# Investigations of $p$+Pb Collisions at Perturbative and Non-Perturbative QCD Scales

by

**Kurt Keys Hill**

B.A. Physics, University of California Davis, 2011

B.S. Mathematics, University of California Davis, 2011

A thesis submitted to the

Faculty of the Graduate School of the

University of Colorado in partial fulfillment

of the requirements for the degree of

Doctor of Philosophy

Department of Physics

2020

This thesis entitled:
Investigations of $p$+Pb Collisions at Perturbative and Non-Perturbative QCD Scales
written by Kurt Keys Hill
has been approved for the Department of Physics

Prof. Jamie Nagle

Dennis V. Perepelitsa

Date

The final copy of this thesis has been examined by the signatories, and we find that both the content and the form meet acceptable presentation standards of scholarly work in the above mentioned discipline.

Hill, Kurt Keys (Ph.D., Physics)

Investigations of $p$+Pb Collisions at Perturbative and Non-Perturbative QCD Scales

Thesis directed by Prof. Jamie Nagle

High energy nuclear collisions manifest a variety of interesting phenomena over a broad range of energy scales. Many of these phenomena are related to the formation of a hot and dense state of deconfined quarks and gluons known as the quark gluon plasma (QGP). Chief among these are the observed near inviscid hydrodynamic expansion of the QGP, as measured through azimuthal anisotropy coefficients of low-$p_{\mathrm{T}}$ final state hadrons $v_n$, and the loss of energy of high-$p_{\mathrm{T}}$ color charges as they traverse the QGP which is observed as a quenching of strongly-interacting final state objects like jets. Observations of these phenomena in Au+Au and Pb+Pb collisions at RHIC and the LHC provide compelling evidence of QGP formation. Small collision systems, like $p$+Pb, also show evidence for the creation droplets of QGP through the observation of anisotropic flow; however, measurements of high-$p_{\mathrm{T}}$ jet and particle spectra show no signs of the energy loss observed in large collision systems. Thus, small systems are an ideal venue to explore the relationship between high- and low-$p_{\mathrm{T}}$ QGP phenomena. Furthermore, the low ambient energy of $p$+Pb compared to A+A collisions allow for the precise determination of perturbative process rates which can be used to understand the nuclear modification of nucleon parton densities.

This dissertation explores 165 nb$^{-1}$ of 8.16 TeV $p$+Pb data collected in 2016 with the ATLAS detector at the LHC through the exposition of two measurements. The cross section and nuclear modification factor for prompt, isolated photons are studied over a broad range of transverse energies and rapidities. These results are compared to predictions from perturbative QCD calculations and a model of initial state energy loss. Additionally, the charged hadron azimuthal anisotropy coefficients, $v_2$ and $v_3$, are measured via two-particle correlations as a function of particle $p_{\mathrm{T}}$ and event centrality. Results are shown from minimum bias events and events selected because of the presence of a high-$p_{\mathrm{T}}$ jet. The elliptic flow coefficient is observed to be non-zero for $0.5 < p_{\mathrm{T}} < 50$ GeV. These results are discussed in the context of hydrodynamic and energy loss models.

## Dedication

For Margarite. Thank you for accompanying me on this journey. I know I would not be here without you.

## Acknowledgements

First, I would like to acknowledge my advisors at CU for structuring an open and respectful environment in which to learn, collaborate, and work. I am grateful that I landed in this group; this effort might have seemed insurmountable had the lab not been such an enjoyable and stimulating place to be. To my advisor Jamie Nagle, thank you for your leadership, honesty, and compassion, and for teaching me to keep the greater context in sight. To my advisor Dennis Perepelitsa, thank you for the enthusiasm and curiosity you brought to the group. It was truly difficult to not be excited when working with you. Thank you for mentoring and including me.

I would like to highlight the support and guidance I received from the wonderful students and postdocs in the group. To Qipeng Hu and Sanghoon Lim, it was such a pleasure collaborating with and learning from you. Thank you for your strong technical and intellectual support on the high $p_T$ flow result. I will always remember our problem solving sessions fondly. To Darren McGlinchey and Ron Belmont, thank you for teaching and enduring me as I became more independent. Thank you for mentoring me on PHENIX. Thank you Javier Orjuela Koop and Theo Koblesky for bringing me up to speed. Thanks to Javier in particular for sharing his graphic design sensibilities and vast institutional knowledge of student life at CU. Thank you Blair Seidliz and Jeff Ouellette for companionship while exploring the world of ATLAS

I would also like to thank my collaborators on ATLAS for keeping such an amazing piece of machinery running. Particularly, I thank my friends in ATLAS heavy ions that have been crucial support for the work presented here: Peter Steinberg, Zvi Citron, Iwona Grabowska-Bold, Sasha Milov, Anne Sickles, Soumya Mohapatra, Aaron Angerami, Brian Cole, Martin Spousta, Martin Rybar, and Jiangyong Jia. Thank you Witold Kozanecki and Richard Hawkings for your help and guidance with the 2016 $p$+Pb luminosity length scale calibration.

I want to acknowledge some of my mentors and educators that introduced me to this path. To my undergrad advisors Manuel Calderón and Daniel Cebra, thank you for introducing me to heavy ions and physics research in general. Thank you for taking me on and giving me a chance; I would not have considered grad school without that exposure. Thank you Rosi Reed for bringing me up to speed when I knew nothing. Thank you to my instructors Larry Green and Bruce Armbrust at LTCC for introducing me to math and physics during a particularly formative time in my life. Thank you Bruce for stepping up to teach the physics sequence when they almost canceled the program.

Finally I would like to offer my deepest gratitude to my family. Thank you for giving me the foundation of love and determination on which my education rests. To my parents, thank you for your open minds and hearts throughout my continued upbringing. You are the most generous people in my life, and I only hope I can live up to your example. To Christine, I have always admired your moral character. Thank you for teaching me strength, leadership, and humor. It was always easier following in your footsteps. To Grammy, thank you for your love and friendship. Thank you for teaching me to appreciate the value of craftsmanship. To Marcus, you have always inspired me with your determination and commitment. Thank you for always pushing to get out there. To Rick and Susan, thank you for your love and support. Thank you for distracting me from physics with style and art and refining my ability to have a good time. To Karen, Holly, and Emma, it is such a pleasure having you in my life. Thank you for being there for me, especially during my time at Brookhaven.

# Contents

**Chapter**

**Appendix**

# Tables

Table

# Figures

**Figure**

# Chapter 1

## Introduction

Until about 10 micro-seconds after the Big Bang, the universe was too hot and dense for the nuclei of atoms to be able to bind together [1]. In fact, during this epoch, even the constituent protons and neutrons could not exist, and the universe was filled with a liquid composed of the fundamental quarks and gluons that make up nuclear matter. This form of matter is called the *quark gluon plasma* (QGP) [2]. At this point the universe was very dense and trillions of Kelvins, or millions of times hotter than the core of a star. As the universe expanded, it cooled and the quarks and gluons crystallized into composite protons and neutrons (together *nucleons*) which then, around three minutes later, came together to form the nuclei of atoms. Around 370,000 years after the Big Bang, the universe was cool enough that electrons could bind to the nuclei, neutralizing the matter and effectively removing the scattering centers for photons [1]. Thus, around this point, light in the universe decoupled from matter and exists today as the Cosmic Microwave Background (CMB) radiation. Telescopes are not able to see beyond this primordial wash of light, limiting the ability to study earlier periods in the universe.[1]

The situation in the universe today is significantly different than the hot and dense periods just described. The universe has expanded such that the CMB photons, that were liberated with a black-body spectrum around 3000 K, have been red-shifted to less than 3 K [1]. Though quarks and gluons are normally confined to nucleons (or more generally *hadrons*), at particle accelerators it is possible to collide the nuclei of large atoms at close to the speed of light, and for a brief moment in time, create tiny droplets of the extremely hot and dense matter of liberated quarks and gluons that filled the early universe. Thus, though it is impossible to observe the primor-dial QGP phase of the universe, the QGP and the interactions that define it can be studied systematically in a laboratory setting.

[1] It is possible that future gravitational wave observatories, like the planed LISA mission [3], will be able to detect gravitational waves from periods before the CMB limit.

The fundamental interaction between quarks and gluons is that of the strong force whose microscopic dynamics are described by *Quantum Chromodynamics* (QCD). The properties of QCD depend on the energy scale of the interaction. At high energies, its coupling strength is weak, and perturbative calculation methods are highly successful. At low energies, however, the coupling is strong and non-perturbative. The confining scale of quarks and gluons corresponds naturally to the transition between these regimes. The complexity of the QCD interaction gives rise to a wealth of rich emergent phenomena that are not obvious from the fundamental degrees of freedom of the theory.

Nuclear collision experiments at CERN in Europe and Brookhaven National Laboratory in the USA provide an opportunity to study QCD matter across a broad range of energy scales. The discovery that the QGP flows as a collective fluid in large collision systems, such as Au+Au and Pb+Pb, has led to the development of effective models describing the dynamics of the deconfined but strongly coupled regime [4–6]. These describe the entire collision evolution from initial spatial energy depositions to final state hadron distributions through the nearly inviscid hydrodynamic expansion of the QGP. In this well established framework, anisotropic flow, as measured by the azimuthal modulation of particle momenta, are a gradient response of the fluid to the spatial geometry anisotropies of the interacting nucleons. In data, this nuclear geometry can be inferred through *centrality*, measured via the overall event activity, and a model of the geometry, such as the Monte Carlo Glauber [7] model. The success of these hydrodynamic descriptions at all centralities has led to this model becoming the standard paradigm. However, the average transverse momentum ($\langle p_{\mathrm{T}} \rangle$) of the expanding QGP is of order a few hundred MeV, and therefore, this picture is only applicable to the bulk of the particles produced with low $p_{\mathrm{T}}$. That said, measurements of particle azimuthal anisotropy show significant non-zero results extending well into the high $p_{\mathrm{T}}$ regime.

High $p_{\mathrm{T}}$ particles originate from rare, high-energy scatterings during the earliest times in the collision. These perturbative processes free quarks and gluons from the initial colliding nucleons that, given sufficient energy, fragment into collimated clusters of particles called *jets*. Since these scatterings occur early in the collision evolution, jets can act as probes of the bulk QGP matter as they must traverse the expanding medium. Mea-surements of the rates of these processes show that the QGP is largely opaque to jets; i.e., jets lose energy as they pass through the medium through a process called *jet quenching* [8–10]. Furthermore, the level of jet energy loss is observed to be smaller for less central (more *peripheral*) events in which the transverse size of the QGP is smaller. Given a path length dependent effect, it is plausible that the high $p_{\mathrm{T}}$ azimuthal anisotropy is due to the jet traversing different paths through the QGP. Both the signal at low and high $p_{\mathrm{T}}$ would then be due to the same geometric pattern of the interacting nucleons; however, the mechanism to transfer this pattern to the final state particle momentum distribution would be different in each case.

Small collision systems, where one or both of the colliding bodies consist of at most a few nucleons, demonstrate flow signatures similar to those in large systems [11–17]. These observations were unexpected because it was thought that partons in such small regions would not interact enough to behave collectively. Nevertheless, the ability of hydrodynamic models to accurately describe the data has led to the conclusion that a QGP, with its characteristic evolution as a response to the initial interacting nucleon geometry, is being created in these systems. However, the jet quenching phenomena seen in large systems is notably absent [18–21]. This raises the question of whether it is possible for collective QGP droplets to form in such a way that they don't modify the jet energy spectrum. Or perhaps there is some jet modification, but the current measurements are not sensitive enough to detect it. Thus, small systems provide a testing ground to study QGP phenomena in relation to the size of the QGP droplet. There is an opportunity to test theories that relate the flowing medium to energy loss mechanisms in a situation in which one exists without the other [22–24].

The modification to measured perturbative process rates from the interaction with the QGP is known as a *final state* effect, since the mechanism modifies the parton *after* the high energy scattering. However, there also are *initial state* effects that modify process rates in nuclear collisions. These modifications are usually interpreted as changes to the parton densities of nucleons when they are bound in nuclei. Thus, in large collision systems, strongly interacting final states, like jets, will contain both effects creating a potential ambiguity. Final state particles that don't interact via QCD, like electroweak bosons, are unmodified by the presence of the QGP [25–28]. Thus, studying electroweak (EW) processes in these collisions can provide a "standard candle" from which to measure QGP phenomena and map out the parton densities in the initial colliding nucleons. Furthermore, asymmetric systems like $p$+Pb allow the study of these processes in a "clean" nuclear environment with limited ambient particle production.

This thesis details two novel measurements in $\sqrt{s_{_\text{NN}}} = 8.16$ TeV $p$+Pb collisions from the Large Hadron Collider (LHC) and using the ATLAS detector published as Refs. [29, 30]: a measurement of the production cross section of direct, high $p_\text{T}$ photons, and a measurement of the azimuthal anisotropy of charged particles over a broad range of $p_\text{T}$. Chapter 2 gives an overview of the physics of ultra relativistic nuclear collisions to provide context for these measurements. Chapter 3 presents the experimental apparatus and data used in the analyses. An analysis to calibrate event-by-event geometric properties of the colliding bodies is given in Chapter 4. Chapters 5 and 6 detail a measurement of the production cross section of high energy photons. The results are compared to theoretical predictions from perturbative calculations, with and without nuclear parton distribution modification, and an expectation from a model predicting energy loss of the scattering partons before the high energy scatter. Chapters 7 and 8 present a measurement of flow signals in these collisions as they depend on the transverse momentum of the particles produced and the underlying process which creates them. Finally, conclusions are given in Chapter 9.

# Chapter 2

## Ultra-Relativistic Heavy Ion Collisions

During the ultra-relativistic collisions of heavy ions, nuclei act as sources of many microscopic scattering centers that interact via the strong, weak, and electromagnetic interactions as the nuclei move through each other. Initial kinetic energy from the impacting particles is converted into, sometimes, thousands of final state particles that expand in all directions. These collision fragments can be captured in detectors and used to infer conclusions about the whole evolution of the collision.

It is helpful to consider a given collision in a probabilistic sense as a single sample from the set of all possible outcomes. To paraphrase the totalitarian principle (its modern incarnation attributed to Gell-Mann), any outcome that is not strictly forbidden by natural conservation laws is compulsory; i.e., it must happen with some non-zero probability. Particle colliders like the Large Hadron Collider (LHC) and the Relativistic Heavy Ion Collider (RHIC) are capable of generating millions of collisions per second, enabling experiments like ATLAS to record enormous sets of data and capture all but the rarest of outcomes.

This thesis is concerned, primarily, with phenomena associated with the strong interaction, so-called due to its dominance at a certain length scale with respect to the other three fundamental interactions. In this chapter, some time is spent reviewing QCD within the context of the Standard Model of Particle Physics. Then an overview of QCD matter is presented, including discussions of hadrons and deconfined media. Finally, the geometry and evolution of nuclear collisions are explored in the context of experimental results.

## 2.1 The Standard Model

High energy physics seeks to understand the fundamental building blocks and dynamics of matter in the universe. The current best understanding is encapsulated in the so-called Standard Model of Particle Physics (SM).

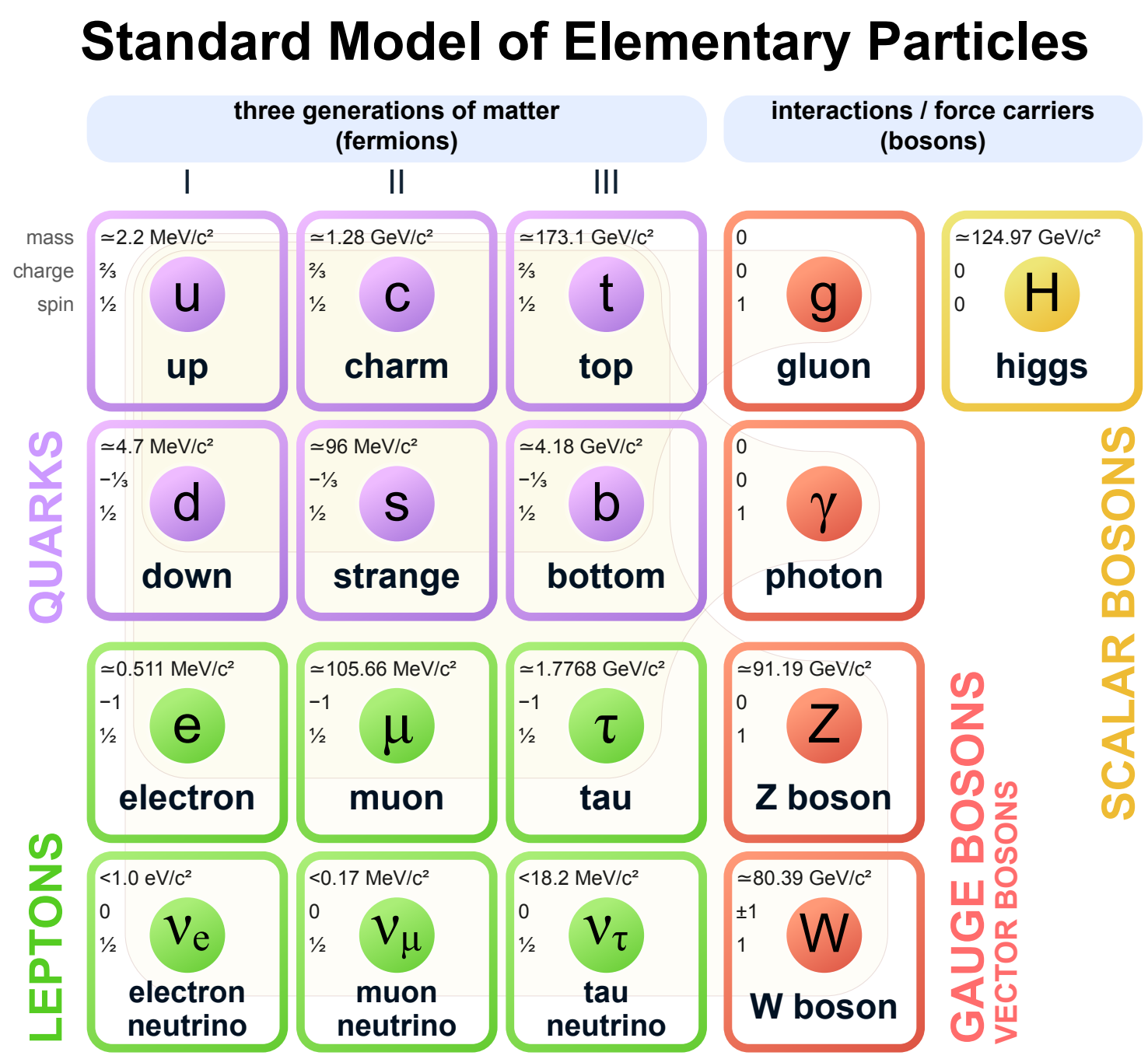

Figure 2.1: A chart of the particles making up the Standard Model [31].

The dynamics of matter are governed by four fundamental interactions: gravity, electromagnetism, and the strong and weak forces. Gravity is a factor of about $10^{30}$ weaker than the other three [32], and is not included in the SM. Interactions occur between bodies that carry a property that is specific to each type of interaction. This property is called electric charge, color charge, or weak isospin for electromagnetic, strong, or weak interactions respectively. All matter is composed of twelve fermions (spin 1/2 particles), shown in the first three columns of Fig. 2.1. These are divided into quarks which hold color, and leptons that do not. Quarks and the upper row of leptons carry electric charge; the bottom row of leptons, the neutrinos, are neutral. All of the fermions carry weak isospin. Each row in the chart corresponds to the three generations of mass hierarchy, the left being lightest and the right being heaviest.[1] Finally, each of the fermions has a corresponding anti-particle with exactly the same properties, but opposite electric charge.[2]

The SM is represented mathematically as a quantum field theory with $\mathrm{SU}(3) \times \mathrm{SU}(2)_L \times \mathrm{U}(1)$ gauge symmetry. The $\mathrm{SU}(3)$ component gives rise to QCD along with eight gauge bosons, known collectively as gluons, corresponding to the eight generators of the symmetry. The remaining $\mathrm{SU}(2)_L \times \mathrm{U}(1)$ symmetry defines the electroweak theory and the three weak bosons, $W^{\pm}$ and $Z^0$, and the photon of electromagnetism. The gauge bosons act to mediate the interactions between the fermions, via particle exchange, and are thus known as force carriers. All of the gauge bosons are required to be massless in order to preserve local gauge symmetry; however, the weak bosons acquire mass through interaction with an additional scalar field via the Higgs mechanism. The gauge symmetries of the SM are continuous, and therefore, Noether's theorem states there are conservation laws for each symmetry. In this case, the corresponding charges are conserved under those interactions.

## 2.2 Quantum Chromodynamics

### 2.2.1 Basic Formulation

QCD is a quantum field theory with $\mathrm{SU}(3)$ gauge symmetry that describes the dynamics of quarks and gluons in the SM. The generators, $T^C$, of the $\mathrm{SU}(3)$ Lie group are the eight elements with Lie algebra,

$$[T^A, T^B] = i f_{ABC} T^C \tag{2.1}$$

defined by the totally anti-symmetric structure constants of $\mathrm{SU}(3)$, $f_{ABC}$. The elements, $T^C$, are represented by the $3 \times 3$ Gell-Mann matrices. A local gauge transformation from the group can then be constructed via exponential mapping of the generators to yield the unitary transformation

$$U(x) = e^{i\alpha_C(x) T^C} \ , \tag{2.2}$$

where the functions $\alpha_C(x)$ parameterize the group operation. The fundamental representation of the group is three dimensional corresponding to color charges red, green, and blue. Quark fields, $\psi(x)$, take the form of

[1] Except for, perhaps, the neutrinos, whose masses are still uncertain.

[2] It is interesting to note that the existence of anti-particles (first proposed by Dirac in his development of relativistic quantum mechanics) and the fermionic nature of matter particles are not a consequences of quantum mechanics *per se*. These two properties follow simply from the algebraic structure of the Poincaré group comprising the symmetries of space-time under relativity. See Refs. [33, 34].

color vectors in this space and transform under gauge rotation as

$$\psi(x) \to U_{ab}(x)\psi^b(x) = e^{i\alpha_A(x)T^A_{ab}}\psi^b(x). \tag{2.3}$$

The color vector indices will be suppressed moving forward.

Ordinary space-time derivatives acting on transformed vectors will receive an extra term, thus breaking the symmetry. This is solved by introducing a covariant derivative

$$D_\mu = \partial_\mu - ig_s T_C A^C_\mu \,, \tag{2.4}$$

where $g_s$ is the QCD coupling constant, and $A^C_\mu$, is a new set of dynamical gauge fields, one for each generator. These are the gluon fields, whose transformation rules can be determined by requiring the gauge invariance of $\bar{\psi} D_\mu \psi$.

Geometrically, imposing $\mathrm{SU}(3)$ gauge symmetry defines a principal $\mathrm{SU}(3)$-bundle, $\mathcal{P}$ on the space-time manifold. To generalize the derivative, a Lie algebra valued 1-form, $\mathcal{A}$ (a connection) is defined on $\mathcal{P}$ that connects the tangent spaces (fibers) of different points in space-time. Then $ig_s T_C A^C_\mu$ is the pullback of $\mathcal{A}$ from $\mathcal{P}$ to space-time. The curvature on $\mathcal{P}$ is the field strength, a Lie algebra valued 2-form that, when pulled back, can be determined as

$$[D_\mu, D_\nu] = -ig_s T_A F^A_{\mu\nu} \tag{2.5}$$

$$\implies F^A_{\mu\nu} = \partial_\mu A^A_\nu - \partial_\nu A^A_\mu - g_s f_{ABC} A^B_\mu A^C_\nu \,. \tag{2.6}$$

The third term on the RHS implies a coupling between the gluon fields themselves. This is a direct consequence of the non-Abelian nature of the symmetry group.

Finally, the fully covariant QCD Lagrangian density is given by

$$\mathcal{L} = \sum_q \bar{\psi}_q (i\gamma^\mu D_\mu - m)\psi_q - \frac{1}{4} F^A_{\mu\nu} F^{A\,\mu\nu} \,, \tag{2.7}$$

where the sum runs over all quark flavors. From this, the dynamics of QCD systems can be derived via the principle of least action.

### 2.2.2 Asymptotic Freedom

Interactions in renormalizable quantum field theories (such as those in the SM) depend on the energy scale of the interaction. If one imagines a particle (probe) scattering off a target, this scale acts as the resolving power of the probe. A higher momentum (energy) probe resolves shorter distance (time) scales. Quantum uncertainty implies the breaking of energy/momentum conservation to a certain degree, allowing fluctuations that take the form of short-lived particles. These fluctuations change the effective values the parameters of the theory (e.g. coupling constants and particle masses) in a scale dependent way. To use the more formal language of Feynman diagrams, loop corrections to vertices and propagators involve off-shell integrals that diverge. A strategy for handling these divergences is to augment the theory parameters in a way as to cancel the divergence at the given scale. This renormalization procedure introduces a scale dependence to the parameters.

The renormalization group equation encodes the relationship between a coupling constant, $g$, and the energy scale, $\mu$,

$$\beta(g) = \frac{\partial g}{\partial \log \mu}\,, \tag{2.8}$$

where the beta function, $\beta$, can be found by studying the divergences perturbatively in successive loop corrections. In QCD the equation reads

$$\beta(\alpha_s) = \frac{\partial \alpha_s}{\partial \log Q^2} = -(b_0\alpha_s^2 + b_1\alpha_s^3 + b_2\alpha_s^4 + \cdots)\,, \tag{2.9}$$

where $b_n$ is the $n$-loop beta function, $\alpha_s = g_s^2/4\pi$, and $Q$ is the momentum transfer of the interaction. This can be solved analytically for the first correction, $b_0 = (33 - 2n_f)/(12\pi)$, where $n_f = 3$ is the number of light quarks ($u$, $d$, and $s$). With an appropriate choice of $\Lambda_{QCD}$, the solution can be expressed as

$$\alpha_s(Q^2) = \frac{1}{b_0 \log(Q^2/\Lambda_{QCD}^2)}\,. \tag{2.10}$$

Eqn. 2.10 implies that the QCD coupling strength actually decreases with the momentum scale, $Q$, approaching zero for vary large $Q$, or equivalently, very short distance scales. Thus, QCD is said to be asymptotically free. This discovery earned the authors, Gross, Wilczek, and Politzer, the 2004 Nobel Prize in physics [35]. The momentum scale, $\Lambda_{QCD}$, determines a threshold between strong and weak coupling regimes. Fig. 2.2 gives a summary of $\alpha_s$ measurements.

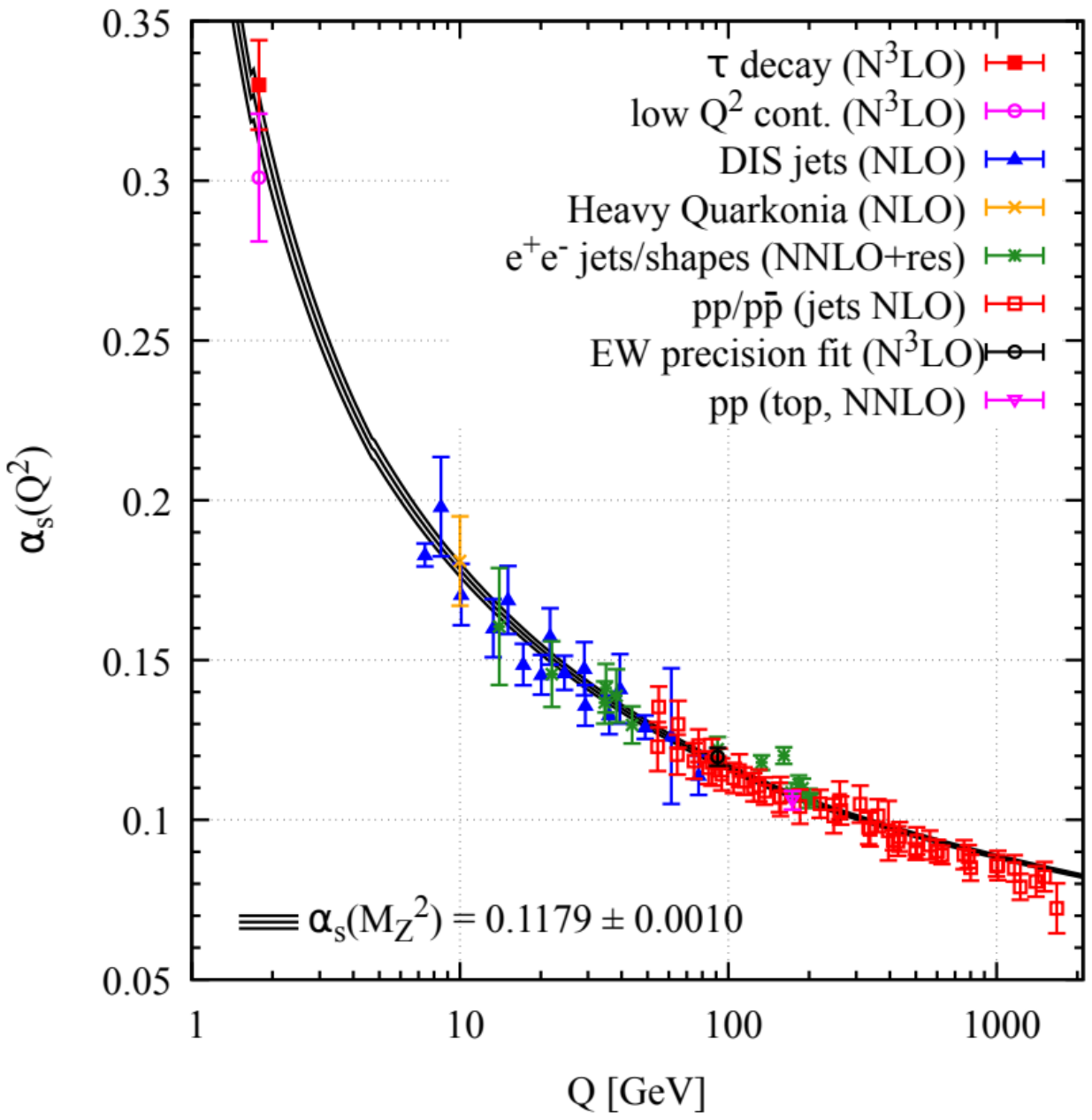

Figure 2.2: A summary of measurements of $\alpha_s$ as a function of momentum transfer $Q$ [1].

### 2.2.3 Color Confinement

Asymptotic freedom describes the QCD coupling at short distance scales, where the theory can be handled with perturbative methods (pQCD). However, the Eqn. 2.10 has a divergence in the coupling as $Q^2 \to \Lambda^2_{QCD}$. This behavior indicates a breakdown in the perturbative assumptions, thus implying the need for new methodology at large distance scales.

Experimentally, the fundamental particles participating in QCD have only been observed in color singlet bound states, a phenomenon known as color confinement. Quarks cannot be isolated in vacuum and are inferred indirectly from hadron spectra and deep inelastic scattering data (see Sec. 2.3.1). The typical hadron mass then defines a natural resolving scale of QCD charges (i.e. $\Lambda_{QCD}$).

A non-perturbative approach to QCD exists in which space-time is discretized into a lattice. In this Lattice QCD (LQCD), quarks fields are located at lattice points and are connected by gluon fields. The infinite dimensional integrals from the path integral formulation become finite dimensional and can be computed directly. The lattice spacing sets a high frequency cutoff that regulates the integrals, and the discrete nature of the problem lends itself to computational methods. Final results are typically extrapolated to continuous space.

Being non-perturbative, LQCD calculations can be valid in the strong coupling regime but are limited in that system dynamics cannot be modeled. Thus, properties related to hadron structure, like hadron masses, form factors, and the equilibrium QCD equation of state, can be calculated from first principles, but time dependant phenomena like hydrodynamic transport coefficients cannot.

Considering the simplest problem of a heavy quark–anti-quark bound state, in which the motion of the quarks is negligible, the effective QCD potential has been shown to match the Cornell-type potential [36, 37],

$$V(r) = -\frac{4}{3}\frac{\alpha_s}{r} + \sigma r\,. \tag{2.11}$$

The first term, dominant at short distances, is a perturbative Coulomb-like contribution, and the second, string-like potential with tension $\sigma$[3] , is dominant at long distances [38]. This string-like behavior is interpreted as a color flux tube in which transverse gluons are suppressed, due to their self interactions, leaving a collimated color field. This is then responsible for the confining behavior; as the quarks are pulled apart, the potential energy grows to a point in which the string "breaks", i.e. a new quark–anti-quark pair is created, leaving two hadrons in the final state.

This picture of color flux tubes is well established in LQCD for heavy quarks, but is not rigorous for light quark bound states. In fact, color confinement in general has no analytic proof. It is seen as an emergent phenomenon produced via dynamical chiral symmetry breaking [39]. Nevertheless, it is helpful to keep this picture in mind.

## 2.3 QCD Matter

### 2.3.1 Hadrons

As described in the previous Section 2.2.3, low-energy QCD matter consists of quarks bound in color singlet hadrons. The constituent quarks and guons are known collectively as partons. The observed states take the form of mesons, bound quark–anti-quark, and baryons, bound three-quark states. Before the Quark Model, the net number of baryons minus anti-baryons was observed to be conserved in scattering experiments and was proposed as an explanation for why the proton was stable. This led to a proposed quantum number, $B$, called baryon number. With the discovery of the Quark Model and QCD, quarks were assigned $B = 1/3$ and anti-quarks $B = -1/3$.[4] Thus, mesons, which can decay purely leptonically, are composed of a quark and anti-quark ($B = 0$).

[3] This differs from an actual string system in that in this case, $\sigma$ is distance independent.

Hadronic states are, then, the outcome of a quark chemistry obeying a set of quantum number conservation rules: color, baryon number, and spin. Though the fundamental theory describing the structure of these particles is known, the energy scale is non-perturbative and the dynamics are not fully understood. However, LQCD methods are able to calculate hadron masses with good accuracy. Fig. 2.3 shows a plot of hadron masses from world data overlaid with calculations from LQCD.

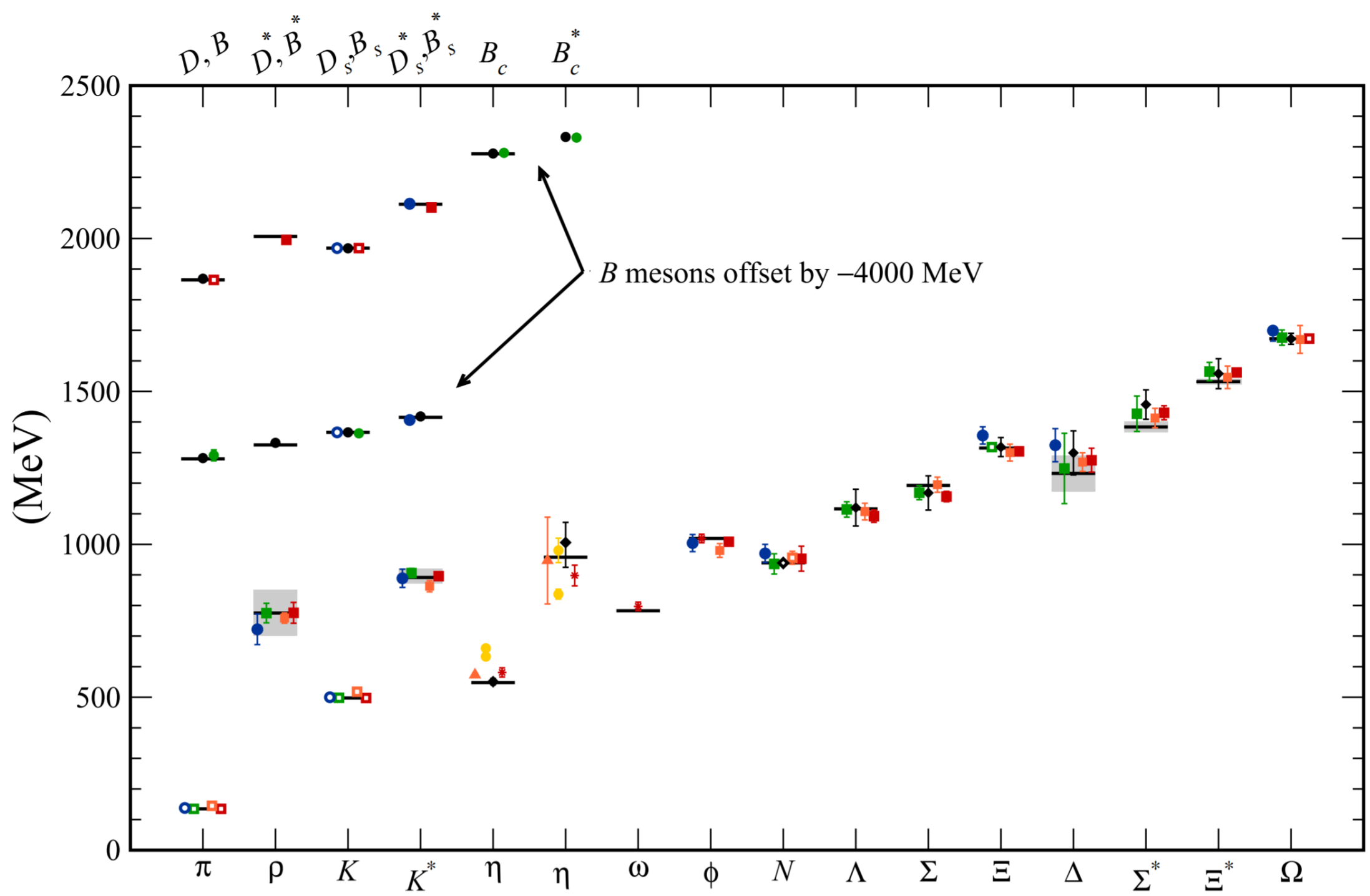

Figure 2.3: A summary of hadron mass calculations from LQCD (colored points), overlaid with measurements from world data (black lines) [1].

[4] It is notable that baryon number conservation is broken in perturbative QCD via Adler–Bell–Jackiw anomalies in higher than tree level diagrams. However, it is believed that the quantity $B - L$, where $L$ is lepton number, is exactly conserved [40].

### 2.3.2 Fragmentation and Jets

Measurements of hadrons from $e^+e^-$ annihilation experiments provide a particularly clear test of the Quark Model and QCD. If one assumes the existence of charged quarks, then the process of $e^+e^- \rightarrow q\bar{q}$, mediated by a photon, is expected. However, given confinement, there must be a subsequent step in which the quarks become hadrons, which is non-perturbative. Nevertheless, if all hadronic final states are integrated, properties of the fundamental process can be inferred.[5] In particular, one can compare the inclusive cross section for having a hadronic final state [41], $\sigma(e^+e^- \rightarrow \text{hadrons})$, to that of having muons in the final state, $\sigma(e^+e^- \rightarrow \mu^+\mu^-)$. One might naively expect the following counting relation

$$\sigma(e^+e^- \rightarrow \text{hadrons}) = 3\sum_q e_q^2 \sigma(e^+e^- \rightarrow \mu^+\mu^-)\,, \tag{2.12}$$

where $e_q$ is the fraction of the unit charge for the quark and the factor of three is for the three colors in QCD. The ratio reduces to

$$R = \frac{\sigma(e^+e^- \rightarrow \text{hadrons})}{\sigma(e^+e^- \rightarrow \mu^+\mu^-)} \tag{2.13}$$

$$= 3\sum_q e_q^2\,. \tag{2.14}$$

Thus, $R = 2$, 10/3, or 11/3 for 3, 4, or 5 quark flavors respectively. Fig. 2.4 shows $R$ as a function of the scattering momentum scale from world data compared to the naive model. The steps in the predictions correspond to the scale at which there is enough total energy to create heavier quark pairs. The vertical arrows correspond to hadronic resonances going off scale. The agreement validates the three quark colors and the Quark Model in general. The modest disagreement can be attributed to the fact that higher order processes involving gluon vertices were ignored.

The scattered partons in these reactions become hadrons through a process called fragmentation. At sufficiently high energy, the color charges can become dozens of hadrons emanating from the collision in a roughly conical pattern called a *jet*. This non-perturbative hadronization process is parameterized by fragmentation functions, probability densities in terms of a momentum scale and the fraction of the initial parton's momentum going to each hadron [1]. These fragmentation functions can be used to compute structure functions that contribute directly to individual hadron production cross sections. Though $e^+e^-$ reactions were used for this example, this process of fragmentation is universal and applies to parton scattering in hadronic collisions as well.

[5] The ability to separate the quark scattering properties from the hadronization by a difference of scales is called factorization.

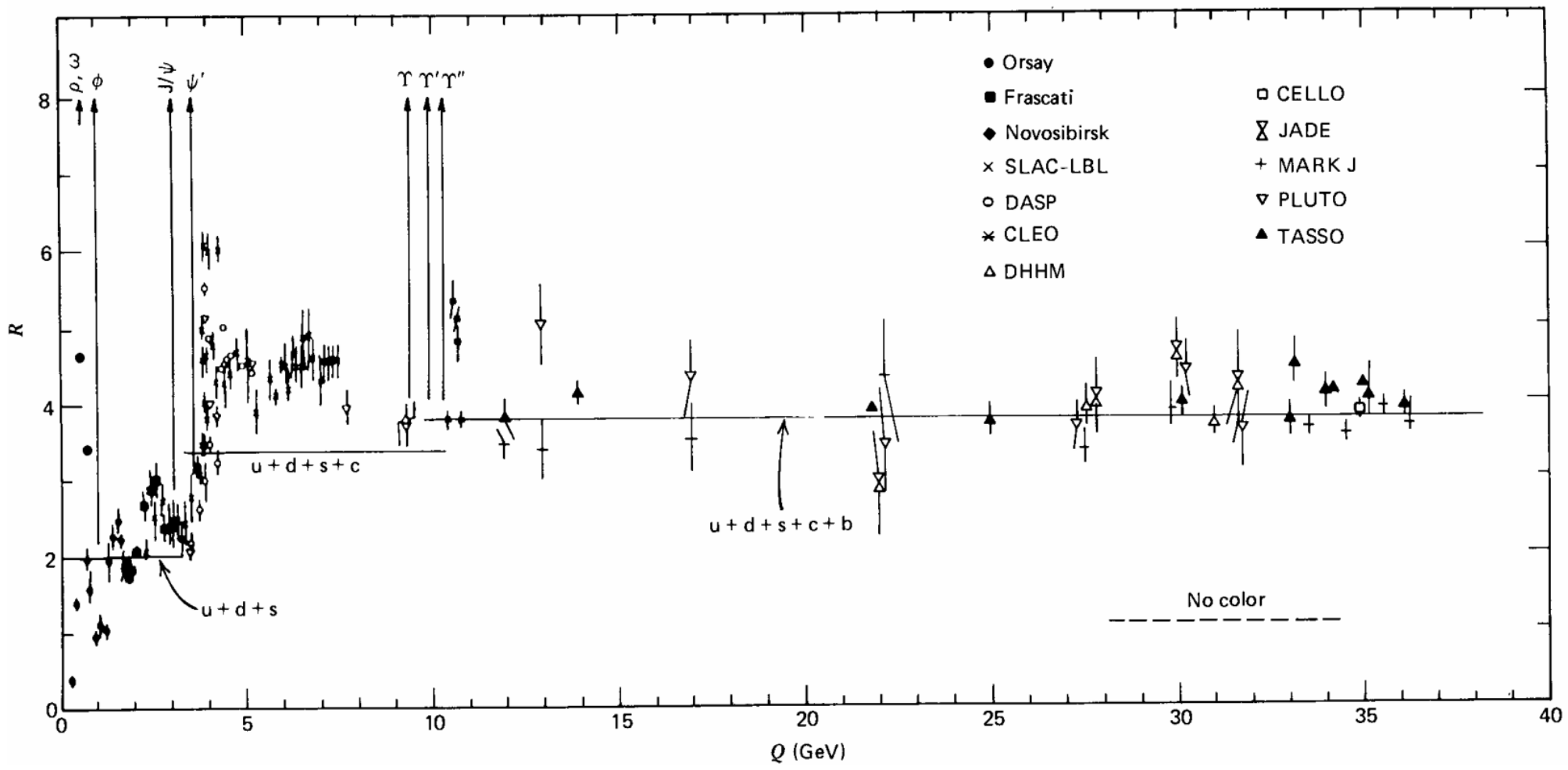

Figure 2.4: The cross section ratio, R, of $\sigma(e^+e^- \to \text{hadrons})$ to that of $\sigma(e^+e^- \to \mu^+\mu^-)$ as a function of momentum scale from world data (points) compared to a naive quark and color scaling model [41].

The phenomenological Lund string model [42], represented in Fig. 2.5, considers fragmentation as an iterative breaking of the color strings mentioned in Sec. 2.2.3. In this picture, a fragmenting quark–anti-quark pair is placed on a light cone, spreading in opposite $(x)$ directions with gluon string connections. The quarks progress until the potential energy in the string grows to the point that it "breaks" by creating a new quark–anti-quark pair (hadron) at a space-time point called a vertex. Each vertex is spacelike separated and the hadron carries away some fraction of the total energy. The probability to tunnel out a hadron of mass $m$ is given by the Lund Symmetric Fragmentation Function $f(x)$ of the form

$$f(x) = N\frac{1}{x}(1-x)^a e^{\frac{-bm^2}{x}}\,. \tag{2.15}$$

The process continues until there is not enough energy to further break the string. Vertices will lie on a hyperbola in the space-time diagram on average. The model represents gluons as excitations on the string. Hadrons are considered as quark pairs without enough energy to break the string connecting them; the quarks switch directions back and forward in a confined "*yo-yo* mode". The Lund model is employed in a number of Monte Carlo (MC) event generators, and has enjoyed great success representing jet fragmentation to hadrons. These generators (the most popular being the Pythia event generator [43]) have become an indispensable tool for experimental efforts.

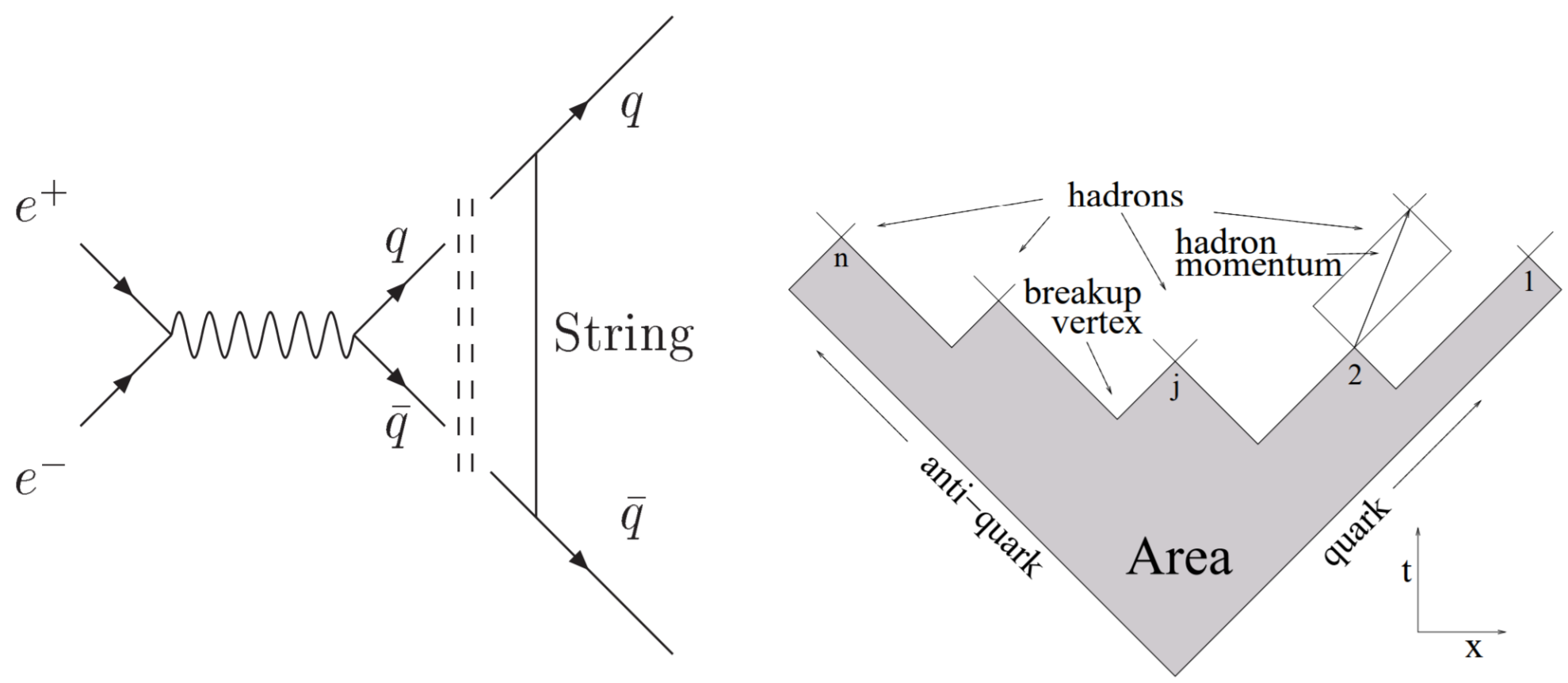

Figure 2.5: A graphical representation of the Lund string fragmentation model. The left figure shows the Feynman diagram of the $e^+e^- \rightarrow q\bar{q}$ process, and the right plot shows the space-time diagram of the fragmentation [44].

In principle, summing the four-momenta of all the hadrons in a jet will yield the kinematics of the initial parton. However, in practice, it is impossible to know the origin of any given hadron in an event, particularly in hadronic collisions where there are potentially many partons interacting simultaneously. Therefore, clustering algorithms were developed to collect nearby hadrons into jet objects. Presently, the most popular method is the so-called anti-$k_t$ clustering algorithm [45].

The anti-$k_t$ algorithm is part of class of sequential clustering algorithms differentiated by a distance metric between particles $i$ and $j$ (or detector objects like reconstructed tracks or calorimeter towers), $d_{ij}$, and between particles $i$ and the beam, $d_{iB}$. The process is to find the smallest distances, and if it is $d_{ij}$, combine the particles; if it is $d_{iB}$, label it a jet and remove it from the list. Continue the process until the list of particles is empty. The distances take the form:

$$d_{ij} = \min(k_{Ti}^{2p}, k_{Tj}^{2p}) \frac{\Delta_{ij}^2}{R^2} \tag{2.16}$$

$$d_{iB} = k_{Ti}^{2p}\,, \tag{2.17}$$

where $\Delta_{ij}^2 = (y_i - y_j)^2 + (\phi_i - \phi_j)^2$ and $k_{Ti}$, $y_i$, and $\phi_i$ are the particle's transverse momentum, rapidity, and azimuthal angle respectively. The parameter $R$ sets a scale for the rough size of the clusters. If $p = 1$, the algorithm is called $k_t$, if $p = 0$, it is called Cambridge/Aachen, and if $p = -1$, it is called anti-$k_t$.

Each version has the common geometric distance factor, and thus, clustered jets tend to be composed of constituents close in angle. However, the $k_t$ algorithm is highly susceptible to low momentum, *soft*, fragments that are not necessarily part of a coherent jet, particularly problematic in "messy" hadronic collisions. In the anti-$k_t$ case, soft particles are likely to be clustered with nearby high momentum, *hard*, particles which form the core of a conical jet shape. The overall jet shape is defined, primarily, by the hardest particle, in contrast to the $k_t$ jet case. Fig. 2.6 shows a comparison between $k_t$ and anti-$k_t$ algorithms on a single event with few hard fragments and many random soft "ghost" particles (for details see Ref. [45]). The shape of the jets in the $k_t$ case are affected by the ghost particles, whereas the anti-$k_t$ jets are more circular.

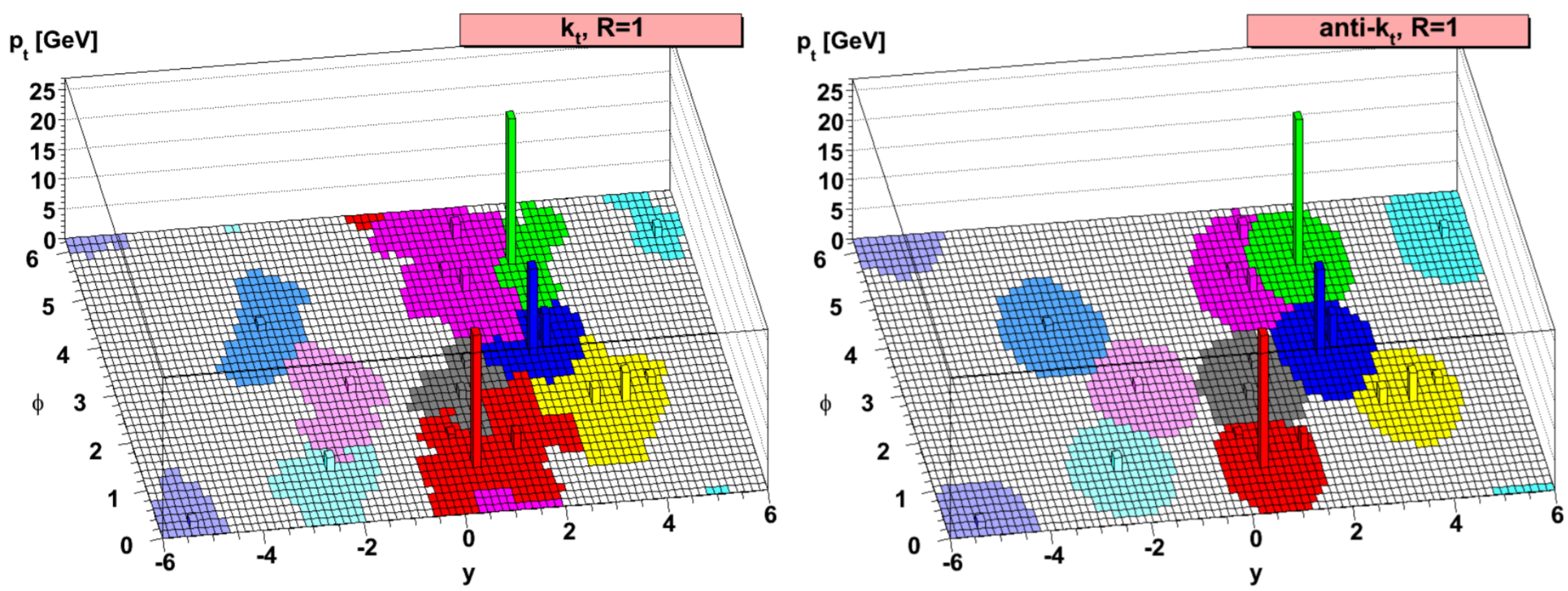

Figure 2.6: A comparison between $k_t$ (left) and anti-$k_t$ (right) jet reconstruction algorithm performance using an event generated with few hard fragments and many random soft "ghost" particles [45].

### 2.3.3 Hadron Structure

In the previous Section 2.3.1, hadrons were defined by their quark content. However, on short timescales, quantum fluctuations make the story much more complicated. The quarks in a hadron radiate gluons that get reabsorbed or connect to the other quarks, and these gluons split to quark anti-quark pairs, etc. Thus, given a high enough resolving power, one not only sees the quarks that contribute to the quantum number rules for a given hadron (valence quarks) but also other quarks in the SM that are fluctuating in and out of existence within the hadron (sea quarks).

Consider lepton-proton scattering in which the interaction is mediated by a virtual electroweak boson as shown in Fig. 2.7. As discussed above, the structure of an object depends on the scale of the probe. At low energies, the long wavelength of the interaction can only resolve the proton as a whole; the internal interactions of the constituents are smoothed over at long time scales, and the body behaves coherently. Calculations in this regime take the form of Mott scattering [46].

On the other hand, consider the case at very high energies, or deep inelastic scattering (DIS), in which the proton breaks apart [47]. To the lepton, the proton is highly Lorentz contracted and its constituents are time dilated. The internal dynamics of the proton are happening at much longer timescales than the interaction, and thus, the boson is imaging a freeze frame of the state of the proton in which the partons have a fixed fraction of the longitudinal momentum, $x$. At a high-scattering-momentum scale, $Q^2$, the short distance of the interaction means the interference with any other interaction between the lepton and proton can be ignored, and the non-perturbative hadronization of the struck parton happens much later and can be factored out of the process. Thus, the proton can be considered a source of approximately incoherent partons, and the scattering between the partons can be considered elastic. The fundamental pieces of information are probability distributions for each possible parton at any given $x$. Furthermore, assuming factorization holds, these Parton Distribution Functions (PDFs) are universal descriptions of the hadron, i.e., they do not depend on the probe.

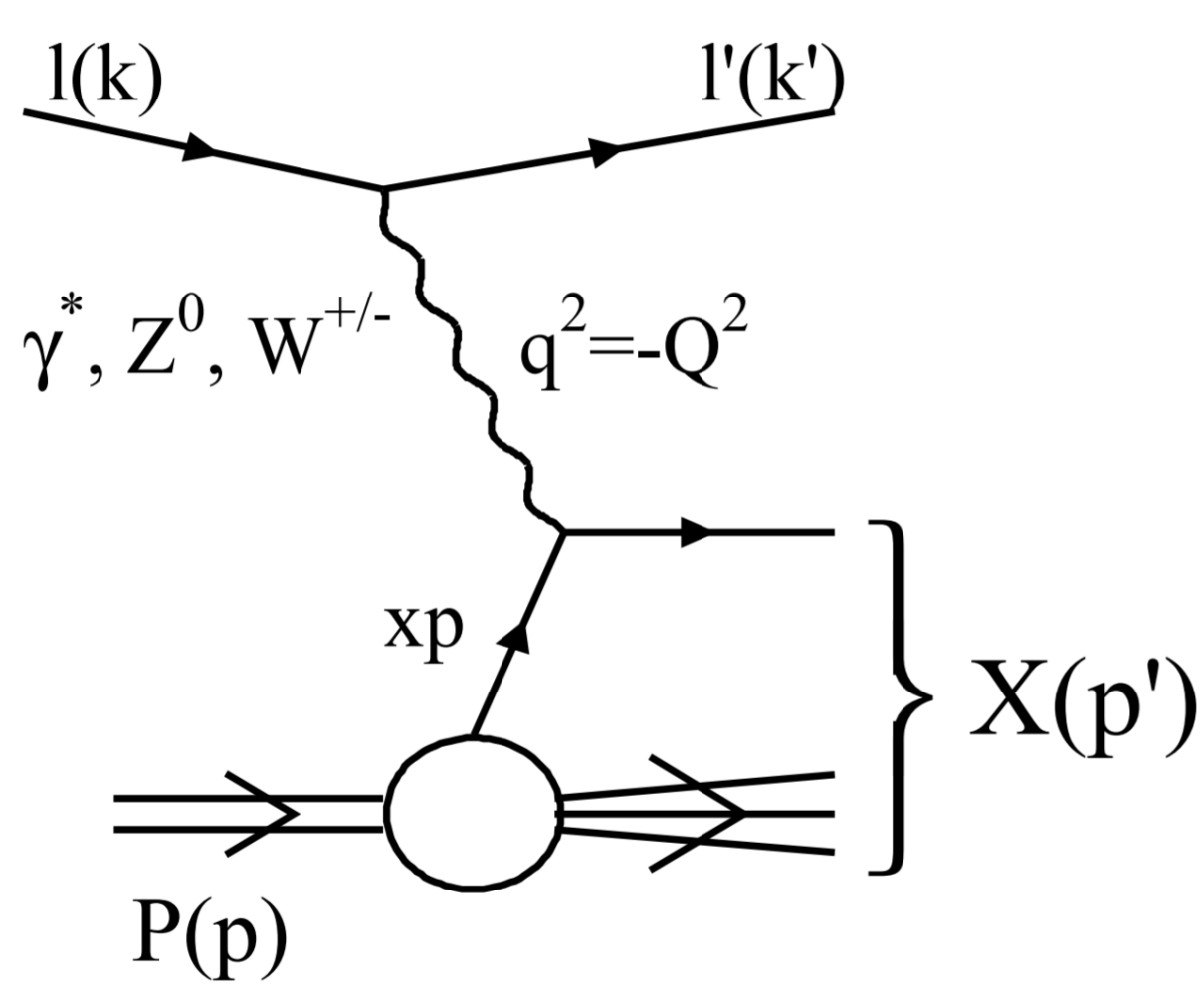

Figure 2.7: Diagram of a DIS process [48].

The scattering cross section for this process, considering photon exchange only, can be expressed in terms of proton structure functions, $F_2$ and $F_L$, as

$$\frac{d\sigma^2}{dxdQ^2} = \frac{4\pi\alpha^2}{2xQ^4}\left[(1+(1-y)^2)F_2(x,Q^2) - y^2F_L(x,Q^2)\right], \tag{2.18}$$

where $y$ is the fraction of the lepton's initial energy lost in the interaction [1]. For elastic scattering on free point-like partons, the structure functions obey so-called *Bjorken scaling*, in which they are independent of $Q^2$, and

$$F_L = F_2 - 2xF_1 = 0\,. \tag{2.19}$$

The structure function can be expressed in terms of the PDFs, $f_q(x)$, for quark $q$ simply as

$$F_2(x) = x\sum_q e_q^2 f_q(x)\,. \tag{2.20}$$

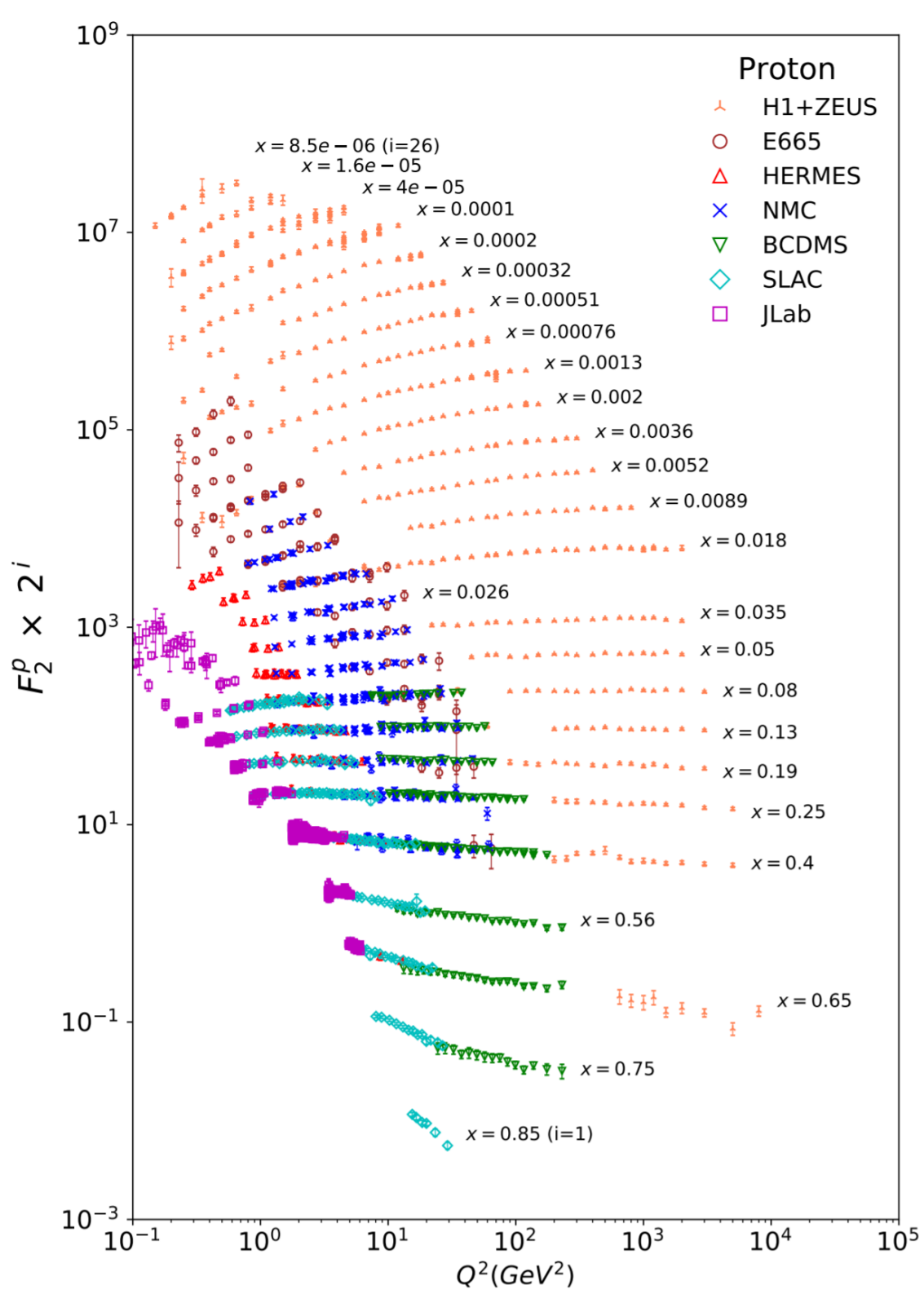

Figure 2.8: Proton structure function, $F_2$, for various $Q^2$ and $x$ values as measured in DIS experiments [1].

Structure functions of the proton, $F_2$, are plotted for various $Q^2$ and $x$ values in Fig. 2.8. One can see the scale independence for a range around $x \approx 10^{-1}$. However the scaling is broken for lower $x$. This breaking can be attributed to corrections from higher orders in pQCD that were not considered above. The parton can split before interacting by emitting a gluon that carries away some momentum. There is an ambiguity between what corrections should be included as part of the hadron structure, $f_q$, and what should be part of the pQCD scattering coefficients. The standard approach is to realize that the emissions should be collinear with the parton to some extent. Thus an arbitrary collinear factorization scale is imposed on the transverse momentum of the emissions, below which, it is categorized with the hadron. The integrals over these processes are divergent but can be renormalized into *running* PDFs, as with the *running* of the QCD coupling. The DGLAP evolution equations govern the evolution of the PDFs with the scale parameter [49–52]. As with the renormalization scale, the collinear factorization scale is typically chosen to be the scattering parameter Q [1].

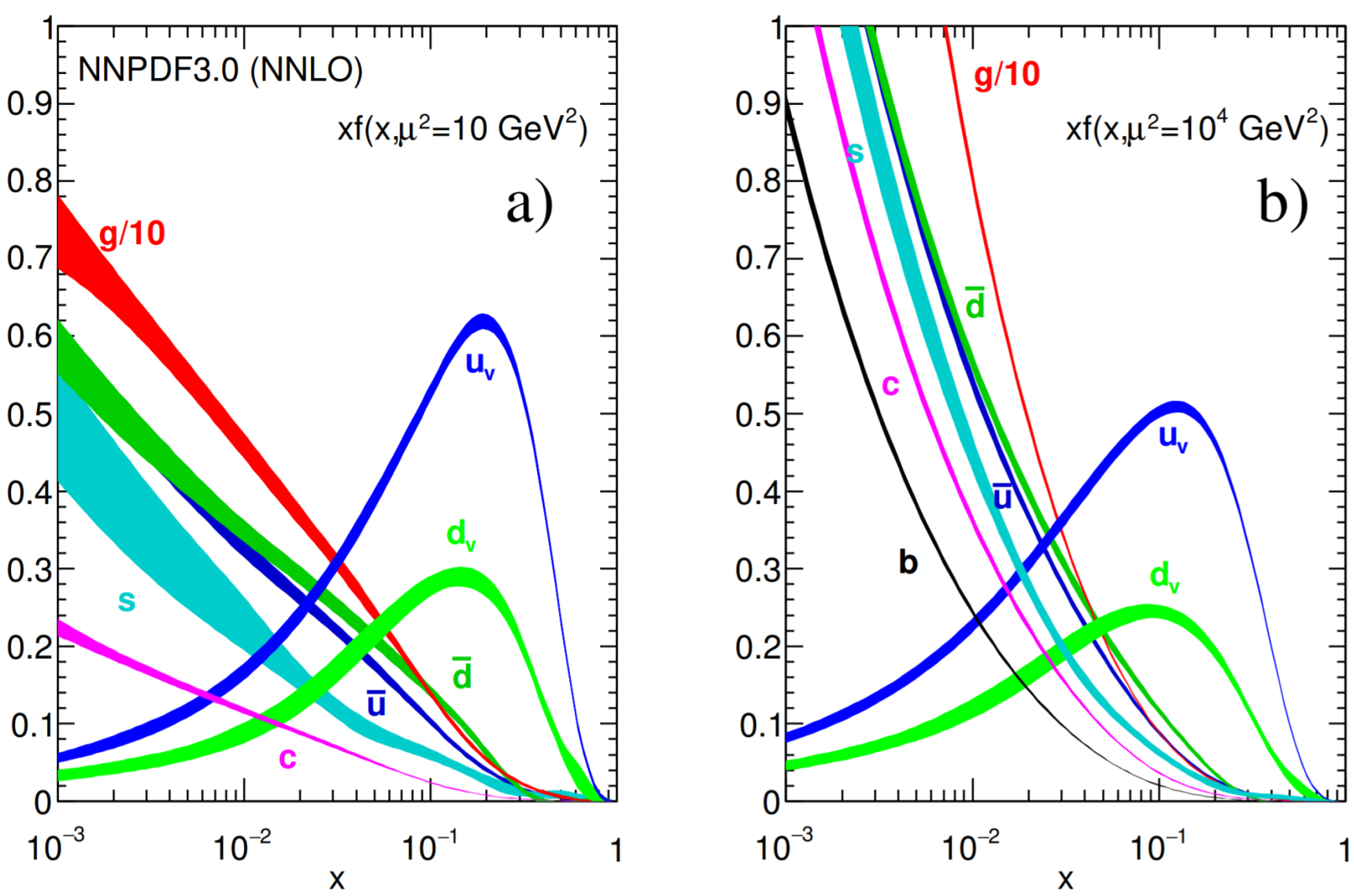

Figure 2.9: Proton PDFs from global fits to world data as a function of parton $x$ for momentum scales 10 GeV (a) and $10^4$ GeV (b) [1, 53].

Fig. 2.9 shows proton PDFs for two different momentum scales. The valence quarks are peaked at significant fractions of the total momentum as one might expect. At lower values of $x$, the probability of finding a sea quark is greater than that of the valence quarks. However, in this region the gluon density is much greater, and even looks divergent at low $x$ and high $Q^2$. It is believed that at low enough $x$, the gluon density saturates as the rate of gluon recombination reaches that of splitting.

Gluon saturation is associated with a scale, $Q_{\mathrm{s}}(Q^2)$, which can be interpreted as the maximum color charge per transverse area of the nucleon at interaction scale $Q^2$ [54]. For sufficient collision energies, $Q_{\mathrm{s}} \gg \Lambda_{QCD}$, and thus the gluon fields my be treated in the weak coupling regime. However, the high densities amplify the interactions such that they cannot be treated perturbatively. The problem is framed in a renormalization group picture where high $x$ partons are viewed as static sources of lower $x$ gluons, and the interactions are resumed in orders of the parton rapidity. The outcome is an effective classical Yang-Mills theory that describe the overlapping color fields as a *Color Glass Condensate* (CGC) [54]. In this theory, coherent low $x$ gluons condense and form effective partons with transverse size on the order of $1/Q_s$. This framework has been used with the IP-Sat model to describe the low $x$ region of DIS data to good accuracy [55].

So far, the discussion has been limited to free protons. However, the internal structure of nucleons bound in nuclei can also be probed in DIS. The PDFs of a nucleus, $f_q^A$, with $A$ total nucleons and $Z$ protons can be written as a linear combination of the individual proton and neutron PDFs, $f_q^{p/A}$ and $f_q^{n/A}$, as

$$f_q^A(x, Q^2) = \frac{Z}{A} f_q^{p/A}(x, Q^2) + \frac{A-Z}{A} f_q^{n/A}(x, Q^2)\,. \tag{2.21}$$

It was observed, first by the European Muon Collaboration (EMC), that the PDFs for protons bound to large nuclei differed from those for free protons, $f_q^p$ [56]. This finding was unexpected because of the vast difference in scales between the nuclear and nucleon binding energies. The observation was that there was a reduction in the scattering rate for $x \approx 0.6$, and an enhancement for $x \approx 0.1$. Further studies to lower $x$ showed a reduction for $x < 0.05$ [57]. These findings prompted the development of nuclear PDFs (nPDF) to be used in the calculations of scattering cross sections in nuclear collisions. These nPDFs are determined from global fits to world data. Fig. 2.10 gives an example of nucleon nPDFs represented as ratios

$$R_X^{Pb}(x, Q^2) = \frac{f_X^{p/Pb}(x, Q^2)}{f_X^p(x, Q^2)}\,, \tag{2.22}$$

for valence quarks, $X = V$, sea quarks, $X = S$, and gluons, $X = G$. One can see the reduction for $x \approx 0.6$, enhancement for $x \approx 0.1$, and depletion for $x < 0.02$, now known as the EMC-effect, anti-shadowing, and shadowing regions, respectively. One also sees a sharp increase in the valence quark case at the highest values of $x$ due to the Fermi motion of the nucleons within the nucleus.

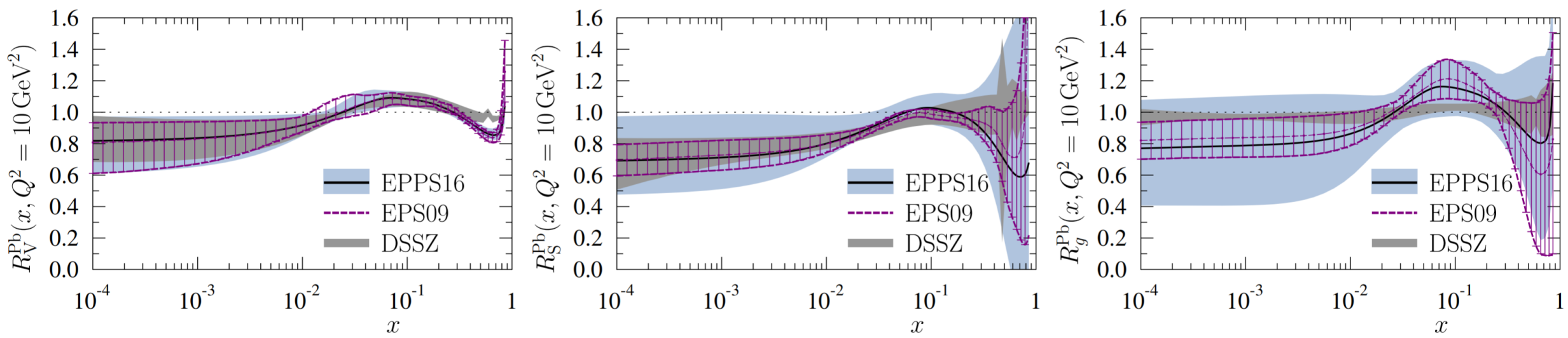

Figure 2.10: Proton nPDFs from global fits to world data as a function of parton $x$ for momentum scale 10 GeV. The results are presented as the ratio of PDFs in Pb nuclei to those from free protons for valence quarks, left, sea quarks, middle, and gluons, right [58].

### 2.3.4 Quark Gluon Plasma

Given that hadrons are composite objects with well defined binding energies, it is natural to assume that there could be a form of matter with high enough energy density such that hadrons could not exist. In this state, quarks and gluons would be deconfined and free to move about independently. This is analogous to increasing the temperature of ice to make water or to ionize gas into a plasma. Asymptotic freedom implies the existence of a weakly interacting gas of quarks and gluons at extremely high temperatures ($T \gg \Lambda_{QCD}$). However, in the strongly coupled regime, it is not obvious what to expect.

The first work suggesting a limit to the temperature of hadronic matter is that of Hagedorn in his statistical bootstrap model [59] which, interestingly, came before the establishment of the Quark Model and QCD as the theory of the strong interaction. In this case, he was trying to understand the origin of the exponential spectrum of hadronic masses by applying simple equilibrium statistical mechanics. The model postulated that hadronic states are self similar; each state is composed of lighter states *ad infinitum*. A limiting temperature naturally arises due to the increasing availability of higher mass states with increasing energy, thus limiting the average kinetic energy of per particle. By constraining this model with the measured hadron spectrum, a temperature of $T_c \approx 160$ MeV was extracted. With the acceptance of the Quark Model, it was proposed that this critical temperature is instead the result of a second order phase transition from hadronic matter to matter composed of deconfined quarks and gluons called quark-gluon plasma (QGP) [60, 61].

In the decades since these initial predictions, deconfinement has been explored in LQCD simulations in which the QCD equation of state can be determined. Calculations in this framework are limited to low net baryon densities due to due to the so-called "sign problem" in lattice quantum field theories [62]. However, results indicate a smooth crossover between quark and hadron matter at $T \approx 150$ MeV in remarkable agreement with the Hagedorn temperature. Fig. 2.11 shows temperature dependent pressure, energy, and entropy densities of quark matter from lattice simulations [63]. At high temperature, the values slowly approach the non-interacting particle limit as expected from asymptotic freedom. Around $T = 150$ MeV, the values vary rapidly with temperature as the phase space grows from the added color degrees of freedom.

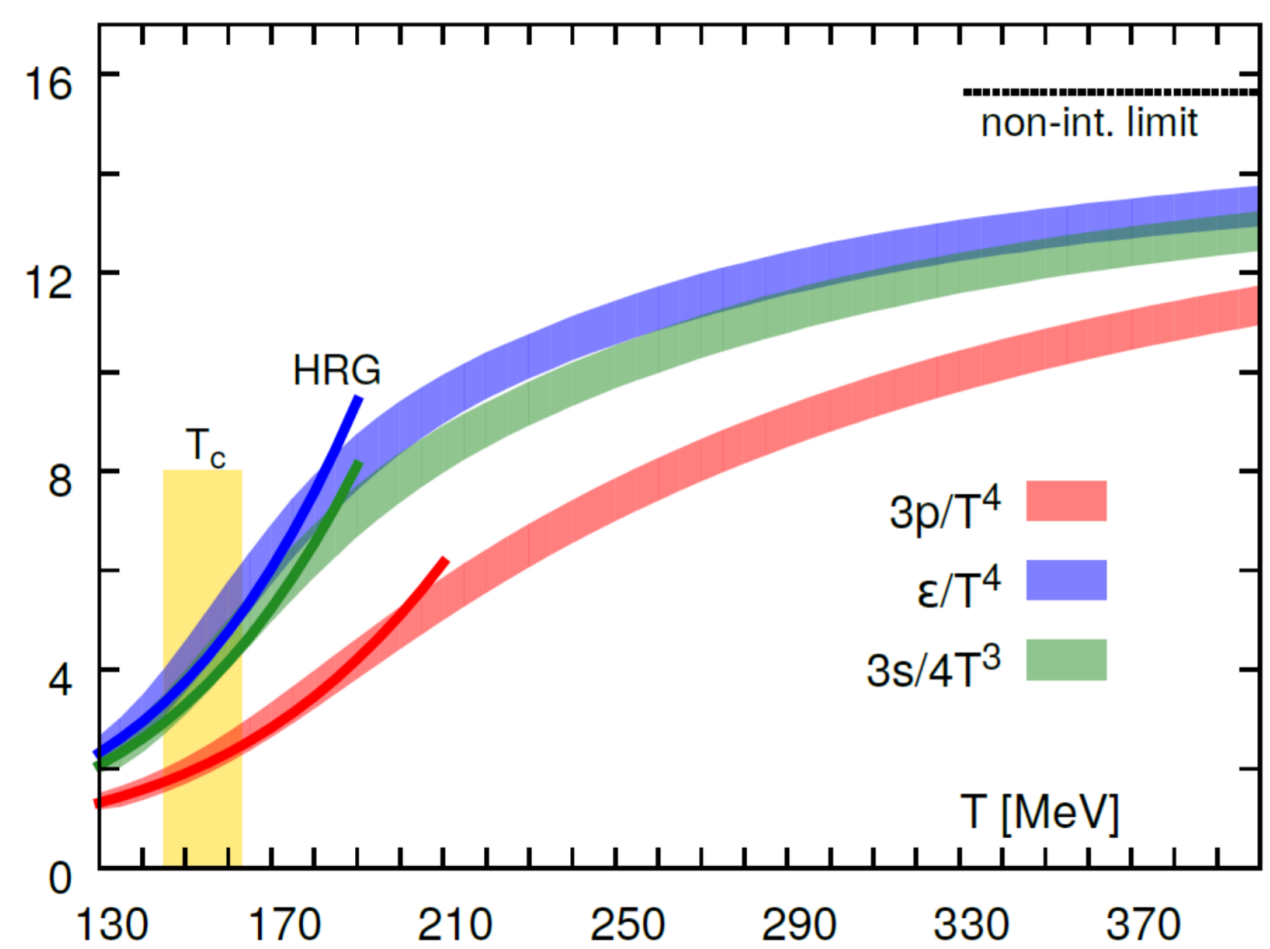

Figure 2.11: Temperature dependent pressure, energy, and entropy densities of quark matter from lattice simulations [63].

The phases of QCD matter are typically depicted on a scale of temperature and net baryon density (or baryon chemical potential) as shown in Fig.2.12. The nuclei of atoms exist at zero temperature and baryon density of one. For small baryon densities, the transition from hadrons to QGP is around $T = 150$ MeV, where LQCD simulations indicate a smooth cross over between QGP and hadronic phases. Alternatively, at low temperature and high baryon density are neutron stars, and as the density grows, long range color Cooper pairs are believed to form creating a superconductor type material [64]. It is presently unknown whether a first order phase transition and associated critical point occurs at moderate temperature and baryon density, but experimental search efforts are underway at RHIC, FAIR and NICA.

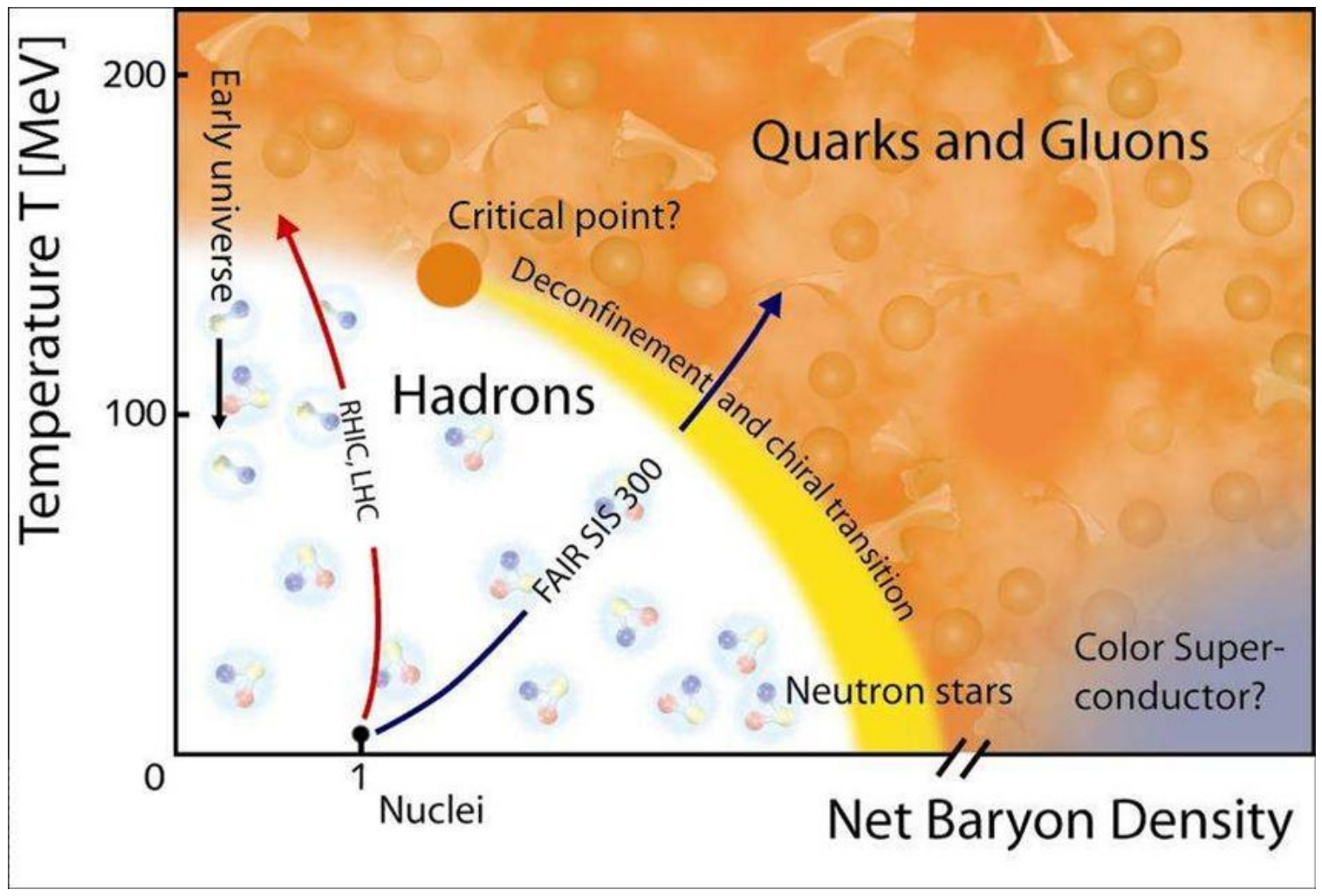

Figure 2.12: A qualitative description of the QCD phase diagram for any given temperature and net baryon density.

The experimental evidence for the creation of the QGP in heavy ion collisions at RHIC and the LHC are plentiful. Aspects of which will be discussed in detail in Sec. 2.4.

## 2.4 Nuclear Collisions

At RHIC and the LHC, respectively, Au and Pb nuclei are accelerated to velocities close to the speed of light and then allowed to collide. The nuclei are Lorentz contracted into disks in the center-of-mass (lab) frame by factors of 100 and 2500. The overlapping disks form a geometry dependent on the impact parameter and the stochastic distribution of the nucleons within each nucleus; nucleons in the overlap region are said to *participate* and the rest *spectate*. The spectator nucleons shear off and continue at beam rapidity which is roughly 5.3 and 8.5 for the two cases. The participating nucleons act as sources of longitudinal quarks and gluons that form a continuum of interacting color fields as the two nuclei move through each other. Most interactions are soft, with little exchanged momentum (i.e. $Q^2 \lesssim 1\text{GeV}^2$), and most of the energy ends up at large absolute rapidities near the beam line. However, a significant amount of energy is distributed in the final state amongst low momentum hadrons at mid-rapidity ($|y| < 1$). For example [63], in head on Pb+Pb collisions with total energy per nucleon pair of $\sqrt{s_{\rm NN}} = 2.76$ TeV, the total transverse energy in $|y| < 0.5$ can be around $1.65$ TeV. The energy density can be estimated as

$$\epsilon = \frac{1}{\mathcal{A}t}\frac{dE_{\rm T}}{dy}\,, \tag{2.23}$$

where $\mathcal{A}$ is the area of the overlap and $t$ is the time since the initial contact [65]. In this case, about 1 fm/c after the collision gives $\epsilon \approx 12$ GeV/fm$^3$, or about 20 times that normal nuclear matter.

As will be shown in Sec. 2.4.3, the QGP that carries this immense energy density acts as a strongly coupled liquid, not a weakly interacting gas as would be expected at much higher energies. This liquid has been observed to flow in response to the initial geometry of the colliding nuclei, a process described with good accuracy by relativistic hydrodynamic simulations [4–6]. Additionally, the medium has shown a high level of opacity to high energy color charges created from scatterings in the earliest times of the collisions [8–10]. This phenomenon of jet quenching will be discussed in more detail in Sec. 2.4.4. Finally, a growing body of evidence points to the creation of a QGP in small collision systems like $p$+Pb and even $pp$ [11–17]. This will be discussed in Sec. 2.4.5

### 2.4.1 The Glauber Model

In nuclear collisions (A+A), one often wants to study the rates of processes in comparison to those in $pp$ collisions. In this way, effects from the nucleons being part of larger nuclei can be separated from effects of individual nucleon-nucleon scattering. However, there will be a trivial scaling just because A+A collisions have more binary nucleon interactions and, thus, more opportunities to include any given process. Therefore, to make these comparisons, one would want to know the total number of $pp$-like collisions, $N_{\rm coll}$, or considering that some quantities might instead scale with the total initial matter density, one would want to know the total number of participating nucleons, $N_{\rm part}$. The scaling factor for hard processes is the *nuclear thickness function*, $T_{\rm AB}$, defined as

$$T_{\rm AB} = \int d^2\vec{b} \int d^2\vec{x}\rho_{\rm A}(\vec{x})\rho_{\rm B}(\vec{x}-\vec{b})\,, \tag{2.24}$$

for nuclei A and B with transverse nucleon density profiles $\rho_{\rm A}(\vec{x})$ and $\rho_{\rm B}(\vec{x})$ respectively. This acts as an effective nucleon-nucleon luminosity and, given the nucleon-nucleon inelastic cross section, $\sigma_{\rm NN}$, can be related to $N_{\rm coll}$

as

$$T_{\mathrm{AB}} = \frac{N_{\mathrm{coll}}}{\sigma_{\mathrm{NN}}} \,. \tag{2.25}$$

The problem is that there is no way of knowing these quantities directly since the nuclei collide with random, femtometer scale, impact parameter and have fluctuating distributions event-to-event.

The strategy is to measure *centrality* as percentiles of some detector quantity as a proxy of overall event activity such as multiplicity or total transverse energy, and then map this distribution onto the distribution of possible $N_{\mathrm{part}}$ or $N_{\mathrm{coll}}$. Thus, these quantities can be inferred on average from the event activity. An illustration of centrality as measured with mid-rapidity charged particle multiplicity is shown in Fig. 2.13. So-called *central* collisions have a large $N_{\mathrm{part}}$, small impact parameter, $b$, and produce a large number of charged particles; *peripheral* collisions have a small $N_{\mathrm{part}}$, large $b$, and produce a small number of charged particles. The process of determining centrality will be discussed in detail in Ch. 4. Here we present an overview of determining the distributions of $N_{\mathrm{part}}$ and $N_{\mathrm{coll}}$ with a Monte Carlo (MC) Glauber model [7].

The Glauber model works in the short wavelength limit where nucleons can be considered semi-classical objects that have straight line trajectories, undisturbed by interactions. Nucleon positions are chosen randomly event-by-event according a realistic 3-d distribution. For spherical nuclei, like Pb and Au, the distribution is only radially dependent and takes the form of a Fermi distribution

$$\rho(r) = \rho_0 \frac{1}{1 + e^{\frac{r-R}{a}}} \,. \tag{2.26}$$

The radius, $R$, and skin depth, $a$, are chosen to match measured charge distributions from low energy electron scattering experiments [66]. To model inter-nucleon repulsion, a minimum separation (typically $d_{\mathrm{min}} > 0.4$ fm) is imposed by re-throwing any given nucleon if it overlaps with another. The parameters are scaled through an iterative procedure to account for this distortion on the nucleon distribution function [66].

Given the nucleon positions, the general assumptions allow for an entirely 2-d collision process in the transverse plane. Impact parameters, $b$, with-respect-to the calculated nuclei centers are chosen randomly such that $dN/db \propto b$ in a range significantly larger than the average diameter of the nucleus. Nucleons are given a ball diameter, $D = \sqrt{\sigma_{\mathrm{NN}}/\pi}$, dependent on the measured $pp$ cross section, $\sigma_{\mathrm{NN}}$ for the given collision energy.

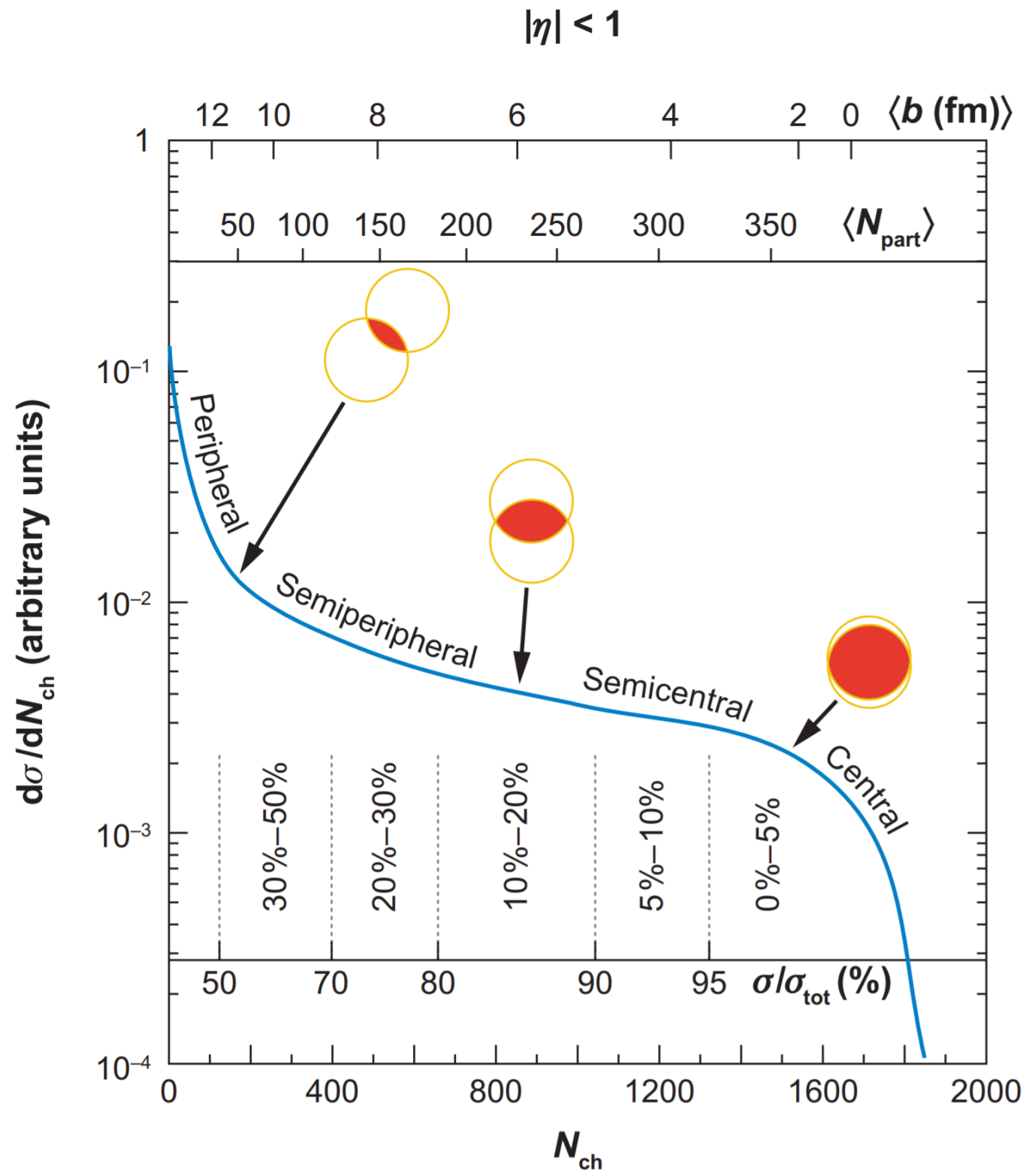

Figure 2.13: An illustration of centrality as measured with mid-rapidity charged particle multiplicity [7].

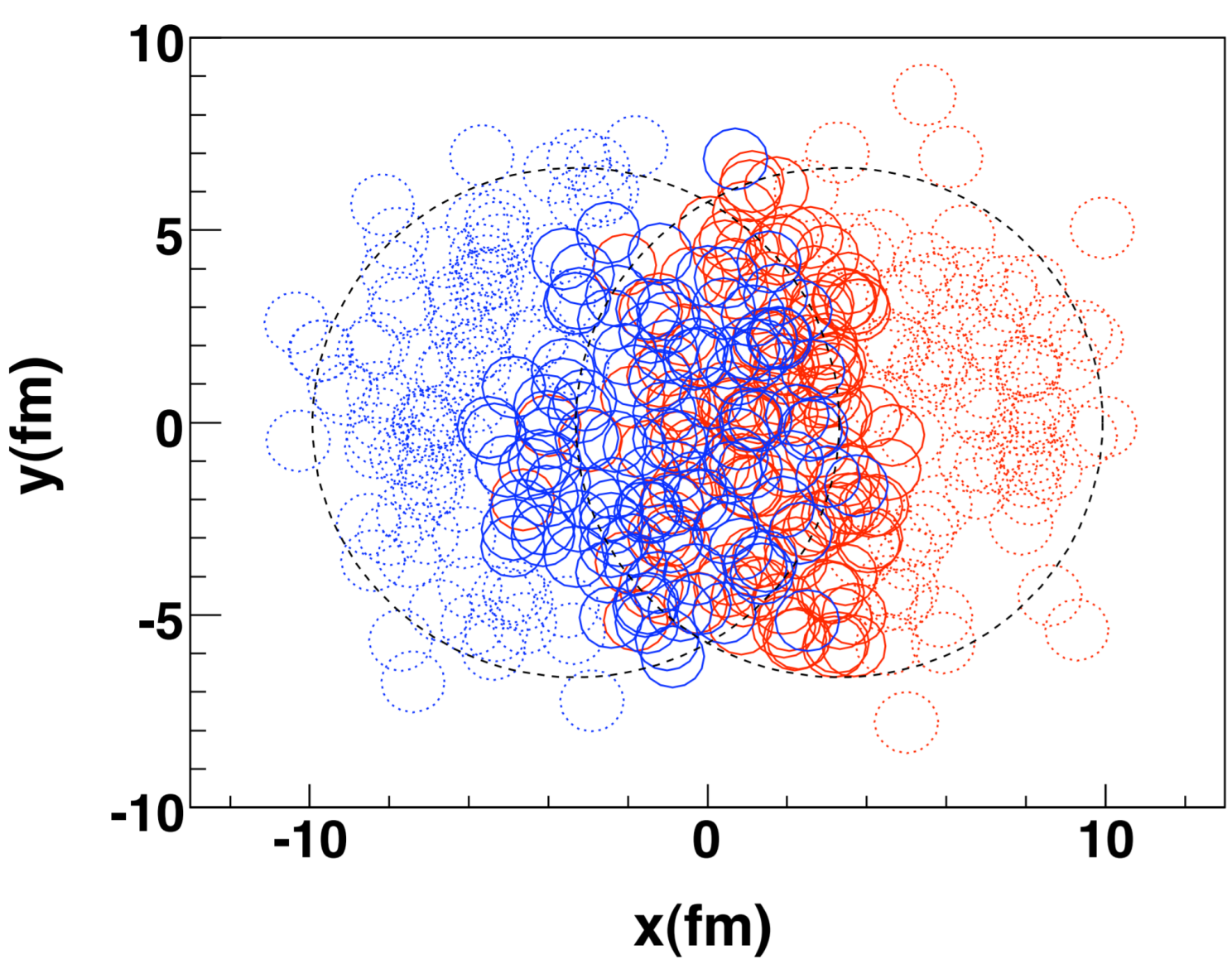

Figure 2.14: An example of a Pb+Pb collision from the PHOBOS MC Glauber simulation [67]. The nucleons are represented as circles, where their color marks which nucleus they originate from. The closed circles are participating nucleons and the dashed circles are spectating nucleons.

Individual nucleons are tagged as wounded (or participating) if they overlap with any nucleons from the other nucleus. The nuclei collide if there is at least one nucleon-nucleon interaction. Fig. 2.14 shows an example Pb+Pb collision from the PHOBOS MC Glauber simulation [67].

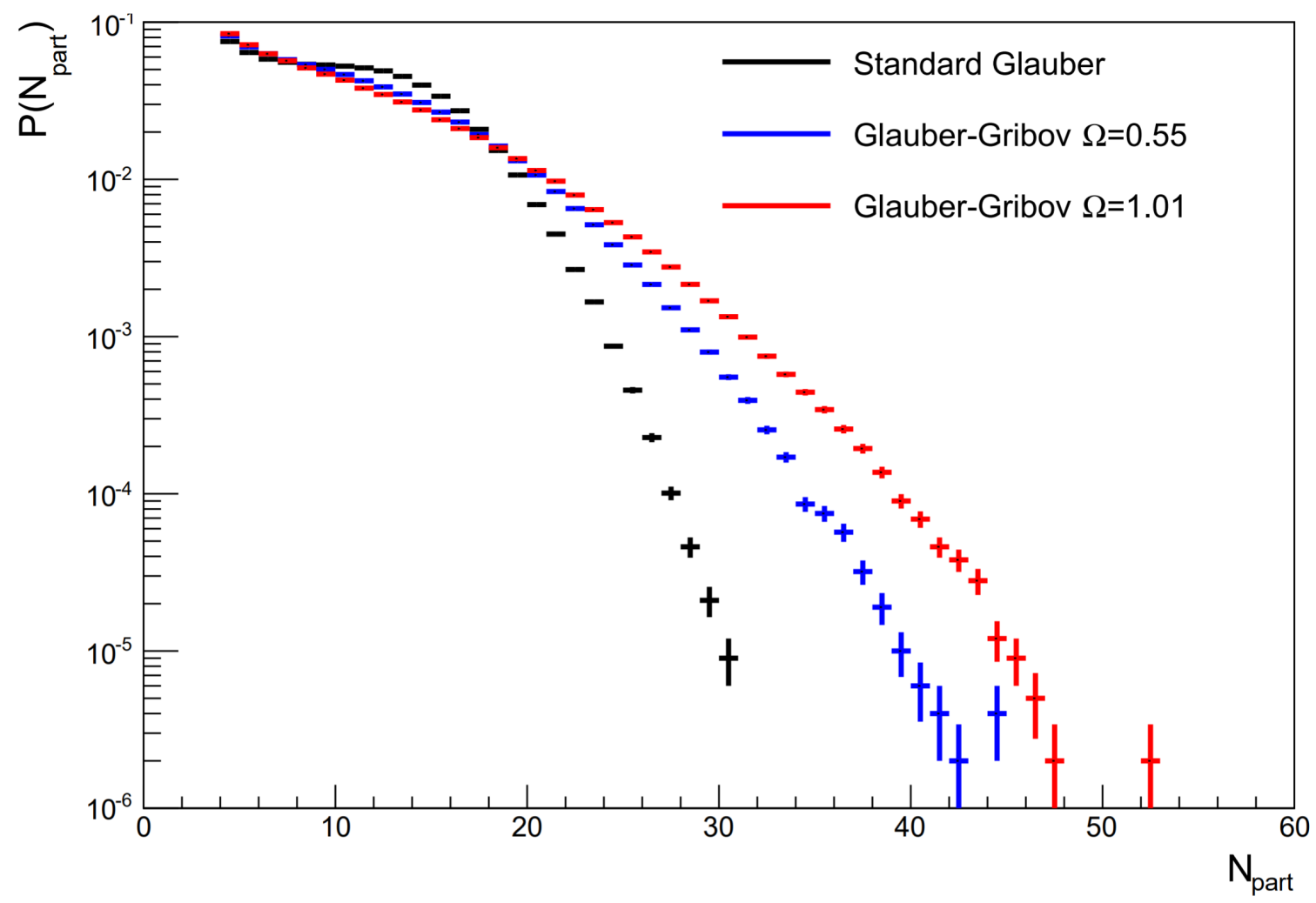

Figure 2.15: $N_{\text{part}}$ distributions from $p$+Pb collisions from the PHOBOS MC Glauber simulation [67]. The black points are from the standard Glauber implementation, and the blue and red points are from the Glauber-Gribov extension with two different values of the fluctuation parameter $\Omega$.

The Glauber-Gribov extension of the above simulation incorporates nucleon size fluctuations. In this case, $\sigma_{\text{NN}}$ is distributed as $(1/\lambda)P(\sigma/\lambda)$ where

$$P = \rho\frac{\sigma}{\sigma+\sigma_0}e^{-(\frac{\sigma-\sigma_0}{\sigma_0\Omega})^2}. \tag{2.27}$$

Here, $\sigma_0$ and $\Omega$ are the mean and width of $\sigma_{\text{NN}}$. Fig.2.15 shows $N_{\text{part}}$ distributions from $p$+Pb collisions for simulations using both standard Glauber and the Glauber-Gribov extension for proton size fluctuations (only the projectile proton size is allowed to fluctuate in this case). The added fluctuations act to broaden the distribution, giving access to larger values of $N_{\text{part}}$ in the tail corresponding to larger sizes of the proton.

These Glauber models are important, not only to determine centrality as a whole, but because they can provide geometrical information about the initial energy densities that are used as initial conditions for transport models like hydrodynamics. The spatial distribution of participating nucleons can be decomposed into

eccentricity moments as

$$\varepsilon_n = \frac{\sqrt{\langle r^2 \cos(n\phi)\rangle^2 + \langle r^2 \sin(n\phi)\rangle^2}}{\langle r^2 \rangle}\,, \tag{2.28}$$

where the brackets indicated averages over nucleons and events [68]. As will be discussed in more detail in the next Section, non-zero eccentricities give rise to momentum anisotropies in the final state hadron spectra via hydrodynamic expansion.

### 2.4.2 The Evolution Model

In the previous section, the Glauber model was described as a method to determine the number and spatial geometry of the participating nucleons in heavy ion collisions. How the energy deposited by these overlapping nucleons evolves into the hadrons seen in detectors is the topic of this Section. The standard paradigm for this multi stage process proceeds in order: 1) initial energy depositions, 2) pre-hydrodynamics, 3) hydrodynamic evolution, 4) hadronization, 5) hadronic scattering, and finally, free streaming of hadrons to the detector. There are quite a few packages on the market that simulate this process, and as a whole, they have been highly successful in describing the experimental results. This is not meant to be an exhaustive review of the topic, but a brief overview of the techniques typically used. For a detailed treatment of the subject, see Refs. [5, 69].

Hydrodynamics is an effective field theory of dynamical matter appropriate at large distance and time scales relative to the microscopic constituents and their dynamics but small with respect to the entire system. The dynamics of the theory are determined by tracking conserved currents described by the energy-momentum tensor, $T^{\mu\nu}$, a symmetric space-time tensor constructed from the effective energy density and velocity fields, $\epsilon$ and $u^\mu$, the metric tensor, $g_{\mu\nu}$, and all gradients that respect the symmetries. If the system is close to equilibrium, the gradients will be small. Thus, $T^{\mu\nu}$ is often represented as a gradient expansion

$$T^{\mu\nu} = T^{\mu\nu}_{(0)} + T^{\mu\nu}_{(1)} + T^{\mu\nu}_{(2)} + \cdots\,, \tag{2.29}$$

where $T^{\mu\nu}_{(0)}$ contains no gradients, $T^{\mu\nu}_{(1)}$ has single gradient terms, etc. In Minkowski space, conservation rules are imposed by requiring

$$\partial_\mu T^{\mu\nu} = 0\,, \tag{2.30}$$

which serve as the equations of motion. Similar rules can be made to include conserved charges, like Baryon number, but these are excluded from the present conversation. The zeroth order approximation gives rise to *ideal hydrodynamics* such that

$$T^{\mu\nu}_{(0)} = (\epsilon + P)u^\mu u^\nu + Pg^{\mu\nu} = \epsilon u^\mu u^\nu + P\Delta^{\mu\nu}\,, \tag{2.31}$$

where the two $P$ factors have been grouped into the projection tensor, $\Delta^{\mu\nu}$. The coefficient $P$ is identified as the pressure which must be related to $\epsilon$ through an equation of state which is typically calculated at equilibrium in LQCD simulations.

In practical scenarios, like high energy nuclear collisions, systems are not in equilibrium and there are significant gradients. Including higher orders introduces dissipative terms that are typically grouped into scalar, $\Pi$, and traceless, $\pi^{\mu\nu}$, components representing bulk and shear stress respectively. The energy-momentum tensor takes form

$$T^{\mu\nu} = T^{\mu\nu}_{(0)} + \pi^{\mu\nu} + \Pi\Delta^{\mu\nu}\,. \tag{2.32}$$

To first order, this becomes

$$T^{\mu\nu} = T^{\mu\nu}_{(0)} + \eta\sigma^{\mu\nu} + \zeta\partial_\rho u^\rho \Delta^{\mu\nu}\,, \tag{2.33}$$

where

$$\sigma^{\mu\nu} = \Delta^{\mu\rho}\partial_\rho u^\nu + \Delta^{\nu\rho}\partial_\rho u^\mu - \frac{2}{3}\Delta^{\mu\nu}\partial_\rho u^\rho\,. \tag{2.34}$$

The transport coefficients, $\eta$ and $\zeta$, are the shear and bulk viscosities which depend on the internal dynamics of the microscopic theory, and can, therefore, provide a connection from hydrodynamic simulations to fundamental properties of QCD. [6]

Given an initial $T^{\mu\nu}$, these equations can then be used to evolve fluid cells in simulation. The initial conditions are generated by first determining an energy density, typically by parameterizing the overlap of participating nucleons in a Glauber model. For example, one can add a simple Gaussian at the location of each nucleon [70], or add entropy to the system for each nucleon and use an equation of state to transform the entropy to energy density [71]. In all cases, the parameters are tuned to match the measured particle multiplicities.

[6] If second order corrections are included, there emerge 15 separate coefficients, and third order corrections bring this number to 68 [5].

The initial energy density of the system is then evolved non-hydrodynamically during a pre-hydrodynamic phase. This can be interpreted as the point that the energy is converted into dynamical matter and entropy is created. The matter is evolved in time and accumulates velocities given a model defined prescription. In perhaps the simplest case, energy cells are interpreted as isotropic momentum sources and are evolved as non-interacting expanding shells. In the more sophisticated approach of the IP-Glasma model [72, 73], color fields are distributed within the colliding nucleons according to the IP-Sat model [55] (mentioned in Sec 2.3.3) and then evolved in the weak coupling CGC framework according to the classical Yang-Mills equations. Alternatively, in the strong coupling limit, one can apply the so-called AdS-CFT correspondence between weakly interacting fields in gravitational theory and strongly coupled fields in certain quantum field theories that are similar to QCD [74]. Then, intractable calculations in the strongly interacting field theory are mapped to tractable calculations in general relativity. This is the approach taken, for example, by the superSONIC model [75].

After some small amount of time (typically 0.1-0.5 fm/c), the system is packaged into $T^{\mu\nu}$ and passed off to the hydrodynamic simulation. This pre-hydrodynamic stage has the effect of smoothing out the initially large energy gradients and bringing the system into the range of applicability of hydrodynamics. There is a significant uncertainty in these early stage models and many seem rather *ad hoc*. Notice the broad range of techniques in described above, from the zero-coupling free streaming case, to the weak coupling IP-Glasma, and finally to the strong coupling superSONIC. In large collision systems, the hydrodynamic phase is so dominant that results are relatively insensitive to the particular model or start time. However, for small systems, it can have a significant impact. Therefore, small collision systems could serve as a testing ground for pre-hydrodynamic models.

Once $T^{\mu\nu}$ has been passed to the hydrodynamics code, the system is evolved until individual fluid cells reach the critical temperature, at which point hadronization takes over. A contiguous transition hyper-surface, $\sigma$, is constructed by connecting cells below the critical temperature. Then the phase change to hadrons is handled by the Cooper-Frye prescription [76] such that energy and momentum is conserved; i.e. $T^{\mu\nu}$ is made to be continuous across the boundary. If the fluid cells on the transition hyper-surface are assumed to be in local equilibrium, the momentum distribution of hadron resonance $i$ is

$$E\frac{dN_i}{d^3p} = \int f_i(x^\mu, p^\mu)p^\mu d\sigma_\mu \,, \tag{2.35}$$

where $f_i(x, p)$ are distribution functions in phase space. The hadrons then undergo a gas scattering phase until becoming so diffuse that they free-stream to the detectors.

### 2.4.3 Bulk Observables

All observables in particle collisions are derived from measurements of final state particles that end up in detectors. As discussed in briefly in Sec 2.3.2, QCD processes are separated by differences in scale. Particles with large $p_\mathrm{T}$ originate from perturbative processes that show different behavior of production rates for low $p_\mathrm{T}$ particles. In nuclear collisions, low $p_\mathrm{T}$ particles predominantly come from the hadronization of the bulk QGP medium and are thus emitted at later times, whereas high $p_\mathrm{T}$ particles are created in hard scatterings in the initial overlap of the colliding discs. Observables relating mostly to the soft, bulk interactions are discussed here, and those relating to hard processes are detailed in the next Section 2.4.4.

The $p_\mathrm{T}$ spectrum of particles can then be divided into low $p_\mathrm{T}$ ($p_\mathrm{T} \lesssim 3$ GeV) and high $p_\mathrm{T}$ ($p_\mathrm{T} \gtrsim 3$ GeV) regions. The left panel of Fig. 2.16 shows the $p_\mathrm{T}$ spectrum of charged particles from central Au+Au collisions at RHIC. The low $p_\mathrm{T}$ behavior is roughly exponential and in good agreement with hydrodynamic simulations. At high $p_\mathrm{T}$, the spectrum changes to a power law shape that can be described by pQCD. The collective hydrodynamic expansion of the QGP affects the shape of the particle $p_\mathrm{T}$ spectrum. The fluid undergoes rapid radial expansion in the transverse plane, giving particles a large velocity boost. The right panel of Fig. 2.16 shows the $p_\mathrm{T}$ spectrum for identified hadrons overlaid with results from hydrodynamic simulations. The spectrum for heavier hadrons are pushed out to higher $p_\mathrm{T}$ as one would expect if the particles gained a common velocity boost, providing evidence for the presence of strong radial flow.

Hydrodynamic expansion converts initial energy density gradients from overlapping nucleus geometry anisotropies into hadron momentum distributions. Thus, an imprint of the initial transverse distribution of the nucleons is left on the distribution of final state hadrons. This is in striking contrast to what one would expect if there was no collectivity; if each nucleon-nucleon interaction were independent, the total distribution of the produced particles would be azimuthally symmetric. These azimuthal anisotropies, are measured relative to the so-called reaction plane, $\Psi_r$, the plane spanned by the impact parameter and the beam axis. The distribution of particle azimuthal angles can then be measured relative to $\Psi_r$ and decomposed in a Fourier series

$$E\frac{d^3N}{d^3p} = \frac{1}{2\pi}\frac{d^2N}{p_{\rm T}dp_{\rm T}dy}(1 + \sum_{n=1}^{\infty} 2v_n \cos\left[n(\phi - \Psi_r)\right])\,, \tag{2.36}$$

where the coefficients $v_n$ quantify the magnitude of the anisotropy. The coefficients $v_2$ and $v_3$ quantify elliptic and triangular flow respectively.

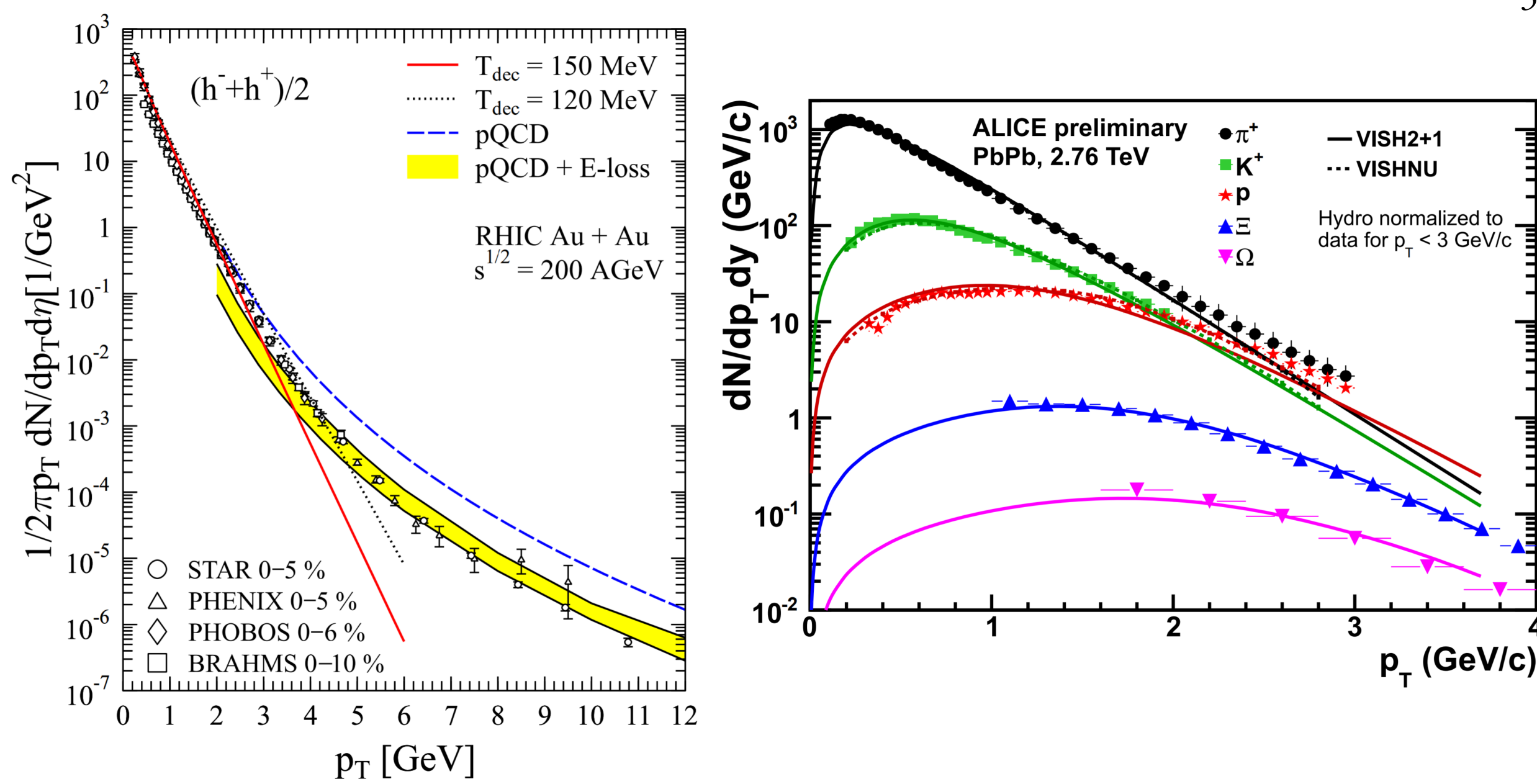

Figure 2.16: **Left:**The $p_{\rm T}$ spectrum of charged hadrons from 200 GeVAu+Au collisions measured with a variety of RHIC experiments. The data is overlaid with theoretical calculations from both hydrodynamics and pQCD [6]. **Right:** The $p_{\rm T}$ spectrum of identified charged hadrons from 2.76 GeVPb+Pb collisions. The data is overlaid with theoretical calculations from the VISHNU hydrodynamic model [77].

The reaction plane must be estimated event-by-event. This can be done using the anisotropy of the particles themselves by calculating the *event plane* as

$$\Psi_n = \Big(\tan^{-1}\frac{\Sigma_i \sin(n\phi_i)}{\Sigma_i \cos(n\phi_i)}\Big)/n\,, \tag{2.37}$$

where the sums run over the particles, and $n$ is the harmonic number. In this way, each harmonic estimates a different event plane. In practice, the event plane is determined in a different region of pseudorapidity than where the flow is measured in order to remove auto-correlations. Up to a correction from the event plane resolution, the flow coefficients can be determined simply as

$$v_n = \langle \cos[n(\phi - \Psi_n)] \rangle \,, \tag{2.38}$$

where the angle brackets represents an average over all particles and over all events.

Complementary methods to determine $v_n$ have been invented using so-called *multi-particle cumulants*. Instead of estimating the reaction plane, these methods correlate particles in various combinations. For example [78], $v_n$ can be calculated using 2- and 4-particle correlations as

$$v_n\{2\} \equiv \sqrt{\langle \cos[n(\phi_1 - \phi_2)] \rangle} \tag{2.39}$$

and

$$v_n\{4\} \equiv \left(2\langle \cos[n(\phi_1 - \phi_2)] \rangle^2 - \langle \cos[n(\phi_1 + \phi_2 - \phi_3 - \phi_4)] \rangle\right)^{1/4} . \tag{2.40}$$

The different multi-particle estimators of $v_n$ are sensitive to fluctuations in different ways, and can therefore be used to learn more about the underlying $v_n$ distributions [79]. Using the previous example,

$$v_n\{2\} = \sqrt{\langle v_n \rangle^2 + \sigma_{v_n}^2} \tag{2.41}$$

and

$$v_n\{4\} \approx \sqrt{\langle v_n \rangle^2 - \sigma_{v_n}^2} \,, \tag{2.42}$$

where the approximation is exact in the limit of Gaussian fluctuations.

Figure 2.17 shows a distribution of 2-particle $\Delta\eta$-$\Delta\phi$ correlations in mid-central Pb+Pb collisions. There are two ridge structures at $\Delta\phi = 0$ and $\Delta\phi = \pi$ which extent across the whole acceptance in $\Delta\eta$. These ridges indicate a strong elliptical modulation to the momentum distribution of particles in the transverse plane. The fact that the structure is almost independent of the particle separation in $\eta$ suggests that this is a global event phenomenon that fits with the collective expansion picture of the hydrodynamic model. The strong elliptic flow is a response to the highly elliptical shape of the nuclear overlap in mid-central collisions. Additionally, there is a peak near $\Delta\eta = \Delta\phi = 0$ from short range correlations created by particle decays and jets. If *flow* correlations are defined to be those corresponding with the event geometry, then these jet-type correlations can be considered *non-flow*. Because a large component of non-flow comes from "small number" particle correlations like decays, multi-particle correlation methods using larger numbers of particles tend to be less sensitive to non-flow. In large collision systems, flow signals are strong enough to dominate non-flow; however, as will be discussed in more detail in Sec.2.4.5, care must be taken to separate non-flow from flow signals when studying small collision systems.

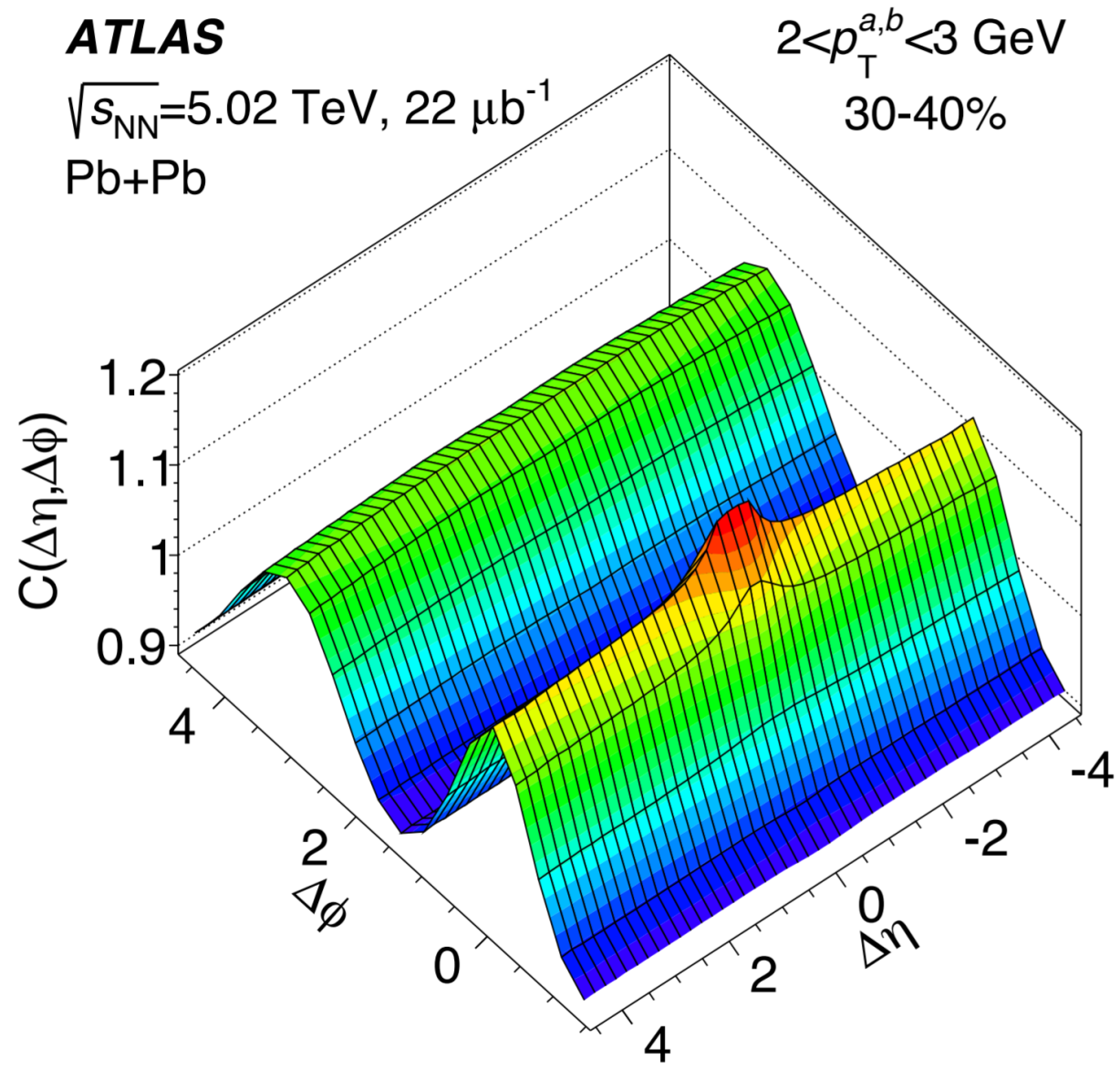

Figure 2.17: Distribution of 2-particle correlations in $\Delta\eta$ and $\Delta\phi$ from 30-40% central Pb+Pb collisions at $\sqrt{s_{\text{NN}}} = 5.02$ TeV [80]. Both particles are required to have $p_{\text{T}} = 2 - 3$ GeV.

Fig. 2.18 shows the elliptic flow, $v_2$, from various methods plotted as a function of event centrality in Pb+Pb collisions. The $v_2$ is smallest in the most central collisions where the impact parameter is small and the overlap region is more circular. The distribution peaks between 30% and 50% central where the overlap geometry is most elliptical. The $v_2\{2\}$ is everywhere greater than that of the event-plane method (EP), and the $v_2\{4\}$ is always less than that of the EP. This makes sense given the way the fluctuations influence $v_2\{2\}$ and $v_2\{4\}$; the separation between the two methods is larger in the most central and most peripheral cases, where the geometric anisotropy is driven by the event-to-event distributions of nucleons and less by the impact parameter. On the other hand, $v_2\{4\}$ should be more insensitive to non-flow correlations than $v_2\{2\}$, so perhaps there is some contamination.

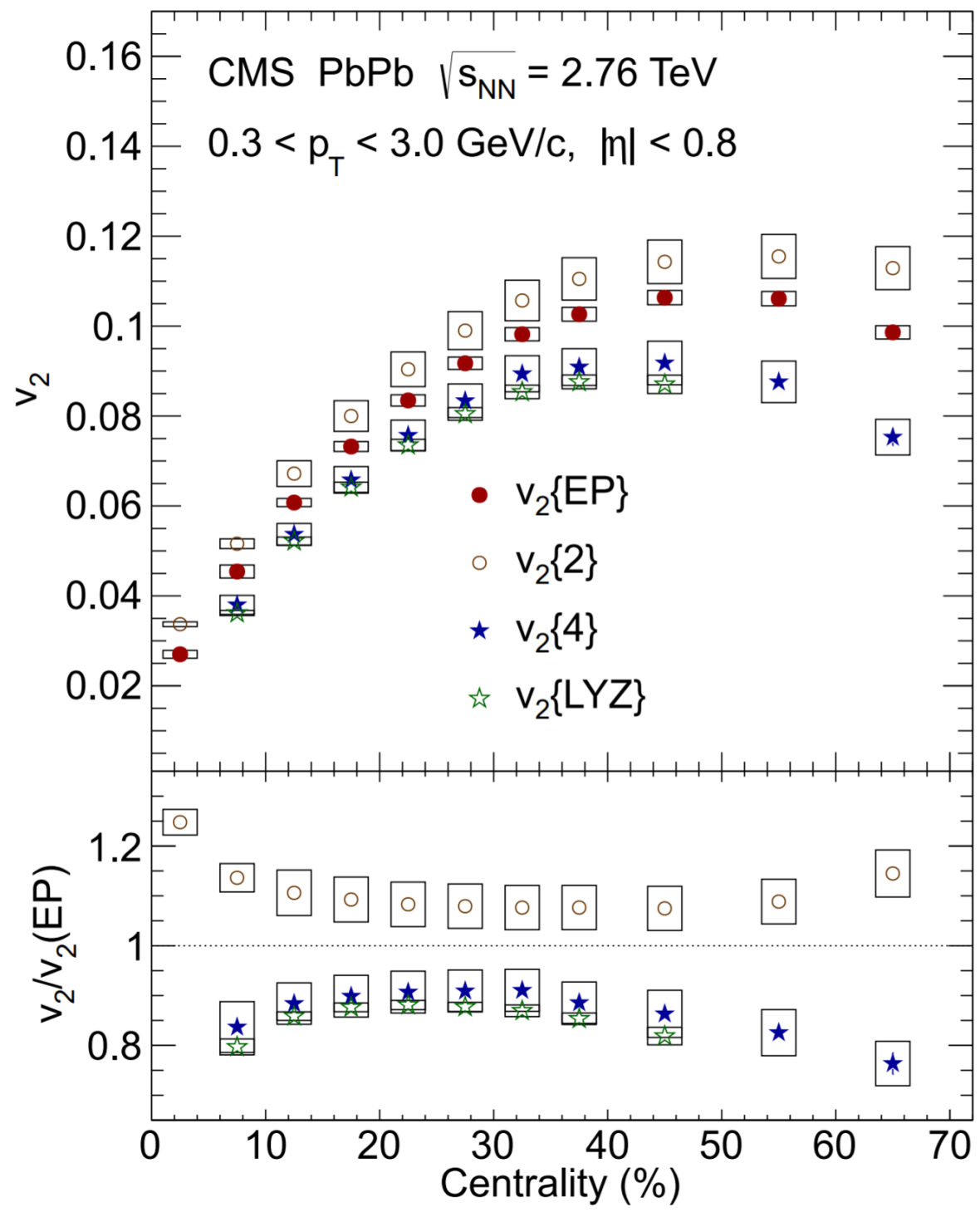

Figure 2.18: Elliptic flow $v_2$ as a function of centrality from the event-plane method, 2- and 4-particle correlations, and Lee-Yang zeros methods$\sqrt{s_{\mathrm{NN}}}$ [81].

Simulations using viscus hydrodynamic expansion are able to reproduce the distributions of $v_n$ to a high degree of precision at low $p_{\mathrm{T}}$. Fig.2.19 shows $v_n$ as a function of particle $p_{\mathrm{T}}$ using the EP method [82]. The $v_2$ is larger than the other harmonics as expected from the large ellipticity of the overlapping nuclei in this centrality. Additionally, the viscosity dampens higher harmonics to a greater degree; this produces lower values for higher order $v_n$ given the same initial spatial eccentricities. The points for each harmonic increase with $p_{\mathrm{T}}$ over the plotted range as one might naively expect from anisotropic expansion; i.e., fluid cells with higher velocity felt greater expansion forces from gradients and, thus, have a greater anisotropy. The theoretical curves match the data points almost perfectly. Anisotropic flow can also be studied with identified hadrons. Fig. 2.20 shows $v_2$ for identified hadrons; the heavier hadron $v_2$ distributions are pushed out in $p_{\mathrm{T}}$ in good agreement with hydrodynamic simulations, again providing evidence for a collective velocity field driving the expansion.

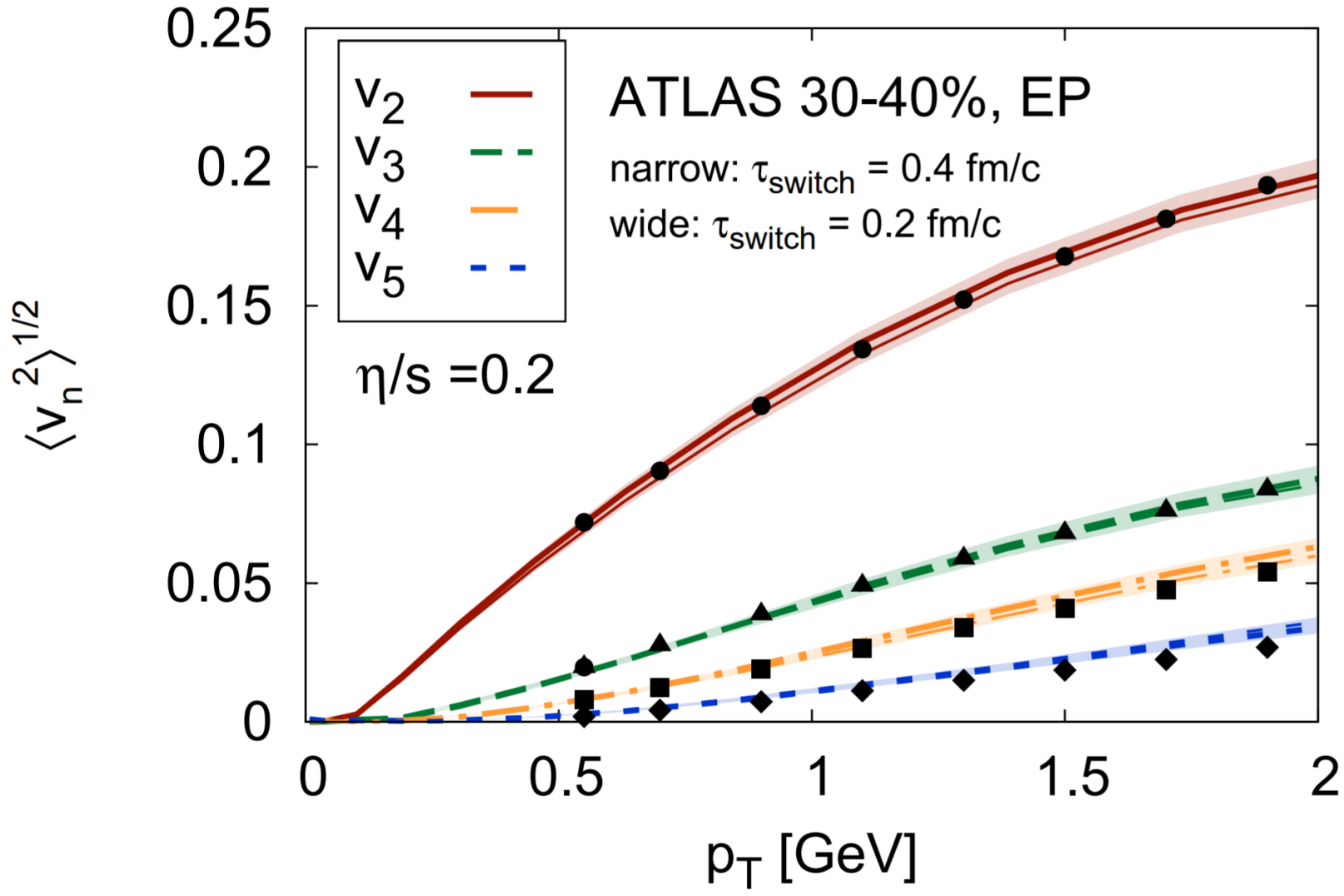

Figure 2.19: Coefficients $v_n$ from the EP method from 30-40% central Pb+Pb events plotted as a function of $p_T$ with comparisons to theoretical simulations using viscus hydrodynamic expansion [82].

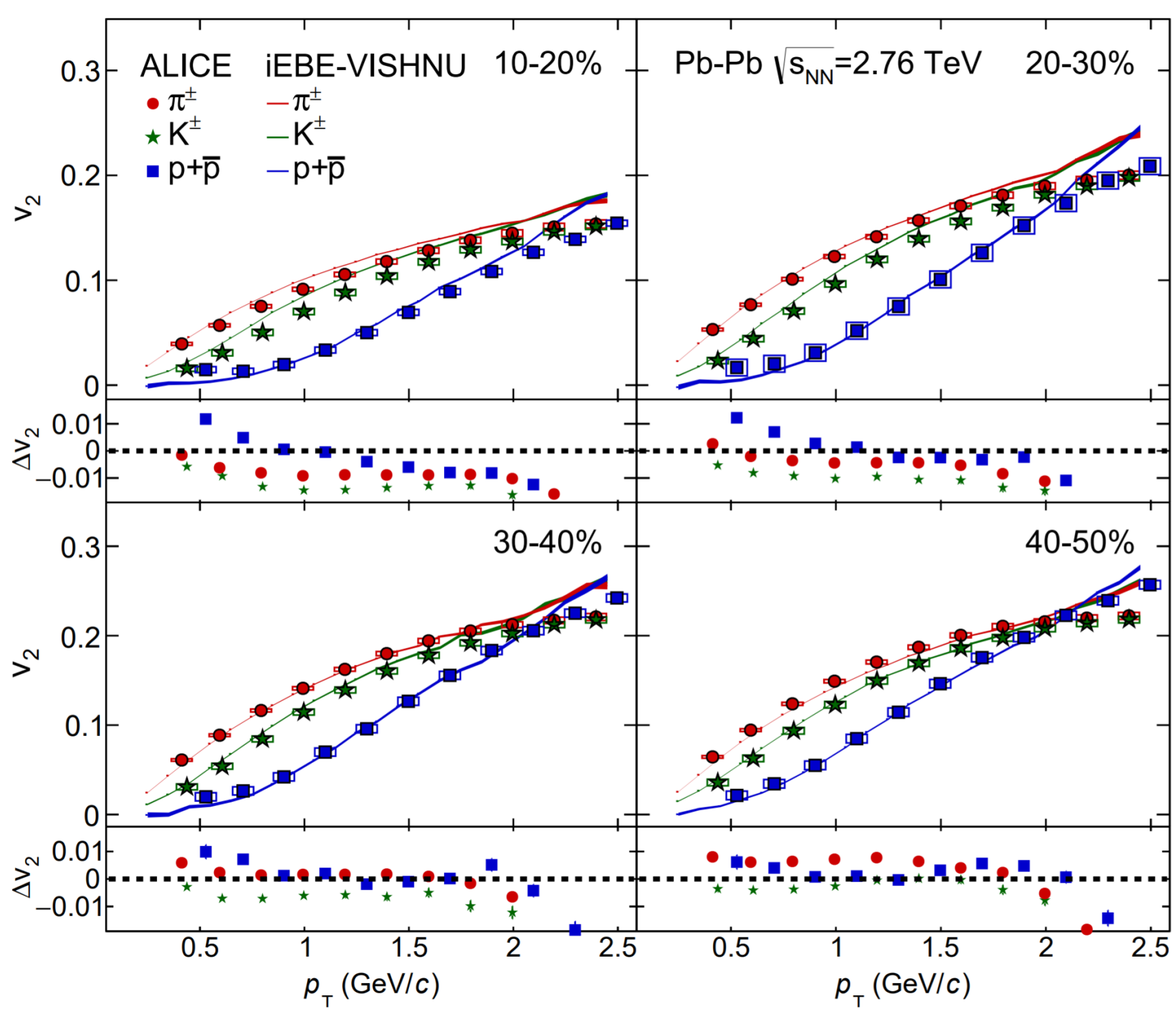

Figure 2.20: Coefficient $v_2$ for identified hadrons measured at various centralities in Pb+Pb collisions with $\sqrt{s_{NN}} = 2.76$. The data are overlaid with curves from hydrodynamic simulations TeV [83].

If the flow coefficients are plotted in a broader $p_T$ range, as in Fig. 2.21, one sees a much more complicated structure. As before, $v_n$ rises with $p_T$ until about 3 GeV, then decreases for higher $p_T$. At $p_T \approx 7$ GeV there is a change in slope indicating, perhaps, a change in the origin of the correlation. Hydrodynamic expansion predicts the low $p_{\mathrm{T}}$ rise of $v_n$, but predicts only monotonically increasing behavior with $p_{\mathrm{T}}$. As discussed above, about 3 GeV is where the slope in the particle $p_{\mathrm{T}}$ spectrum changes indicating a change from bulk to hard scattering particle production. Thus it fits that this is where the hydro-like behavior of the $v_n$ stops. After 3 GeV, there are particles from jets and other hard scattering processes entering these correlations in increasingly higher proportions. Thus, one might assume there is a mixture of particles from QGP hadronization and from jet fragmentation at any given $p_{\mathrm{T}}$; the lower $p_{\mathrm{T}}$ region is dominated by particles from the bulk, and the high $p_{\mathrm{T}}$ region is dominated by jet particles (almost by definition at say $p_{\mathrm{T}} \approx 50$ GeV). The possible origin of flow like correlations at high $p_{\mathrm{T}}$ will be discussed in the next Section 2.4.4.

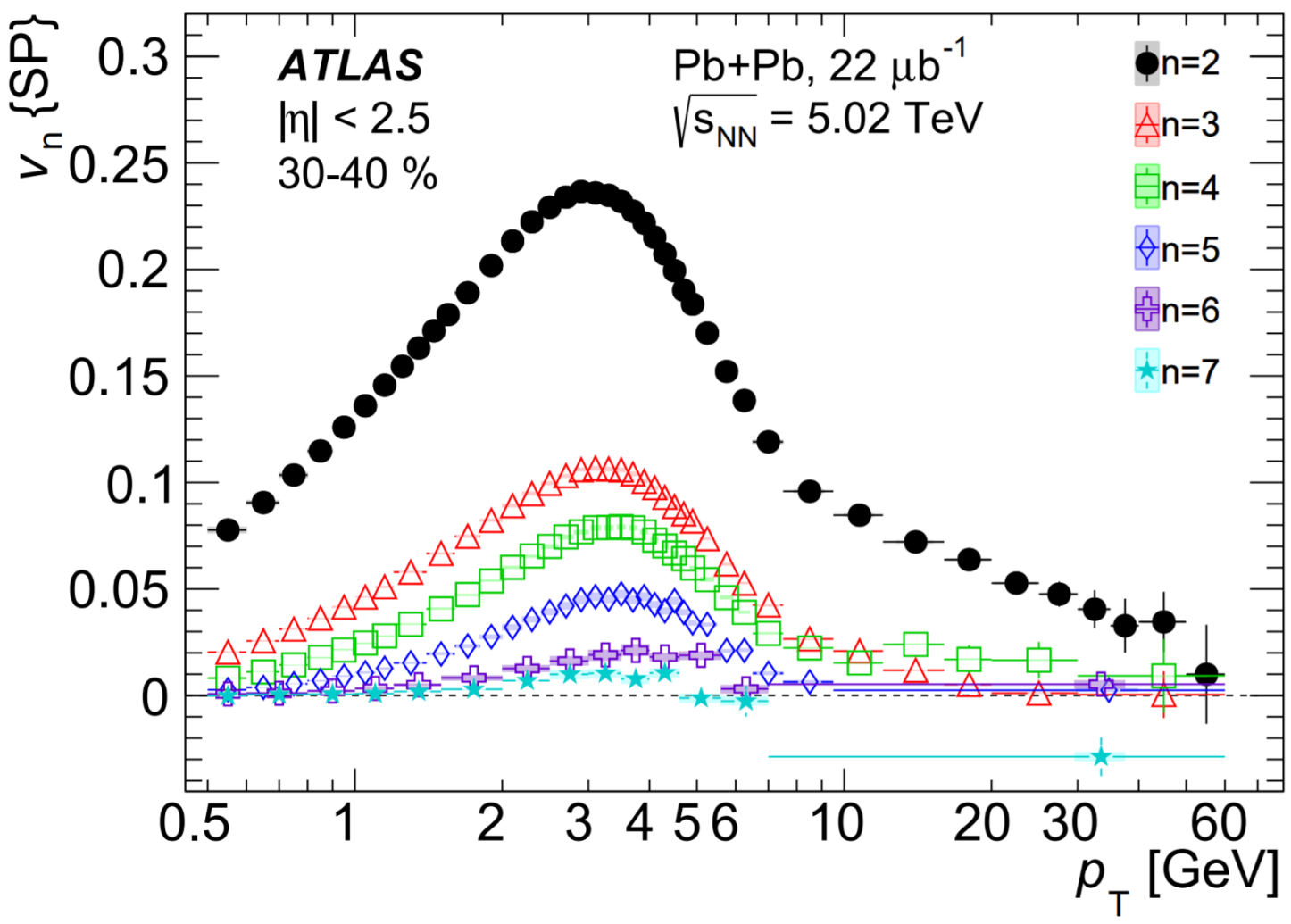

Figure 2.21: Coefficients $v_n$ from 30-40% central Pb+Pb events plotted as a function of $p_{\mathrm{T}}$ [80].

### 2.4.4 Hard Probes

In contrast to bulk particle production, hard scattering events are rare. These processes can be used to determine information about nPDFs and may be considered probes to study the QGP. Particle $p_{\mathrm{T}}$ distributions can be compared to expectations from $pp$ collisions scaled by $T_{\mathrm{AB}}$. Define the nuclear modification factor, $R_{\mathrm{AB}}$, to measure process $X$ in A+B collisions

$$R_{\mathrm{AB}}(p_{\mathrm{T}}) = \frac{\frac{1}{N_{\mathrm{evnt}}} dN^X_{\mathrm{AB}}/dp_{\mathrm{T}}}{\langle T_{\mathrm{AB}} \rangle d\sigma^X_{pp}/dp_{\mathrm{T}}}\,, \tag{2.43}$$

where $\frac{1}{N_{\text{evnt}}} N_{\text{AB}}^{X}$ is the per-event-yield of observations of $X$ in A+B collisions, and $\sigma_{pp}^{X}$ is the cross section of $X$ in *pp* collisions. Thus, $R_{\text{AB}}$ quantifies the comparison of process rates between A+B and *pp* collisions, where $R_{\text{AB}} = 1$ indicates that the nuclear collision can be considered an incoherent superposition of $N_{\text{coll}}$ different *pp* collisions. Deviation from unity can be due to a number of possibilities:

- First, the process could not scale with $T_{\text{AB}}$. Consider particle production at mid-rapidity for collisions of a fixed energy where $x \approx p_{\text{T}}/(\sqrt{s}/2)$. At low $x$, where gluon densities are high, nucleons can be considered opaque disks of dense color fields.[7] Thus, one would expect soft particle production to scale with simply with the overall number of overlapping nucleons ($N_{\text{part}}$). On the other hand, hard processes occur between high $x$ partons that are relatively diffuse within the nucleons. Therefore, these processes should scale with the number of opportunities for the high $x$ partons to find each other in a combinatorial sense ($N_{\text{coll}}$). Models of particle production assuming a two component soft and hard linear scaling with $N_{\text{part}}$ and $N_{\text{coll}}$ are generally very successful [84–86].

- Second, the process could have a modified initial state production rate, as expected to some degree from nuclear PDF modification and the proton–neutron (*isospin*) asymmetry.

- Finally, the object's kinematics could be modified due to interaction with the QGP; i.e., particles lose or gain energy on average, and, because the spectrum is steeply falling this causes an apparent increase or decrease in the final state production rate respectively.

Figure 2.22 shows the $R_{\text{AA}}$ for direct photons and $Z$ bosons in Pb+Pb collisions. In this context, direct photons are defined as those created initially in hard scattering between partons and not from the decay of hadrons like $\pi_0$. Because these are not colored objects, they do not interact strongly with the QGP and, therefore, provide a clean test of $T_{\text{AB}}$ scaling and nPDF effects. Both figures show an $R_{\text{AA}}$ around unity within uncertainties indicating no modification over a broad range of centrality. The left plot provides a prediction from a next-to-leading order pQCD calculation (Jetphox [87]) including nPDF modifications. The predicted modification is small compared to the uncertainties in the data.

[7] At sufficient beam energy, it is believed the gluon fields condense and saturate, and the nucleons can be considered disks of CGC.

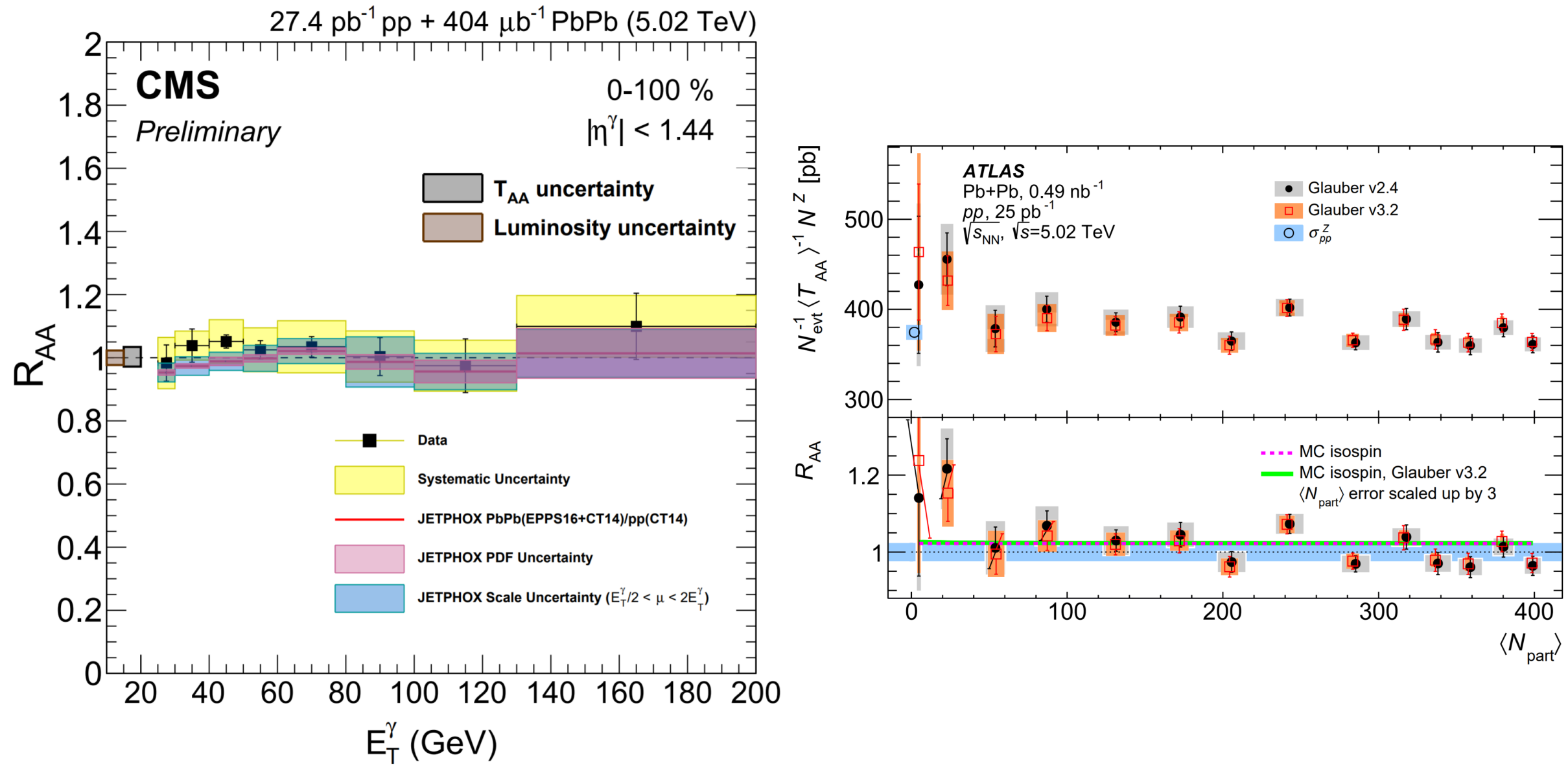

Figure 2.22: Left: prompt and isolated photon $R_{AA}$ plotted as a function of the photon's transverse energy [88]. Right: $Z$ boson $R_{AA}$ as a function of $N_{part}$ from Pb+Pb collisions at $\sqrt{s_{NN}} = 5.02$ TeV [28].

The left panel of Fig. 2.23 shows the $R_{AA}$ of charged hadrons in Pb+Pb collisions. The $p_T$ ranges from 0.5 to almost 200 GeV and shows a significant *suppression* of the Pb+Pb spectra that is stronger in more central collisions. There is a rising behavior at low $p_T$ that peaks around 2 GeV, continuing to a decreasing trend until about 7 GeV, where it starts rising again until the statistics run out. It is worth pointing out that the extrema in these distributions correspond to the points where the behavior in the flow coefficients also changes, shown in Fig. 2.21 changed. As before, the low $p_T$ region is dominated by particles from the bulk; the rising behavior may be due to the radial flow of the medium pushing particles to higher $p_T$ on average, while the overall suppression is due to the different scaling behavior. At high $p_T$, $T_{AA}$ scaling should hold and nPDF modification could not cause such a strong effect, thus, the suppression is an effect of the energy loss of particles traversing the QGP. This phenomenon is known as *jet quenching*.

Microscopically, jets emerge from scattered asymptotically free partons; this happens at the earliest times in the collisions. The high $p_T$ parton then must move through the developing QGP fire-ball while fragmenting. The parton and its fragments are colored and interact strongly with the plasma, radiating gluons and losing energy as they move through. Single hadron spectra give a piece of the picture, but the physics of in-medium energy loss is convoluted with the fragmentation process. Thus, jet quenching can be studied more directly with fully reconstructed jets. The right panel of Fig. 2.23 shows the inclusive jet $R_{\mathrm{AA}}$. As with the single hadrons there is a clear ordering with centrality, and suppression is less for higher $p_{\mathrm{T}}$ jets. It is remarkable that even jets with $p_{\mathrm{T}}$ near 1 TeV exhibit such large modification.

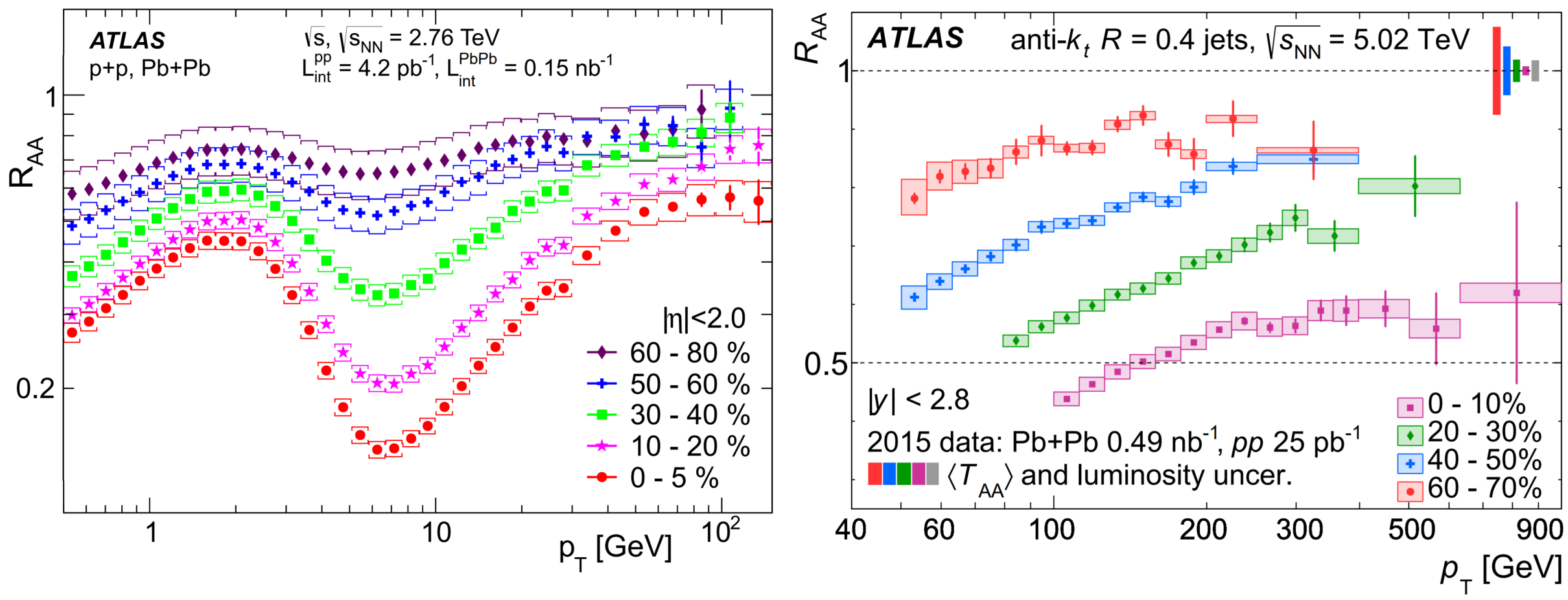

Figure 2.23: Left: Charged particle $R_{\mathrm{AA}}$ for various centralities of Pb+Pb collisions at $\sqrt{s_{\mathrm{NN}}} = 2.76$ TeV as a function of particle $p_{\mathrm{T}}$ [89]. Right: inclusive jet $R_{\mathrm{AA}}$ for various centralities of Pb+Pb collisions at $\sqrt{s_{\mathrm{NN}}} = 5.02$ TeV as a function of jet $p_{\mathrm{T}}$ [90].

At tree level in pQCD diagrams, quark and gluon final states are most often balanced by other quarks and gluons. Thus, jets tend to come in pairs balancing each other in $p_{\mathrm{T}}$ and azimuth in a *di-jet* configuration. This provides another way to study jet quenching by correlating reconstructed jets in any given event. These di-jet systems are created anywhere in the nuclear overlap region and with isotropic orientation. One would assume the amount of energy loss is a function of the distance traveled in the medium. Therefore, the level of quenching from either is random event-to-event but correlated between the jets; i.e., if the jets were created near the surface of the medium, one jet is likely to experience more quenching than the other. Fig. 2.24 shows a representation of the calorimeter energy depositions in a single Pb+Pb di-jet event. The jets are balanced in azimuth, but there is a large asymmetry in their $p_{\mathrm{T}}$.

Figure 2.25 shows the di-jet $p_{\mathrm{T}}$ balance, $x_J = p_{\mathrm{T}}^{\mathrm{jet2}}/p_{\mathrm{T}}^{\mathrm{jet1}}$, where jets 1 and 2 are the leading and sub-leading jets respectively. The most probable value in the $pp$ case is near unity showing that most di-jet systems are balanced. There is a roughly linear decrease in the yield as one moves to lower values of $x_J$. This is due to higher order multi-jet processes that produce a significant imbalance for the two leading jets. The case for central Pb+Pb collisions is very different where the most probable value is $x_J \approx 0.5$. In more peripheral collisions, the Pb+Pb values seem to smoothly converge to the *pp* values, showing that the effect is controlled by the transverse size of the QGP.

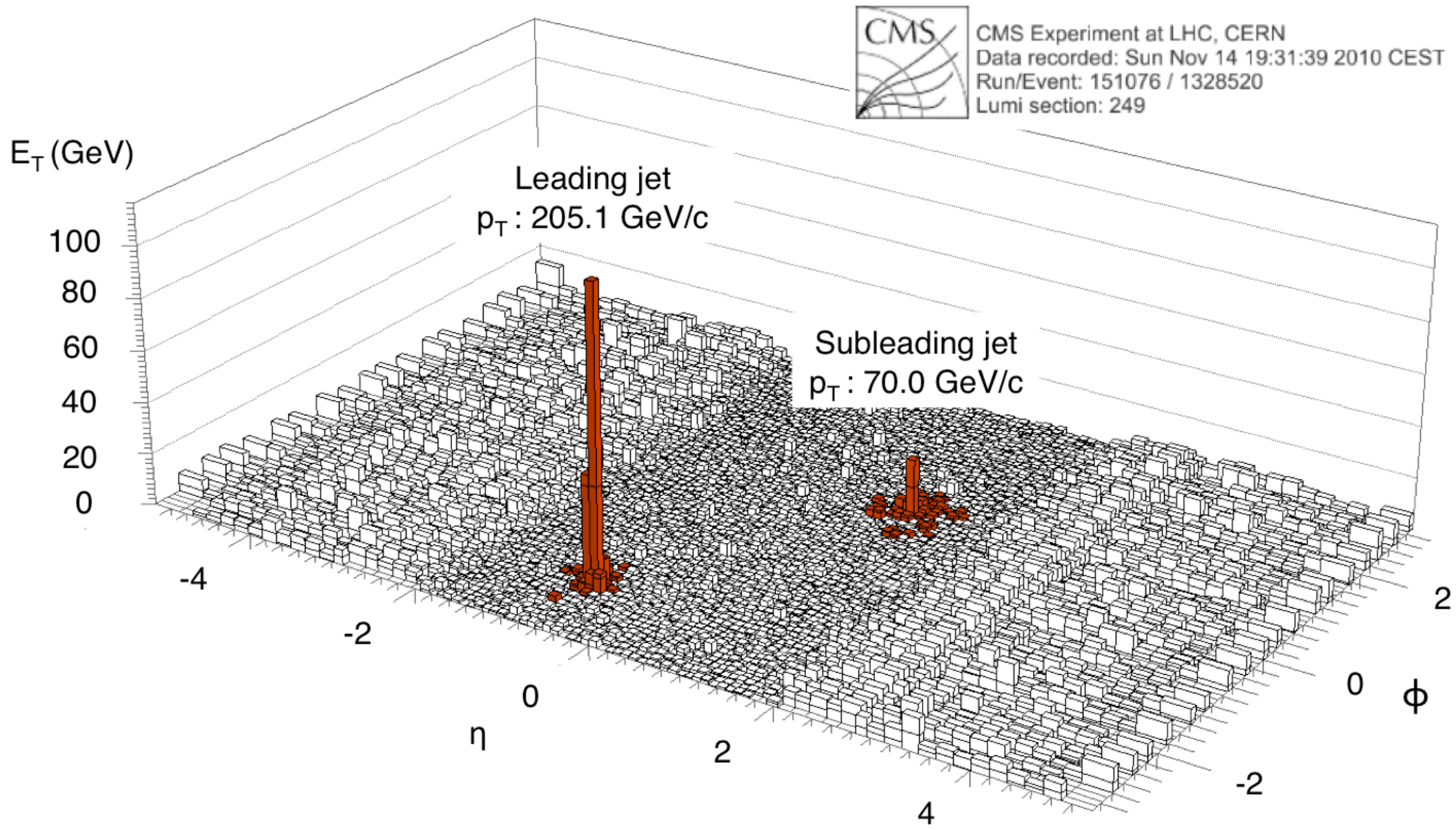

Figure 2.24: A representation of energy depositions in the CMS calorimeter system in a Pb+Pb di-jet event [91].

Returning now to the non-zero flow coefficients observed at high $p_{\mathrm{T}}$ in Pb+Pb collisions (Fig.2.21). Above 10 GeV is in the hard scattering regime, and as is seen in Fig. 2.23, hadrons are significantly affected by quenching. It has been predicted that the quenching phenomena is what is causing the flow signal [93, 94]. In this case the jet energy loss has a path-length dependence, and the shape of the QGP is elliptical at early times; thus the jet loses more energy traversing the major axis than if it was traversing the minor axis. This geometry yields a modulation in the jet energy loss that gives rise to the "flow" signal. Consistently modeling the $R_{\mathrm{AA}}$ and $v_n$ in a single framework has been a long standing challenge to the field (known as the "$R_{\mathrm{AA}} + v_2$ puzzle"). However, progress has been made, particularly by Ref. [22]. Figure 2.26 shows data of $R_{\mathrm{AA}}$, $v_2$, and $v_3$ compared to theory curves able to match the data quite well above about 10 GeV.

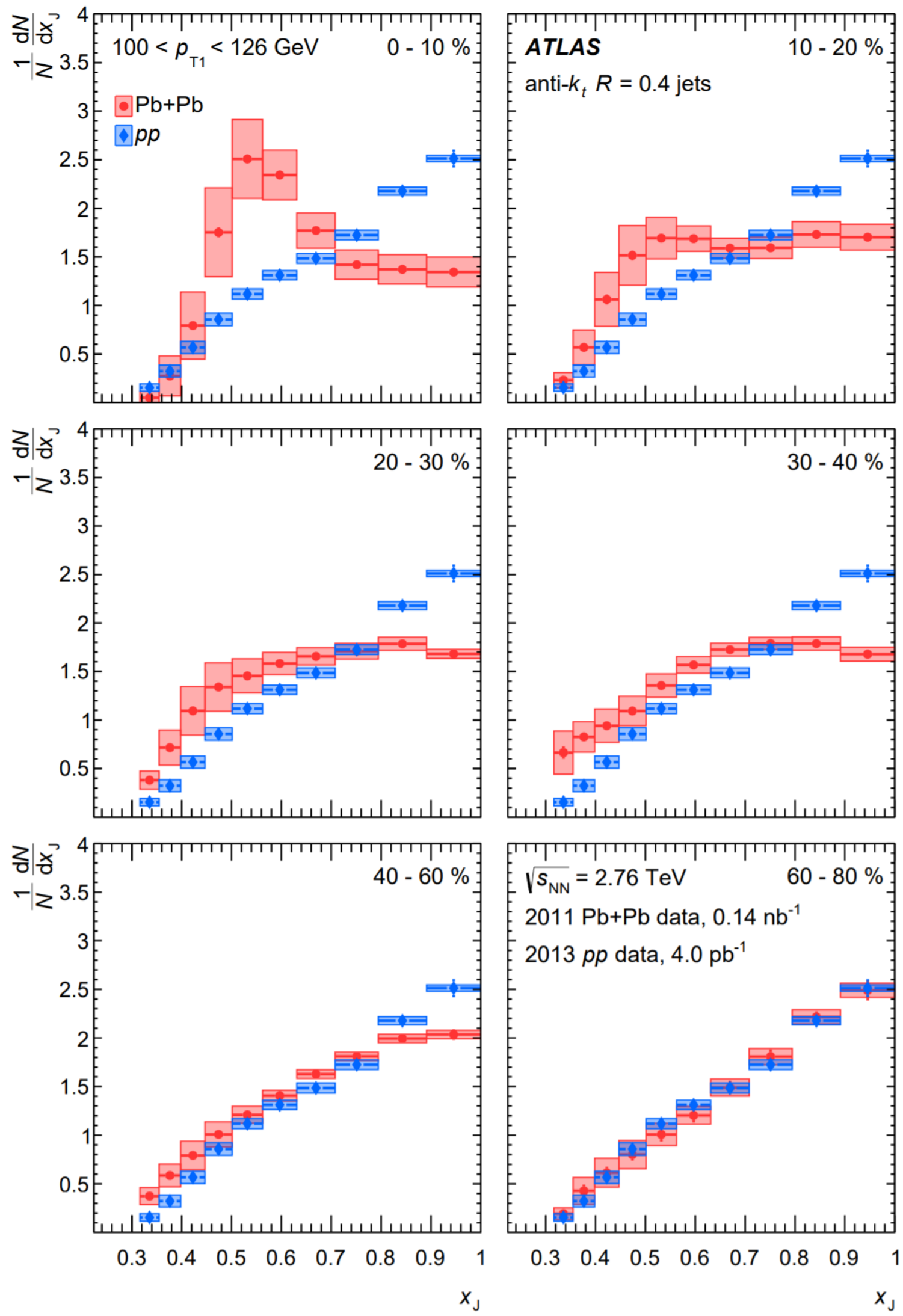

Figure 2.25: Di-jet yields from Pb+Pb collisions at $\sqrt{s_{\mathrm{NN}}} = 2.76$ TeV plotted as a function of their $p_{\mathrm{T}}$ balance, $x_J$, for several centrality classes [92].

### 2.4.5 Small Systems

Small collision systems are those in which at least one of the colliding nuclei comprises only a few nucleons or even just one. Asymmetrical systems like $d$+Au and $p$+Pb were originally proposed to study nuclear effects not related to the formation for the QGP, and today, still provide the best hadron-hadron collision data for nPDF global analyses. However, relatively recently it was discovered that these small systems exhibit signs of collective expansion similar to large systems [17]. In the years since, small systems have become a testing ground for QGP related phenomena.

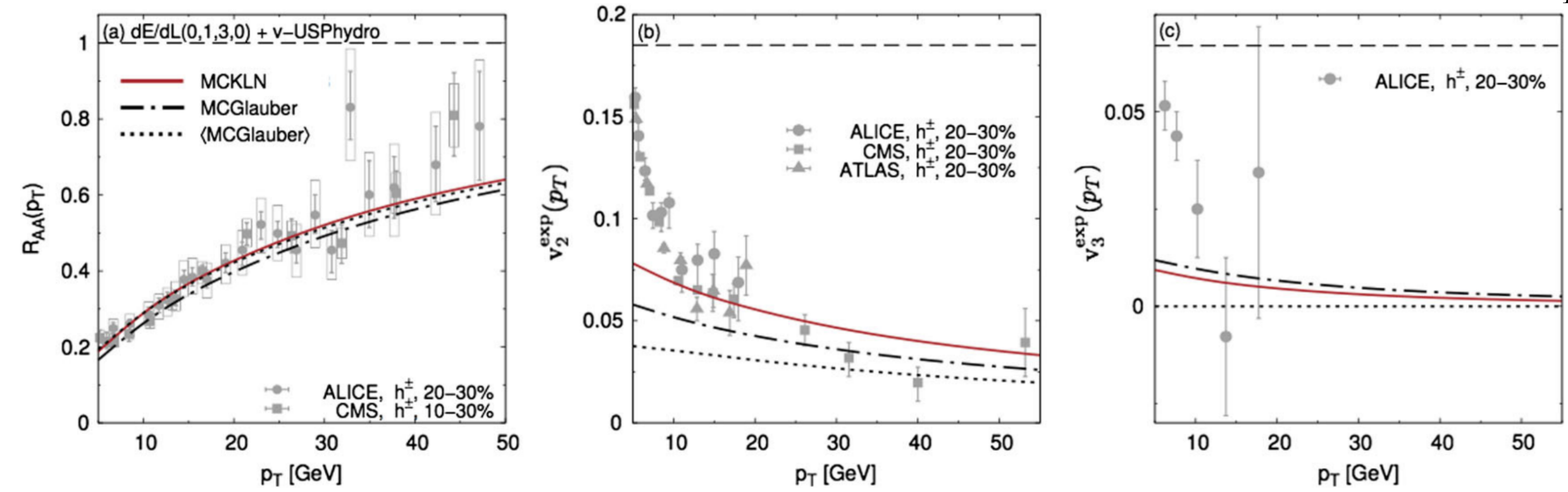

Figure 2.26: $R_{\mathrm{AA}}$ (left), $v_2$ (center), and $v_3$ (right) from a theoretical jet quenching calculation showing good agreement, at high $p_{\mathrm{T}}$, with the measured data [22].

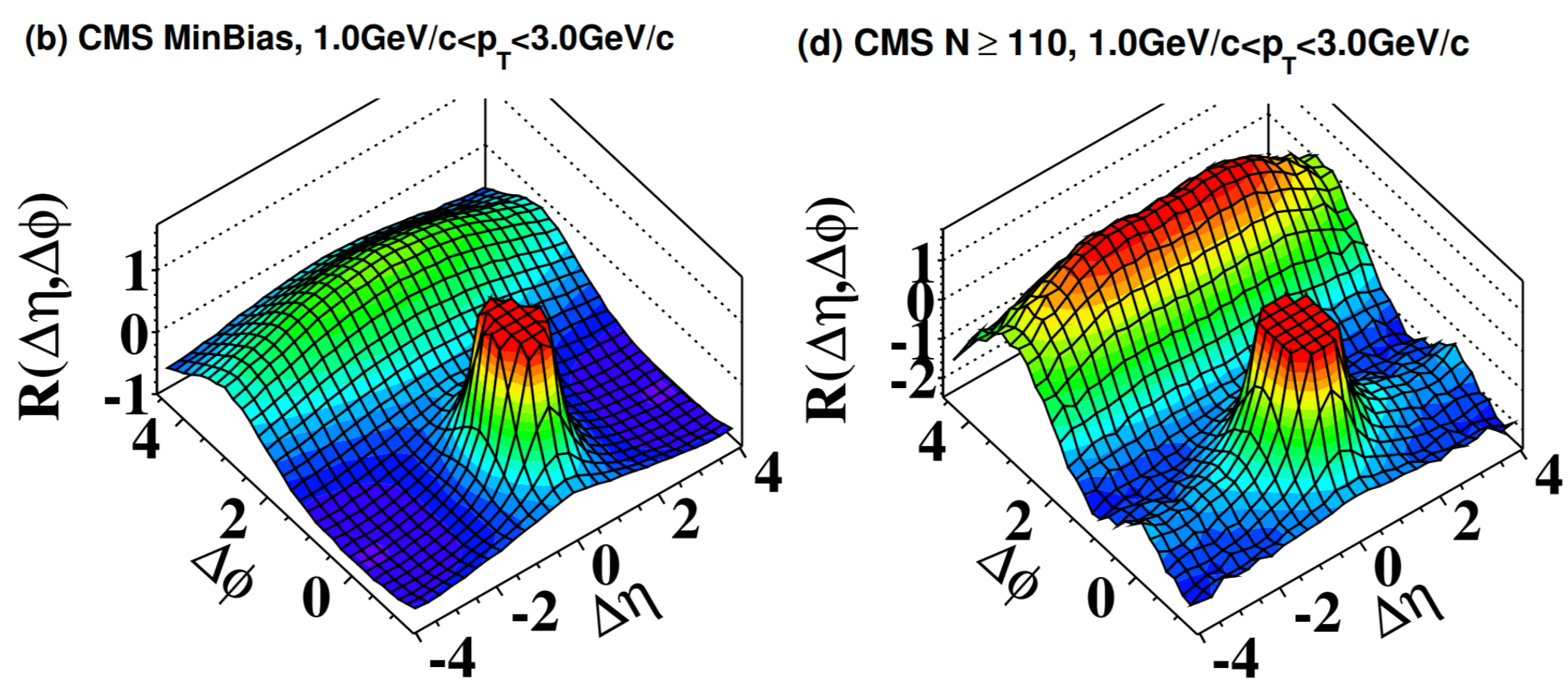

Figure 2.27: Two particle correlation functions in $\Delta\eta$ and $\Delta\phi$ from minimum bias (left) and high multiplicity (right) selected events [14].

Fig. 2.27 shows two-particle correlation functions from inclusive minimum bias (left) and high multiplicity (right) $pp$ collisions. The left plot is what one would expect, there is a strong peak at $\Delta\eta = \Delta\phi = 0$ from short range jet and resonance correlations, and there is a broad away-side ridge from momentum balancing particles (e.g. di-jet pairs). The high multiplicity plot on the right is similar except for the surprising addition of a noticeable near-side ridge extending the full range in $\Delta\eta$. This near-side ridge calls into question earlier assumptions that these small systems are too small to create the QGP. Note the difference between the right panel of Fig. 2.27 and Fig. 2.17 for the Pb+Pb case; the flow signal-to-background is much stronger in Pb+Pb than $pp$. Methods have since been developed to subtract contributions from non-flow correlations that would contaminate the extracted $v_n$. This will be discussed in detail in Ch.7.

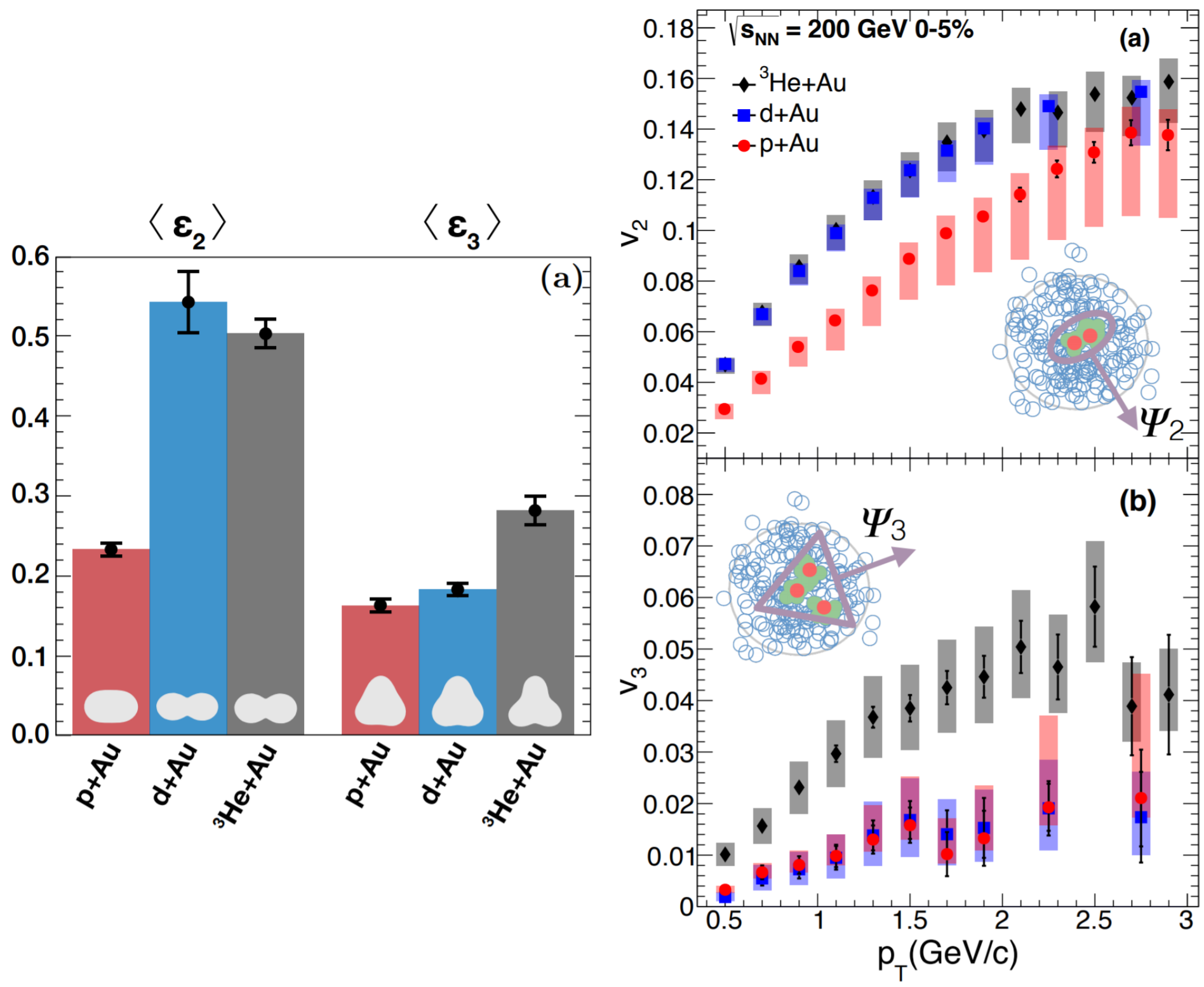

Figure 2.28: Left: average eccentricities $\varepsilon_2$ and $\varepsilon_3$ for each system in the geometry scan from MC Glauber simulations. Right: Flow coefficients $v_2$ and $v_3$ from each system plotted as a function of particle $p_{\mathrm{T}}$ [17].

Since it's initial discovery, flow signals have been systematically studied in many small systems at RHIC and the LHC. Perhaps the most striking of these comes from the so-called small systems geometry scan at RHIC to test whether the flow signal is correlated with the initial eccentricities, as with large systems. In this case, small nuclei with different spatial geometries, $p$, $d$, and $^3$He, were collided on Au target nuclei. The left panel of Fig. 2.28 shows the average ellipticity, $\varepsilon_2$, and triangularity, $\varepsilon_3$, for the three species from Glauber simulations. As one might expect $p$+Au has significantly smaller $\varepsilon_2$ than $d$+Au and $^3$He+Au, and $^3$He+Au has significantly larger $\varepsilon_3$ than the other two. The right panel of Fig. 2.28 shows the resulting $v_2$ and $v_3$ from the scans. The ordering is precisely what is expected given collective fluid expansion translating the spatial anisotropies into momentum anisotropies. Ref. [17] also shows hydrodynamic models are in good agreement with the data. Furthermore, Fig. 2.29 shows that a single hydrodynamic model can describe the measured flow signal from $pp$, $p$+Pb, and Pb+Pb without tuning model parameters between the systems, providing confidence that the flow signal observed in these small systems is of the same origin as that of large systems.

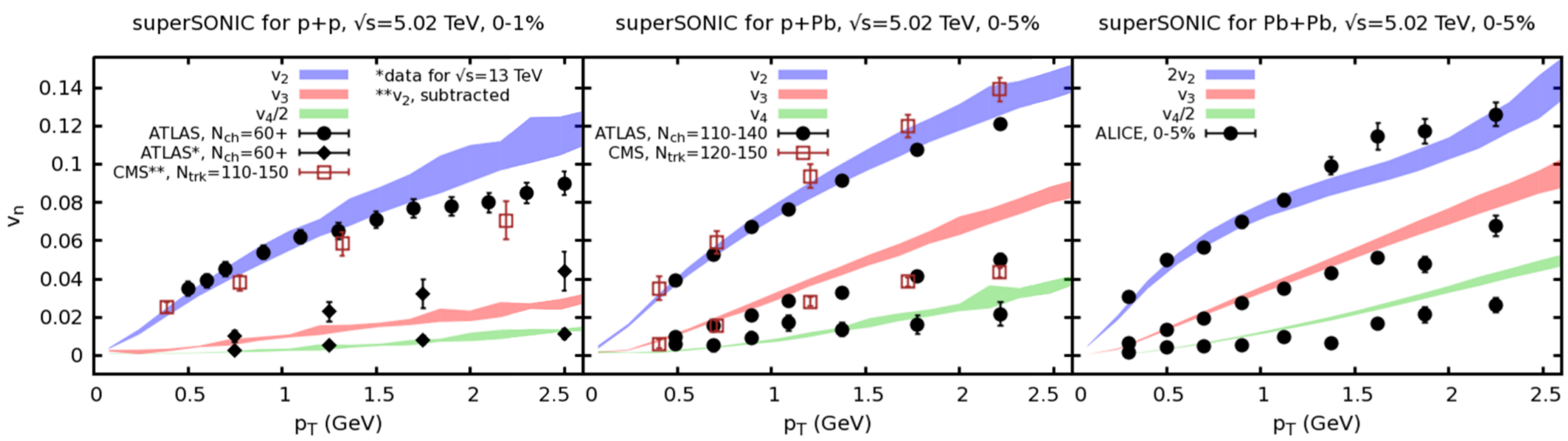

Figure 2.29: Flow coefficients, $v_n$, plotted as a function of particle $p_{\mathrm{T}}$ for $pp$ (left), $p$+Pb (center), and Pb+Pb (right). Measured data are shown as black markers and are overlaid with curves from the superSONIC hydrodynamic simulation model [71].

Fig. 2.30 shows a comparison to the charged hadron nuclear modification factor between inclusive Pb+Pb and $p$+Pb systems. The $p$+Pb results show no sign of the suppression observed in Pb+Pb. If the flow results, discussed above, really do imply the creation of a QGP droplet, one might expect it to be accompanied by energy loss. One possibility is that these small systems create too small a droplet to affect the high $p_{\mathrm{T}}$ parton; i.e. there is some path-length dependence, but the path length is too short. Using the same reasoning as was used for the Pb+Pb case, the fact that small systems like $p$+Pb do not exhibit jet quenching signals means there should be no mechanism to create a flow pattern at high $p_{\mathrm{T}}$.

The right panel of Fig. 2.30 shows the flow coefficient $v_2$ for these same systems, where the $p$+Pb values are scaled to match the Pb+Pb values at low $p_{\mathrm{T}}$. The shapes of the distributions are remarkably similar up until the $p$+Pb values run out of statistics at $p_{\mathrm{T}} \approx 10 - 12$ GeV. The data are overlaid with a result from a jet quenching calculation showing good agreement with the Pb+Pb data for $p_{\mathrm{T}} \gtrsim 10$ GeV. The fact that the $p$+Pb points agree in this range (though with significant uncertainty) raises the question of whether the same high $p_{\mathrm{T}}$ flow signal exists in a system in which there is no observed jet quenching. The results in Ch. 7 will extend the $p_{\mathrm{T}}$ range of this measurement and provide insight to the long standing $R_{\mathrm{AA}} + v_2$ puzzle.

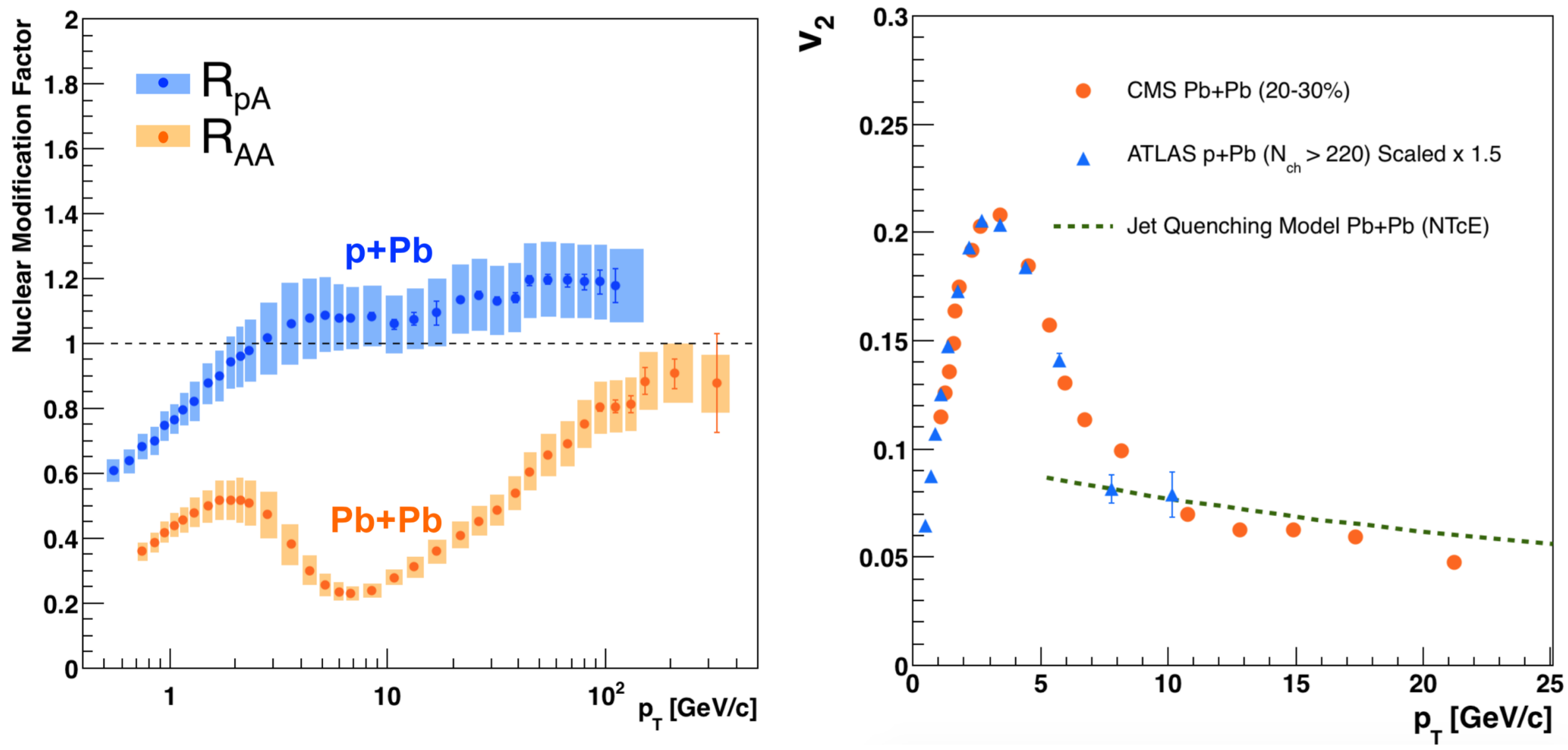

Figure 2.30: **Left:** Charged hadron nuclear modification factor $R_{\mathrm{AA}}$ from Pb+Pb collisions compared to $R_{p\mathrm{Pb}}$ as measured by the CMS collaboration. **Right:** Azimuthal anisotropy coefficient $v_2$ from Pb+Pb and $p$+Pb scaled to match at low $p_{\mathrm{T}}$ [95].

# Chapter 3

# The Experiment

It is hard to overstate how important particle accelerators have been in the development of humanity's current understanding of the universe. Every new accelerator has led to discoveries that inform us of the most basic properties of matter and how it came to be in the cosmological record. This dissertation is based on data taken at the state-of-art and highest energy accelerator facility in the world, the LHC at CERN. The experimental apparatus consists of the LHC accelerator chain that accelerates and collides hadrons, and the ATLAS detector that measures the particles emanating from the collision point. In this Chapter, an overview is given of the LHC and its injector chain, the ATLAS detector, with details provided for the most relevant aspects, and the dataset used for the analyses.

## 3.1 The Large Hadron Collider

The LHC [96] is the largest particle collider in the world. It is a 26.7 km circular synchrotron that inhabits the tunnel that was built for the, now decommissioned, Large Electron-Positron collider (LEP). The tunnel lies about 100 m below the French and Swiss countryside and has eight straight sections, and eight arc sections. Due to the limited space in the LEP tunnel, the two rings that store the two counter-rotating beams share the same twin-bore superconducting magnet system. The twin-bore design is more compact but less flexible due to magnetic coupling between the rings. There are four detector experiments located on the ring at the four interaction points (IP) where the beams are crossed and switch magnet bores. The two general purpose and high rate detectors are ATLAS at IP1 and CMS at IP5. The purpose built heavy ion physics experiment, ALICE, lives at IP2, and LHCb, the experiment focusing on physics relating to the bottom quark, is located at IP8.

The beams are steered in the eight arc sections by 1232 8.33 T dipole magnets and focused with 392 quadrupole magnets. To create such immense fields, the magnets are cabled with NbTi and held in the superconducting regime at 1.9 K with superfluid helium. The eight straight sections are used for utility insertions. Four of these sections are used for the IPs, where IP2 and IP8 also store the beam injector systems. There are two sections used for beam cleaning and collimation: one section houses the beam dumping system, and one section contains the two radio frequency (RF) systems - one for each beam. The RF cavities accelerate the beam by applying an alternating gradient potential that attracts the beam particles as they enter, and repulses the particles as they exit. Thus, the particles are clustered longitudinally in bunches with dimension defined by the frequency and the speed of the bunches (the speed of light, $c$). In the case of the LHC, the bunch length must be similar to that of the injector of 1.7 ns$\cdot c$. Therefore, the 400.8 MHz RF system (creating 2.5 ns$\cdot c$ bunches) in the LHC can accommodate the injected beam with minimal losses. The bunches are spaced by a minimum of 25 ns which allows for 3564 bunch places on the rings.

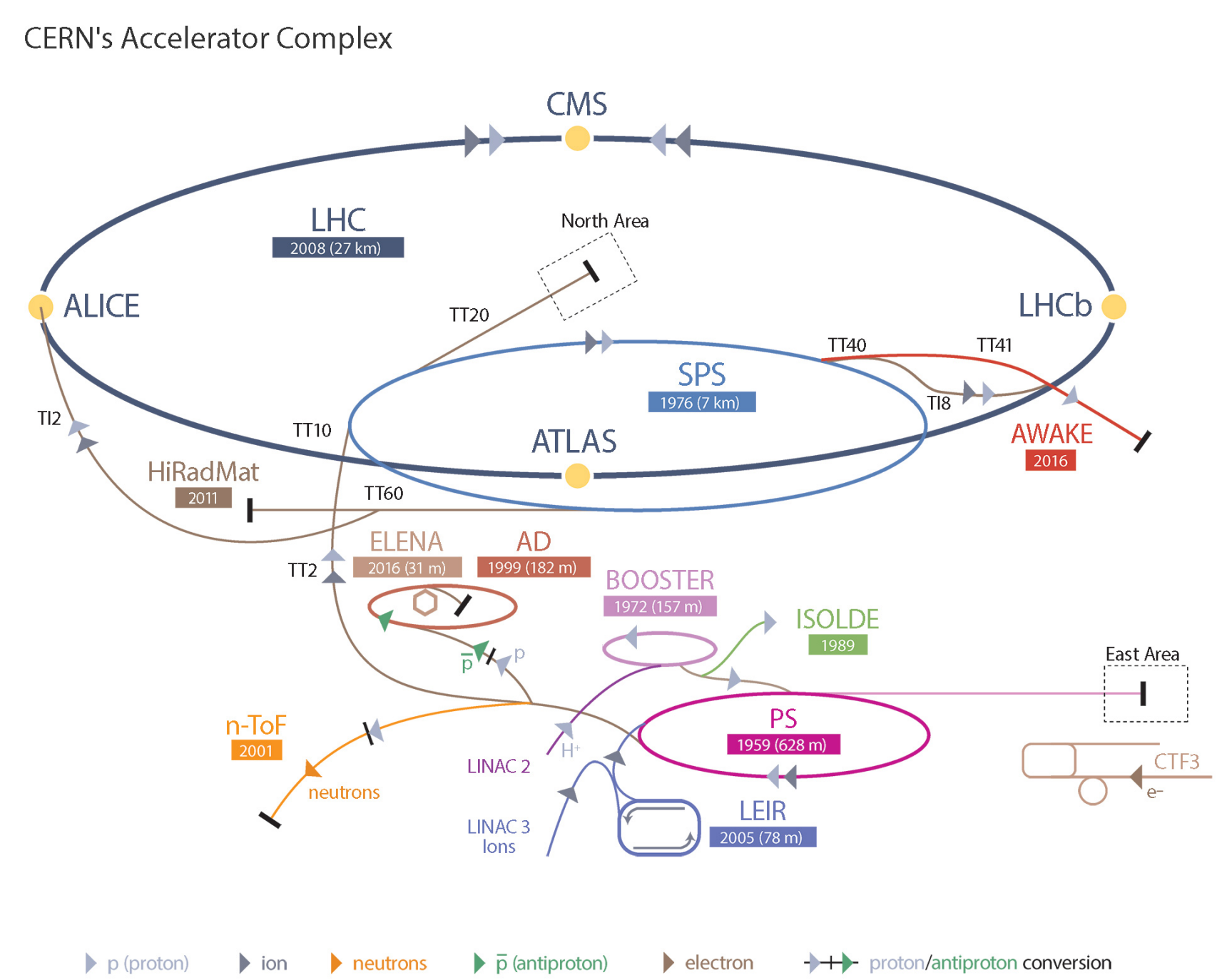

Figure 3.1: A diagram of the CERN accelerator complex, including the injection chain for the LHC [97].

The full injection chain connects some of the oldest and newest accelerator systems at CERN. A diagram can be viewed in Fig. 3.1. The protons and Pb atoms are ionized and accelerated in Linacs up to 50 MeV for protons and 4.2 MeV/n for Pb. The protons then enter the Booster at 1.4 GeV, the Proton Synchrotron (PS) at 25 GeV, the Super PS (SPS) at 450 GeV, before being injected into the LHC at 7 TeV. For Pb ions, they go from the Linac to the Low Energy Ion Ring (LEIR) at 72 MeV/n, the PS at 6 GeV/n, the SPS at 177 GeV/n, before entering the LHC at 2.76 TeV/n.[1] Each beam of the LHC is filled in 12 steps from the SPS, and each SPS fill takes 3 to 4 steps from the PS in a process that can take close to an hour. The filled bunch positions form a pattern such that bunch crossings occur at the IPs.

The principal performance metric of a collider is *luminosity* ($L$). This is the beam-beam parameter that when combined with a process cross section, gives the rate of said process as

$$\frac{dN_p}{dt} = L\sigma_p \,. \tag{3.1}$$

For Gaussian beam profiles colliding head on, the instantaneous luminosity can be expressed as

$$L = f\frac{n_1 n_2}{4\pi\sigma_x\sigma_y} \,, \tag{3.2}$$

where $n_1$ and $n_2$ are the number of particles in the colliding bunches 1 and 2, $f$ is the frequency at which they collide, and $\sigma_x$ and $\sigma_y$ are the beam widths in $x$ and $y$. The transverse trajectory of the particles in a synchrotron is approximated for each coordinate $x$ and $y$ as

$$x(s), y(s) = A_{x,y}\sqrt{\beta_{x,y}}\cos(\psi_{x,y}) \,, \tag{3.3}$$

where $\psi$ and the amplitude function $\beta$ are implicit functions of the longitudinal distance parameter $s$ [1]. In this case, the $\beta$ functions modulate the envelope of the beam in the transverse planes. Thus, for higher luminosity, the beam is squeezed so that its transverse size reaches a minimum at the IP; the $\beta$ function at this minimum is defined to be $\beta^*$ or *beta-star*. The beams are sent through a special pair of quadrupole triplet focusing magnets at each IP to lower the $\beta^*$ to as small as possible. At IP1 and IP5 beam sizes are squeezed to about a factor of 12.5 smaller. Fig. 3.2 shows a rendering of the beam optics in the ALICE detector at IP2 during the 2018 Pb+Pb running. One can see the beam envelopes (represented as colored tubes) get compressed as the beams are crossed at the interaction point.

[1] The difference in the maximum energy between the protons and ions is due to the different mass-to-charge ratio. Magnetic fields bend charged particles with the same momentum-to-charge ratio in the same arcs. Thus, the maximum momentum for the Pb ions is $Z/A = 82/208$ times the momentum of the proton in the same system.

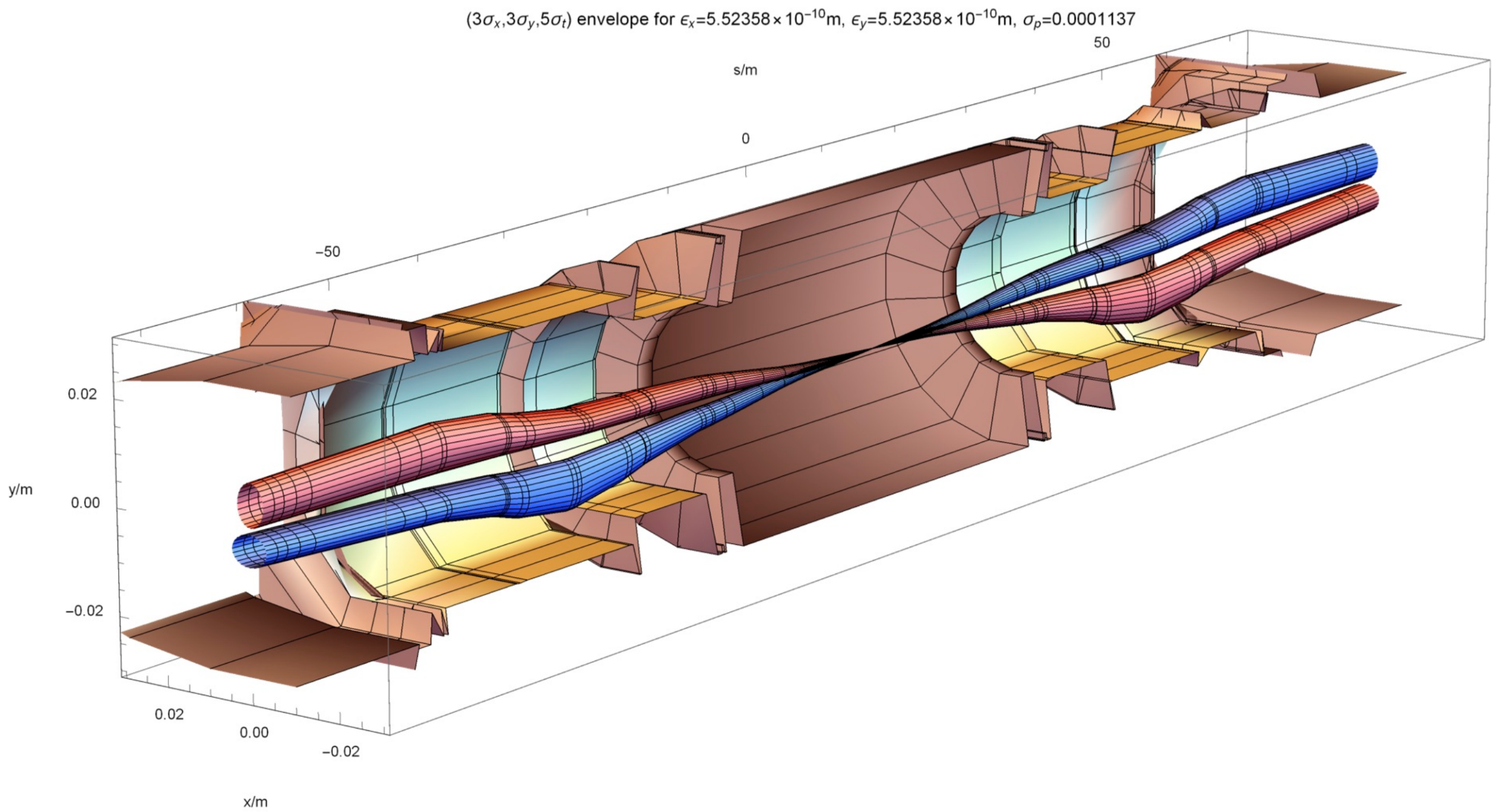

Figure 3.2: A rendering of the LHC beam optics at IP2 during the 2018 Pb+Pb running period [98].

Because the luminosity is a direct factor in any cross section measurement, it is important to have a precise measurement of it and good understanding of its uncertainty for any given running period. Beam profiles are determined with so-called Van der Meer scans at each IP. The scans start with the beams separated, then they are swept through each other separately for each transverse dimension. Luminometers measure the collision rate through this process and the spatial extent of the beams can be mapped out. A separate length scale scan is performed by offsetting a given beam by a fixed amount and using the other beam to scan over it to find its center. Different sets of very precise beam position monitors can be used to study time dependent drifts in the beam orbits that can distort the scale estimation.

The beam luminosity degrades over time due to stochastic losses from inter-beam scattering or RF noise as well as beam loss through collisions, which is the main source of loss of luminosity at the LHC. The decay time of the bunch intensity due to normal collisional losses can be calculated as

$$\tau = \frac{N_0}{L\sigma k}\,, \tag{3.4}$$

where $N_0$ is the initial beam intensity, $L$ is the initial luminosity, $\sigma$ is the total scattering cross section for the particles comprising the beam, and $k$ is the number of IPs. At the LHC for $pp$ running, given the design specified luminosity of $L = 10^{34}\ \mathrm{cm}^{-2}\mathrm{s}^{-1}$, this gives the time to reach $1/e$ of the initial luminosity as about 29 hours [96]. Adding in other corrections shortens the life time to about 15 hours. The scattering cross section is larger and inter-beam forces are stronger for heavy ion beams, and therefore, lifetimes tend to be shorter. Because of this decay behavior, most of the luminosity recorded at the experiments is from the earlier period in a given fill. Fig. 3.3 shows a representation of beam luminosity and intensity over time. One can see the stepwise increases in intensity as the beam are filled from the SPS. Then the magnetic field ramps up as the bunches are brought to full energy. The decay slope kinks at the point the beams are brought into collision. Finally the beam is dumped about 10 hours later, and the magnets are ramped down.

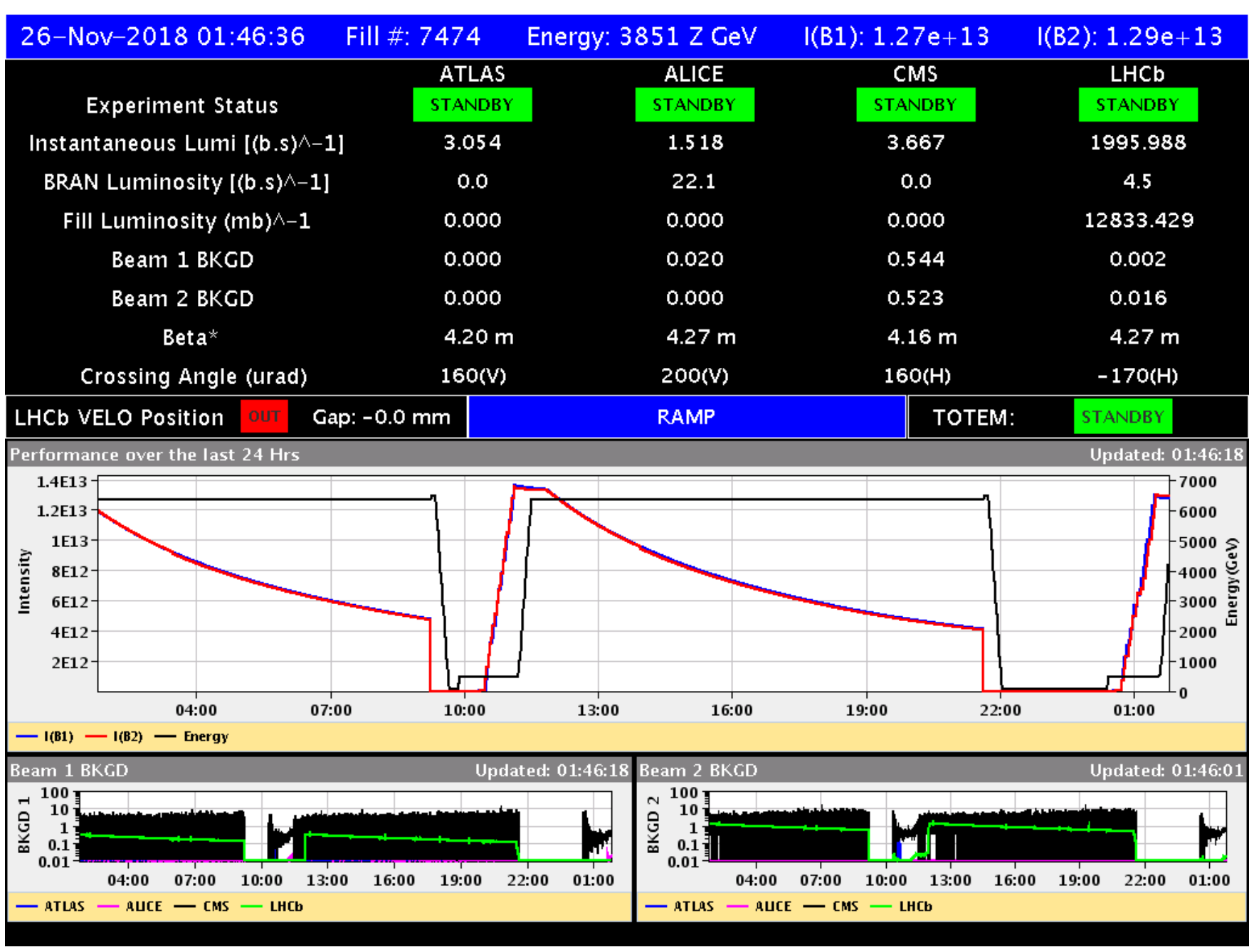

Figure 3.3: An example from the online Vistar LHC monitoring showing the beam current, dipole magnetic field, and instantaneous luminosity as a function of time spanning a couple fills.

## 3.2 The ATLAS Detector

The ATLAS experiment [99] is a large acceptance general purpose detector designed for high luminosity running at the LHC. The detector comprises many different specialized sub-systems in a cylindrical format with 25 m diameter and 44 m length. The inner detector houses layers of silicon pixel and strip tracking detectors spanning $|\eta| < 2.5$ and is surrounded by a thin superconducting solenoid creating a 2 T magnetic field. Outside of the solenoid is the high-granularity Pb-liquid-argon (LAr) electromagnetic calorimeter covering the range $|\eta| < 3.2$. The hadronic calorimeter is composed multiple technologies spanning $|\eta| < 4.9$. Surrounding the calorimeter is the three layer muon spectrometer amid a toroid magnet system. The trigger has a hardware Level-1 (L1) system capable of a rejection of about $5 \times 10^6$ and a software High Level Trigger (HLT) providing further rejection based on the full detector information. A rendering of the detector is shown in Fig. 3.4.

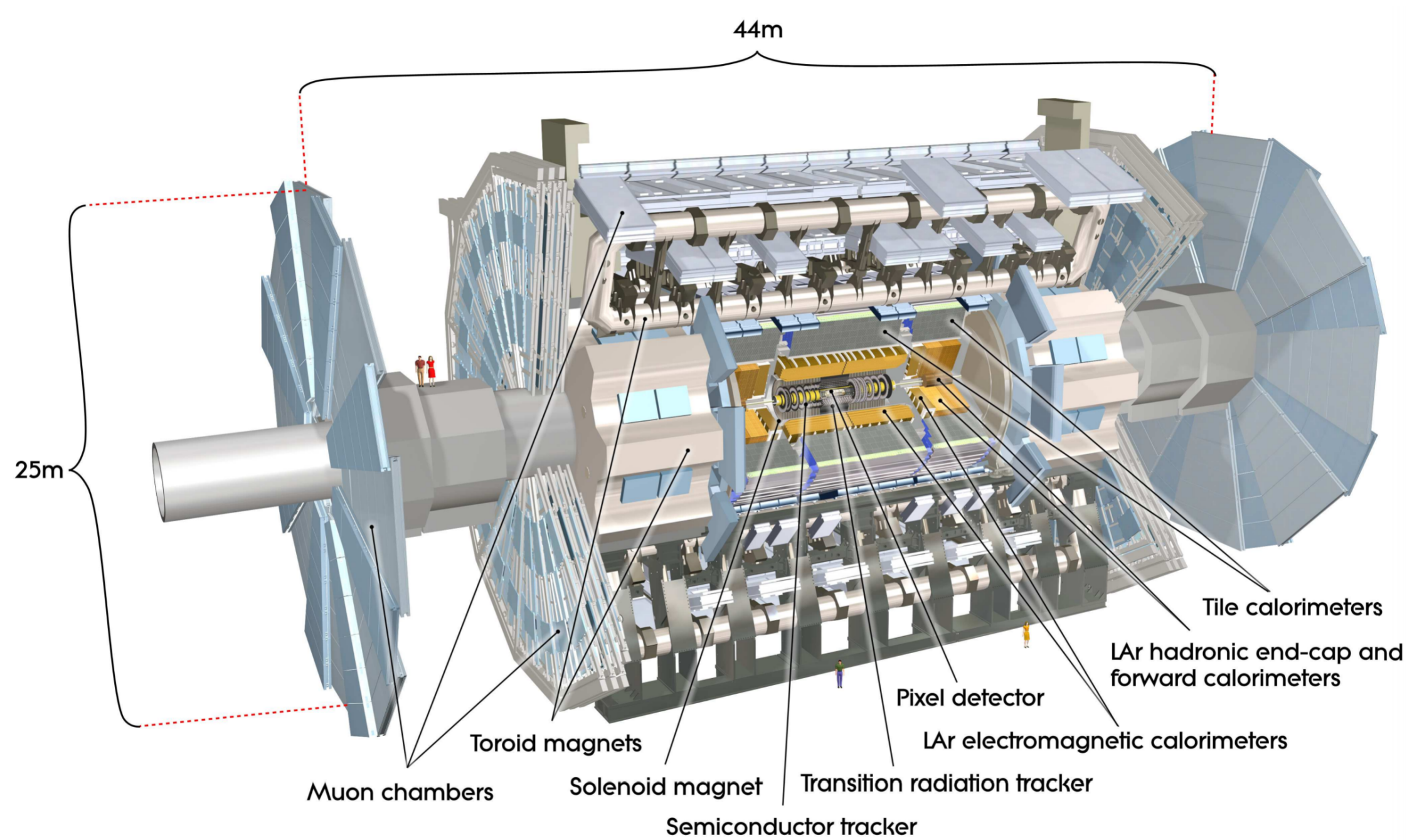

Figure 3.4: A rendering of the ATLAS detector with a cut away exposing the inner detector [99].

ATLAS uses a right-handed coordinate system with its origin at the nominal interaction point (IP) in the centre of the detector and the $z$-axis along the beam pipe. The $x$-axis points from the IP to the centre of the LHC ring, and the $y$-axis points upward. Cylindrical coordinates $(r, \phi)$ are used in the transverse plane, $\phi$ being the azimuthal angle around the $z$-axis. The pseudorapidity is defined in terms of the polar angle $\theta$ as $\eta = -\ln\tan(\theta/2)$ and the rapidity of the components of the beam, $y$, are defined in terms of their energy, $E$, and longitudinal momentum, $p_z$, as $y = 0.5\ln\frac{E+p_z}{E-p_z}$.

### 3.2.1 The Inner Detector

The inner detector is responsible for charged particle tracking and pattern recognition. It accomplishes this with three distinct detector technologies, the silicon pixels in the innermost layers, the silicon strips (SCT) in the middle layers, and the straw-tube Transition Radiation Tracker (TRT) at the outermost. This system is immersed in a 2 T solenoidal field that bends charged particles for high precision momentum measurements down to 0.5 GeV nominally. Fig. 3.5 shows a graphical representation of the inner detector.

The pixel detectors have the highest granularity and sit nearest to the beam-pipe. There are three cylindrical layers in the barrel and three perpendicular disk layers as end-caps on either side of the interaction region. The minimum pixel size in $R-\phi\times z$ is $50\times 400$ $\mu$m$^2$ and a total of 80.4 million readout channels. The insertable B-Layer (IBL) [100, 101] was installed for run 2. This additional pixel tracking layer at the innermost radius adds redundancy and precision for displaced vertex finding for the b-tagging program. In order to fit the extra layer, the beam-pipe was replaced with one of smaller diameter in this region. There are four cylindrical SCT layers in the barrel, each consisting of two perpendicular stereo strip layers. The SCT end-caps are each composed of nine disk layers of radially oriented strips and a set of stereo strips at an angle of 40 mrad. The total SCT readout has about 6.3 million channels.

The TRT covers $|\eta| < 2.0$ and is composed of hollow polyimide drift tubes with 4 mm diameter. The cathode tubes are concentric around gold plated tungsten wires forming the anode. The tubes were designed to be filled with a gas mixture of 70% Xe, 27% $CO_2$, and 3% $O_2$. Charged particles entering the tubes ionize the gas and the electrons are collected producing the signal. Additionally, low energy transition photons from super-fast electrons would be absorbed by the Xe atoms producing larger signals that would be used for particle identification (PID) purposes. However, an irreparable gas leak requires the current use of Ar instead of the much more expensive Xe; this prevents the full PID capabilities from being realized. The barrel is composed of longitudinally oriented 144 cm straw tubes in 73 planes with the wires cut in half around $\eta = 0$, and the end-cap has 37 cm straws oriented radially in wheels making up 160 planes. The TRT system has a total of 351,000 readout channels.

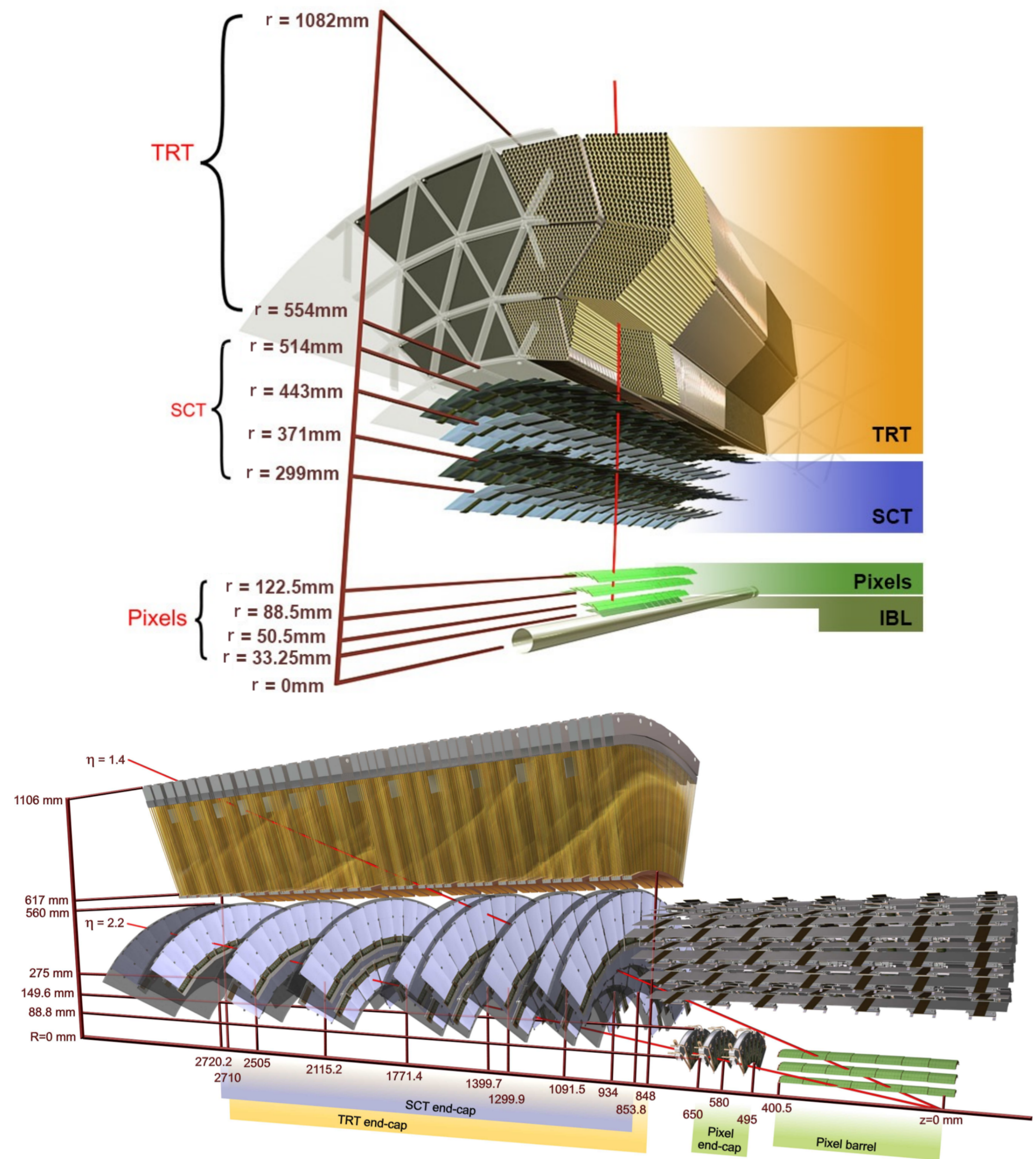

Figure 3.5: A rendering of the ATLAS inner detector barrel (top) and end-cap (bottom) sections [99].

Track finding begins by clustering charge deposition in the pixels and SCT [102, 103]. These clusters form three-dimensional space-points that are used in the track finding algorithm. Combinatorial tracking seeds are first created from sets of three space-points fit to helices. From these the impact parameter with-respect-to the center of the detector is estimated. A Kalman filter builds track candidates from estimated high purity seeds by incorporating space points from the other layers. Track candidates are then assigned a quality score dependent on how many clusters were used, how many holes were found and the fit $\chi^2$. The ambiguity solver then processes the tracks in descending order of their quality trying to disambiguate clusters shared between candidates. Clusters can be shared by no more than two tracks, and tracks with the higher score get preference. Tracks are removed form contention if they fail to meet a set of basic criteria defining a certain *working point*. The `MinBias` track selection working point, that is used in this thesis, is defined by the following quality cuts:

- $p_{\mathrm{T}} > 400$ MeV
- $|\eta_{\mathrm{trk}}| < 2.5$
- $N_{\mathrm{Pix}} \geq 1$
- $N_{\mathrm{SCT}} \geq 6$
- $\chi^2$ probability $> 0.01$ for $p_{\mathrm{T}} > 10$ GeV
- $N_{\mathrm{IBL}} + N_{\mathrm{B-Layer}} > 0$, if both IBL hit and B-layer hit are expected
- $N_{\mathrm{IBL}} + N_{\mathrm{B-Layer}} \geq 0$, if either IBL hit or B-layer hit is not expected
- $|d_0|$ with-respect-to primary vertex less than 1.5 mm
- $|z_0 \sin\theta|$ with-respect-to primary vertex less than 1.5 mm

where $d_0^{\mathrm{BL}}$ and $z_0^{\mathrm{BL}}$ are the transverse impact parameter with-respect-to the beam axis and the longitudinal distance of closest approach to the primary vertex respectively.

### 3.2.2 The Calorimeter Systems

The electromagnetic (EM) calorimeter, shown in Fig. 3.6 consists of a cylindrical barrel region covering $|\eta| < 1.475$ and two wheel end-caps covering $1.375 < |\eta| < 3.2$. The barrel and two end-caps are each kept

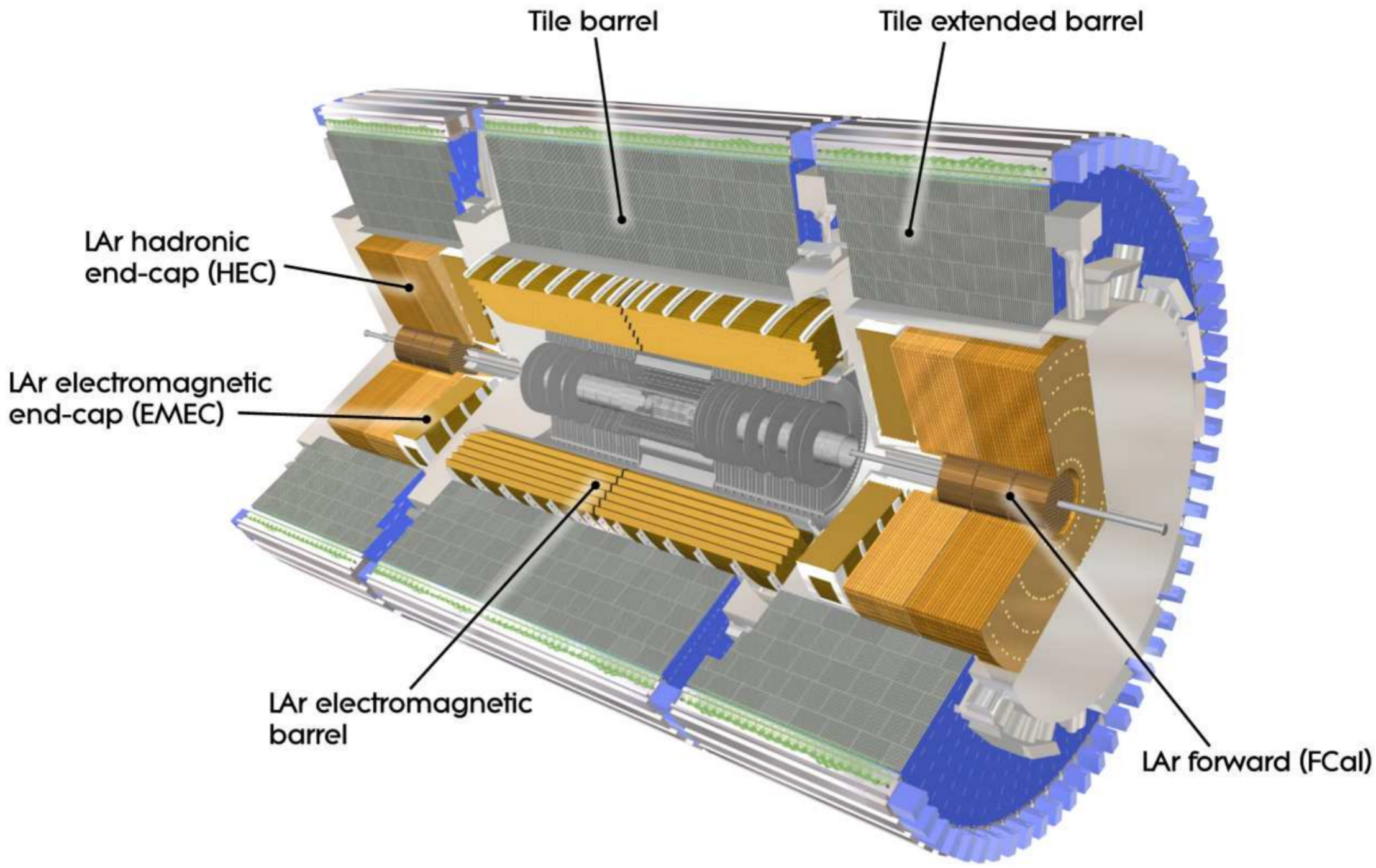

Figure 3.6: A rendering of the ATLAS calorimeter systems [99].

in independent cryostats that maintain a temperature of 80 K, sufficient to liquefy argon, with a liquid nitrogen refrigeration system. The barrel is divided into two symmetric cylinders by a small gap at $\eta = 0$, and the end-caps are divided into two concentric wheels covering $1.375 < |\eta| < 2.5$ and $2.5 < |\eta| < 3.2$ respectively. The Pb absorber has a stacked accordion geometry leaving gaps between the leaves for the LAr and high voltage electrodes. This provides full azimuthal coverage without cracks. The accordion waves are axial in the barrel and vary with radius to maintain a constant gap for the LAr. However, in the end-caps, the waves are radial and the LAr gap is widens, so the wave amplitude and folding angle vary to compensate and yield a linear response. When an electron or photon encounters the calorimeter, it interacts with the charges in the absorber material to create an electromagnetic shower, a cascade of photons pair-producing electrons and electrons radiating photons. Charged particles moving through the LAr will ionize atoms in the liquid and these ionization electrons are collected on the electrodes and readout to electronics as signal.

The EM calorimeter has three radial layers plus a presampler for most of the coverage. A summary of the coverage can be found in Table 3.1 and the cumulative material in Fig. 3.8. The first layer is highly granular in $\eta$ allowing for excellent position resolution and shape information of developing electromagnetic showers. Most of the energy is left in the second layer that has moderate $\eta$ and $\phi$ granularity. The third layer is relatively thin and captures only the tail of high energy showers. The excellent granularity and radial segmentation allows for identification of electrons and photons from calorimeter shape information. In total there are nine different shower shape quantities that enter the decision. Photon reconstruction will be discussed in more detail in Sec. 5.

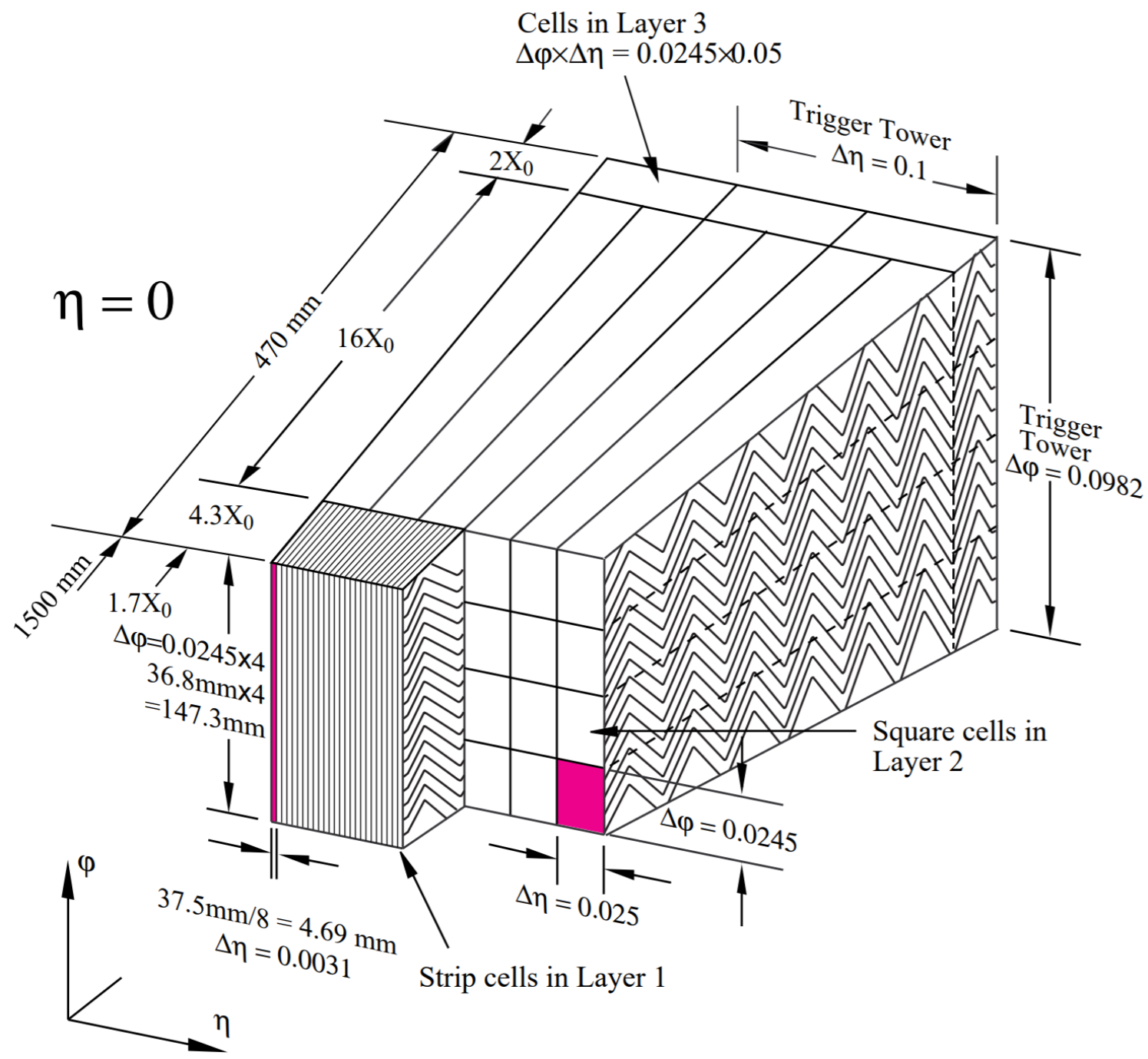

Figure 3.7: A diagram of a barrel EM calorimeter module [99].

The Hadronic calorimeter system is divided into the tile calorimeter (tile) in the barrel covering $|\eta| < 1.7$, the LAr hadronic end-cap calorimeter (HEC) covering $1.5 < |\eta| < 3.2$, and the LAr forward calorimeter (FCal) covering $3.1 < |\eta| < 4.9$. A summary of the coverage and granularity can be found in Table 3.2, and a representation of the total number of interaction lengths taken by each layer is shown in Fig. 3.9.

The tile calorimeter comprises a central barrel region covering $|\eta| < 1.0$ and two extended barrel regions covering $0.8 < |\eta| < 1.7$. The barrel has 64 wedges each spanning $\Delta\phi \approx 0.1$ made of alternating steel and plastic scintillator tiles that are oriented radially and perpendicular to the beam axis. Hadrons moving through the material interact strongly with the nuclei inside. The remnants from these collisions can then have interactions themselves, creating a cascading shower similar to the electromagnetic shower but much more irregular due to the lower cross section and broader range of outcomes for nuclear interactions than electromagnetic ones. Charged particles from these hadronic showers create scintillation light in the tiles that is collected into wavelength-shifting fibers. The fibers are ganged together to form three different readout layers in radius before being coupled to photo-multiplier tubes and read out to electronics.

Table 3.1: A summary of the EM calorimeter coverage and layer granularity.

| | Barrel | | End-cap | |
|---|---|---|---|---|
| Layer | $\eta$ Coverage | Granularity $\Delta\eta \times \Delta\phi$ | $\eta$ Coverage | Granularity $\Delta\eta \times \Delta\phi$ |
| Presampler | $\lvert\eta\rvert < 1.53$ | $0.025 \times 0.1$ | $1.5 < \lvert\eta\rvert < 1.8$ | $0.025 \times 0.1$ |
| Layer 1 | $\lvert\eta\rvert < 1.40$ | $0.00315 \times 0.1$ | $1.375 < \lvert\eta\rvert < 1.425$ | $0.05 \times 0.1$ |
| | $1.40\lvert\eta\rvert < 1.475$ | $0.025 \times 0.025$ | $1.425 < \lvert\eta\rvert < 1.5$ | $0.025 \times 0.1$ |
| | | | $1.5 < \lvert\eta\rvert < 2.5$ | $0.0031\text{-}0.025 \times 0.1$ |
| | | | $2.5 < \lvert\eta\rvert < 3.2$ | $0.1 \times 0.1$ |
| Layer 2 | $\lvert\eta\rvert < 1.40$ | $0.025 \times 0.025$ | $1.375 < \lvert\eta\rvert < 1.425$ | $0.05 \times 0.025$ |
| | $1.40 < \lvert\eta\rvert < 1.475$ | $0.075 \times 0.025$ | $1.425 < \lvert\eta\rvert < 2.5$ | $0.025 \times 0.025$ |
| | | | $2.5 < \lvert\eta\rvert < 3.2$ | $0.1 \times 0.1$ |
| Layer 3 | $\lvert\eta\rvert < 1.35$ | $0.025 \times 0.025$ | $1.5 < \lvert\eta\rvert < 2.5$ | $0.05 \times 0.025$ |

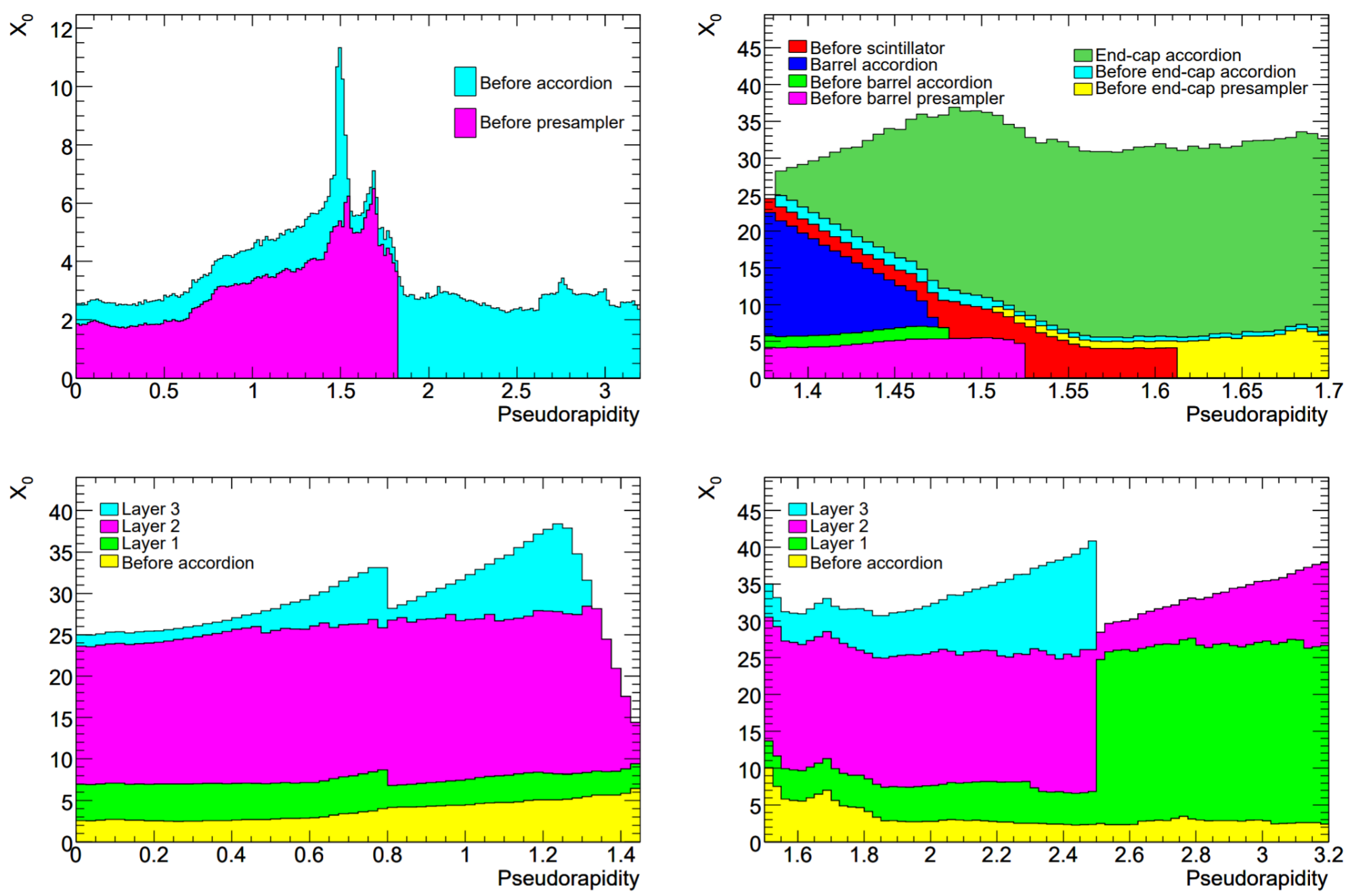

Figure 3.8: The cumulative EM calorimeter material in radiation lengths as a function of absolute pseudorapidity [99].

Table 3.2: A summary of the hadronic calorimeter coverage and layer granularity.

| Layer | Absorber | $\eta$ Coverage | Granularity $\Delta\eta \times \Delta\phi$ |
|---|---|---|---|
| **Scintillator tile calorimeter** | | | |
| Layer 1 & 2 | Steel | $\lvert\eta\rvert < 1.7$ | $0.1 \times 0.1$ |
| Layer 3 | Steel | | $0.2 \times 0.1$ |
| **LAr hadronic end-cap calorimeter** | | | |
| Layers 1 & 2 | Copper | $1.5 < \lvert\eta\rvert < 2.5$ | $0.1 \times 0.1$ |
| Layers 3 & 4 | | $2.5 < \lvert\eta\rvert < 3.2$ | $0.2 \times 0.2$ |
| **LAr forward hadronic calorimeter** | | | |
| Layer 1 | Copper | $3.1 < \lvert\eta\rvert < 4.9$ | approximately $0.2 \times 0.2$ |
| Layer 2 & 3 | Tungsten | | |

Similar to the EM calorimeter, the HEC also uses LAr to collect ionization electrons and shares its cryostats with the EM end-caps and FCals. Each side of the HEC has a front wheel and a rear wheel with each wheel containing two longitudinal layers. The wheels are made of 32 wedge modules, and each module is made of 24 copper absorber plates 25 mm thick. The rear wheels instead have 16 plates at 50 mm each. The gaps for the LAr and high voltage electrodes are each 8.5 mm thick, and thus, the sampling fraction is less for the rear wheels. The gaps contain four electrodes that effectively divide each gap into four drift zones. The configuration of the readout cells defines an approximately projective geometry back to the center of the detector.

Finally, the FCal has one electromagnetic layer and two hadronic layers, each 45 cm thick. Again, the active layers of these are LAr with the absorber for the EM layer being copper and that of the other two layers being tungsten. Additionally, a copper shielding plug covers the outside of the third layer to reduce punch through into the muon system beyond. Similar to the HEC, the EM FCal layer uses a stacked plate design. However, instead of gaps between the plates, each absorber plate has thousands of holes that house the electrodes. The electrodes are copper rods separated from copper tubes by plastic fiber. The LAr permeates the gaps between the rods and tubes. The hadronic layers form a similar structure except using the denser tungsten instead of copper absorber. Additionally, the space between the electrode tubes is filled with tungsten slugs, in order to maximise the absorbing material and interaction lengths.

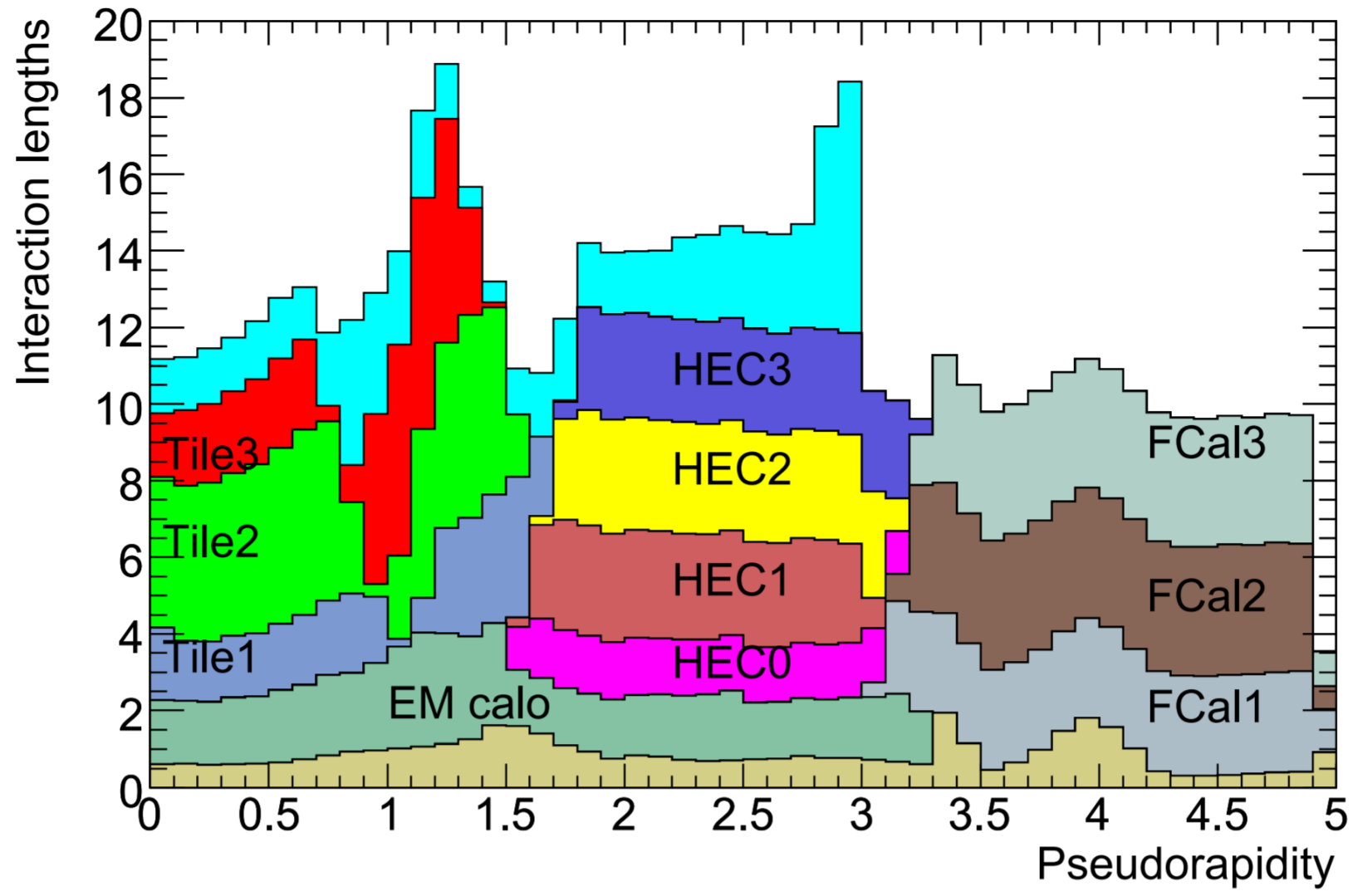

Figure 3.9: The cumulative hadronic calorimeter material in nuclear interaction lengths as a function of absolute pseudorapidity [99].

### 3.2.3 The Minimum Bias Trigger Scintillators

The Minimum Bias Trigger Scintillators (MBTS) provides a low bias and fast readout signal necessary to trigger on MB events in low luminosity running. The detectors are two 2 cm thick polystyrene scintillator disks mounted 3.6 m to either side of the interaction region. The disks are composed of eight sectors where each sector has two sections in radius. A representation of one side can be seen in Fig. 3.10. Wavelength shifting fibers are embedded into each sector extending radially. The outer sections cover $2.08 < |\eta| < 2.78$, and the inner section covers $2.78 < |\eta| < 3.75$. The fibers are coupled to PMTs that read out to the same electronics used in the tile calorimeter.

### 3.2.4 The Trigger

The LHC is capable of colliding bunches at 40 MHz, and each event is on the order of 1 MB. If ATLAS wanted to record every event, the data acquisition system (DAQ) would need to write TBs of data every second. This is unfeasible, so it is necessary for a triggering system to select events to write to disk. The trigger is divided into two steps: a fast but coarse hardware based Level-1 (L1), and a slower but more precise software based High Level Trigger (HLT).

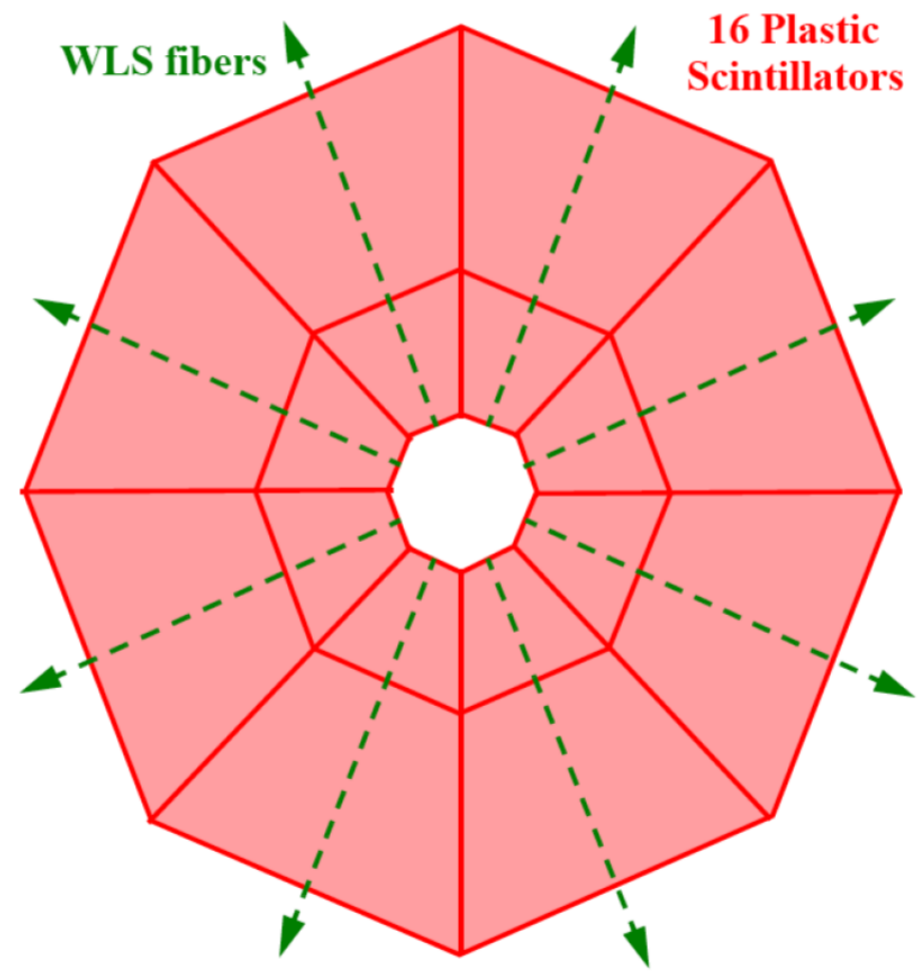

Figure 3.10: A representation of one side of the Minimum Bias Trigger Scintillators [104].

The L1 system is capable of processing the full 40 MHz bunch crossing rate and paring it down to about a 100 kHz accept rate that gets passed off to the HLT. Simply due to the size of the detector, the time of flight of the particles on which L1 decisions are based is longer than the bunch crossing interval. Thus, detector signals must be stored in pipeline buffers while the decision is being made. It is of interest in terms of cost and reliability to minimize this buffer storage, so the L1 is designed to make decisions with less than 2.5 $\mu$s latency. This is accomplished using custom electronic processors. The decisions are based on coarsened information from the calorimeters and muon detectors looking for high $p_{\mathrm{T}}$ muons, EM calorimeter clusters, jets, $\tau$-leptons, large missing transverse energy, and total transverse energy. The Central Trigger Processor (CTP) takes in information from all the L1 object types and is responsible for making the final decision. In total, 256 accept modes are possible, being combinations of, e.g., flags specifying thresholds that have been met on the L1 objects. Geometric information from flagged L1 objects is then passed to the HLT as Regions-of-Interest (ROI) to seed the HLT trigger decision. A diagram of the L1 flow is plotted in Fig. 3.11.

The HLT and DAQ systems interface the detector readout electronics and L1 decisions with the CERN Tier-0 mass storage and computing resources. Limitations and requirements are based not just on the capabilities of the DAQ, but on the ability of the computer farms at CERN to store and process the volume of data that is output. The HLT is divided historically into two sections, the Level-2 (L2) and Event Filter (EV), though both are implemented in software on servers on site. The L2 takes detector information around ROIs from the L1 and reconstructs these limited regions in detail. In this way L2 decision can be made using a small fraction (∼5%) of the total HLT bandwidth. After the L2 decision, the event is passed to the EV where events are reconstructed with offline algorithms and filtered and classified into distinct *streams* to be written to disk. The final output bandwidth of the HLT streams tend to be around 1 GB/s.

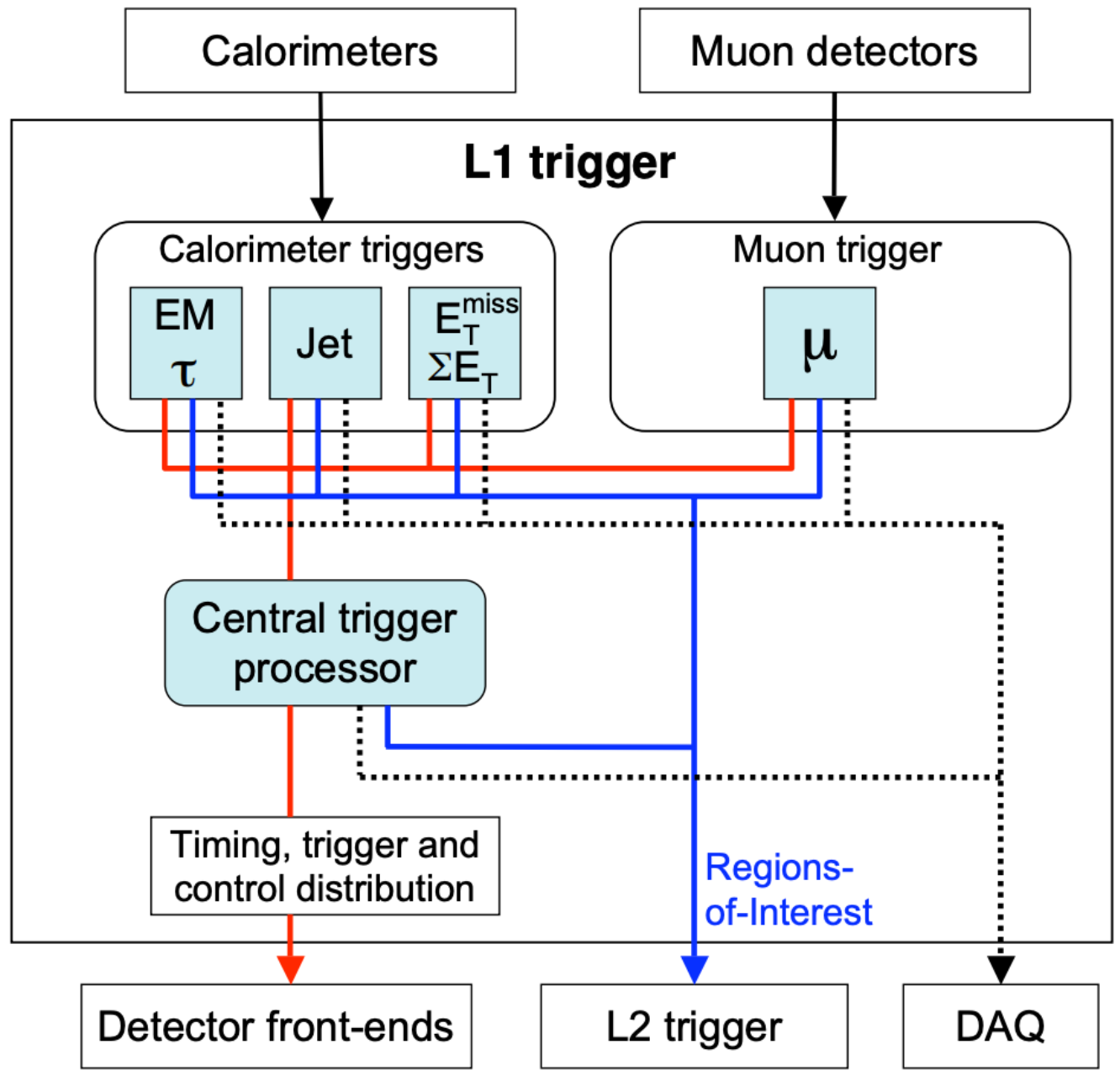

Figure 3.11: A diagram of the ATLAS L1 trigger flow [99].

## 3.3 The Dataset

This dissertation is based on data recorded from the $p$+Pb $\sqrt{s_{\mathrm{NN}}} = 8.16$ TeV collisions delivered by the LHC to ATLAS in November and December 2016. In these collisions, the proton has $E = 6.5$ TeV, while the Pb nucleus has $E = 6.5$ TeV per unit charge, resulting in an energy per nucleon of $(Z/A) \times 6.5$ TeV$=$ $82/208 \times 6.5$ TeV$=2.56$ TeV. The resulting nucleon–nucleon collision system has $\sqrt{s_{\mathrm{NN}}} = 8.16$ TeV and the center of mass of this system has a rapidity shift with respect to the ATLAS laboratory system of $\Delta y = \pm 0.465$ (where the sign is dependent on the direction of the proton). The data comprises two distinct beam orientations: (1) in the first period, called the "$p$+Pb" orientation, the Pb beam is moving from negative to positive rapidity, and the center of mass is boosted by $\Delta y = -0.465$, and (2) the reverse orientation, with the Pb beam moving from positive to negative rapidity.

The dataset is divided into *runs* of continuous DAQ operation that usually correspond to whole fills of the LHC. The runs are then broken into small chunks called *lumi-blocks*. Analyzed events are required to pass a set of *good run* criteria based on the performance of the detector subsystems and the LHC; for this dataset, there are a total of 30 good runs containing 165 nb$^{-1}$± 2.4% of integrated luminosity. The run numbers and luminosities for each run period are:

- $p$+Pb orientation: 11 runs, run numbers 313063-313435, $L_{\text{int}} = 56.76$ nb$^{-1}$
- Pb+$p$ orientation: 19 runs, run numbers 313572-314170, $L_{\text{int}} = 107.80$ nb$^{-1}$

For consistency when presenting physics results, the proton-going direction defines the positive rapidity direction.

One important detail of the detector operation during data-taking is that one quadrant of the HEC on the A side was disabled. The affected region is approximately $+1.56 < \eta^{\text{lab}} < 2.37$, $-\pi < \phi < -\pi/2$ (25% of the total A-side HEC acceptance). Specific analysis choices were made to either eliminate this problem or to analyse any bias this might have caused.

# Chapter 4

## Centrality Determination

A central concept in heavy ion collisions, centrality is a measure of an event's overall activity or soft particle production. With the use a model such as Glauber, the event activity is translated to geometric quantities characterizing the collision, e.g. the impact parameter, $N_{\mathrm{part}}$, etc.. Thus, centrality provides not only a handle on the scaling of hard and soft particle production mechanisms, but also the shape and geometry of the collision region which can be correlated to the final state particle distributions.

In ATLAS, the total transverse energy deposited into the forward calorimeter sections (FCal$_{\mathrm{A}}$ : $3.1 < \eta < 4.9$ and FCal$_{\mathrm{C}}$ : $-4.9 < \eta < -3.1$) is used to sample the event activity. In symmetric collision systems, like Pb+Pb, the sum of the transverse energy in both FCals is used. However, for $p$+Pb, the intrinsic asymmetry of the projectiles as well as the boost of the center of mass frame relative to the detector lead to an asymmetry in the lab pseudorapidity distribution of soft particles. Figure 4.1 shows the correlation between total transverse energy deposited in the $p$-going FCal ($\Sigma E_{\mathrm{T}}^{p}$) versus that in the Pb-going FCal ($\Sigma E_{\mathrm{T}}^{\mathrm{Pb}}$). At low total energy in both FCals, the correlation is roughly one-to-one, but as the energy in the Pb-going direction increases, the response in the $p$-going direction seems to saturate. This behavior indicates a lack of sensitivity of the $p$-going direction to the overall soft particle production, and therefore, this analysis uses solely $\Sigma E_{\mathrm{T}}^{\mathrm{Pb}}$.

The $\Sigma E_{\mathrm{T}}^{\mathrm{Pb}}$ distribution is connected with MC Glauber models to correlate the event activity to geometric properties of the collision. As discussed in Sec. 2.4.1, MC Glauber simulates a nucleus-nucleus collision as a collection of nucleon-nucleon collisions. The event activity distribution is assumed to be parameterized with a scaling in $N_{\mathrm{part}}$ (in $p$+Pb, $N_{\mathrm{coll}} = N_{\mathrm{part}} - 1$). With this parameterization, the $N_{\mathrm{part}}$ distribution is fit to the $\Sigma E_{\mathrm{T}}^{\mathrm{Pb}}$ distribution as an $N_{\mathrm{part}}$ convolution of the parametric function.

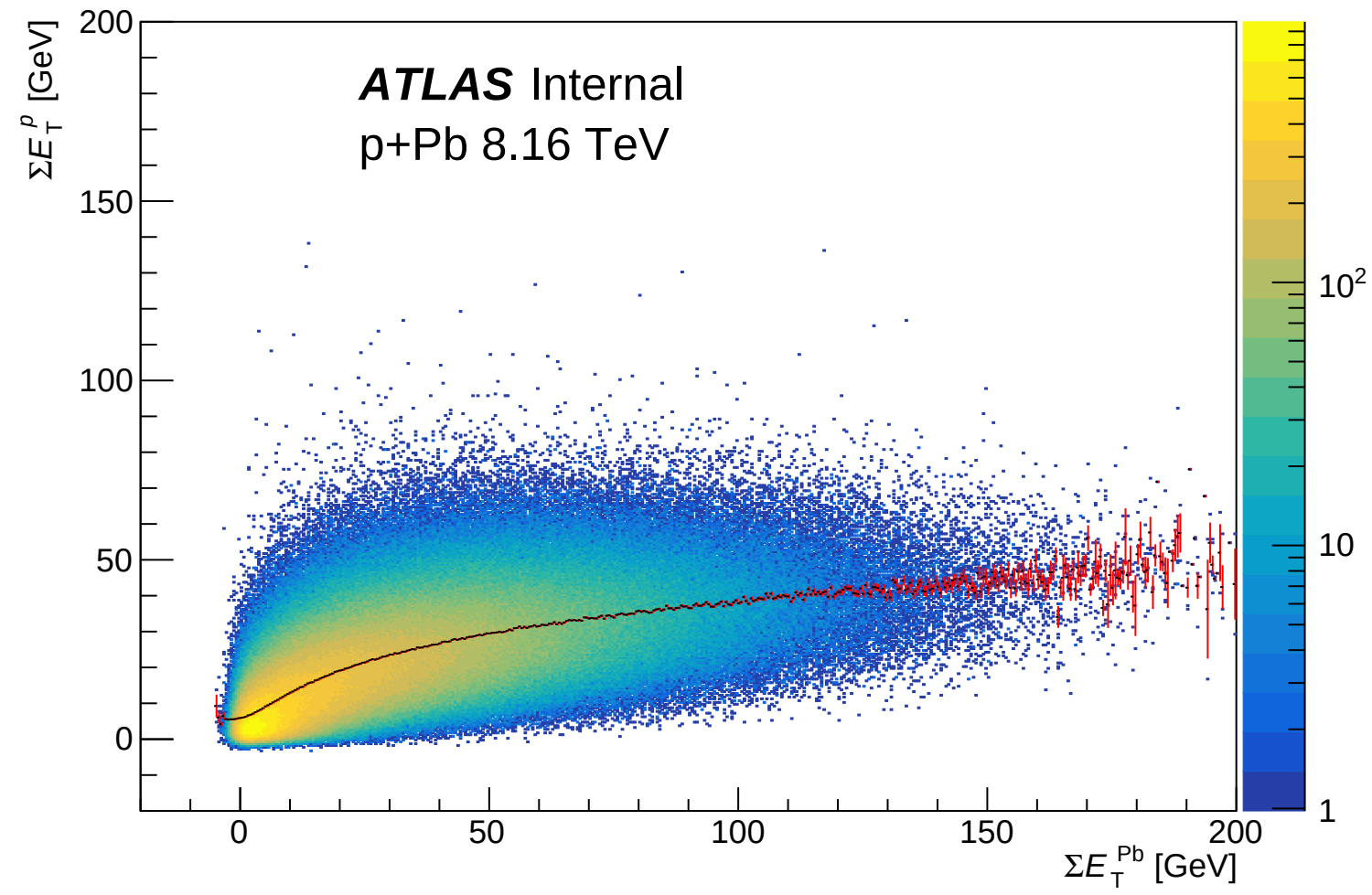

Figure 4.1: Correlation of the sum of the energy deposited in the $p$-going (y-axis) versus Pb-going (x-axis) forward calorimeters. The red histogram shows the average of the $p$-going energy for each Pb-going energy.

## 4.1 Datasets

The analysis is based on the $p$+Pb data described in Sec. 3.3. Simulated events (MC data) are used to study the detector response from both inelastic and elastic events.

### 4.1.1 Monte Carlo Data

Two sets of minimum bias MC data were used in this analysis. MC event generators, Pythia [43] and Hijing [105] were used to create simulated data. As mentioned in Sec. 2.3.2, the soft particle production in Pythia is based on the Lund string fragmentation model. Hard processes are leading order perturbative calculations. Until very recently, Pythia could not simulate heavy ion collisions, and therefore, there was a need for an extension. The most common extension, Hijing uses a version of Pythia for its nucleon-nucleon collision sub routines and adds additional modeling of color strings to match measured bulk particle production in large collision systems.

Both generators were used at $\sqrt{s_{\mathrm{NN}}} = 8.16$ TeV with the correct center of mass boost with respect to the lab frame to reflect the real data. 100 k Pythia $pp$ events each were generated using non-diffractive, single-diffractive, and double-diffractive processes, and 1 M Hijing $p$+Pb events were generated in the same configuration but 5 different $z$-vertex positions. Diffractive events are characterized by the exchange of color neutral objects, and thus, one or both of the colliding particles remain intact; These Events account for a large fraction of the total scattering cross section and must be studied as part of the total energy production. The ATLAS detector response to the generated events was determined through a full GEANT4 simulation [106, 107], and the simulated events were reconstructed in the same way as the data.

## 4.2 Event Selection

The data for this analysis was selected using the minimum bias trigger, `HLT_mb_sptrk_L1MBTS_1`, that required a hit in one of the MBTS counters and one space point in the tracker as well a one HLT track. This centrality analysis relies on the Glauber geometric model which does not naturally take into account diffractive or photo-nuclear events. Therefore, it is beneficial to remove them. Photo-nuclear events are such that the boosted electric fields emanating from a colliding particle strike the other as a nearly non-virtual photon. They are characterized by the disassociation of one of the projectiles and not the other. Because of the significant charge enhancement of the Pb nucleus relative to the proton, the photon is much more likely to emanate from the Pb nucleus to strike the proton leaving the nucleus intact. To reject these events, the L1 ZDC trigger bit in the Pb going direction is required to have been set, indicating the breakup of the nucleus.

Diffractive events are associated with significant pseudorapidity gaps in the particle production. Figures 4.2 and 4.3 show the gap distributions measured from the forward edge of the Pb-going and $p$-going FCal, using truth particles and clusters above 200 MeV from the entire calorimeter system, from non-diffractive, single-diffractive, and double-diffractive PYTHIA data. The separation between non-diffractive and diffractive distributions has a larger magnitude and occurs at a smaller gap size, $\Delta\eta_{\mathrm{Gap}}$, in the Pb-going direction. In this analysis, we reject events with $\Delta\eta_{\mathrm{Gap}}^{\mathrm{Pb}} > 1.8$.

In-time pileup is when more than one collision occur in a given bunch crossing. To minimize the contamination of these events, a reconstructed vertex based rejection is implemented. At least one vertex is required in the event and any additional vertices are required to have six or fewer tracks associated with it. Motivating this cut, Fig. 4.4 shows a histogram of the number of tracks associated with the secondary vertex, in events with more that one vertex, from both data and HIJING MC with no simulated pileup. As a check, Fig. 4.5 shows the distribution of vertex $z_1 - z_2$ in events with more than one vertex (black). The left plot selects events in which the second vertex has greater than 6 tracks, to choose pileup events, (green), and it agrees well with simulated pileup events generated by mixing HIJING events (red). Conversely, the right plot selects events passing the pileup rejection (green) compared to the distribution from no pileup HIJING events (red) showing, again, good agreement.

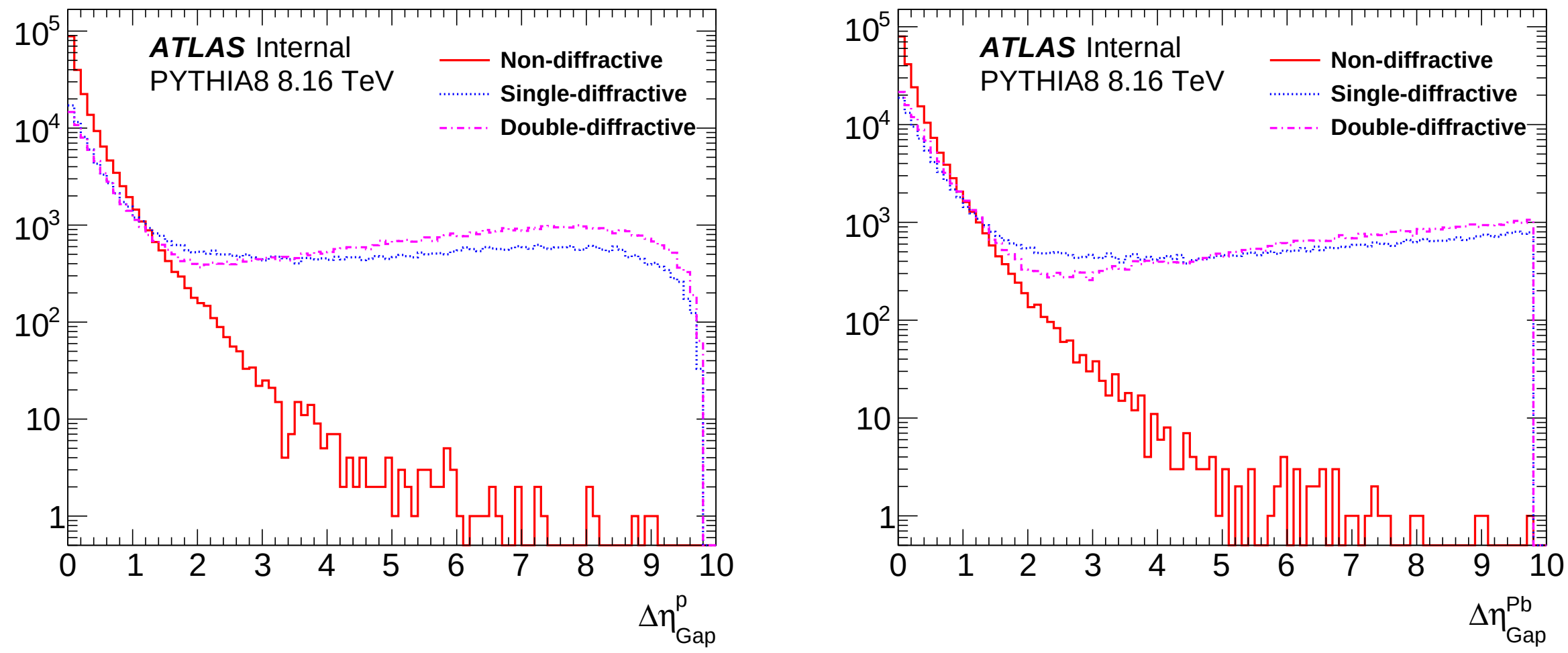

Figure 4.2: Edge gaps measured from the forward edge of the calorimeter system ($\eta = 4.9$) on the $p$-going (**Left**) and Pb-going (**Right**) sides of truth particles above 200 MeV.

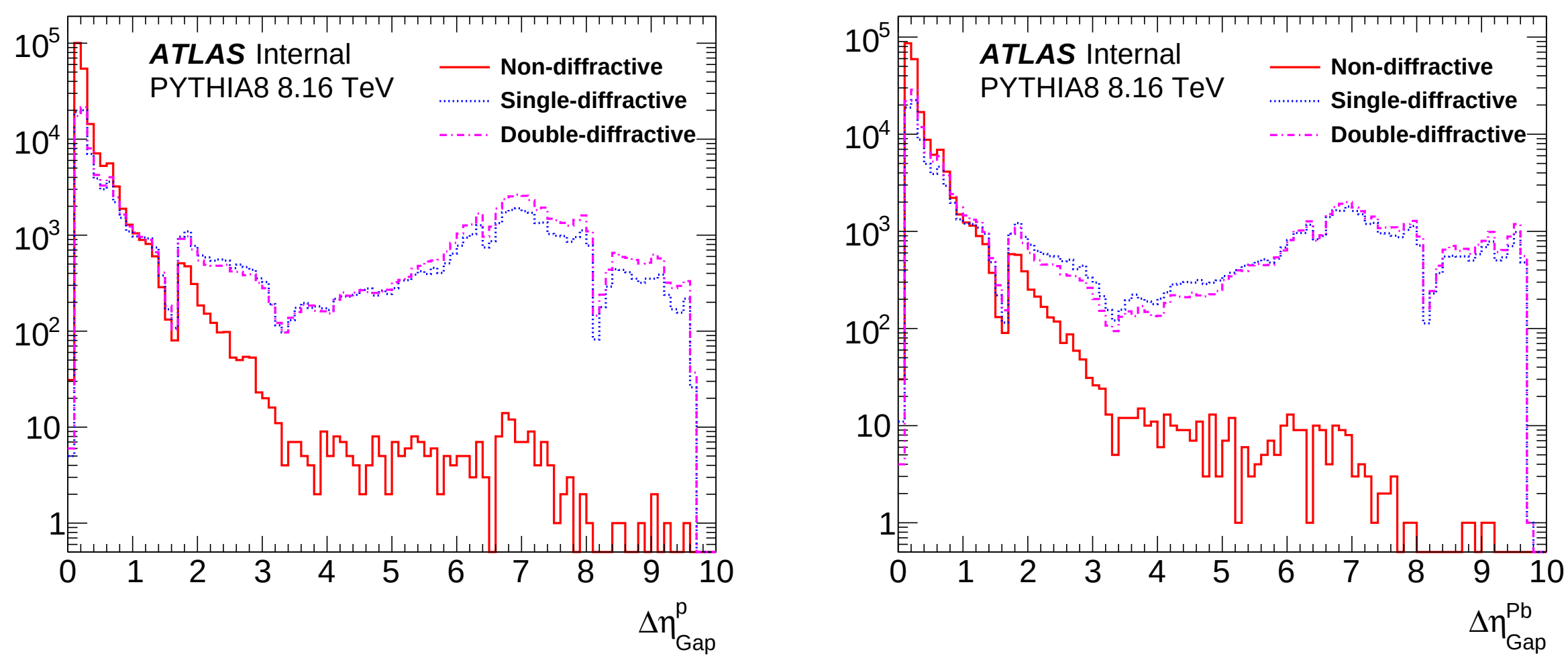

Figure 4.3: Edge gaps measured from the forward edge of the calorimeter system ($\eta = 4.9$) on the $p$-going (**Left**) and Pb-going (**Right**) sides of calorimeter clusters above 200 MeV.

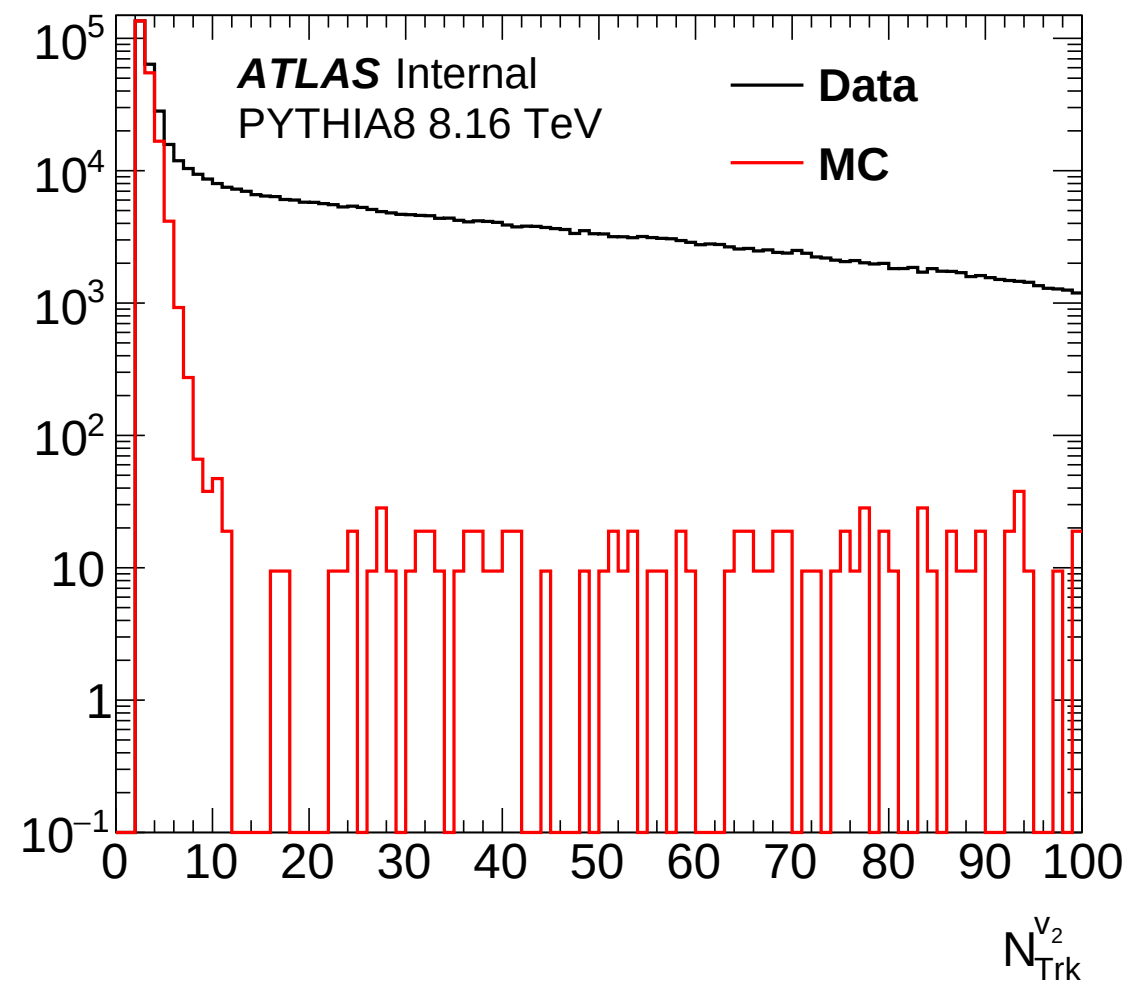

Figure 4.4: Histogram of the number of tracks associated with the second vertex in data (black) and HIJING MC with no pileup (red).

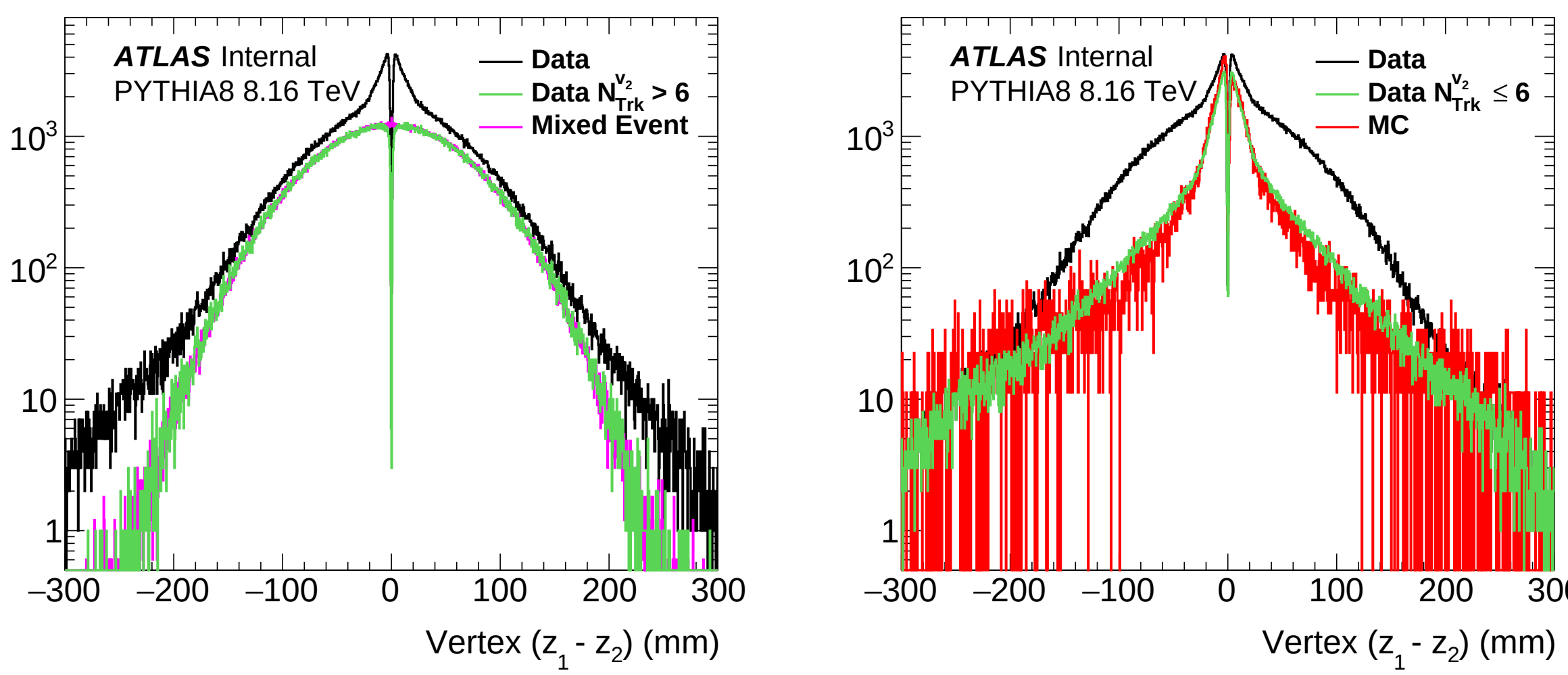

Figure 4.5: Distribution of vertex $z_1 - z_2$ in events with more than one vertex (black). **Left:** Events in which the second vertex has greater than 6 tracks to choose pileup events (green) and agrees well with simulated pileup events generated by mixing Hijing events (purple). **Right:** Events passing the pileup rejection (green) compared to the distribution from no pileup Hijing events (red) showing, again, good agreement.

Efficiencies for these requirements are found in Pythia *pp* simulations and is shown in Fig. 4.6. In total, they exclude 50.8% of the diffractive events while being 98.7% efficiency for the non-diffractive portion.

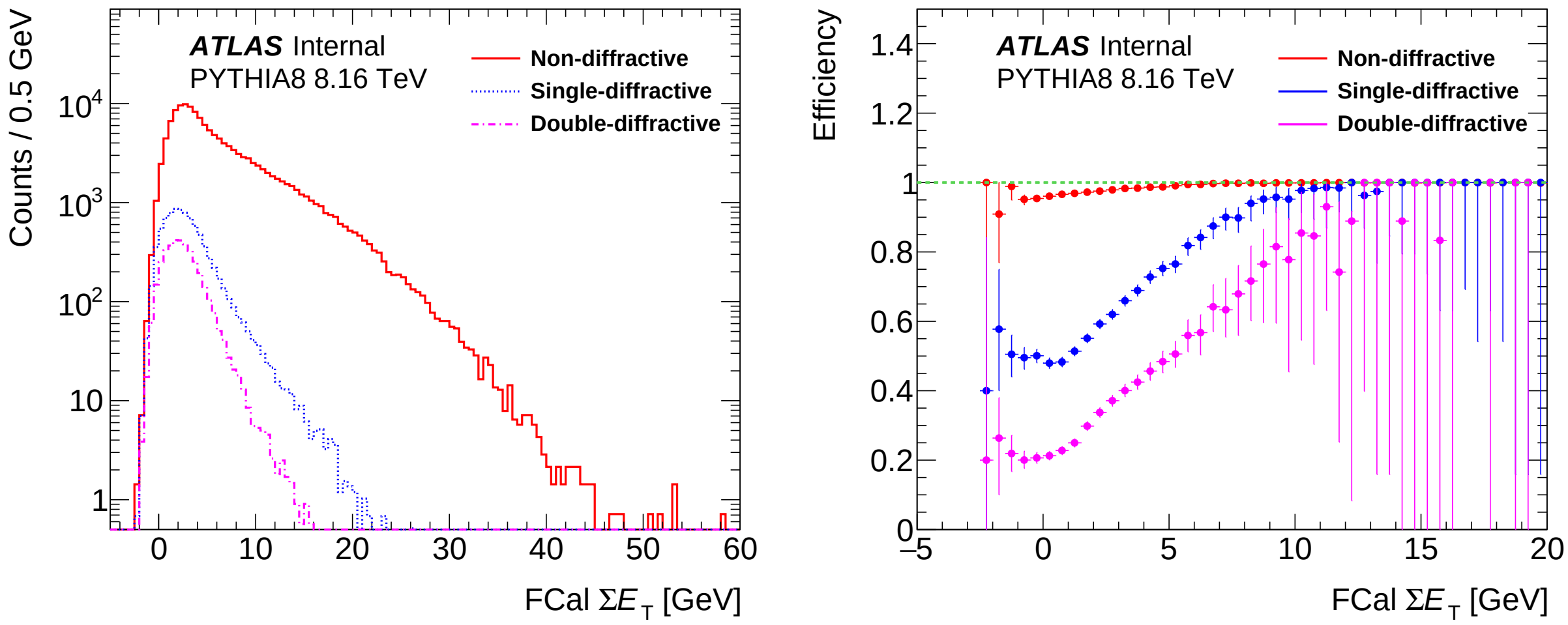

Figure 4.6: **Left:** Pb-going FCal total transverse energy distributions from the non-diffractive, single-diffractive, and double-diffractive components from Pythia $pp$ simulations after applying the event selection requirements. **Right:** Event selection efficiency plotted as a function of FCal $\Sigma E_{\mathrm{T}}^{\mathrm{Pb}}$ for non-diffractive, single-diffractive, and double-diffractive components.

## 4.3 Forward Calorimeter Energy Distributions

Data taking was separated into two periods: the first in which the Pb nucleus was moving towards $\mathrm{FCal}_{\mathrm{A}}$ and the proton towards $\mathrm{FCal}_{\mathrm{C}}$, the second in which the beam directions were reversed and therefore the Pb nucleus was moving in the directions of $\mathrm{FCal}_{\mathrm{C}}$. Noise contributions are determined by examining triggers from empty beam crossings. These noise distributions are fit to Gaussian functions (Fig. 4.7) from which the mean and width are extracted for each run. The means and widths are plotted in Figure 4.8. The noise widths show a clear separation between the first and second periods showing a characteristic difference between $\mathrm{FCal}_{\mathrm{A}}$ and $\mathrm{FCal}_{\mathrm{C}}$. A slight run dependence is found for the pedestal means and is corrected run-by-run for the rest of the analysis. Figure 4.9 shows the means after this correction.

As has been seen in previous analyses, $\Sigma E_{\mathrm{T}}^{\mathrm{Pb}}$ distributions, integrated over entire runs, show a significant negative energy tail. This tail can be explained by the existence of out-of-time pileup, in which the pulses from previous beam crossings overlap in the electronics. The pulses are shaped in a way that cancels this effect on average, but can give rise to significant negative energy for events closely following others in time. Out-of-time-pileup has little affect on the positive side of the distribution; However, to prevent the negative side from biasing the global fit, we select only events first in the bunch train, thus ensuring a significant time gap from the preceding event and no out-of-time pileup. Figure 4.10 shows the $\Sigma E_{\mathrm{T}}^{\mathrm{Pb}}$ distribution from all bunch crossings and after selecting only the first in the train, where the modest negative contribution can be explained by the width of the electronic noise.

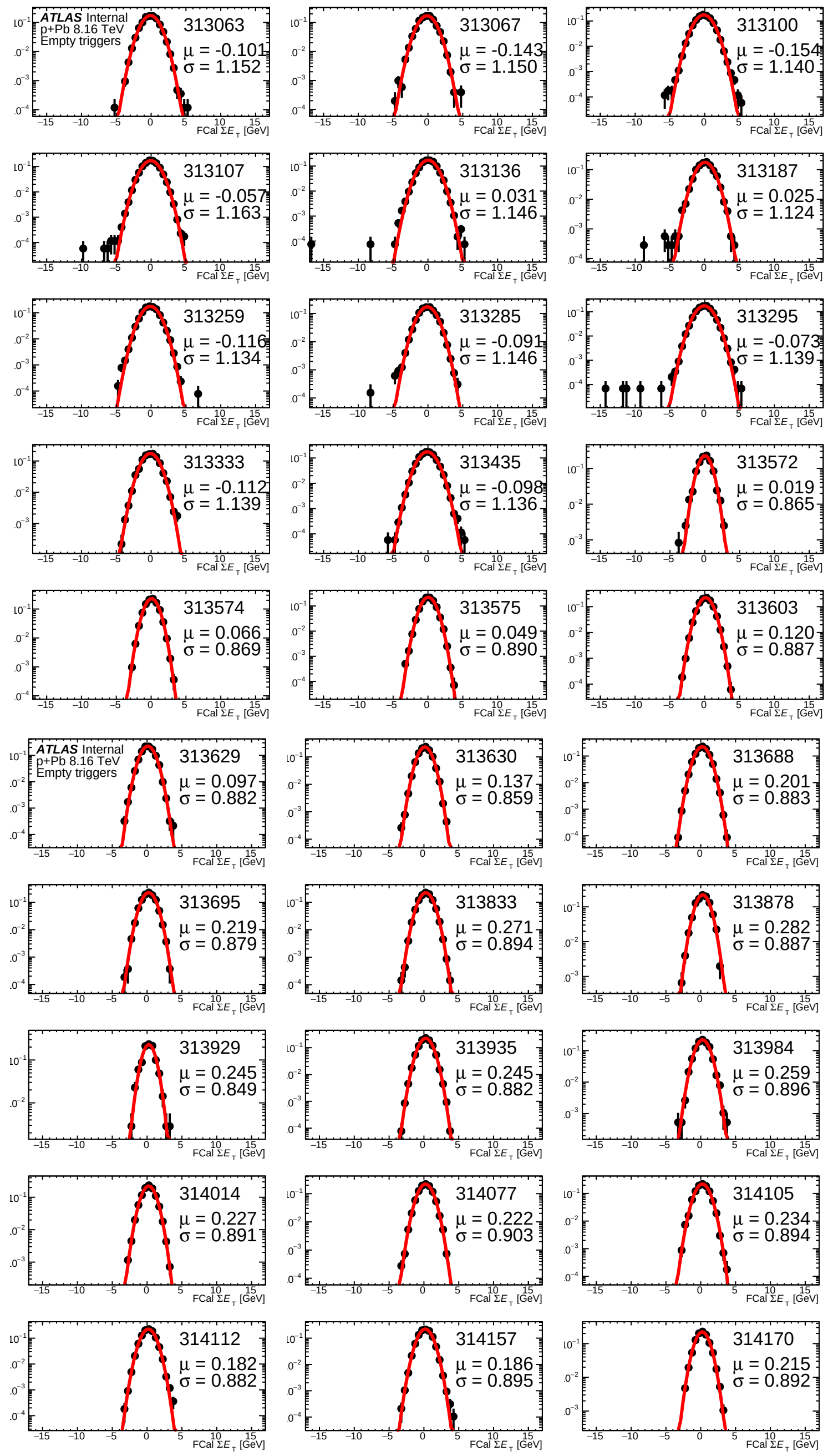

Figure 4.7: Gaussian fits to FCal noise $\Sigma E_{\mathrm{T}}^{\mathrm{Pb}}$ distributions from empty triggered events for each run.

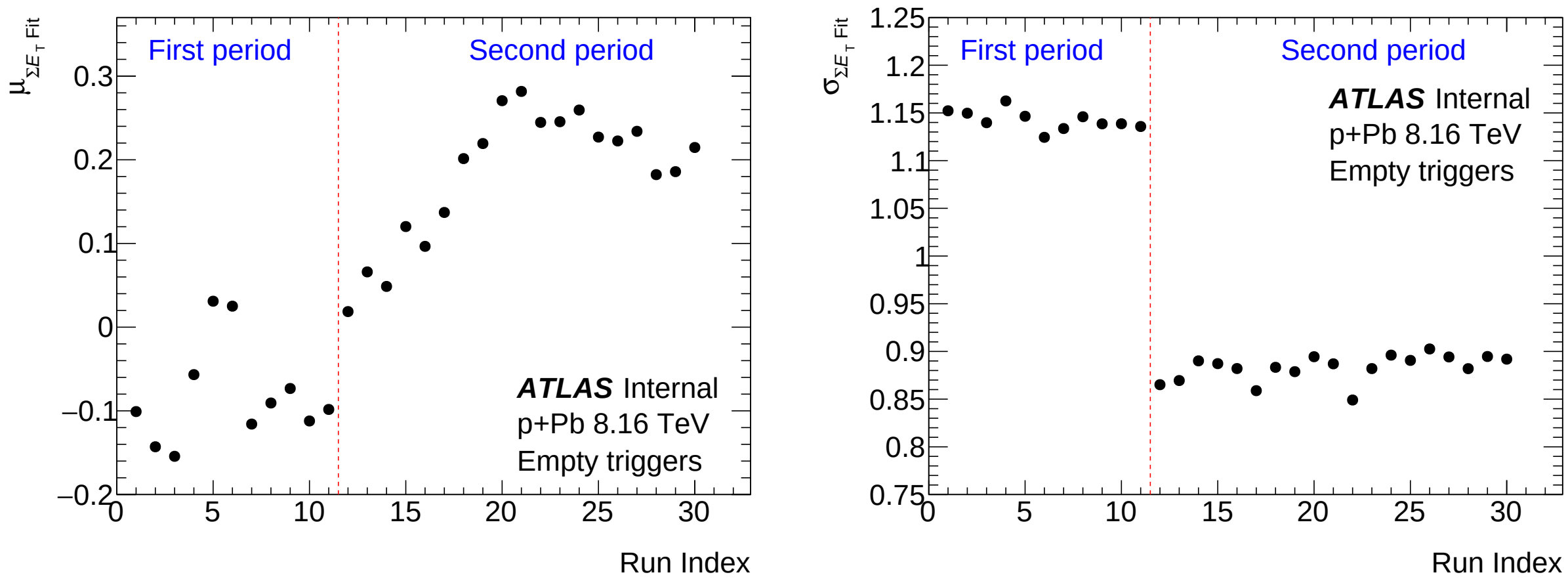

Figure 4.8: Extracted means (**Left**) and widths (**Right**) from Gaussian fits to $\Sigma E_{\mathrm{T}}^{\mathrm{Pb}}$ distributions from empty triggers. The parameters are plotted for each run in chronological order (run index).

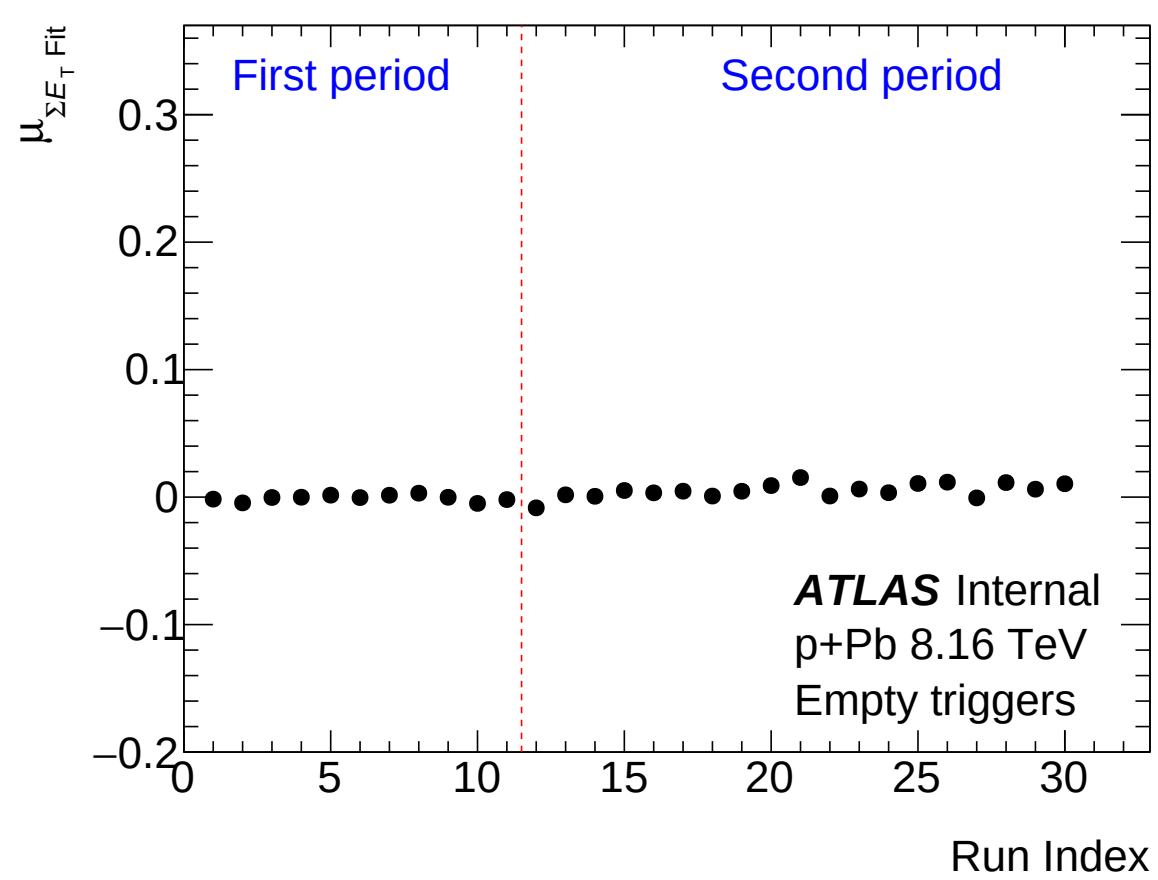

Figure 4.9: Extracted means from Gaussian fits to $\Sigma E_{\mathrm{T}}^{\mathrm{Pb}}$ distributions from empty triggers after correction.

To form the final $\Sigma E_{\mathrm{T}}^{\mathrm{Pb}}$ distribution used in this analysis, the Pb-going distributions from each period are added (i.e. $\mathrm{FCal_A}$ from period 1 is added to $\mathrm{FCal_C}$ from period 2). It is therefore necessary that the energy scale from the two calorimeters be consistent. Figure 4.11 shows a comparison of the $\Sigma E_{\mathrm{T}}^{\mathrm{Pb}}$ distributions from each FCal. It is clear from the ratio that there is a slight discrepancy in the energy scales that is corrected by a constant scaling factor of 0.989 applied to the energies from $\mathrm{FCal_C}$. This is adjusted from this point on in the analysis.

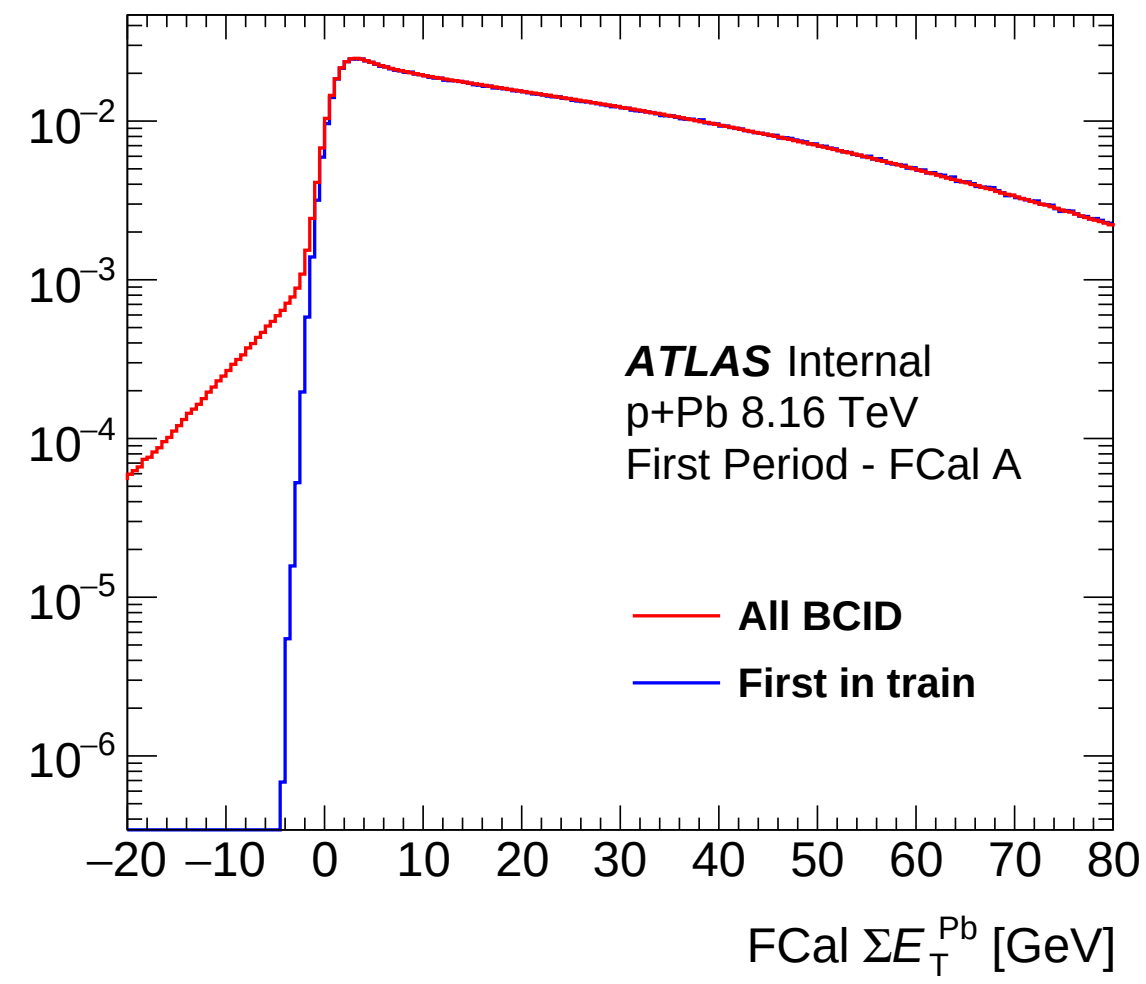

Figure 4.10: Mean $\Sigma E_{\mathrm{T}}^{\mathrm{Pb}}$ from all bunch crossing positions (red) and from only the first position in the bunch train.

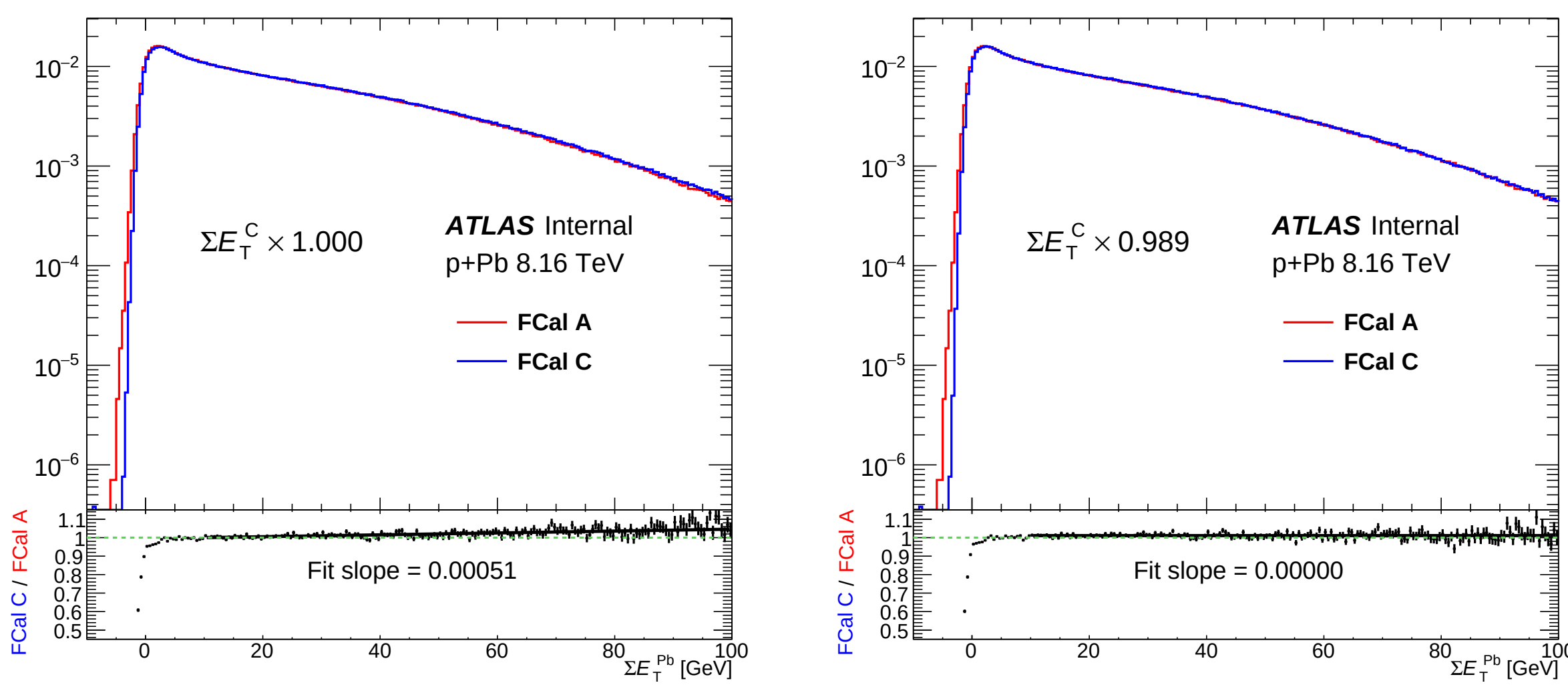

Figure 4.11: **Left:** Comparison between the $\Sigma E_{\mathrm{T}}^{\mathrm{Pb}}$ distributions from each running period, both scaled to unit integral. **Right:** The same comparison after applying a scale factor of 0.989 to each entry in the $\mathrm{FCal}_{\mathrm{C}}$ histogram.

The shape of the pseudorapidity distribution of produced particles in these asymmetrical collisions combined with the acceptance of the Pb-going FCal lead to a vertex $z$ position dependence on $\Sigma E_{\mathrm{T}}^{\mathrm{Pb}}$ distributions. This is simply due to the discrepancy between the fixed (lab) pseudorapidity position relative to the $z$-translated physical pseudorapidity distribution of the produced particles. To quantify this effect, we consider the mean total $\Sigma E_{\mathrm{T}}^{\mathrm{Pb}}$ in bins of vertex $z$ position, shown in Fig. 4.12 for each period (and thus each FCal) separately. The vertex $z$ dependence is well modeled by a linear function, the slope of which is given in the plot and can be used to correct for this dependence with a scaling event-by-event. The $\Sigma E_{\mathrm{T}}^{\mathrm{Pb}}$ distributions contain this correction from now on. The final $\Sigma E_{\mathrm{T}}^{\mathrm{Pb}}$ distribution used in this analysis is given in Fig. 4.13.

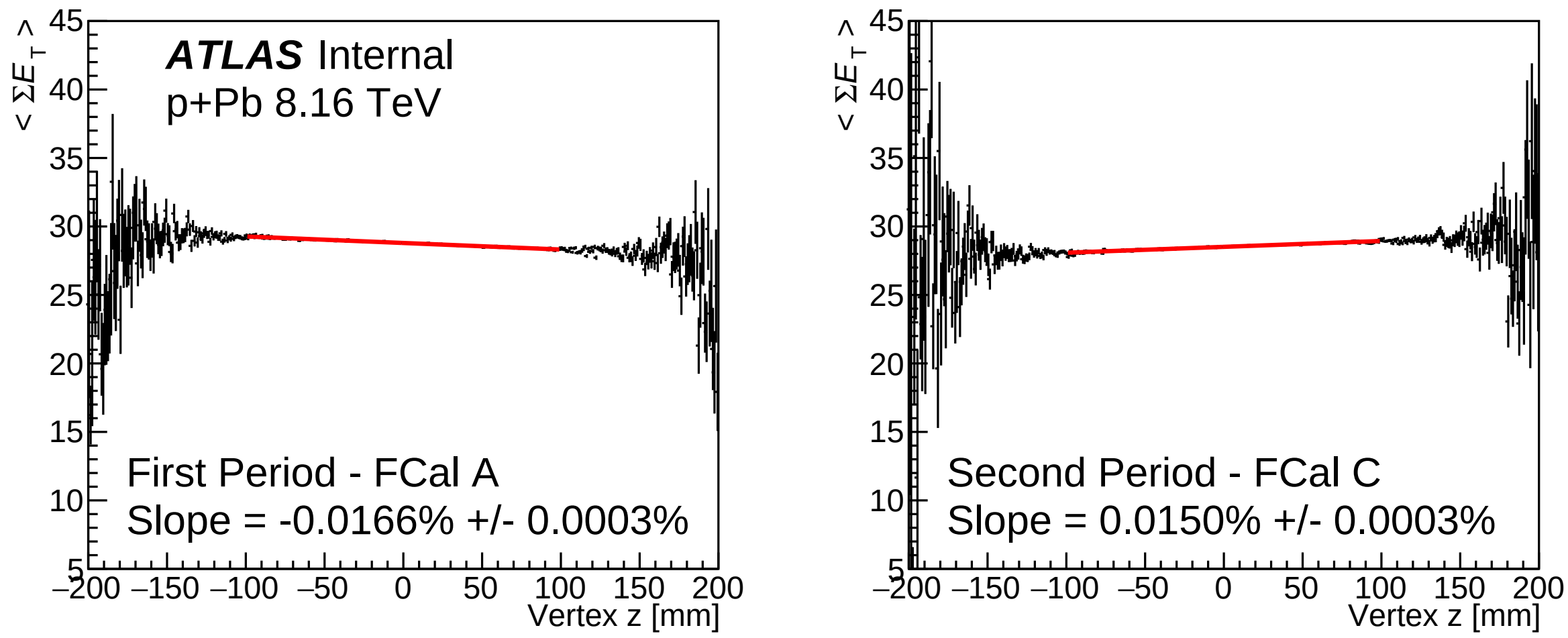

Figure 4.12: Mean $\Sigma E_{\mathrm{T}}^{\mathrm{Pb}}$ as a function of vertex $z$ from period 1 (**Left**) and period 2 (**Right**). The plots are fit to a line, the parameters of which are used to correct for this effect in the data.

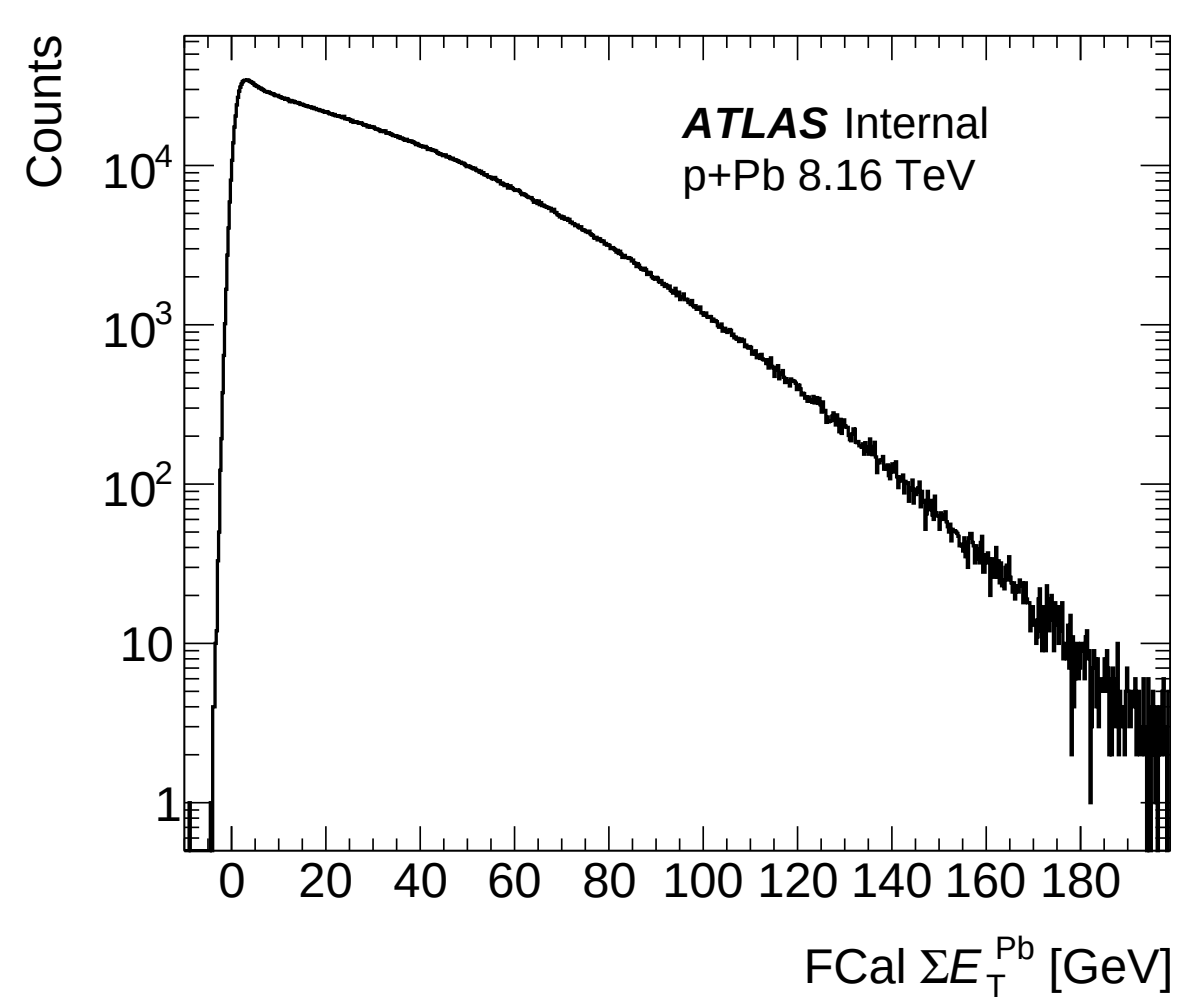

Figure 4.13: $\Sigma E_{\mathrm{T}}^{\mathrm{Pb}}$ distribution after all event selection and corrections and integrated over all runs.

## 4.4 Glauber Models

To map the $\Sigma E_{\mathrm{T}}^{\mathrm{Pb}}$ distributions to geometric properties of the collisions, it is necessary to introduce models of individual nucleus-nucleus interactions. Events can then be classified by the $N_{\text{part}}$. The MC Glauber and Glauber-Gribov models used here were described in Sec. 2.4.1. The nucleon-nucleon cross section used for this analysis corresponds to an extrapolation from the current measurement of the total $pp$ inelastic cross section at $\sqrt{s_{NN}}$ = 8 TeV from the TOTEM collaboration [108]

$$\sigma_{tot}^{pp} = (75 \pm 2)\ \text{mb}. \tag{4.1}$$

To review, the Glauber-Gribov model builds on base MC Glauber by incorporating interaction strength fluctuations. It accomplishes this by introducing a fluctuation in the total cross section ($\sigma_{tot}$) such that it is drawn from the probably density function:

$$P(\sigma_{\text{tot}}) = N\frac{\sigma_{\text{tot}}}{\sigma_{\text{tot}} + \sigma_0}\, exp\Big( - \frac{\sigma_{\text{tot}}/\sigma_0 - 1}{\Omega^2}\Big), \tag{4.2}$$

where $\sigma_0$ controls the nominal mean nucleon-nucleon cross-section $\langle\sigma_{tot}\rangle$, $\Omega$ is a dimensionless parameter that describes the magnitude of the cross-section fluctuations, and $N$ is a normalization to ensure unit integral. The inelastic nucleon-nucleon cross-section $\sigma_{\text{NN}}$ is taken to be a fixed fraction of the total cross-section $\sigma_{\text{tot}}$ according to $\sqrt{s_{\text{NN}}} = \lambda\sigma_{\text{tot}}$, so that $P(\sqrt{s_{\text{NN}}}) = (1/\lambda)P(\sigma_{\text{tot}}/\lambda)$. For each choice of the parameters $\sigma_0$ and $\Omega$, we choose $\lambda$ such that $\langle\sigma_{\text{tot}}\rangle = (75 \pm 2)$ mb, as in the default Glauber, and $\Omega$ = 0.55.

## 4.5 Global Fits

In this section, the Glauber model derived $N_{\text{part}}$ distributions are connected to the observed data FCal $\Sigma E_{\mathrm{T}}^{\mathrm{Pb}}$ distribution by introducing $N_{\text{part}}$ scaling parameterizations and performing global fits to the data. As has been assumed in previous analyses, the $\Sigma E_{\mathrm{T}}^{\mathrm{Pb}}$ response distribution for any given $N_{\text{part}}$ is assumed to follow a gamma distribution with parameters $k$ and $\theta$ that may depend on $N_{\text{part}}$. The modeled $\Sigma E_{\mathrm{T}}^{\mathrm{Pb}}$ distribution is then computed as the $N_{\text{part}}$ convolution of the gamma distributions. In this analysis, we explore two different $k(N_{\text{part}})$ and $\theta(N_{\text{part}})$ parameterizations:

- Model 1:

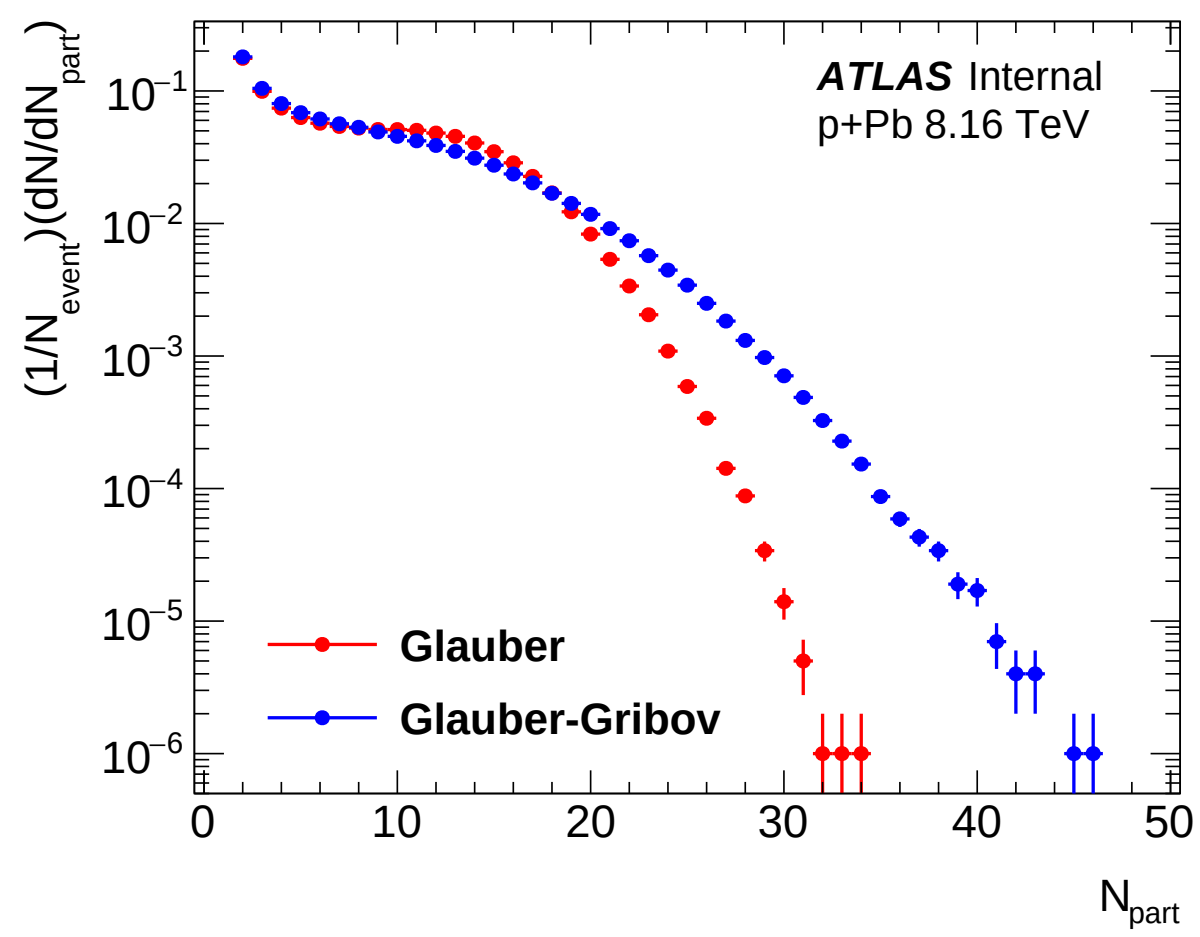

Figure 4.14: Distribution of $N_{\text{part}}$ values from Glauber (with $\sigma_{\text{NN}} = (75 \pm 2)$ mb) and Glauber-Gribov (with $\sigma_{\text{NN}} = (75 \pm 2)$ mb and $\Omega = 0.55$) models at $\sqrt{s_{\text{NN}}}$ 8.16 TeV $p$+Pb.

  - $k = k_0\, N_{\text{part}}$
  - $\theta = \theta_0$

- Model 2:
  - $k = k_1 + k_0(N_{\text{part}} - 2)$
  - $\theta = \theta_1 + \theta_0\, Log(N_{\text{part}} - 1)$

where, in both models, $k_0$ and $\theta_0$ are free parameters to be determined from the fit. Model 2 attempts to get a handle on the $pp$ ($N_{\text{part}} = 2$) behavior by fixing the parameters $k_1$ and $\theta_1$ to those obtained from a fit to the $\Sigma E_{\text{T}}^{\text{Pb}}$ from Pythia $pp$ simulations.

### 4.5.1 Fit to MC Data

Pythia $pp$ simulations, with the same center of mass boost found in $p$+Pb data, from representative admixture non-diffractive, single-diffractive, and double-diffractive events are used to generate a simulated $\Sigma E_{\text{T}}^{\text{Pb}}$ distribution, which includes all event selection used in the data except the Pb-going ZDC requirement (The ZDC was not simulated). Additionally, to exclude events in which ATLAS would not detect or trigger on (and are thus not included in this analysis), at least one truth particle is required to fall within the Pb-going FCal acceptance.

Fig. 4.15 shows the fit to the noise distribution measured in events in which no truth particles struck the Pb-going FCal. Because this level of noise is slightly less that found in data (Fig. 4.8), the PYTHIA $\Sigma E_{\mathrm{T}}^{\mathrm{Pb}}$ data is smeared with a Gaussian to match the average value found in data. This distribution was then fit to a gamma function convolved with a Gaussian with a width fixed to this value. Fig. 4.16 gives this distribution and fit. It is apparent that this functional form is not perfect in describing the data. However it's simple form and analytic convolutional properties make it a convenient choice.

From this fit, we obtain $k_0$ and $\theta_0$ used to set the $N_{\text{part}} = 2$ parameters in model 2.

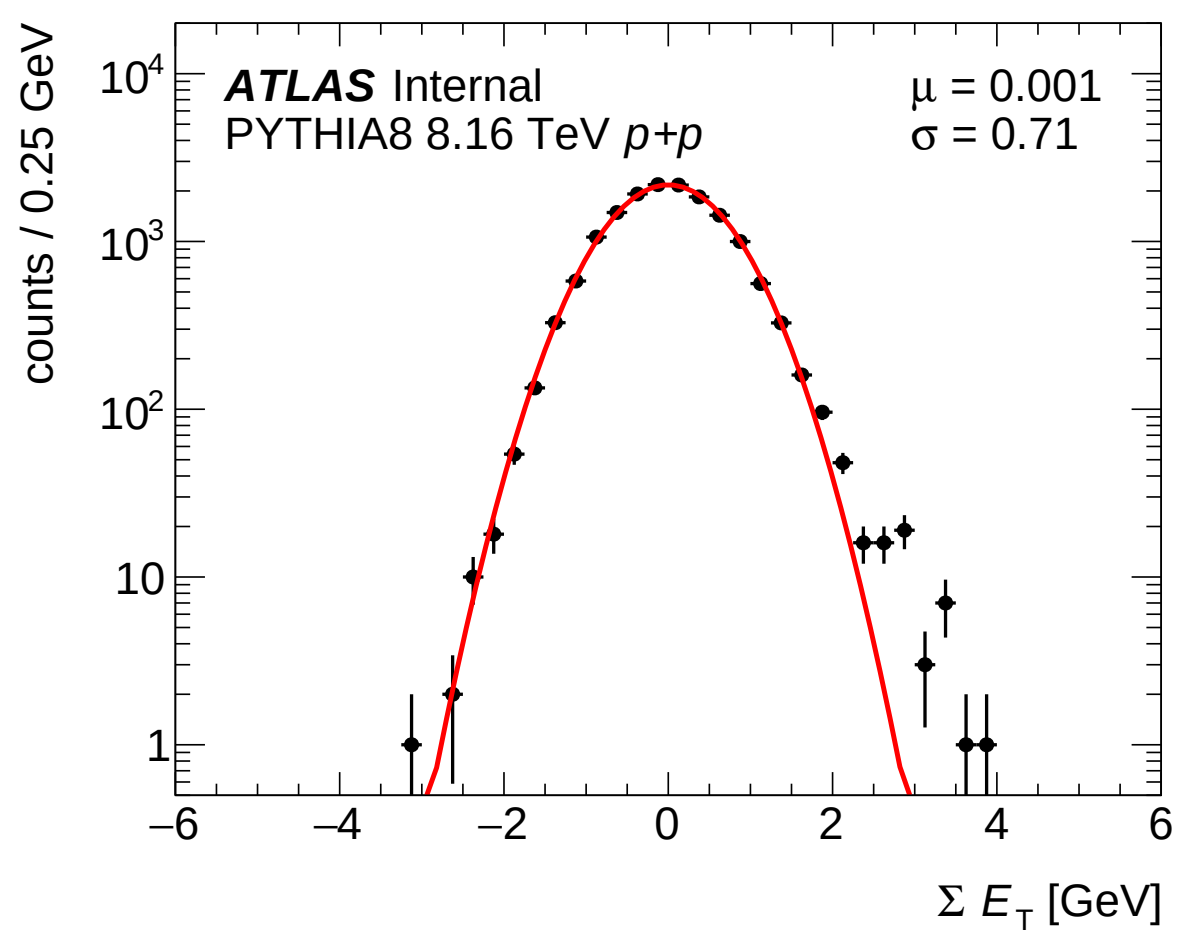

Figure 4.15: Gaussian fit to the $\Sigma E_{\mathrm{T}}^{\mathrm{Pb}}$ noise distribution.

### 4.5.2 Fits to Full Data Distribution

Global fits are performed using, as a functional form, the convolution of the $N_{\text{part}}$ distributions from the Glauber and Glauber-Gribov models (Fig. 4.14) with gamma distributions who's parameters are generated by the two stated scaling models. The fits are made to points above 10 GeV to ensure that any poorly understood inefficiency or diffractive elements, contributing at low $\Sigma E_{\mathrm{T}}^{\mathrm{Pb}}$, would not influence the fits. Figure 4.17 shows the fits using traditional Glauber and each scaling model. Figure 4.18 gives the fits using Glauber-Gribov. The results of these fits are summarized in Table 4.1. As a check, the fits are computed again, this time fitting above 5 GeV instead of 10 GeV. The results of which are summarized in Table 4.2, and the fits are given in Figures 4.19 and 4.20.

Table 4.1: Summary of fits to $\Sigma E_{\mathrm{T}}^{\mathrm{Pb}}$ data > 10 GeV

| Glauber model | Response model | $k_0$ | $\theta_0$ | Efficiency |
|---|---|---|---|---|
| Traditional Glauber | model 1 | 0.361 | 9.733 | 0.972 |
| Traditional Glauber | model 2 | 0.413 | 1.413 | 0.971 |
| Glauber-Gribov | model 1 | 0.620 | 5.788 | 0.990 |
| Glauber-Gribov | model 2 | 0.867 | 0.014 | 0.987 |

Table 4.2: Summary of fits to $\Sigma E_{\mathrm{T}}^{\mathrm{Pb}}$ data > 5 GeV

| Glauber model | Response model | $k_0$ | $\theta_0$ | Efficiency |
|---|---|---|---|---|
| Traditional Glauber | model 1 | 0.351 | 9.997 | 0.976 |
| Traditional Glauber | model 2 | 0.483 | 1.112 | 0.999 |
| Glauber-Gribov | model 1 | 0.622 | 5.820 | 1.005 |
| Glauber-Gribov | model 2 | 0.802 | 0.132 | 1.004 |

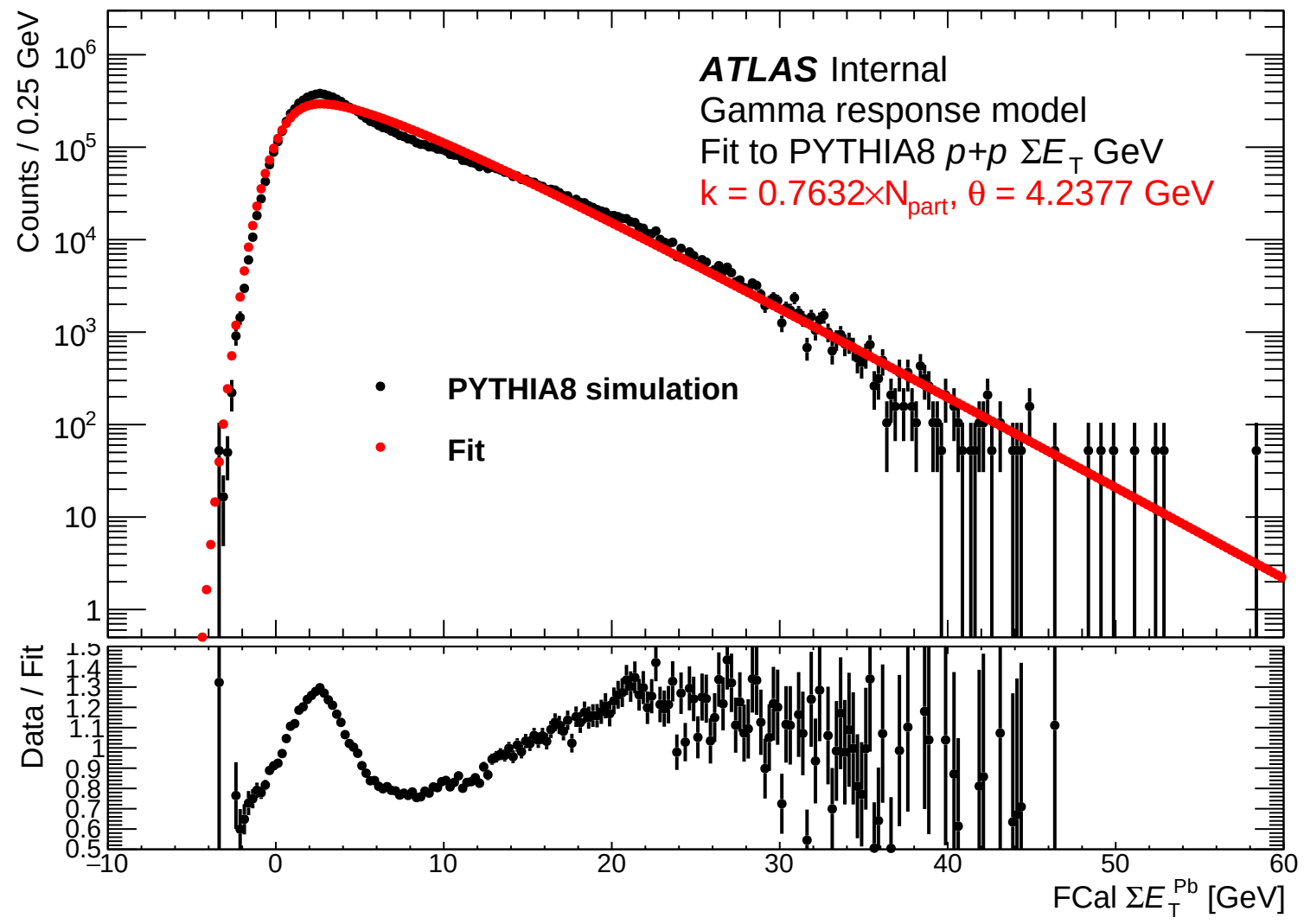

Figure 4.16: $\Sigma E_{\mathrm{T}}^{\mathrm{Pb}}$ distribution from simulated Pythia events (black) fit with a gamma distribution convolved with a Gaussian with width fixed by the noise fit (red). The lower panel gives the ratio of data to fit.

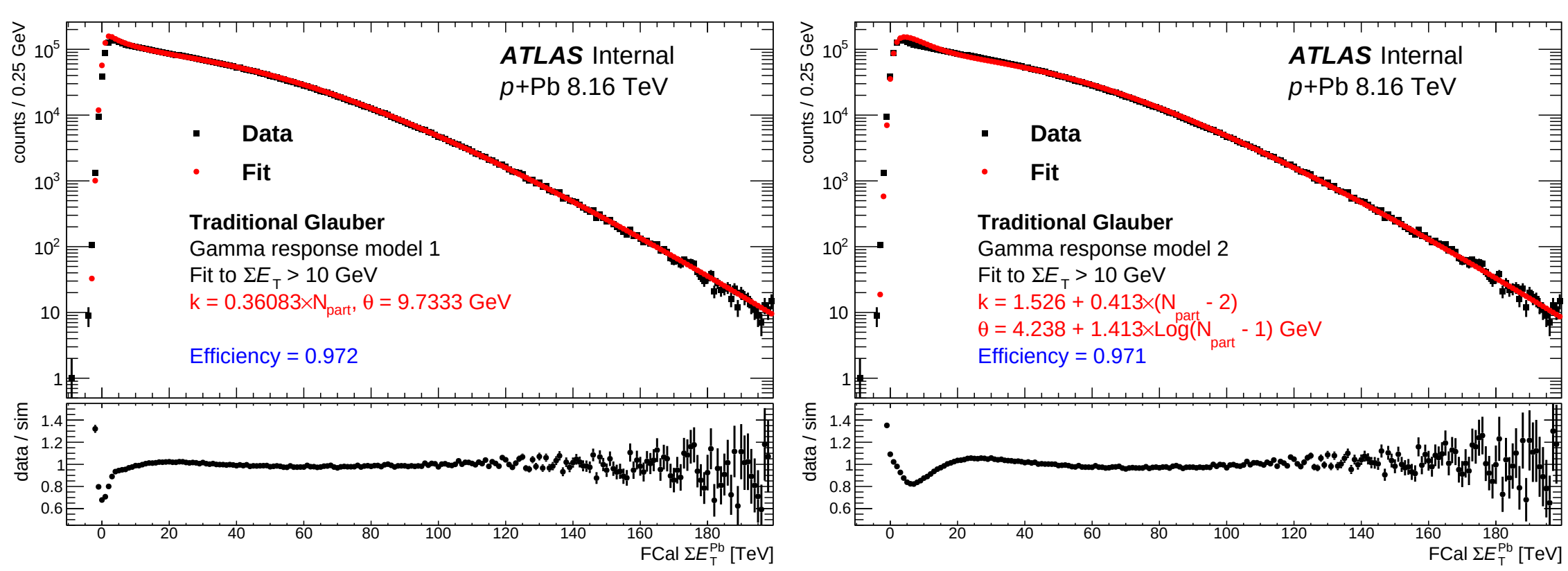

Figure 4.17: Fit to $\Sigma E_{\mathrm{T}}^{\mathrm{Pb}}$ distribution using the Glauber $N_{\mathrm{part}}$ distribution and gamma distribution scaling model 1 (**Left**) and model 2 (**Right**).

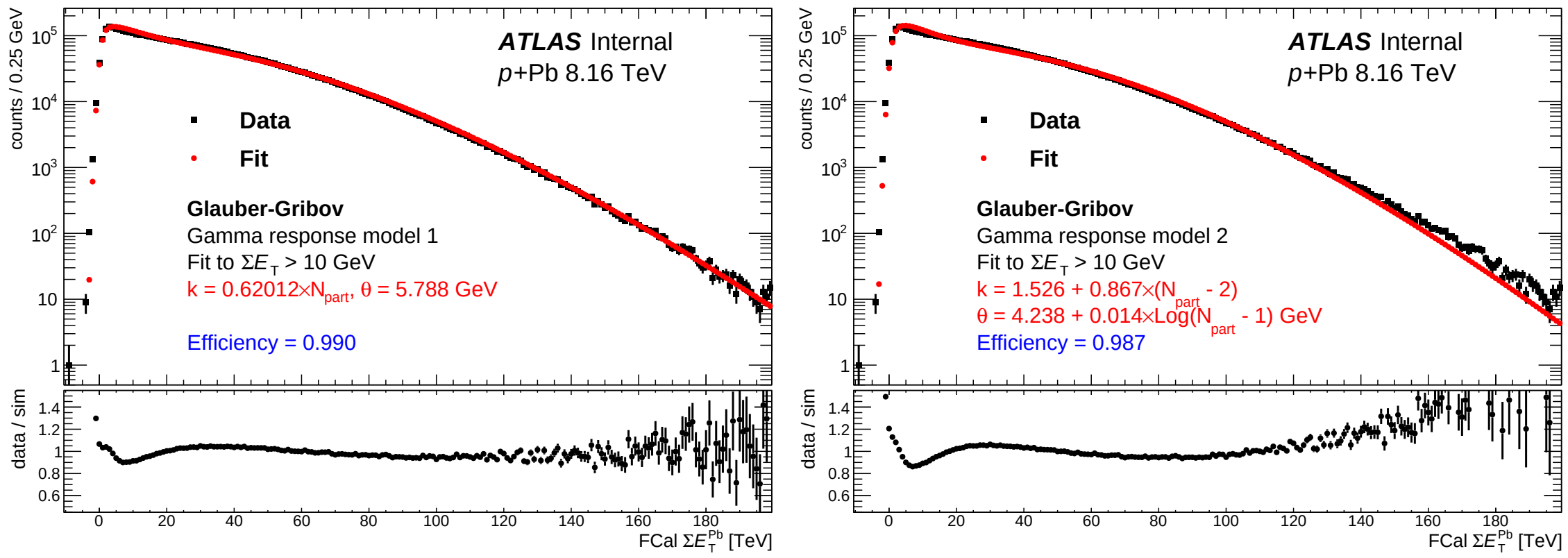

Figure 4.18: Fit to $\Sigma E_{\mathrm{T}}^{\mathrm{Pb}}$ distribution using the Glauber-Gribov $N_{\mathrm{part}}$ distribution and gamma distribution scaling model 1 (**Left**) and model 2 (**Right**).

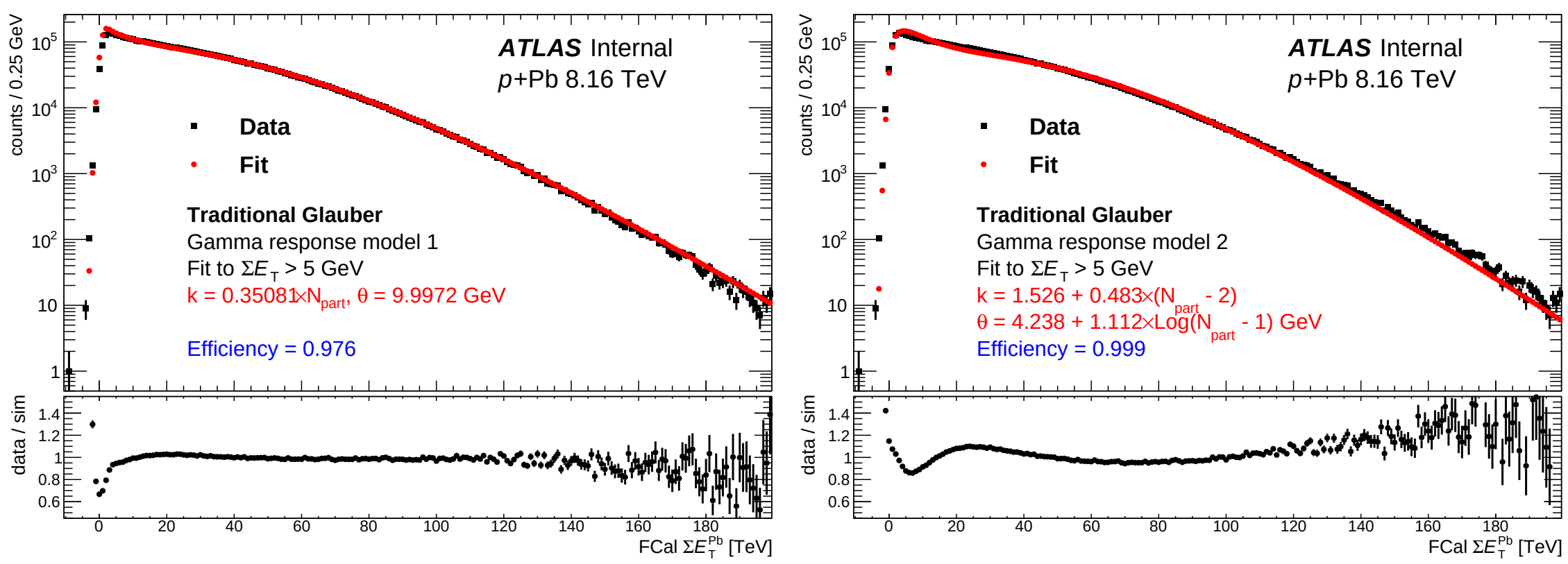

Figure 4.19: Fit to $\Sigma E_{\mathrm{T}}^{\mathrm{Pb}}$ distribution using the Glauber $N_{\mathrm{part}}$ distribution and gamma distribution scaling model 1 (**Left**) and model 2 (**Right**).

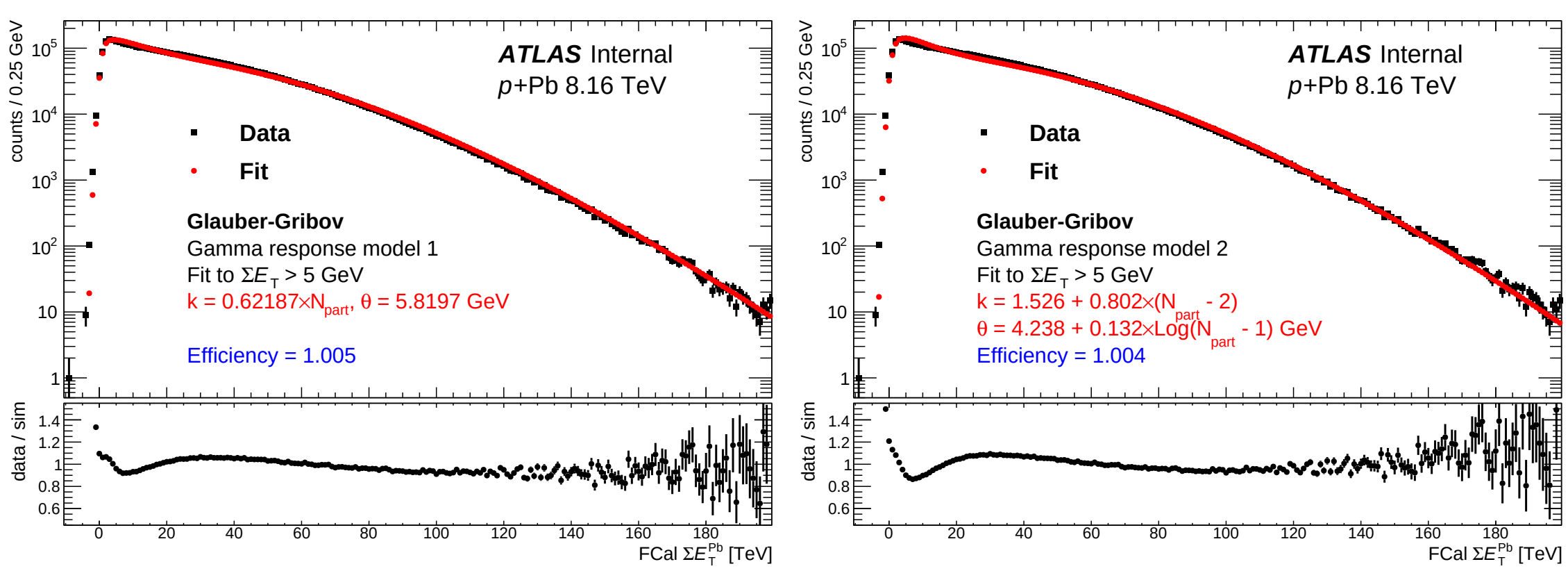

Figure 4.20: Fit to $\Sigma E_{\mathrm{T}}^{\mathrm{Pb}}$ distribution using the Glauber-Gribov $N_{\mathrm{part}}$ distribution and gamma distribution scaling model 1 (**Left**) and model 2 (**Right**).

## 4.6 Results

### 4.6.1 Centrality Selections

Comparing the simulated distribution for each model to the data provides a data-driven method for estimating the event selection efficiency. Pythia simulations in Sec. 4.2 indicate that efficiency losses and any enhancement due to residual diffractive/photo-nuclear events are primarily constrained to the region $\Sigma E_{\mathrm{T}}^{\mathrm{Pb}} < 5$. However, the precise centrality designations are sensitive to the total efficiency. Given the variation in efficiency estimations for each model, a nominal value 98% with variation +2% -1% is assigned, and the centrality designations for the nominal efficiency and variations are given in Table 4.3.

### 4.6.2 $\langle N_{\mathrm{part}} \rangle$ and $\langle T_{\mathrm{AB}} \rangle$

Mean $N_{\mathrm{part}}$ and $T_{\mathrm{AB}}$ are extracted for the regions between each centrality designation (0-1%, 1-5%, 5-10%, 10-20%,20-30%, 30-40%, 40-50%, 50-60%, 60-70%, 70-80%, 80-90%, as well as the whole range 0-90%). The values are plotted for both traditional Glauber and Glauber-Gribov in Fig. 4.21 and tabulated in Tables 4.4 and 4.5.

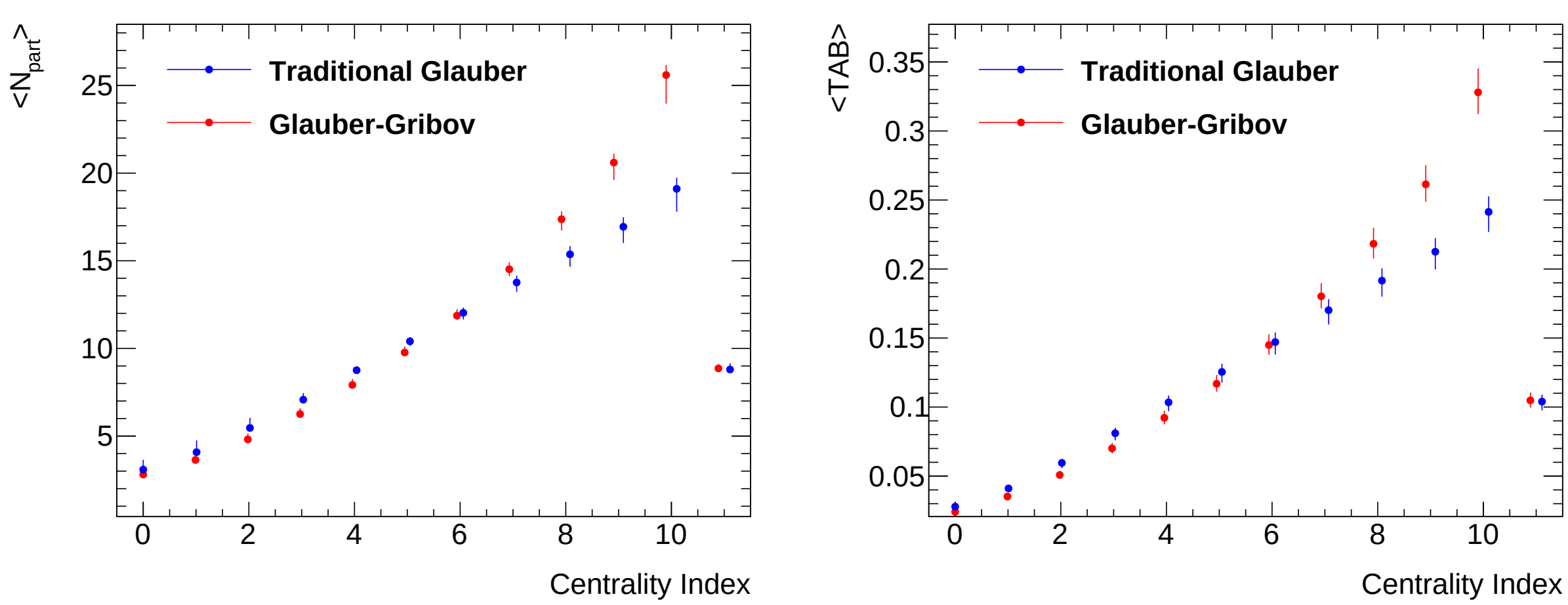

Figure 4.21: Mean $N_{\mathrm{part}}$ (**Left**) and $T_{\mathrm{AB}}$ (**Right**) calculated from both Glauber and Glauber-Gribov models and plotted as a function of centrality class.

Table 4.3: Summary of fits to $\Sigma E_{\mathrm{T}}^{\mathrm{Pb}}$ data > 5 GeV.

| Centrality | $\Sigma E_{\mathrm{T}}^{\mathrm{Pb}}$ cut (98%) | $\Sigma E_{\mathrm{T}}^{\mathrm{Pb}}$ cut (97%) | $\Sigma E_{\mathrm{T}}^{\mathrm{Pb}}$ cut (100%) |
|---|---|---|---|
| 1% | 108.51 GeV | 108.33 GeV | 108.85 GeV |
| 5% | 78.99 GeV | 78.79 GeV | 79.39 GeV |
| 10% | 64.57 GeV | 64.35 GeV | 65.01 GeV |
| 20% | 48.50 GeV | 48.25GeV | 49.00 GeV |
| 30% | 37.96 GeV | 37.68 GeV | 38.51 GeV |
| 40% | 29.79 GeV | 29.48 GeV | 30.38 GeV |
| 50% | 22.96 GeV | 22.64 GeV | 23.60 GeV |
| 60% | 17.04 GeV | 16.69GeV | 17.71 GeV |
| 70% | 11.80 GeV | 11.44 GeV | 12.49 GeV |
| 80% | 7.13 GeV | 6.77 GeV | 7.84 GeV |
| 90% | 3.07 GeV | 2.71 GeV | 3.75 GeV |

Table 4.4: Summary of mean $N_{\mathrm{part}}$ values for Glauber and Glauber-Gribov for each centrality class.

| Centrality | Traditional Glauber | Glauber-Gribov |
|---|---|---|
| 0-1% | $19.10^{+0.61}_{1.28}$ | $25.60^{+0.55}_{1.61}$ |
| 1-5% | $16.94^{+0.52}_{0.90}$ | $20.60^{+0.47}_{0.98}$ |
| 5-10% | $15.37^{+0.45}_{0.69}$ | $17.37^{+0.42}_{0.62}$ |
| 10-20% | $13.76^{+0.37}_{0.51}$ | $14.52^{+0.38}_{0.37}$ |
| 20-30% | $12.03^{+0.28}_{0.36}$ | $11.87^{+0.34}_{0.21}$ |
| 30-40% | $10.41^{+0.21}_{0.24}$ | $9.77^{+0.32}_{0.15}$ |
| 40-50% | $8.76^{+0.20}_{0.18}$ | $7.92^{+0.31}_{0.11}$ |
| 50-60% | $7.08^{+0.36}_{0.14}$ | $6.25^{+0.32}_{0.10}$ |
| 60-70% | $5.46^{+0.55}_{0.13}$ | $4.81^{+0.31}_{0.09}$ |
| 70-80% | $4.08^{+0.64}_{0.10}$ | $3.64^{+0.30}_{0.07}$ |
| 80-90% | $3.09^{+0.54}_{0.07}$ | $2.81^{+0.24}_{0.06}$ |
| 0-90% | $8.80^{+0.33}_{0.18}$ | $8.86^{+0.22}_{0.18}$ |

Table 4.5: Summary of mean $T_{\mathrm{AB}}$ values for Glauber and Glauber-Gribov for each centrality class.

| Centrality | Traditional Glauber | Glauber-Gribov |
|---|---|---|
| 0-1% | $0.2414^{+0.0109}_{0.0143}$ | $0.3280^{+0.0170}_{0.0155}$ |
| 1-5% | $0.2125^{+0.0096}_{0.01255}$ | $0.2614^{+0.0136}_{0.0123}$ |
| 5-10% | $0.1916^{+0.0087}_{0.0113}$ | $0.2183^{+0.0114}_{0.0103}$ |
| 10-20% | $0.1702^{+0.0077}_{0.0100}$ | $0.1802^{+0.0094}_{0.0085}$ |
| 20-30% | $0.1471^{+0.0067}_{0.0087}$ | $0.1450^{+0.0075}_{0.0068}$ |
| 30-40% | $0.1254^{+0.0057}_{0.0074}$ | $0.1169^{+0.0061}_{0.0055}$ |
| 40-50% | $0.1034^{+0.0047}_{0.0061}$ | $0.0922^{+0.0048}_{0.0043}$ |
| 50-60% | $0.0810^{+0.0037}_{0.0048}$ | $0.0701^{+0.0036}_{0.0033}$ |
| 60-70% | $0.0595^{+0.0027}_{0.0035}$ | $0.0508^{+0.0026}_{0.0024}$ |
| 70-80% | $0.0411^{+0.0019}_{0.0024}$ | $0.0352^{+0.0018}_{0.0017}$ |
| 80-90% | $0.0279^{+0.0013}_{0.0016}$ | $0.0241^{+0.0013}_{0.0011}$ |
| 0-90% | $0.1040^{+0.0047}_{0.0061}$ | $0.1048^{+0.0055}_{0.0049}$ |

### 4.6.3 Systematic Uncertainties

There are three main sources of systematic uncertainty in this analysis. These sources are varied, and the response on $N_{\text{part}}$ and $T_{\text{AB}}$ is measured relative to the nominal values. The variations are as follows:

- **Uncertainty in efficiency:** The efficiency is varied to 97% and 100% corresponding to observed variations in estimates from the fits.
- **Glauber parameters:**
  * Nucleon-Nucleon cross section ($\sigma_{NN}$) is varied $\pm$ 2 mb corresponding to the uncertainty in the TOTEM measurement (Fig. 4.22, 4.25)
  * Woods-Saxon radius and skin depth (R, a) are varied according to their uncertainty and in a way that preserves their anti-correlation ( (6.62,0.546) $\rightarrow$ (6.68,0.536) $\rightarrow$ (6.56,0.556) ) (Fig. 4.23, 4.26)
  * Hard core radius (d$_{min}$ $\pm$ 0.2 fm from the nominal 0.4 fm
- **Fit model:** Alternate fit model 1 is used and relative difference assigned as uncertainty (Fig. 4.24, 4.27)

At most peripheral and most central the model variation dominates the uncertainty, whereas in the mid centrality range, the cross section uncertainty dominates.

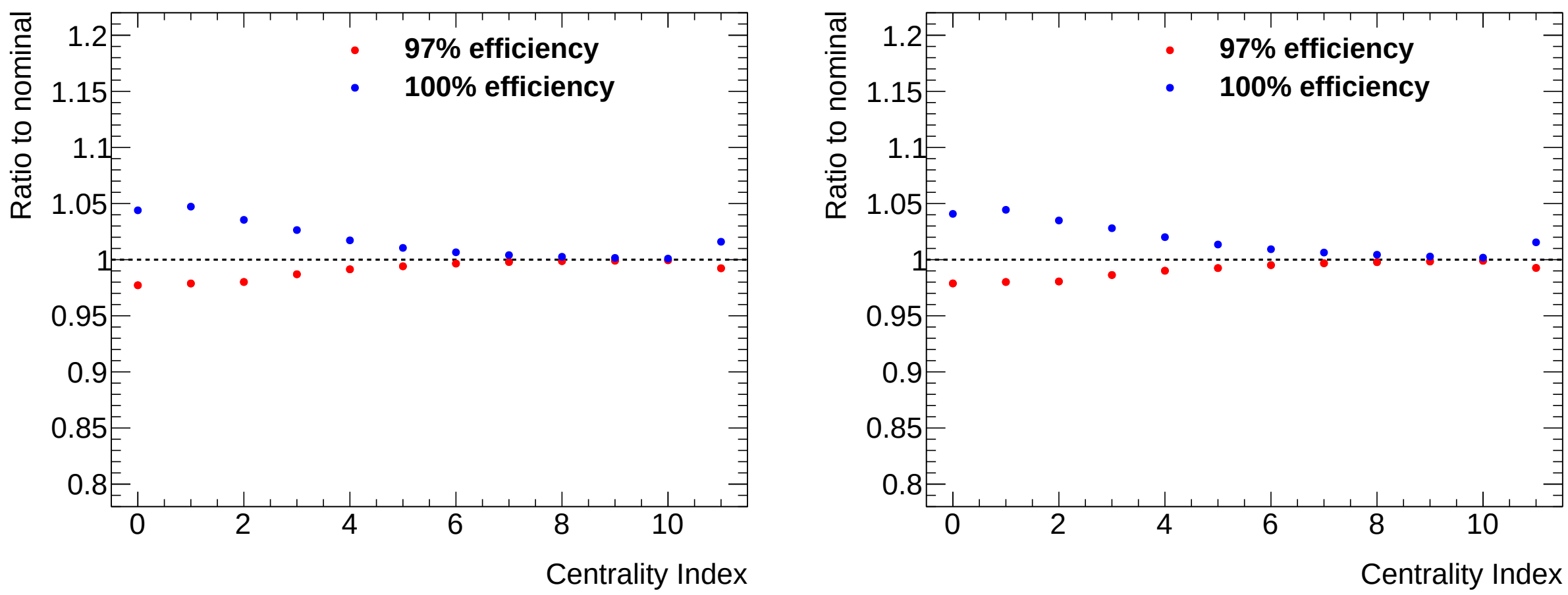

Figure 4.22: Systematic variation of $N_{\text{part}}$ generated by setting efficiencies to 97% and 100%, and calculated using Glauber (**Left**) and Glauber-Gribov (**Right**) of each centrality class.

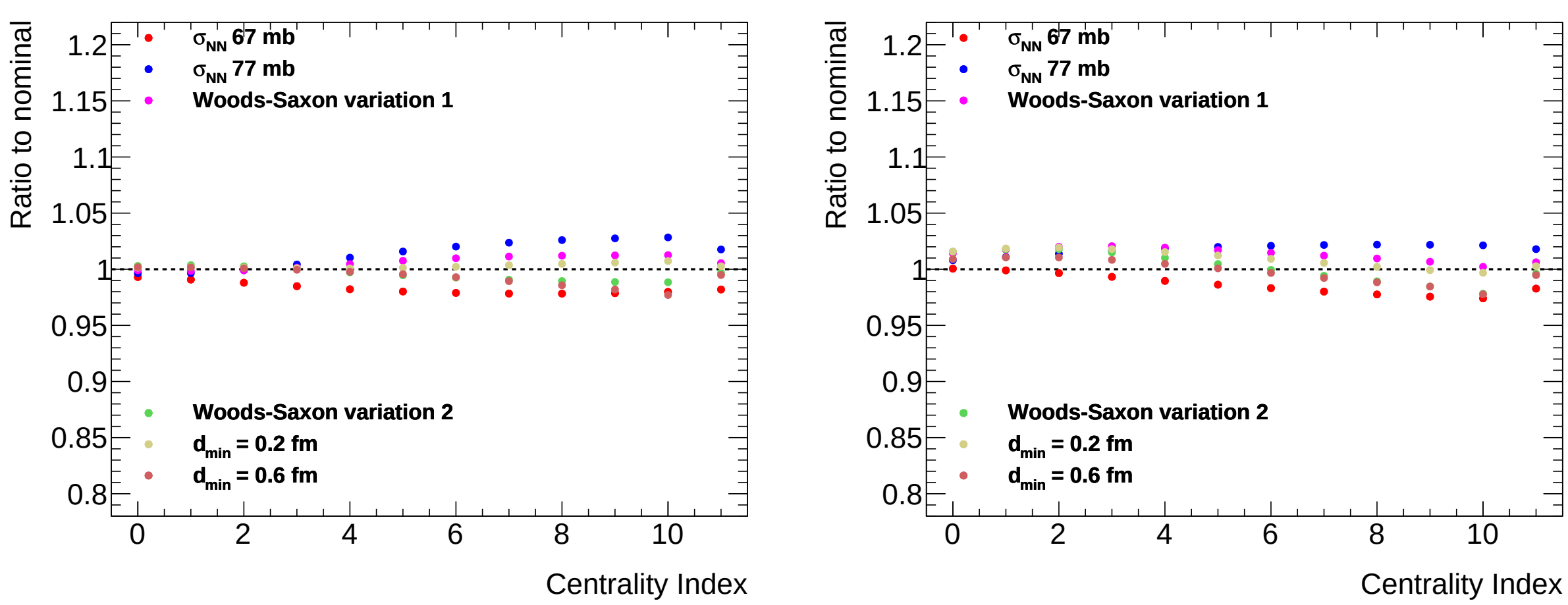

Figure 4.23: Systematic variation of $N_{\text{part}}$ generated by varying Glauber parameters, and calculated using Glauber (**Left**) and Glauber-Gribov (**Right**) of each centrality class.

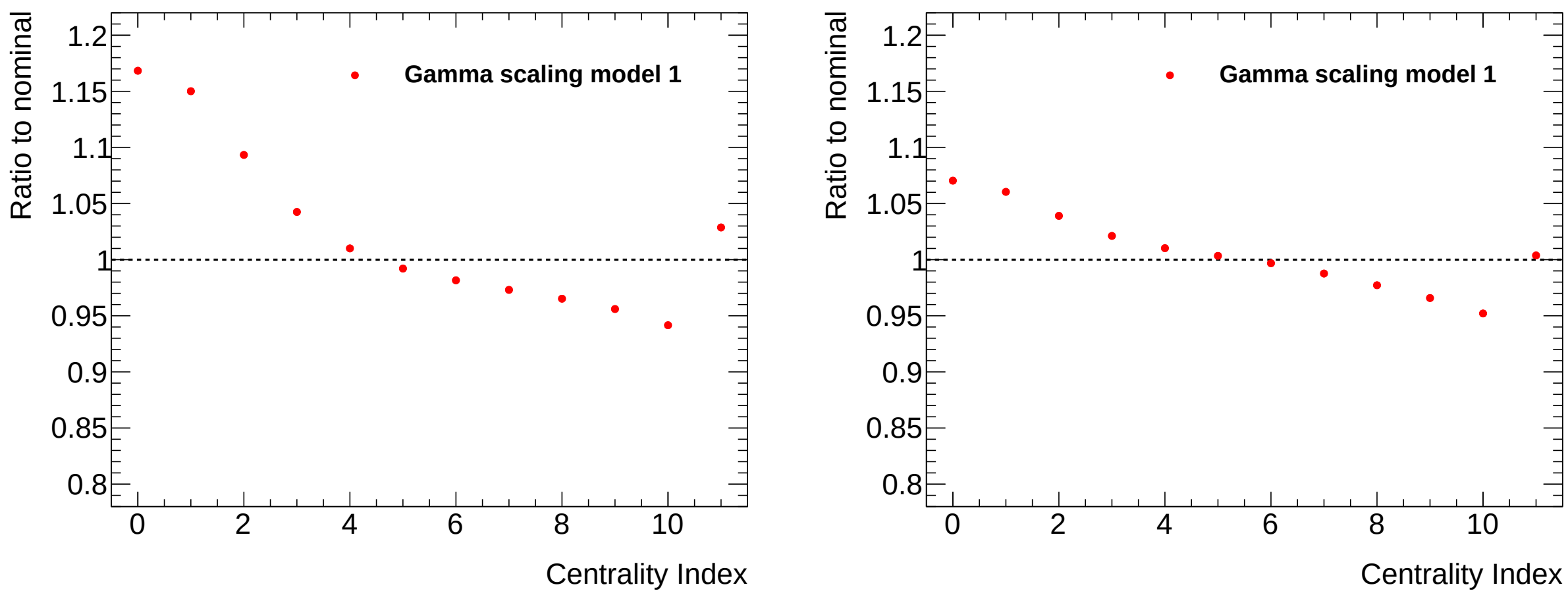

Figure 4.24: Systematic variation of $N_{\text{part}}$ generated by varying the scaling model to model 1 compared to the nominal model 2, and calculated using Glauber (**Left**) and Glauber-Gribov (**Right**) of each centrality class.

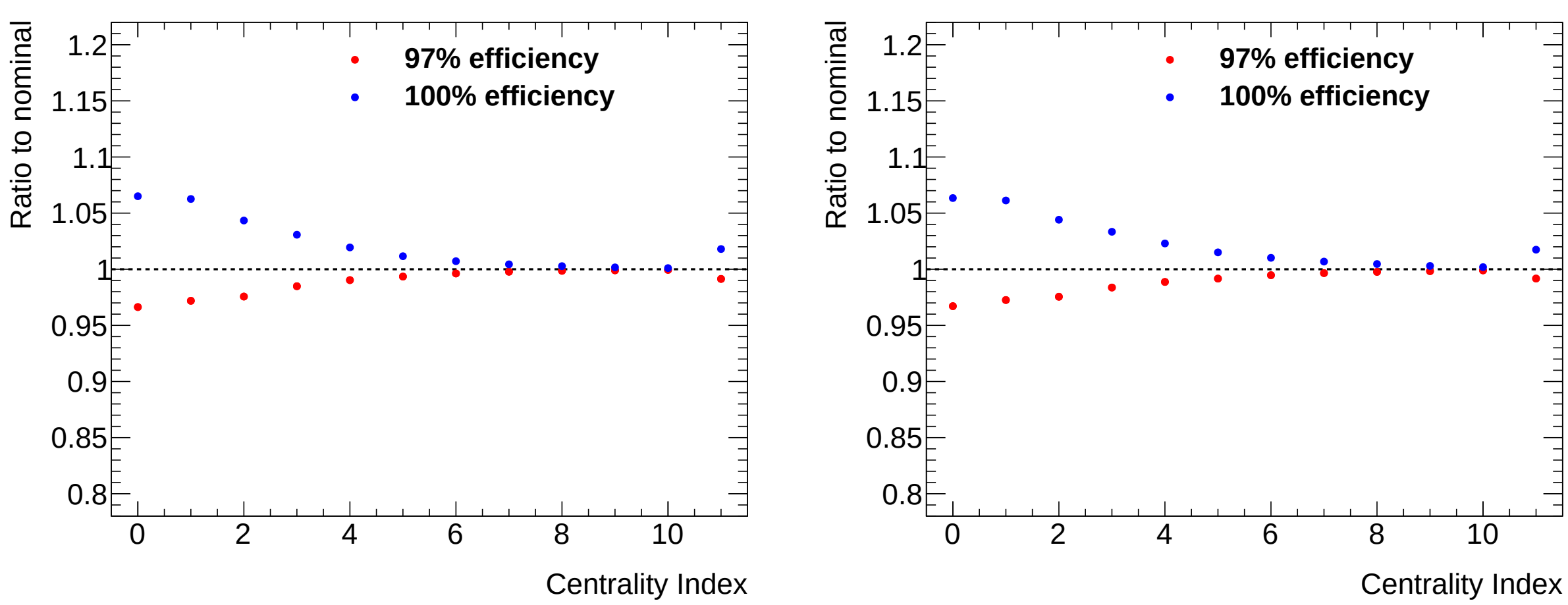

Figure 4.25: Systematic variation of $T_{\text{AB}}$ generated by setting efficiencies to 97% and 100%, and calculated using Glauber (**Left**) and Glauber-Gribov (**Right**) of each centrality class.

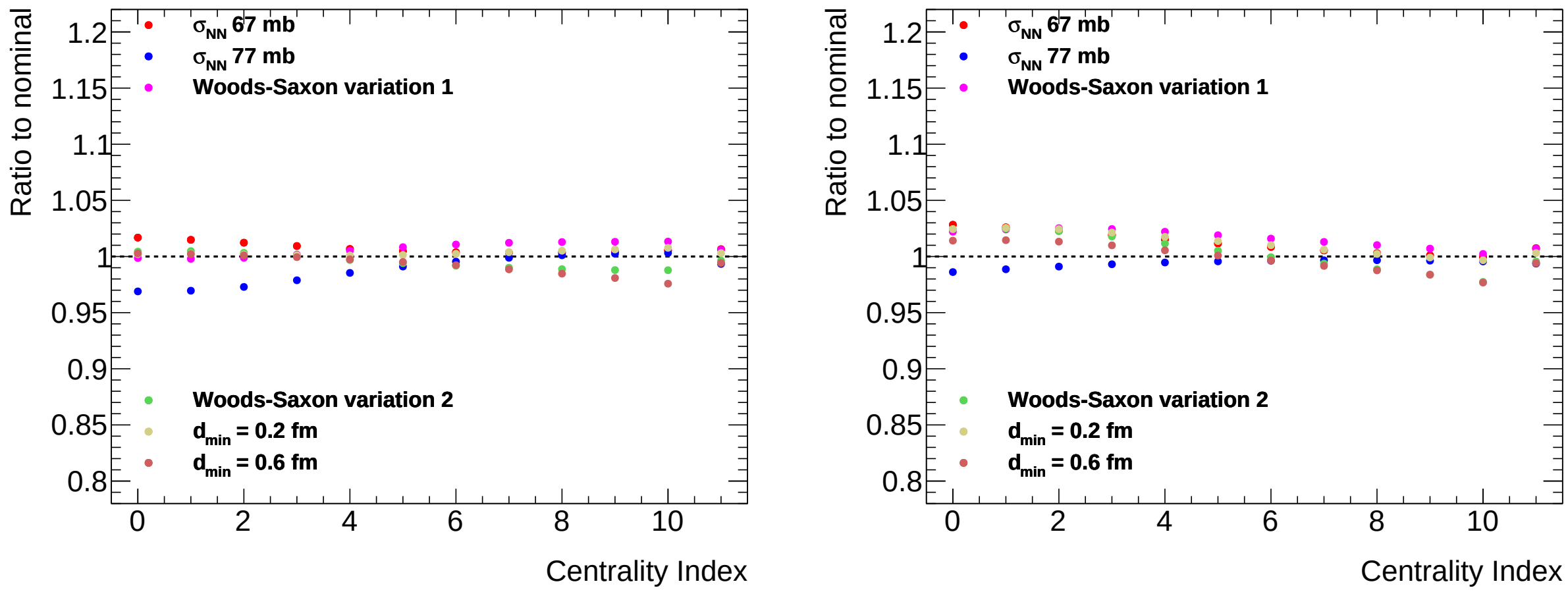

Figure 4.26: Systematic variation of $T_{AB}$ generated by varying Glauber parameters, and calculated using Glauber (**Left**) and Glauber-Gribov (**Right**) of each centrality class.

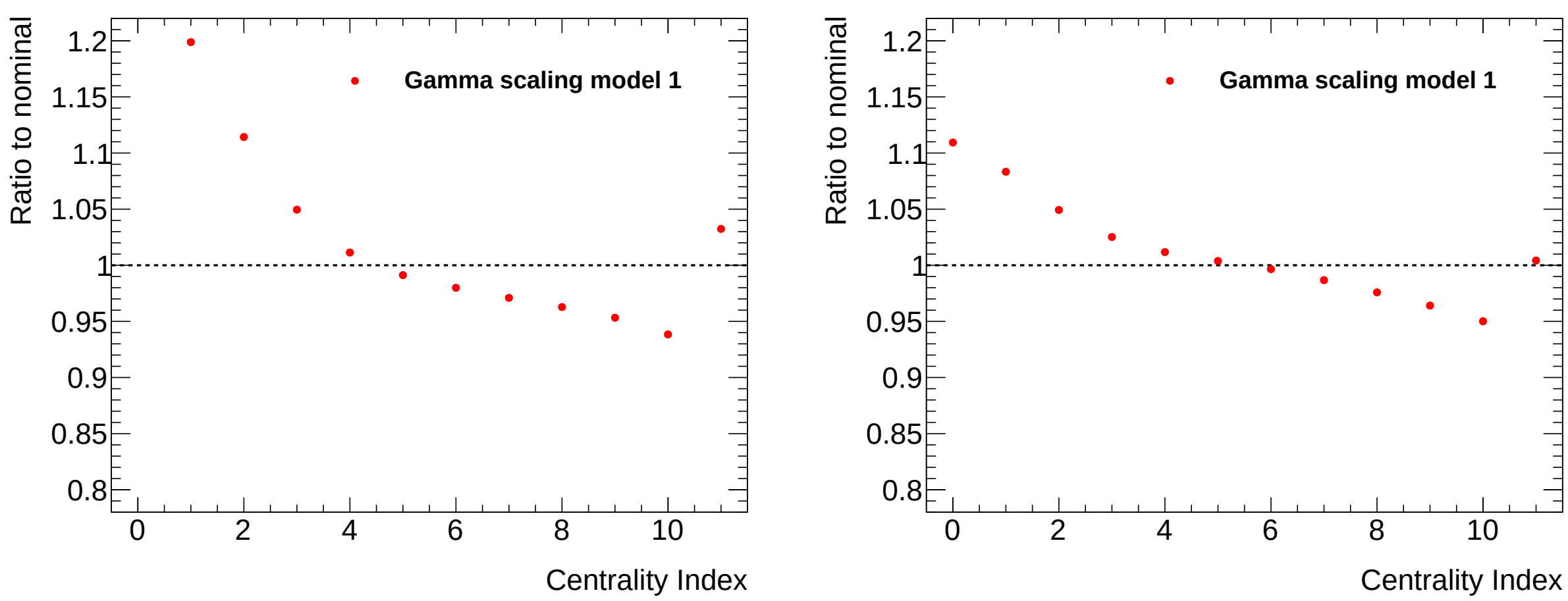

Figure 4.27: Systematic variation of $T_{AB}$ generated by varying the scaling model to model 1 compared to the nominal model 2, and calculated using Glauber (**Left**) and Glauber-Gribov (**Right**) of each centrality class.

# Chapter 5

## Measurement of Direct Photon Production

Measurements of particle and jet production rates at large transverse energy are a fundamental method of characterising hard-scattering processes in all collision systems. In collisions involving large nuclei, production rates are modified from those measured in proton+proton (*pp*) collisions due to a combination of initial- and final-state effects. The former arise from the dynamics of partons in the nuclei prior to the hard-scattering process, while the latter are attributed to the strong interaction of the emerging partons with the hot nuclear medium formed in nucleus–nucleus collisions. As discussed in Sec.2.4.4, modification due to the nuclear environment is quantified by the nuclear modification factor, $R_{\mathrm{AA}}$, defined as the ratio of the cross-section measured in A+A to that in *pp* collisions, scaled by the expected difference in the degree of geometric overlap between the systems.

Measurements of prompt photon production rates offer a way to isolate the initial-state effects because the final-state photons do not interact strongly. These initial-state effects include the degree to which parton densities are modified in a nuclear environment [109–111], as well as potential modification due to an energy loss arising through interactions of the partons traversing the nucleus prior to the hard scattering [112, 113]. Constraints on such initial-state effects are particularly important for characterising the observed modifications of strongly interacting final states, such as jet and hadron production [89, 114], since they are sensitive to effects from both the initial and final state. Due to the significantly simpler underlying-event conditions in proton–nucleus collisions, measurements of photon rates can be performed with better control over systematic uncertainties than in nucleus–nucleus collisions, allowing a more precise constraint on these initial-state effects.

Prompt photon production has been extensively measured in *pp* collisions at a variety of collision energies [115–119] at the LHC. It was also measured in lead–lead (Pb+Pb) collisions at a $\sqrt{s_{\mathrm{NN}}} = 2.76$ TeV [25, 26] at the LHC, and in gold–gold collisions at $\sqrt{s_{\mathrm{NN}}} = 200$ GeV at RHIC [27], where the data from both colliders indicate that photon production rates are unaffected by the passage of the photons through the QGP. At RHIC, photon production rates were measured in deuteron–gold collisions at $\sqrt{s_{\mathrm{NN}}} = 200$ GeV [120, 121] and were found to be in good agreement with pQCD calculations. Additionally, jet production [18, 19] and electroweak boson production [122–124] were measured in 28 nb$^{-1}$ of $p$+Pb collision data at $\sqrt{s_{\mathrm{NN}}} = 5.02$ TeV recorded at the LHC; the former is a strongly interacting final state, while the latter is not. All measurements provided some constraints on initial-state effects.

The data used in this measurement are described in Sec. 3.3. By convention, the results are reported as a function of photon pseudorapidity in the nucleon–nucleon collision frame, $\eta^*$, with positive $\eta^*$ corresponding to the proton beam direction, and negative $\eta^*$ corresponding to the Pb beam direction. Because photons are massless, their pseudorapidity is equal to their rapidity. Furthermore, when discussing detector level quantities, $\eta^{\mathrm{lab}}$ will sometimes be used to refer to the *lab* coordinates.

At leading order, the process $p+\mathrm{Pb} \to \gamma + X$ has contributions from direct processes, in which the photon is produced in the hard interaction, and from fragmentation processes, in which it is produced in the parton shower. Beyond leading order the direct and fragmentation components are not separable and only their sum is a physical observable. To reduce contamination from the dominant background of photons mainly from light-meson decays in jets, the measurements presented here require the photons to be isolated from nearby particles. This requirement also acts to reduce the relative contribution of fragmentation photons in the measurement, and thus, the same fiducial requirement must be imposed on theoretical models when comparing with the data. Specifically, as in previous ATLAS measurements [116, 117], the sum of energy transverse to the beam axis within a cone of $\Delta R \equiv \sqrt{(\Delta\eta)^2 + (\Delta\phi)^2} = 0.4$ around the photon, $E_{\mathrm{T}}^{\mathrm{iso}}$, is required satisfy

$$E_{\mathrm{T}}^{\mathrm{iso}} < 4.8 + 4.2 \times 10^{-3} E_{\mathrm{T}}^{\gamma}\ [\mathrm{GeV}] \qquad (5.1)$$

where $E_{\mathrm{T}}^{\gamma}$ is the transverse energy of the photon. At particle level in simulations and calculations, $E_{\mathrm{T}}^{\mathrm{iso}}$ is calculated as the sum of transverse energy of all particles with a decay length above 10 mm, excluding muons and neutrinos. This sum is corrected for the ambient contribution from uncorrelated soft (underlying-event) particles, consistent with the previous measurements [116, 117].

The results report a measurement of the cross-section for prompt, isolated photons in $p$+Pb collisions. Photons are measured with $E_{\mathrm{T}}^{\gamma} > 20$ GeV, the isolation requirement detailed above, and in three nucleon–nucleon centre-of-mass pseudorapidity ($\eta^*$) regions, $-2.83 < \eta^* < -2.02$, $-1.84 < \eta^* < 0.91$, and $1.09 < \eta^* < 1.90$. In addition to the cross-section, the data are compared to a $pp$ reference cross-section derived from a previous measurement of prompt photon production in $pp$ collisions at $\sqrt{s} = 8$ TeV that used the identical isolation condition [116]. The nuclear modification factor $R_{p\mathrm{Pb}}$ is derived in each pseudorapidity region, using an extrapolation for the different collision energy and center-of-mass pseudorapidity selection, and is reported in the region $E_{\mathrm{T}}^{\gamma} > 25$ GeV where reference data is available. Furthermore, the ratio of $R_{p\mathrm{Pb}}$ in the forward region to that in the backward region is presented to take advantage of the strong cancellation of systematic uncertainties. The measurements are compared with next-to-leading-order (NLO) pQCD predictions from JETPHOX [87] using parton distribution functions (PDF) extracted from global analyses that include nuclear modification effects [58, 125]. Additionally, the data are compared with predictions from a model of initial-state energy loss [112, 113, 126].

## 5.1 Data and Event Selection

### 5.1.1 Trigger and Event Selection

Events are selected for analysis from multiple high-level triggers (HLT) which require a reconstructed photon at the trigger level with $E_{\mathrm{T}}^{\gamma}$ above some minimum threshold. At the trigger level, only a very loose identification requirement (detailed in Sec. 5.3.2) is imposed to reject hadron backgrounds. These are the `HLT_gX_Loose` triggers, with the online thresholds `X`$= 15, 20, 25, 30, 35$ (in GeV). All triggers with thresholds less than 35 are prescaled, meaning an event selected by the trigger is recorded only after some number of that same trigger has fired; this is done in cases in which the requirement would accept too many events and swamp the bandwidth. The five triggers then span the range of the measurement.

Events selected by each trigger are used to populate a specific kinematic region of the spectrum measurement. Specifically, each $E_{\mathrm{T}}^{\gamma}$ region is filled by the trigger containing the largest number of photons in the kinematic range and which is on its efficiency plateau in the region. In practice, this means that each HLT trigger solely populates the photon $E_{\mathrm{T}}^{\gamma}$ region which begins 5 GeV above its online threshold value, until a higher-threshold trigger takes over at higher-$E_{\mathrm{T}}^{\gamma}$ via explicit cuts. In addition, at least one reconstructed vertex is required. Table 5.1 summarizes the triggers used, as well as the kinematic regions they populate, and the luminosity sampled by each trigger in each running period.

Table 5.1: Photon triggers used in analysis, with the corresponding offline $E_{\mathrm{T}}^{\gamma}$ range where they are used, and sampled luminosities in both running periods.

| Trigger | offline $E_{\mathrm{T}}$ region | $L_{\mathrm{int}}$ ($p$+Pb period) | $L_{\mathrm{int}}$ (Pb+$p$ period) |
|---|---|---|---|
| HLT_g15_Loose | 20-25 GeV | 0.691 $\mathrm{nb}^{-1}$ | 0.690 $\mathrm{nb}^{-1}$ |
| HLT_g20_Loose | 25-30 GeV | 1.515 $\mathrm{nb}^{-1}$ | 1.913 $\mathrm{nb}^{-1}$ |
| HLT_g25_Loose | 30-35 GeV | 3.237 $\mathrm{nb}^{-1}$ | 4.592 $\mathrm{nb}^{-1}$ |
| HLT_g30_Loose | 35-40 GeV | 5.935 $\mathrm{nb}^{-1}$ | 8.791 $\mathrm{nb}^{-1}$ |
| HLT_g35_Loose | $> 40$ GeV | 59.88 $\mathrm{nb}^{-1}$ | 110.62 $\mathrm{nb}^{-1}$ |

The efficiency of the high-level trigger for selecting offline, loose photons is determined in data. Since there is not sufficient minimum bias data to directly determine the HLT efficiencies with high precision, the trigger efficiency is instead determined separately for each part of the trigger chain, using supporting HLT, L1 pass-through and MB triggers in data. The efficiency is always reported as a function of offline photon $E_{\mathrm{T}}^{\gamma}$, for loose-identified photons.

(1) Using minimum bias triggered data (`HLT_mb_sptrk_L1MBTS_1`), detailed in Sec. 4.2, the efficiency of the Level-1 EM trigger is determined for different thresholds. Offline loose photons in these events are considered to fire the trigger if they are geometrically matched with an L1 EM RoI above the given threshold. This efficiency is estimated separately for all of the L1 EM thresholds which seed the HLT triggers used in this analysis.

(2) Using supporting L1 pass-through triggers (`HLT_noalg_L1EM`$X$ for different thresholds $X$), the efficiency of the HLT trigger without any identification requirement (e.g. `etcut`) is evaluated. This condition is checked for different HLT $E_{\mathrm{T}}$ thresholds. Offline loose photons in these L1-triggered events are geometrically matched to an HLT photon (without any ID requirement) above the given threshold.

(3) Using supporting HLT photon triggers without online identification (HLT_g$X$_etcut, for different thresholds $X$), the HLT photon efficiency with loose identification is determined. This condition is checked for different HLT thresholds separately. Offline loose photons in these HLT-etcut triggered events are checked to see if the HLT photon they are matched pass the online loose ID selection.

In this way, the total efficiency is factorized into three schematic components and examined separately in each: (1) the efficiency for an offline photon to be identified at L1; (2) the efficiency for the photon to be identified at the HLT level; (3) the efficiency of photon to pass online identification cuts. Moreover, efficiencies are evaluated separately for photons in the barrel ($|\eta| < 1.37$) and end-cap ($1.56 < |\eta| < 2.37$).

Figure 5.1 summarizes the three efficiency components for each of the four HLT triggers used in the analysis. In the $E_{\mathrm{T}}^{\gamma}$ region where each trigger is used (5 GeV above the HLT online threshold), the L1 and HLT without ID efficiencies saturate at 100%. However, the identification at the online level introduces a small inefficiency. For offline loose identified photons, this last efficiency saturates at 99 % at mid-rapidity and 98% at forward rapidity. However, when this is evaluated for offline tight identified photons (the signal selection used in the analysis), these efficiencies are greater than 99.5% in both regions. The resulting inefficiency has a much smaller effect on the experimental measurement than the dominant systematic uncertainties, thus, no correction for the trigger efficiency is applied or systematic uncertainty assigned.

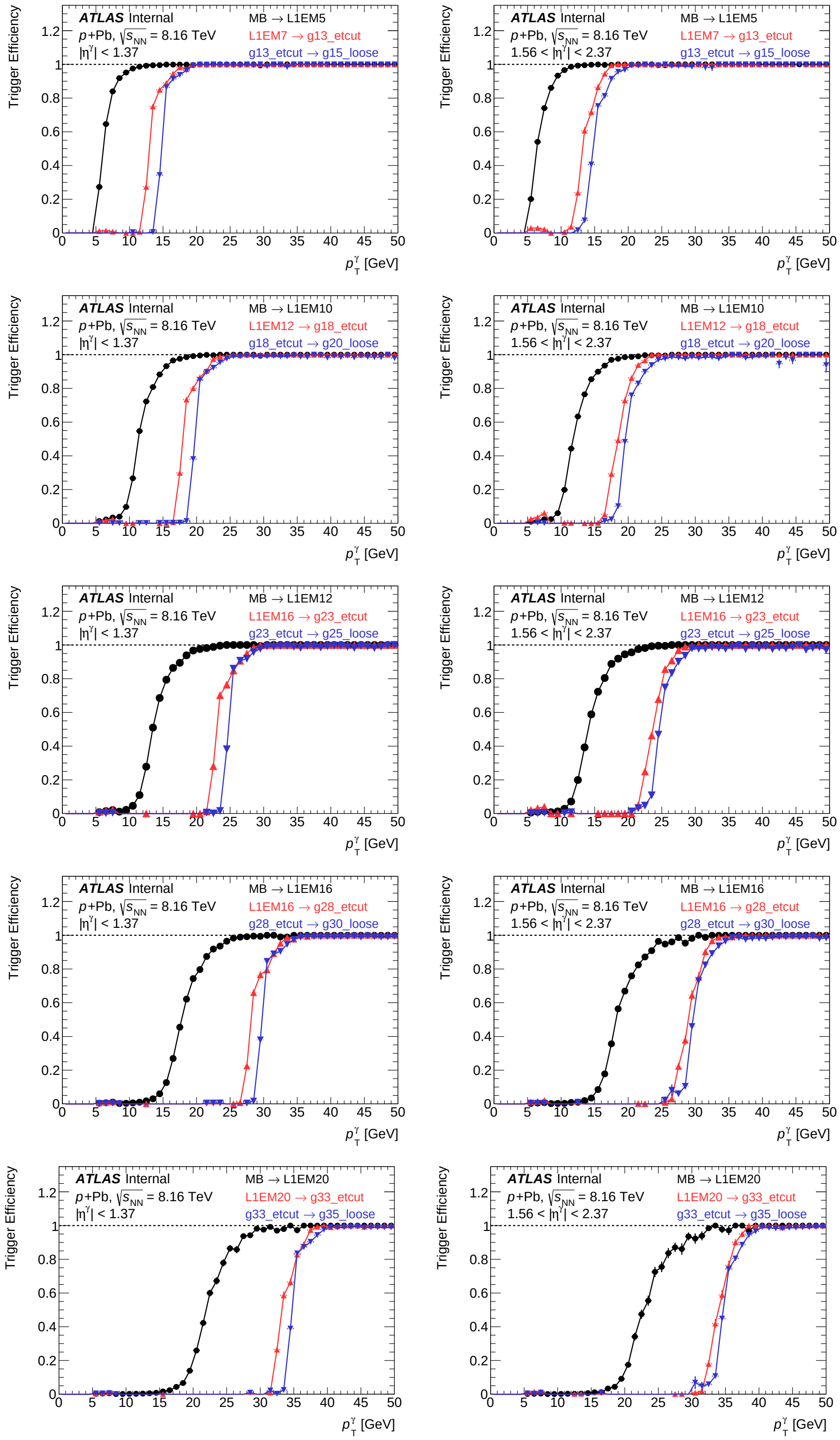

Figure 5.1: Trigger efficiency for each "stage" of the total efficiency for HLT loose triggers. Each row shows the stages for a particular HLT loose threshold, while the left and right columns show the efficiency in the barrel and end-cap regions. Efficiencies are always shown with respect to offline loose photon. Separate efficiencies are shown for the L1 efficiency in MB events (black), the HLT efficiency without ID in L1 pass-through triggered events (red), and the HLT efficiency with online loose ID requirement in HLT trigger photon events without ID (blue).

## 5.2 Simulated Event Samples

Samples of MC simulated events were generated to study the detector performance for signal photons. Proton–proton generators were used as the source of events containing photons. To include the effects of the $p$+Pb underlying-event environment, these simulated $pp$ events were combined with minimum bias $p$+Pb events from data before reconstruction in a process called *data overlay*. In this way, the simulated events contain the effects of the $p$+Pb underlying-event identical to those observed in data.

The Pythia 8.186 [127] generator was used to generate the nominal set of MC events, with the NNPDF23LO PDF set [128] and a set of generator parameters tuned to reproduce minimum-bias $pp$ events with the same collision energy as that in the $p$+Pb data ("A14" tune) [129]. A centre-of-mass boost was applied to the generated events to bring them into the same laboratory frame as the data. The generator simulates the direct photon contribution and, through final-state QED radiation in $2 \to 2$ QCD processes, also includes the fragmentation photon contributions; these components are defined to be signal photons. Events were filtered in six exclusive $E_{\mathrm{T}}^{\gamma}$ ranges from 17 GeV to 500 GeV.

An additional MC sample was used to assess the sensitivity of the measurement to this choice of generator. The Sherpa 2.2.4 [130] event generator produces fragmentation photons in a different way from Pythia and was thus chosen for the comparison. The NNPDF3.0NNLO PDF set [53] was used, and the events were generated in the same kinematic regions as the Pythia events. These events were generated with leading-order matrix elements for photon-plus-jet final states with up to three additional partons, which were merged with the Sherpa parton shower.

The Pythia and Sherpa $pp$ events were passed through a full Geant4 simulation of the ATLAS detector [106, 107]. To model the underlying event (UE) effects, each simulated event was combined with a minimum-bias $p$+Pb data event and the two were reconstructed together as a single event, using the same algorithms as used for the data. These events were split between the two beam configurations in a proportion matched to that in data-taking. The UE activity levels, as characterized by the FCal $\Sigma E_{\mathrm{T}}^{\mathrm{Pb}}$, are different in the photon-containing data events from the minimum-bias data events used in the simulation. Thus, the simulated events were weighted on a per-event basis to match the UE activity distribution in data.

Generally, distributions in the MC are constructed via a weighted sum of the distribution in each sub-sample, where the weighting accounts for differences in the cross-section ($\sigma_{\rm DP}$), generator-level filter efficiency ($\epsilon_{\rm DP}$), and total sample statistics ($N_{\rm evt}^{\rm DP}$), e.g.

$$\mathcal{O} = \sum_{\rm DP} \mathcal{O}_{\rm DP} \left(\sigma_{\rm DP}/\epsilon_{\rm DP} N_{\rm evt}^{\rm DP}\right) \tag{5.2}$$

Tables A.1 and A.2 in the Appendix summarize the MC samples used in this analysis.

### 5.2.1 Generator-Level Definition and Spectra

At the generator level, prompt isolated photons are defined as those which are photons (Pythia particle code = 22), final-state, and primary particles. Additionally, the generator-level transverse isolation energy, $E_{\rm T}^{\rm iso}$, is defined as the scalar sum of the transverse energies of final-state, primary particles (excluding muons and neutrinos) within a $\Delta R < 0.4$ cone around the photon. To account for the effects of particle level UE on the truth isolation energy, the ambient-energy-density is determined event-by-event using the jet area method [131], and subtracted from the truth isolation energy before imposing the isolation requirement. Figure 5.2 shows the average ambient-energy-density as a function of truth photon $E_{\rm T}^{\gamma}$ in each pseudorapidity region for both Pythia and Sherpa. This correction ends up having only a sub-percent level effect on the resulting cross sections as can be seen in the ratios in Figure 5.3.

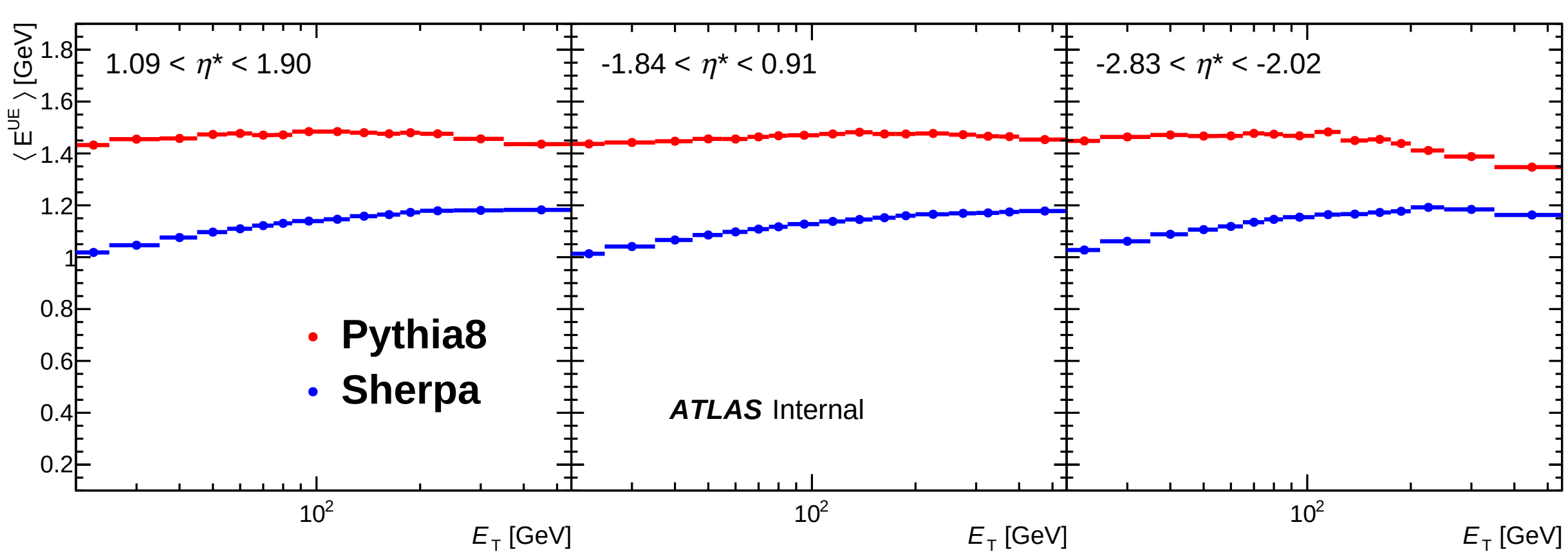

Figure 5.2: Average ambient-energy-density as determined, event-by-event at particle level using the jet area method, and plotted as a function of truth photon $E_{\rm T}^{\gamma}$ in each pseudorapidity region for both Pythia and Sherpa.

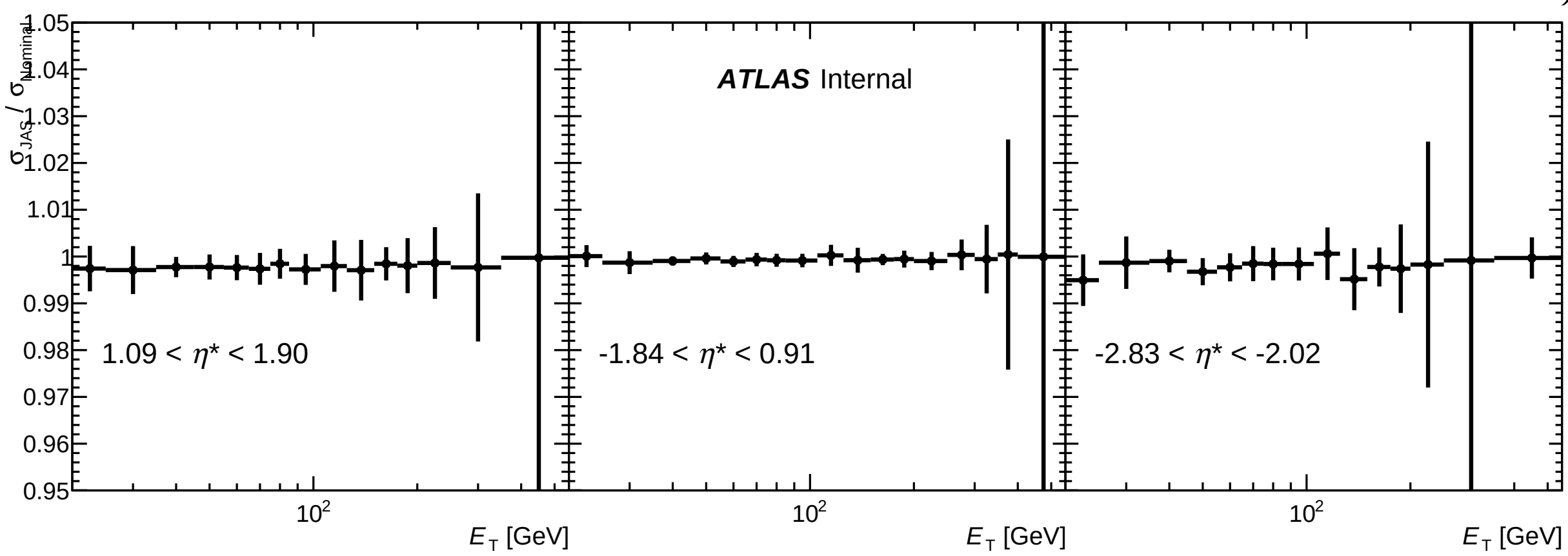

Figure 5.3: Ratio of resulting photon cross section measurements with and without jet-area subtraction of the UE.

Figure 5.4 shows a scatterplot of the generator-level $E_T^{iso}$ as a function of $E_T^{\gamma}$, indicating the region in which photons are defined to be isolated. Additionally, it gives an example of how the full isolated photon cross-section is built out of the spectra in individual MC sub-samples.

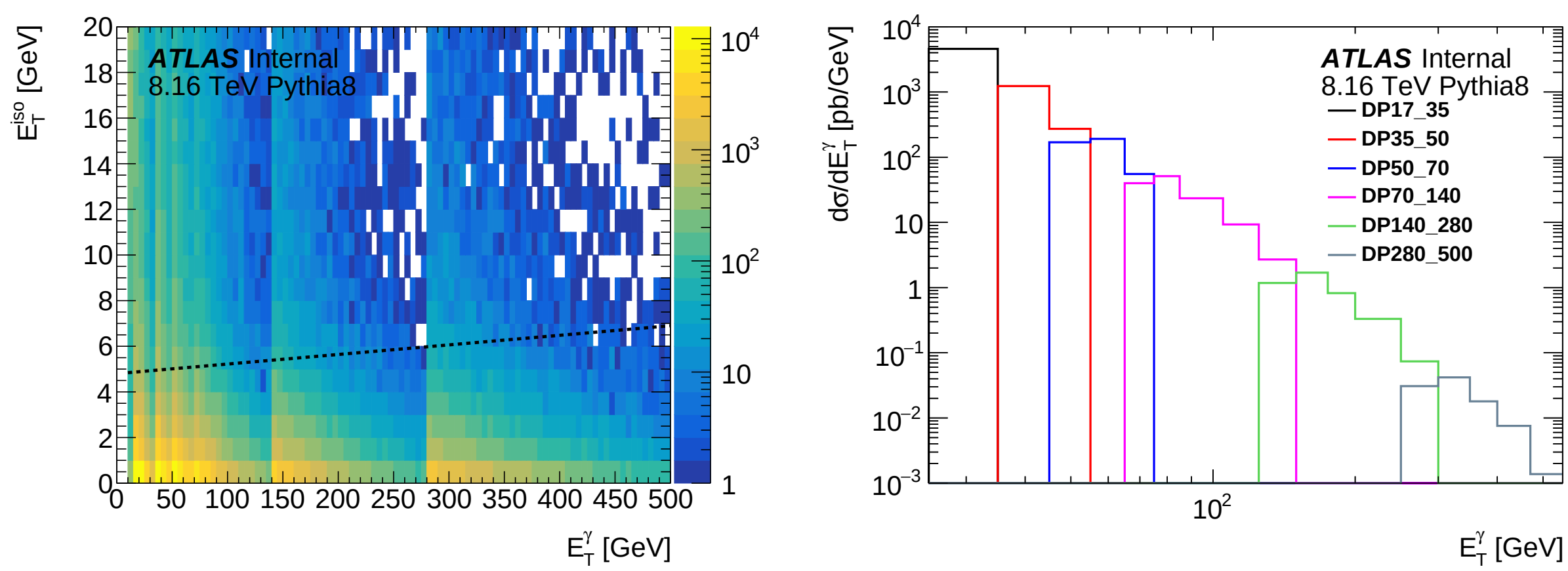

Figure 5.4: **Left:** Distribution of generator-level isolation values as a function of generator-level $E_T^{\gamma}$. All photon $E_T^{\gamma}$ slices are included without cross-section weighting. The dashed line indicates the position of the isolation cut. **Right:** Isolated photon cross-section at the generator-level in Pythia events. The binning matches the binning used in the analysis, with the (weighted) contribution from each DP sample shown as a different color.

## 5.3 Photon Reconstruction and Identification

Photons are reconstructed and identified following a procedure used extensively in previous ATLAS measurements [117], of which only the main features are summarised here. Efficiencies for each step of the process are calculated in each period separately to check for period differences, and the final values are combined as the luminosity weighted average.

### 5.3.1 Photon Reconstruction

As detailed in Refs. [132, 133], photon candidates are reconstructed from clusters of energy deposited in the electromagnetic calorimeter in three regions corresponding to the laboratory-frame ($\eta^{\mathrm{lab}}$) positions of the barrel and forward and backward end-caps $\left|\eta^{\mathrm{lab}}\right| < 2.37$. The transition region between the barrel and end-cap calorimeters, $1.37 < \left|\eta^{\mathrm{lab}}\right| < 1.56$, is excluded due to its higher level of inactive material. The measurement of the photon energy is based on the energy collected in calorimeter cells in an area of size $\Delta\eta \times \Delta\phi = 0.075 \times 0.175$ in the barrel and $\Delta\eta \times \Delta\phi = 0.125 \times 0.125$ in the end-caps. It is corrected via a dedicated energy calibration [134] which accounts for losses in the material before the calorimeter, both lateral and longitudinal leakage, and for variation of the sampling-fraction with energy and shower depth.

The reconstruction efficiency is calculated in MC by computing the fraction of truth isolated photons, $N^{\gamma}_{\mathrm{TIso}}$, which are reconstructed, $N^{\gamma}_{\mathrm{TIso,reco}}$, in each $E^{\gamma}_{\mathrm{T}}$ and $\eta$ bin. The reconstructed photons are required to match the truth level photon geometrically within an $\Delta$R distance of 0.2.

$$\epsilon_{\mathrm{reco}} = \frac{N^{\gamma}_{\mathrm{TIso,reco}}}{N^{\gamma}_{\mathrm{TIso}}} \tag{5.3}$$

The efficiency is plotted in Fig. 5.5 showing 99% in the barrel, and 98% in the end caps and similar values for $p$+Pb and Pb+$p$ periods.

### 5.3.2 Photon Identification

The photons are identified using two sets of calorimeter shower shape requirements corresponding to *loose* and *tight* selections described in detail in Ref. [133]. The tight requirements select clusters which are compatible with originating from a single photon impacting the calorimeter. Photon identification criteria is designed to minimize background from non-prompt sources and is based on the classification of the full shower shape in the calorimeters. Nine shower shape variables are created using information from the hadronic calorimeter, the lateral shower shape in the second layer of the electromagnetic calorimeter, and the detailed shower shape in the finely segmented first layer. A pictorial representation of these variables is given in Fig. 5.6. Criteria on each of them together form the tight identification category. The loose selection class is designed to be less discriminating and accept a larger number of background and decay photons. This is done by relaxing some of the second EM and hadronic layer's shower shape criteria and not using the information from the first layer. In this analysis, the loose selection is only used at the triggering level to select both direct and background photons (which are needed for the purity determination).

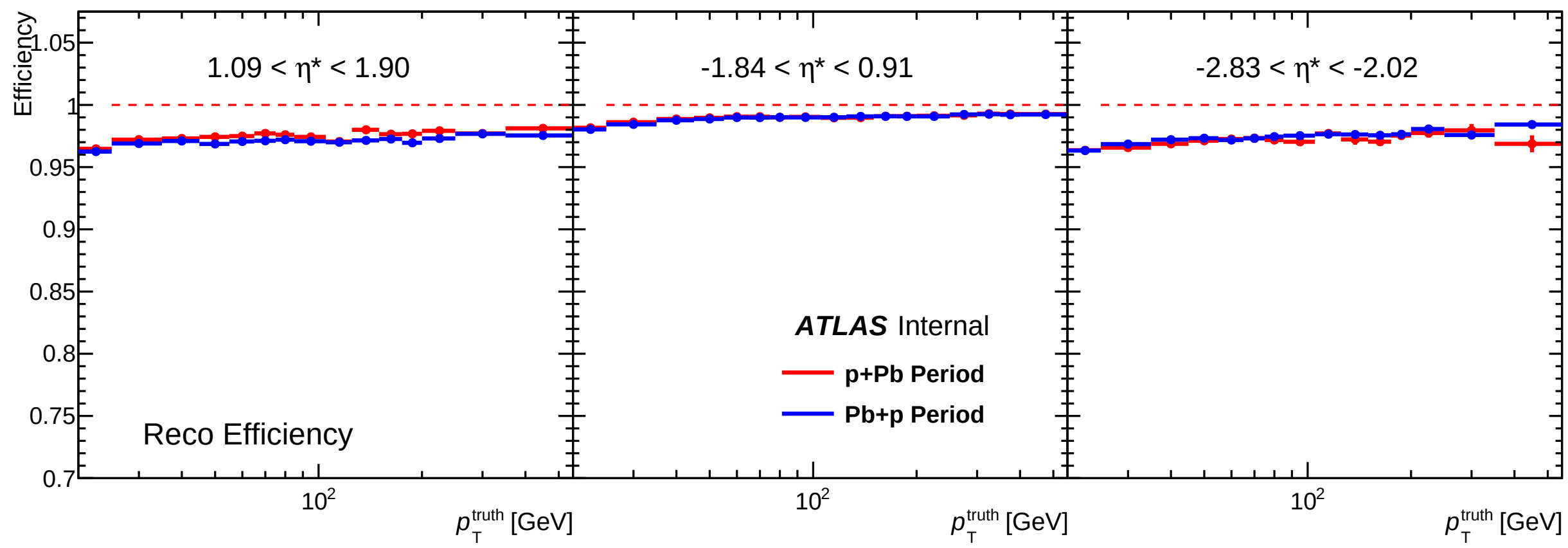

Figure 5.5: Reconstruction efficiency for truth isolated photons plotted as a function of $E_T^{\gamma}$ in each center of mass pseudorapidity slice for both running periods.

The same selection scheme is applied to MC after corrections or *fudge factors* are imposed to make the distributions better agree with data. Fig. 5.7 gives an example of one shower shape variable, $R_\eta$, in data as compared with MC before and after fudging for each pseudorapidity region and a representative $E_T^{\gamma}$ bin. Examples of the rest of the variables can be found in Appendix A.7. Furthermore, when plotting the MC photons, a small additional correction is applied as a weight to correct an observed discrepancy between the tight I.D. efficiencies from MC and those derived from data driven studies.

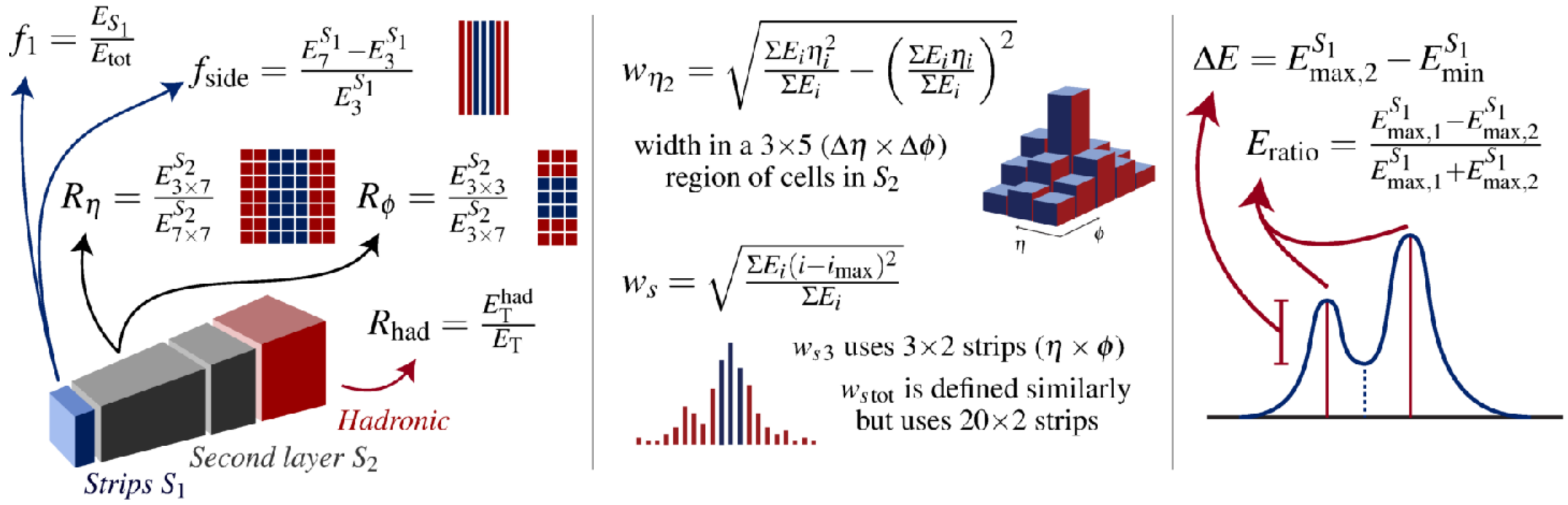

Figure 5.6: A pictorial representation of the shower shape variables used for photon identification.

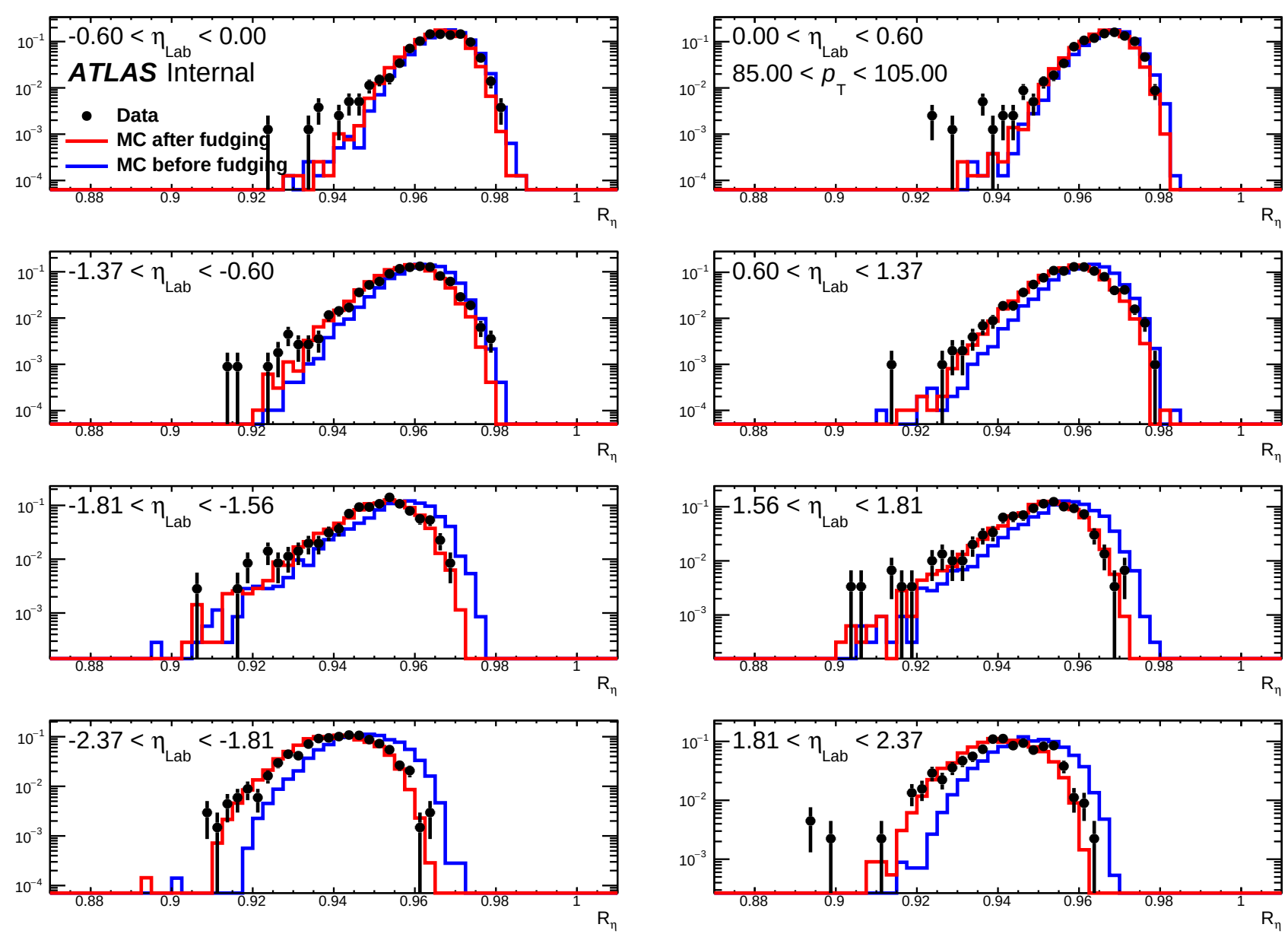

Figure 5.7: Shower shape parameter, $R_\eta$, in each pseudorapidity slice from representative $E_{\mathrm{T}}^{\gamma}$ bin ($85\,\mathrm{GeV} < E_{\mathrm{T}}^{\gamma} < 105\,\mathrm{GeV}$). Reconstructed data plotted as black points overlaid with MC before (blue histogram) and after (red histogram) fudging.

The efficiency for tight selection of reconstructed and truth isolated photons is calculated in MC via the equation

$$\epsilon_{\mathrm{tight}} = \frac{N^{\gamma}_{\mathrm{TIso,reco,tight}}}{N^{\gamma}_{\mathrm{TIso,reco}}} , \tag{5.4}$$

and plotted as a function of truth $E_{\mathrm{T}}^{\gamma}$ in each $\eta$ bin in Fig. 5.8.

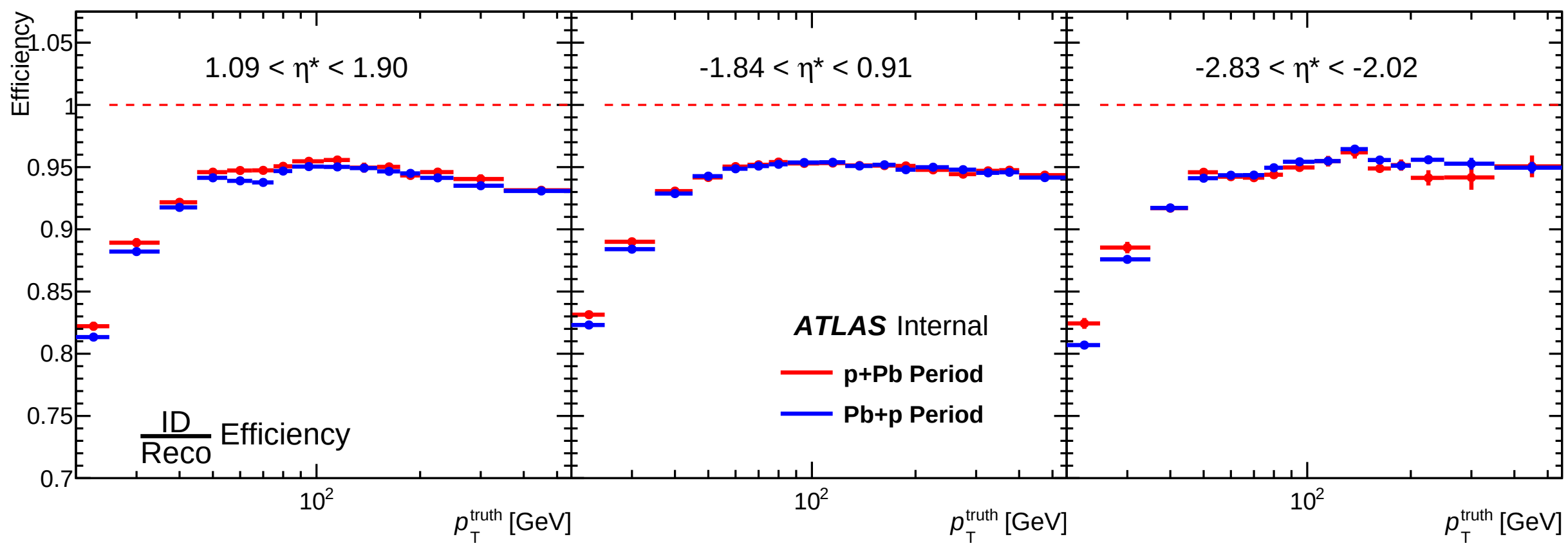

Figure 5.8: Tight identification cut efficiency for reconstructed and truth isolated photons plotted as a function of $E_{\mathrm{T}}^{\gamma}$ in each center of mass pseudorapidity slice for both running periods.

### 5.3.3 Photon Isolation

The $E_{\mathrm{T}}^{\mathrm{iso}}$ is computed from the sum of $E_{\mathrm{T}}$ values in topological clusters of calorimeter cells [135] inside a cone of size $\Delta R = 0.4$ centered on the photon. This cone size is chosen to be compatible with a previous measurement of photon production in $pp$ collisions at $\sqrt{s} = 8$ TeV [116], which is used to construct the reference spectrum for the $R_{p\mathrm{Pb}}$ measurement. This estimate excludes an area of $\Delta\eta \times \Delta\phi = 0.125 \times 0.175$ centred on the photon, and is corrected for the expected leakage of the photon energy from this region into the isolation cone. Photons are then declared isolated if they pass the isolation condition identical to the particle level condition stated in Eqn. 5.1. Isolation energy in MC is shifted to match data for each $E_{\mathrm{T}}^{\gamma}$ and $\eta$ bin. Figure 5.9 shows the relationship between the reconstructed isolation energy to the truth isolation energy, described above. In section 5.6.1.3, this discrepancy is accounted for by deriving an associated systematic uncertainty.

Figure 5.10 shows example $E_{\mathrm{T}}^{\mathrm{iso}}$ distributions for identified and isolated photons, the corresponding distributions for background photons with the normalisation determined by the double-sideband method, discussed in Sec. 5.4, and the resulting signal-photon distributions after background subtraction, compared with those for generator-level photons in MC simulation. The Figure shows the shape of the background distribution within the signal region, and the correspondence between the background-subtracted data and the signal-only Pythia distributions gives confidence that the simulations accurately represent the data.

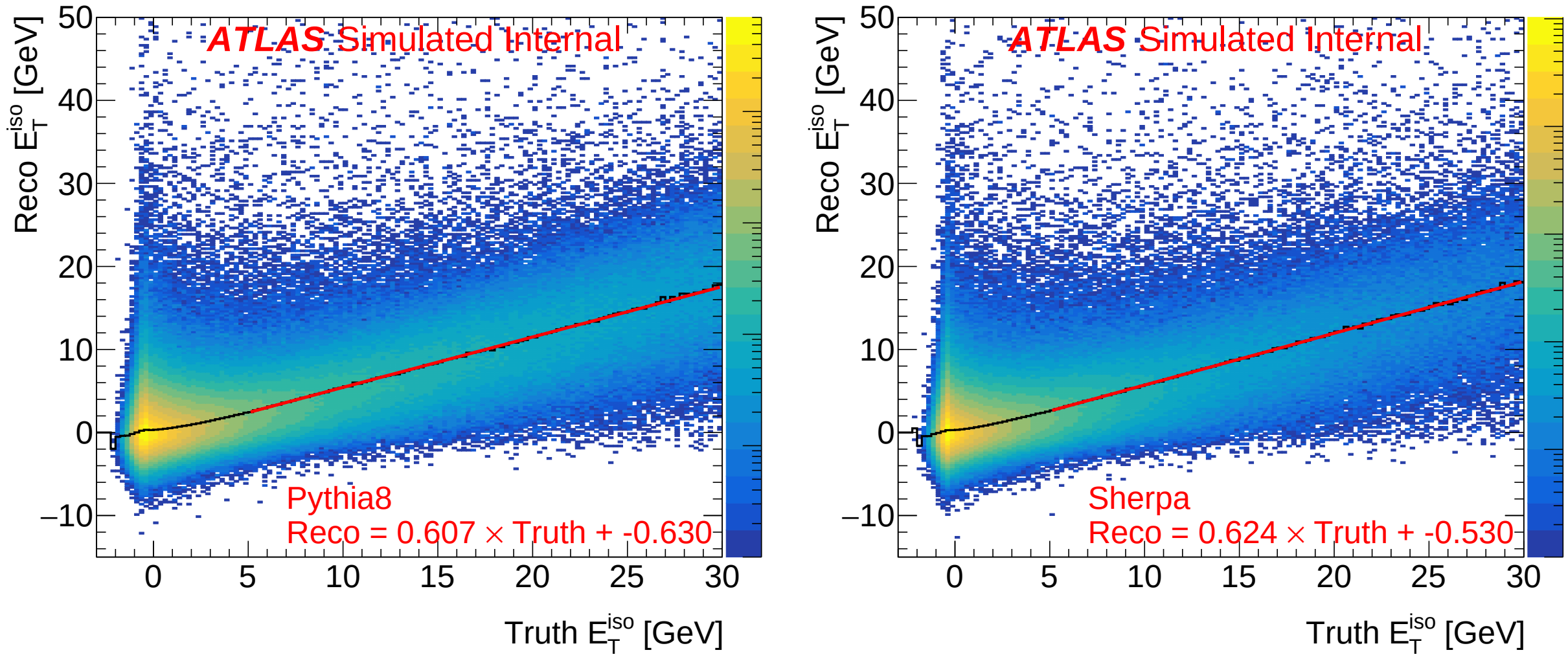

Figure 5.9: A 2d histogram of reconstructed isolation energy vs. the truth transverse energy from final-state, primary particles (excluding muons and neutrinos) particles in a cone of 0.4 around the photon. A profile is plotted as a black line and a linear fit to the profile is plotted in red. **Left:** Pythia data overlay, **Right:** Sherpa data overlay.

The efficiency of the isolation cut is calculated in MC simulations as

$$\epsilon_{\text{Iso}} = \frac{N^{\gamma}_{\text{TIso,reco,tight,Iso}}}{N^{\gamma}_{\text{TIso,reco,tight}}} , \tag{5.5}$$

and plotted as a function of truth $E_{\text{T}}^{\gamma}$ in each $\eta$ bin in Fig. 5.11. The isolation cut efficiency shows a period dependence of about 1% in the backward region. The disabled HEC region can not be the cause of this discrepancy, as it has been removed from the analysis. The root cause of this remains a mystery, and therefore, a flat 1% systematic in the backward region is applied to account for the discrepancy.

This analysis applies no detector level UE corrections to the measured isolation energy. However, the sensitivity to this choice was checked by using the jet-area method [131] with topological calorimeter clusters to estimate the ambient UE energy. Subtraction of this energy from photon isolation cones yields a negligible effect on the measurement.

### 5.3.4 Total Photon Selection Efficiency

The combined reconstruction, tight identification, and isolation efficiency is defined as

$$\epsilon_{\text{total}} = \epsilon_{\text{reco}} \times \epsilon_{\text{tight}} \times \epsilon_{\text{Iso}} , \tag{5.6}$$

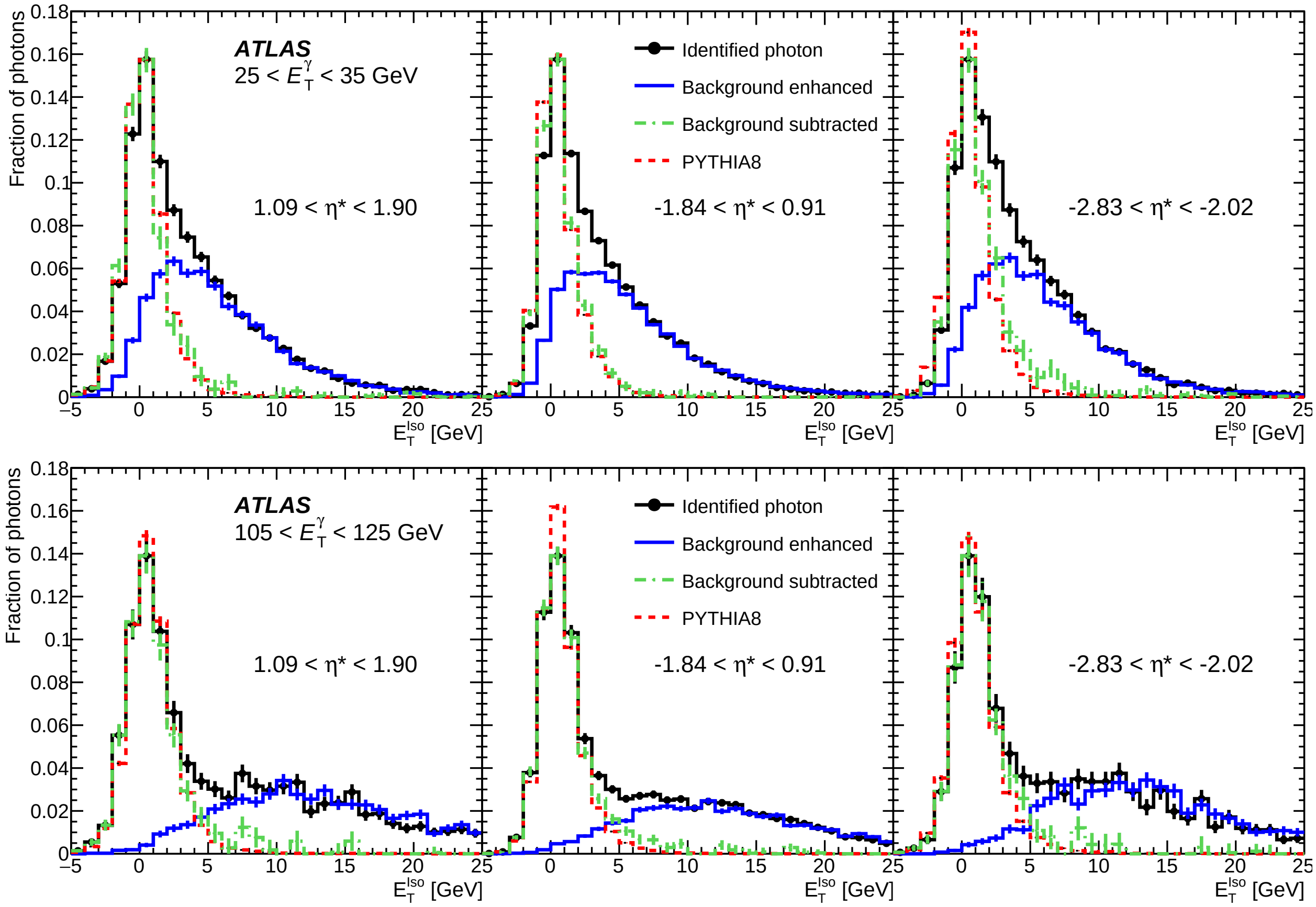

Figure 5.10: Distributions of detector-level photon isolation transverse energy $E_{\mathrm{T}}^{\mathrm{iso}}$ for identified photons in data (black points), background photons scaled to match the data at large $E_{\mathrm{T}}^{\mathrm{iso}}$ (blue solid line), the resulting distribution for signal photons scaled so that the maximum value is the same as that for identified photons (green dot-dashed line), and that for photons in simulation which are isolated at the generator level normalised to have the same integral as the signal photon distribution (red dashed line). Each panel shows a different pseudorapidity region, while the top and bottom panels show the low-$E_{\mathrm{T}}^{\gamma}$ and high-$E_{\mathrm{T}}^{\gamma}$ range respectively. The vertical error bars represent statistical uncertainties only.

and is plotted as a function of truth $E_{\mathrm{T}}^{\gamma}$ for each $\eta$ bin in Fig. 5.12. Tables of efficiency values are printed in Appendix A.3.

A comparison to the total efficiency calculated using the Sherpa data overlay sample is shown for period A in Fig. 5.13 and period B in Fig. 5.14. The Sherpa sample gives an efficiency that is consistent with that from Pythia in all kinematic regions.

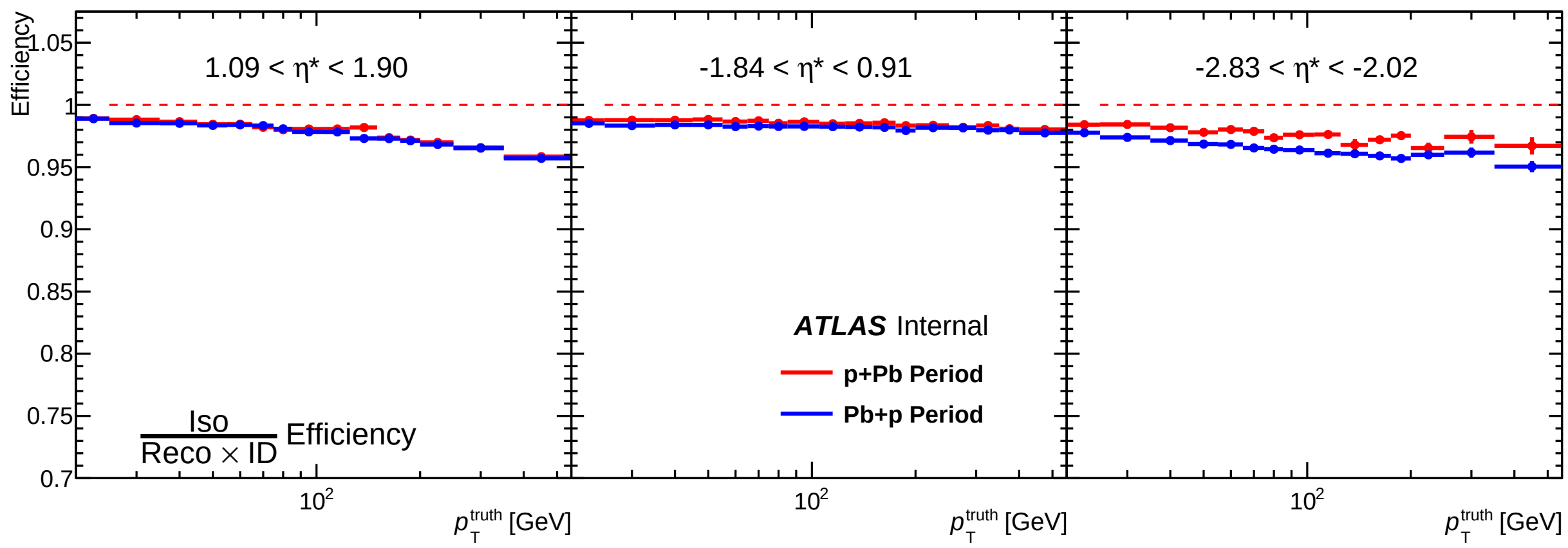

Figure 5.11: Isolation cut efficiency for tight identified, reconstructed, and truth isolated photons plotted as a function of $E_{\mathrm{T}}^{\gamma}$ is each pseudorapidity slice. As a reminder, the photons are only from the region of the detector in which the HEC was live.

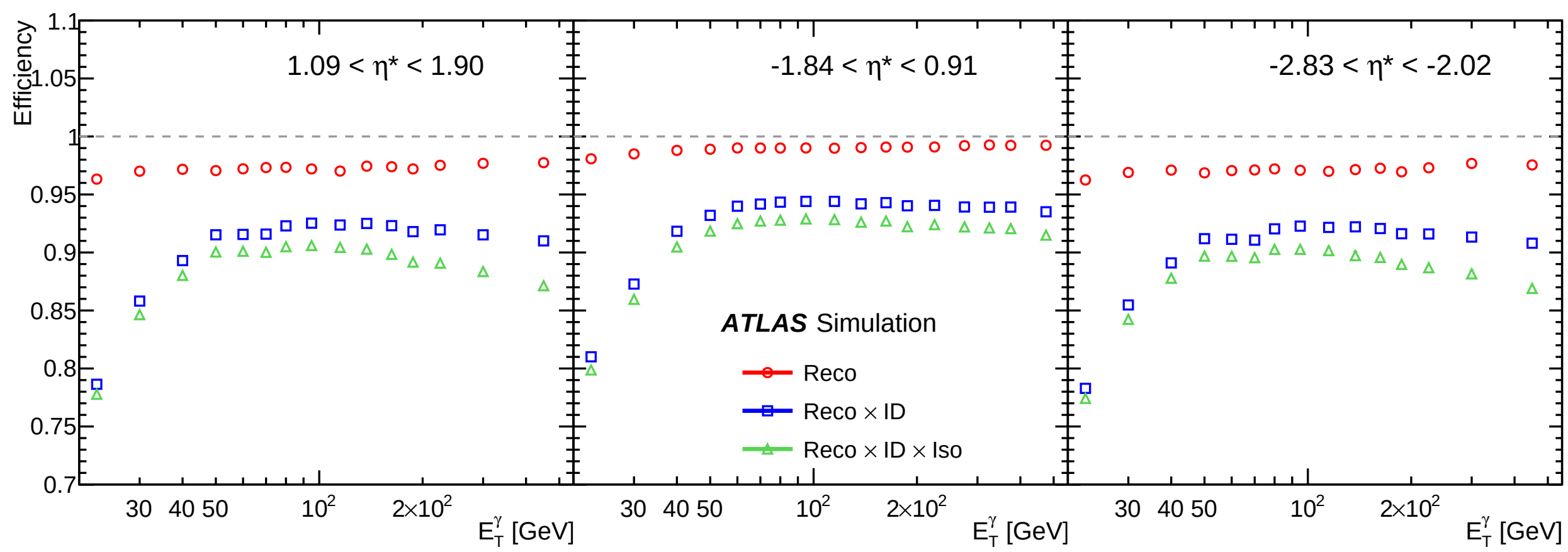

Figure 5.12: Efficiency for simulated photons passing the generator-level isolation requirement, shown as a function of photon transverse energy $E_{\mathrm{T}}^{\gamma}$ with a different pseudorapidity region in each panel. The reconstruction (red circles), reconstruction plus identification (blue squares) and total selection (green triangles) efficiencies are shown separately.

### 5.3.5 Photon Energy Measurement

The photon energy scale and resolution is also determined in MC simulations. In Fig. 5.15, the mean relative energy difference of reconstructed to truth $E_{\mathrm{T}}^{\gamma}$ is plotted for tight identified and isolated photons which quantifies the measured energy scale. The energy scale is within one half of one percent in each region of $\eta$. Fig. 5.16 shows reconstructed photon energy resolution as the width of a Gaussian fit to the energy distribution

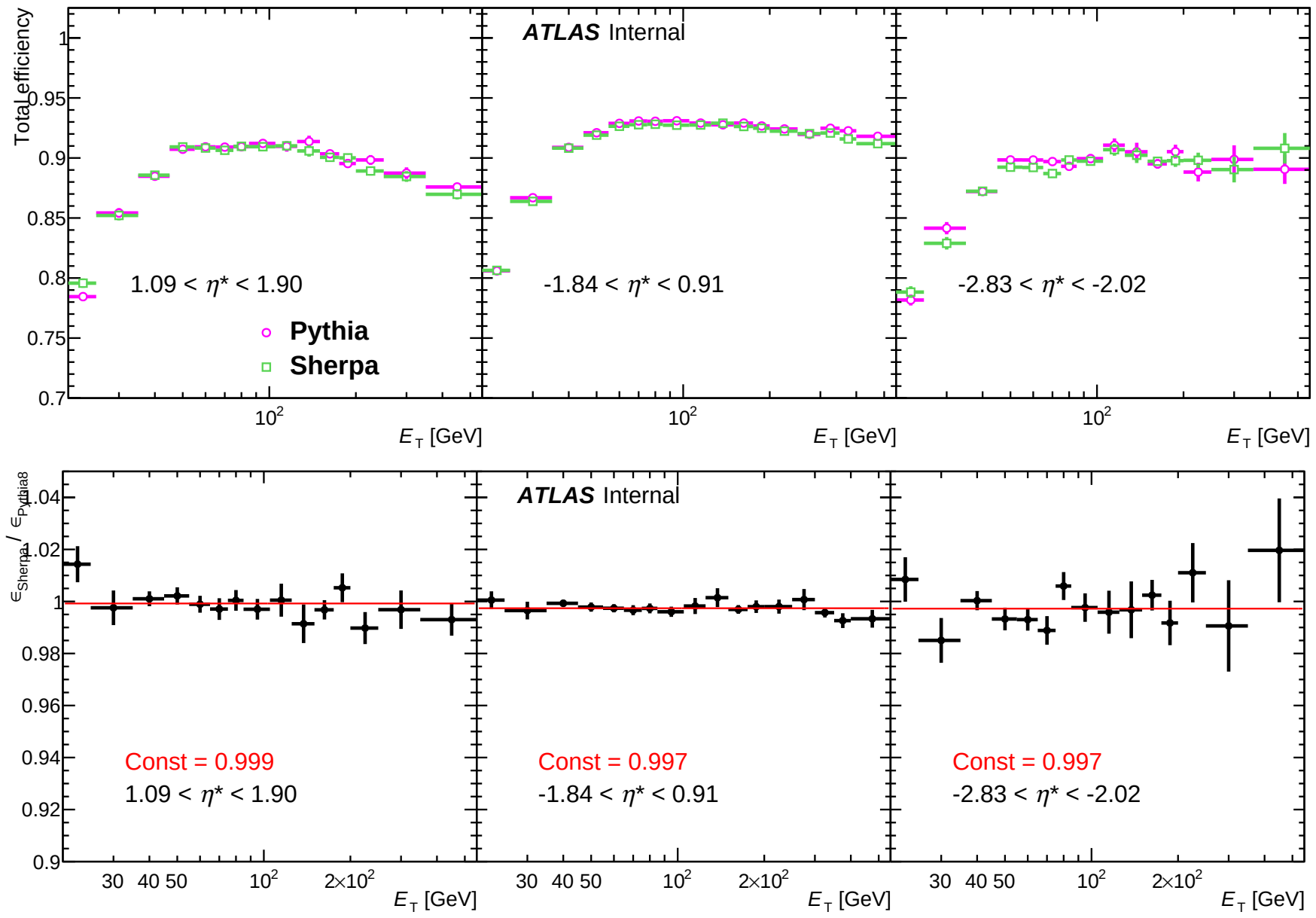

Figure 5.13: Combined reconstruction, tight identification, and isolation cut efficiency for truth isolated photons plotted as a function of $E_{\mathrm{T}}^{\gamma}$ from both Pythia and Sherpa data overlay from the $p$+Pb running period. **Top:** $p$+Pb period, **Bottom:** $p$+Pb period ratio

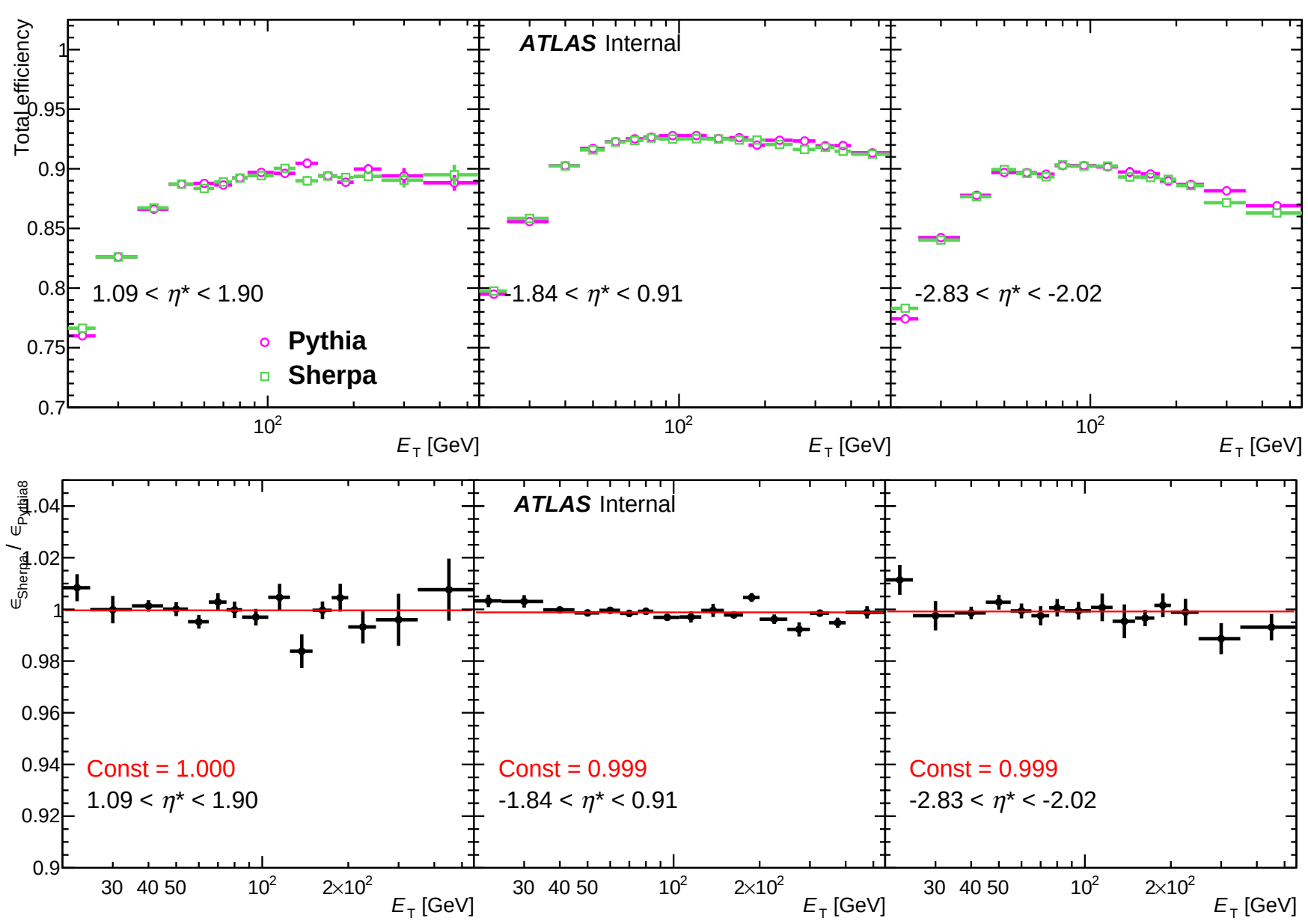

Figure 5.14: Combined reconstruction, tight identification, and isolation cut efficiency for truth isolated photons plotted as a function of $E_{\mathrm{T}}$ from both Pythia and Sherpa data overlay from the Pb+$p$ running period. **Top:** Pb+$p$ period, **Bottom:** Pb+$p$ period ratio

in every given $E_{\mathrm{T}}^{\gamma}$ bin. It is smaller than 3.5% everywhere in the detector. The scale and resolution are as expected, and therefore, we do not make any additional corrections but apply variations as systematic uncertainties.

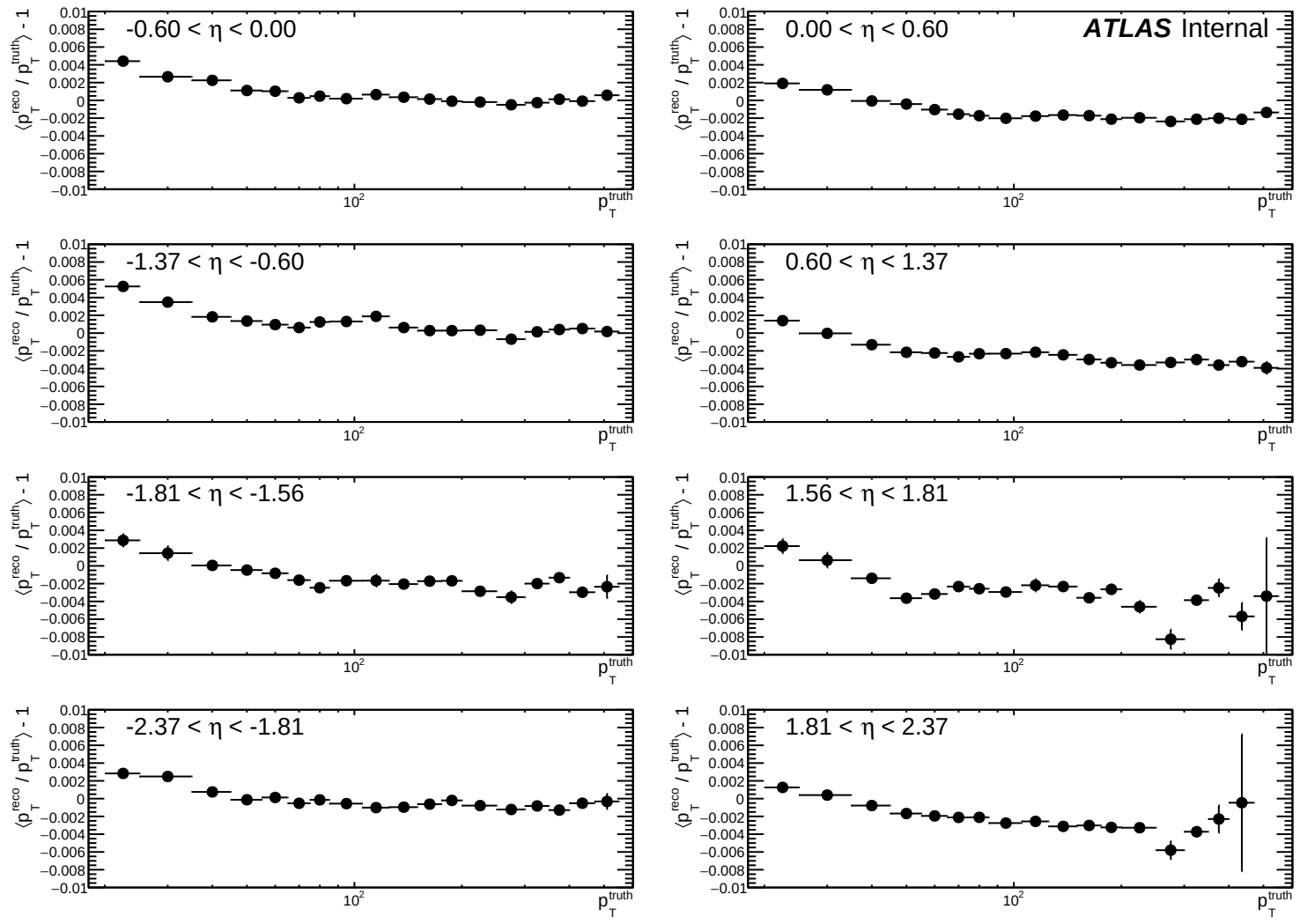

Figure 5.15: Mean relative energy difference of reconstructed to truth $E_{\mathrm{T}}^{\gamma}$ for tight identified and isolated photons plotted as a function of truth $E_{\mathrm{T}}^{\gamma}$ in each pseudorapidity slice.

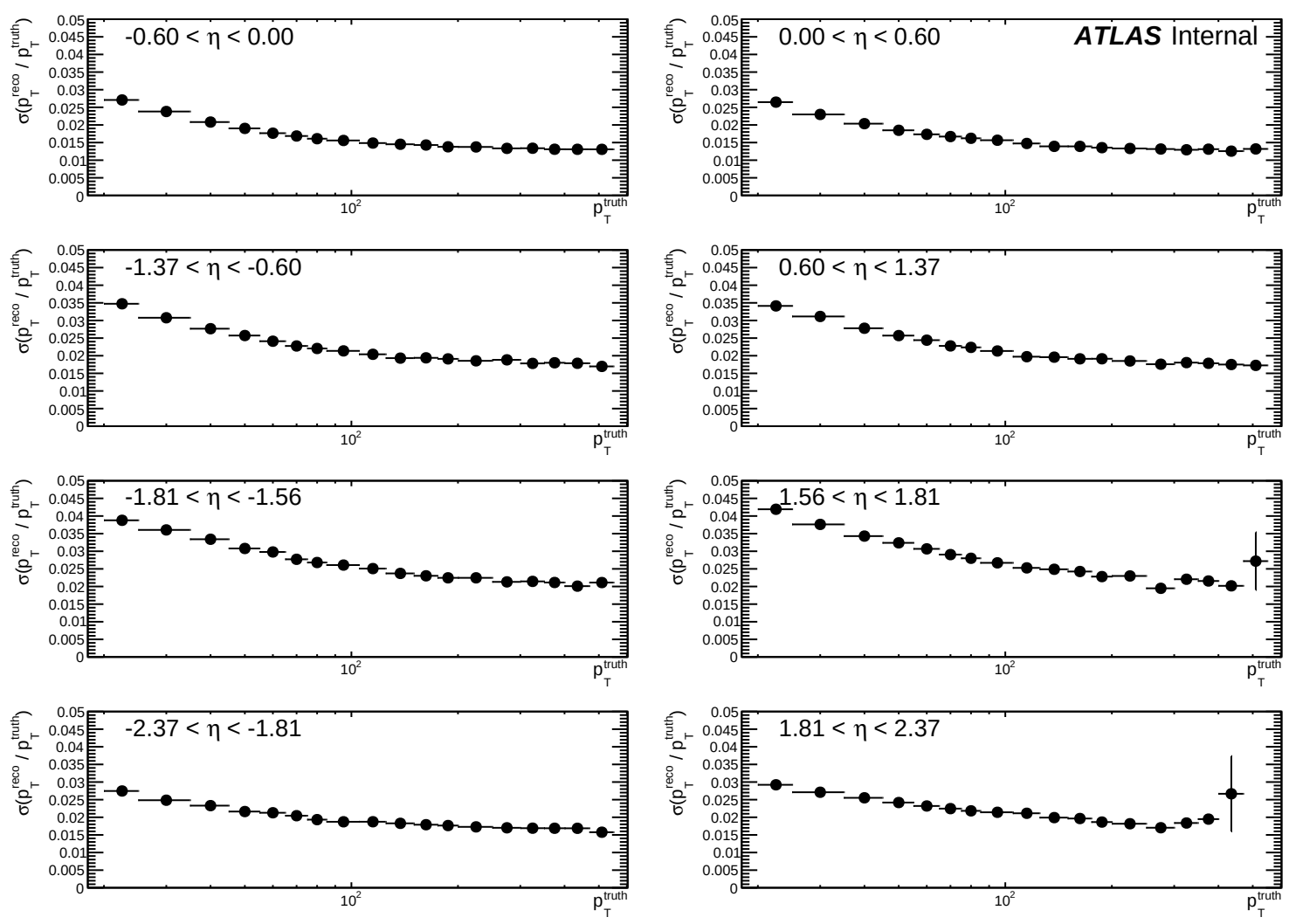

Figure 5.16: Reconstructed photon energy resolution for tight identified and isolated photons plotted as a function of truth $E_{\mathrm{T}}^{\gamma}$ in each pseudorapidity slice.

### 5.3.6 Electron Contamination

The dominant sources of electrons in the present kinematic regions are from heavy flavor meson decays, however, the isolation energy condition strongly suppresses this contribution. The primary source of isolated electrons are from $W^{\pm}$ and $Z$ boson decays as has been noted in previous ATLAS measurements [116, 136]. Mis-reconstructed electrons are may be mis-identified as converted photons with one track. To study contamination in the photon container from these mis-reconstructed electrons, a $Z \rightarrow e^+e^-$ Monte Carlo sample, with full detector simulation, was used to measure the probability to reconstruct electrons as photons.

Reconstructed electrons and photons are geometrically matched to truth electrons from $Z \rightarrow e^+e^-$ decays. The reconstructed photons are required to satisfy the same identification and isolation conditions used for real data in this analysis, and the electrons are required to be pass a loose identification requirement with the recommended track quality cuts. The total per-event-yields of these misidentified photons and electrons as well as the misidentification rate defined as the ratio of the two spectra are plotted in Fig. 5.17. The misidentification rate is about 6% in the forward regions and about 2% at mid-rapidity which is consistent with recent studies at 13 TeV [137].

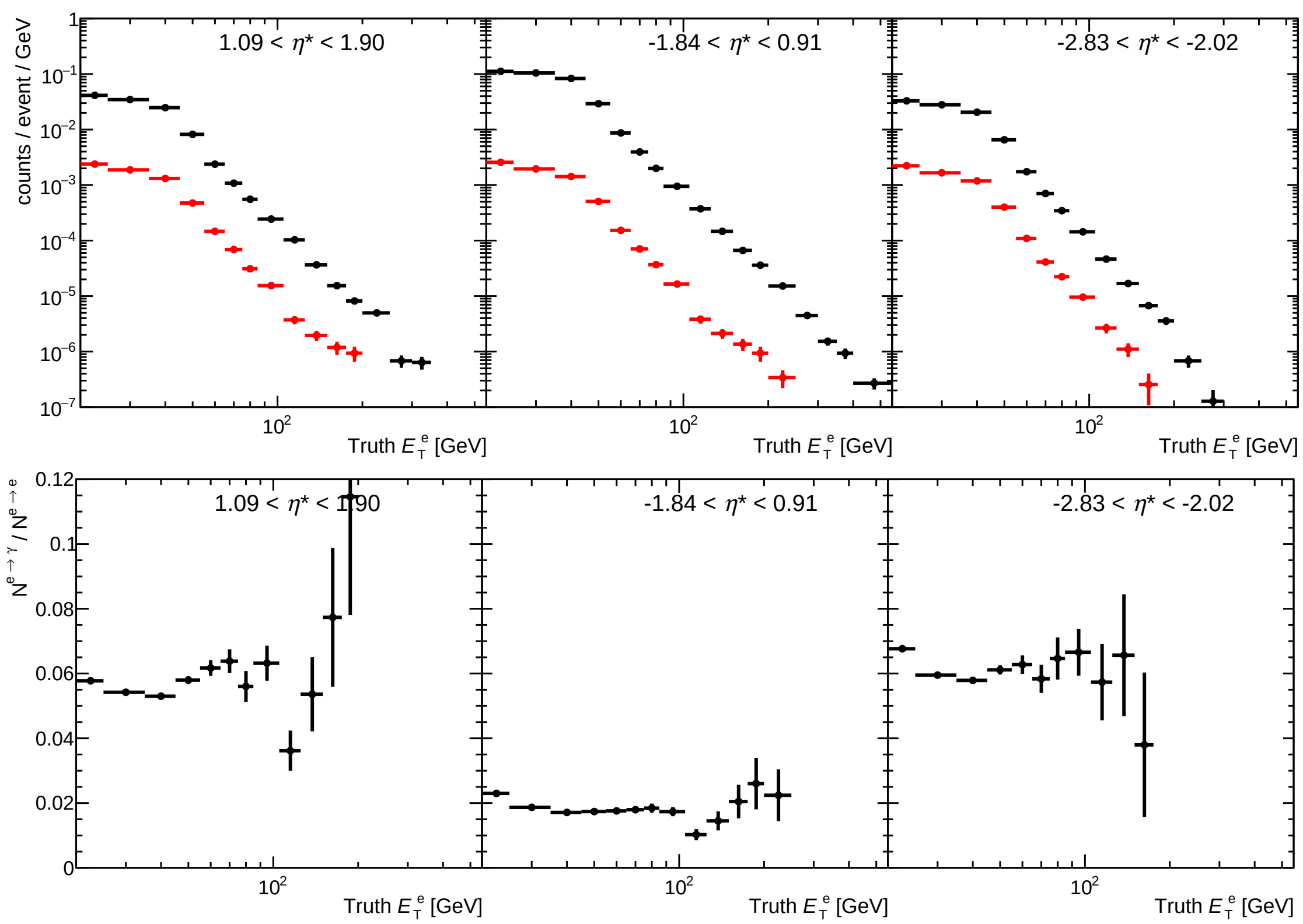

Figure 5.17: Top: The total per-event-yields of reconstructed electrons (black) and misidentified photons (red). Bottom: The photon misidentification rate defined as the ratio of the two spectra.

To estimate the electron contamination, we find the cross section of misidentified electrons in the $Z \to e^+e^-$ sample and take the ratio to the measured photon cross section. The contamination is taken as the ratio of these cross sections which is between 0.05% and 0.33% in the forward regions, and 0.05% and 0.12% at mid-rapidity. Figures of the cross sections and contamination are plotted in Fig. 5.18.

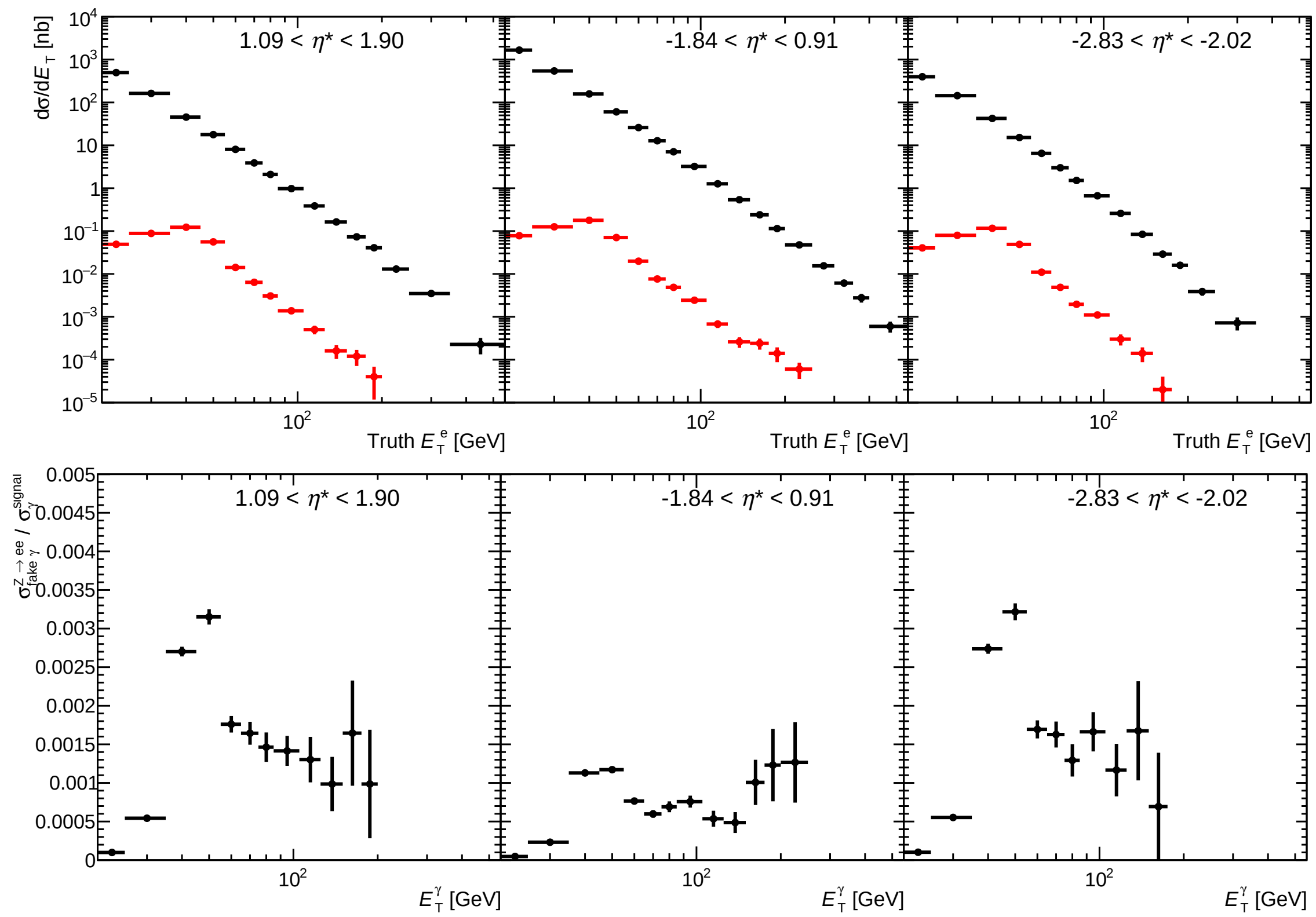

Figure 5.18: Top: Measured photon cross section from this analysis (black) and misidentified photon cross section from electrons from Z decays in MC (red). Bottom: The electron contamination defined as the ratio of the two spectra.

To estimate the contribution from $W^\pm \to e^\pm$ decays, the ratio of mis-identified electrons from $W + Z$ decays to that of only $Z$ decays is calculated using the contamination values found in the internal note accompanying the 13 TeV *pp* inclusive photon results [138], accounting correctly for the factor of two from the $Z$ vs. $W$ decays. These values indicate conservative factors of 4.35 (mid-rapidity) and 4.0 (forward) increases from the contamination from $Z$ decays only. A conservative estimate of the total contamination is $1.3\%$ below 100 GeV and $0.5\%$ above 100 GeV in the forward regions, and $0.5\%$ for all $E_{\mathrm{T}}^{\gamma}$ in mid-rapidity. These values are taken as systematic uncertainties on the final measurement.

### 5.3.7 Electron/Photon Calibration Correction

While investigating $Z$ bosons in the di-electron decay channel in $pp$ collisions at 5.02 TeV, a negative shift in the mass peak was discovered and was found to be a result of an electron energy scale calibration problem. The electron energy scale is parameterized in the following way:

$$E_e^{\mathrm{data}} = E_e^{\mathrm{MC}}(1 + \alpha(cell_\eta, \mu)), \tag{5.7}$$

where $\alpha$ is a function of the pseudorapidity of the calorimeter cell and the pileup rate parameter $\mu$. $\alpha$ is related to $\mu$ through the $OFC(\mu)$ parameters. The source of the problem was that the production was run with the default $OFC(\mu = 20)$ instead of the appropriate $OFC(\mu = 0)$ for the $p$+Pb conditions. Fig. 5.19 shows the ratio of extracted electron energy from an example 13 TeV run using $OFC(\mu = 0)$ to that using $OFC(\mu = 20)$ showing a strong $\eta$ dependence.

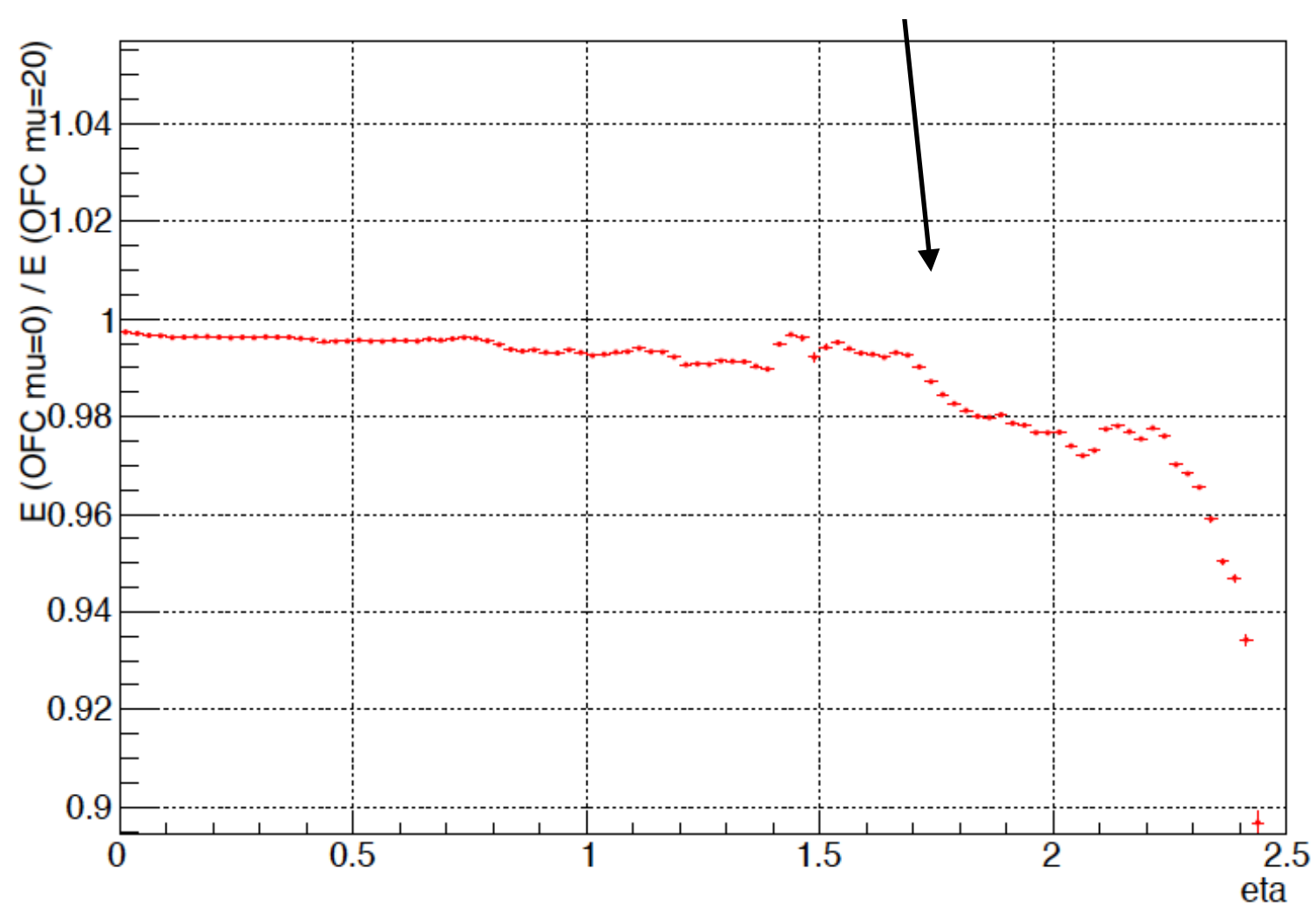

Figure 5.19: The ratio of extracted electron energy from an example 13 TeV run using $OFC(\mu = 0)$ to that using $OFC(\mu = 20)$.

This problem, also affecting photons, is corrected by adding an additional calibration factor to the electron and photon energy scale. Figures 5.20 and 5.21 show this correction applied to a $Z$ mass peak comparison in 5.02 TeV $pp$ data at mid- and forward-rapidity. The same corrections are applied in this photon analysis.

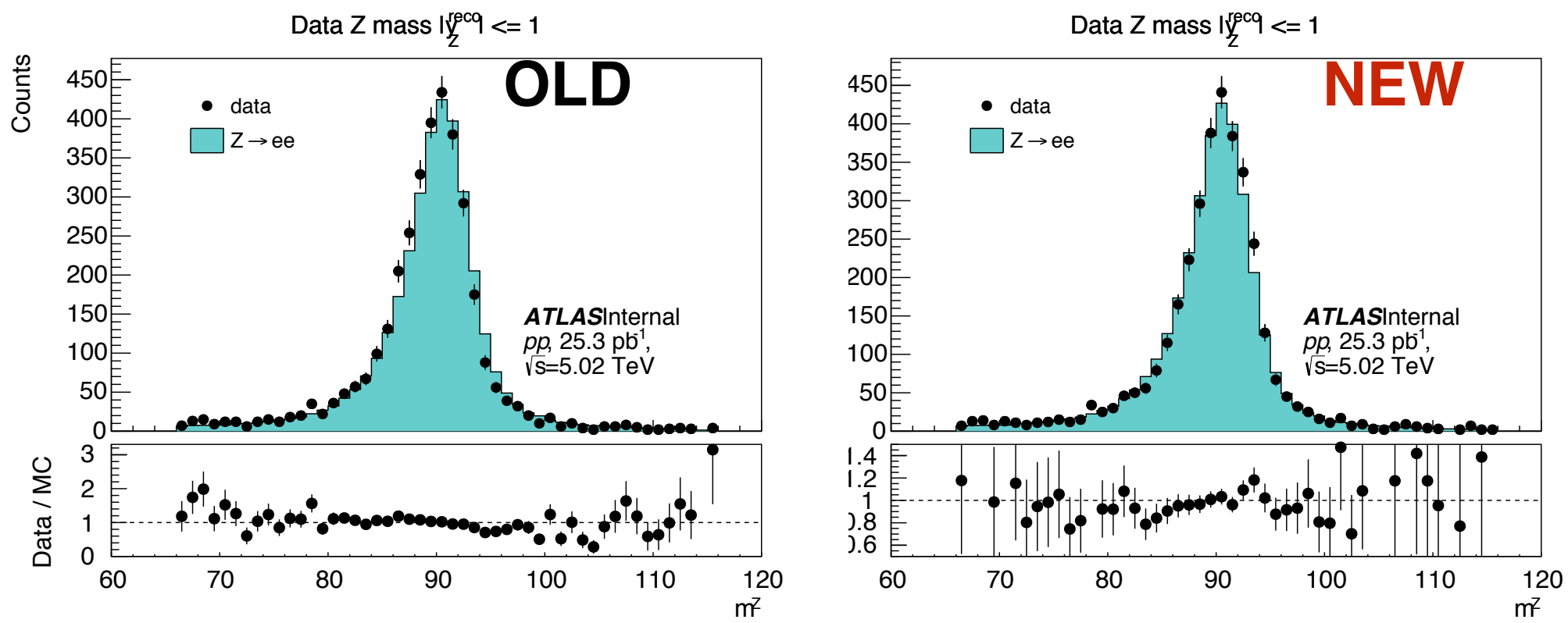

Figure 5.20: Di-electron invariant mass distribution in data compared to MC for uncorrected (left) and corrected(right) electron energy scale data at mid rapidity.

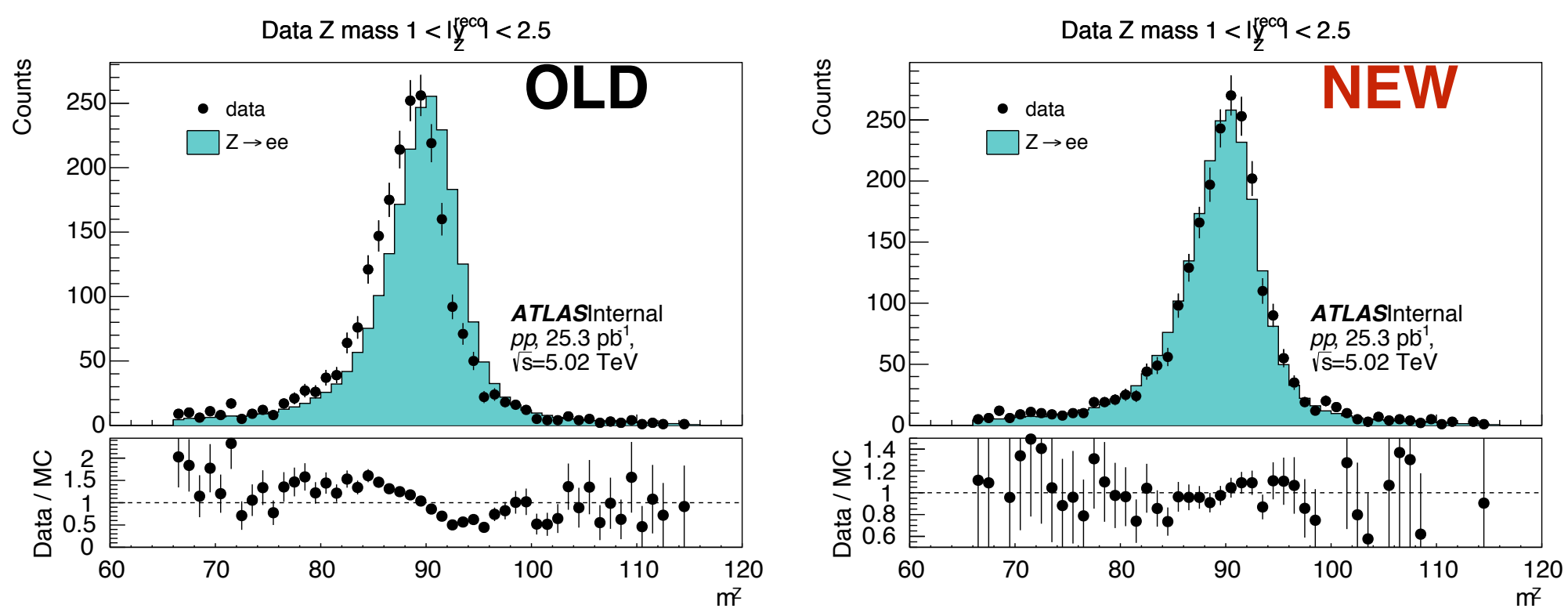

Figure 5.21: Di-electron invariant mass distribution in data compared to MC for uncorrected (left) and corrected (right) electron energy scale data at forward rapidity. Note the different range in y-axis for the ratios in the bottom panels.

## 5.4 Cross Section Signal Extraction

The differential cross-section is calculated for each $E_{\mathrm{T}}^{\gamma}$ and $\eta^{*}$ bin as

$$\frac{\mathrm{d}\sigma}{\mathrm{d}E_{\mathrm{T}}^{\gamma}} = \frac{1}{L_{\mathrm{int}}}\frac{1}{\Delta E_{\mathrm{T}}^{\gamma}}\frac{N_{\mathrm{sig}}P_{\mathrm{sig}}}{\epsilon^{\mathrm{sel}}\epsilon^{\mathrm{trig}}}C_{\mathrm{MC}},$$

where $L_{\mathrm{int}}$ is the integrated luminosity, $\Delta E_{\mathrm{T}}^{\gamma}$ is the width of the $E_{\mathrm{T}}^{\gamma}$ bin, $N_{\mathrm{sig}}$ is the yield of photon candidates passing identification and isolation requirements, $P_{\mathrm{sig}}$ is the purity of the signal selection, $\epsilon^{\mathrm{sel}}$ is the combined reconstruction, identification and isolation efficiency for signal photons, $\epsilon^{\mathrm{trig}}$ is the trigger efficiency, and $C_{\mathrm{MC}}$ is a MC derived bin-by-bin correction for the change in the $E_{\mathrm{T}}^{\gamma}$ spectrum from photons migrating between bins in the spectrum due to the width in the energy response. $C_{\mathrm{MC}}$ is determined after all selection criteria at both reconstruction and particle levels are imposed. The efficiencies have been discussed above. The purity and other factors are detailed below.

### 5.4.1 Photon Purity

The photon purity $P_{\mathrm{sig}}$ is determined via a data-driven double-sideband procedure used extensively in previous measurements of cross-sections for processes with a photon in the final state [116, 117, 139, 140] and summarised here. A matrix of four designations is made: tight, non-tight, isolated, non-isolated. The non-tight selected photons pass the **loose** requirements of Ref. [133] but fail certain components of the tight requirements, designed to mostly select background. The non-iso photons are required to have significant isolation energy such that

$$E_{\mathrm{T}}^{\mathrm{iso}} > 7.8 + 4.2 \times 10^{-3} E_{\mathrm{T}}^{\gamma} \text{ [GeV]}. \tag{5.8}$$

These designations form four groups: A (tight & iso), B (tight & non-iso), C (non-tight & iso), D (non-tight & non-iso). The majority of signal photons are in the tight, isolated region, defined to be the signal region, while the other three regions are dominated by the background. Photon candidates that comprise the background are assumed to be distributed in a way that is uncorrelated along the two axes. The yields in the three non-signal sidebands, with an understanding of the leakage from the signal region into the sidebands, are used to estimate the yield of background in the signal region and this is combined with the yield in the signal region to extract the purity.

Figure 5.22 shows the sideband yields for the $p$+Pb and Pb+$p$ running periods combined (reversing the pseudorapidity of the Pb+$p$ period) as raw yields and as ratios to the signal region A. These ratios are fit to polynomial forms which are used only to smooth the sideband yields for an alternate purity calculation and are not used in the calculation of the cross section.

Leakage rates of signal photons into the sidebands are estimated in MC and are presented in Fig. 5.23. Comparisons between leakages found from the nominal Pythia sample to those from Sherpa can be found in Appendix A.6.

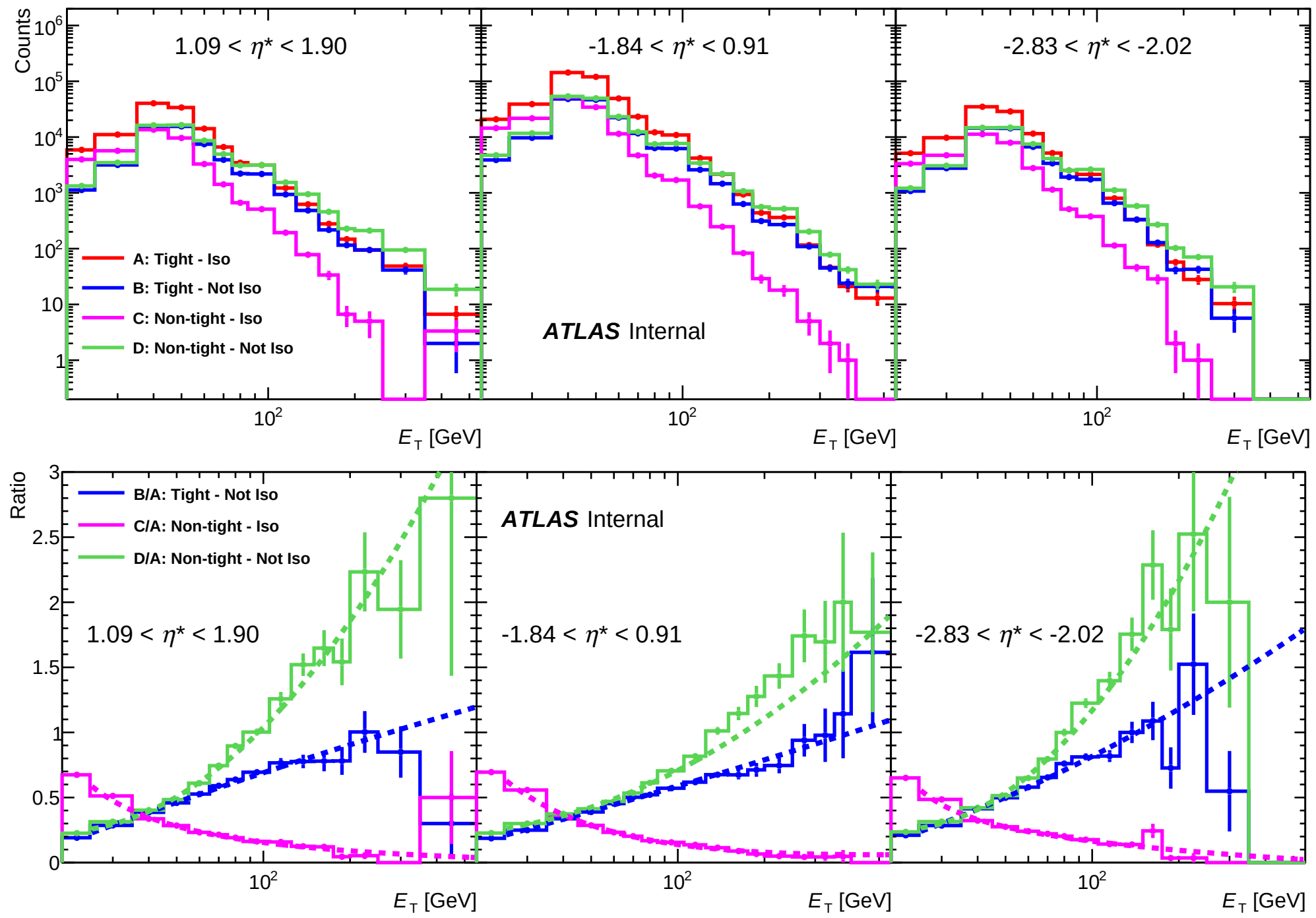

Figure 5.22: Sideband A, B, C, and D yields (Top) and ratios to A (Bottom) from both data taking periods combined. Photon purity is calculated from these ratios together with leakage fractions. The histograms are fit to polynomial functional forms, though the fits are not used in the cross section calculation.

The purity is defined as the ratio of the yield of signal photons to the total photon yield in the tight and isolated region:

$$P = \frac{N^{\mathrm{A,data}}_{\mathrm{signal}}}{N^{\mathrm{A,data}}} \tag{5.9}$$

The yield of signal photons in region A, $N^{\mathrm{A,data}}_{\mathrm{signal}}$, is computed from sideband yields, $N^{\mathrm{X,data}}$, and leakage factors, $f^{\mathrm{X,MC}}$, as

$$N^{\mathrm{A,data}}_{\mathrm{signal}} = N^{\mathrm{A,data}} - R_{\mathrm{bkg}} \cdot \left( (N^{\mathrm{B,data}} - f^{\mathrm{B,MC}} N^{\mathrm{A,data}}_{\mathrm{signal}}) \cdot \frac{(N^{\mathrm{C,data}} - f^{\mathrm{C,MC}} N^{\mathrm{A,data}}_{\mathrm{signal}})}{(N^{\mathrm{D,data}} - f^{\mathrm{D,MC}} N^{\mathrm{A,data}}_{\mathrm{signal}})} \right) . \tag{5.10}$$

Thus, to extract $N^{\mathrm{A,data}}_{\mathrm{signal}}$, we can solve the above quadratic. The factor $R_{bkg}$ quantifies the deviation from perfect factorization and is set to unity to calculate the central purity values and is varied to determine the systematic uncertainty of this assumption on the measurement as described in section 5.6. Because the sideband yields are statistically limited at high $E_{\mathrm{T}}^{\gamma}$, the yields from the two periods are combined, respecting the reverse of the beam direction, before making the calculation directly from the yields and leakage fractions.

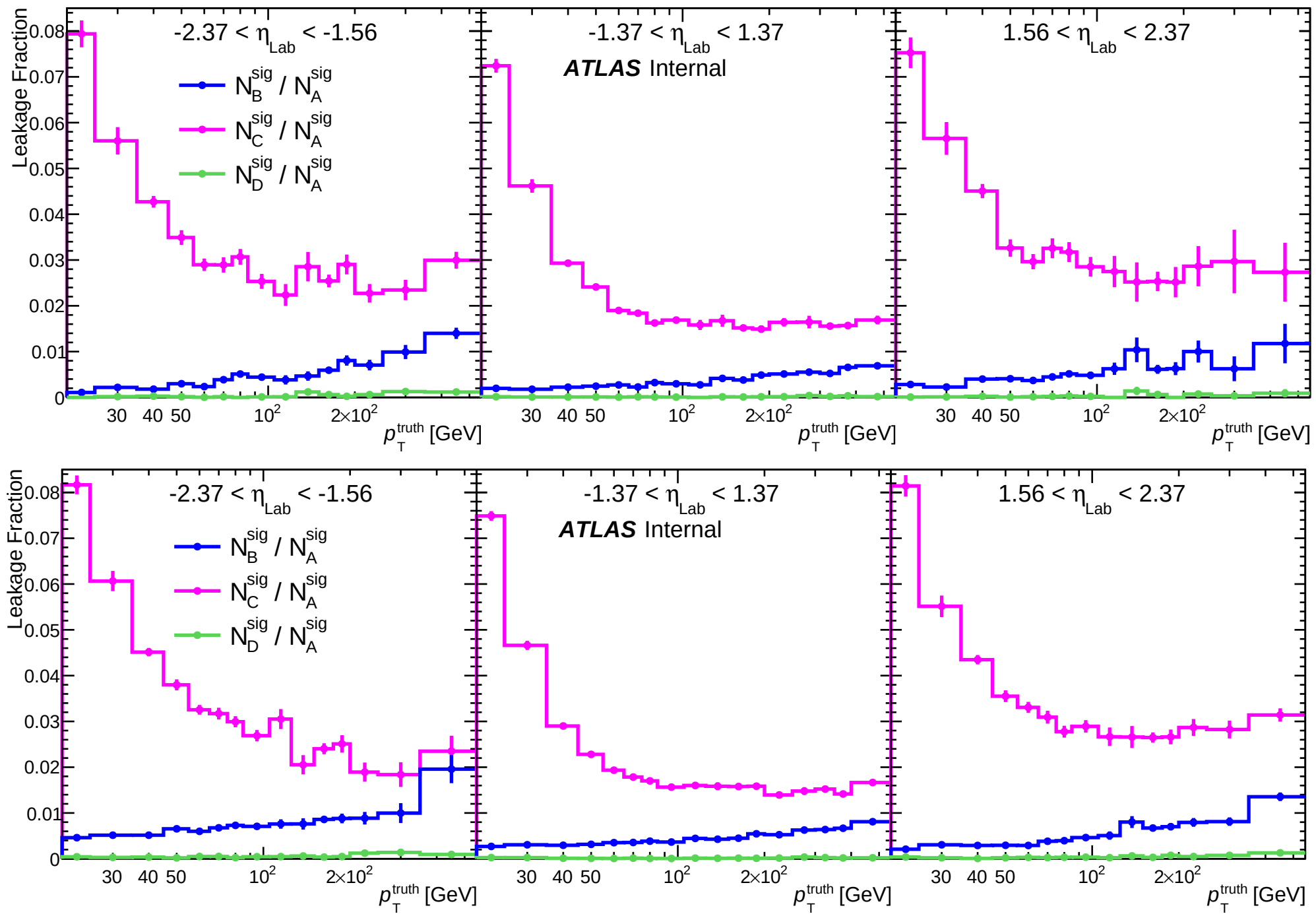

Figure 5.23: Sideband leakage fractions from truth isolated photons in Monte Carlo simulations plotted as a function of $E_{\mathrm{T}}^{\gamma}$ in each pseudorapidity slice from the $p$+Pb (Top) and Pb+$p$ (bottom) running periods.

An alternate calculation is made using polynomial fits to the sideband ratios to extrapolate to the points above 75 GeV, and is used for comparison purposes only. In this case, the observed sideband B, C, and D yields are replaced with the functional forms evaluated at the bin center multiplied by the bin content of A. The bottom panel of Fig. 5.22 show these ratios along with their fits. The uncertainty in this smoothing is calculated by determining the 68% (1 $\sigma$) confidence interval around the fit function for each ratio. The value of the upper and lower band at a point is taken as the error, and is propagated through the purity calculation as a statistical error bar. The statistical error on the direct calculation is discussed in the next section.

The same procedure is carried out using leakages from SHERPA, and the resulting purities, shown in Fig. 5.24, are consistent with the nominal values within 0.5%. It is worth noting here that $N_{\mathrm{signal}}^{\mathrm{A,data}}$ is really what we are after. The purity itself is not used in the cross section measurement, but is a useful performance plot to make.

To estimate the statistical uncertainty on the purity, toy MC simulations are used to capture the asymmetrical nature of the error. Each toy consists of sample sideband yields randomly generated from Poisson distributions

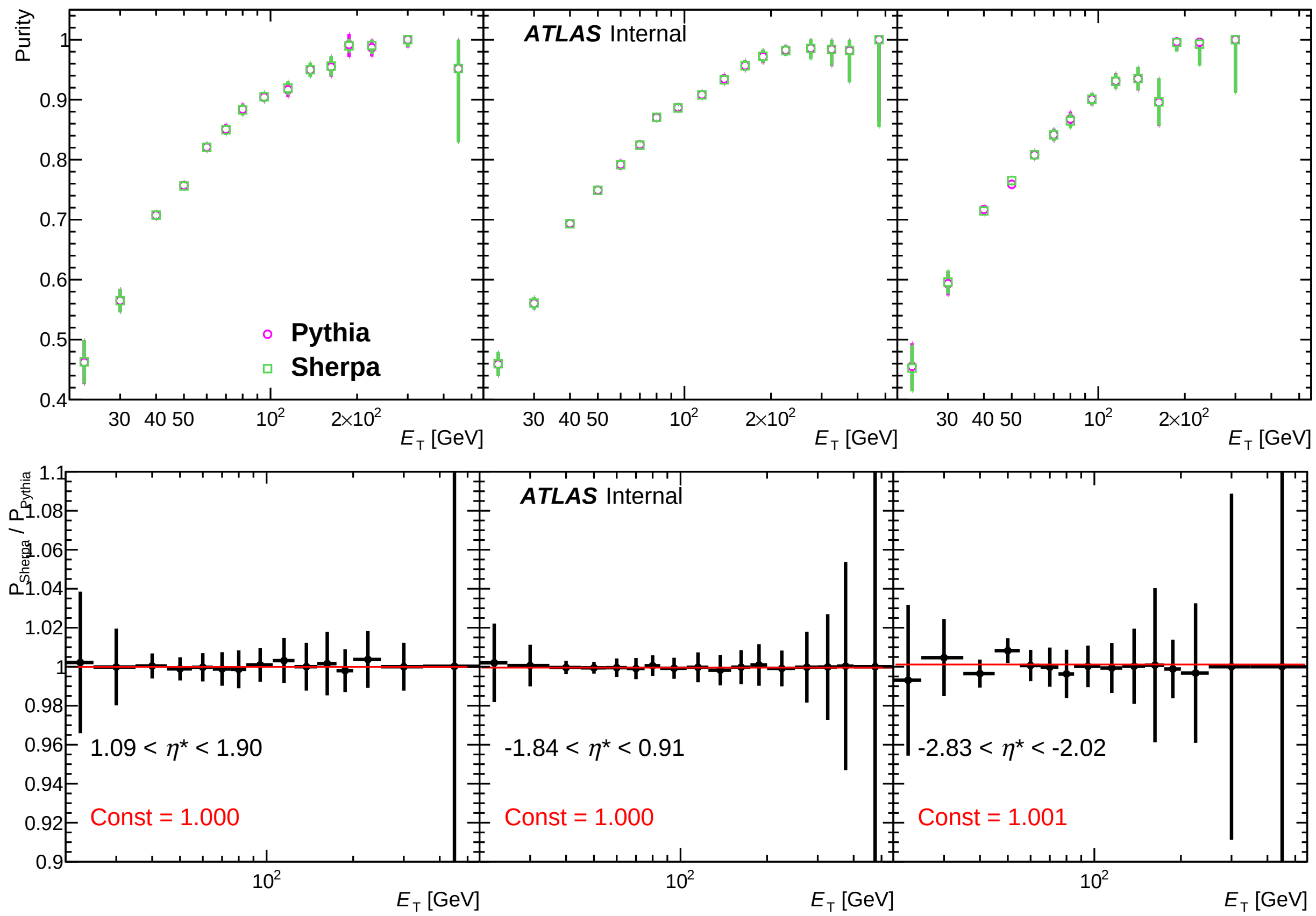

Figure 5.24: Purities for tight identified and isolated photons calculated via the sideband method with leakages from both Pythia and Sherpa data overlay. **Top:** purities; **bottom:** ratios.

with means set to the measured yield in that sideband, and the leakage fractions which are constant and taken from MC. For each toy, the purity is calculated and entered into a finely binned (3000) histogram. 1M toys are thrown for each $E_{\mathrm{T}}^{\gamma}$ and pseudorapidity slice. The error is determined by integrating the distribution to 16% on either side of the mean to approximate one standard deviation. If there is less than 16% above the mean, the top of the error is set to unity. For the case of $N_{\mathrm{signal}}^{\mathrm{A,data}}$, a completely analogous procedure is carried out, however, because it is not constrained like the purity to be less than one, the statistical uncertainties are expected to be symmetric. Thus, the error is simply taken as the RMS of the distribution. A comparison between the direct calculation and the alternate, smoothed version is shown in Fig. 5.25

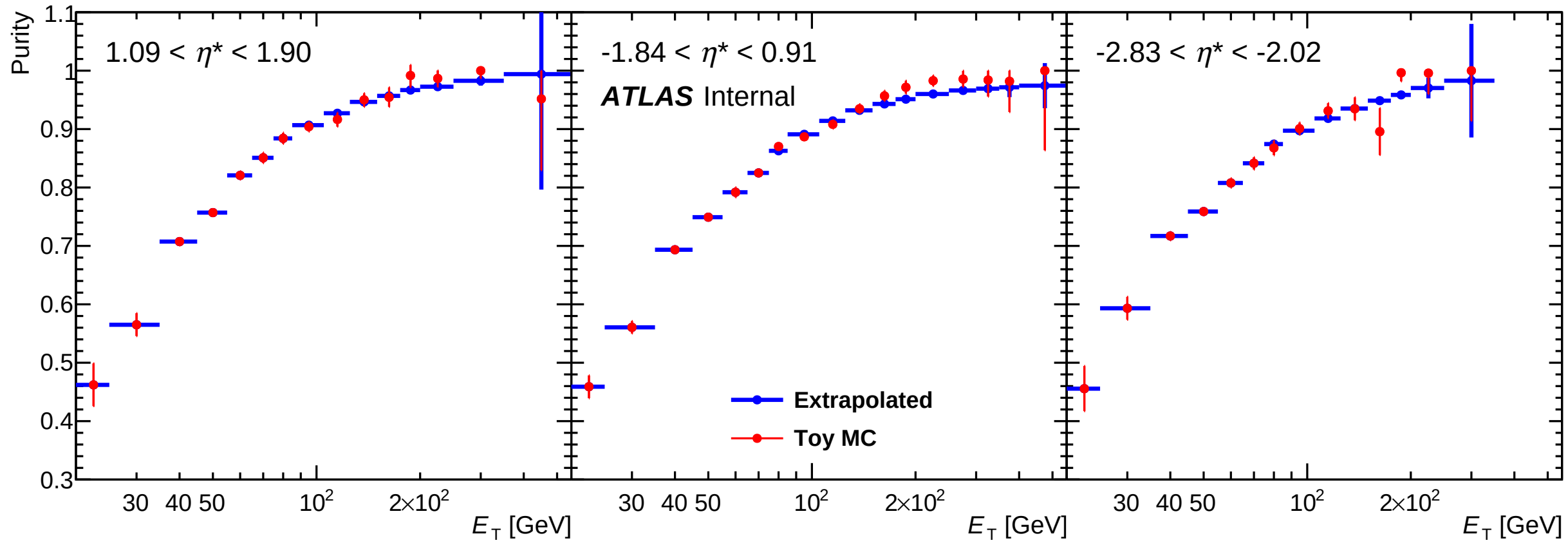

Figure 5.25: Purities for tight identified and isolated photons calculated via the direct method using toy MCs for the statistical error, and those calculated using the smoothed sideband method.

### 5.4.2 Unfolding for Bin Migration

Because of the non-zero energy resolution for photons in the EM calorimeter, some reconstructed photons will end up in a $E_{\mathrm{T}}^{\gamma}$ bin not corresponding to its "truth" $E_{\mathrm{T}}^{\gamma}$. Due to the falling nature of the $E_{\mathrm{T}}^{\gamma}$ spectrum, this bin sharing is not symmetric and there will be an overall migration of counts towards higher $E_{\mathrm{T}}^{\gamma}$ bins as shown in Fig. 5.26. Fig. 5.27 shows the fraction of photon counts that remain in the same bin after reconstruction (do not migrate) to be about 90%. The saw-tooth jumps are due to jumps in the bin size that affect the migration fraction. Thus, this effect is small in the case of photons in ATLAS, and therefore, we employ a simple bin-by-bin correction, $C_{\mathrm{MC}}$, as calculated in MC. The correction is calculated using truth matched tight and isolated photons and taking the ratio of their reconstructed $E_{\mathrm{T}}^{\gamma}$ over their truth $E_{\mathrm{T}}^{\gamma}$

$$C_{\mathrm{MC}} = \frac{N^{\gamma,\mathrm{true}p_{\mathrm{T}}}_{\mathrm{TIso,reco,tight,Iso}}}{N^{\gamma,\mathrm{reco}p_{\mathrm{T}}}_{\mathrm{TIso,reco,tight,Iso}}} \,, \tag{5.11}$$

Fig. 5.28 shows that these corrections for each beam configuration are of order a couple percent at low $E_{\mathrm{T}}^{\gamma}$, but grow to about 5% at high $E_{\mathrm{T}}^{\gamma}$. Note, the error bars on these corrections are not calculated correctly, as they have not taken into account the correlation between the numerator and denominator.

To smooth the statistical fluctuations, the periods are combined (reversing the pseudorapidity of the second period) and the correct errors are calculated in the following way. The population of $i$th bin of the truth ($T$) and reconstructed ($R$) $E_{\mathrm{T}}^{\gamma}$ spectra can be expressed as $T_i = \Sigma_j A_{ji}$ and $T_i = \Sigma_j A_{ij}$ respectively, where $A_{ij}$ are the

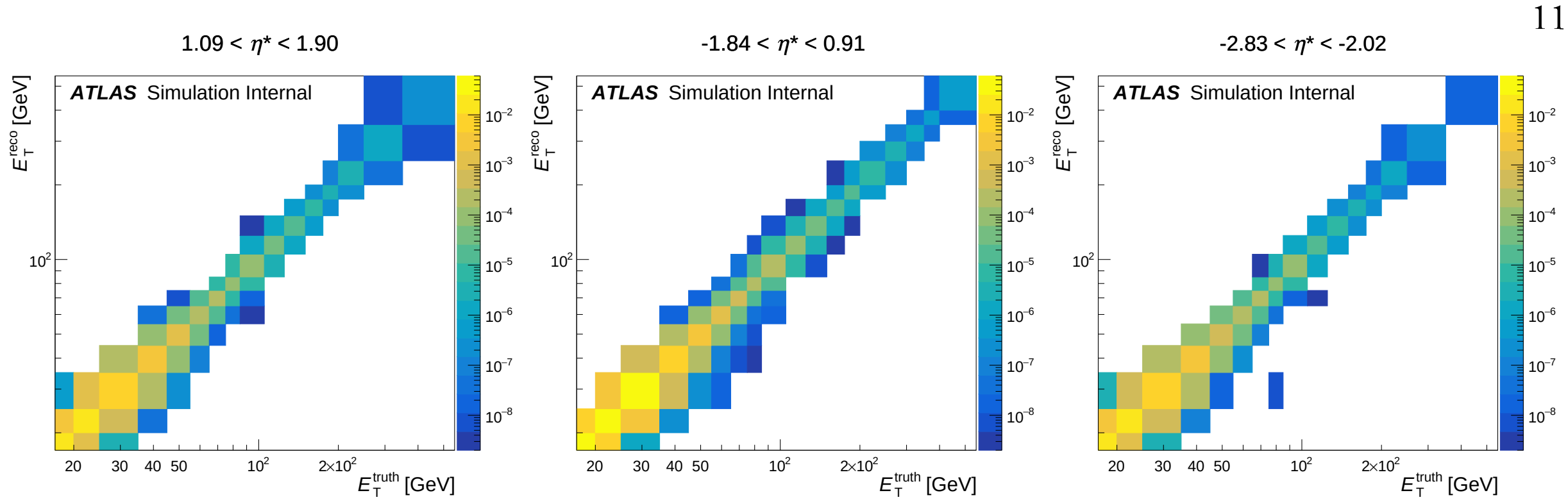

Figure 5.26: Photon $E_{\mathrm{T}}^{\gamma}$ response matrix for tight and isolated truth matched photons from MC data overlay in each pseudorapidity region. The bin from 17-20 GeV acts as an underflow bin.

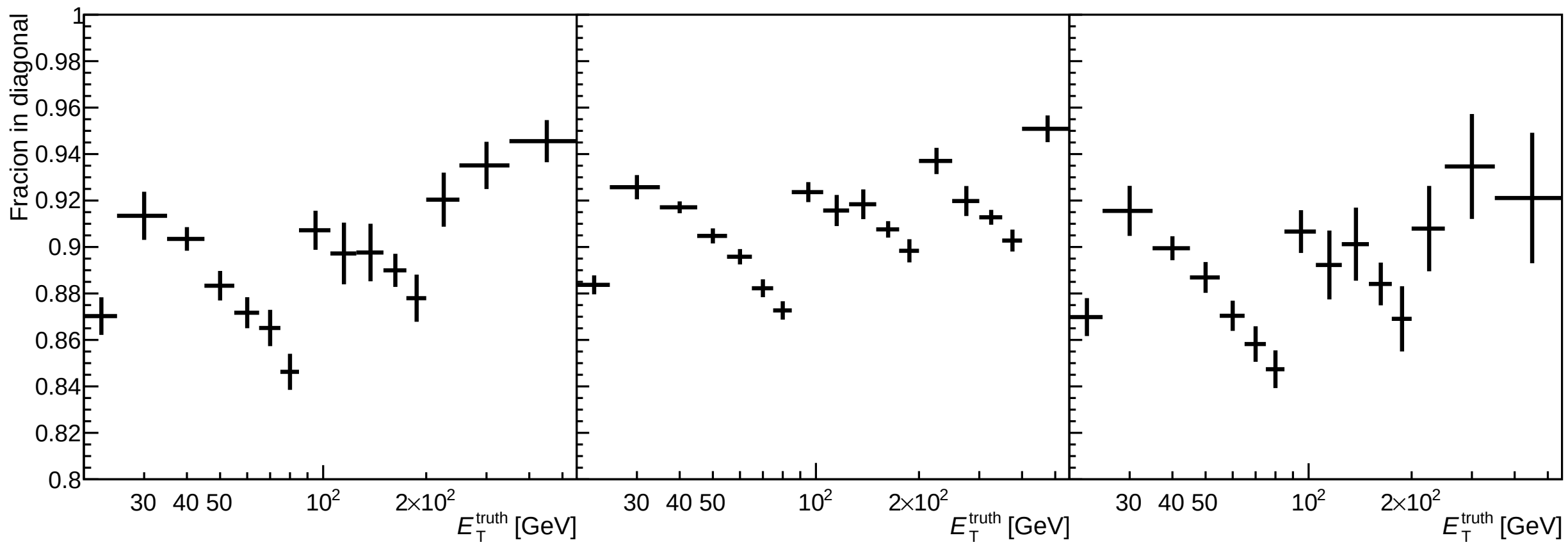

Figure 5.27: The fraction of photon counts that remain in the same $E_{\mathrm{T}}^{\gamma}$ bin after reconstruction as measured in MC data overlay in each pseudorapidity region

matrices given in Fig. 5.26. The correction factors for each bin, $C_i$, can then be expressed as

$$C_i = \frac{R_i}{T_i} = \frac{A_{ii} + \Sigma_{j \neq i} A_{ij}}{T_i} \ . \tag{5.12}$$

The reconstructed bin's contents consist of two populations, (1) those that originated in this bin, and (2) those that migrated into it. Binomial errors are assigned to population (1) using the fractions in Fig. 5.27 as the binomial probability, and Poisson uncertainties are assigned to the counts from (2). The total uncertainty on $R_i$ is then the quadrature sum of the two. $T_i$ is given Poisson uncertainties which is propagated into $C_i$ in the ratio. The corrections, shown in Fig. 5.29, are then fit to a logistic functional form from which the correction factors are taken. The bin-by-bin errors are simply those found in the above exercise.

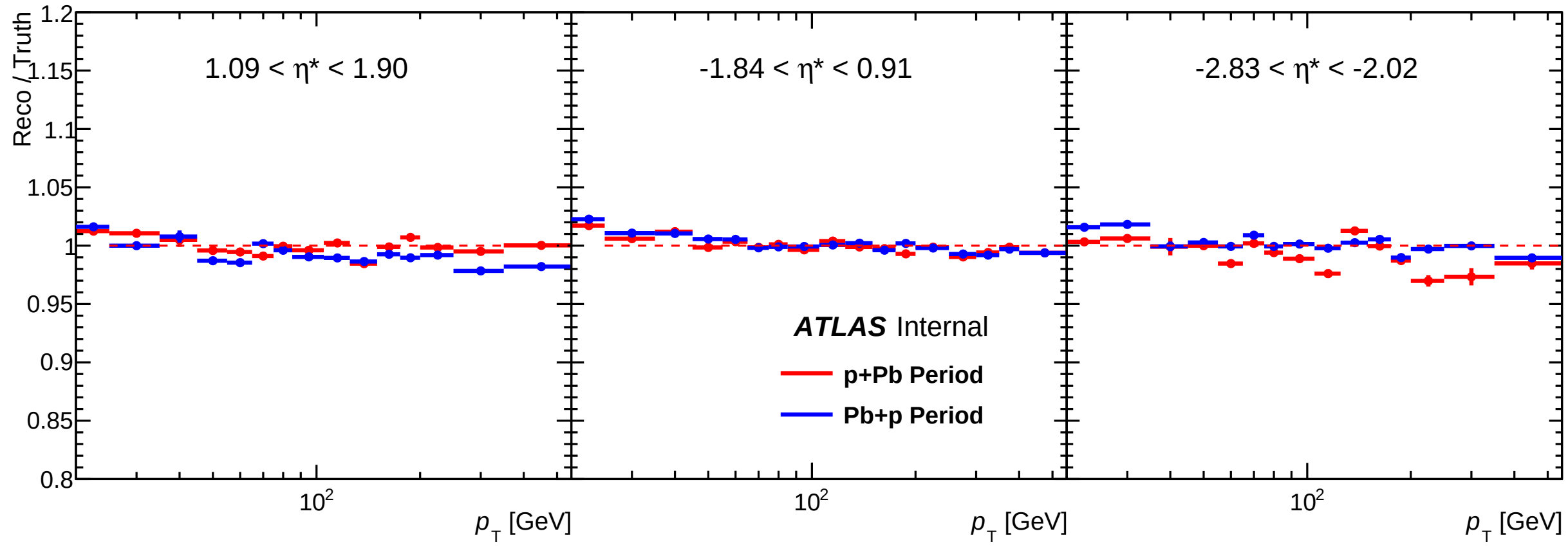

Figure 5.28: Bin migration corrections plotted as a function of $E_T^\gamma$ in each pseudorapidity region for both running periods. Note, the error bars on these data are not calculated correctly, as they have not taken into account the correlation between the numerator and denominator.

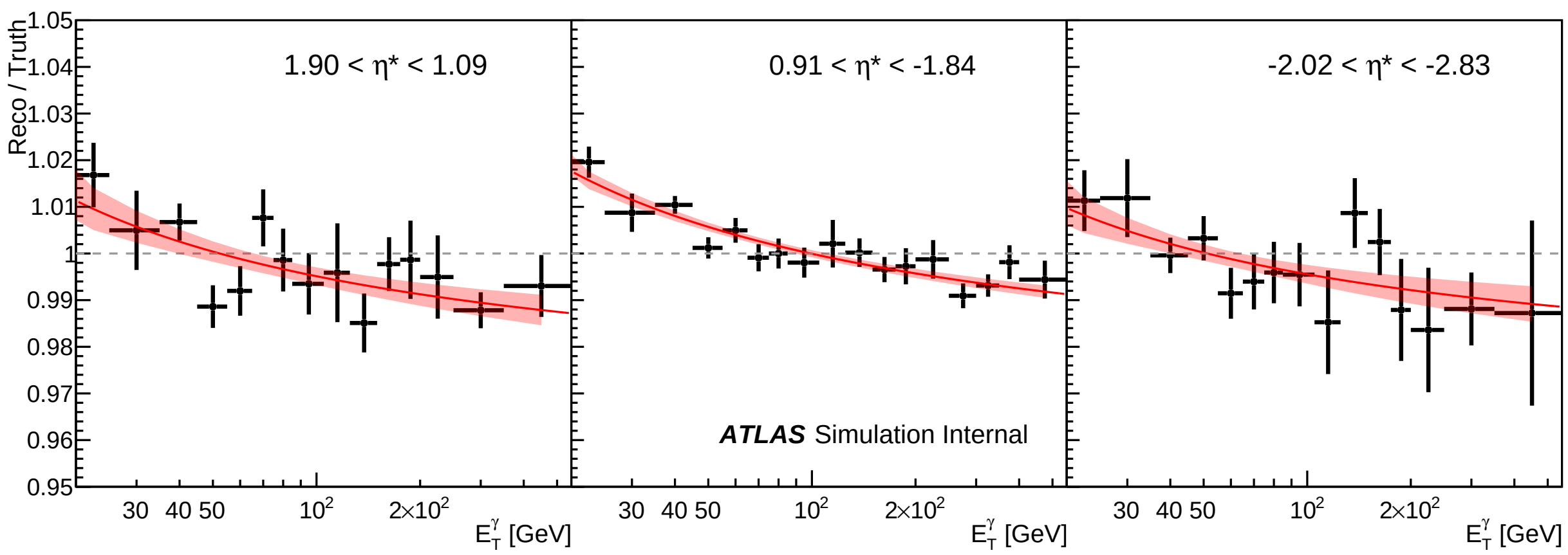

Figure 5.29: Bin migration corrections plotted as a function of $E_T^\gamma$ in each pseudorapidity region. The corrections are fit with a logistic function which is used for the applied corrections.

To test the sensitivity of the corrections to differences in the shape of the $E_T^\gamma$ spectrum between data and MC, the MC is reweighted to match the shape of that in data. Fig. 5.30 gives a comparison between data and MC showing that the data is slightly steeper than MC. The fits to the ratio in the lower panel are used to reweight the MC to match the data. Fig. 5.31 shows the correction factors after this reweighting. The difference to the nominal corrections is $< 0.2\%$ and is considered negligible.

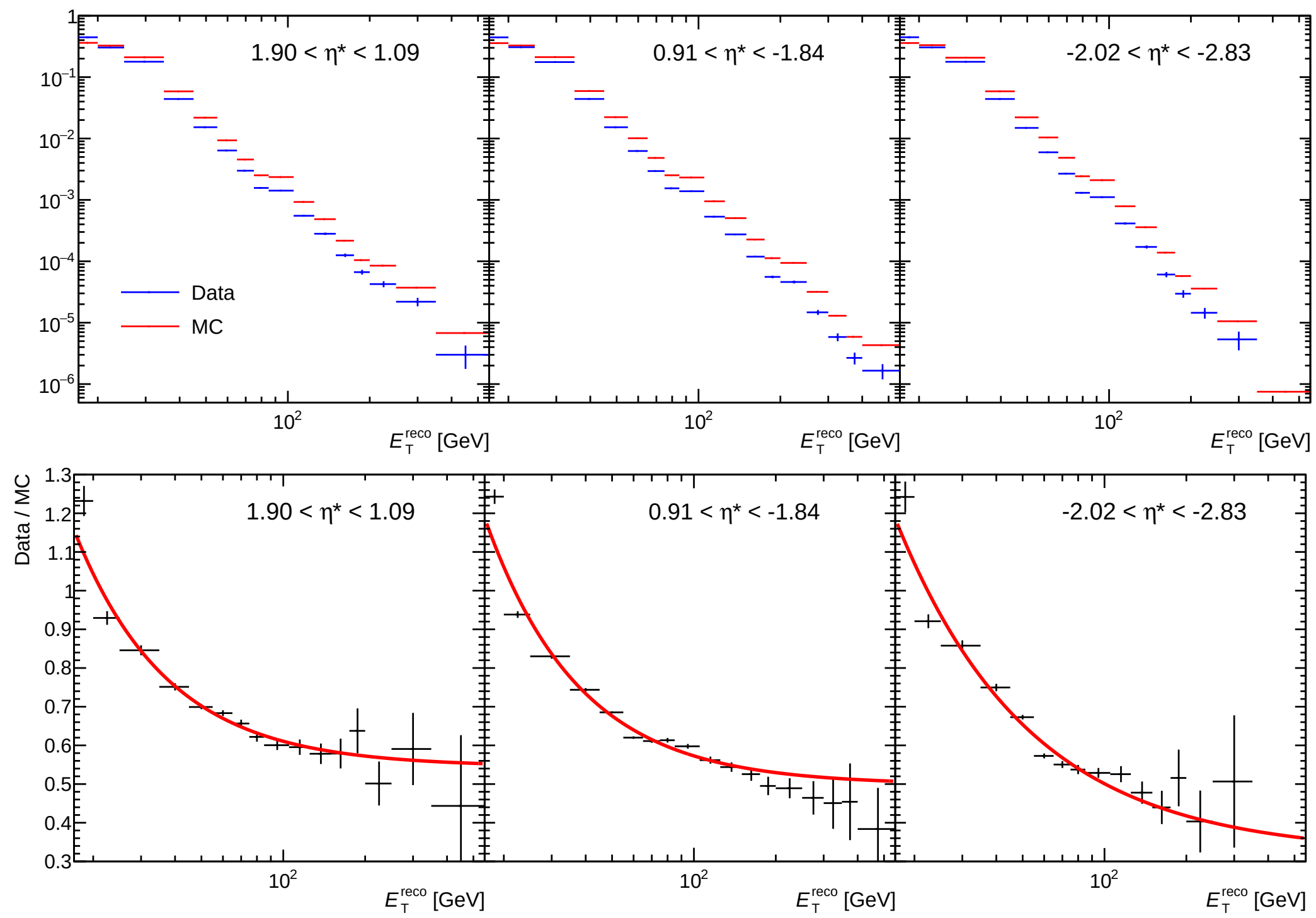

Figure 5.30: (Top) A comparison of the photon $E_{\mathrm{T}}$ spectrum in data (blue) and MC (red), showing that the data spectrum is slightly steeper than that of MC. (Bottom) The ratio of the two including an exponential fit which is used as a factor to reweight the MC to match the data.

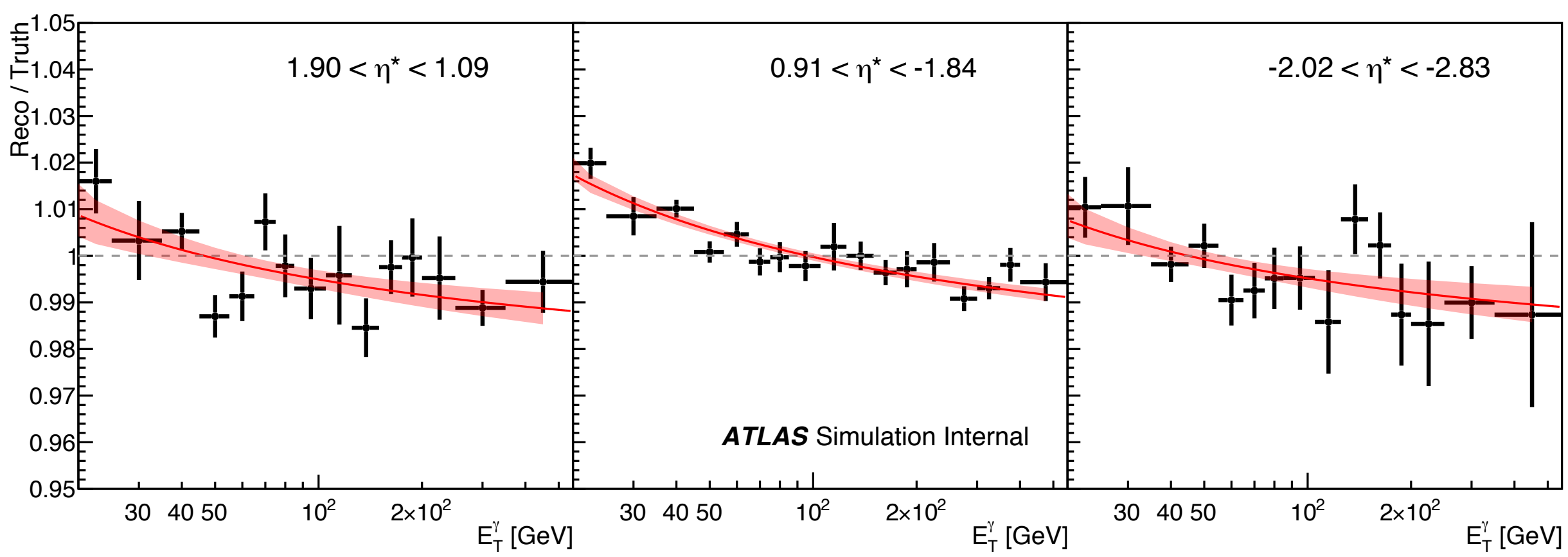

Figure 5.31: Bin migration corrections, determined after MC to data spectrum reweighting, plotted as a function of $E_{\mathrm{T}}^{\gamma}$ in each pseudorapidity region. The corrections show a negligible difference to the nominal values in Fig. 5.29.

## 5.5 Nuclear Modification Factor $R_{p\mathrm{Pb}}$

The nuclear modification factor $R_{p\mathrm{Pb}}$ can be expressed as a ratio of cross-sections in the following way:

$$R_{p\mathrm{Pb}} = (\mathrm{d}\sigma^{p+\mathrm{Pb}\to\gamma+X}/\mathrm{d}E_{\mathrm{T}}^{\gamma})/(A\cdot\mathrm{d}\sigma^{pp\to\gamma+X}/\mathrm{d}E_{\mathrm{T}}^{\gamma}) \; , \tag{5.13}$$

where the geometric factor $A$ is simply the number of nucleons in the Pb nucleus, 208. The reference $pp$ spectrum is constructed by extrapolating a photon spectra measurement from $\sqrt{s} = 8$ TeV $pp$ collisions by ATLAS [116] that used the same particle-level isolation requirement. This extrapolation procedure is discussed in detail below.

### 5.5.1 $pp$ Reference Construction

A reference cross-section for the prompt photon cross-section in $pp$ collisions at the same energy ($\sqrt{s} = 8.16$ TeV) and with the same boost of the center of mass frame with respect to the lab frame ($\Delta y = +0.465$), is constructed by extrapolating previously the measured $\sqrt{s} = 8$ TeV cross-section by ATLAS [116] using MC simulation. For the nominal extrapolation, Pythia simulations were used to generate truth-level photon spectra in 8 TeV (without boost) and 8.16 TeV (with boost) $pp$ collisions. Both samples use the identical tune (A14) and PDF set (NNPDF23LO), and have a consistent definition of truth-level isolation analogous to that in data. The strategy is then to use the ratios of the MC cross-sections as multiplicative extrapolation factors which can be applied to the 8 TeV data to construct an 8.16 TeV $pp$ reference (with matching boost). Specifically,

$$\sigma^{pp,8.16\text{ TeV,boost}}(E_{\mathrm{T}}^{\gamma},\eta^{\mathrm{lab}}) = \sigma^{pp,8\text{ TeV}}(E_{\mathrm{T}}^{\gamma},\eta^{\mathrm{lab}}) \times \frac{\sigma^{\textsc{Pythia},8.16\text{ TeV,boost}}(E_{\mathrm{T}}^{\gamma},\eta^{\mathrm{lab}})}{\sigma^{\textsc{Pythia},8\text{ TeV}}(E_{\mathrm{T}}^{\gamma},\eta^{\mathrm{lab}})} \tag{5.14}$$

Note that the 8 TeV data are reported in symmetric selections in $\eta$, e.g. $|\eta| < 0.6$, whereas the 8.16 TeV reference must be constructed in signed-$\eta$ bins. Therefore, each $|\eta|$ range in 8 TeV $pp$ data is used to construct a $pp$ reference for **two** signed $\eta$ ranges, and the ratios of Pythia distributions are determined in signed $\eta^{\mathrm{lab}}$ ranges.

This approach has several advantages: (1) it is a correction derived from the ratio of MC distributions, meaning that any failure of the MC to describe the $E_{\mathrm{T}}^{\gamma}$, $y$ dependence of the data approximately cancels in the ratios, before it is applied to data; (2) the $p$+Pb to $pp$ comparison ratio is between photon yields measured in the same physical region of the detector, potentially allowing for a cancellation of systematic uncertainties between the data; (3) this factor simultaneously corrects for the different $\sqrt{s}$ and center of mass shift at the same time, rather than separate corrections for both effects.

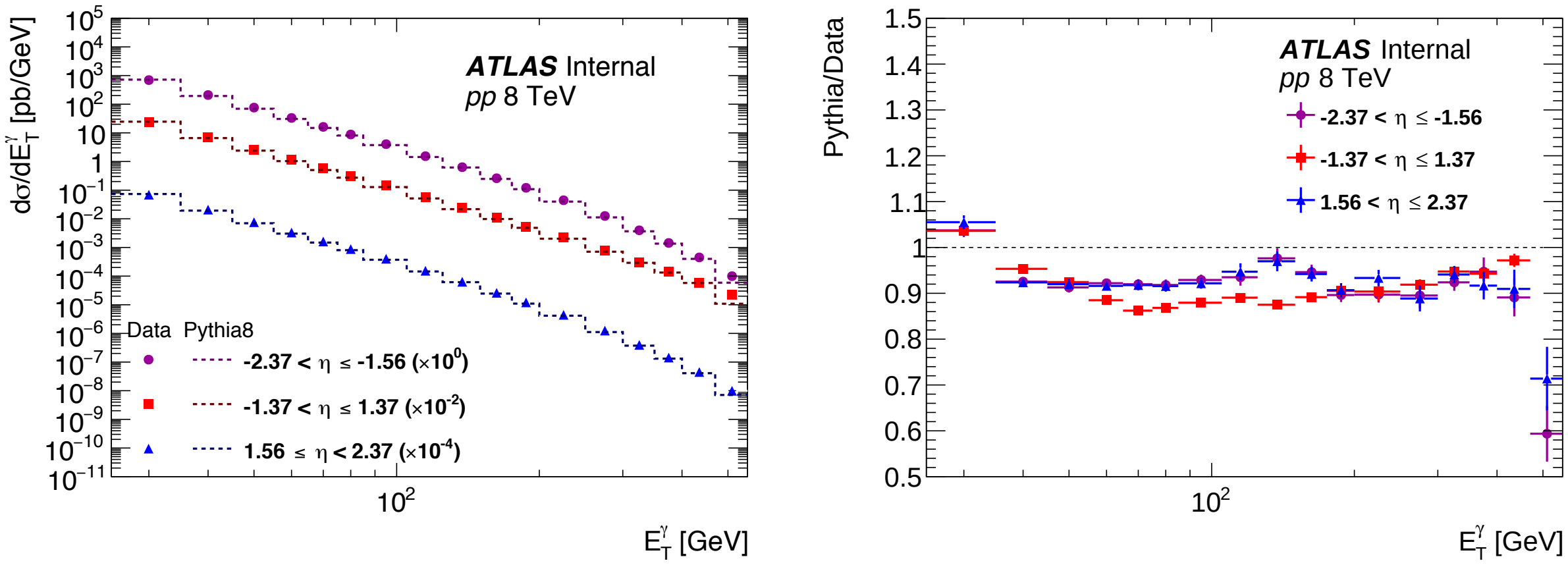

Figure 5.32: **Left**: Comparison of prompt, isolated photon spectrum measured by ATLAS in $\sqrt{s} = 8$ TeV$pp$ data (identical to the points in Fig. 3 in Ref. [116]), to that in Pythia A14 NNPDF23LO at the same energy. **Right**: Pythia/data ratio in each measured $\eta$ selection (systematic uncertainties on data not shown).

Figure 5.32 shows the central values of the measured 8 TeV ATLAS data compared to that in Pythia for each $|\eta^{\mathrm{Lab}}|$ selection in data. It can be seen that Pythia provides a reasonable but not perfect description of the $E_{\mathrm{T}}^{\gamma}$ and $\eta$ dependence in data, and is typically within 15% over the range of interest. (Note that the Pythia comparison in Ref. [116] is a different tune and PDF set – for our purposes we use the same tune and PDF as that to derive the corrections for the 8.16 TeV $p$+Pb analysis.)

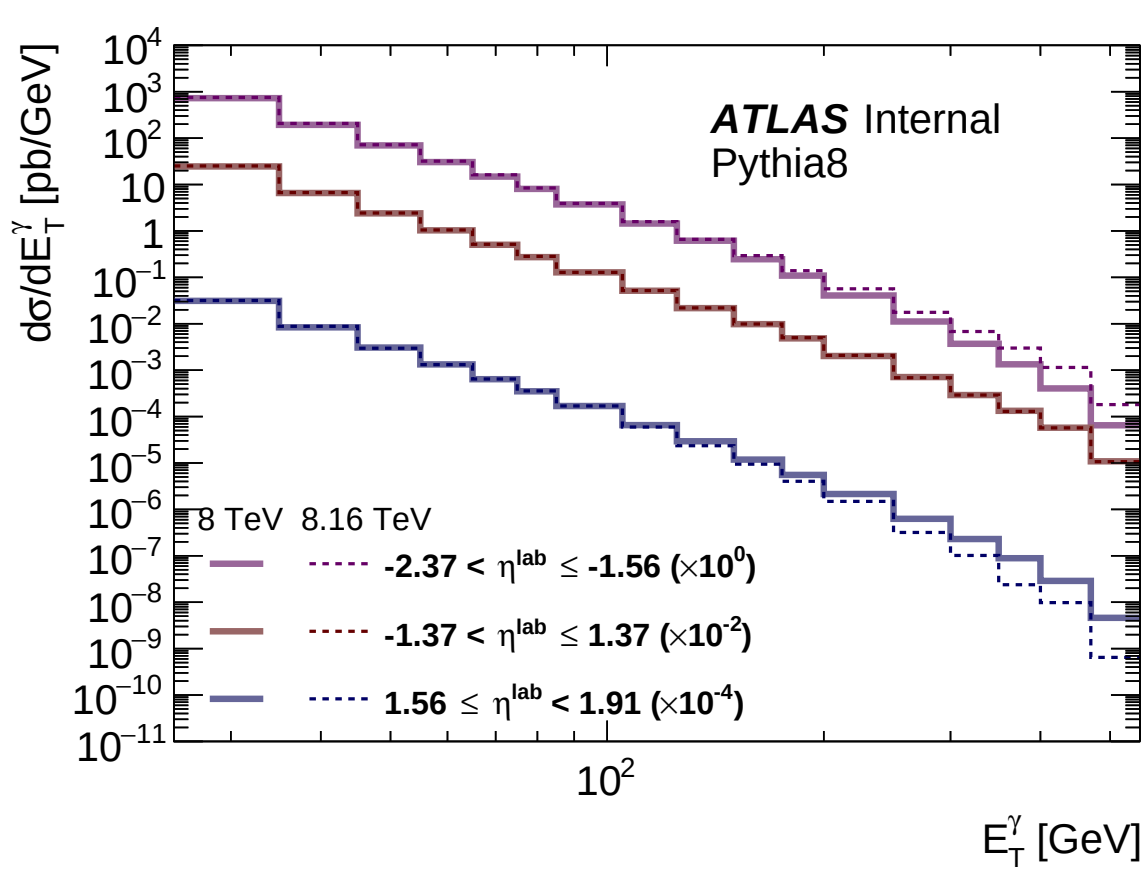

Figure 5.33: Comparison of prompt, isolated photon spectrum in Pythia A14 NNPDF23LO simulation at 8.16 TeV (with $\Delta y = -0.465$ boost) and 8 TeV for each pseudorapidity slice.

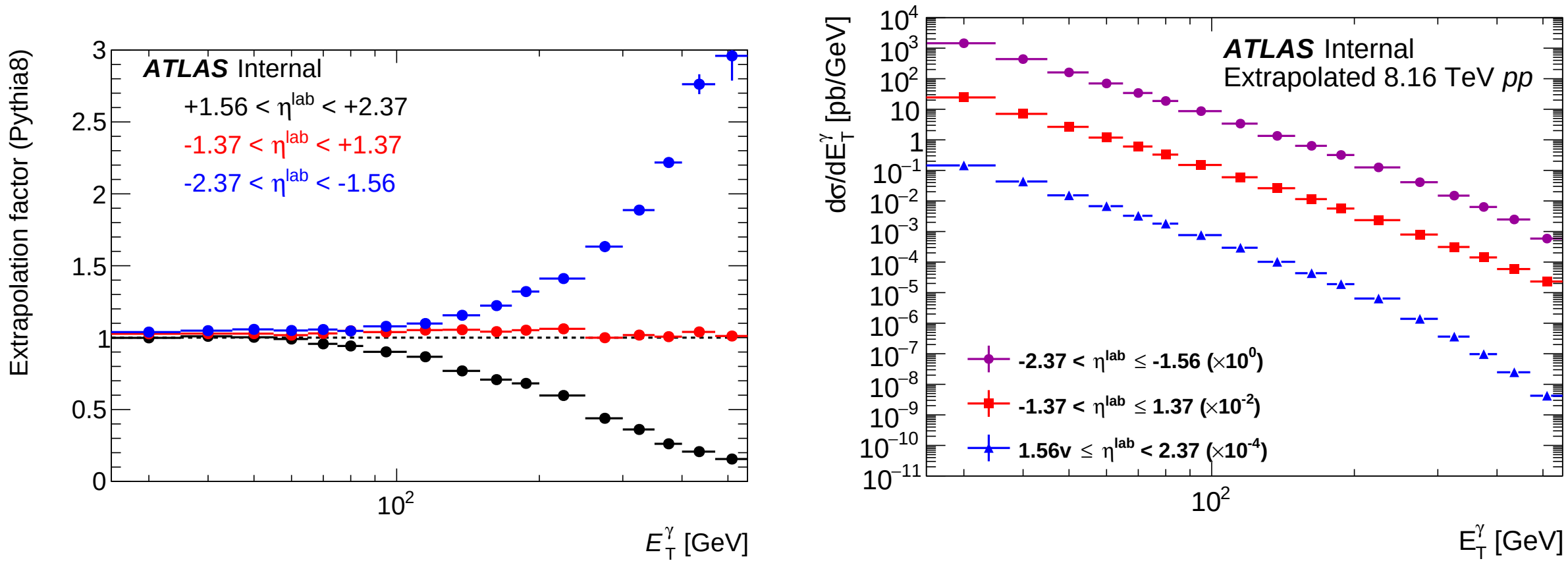

Figure 5.34: **Left**: Ratio of prompt, isolated photon spectrum in Pythia A14 NNPDF23LO simulation between 8.16 TeV (with $\Delta y = -0.465$ boost) and 8 TeV for each pseudorapidity slice. **Right**: Extrapolated $\sqrt{s} =$ 8.16 TeV $pp$ reference spectrum with center of mass boost by $\Delta y = -0.465$. The spectra are shown for each pseudorapidity slice.

Figure 5.33 compares the spectra as a function of signed $\eta^{\mathrm{lab}}$ between Pythia 8.16 TeV (with boost in the $\eta \to -\infty$ direction) and 8 TeV, while Figure 5.34 shows the ratio of the two as in Eqn. 5.14. It can be seen that at mid-rapidity or at moderate-$E_{\mathrm{T}}^{\gamma}$, the ratio of cross-sections is within 5–10% of unity. This is because the $\eta$ distribution at low or moderate-$E_{\mathrm{T}}^{\gamma}$ is fairly flat, meaning that a boost has little effect on the cross-section at mid-rapidity, and because the cross-section grows only slightly from 8 TeV to 8.16 TeV (a 2% difference in $\sqrt{s}$). However, at large-$E_{\mathrm{T}}^{\gamma}$ or at moderate-$E_{\mathrm{T}}^{\gamma}$ and large-$|\eta|$, the ratios deviate from unity since the $\eta$ shape of the photon spectrum becomes steeper. Here the corrections are up to a factor of 2 in the kinematic region reached in the $p$+Pb data.

Figure 5.34 shows the resulting reference $pp$ spectra constructed according to Eq. 5.14. These spectra are the ones used to construct the $R_{p\mathrm{Pb}}$ results. Since the extrapolated $pp$ reference is constructed by applying a multiplicative factor to each measured $E_{\mathrm{T}}^{\gamma}$, $\eta$ bin, the resulting spectrum inherits the relative systematic uncertainties of the original measurements exactly.

## 5.6 Systematic Uncertainties

The sources of systematic uncertainty affecting the measurement are described in this section, which is broken into two parts discussing the uncertainty in 1) the cross-section and 2) the nuclear modification factor $R_{p\mathrm{Pb}}$, including its ratio between forward and backward pseudorapidity regions. Tables including numerical values of all contributions to the systematic uncertainties of the cross section, $R_{p\mathrm{Pb}}$, and forward-to-backward ratio can be found in Appendix A.5.

### 5.6.1 Cross-Section Uncertainty

The major uncertainties in the cross-section can be divided into two main categories: those affecting the purity determination, which are dominant at low $E_{\mathrm{T}}^{\gamma}$ where the sample purity is low, and those affecting the detector performance corrections, which are dominant at high $E_{\mathrm{T}}^{\gamma}$. All other sources tend to be weakly dependent on $E_{\mathrm{T}}^{\gamma}$. A summary is shown in Figure 5.35. In each category, the uncertainty is the sum in quadrature of the individual components; the combined uncertainty is the sum in quadrature of all contributions, excluding those associated with the luminosity. The total uncertainties range from 15% at low and high $E_{\mathrm{T}}^{\gamma}$, where they are dominated by the purity and detector performance uncertainties respectively, to a minimum of approximately 6% at $E_{\mathrm{T}}^{\gamma} \approx 100$ GeV, where both of these uncertainties are modest. Each contribution to these uncertainties are studied in detail below.

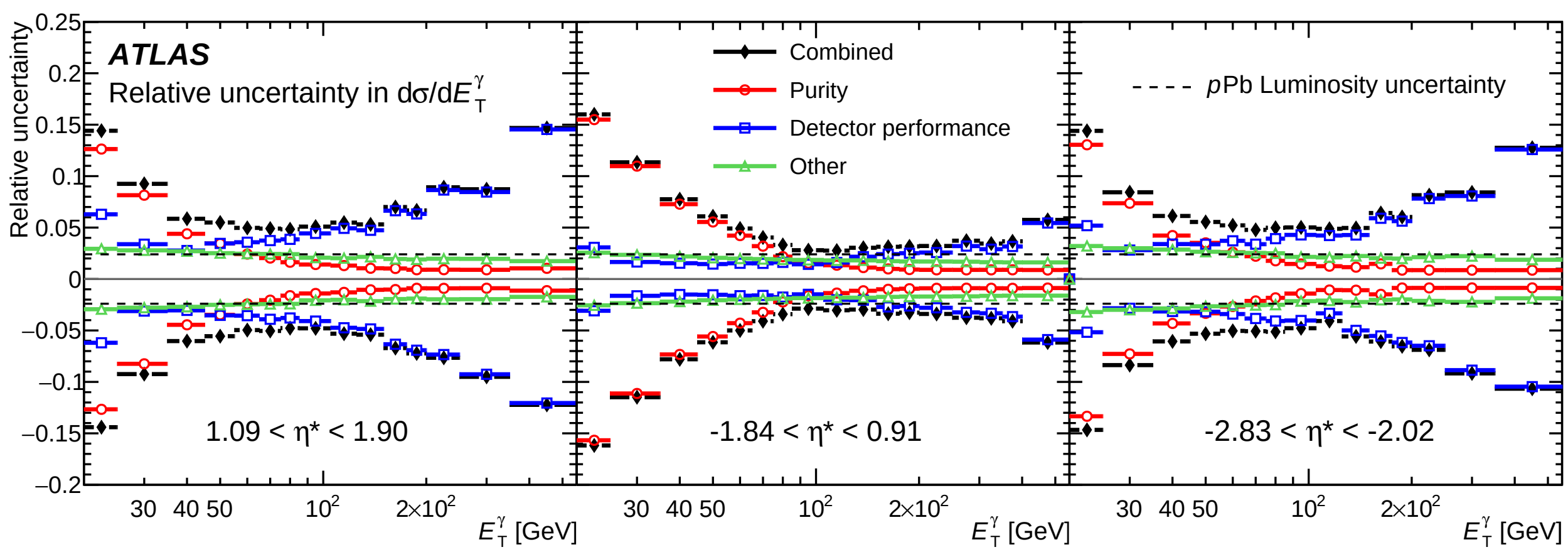

Figure 5.35: Summary of the relative sizes of major sources of systematic uncertainty in the cross-section measurement, as well as the combined uncertainty (excluding luminosity), shown as a function of photon transverse energy $E_{\mathrm{T}}^{\gamma}$.

#### 5.6.1.1 Purity Uncertainty

To assess the uncertainty in the purity determination, each boundary defining the sidebands used in the calculation is varied independently in order to understand the sensitivity of the measurement to the double sideband binning and correlation assumptions. The dominant uncertainty is that from the in the level of sideband correlation. If there is no correlation, one expects the following relationship of the yield of background photons in each sideband region:

$$R_{\text{bkg}} = \frac{N^{\text{A}}_{\text{bkg}} \cdot N^{\text{B}}_{\text{bkg}}}{N^{\text{C}}_{\text{bkg}} \cdot N^{\text{D}}_{\text{bkg}}} = 1 \; . \tag{5.15}$$

This condition has been tested in MC and data in previous studies and was found to be accurate to the level of 10%. This analysis performs the same data-driven check as was done in Ref. [116]. Namely, we divide the not-isolated background into the following separations analogous to the ABCD distinctions in the purity determination:

- B: Tight && (7.8 GeV < $E_{\text{T}}^{\gamma}$ < 17.8 [GeV])
- D: Non-Tight && (7.8 GeV < $E_{\text{T}}^{\gamma}$ < 17.8 [GeV])
- F: Tight && ($E_{\text{T}}^{\gamma}$ > 17.8 [GeV])
- E: Non-Tight && ($E_{\text{T}}^{\gamma}$ > 17.8 [GeV])

$R_{\text{bkg}}$ can now be estimated in the background region as

$$R_{\text{bkg}} = \frac{N^{\text{B}}_{\text{bkg}} \cdot N^{\text{E}}_{\text{bkg}}}{N^{\text{D}}_{\text{bkg}} \cdot N^{\text{F}}_{\text{bkg}}} \; . \tag{5.16}$$

Fig. 5.36 gives this estimation of $R_{\text{bkg}}$, plotted as a function of $E_{\text{T}}^{\gamma}$. The results agree with the a 10% variation found in previous studies, thus we apply $\pm 10\%$ variation on $R_{\text{bkg}}$ and recompute the purities to obtain the systematic variation. Fig. 5.37 shows the effect of this variation on the purity. This $\pm 10\%$ variation in the sideband correlation yields a 13% uncertainty in the cross-section in the lowest $E_{\text{T}}^{\gamma}$ range, decreasing to less than 1% for $E_{\text{T}}^{\gamma} > 100$ GeV.

The inverted photon, non-tight, identification requirement for the background candidates is varied to be less or more restrictive about which shower shapes the background candidates are required to fail. The purities calculated with these variations are shown in Fig. 5.38. As can be seen when plotting the relative deviation of these variations in Fig. 5.39, the purity from these variations is statistically limited. To not introduce significant statistical fluctuations in this systematic variation, the two variations are symmetrized by taking the average of the absolute value of each variation (plotted in black). The average points are further smoothed by fitting to constant functions (blue lines), in the forward and backward regions, and a piece-wise line (in $\log(x)$) between 20 and 90 GeV and constant between 90 and 550 GeV in the mid-rapidity region. The value of the fit functions are used as the symmetric uncertainty for these variations. This yields an uncertainty that is less than 1% for all $E_{\mathrm{T}}^{\gamma}$ in the forward and backward rapidity bins, but is significant at mid-rapidity ($-1.84 < \eta^* < 0.91$) where it is 9% in the lowest $E_{\mathrm{T}}^{\gamma}$ bin and decreases to be less than 1% for $E_{\mathrm{T}}^{\gamma} > 100$ GeV.

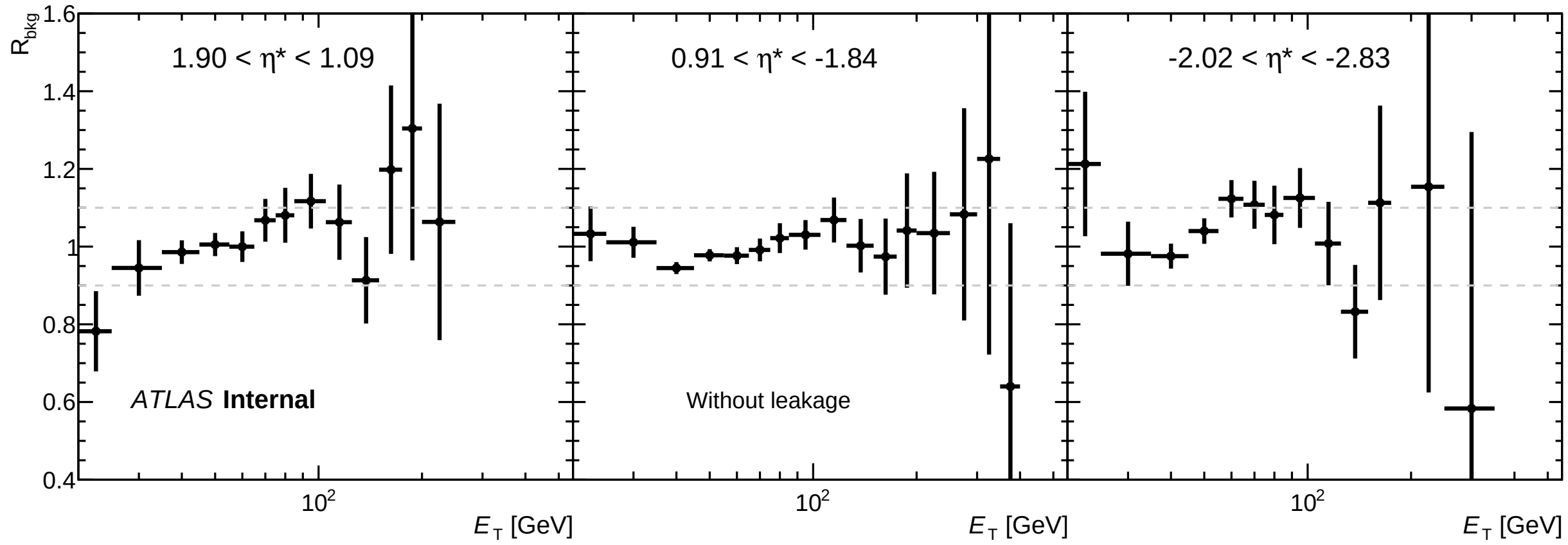

Figure 5.36: $R_{\mathrm{bkg}}$ estimated using the BDFE method plotted as a function of $E_{\mathrm{T}}^{\gamma}$.

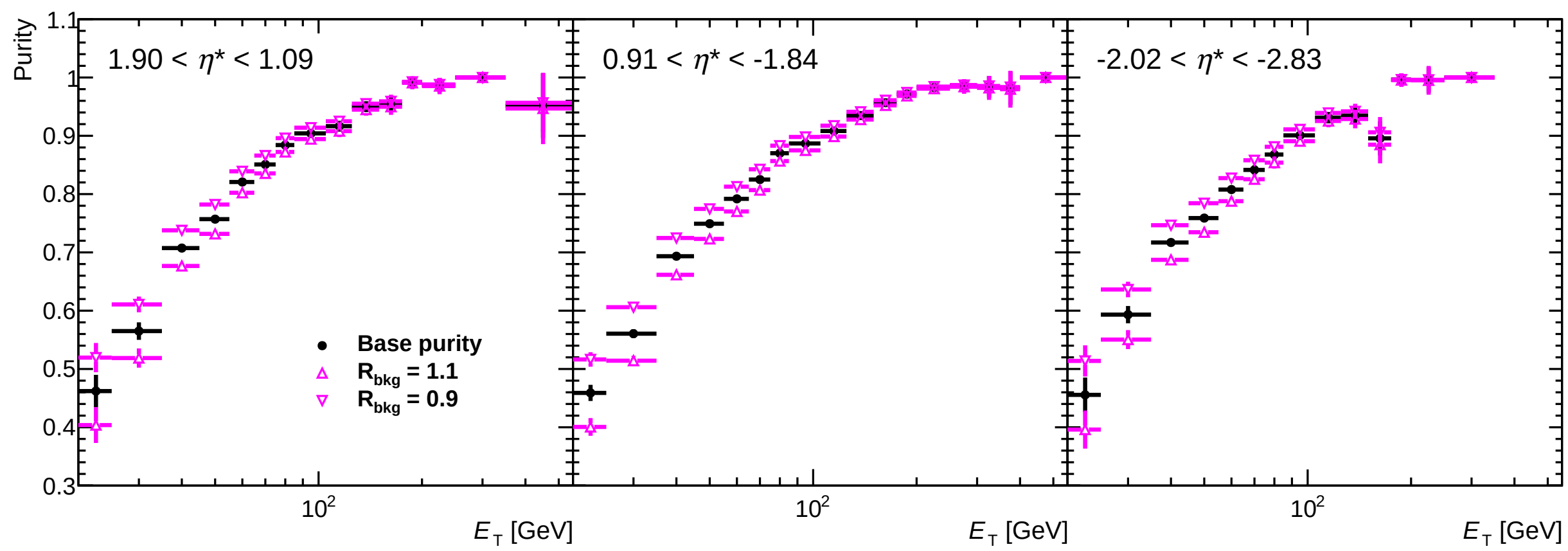

Figure 5.37: The effect of systematic variations in $R_{\mathrm{bkg}}$ when calculating the purity. The base purity is shown as black points and the purity calculated when $R_{\mathrm{bkg}}$ is varied up (and down) are shown as magenta up (down) triangles. The plots are shown in the usual pseudorapidity bins.

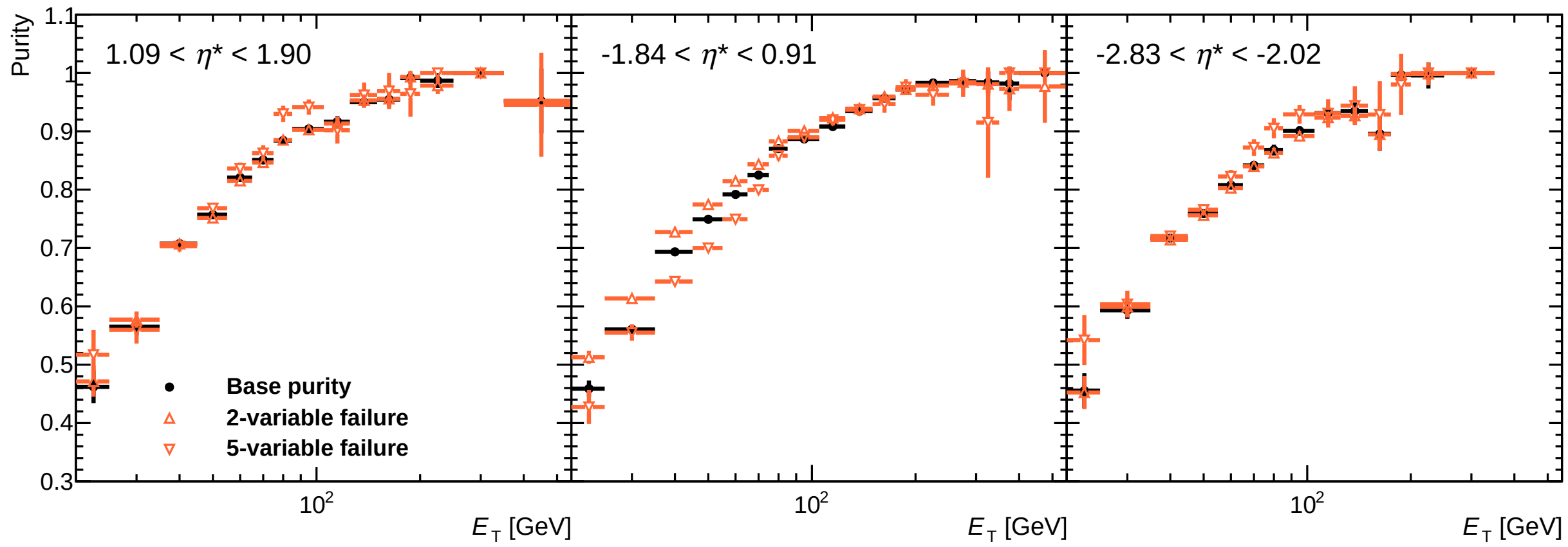

Figure 5.38: The effect of systematic variations in the definition of the non-tight sideband when calculating the purity. The base purity is shown as black points and the purity calculated with 2 (5) shower shape variable failure mode are shown as orange up (down) triangles. The plots are shown in the usual pseudorapidity bins.

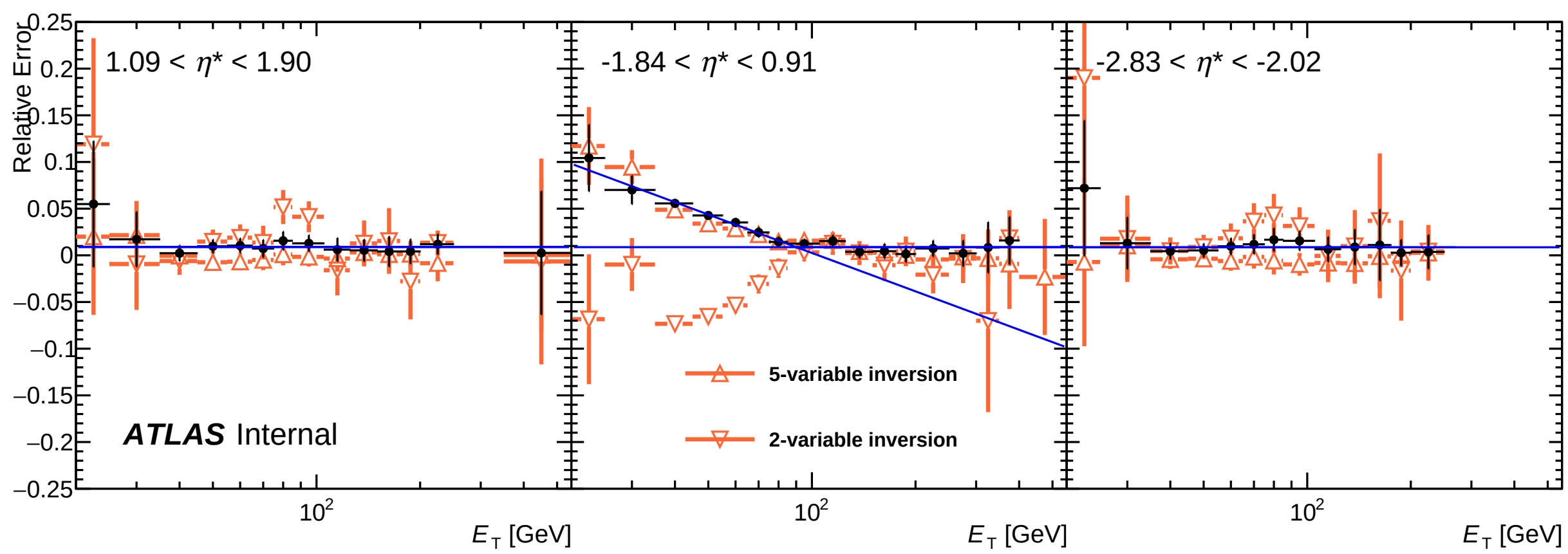

Figure 5.39: The relative deviation in the purity when varying the non-tight definition (orange). The two variations are symmetrized and averaged and plotted as black points. The averaged points are fit to constant functions (blue lines), in the forward and backward regions, and piece-wise line (in $log(x)$) between 20 and 90 GeV and constant between 90 and 550 GeV in the mid-rapidity region. These fits are used as the symmetric error for these variations.

Variations in the isolation energy threshold of $\pm 1$ GeV have been shown to cover any difference between simulations and data [117]. New leakages and purities are then calculated with the new isolation definition. Fig. 5.40 shows the effect of these variations on the purity. The result is a 1–2% effect on the cross-section in the lowest $E_{\mathrm{T}}^{\gamma}$ range and less than 1% at higher $E_{\mathrm{T}}^{\gamma}$. The uncertainty associated with the inverted shower-shape was smoothed and symmetrised; however, this is derived asymmetrically from the positive and negative variations separately.

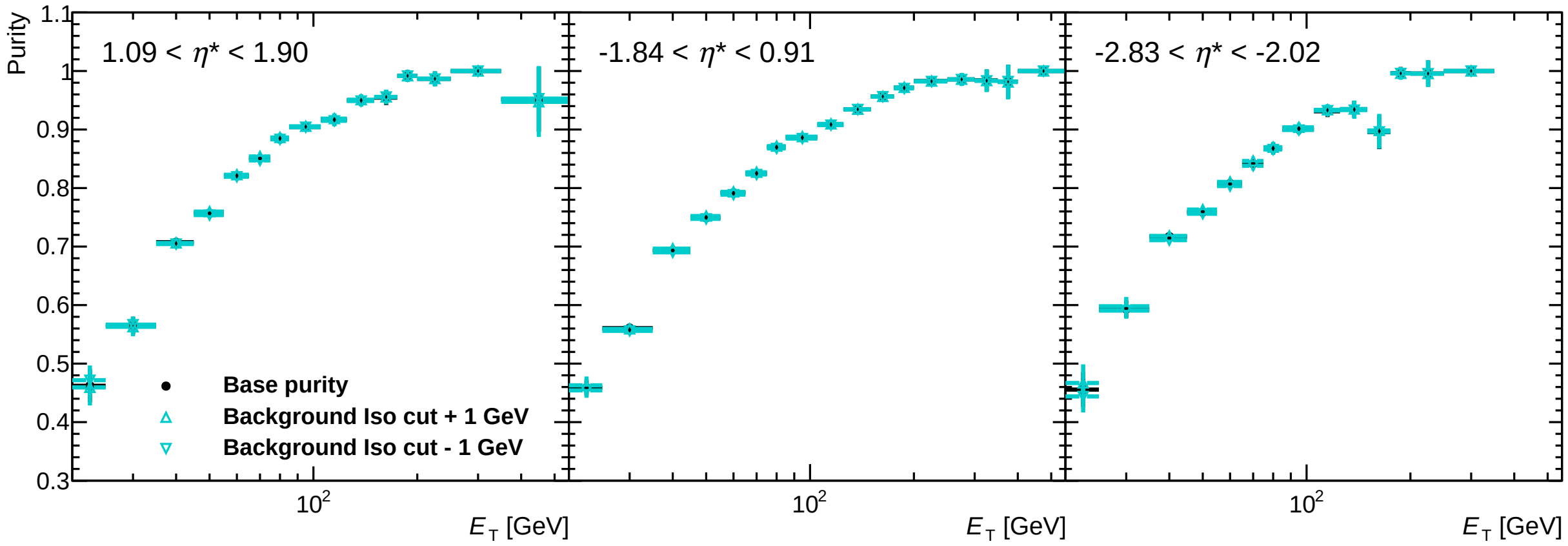

Figure 5.40: The effect of systematic variations in the definition of the non-iso sideband when calculating the purity. The base purity is shown as black points and the purity calculated with +1 GeV (-1 GeV) added to the constant term in the condition are shown as cyan up (down) triangles. The plots are shown in the usual pseudorapidity bins.

The variations affecting the purity are summarized in Fig. 5.41 showing the relative deviations for each variation. The yellow band gives the quadrature sum of each of these contributions as an asymmetric error band.

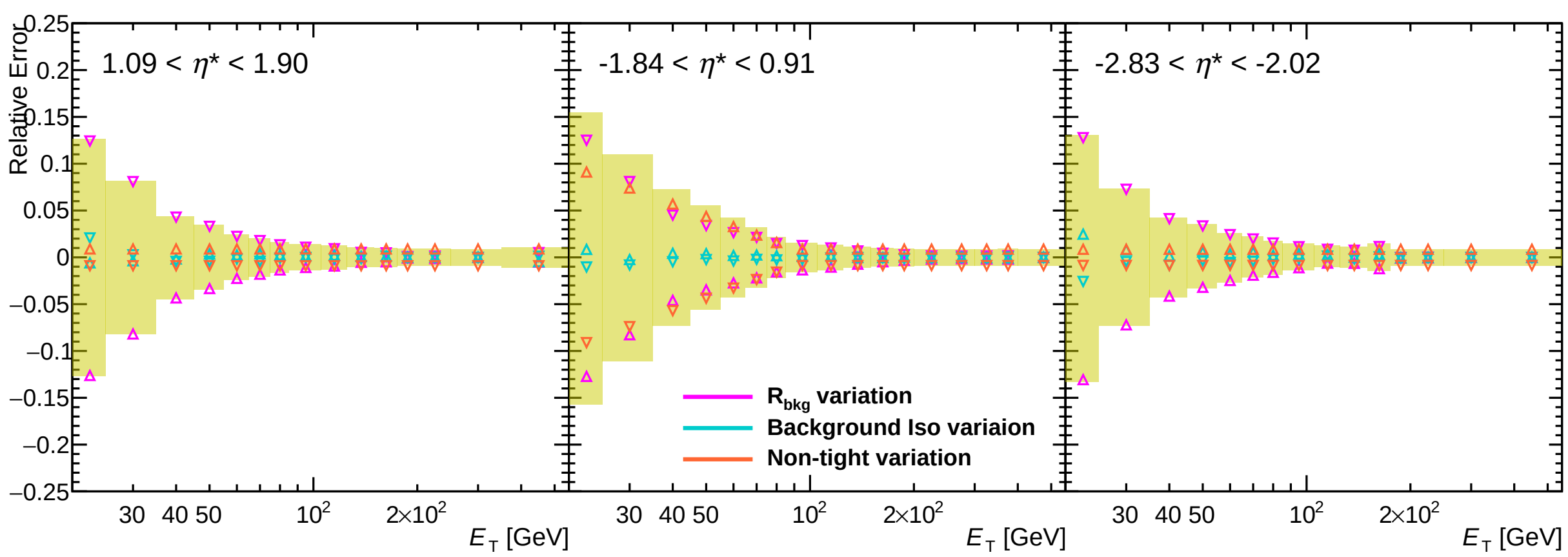

Figure 5.41: The relative deviation in the purity when varying $R_{\mathrm{bkg}}$ (magenta), the non-isolation definition (cyan), non-tight definition (orange), and their quadratic sum (yellow band). The plots are shown in the usual pseudorapidity bins and both.

#### 5.6.1.2 Detector Performance Uncertainty

Uncertainties associated with detector performance corrections are dominant at high $E_{\mathrm{T}}^{\gamma}$. A detailed description of the several components of the photon energy scale and resolution uncertainties are given in Ref. [117]. The impact of these on the measurement is determined by varying the reconstructed photon $E_{\mathrm{T}}^{\gamma}$ in simulation within the energy scale and resolution uncertainties and deriving alternative correction factors for positive and negative variations separately. Additionally, the scale factor variation accounts for uncertainties associated with corrections for small differences in reconstruction, identification and isolation efficiencies observed between data and simulation [133]. Fig. 5.42 shows the relative deviation for each of these systematic uncertainties overlaid. The yellow band gives the quadrature sum of each of these contributions as an asymmetric error band. Of these, the impact of the energy scale variation is dominant, giving a 10–15% contribution at 500 GeV in the forward and backward regions, decreasing to less than 2% at the lowest $E_{\mathrm{T}}^{\gamma}$. In the mid-rapidity region, the energy scale variation gives a 5% uncertainty at high $E_{\mathrm{T}}^{\gamma}$, decreasing to less than 1% at low $E_{\mathrm{T}}^{\gamma}$. The scale factor uncertainties are about 5% in the forward regions and low $E_{\mathrm{T}}^{\gamma}$ and less than 2% elsewhere.

#### 5.6.1.3 Other Uncertainties

Systematic uncertainties related to modelling in simulation, luminosity, electron contamination, and other sources tend to be lower than those previously discussed. However, their combined effect is dominant in the mid-rapidity region and between 90 GeV and 250 GeV.

Figure 5.9 indicates an undesirable discrepancy between the particle level and detector level isolation definitions. Minimizing this gap can potentially reduce model dependencies of the measurement. Recent analyses have accounted for this by adjusting the particle level definition to match at detector level. However, due to this analysis using the 8 TeV $pp$ measurement as a reference, it is important to use the same particle level definition. For this reason, the nominal particle level isolation definition is used for the central values of the measurement, but the effect on the cross section of changing the particle level cut to match the data is added as a systematic uncertainty. That is, the fit from Figure 5.9 is used to derive a mapping to what the particle level cut would be to effectively match the detector level values. Figure 5.43 shows the effect of this redefinition as about 2% at low $E_{\mathrm{T}}^{\gamma}$, and steadily falling to around 1% at the highest $E_{\mathrm{T}}^{\gamma}$.

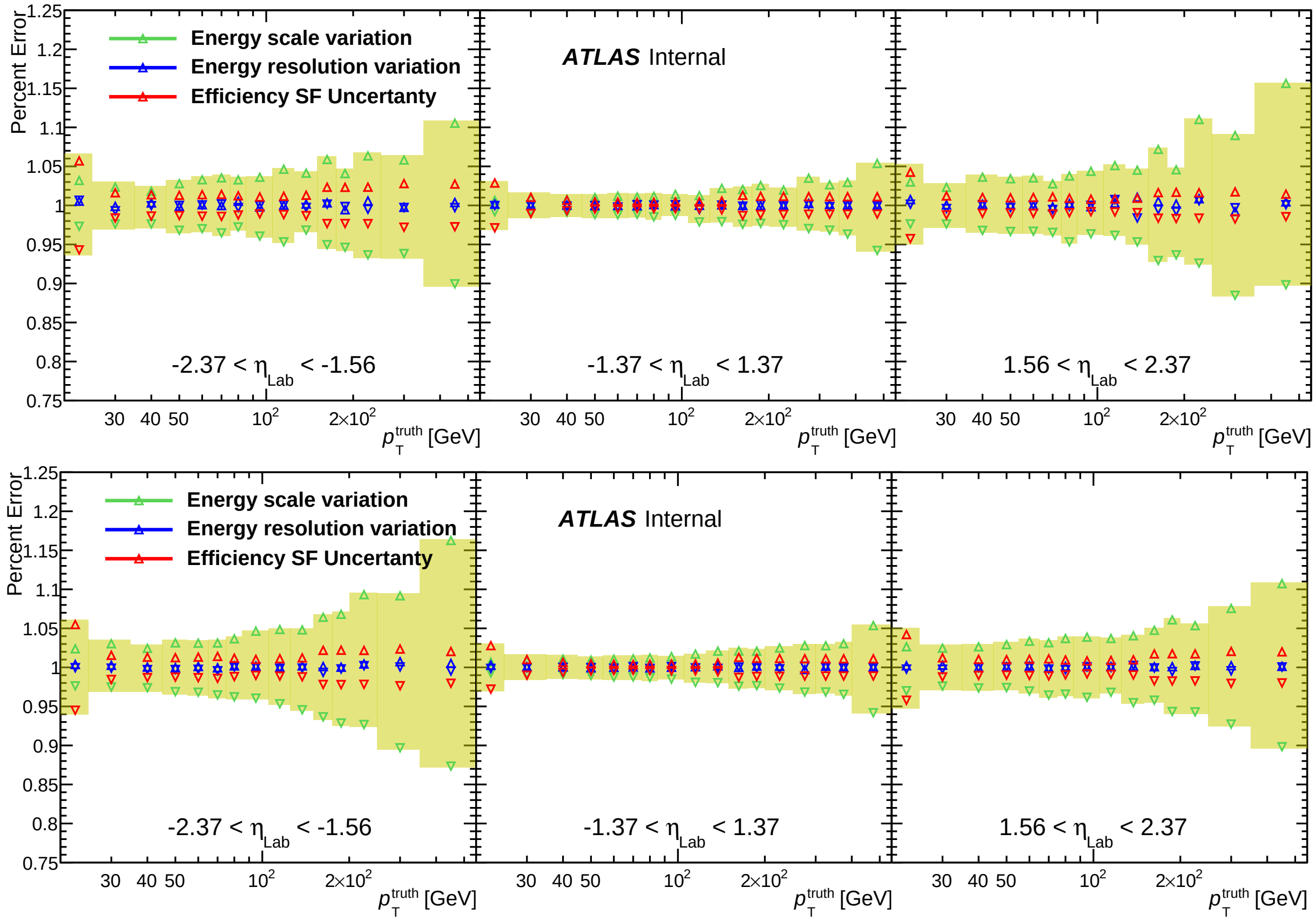

Figure 5.42: The relative deviation in the bin migration correction from the systematic variations in the energy resolution (blue) and scale (green) in MC. Additionally plotted is the error on the efficiency associated with the efficiency scale factors (red). The quadratic sum is overlaid as a yellow band. **Top:** period A; **bottom:** period B.

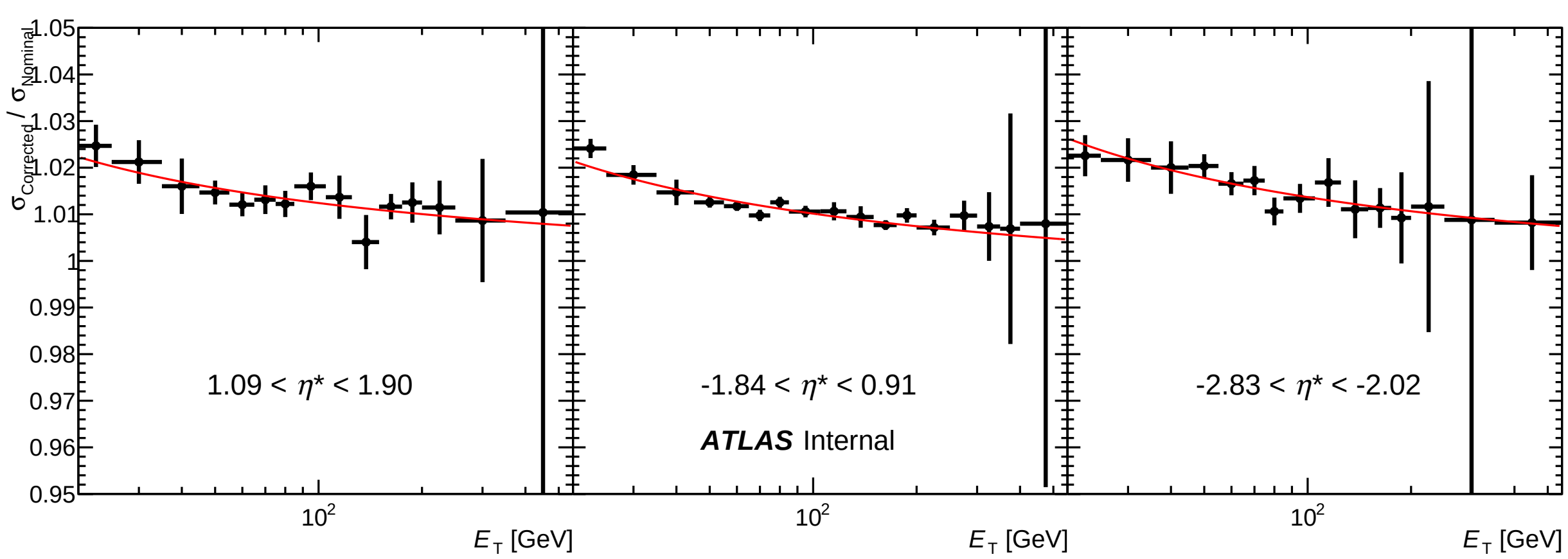

Figure 5.43: Effect on total cross section by changing particle level isolation cut to match the detector level.

As discussed in Sec 5.3.6, an uncertainty is assigned to cover the possible contribution of mis-reconstructed electrons, primarily from the decays of $W^{\pm}$ and $Z$ bosons, to the selected photon yield. Based on simulation studies, and the results of previous measurements [116, 117], this is assigned to be 1.3% for $E_{\mathrm{T}}^{\gamma} < 105$ GeV in forward pseudorapidity regions, and 0.5% everywhere else. To test the beam orientation dependence, the cross-section is measured using the data from each beam configuration separately. The two measurements agree at the level of 1%, well above the statistical uncertainty for most $E_{\mathrm{T}}^{\gamma}$ bins. This difference is taken as a global, symmetric uncertainty in the combined results.

To quantify the sensitivity of this measurement to the admixture of direct and fragmentation photons in the MC samples, we calculate the efficiencies and purities after re-weighting these components in the default Pythia MC sample. It is worth noting again here that the direct and fragmentation photon distinction is not well defined beyond leading order in pQCD calculations and cannot be determined experimentally. The default fraction of direct photons in the Pythia is plotted in Fig. 5.44, where the fraction of fragmentation photons is the complement. As an extreme variation, the MC is re-weighted such that the fraction of direct photons is unity, that is, all photons in the sample are direct. Fig 5.45 shows the effect of this variation as the ratio of cross section with re-weighting to the nominal. The ratios are fit with a constant function and each show a relative variation of approximately 1% which is taken as a systematic uncertainty.

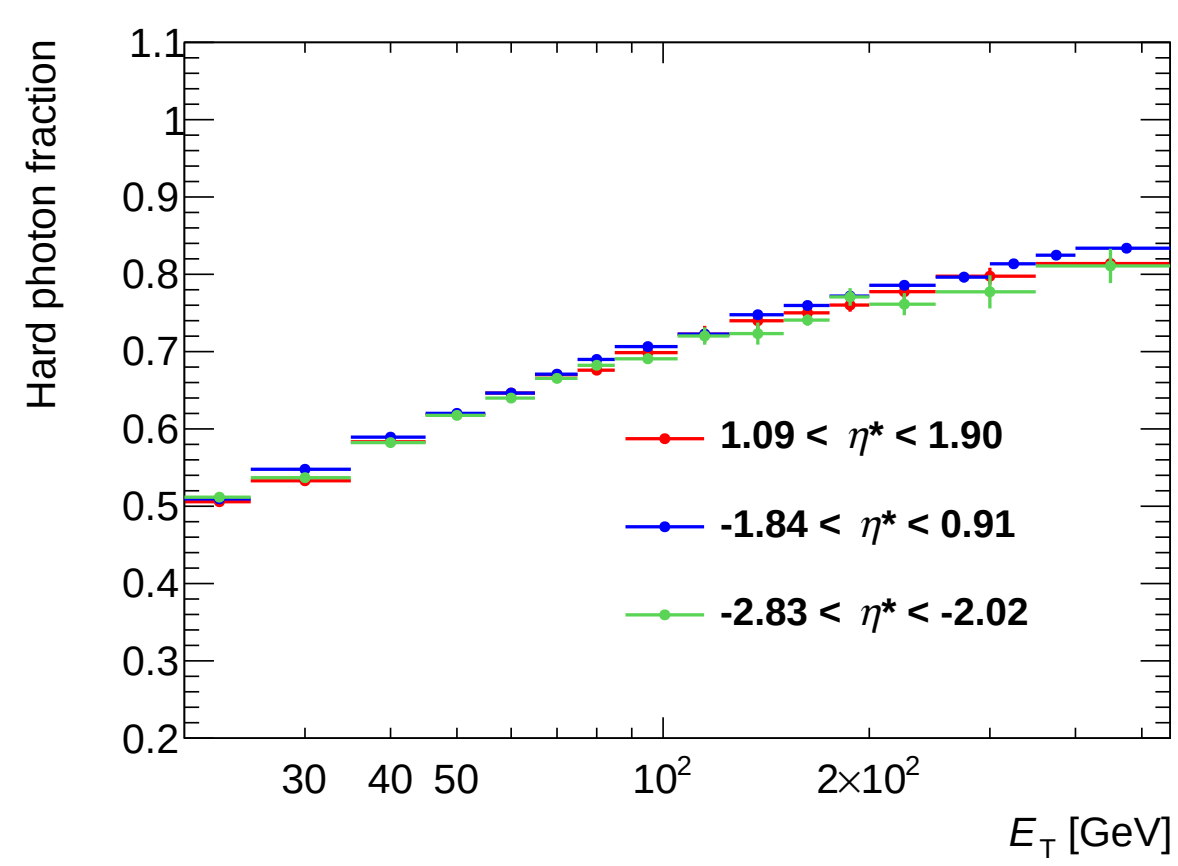

Figure 5.44: The default fraction of hard photons in the Pythia MC sample for each rapidity bin, plotted as a function of photon transverse energy.

Finally, the uncertainty in the integrated luminosity of the combined data sample is 2.4%. It is derived following a methodology similar to that detailed in Ref. [141]. The LUCID-2 detector is used for the baseline luminosity calibrations, and its luminosity scale is calibrated using x-y beam-separation scans [142].

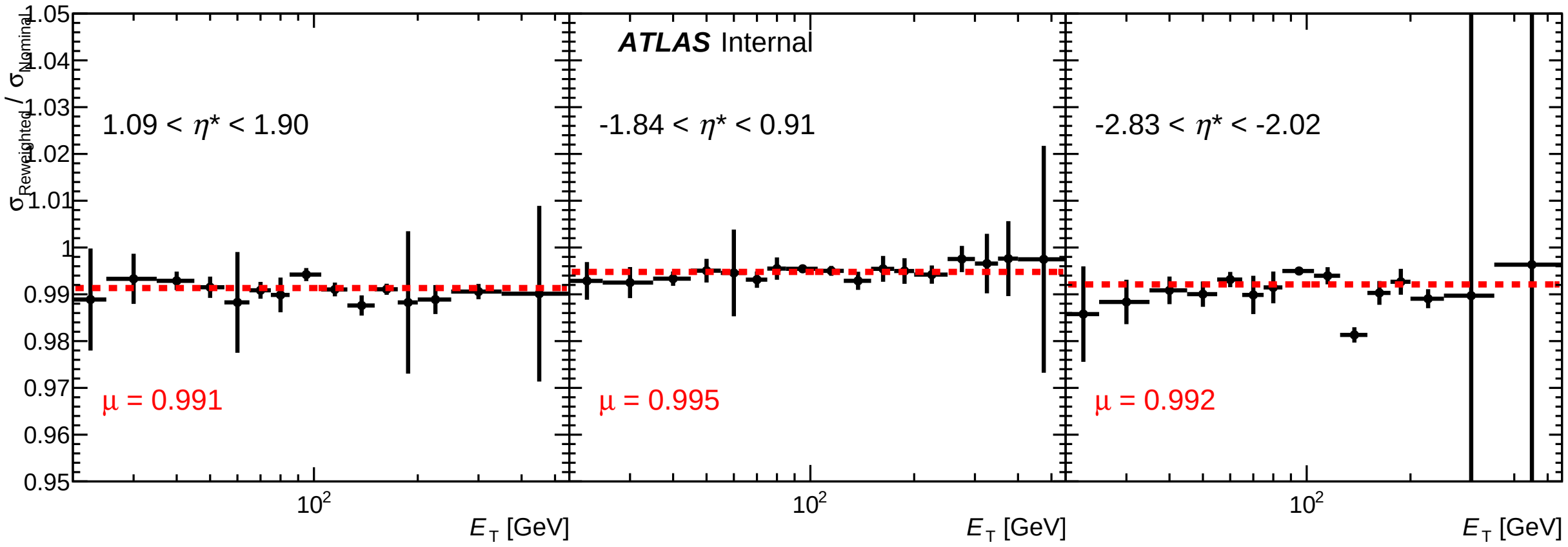

Figure 5.45: Ratios of the resulting cross sections with the direct fraction re-weighting to those from the nominal fraction.

### 5.6.2 $R_{p\text{Pb}}$ Uncertainty

The nuclear modification factor $R_{p\text{Pb}}$ is affected by systematic uncertainties associated with both the $p$+Pb and $pp$ measurements. The uncertainties in the differential cross-section of the $pp$ reference data are obtained directly from Ref. [116] and shown in Fig. 5.46 not including the global luminosity uncertainty of 1.9%. Due to differences in photon reconstruction, energy calibration, isolation and identification procedures between the $pp$ and $p$+Pb datasets, the uncertainties are treated as uncorrelated and added in quadrature.

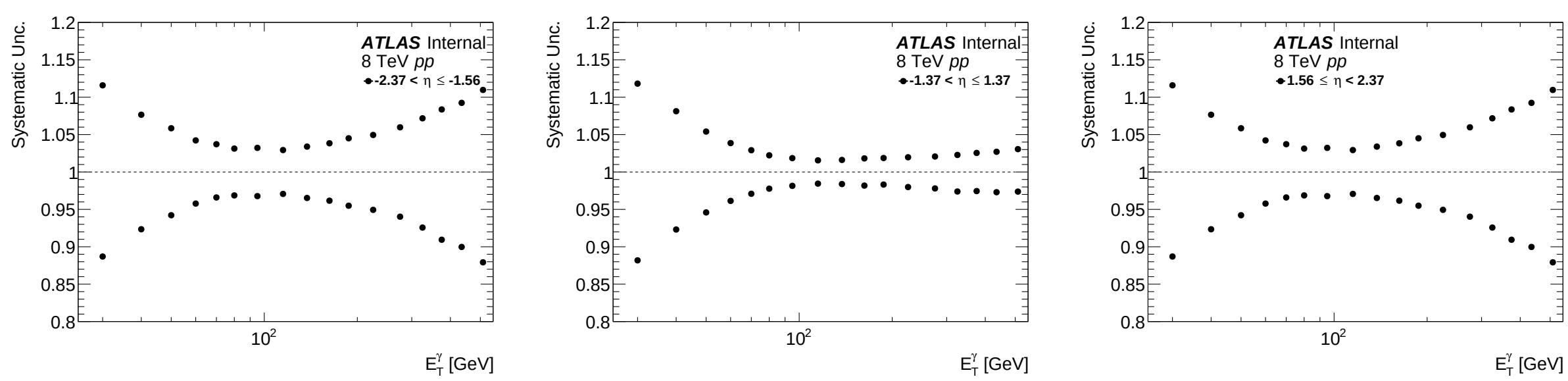

Figure 5.46: Summary of total systematic uncertainties on previously measured 8 TeVphoton spectrum [116], each panel showing a different $|\eta|$ selection. (A global luminosity uncertainty of 1.9% is not included.)

The uncertainty in the extrapolation of the $pp$ $E_{\mathrm{T}}^{\gamma}$ spectrum at 8 TeV is determined by using an alternative method to derive the multiplicative extrapolation factors. Jetphox 1.3.1 is used instead of Pythia to produce the ratio of the boosted 8.16 TeV to (not boosted) 8 TeV$pp$ cross-sections, which are used as the multiplicative factors on the $pp$ data. Jetphox was run in a mode to produce both the direct and fragmentation photon components. Both the Born and higher order contribution were calculated and summed. The Jetphox calculation is run nominally using the CT10 PDF as well as MSTW2008 as an additional source of variation. The variations from Pythia to Jetphox +CT10 are added in quadrature with those from Jetphox +CT10 to Jetphox +MSTW2008.

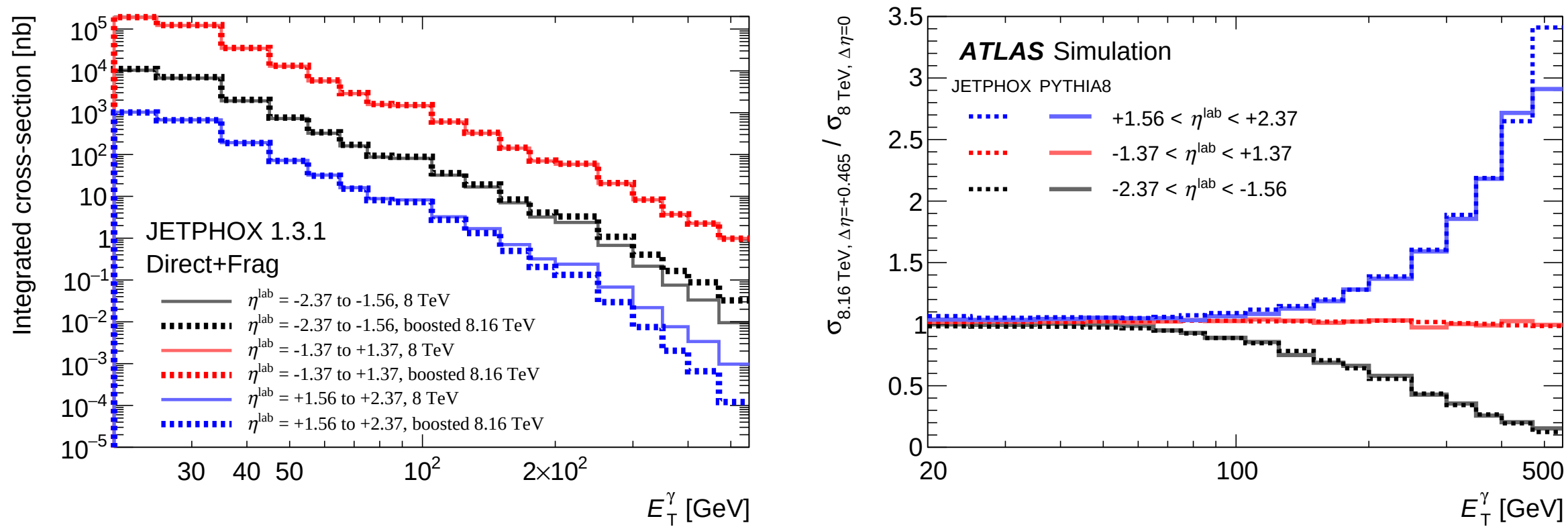

Figure 5.47: **Left:** Comparison of the Jetphox-calculated cross-section for boosted 8.16 TeV $pp$ and non-boosted 8 TeV $pp$ collisions, in the kinematic $E_{\mathrm{T}}^{\gamma}$ and $\eta^{\mathrm{lab}}$ bins used in the analysis. **Right:** Extracted extrapolation from Jetphox compared to the nominal values from Pythia.

The left side of Fig. 5.47 compares the total integrated cross-section for each kinematic bin used in this analysis for (non-boosted) 8 TeV $pp$ and boosted (to $\Delta y = -0.465$) 8.16 TeV $pp$. The right side shows the boosted 8.16 TeV / 8 TeV ratio, which are the extrapolation factors compared to those from Pythia. Clearly, they are qualitatively similar in each case, since the overall shape of the $\eta$ distribution in reasonably well reproduced by multiple theoretical approaches. The difference between these alternate extrapolation factors and the nominal ones are used to set a symmetric systematic uncertainty associated with the $pp$ reference as in Fig. 5.48 (since a different set of extrapolation factors will produce a different reference). Because the values are statistically varying, the histograms are fit to polynomial functions to smooth the variations. The systematic uncertainties up and down are then given by the value of the function at the bin center.

A summary of the total uncertainties is given in Fig. 5.49.

For the measurement of the ratio of $R_{p\mathrm{Pb}}$ values between the forward and backward pseudorapidity regions, each systematic variation affecting the purity and detector performance corrections is applied to the numerator and denominator in a coherent way, allowing them to partially cancel out in the ratio. All uncertainties in the other categories, except those from electron contamination and the beam direction difference, are treated as correlated. For this reason, they cancel out; notably the $p$+Pb luminosity and $pp$ cross-section uncertainties cancel out completely. The extrapolation uncertainties are treated as independent and are added in quadrature to the other uncertainties in $R_{p\mathrm{Pb}}$. The resulting uncertainty ranges from about 5% at the lowest $E_{\mathrm{T}}^{\gamma}$, where it is dominated by the uncertainty in the purity, to about 3% at mid-$E_{\mathrm{T}}^{\gamma}$, and again about 5% at high $E_{\mathrm{T}}^{\gamma}$, where it is dominated by uncertainty in the energy scale. A summary of the uncertainties in the forward-to-backward ratio is shown in Fig. 5.50.

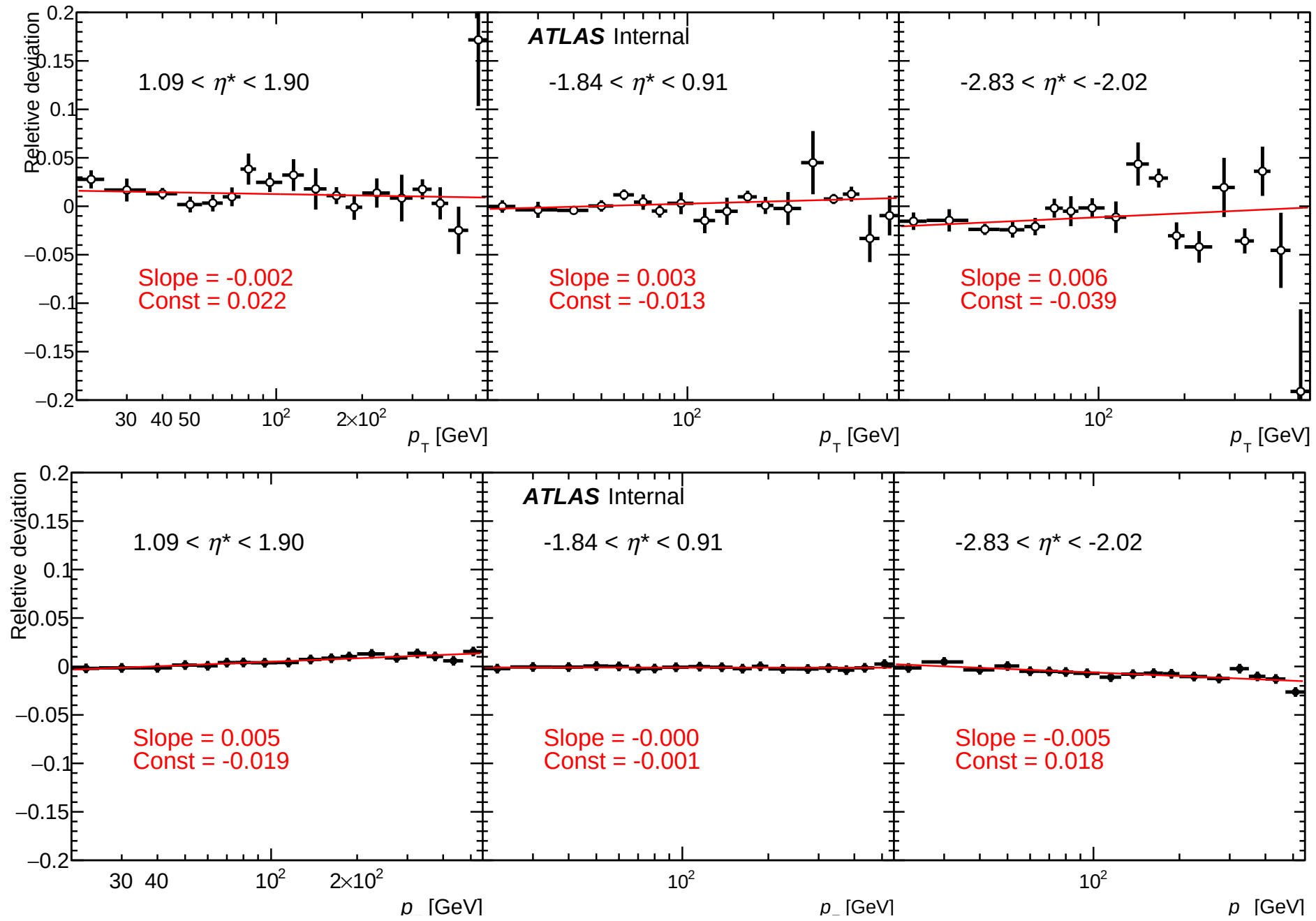

Figure 5.48: Top: Ratio between the constructed $pp$ reference using correction factors from Pythia and that using correction factors from Jetphox, plotted as a percent of the Pythia values. Bottom: Ratio of extrapolation factors calculated with Jetphox +CT10 to Jetphox +MSTW2008. These values are used as asymmetric errors on the extrapolation procedure, and are added in quadrature with the rest of the systematic uncertainties when plotting the $R_{p\mathrm{Pb}}$.

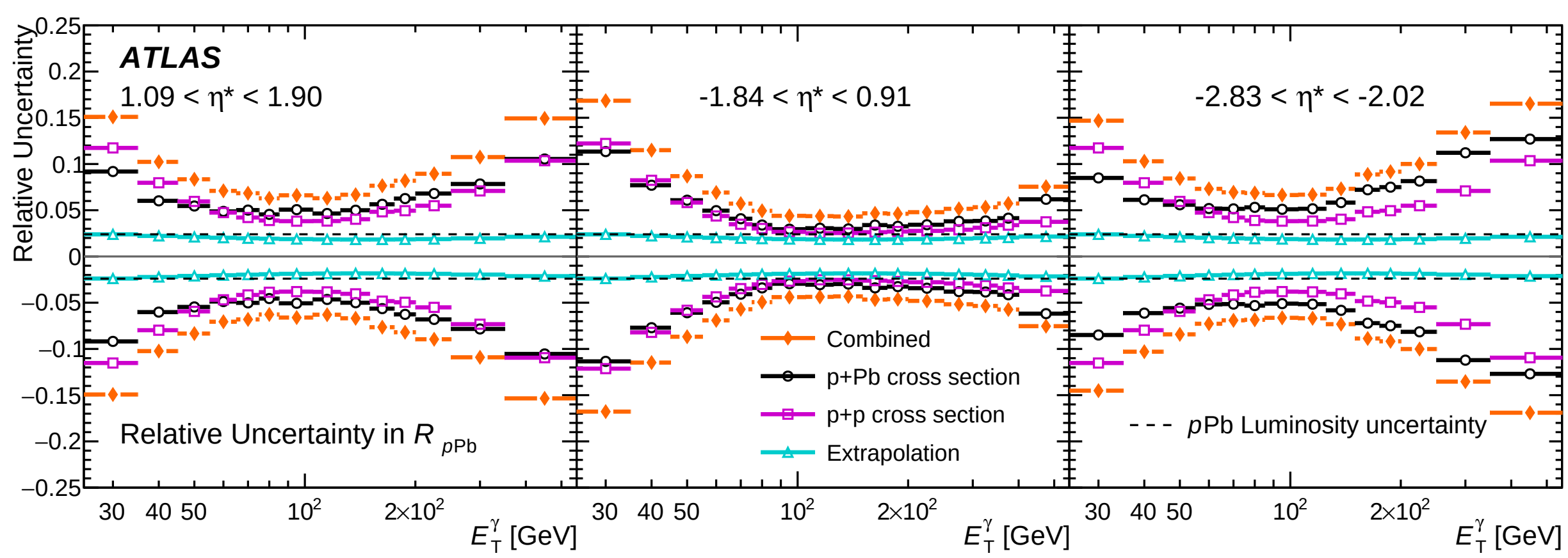

Figure 5.49: A breakdown of all systematic uncertainties on the $R_{p\mathrm{Pb}}$ measurement.

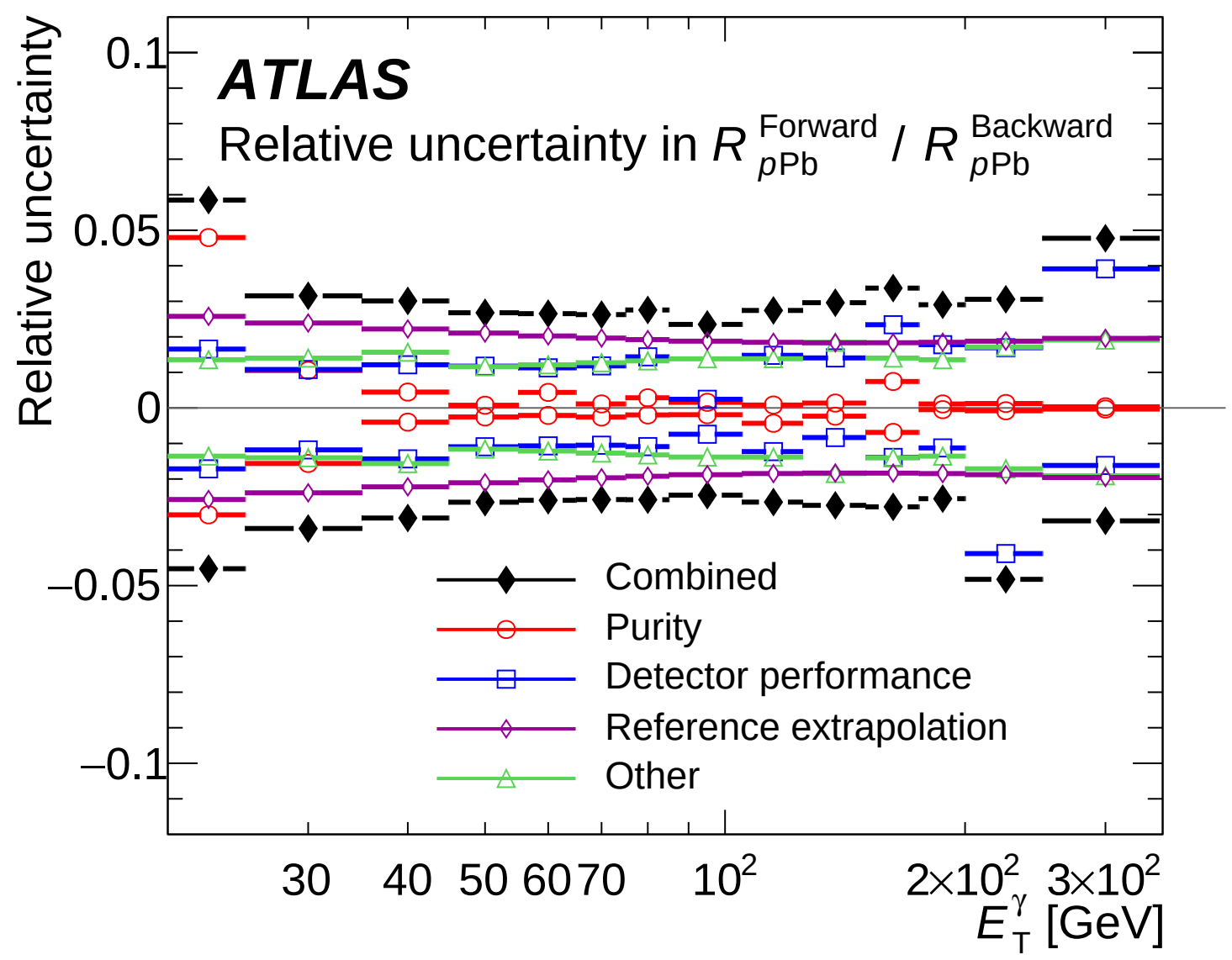

Figure 5.50: Summary of the relative size of major sources of systematic uncertainty in the forward-to-backward ratio of the nuclear modification factor $R_{p\mathrm{Pb}}$, as well as the combined uncertainty, shown as a function of photon transverse energy $E_{\mathrm{T}}^{\gamma}$. The Reference extrapolation refers to the uncertainty related to the extrapolation of the previously measured 8 TeV $pp$ spectrum to 8.16 TeV and boosted kinematics.

# Chapter 6

## Results of the Measurement of Direct Photon Production

Photon production cross-sections are shown in Figure 6.1 for photons with $E_{\mathrm{T}}^{\gamma} > 20$ GeV in three pseudorapidity regions. The measured $\mathrm{d}\sigma/\mathrm{d}E_{\mathrm{T}}^{\gamma}$ values decrease by five orders of magnitude over the complete $E_{\mathrm{T}}^{\gamma}$ range, which extends out to $E_{\mathrm{T}}^{\gamma} \approx 500$ GeV for photons at mid-rapidity. In Pythia, photons in this range typically arise from parton configurations in which the parton in the nucleus has Bjorken scale variable, $x_{\mathrm{A}}$, in the range $3 \times 10^{-3} \lesssim x_{\mathrm{A}} \lesssim 4 \times 10^{-1}$. In the nuclear modified PDF (nPDF) picture, this range probes the so-called shadowing (suppression for $x_{\mathrm{A}} \lesssim 0.1$), anti-shadowing (enhancement for $0.1 \lesssim x_{\mathrm{A}} \lesssim 0.3$), and EMC (suppression for $0.3 \lesssim x_{\mathrm{A}} \lesssim 0.7$) regions [58].

The data are compared with an NLO pQCD calculation similar to that used in Ref. [111] but using the updated CT14 [143] PDF set for the free-nucleon parton densities, where the data is similarly underestimated at low $E_{\mathrm{T}}^{\gamma}$. Jetphox [87] is used to perform a full NLO pQCD calculation of the direct and fragmentation contributions to the cross-section. The BFG set II [144] of parton-to-photon fragmentation functions are used, the number of massless quark flavours is set to five, and the renormalisation, factorisation and fragmentation scales are chosen to be $E_{\mathrm{T}}^{\gamma}$. In addition to the calculation with the free-nucleon PDFs, separate calculations are performed with the EPPS16 [58] and nCTEQ15 [125] nPDF sets. The EPPS16 calculation uses the same free-proton PDF set, CT14, as the free-nucleon baseline to which the modifications are applied. The prediction is systematically lower than the data by up to 20% at low $E_{\mathrm{T}}^{\gamma}$ but is closer to the data at higher $E_{\mathrm{T}}^{\gamma}$, consistent with the results of such comparisons in *pp* collisions at LHC energies [116, 117]. A recent calculation of isolated photon production at NNLO found that the predicted cross-sections were systematically larger at low $E_{\mathrm{T}}^{\gamma}$ than the NLO prediction [145], and thus may provide a better description of the data in this and previous measurements.

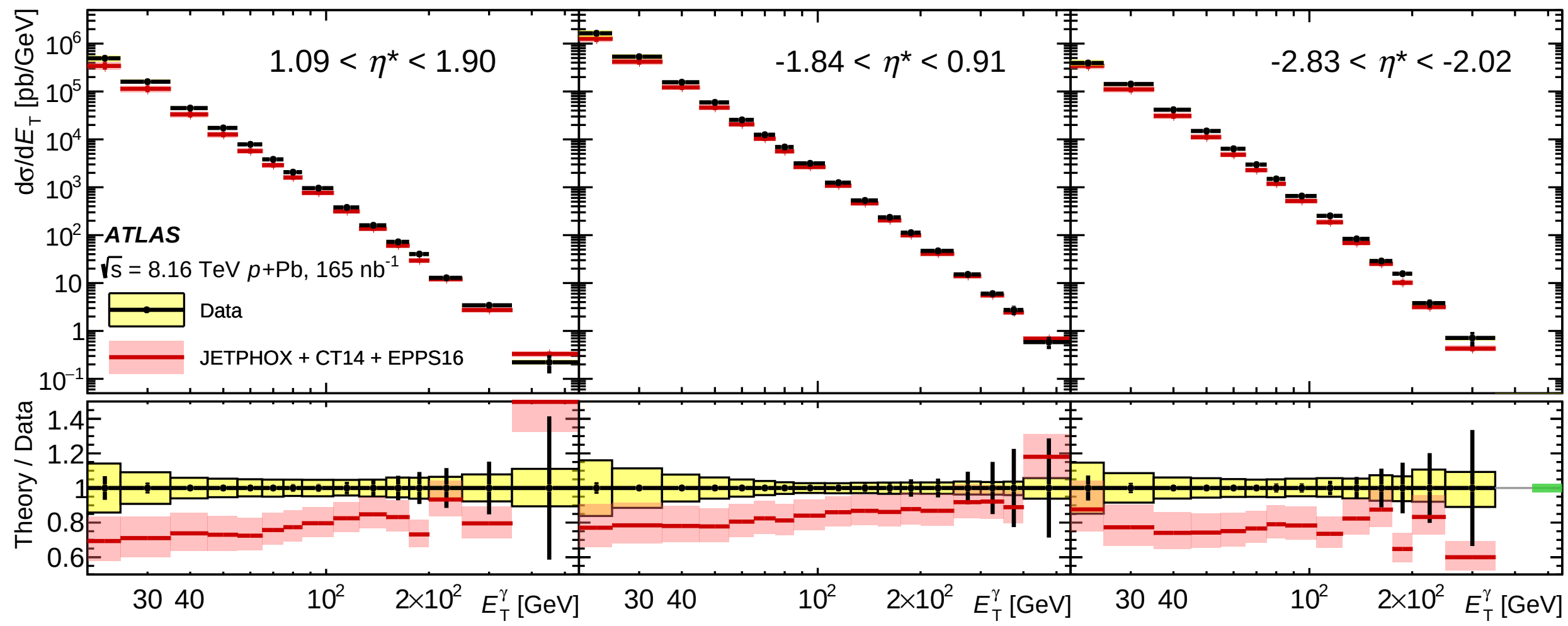

Figure 6.1: Prompt, isolated photon cross-sections as a function of transverse energy $E_{\mathrm{T}}^{\gamma}$, shown for different centre-of-mass pseudorapidity, $\eta^*$, regions in each panel. The data are compared with Jetphox with the EPPS16 nuclear PDF set [58], with the ratio of theory to data shown in the lower panels. Yellow bands correspond to total systematic uncertainties in the data (not including the luminosity uncertainty), vertical bars correspond to the statistical uncertainties in the data, and the red bands correspond to the uncertainties in the theoretical calculation. The green box (at the far right) represents the 2.4% luminosity uncertainty.

Uncertainties associated with these calculations are assessed in a number of ways. Factorisation, renormalisation, and fragmentation scales are varied, up and down, by a factor of two as in Ref. [116]. The uncertainty is taken as the envelope formed by the minimum and maximum of each variation in every kinematic region and is dominant in most regions. PDF uncertainties are calculated via the standard CT14 error sets and correspond to a 68% confidence interval. Again following Ref. [116] the sensitivity to the choice of $\alpha_{\mathrm{S}}$ is evaluated by varying $\alpha_{\mathrm{S}}$ by $\pm 0.002$ around the central value of 0.118 in the calculation and PDF. Uncertainties from nPDFs are calculated from the error sets which correspond to 90% confidence intervals, as described in Ref. [58]. These are converted into uncertainty bands which correspond to a 68% confidence interval. A summary of each variation is shown in Fig. 6.2.

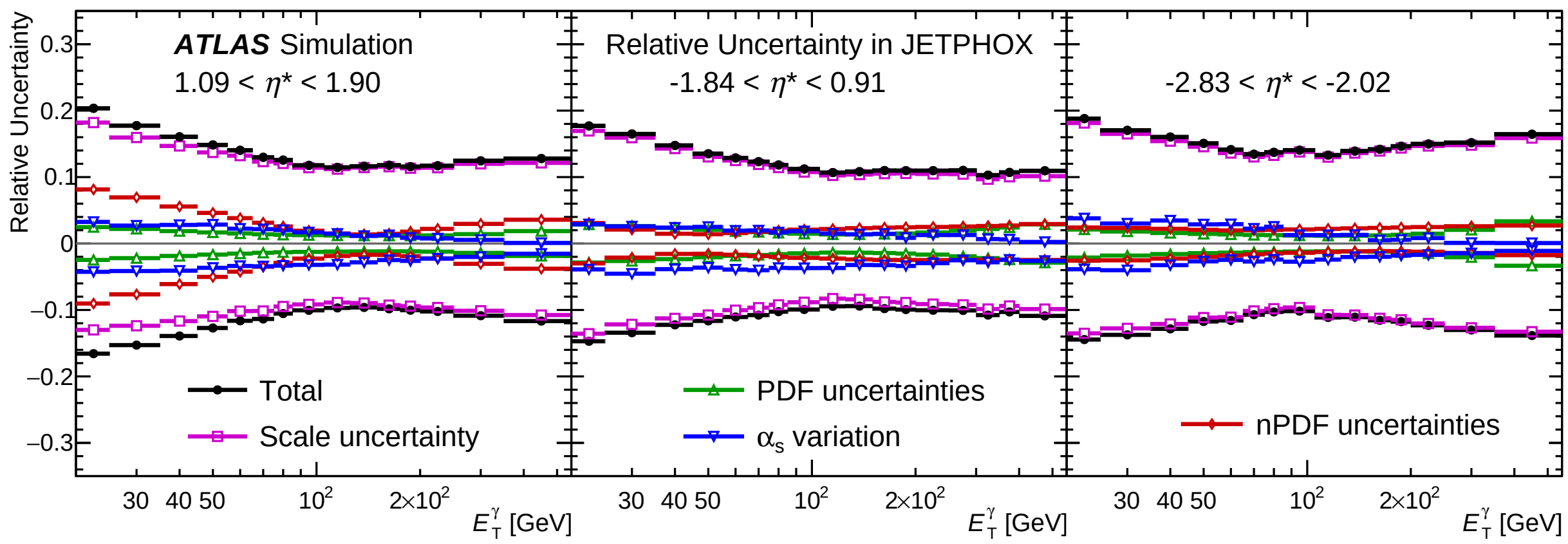

Figure 6.2: A breakdown of all systematic uncertainties in the cross-section prediction from Jetphox with the EPPS16 nPDF set.

## 6.1 $R_{p\mathrm{Pb}}$ Results

Figure 6.3 shows the nuclear modification factor $R_{p\mathrm{Pb}}$ as a function of $E_{\mathrm{T}}^{\gamma}$ in different $\eta^*$ regions. At forward rapidities ($1.09 < \eta^* < 1.90$), the $R_{p\mathrm{Pb}}$ value is consistent with unity, indicating that nuclear effects are small. In Pythia, photons in this region typically arise from configurations with gluon partons from the Pb nucleus with $x_{\mathrm{A}} \approx 10^{-2}$. Nuclear modification pulls the pQCD calculation down slightly for $E_{\mathrm{T}}^{\gamma} < 100$ GeV, above which the modification reverses, indicating a crossover between shadowing and anti-shadowing regions. At mid-rapidity, nuclear effects are similarly small and consistent with unity at low $E_{\mathrm{T}}^{\gamma}$, but at higher $E_{\mathrm{T}}^{\gamma}$, there is a hint that $R_{p\mathrm{Pb}}$ is lower. This feature primarily reflects the different up- and down-quark composition of the nucleus relative to the proton and is more important at larger parton $x$. In this case, the larger relative down-quark density decreases the photon yield. This effect is evident in the Jetphox theory curve in blue dash-dotted line, which includes the proton–neutron asymmetry and the free-nucleon PDF set CT14. This effect is most pronounced at backward pseudorapidity where, in Pythia, the nuclear parton composition is typically a valence quark with $x_{\mathrm{A}} \approx 0.2$. Here, nPDF modification moves $R_{p\mathrm{Pb}}$ above the free-nucleon PDF calculation at low $E_{\mathrm{T}}^{\gamma}$ but below at high $E_{\mathrm{T}}^{\gamma}$, indicating the crossover from the anti-shadowing to the EMC region.

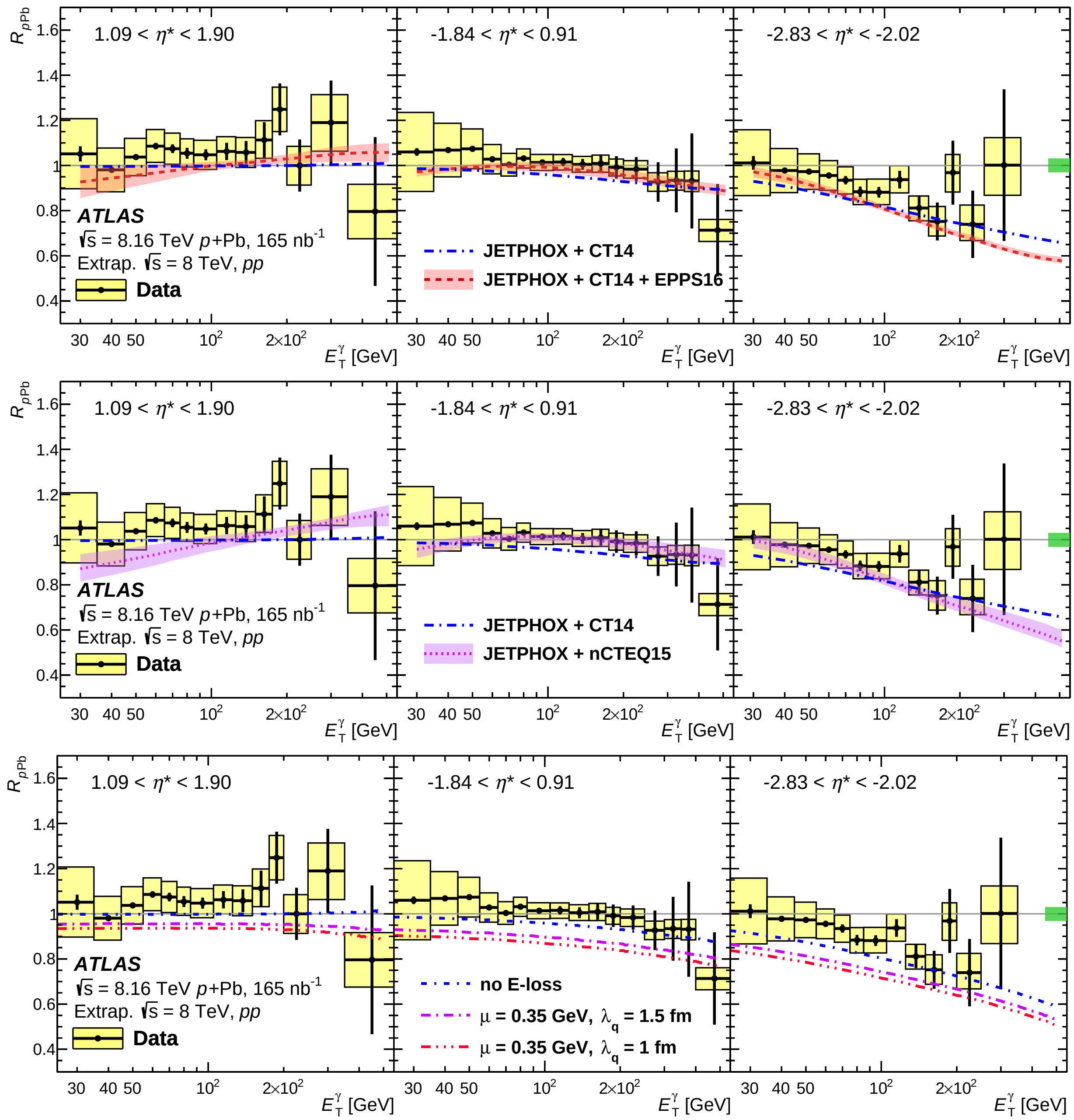

Figure 6.3: Nuclear modification factor $R_{p\mathrm{Pb}}$ for isolated, prompt photons as a function of photon transverse energy $E_{\mathrm{T}}^{\gamma}$, shown for different centre-of-mass pseudorapidity, $\eta^{*}$, regions in each panel. The $R_{p\mathrm{Pb}}$ is measured using a reference extrapolated from $\sqrt{s} = 8$ TeV $pp$ data (see text). The data are identical in each row, but show comparisons with the expectations based on Jetphox with the EPPS16 nuclear PDF set (top) [58], with the nCTEQ15 nuclear PDF set (middle) [125], and with an initial-state energy-loss calculation (bottom) [112, 113, 126]. In all plots, the yellow bands and vertical bars correspond to total systematic and statistical uncertainties in the data respectively. In the top and middle panels, the red and purple bands correspond to the systematic uncertainties in the theoretical calculations. The green box (at the far right) represents the combined 2.4% $p$+Pb and 1.9% $pp$ luminosity uncertainties.

The $R_{p\mathrm{Pb}}$ calculations including nPDFs consider only the nPDF uncertainty, since previous calculations have shown that the scale and PDF uncertainties cancel out almost completely in the kinematic region of the measurement [111], and no non-perturbative corrections are applied. Within the present uncertainties, the data are consistent with both the free-proton PDFs and with the small effects expected from a nuclear modification of the parton densities.

The $R_{p\mathrm{Pb}}$ measurements are also compared with an initial-state energy-loss prediction that is calculated within the framework described in Refs. [112, 113, 126]. In this model, the energetic partons undergo multiple scattering in the cold nuclear medium, and thus lose energy due to this medium-induced gluon bremsstrahlung, before the hard collision. The calculation is performed with a parton–gluon momentum transfer $\mu = 0.35$ GeV and mean free path for quarks $\lambda_q = 1.5$ fm. Alternative calculations with a shorter path length ($\lambda_q = 1$ fm), and a control version with no initial-state energy loss, are also considered. The data disfavour a large suppression of the cross-section from initial-state energy-loss effects.

The ratio of the $R_{p\mathrm{Pb}}$ values between forward and backward pseudorapidity, shown in Figure 6.4, is studied as a way to reduce the effect of common systematic uncertainties and better isolate the magnitude of nuclear effects [146]. The remaining systematic uncertainty, discussed in Sec. 5.6.2, is dominated by the reference extrapolation and treated as uncorrelated between points. Below $E_{\mathrm{T}}^{\gamma} \approx 100$ GeV, this corresponds roughly to the ratio of $R_{p\mathrm{Pb}}$ from photons from gluon nuclear parton configurations in the shadowing $x_{\mathrm{A}}$ region to that from quark partons in the anti-shadowing region. This can be seen in the top two panels of Figure 6.4, where the nuclear modification (red/purple bands) brings the JETPHOX calculation below that of the free-nucleon PDF (blue curve), though the effect from EPPS16 is less significant. In contrast, the behaviour is reversed at higher $E_{\mathrm{T}}^{\gamma}$ where the numerator probes the shadowing/anti-shadowing crossover region and the denominator moves deeper into the EMC region [58]. The data are consistent with the pQCD calculation before incorporating nuclear effects, except possibly in the region $E_{\mathrm{T}}^{\gamma} < 55$ GeV, which is sensitive to the effects from gluon shadowing. At low $E_{\mathrm{T}}^{\gamma}$, the data are systematically higher than the calculations which incorporate nPDF effects, but approximately within their theoretical uncertainty. Additionally, in the lower plot of Figure 6.4, the forward-to-backward ratios are compared with predictions from a model incorporating initial-state energy loss. The data show a preference for no or only a limited amount of energy loss.

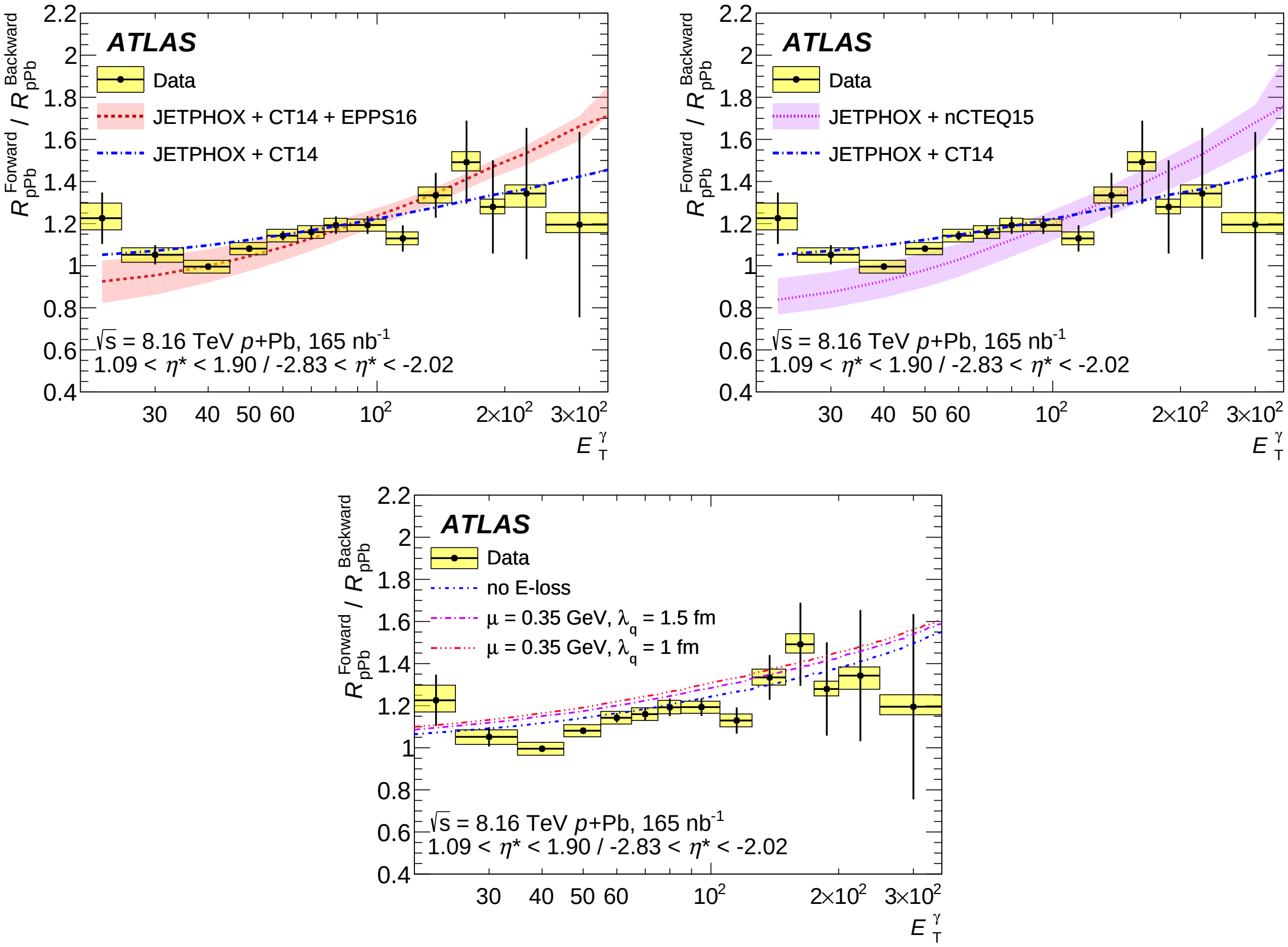

Figure 6.4: Ratio of the nuclear modification factor $R_{p\mathrm{Pb}}$ between forward and backward pseudorapidity for isolated, prompt photons as a function of photon transverse energy $E_{\mathrm{T}}^{\gamma}$. The data are identical in each panel, but show comparisons with the expectations based on Jetphox with the EPPS16 nuclear PDF set (top, left) [58] or with the nCTEQ15 nuclear PDF set (top, right) [125], and with an initial-state energy-loss calculation (bottom) [112, 113, 126]. The strength of the initial-state energy-loss effect is parameterized by $\lambda_{\mathrm{q}}$, which represents the mean free path of partons in the nuclear medium and is smaller for a larger degree of energy loss. In all plots, the yellow bands and vertical bars correspond to total systematic and statistical uncertainties in the data respectively. In the left and right panels, the red and purple bands correspond to the systematic uncertainties in the calculations.

# Chapter 7

## Measurement of Azimuthal Anisotropy

### 7.1 Introduction

Small collision systems provide a way to test the interpretation that the $v_n$ flow signal at high $p_{\mathrm{T}}$ in large systems is a result of the path-length differential energy loss of hard partons traversing the QGP, as discussed in Sec 2.4.4. In small systems, low-$p_{\mathrm{T}}$ flow signals match well with expectations from a nearly inviscid hydrodynamic flow of the QGP [5]; however, measurements aimed at observing signatures of jet quenching in have found no such effect. The ATLAS experiment has published results on the charged hadron azimuthal anisotropy up to $p_{\mathrm{T}} \approx$ 12 GeV in $p$+Pb that hint at a non-zero anisotropy extending into the region beyond the usual hydrodynamic interpretation and into the regime of jet quenching [147]. However, it seems unlikely that there can be differential jet quenching as a function of orientation relative to the QGP geometry if there is no jet quenching in $p$+Pb collisions as observed in the spectra. Thus, there are two related outstanding puzzles, one being the lack of jet quenching observed in the spectra, if indeed small droplets of QGP are formed, and the other being what mechanism can lead to high-$p_{\mathrm{T}}$ hadron anisotropies other than differential jet quenching.

This Chapter presents a measurement of the azimuthal anisotropy of unidentified hadrons as a function of $p_{\mathrm{T}}$ and centrality in $\sqrt{s_{\mathrm{NN}}} = 8.16$ TeV $p$+Pb collisions with the ATLAS detector. The measurement is made using two-particle correlations, measured separately for minimum-bias triggered events and events requiring a jet with $p_{\mathrm{T}}$ greater than either 75 GeV or 100 GeV. A standard template fitting procedure [12, 13] is applied to subtract the non-flow contributions to the azimuthal correlations from particle decays, jets, dijets, and global momentum conservation. To decrease the residual influence of the non-flow correlation in the jet events, a novel procedure is used to restrict the acceptance of particles according to the location of jets in the event. Assuming that the two-particle anisotropy coefficients are the products of the corresponding single-particle coefficients (factorisation), the elliptic and triangular anisotropy coefficients, $v_2$ and $v_3$, are reported as a function of $p_{\mathrm{T}}$. Additionally, $v_2$ results are presented as a function of centrality in three different $p_{\mathrm{T}}$ ranges. Finally, the fractional contribution to the correlation functions from jet particles is determined as a function of $p_{\mathrm{T}}$.

## 7.2 Data and Event Selection

### 7.2.1 Trigger and Event Selection

The events used in this analysis fall into two categories: minimum bias and jet events. The minimum bias triggered (MBT) selection is composed of a single trigger requiring at least one high level space-point track and is seeded by an L1 requirement of a hit in either MBTS. A previous iteration of the analysis combined this trigger with a set of high multiplicity triggers (HMT), to enhance the statistics at high multiplicity. However, the final results are restricted to the single trigger because, though the added statistics reduced the statistical uncertainty, combining the triggers introduced systematic uncertainty that negated the statistical gains. A comparison to the combined trigger approach is given later in the Chapter, and therefore, the HMTs are discussed in this section.

The four HMTs used for comparison have varying space point, online track, and pileup suppression requirements and are seeded from various L1 total energy requirements. Each trigger is used to populate an exclusive offline range in the number of charged tracks ($N_{\mathrm{trk}}$) for which it has recorded the most events. Because some triggers were not active during all runs, this optimization is done on a run-by-run basis. The multiplicity distributions from each MBT trigger for each run are shown in Figures B.1 and B.2 in the Appendix.

The efficiency of L1 trigger `L1_MBTS_1` can be determined in $p$+Pb data with respect to `HLT_mb_sptrk` which is seeded by a random trigger at L1. The efficiency is measured in two different ways: one is the `L1_MBTS_1` efficiency, before the trigger prescale is applied, with respect to `HLT_mb_sptrk`; the other is the prescale corrected efficiency of `HLT_mb_sptrk_L1MBTS_1` with respect to `HLT_mb_sptrk`. Data collected in one single run at the beginning of 2016 $p$+Pb collisions with most of the statistics of HLT_mb_sptrk triggered sample is used to determine the efficiency. The measured efficiency is shown as a function of multiplicity as shown on the left panel in Figure 7.1. The two measurements show very good consistency indicating that there is no additional inefficiency at HLT for `HLT_mb_sptrk_L1MBTS_1` with respect to `L1_MBTS_1`. Due to radiation damage in 2015 and 2016, the MBTS is not fully efficiency until $N_{ch} > 120$. Trigger efficiency corrections for `HLT_mb_sptrk_L1MBTS_1` are applied as a function of multiplicity to its triggered sample. The efficiency of the HLT requirement of `HLT_mb_sptrk` is measured as the relative efficiency of `HLT_mb_sptrk_L1MBTS_1` with respect to `L1_MBTS_1` and shown as a function of multiplicity on the right panel in Figure 7.1. The measured relative efficiency is always 100% which means the HLT requirement of `HLT_mb_sptrk` is fully efficient in events that satisfy `L1_MBTS_1`.

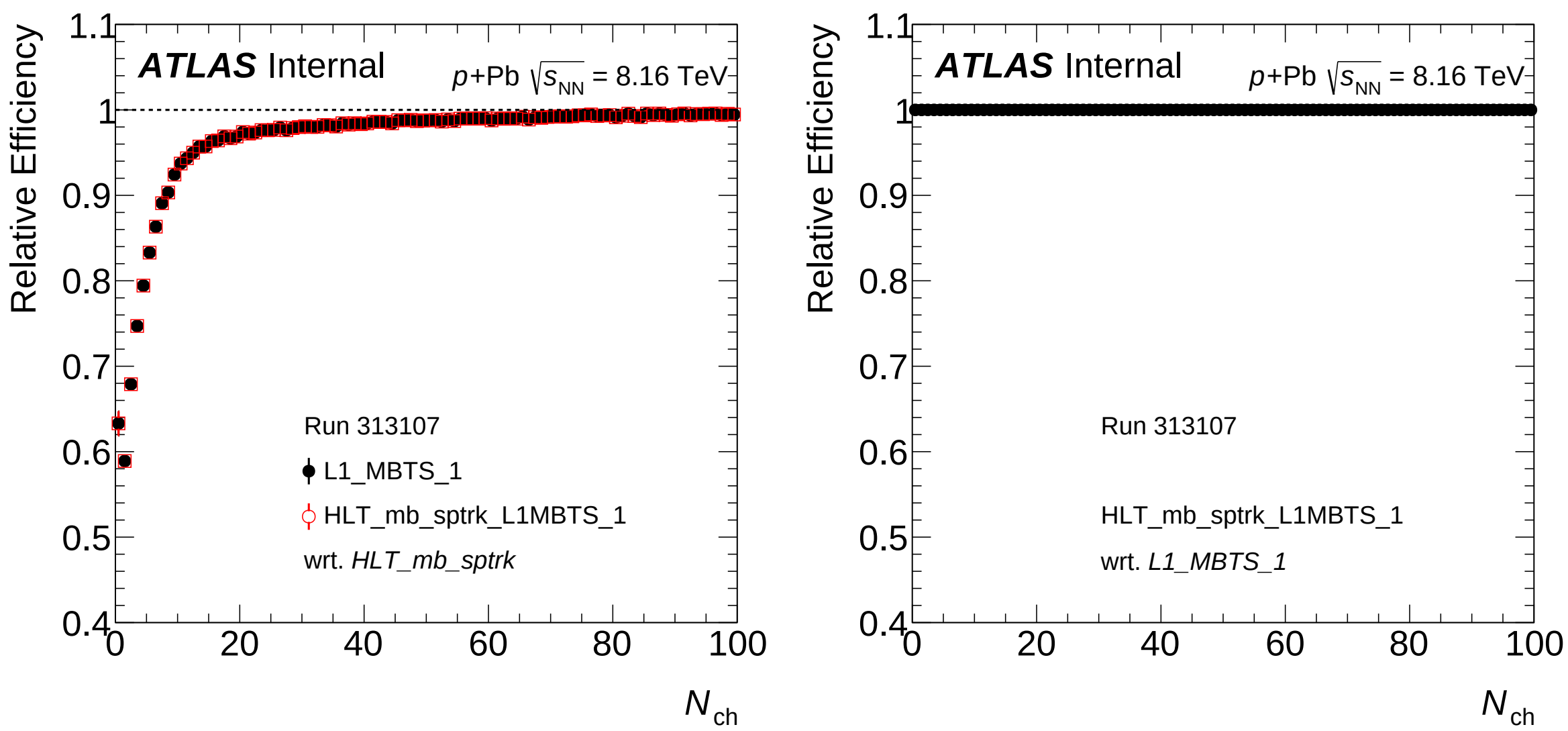

Figure 7.1: **Left:** `L1_MBTS_1` trigger efficiency as a function of multiplicity in 2016 $p$+Pb collisions, measured as the relative efficiencies of `L1_MBTS_1` and `HLT_mb_sptrk_L1MBTS_1` with respect to `HLT_mb_sptrk`. Due to radiation damage in 2015 and 2016, the MBTS is not fully efficiency until $N_{ch} > 120$. **Right:** `HLT_mb_sptrk_L1MBTS_1` trigger efficiency with respect to `L1_MBTS_1` as a function of multiplicity in 2016 $p$+Pb collisions. No inefficiency observed for the HLT requirement `HLT_mb_sptrk` with respect to `L1_MBTS_1` since the relative efficiency is always 100%.

Due to tighter quality cuts applied to tracks offline than those applied online, the track counting at HLT in HMT's is expected to be fully efficient. The efficiencies of various L1 total energy triggers (`L1_TE`), used to seen the HMTs, are studied as a function of multiplicity as shown in the left panel of Fig. 7.2. The efficiencies are measured as the relative efficiency to MB trigger, `HLT_mb_sptrk_L1MBTS_1`. As one can see in the figure, most of the `L1_TE` triggers are very efficient where a multiplicity cut is applied offline.

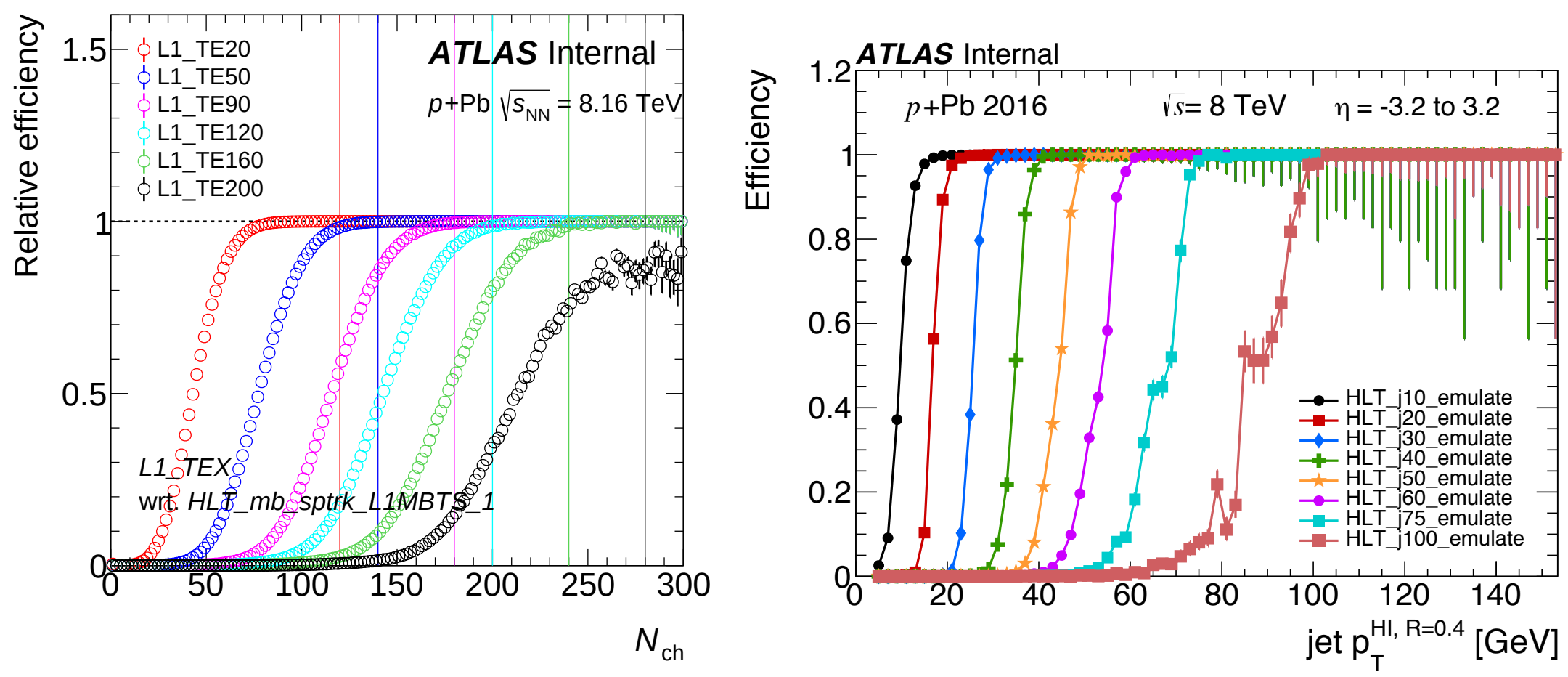

Figure 7.2: **Left:** `L1_TEX` trigger efficiency as a function of multiplicity in 2016 $p$+Pb data, measured with respect to `HLT_mb_sptrk_L1MBTS_1`. The colored lines correspond to the multiplicity threshold applied offline to each of the trigger with the same color. **Right** Jet trigger efficiencies as a function of offline jet $p_{\mathrm{T}}$.

The second dataset is selected by a jet trigger requiring a high level jet to pass a 75 GeV or 100 GeV online threshold and is seeded by a L1 jet 20 GeV requirement. The 100 GeV trigger is the lowest threshold unprescaled jet trigger, and thus samples the full luminosity delivered to ATLAS during the running period. Offline, the events are required to contain a calibrated anti-$k_t$, $R = 0.4$ calorimeter jet with transverse momentum greater than 75 GeV or 100 GeV. Jet trigger efficiencies were studied as a function of offline jet $p_{\mathrm{T}}$ and shown in the right panel of Fig. 7.2. These studies indicate that the J75 and J100 triggers have greater than 97% efficiency for offline jets with $p_{\mathrm{T}} > 75$ GeV and $p_{\mathrm{T}} > 100$ GeV respectively. No jet trigger corrections are applied.

To minimize bias on the track $\phi$ distribution due to the rejection of jets in the disabled sector of the HEC, the offline jet satisfying the threshold is required to have an $\eta < 1.16$. Thus, the whole $\eta$ slice containing the disabled HEC, including an $R = 0.4$ buffer, is removed from the jet selection. Fig. 7.3 shows the effect of this jet restriction on the track $\phi$ distributions. The first row shows the $\eta - \phi$ distribution of tracks, where each vertical $\eta$ slice is self normalized to show the relative structure in $\phi$, and the second row shows the self normalized $\phi$ distributions. The left column is from MB+HMT events and acts as a baseline, as there can be no jet selection bias. The center shows jet events with no offline jet condition; this shows an obvious bias on the track distributions due to the jet trigger inefficiency in the disabled HEC region. The right plots are jet events after making the offline jet selection described in the previous paragraph; in this case, the bias on the track distributions is drastically reduced.

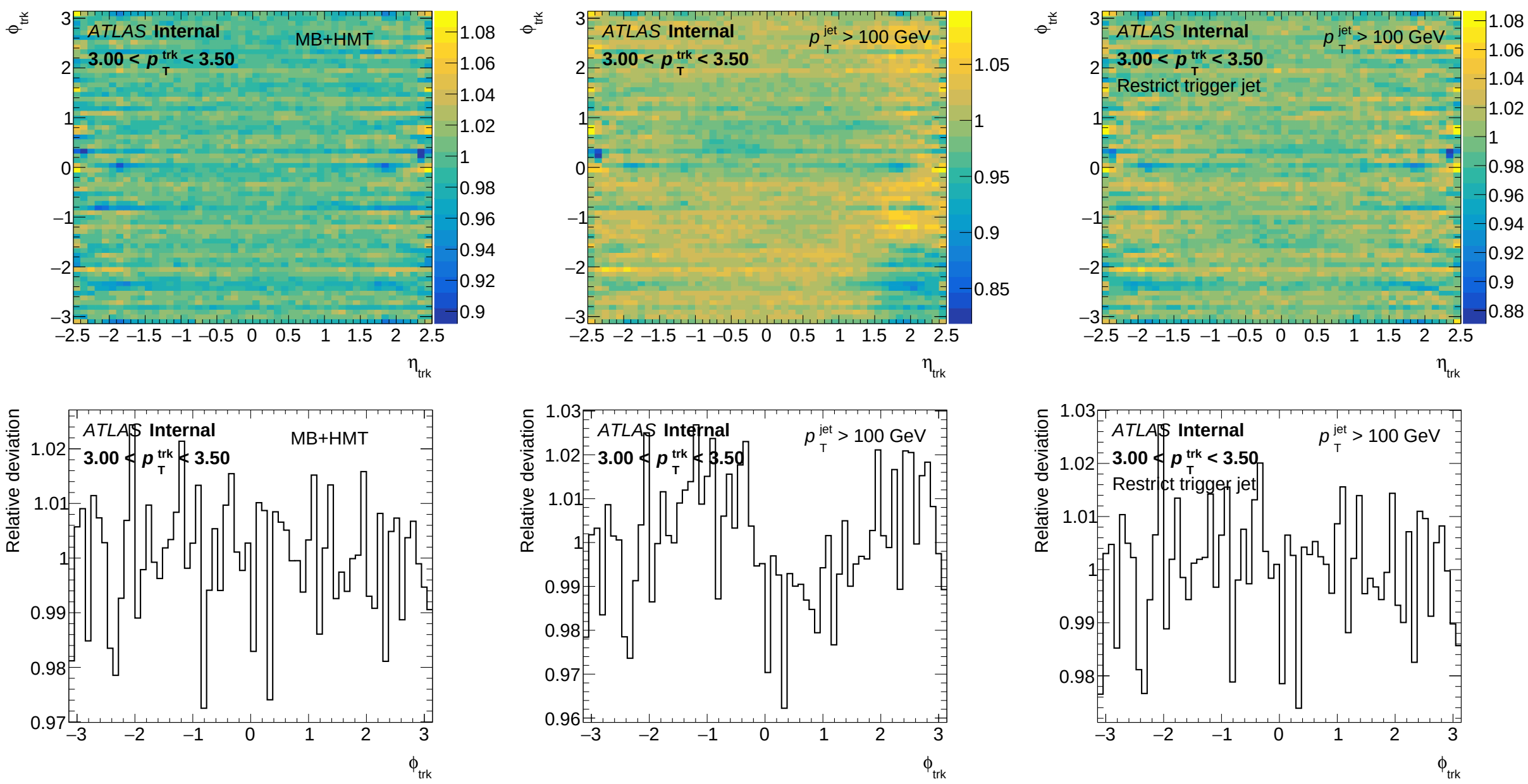

Figure 7.3: The $\eta - \phi$ distributions (top) and $\phi$ projections (bottom) of tracks in which each $\eta$ slice is self normalized such that the values represent the relative deviation from the mean in $\phi$. The left is from MB+HMT events, the center is from jet events, and the right is from jet events after the offline jet selection is made.

Table 7.1 summarizes the triggers used, offline conditions, and luminosity sampled by each trigger in each running period. Fig. 7.4 shows the multiplicity and track $p_{\mathrm{T}}$ distributions for each set of events. Fig. 7.5 shows the per-event normalized $\eta$ distributions of tracks for central and peripheral events for each set of trigger selected events. In these coordinates, the Pb nucleus is moving towards positive $\eta$, the proton toward negative $\eta$, and there is a center-of-mass boost of 0.465 units to the negative direction. In the peripheral case, jet events show an enhancement in particle production due to the presence of jets and the shift of the peak in the direction of the boost, whereas the MB+HMT events give a broad and flat distribution. In central events, the fragmenting nucleus dominates the particle production, and there is an enhancement in the nuclear-going direction. In this case, the MB+HMT events show a larger yield due to the HMTs.

Table 7.1: Triggers used in analysis, with the sampled luminosities in both running periods.

| Trigger | Offline requirement | $L_{\mathrm{int}}$ ($p$+Pb period) | $L_{\mathrm{int}}$ (Pb+$p$ period) |
|---|---|---|---|
| HLT_mb_sptrk_L1MBTS_1 | $N_{\mathrm{trk}} < 120$ | 0.039 $\mathrm{nb}^{-1}$ | 0.040 $\mathrm{nb}^{-1}$ |
| HLT_mb_sp2400_pusup500_trk120_hmt_L1TE20 | $120 \leq N_{\mathrm{trk}} < 140$ | 0.048 $\mathrm{nb}^{-1}$ | 0.059 $\mathrm{nb}^{-1}$ |
| HLT_mb_sp2800_pusup600_trk140_hmt_L1TE50 | $140 \leq N_{\mathrm{trk}} < 200$ | 0.162 $\mathrm{nb}^{-1}$ | 0.503 $\mathrm{nb}^{-1}$ |
| HLT_mb_sp4100_pusup900_trk200_hmt_L1TE120 | $200 \leq N_{\mathrm{trk}} < 240$ | 1.434 $\mathrm{nb}^{-1}$ | 2.805 $\mathrm{nb}^{-1}$ |
| HLT_mb_sp4800_pusup1100_trk240_hmt_L1TE120 | $240 \leq N_{\mathrm{trk}}$ | 2.576 $\mathrm{nb}^{-1}$ | 5.960 $\mathrm{nb}^{-1}$ |
| HLT_j75_ion_L1J20 | $p_{\mathrm{T}}^{\mathrm{jet}} > 75\mathrm{GeV}$ | 3.47 $\mathrm{nb}^{-1}$ | 0 $\mathrm{nb}^{-1}$ |
| HLT_j75_L1J20 | $p_{\mathrm{T}}^{\mathrm{jet}} > 75\mathrm{GeV}$ | 0 $\mathrm{nb}^{-1}$ | 22.53 $\mathrm{nb}^{-1}$ |
| HLT_j100_ion_L1J20 | $p_{\mathrm{T}}^{\mathrm{jet}} > 100\mathrm{GeV}$ | 56.76 $\mathrm{nb}^{-1}$ | 0 $\mathrm{nb}^{-1}$ |
| HLT_j100_L1J20 | $p_{\mathrm{T}}^{\mathrm{jet}} > 100\mathrm{GeV}$ | 0 $\mathrm{nb}^{-1}$ | 107.80 $\mathrm{nb}^{-1}$ |

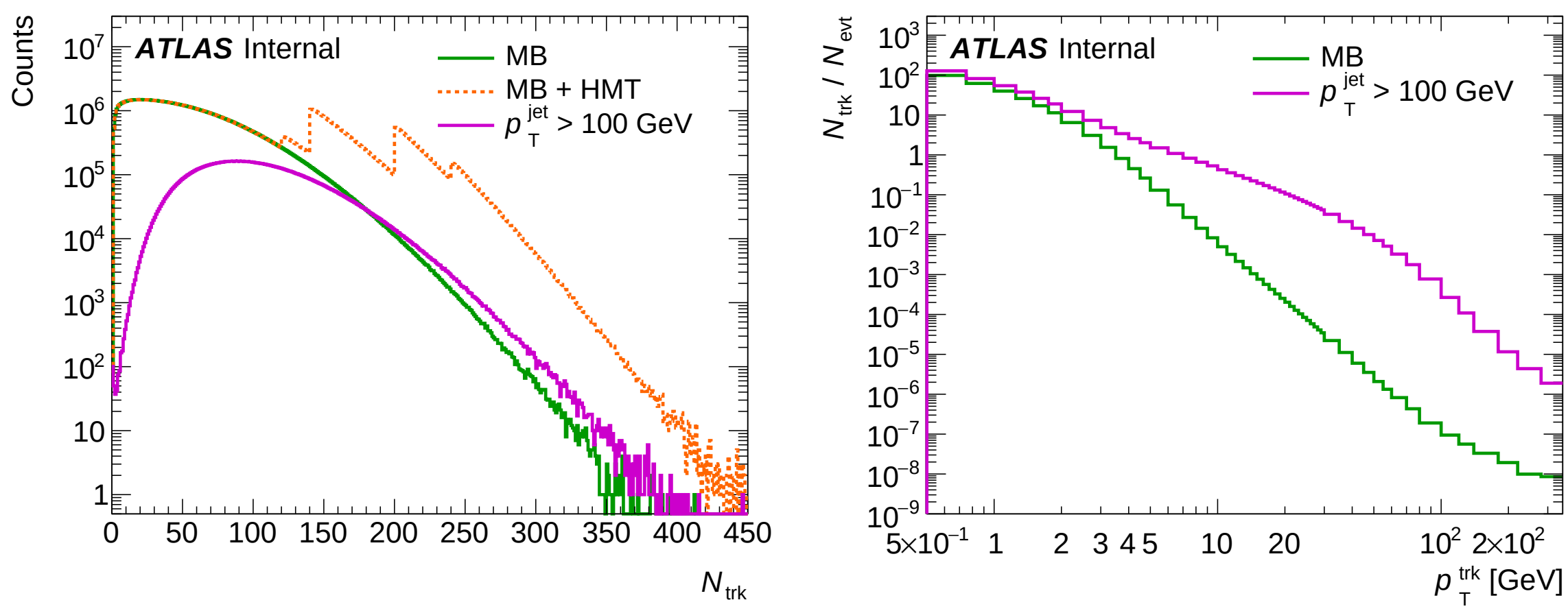

Figure 7.4: Multiplicity (**left**) and track $p_{\mathrm{T}}$ (**right**) distributions from the two sets of events considered: minbias with high multiplicity triggers in green, and those with 100 GeV jets in magenta.

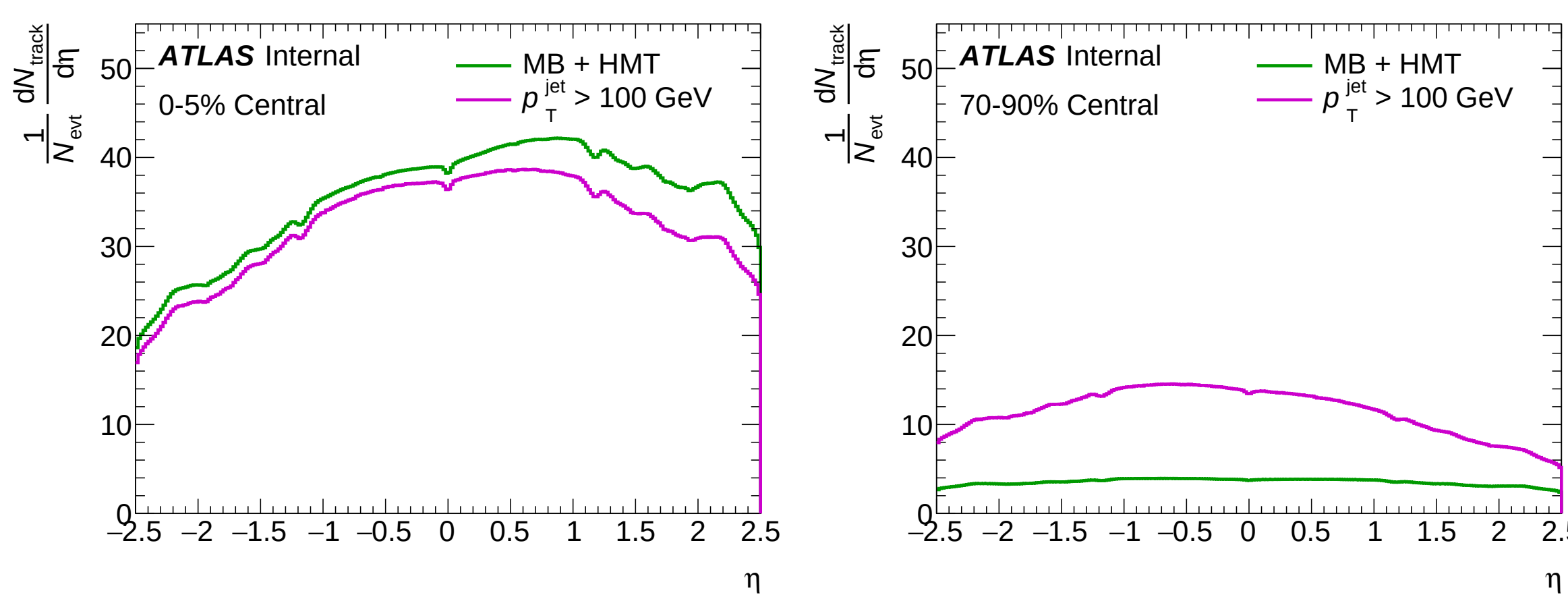

Figure 7.5: $dN_{\mathrm{ch}}/d\eta$ vs $\eta$, uncorrected for detector inefficiencies, for central (**left**) and peripheral events (**right**) from the two sets of events considered: minbias with high multiplicity triggers in green, and those with 100 GeV jets in magenta.

In-time pileup events are reduced in the same way for the centrality analysis described in Sec. 4.2. Events are required to contain at least one vertex, and are rejected if there exists any other vertices with greater than six tracks associated with it.

### 7.2.2 Simulation Samples

Contrary to the photon analysis, this measurement is primarily self normalized and corrected and is, therefore, insensitive to corrections derived from MC simulations. However, reconstruction and selection efficiencies for *primary* [148] charged hadrons to meet the track quality criteria, described below, were determined using a sample of 3 million minimum-bias $p$+Pb events simulated by the Hijing generator [84]. Events were generated with both beam configurations. The ATLAS detector response to the generated events was determined through a full Geant4 simulation [106, 107], and the simulated events were reconstructed in the same way as the data.

Additionally, a Pythia di-jet data overlay sample is used to check the bias on the jet selection from the UE. In this case, hard QCD Pythia events are filtered at truth level on the requirement that they contain a truth jet with $p_T$ > 60 GeV. The detector response is simulated in GEANT4, and the events are mixed with minimum bias data events before being reconstructed in the same way as was described in Sec. 5.2.

### 7.2.3 Object Selection

The correlations studied in this analysis are made between inclusive charged hadrons. These are defined to be any reconstructed track in the inner detector passing the `MinBias` track selection working point quality cuts that are described in Sec. 3.2.1. Charged-particle tracks and collision vertices are reconstructed in the ID using the algorithms described in Sec. 3.2.1 and Refs. [102, 103]. Only inner detector tracks with $p_{\mathrm{T}} > 0.4$ GeV and $|\eta| < 2.5$ are used in this analysis. The total number of reconstructed ID tracks satisfying these selection criteria in a given event is called the multiplicity or $N_{\mathrm{ch}}^{\mathrm{rec}}$. The `MinBias` tracking efficiency is measured as the probability of a truth primary charged pion to be reconstructed as a `MinBias` track by the ID in the Hijing data samples described above. The measured efficiencies are shown in Figure 7.6 as a function of prompt pion $p_{\mathrm{T}}$ and $\eta$. Over the measured kinematic range, the track reconstruction efficiency varies from approximately 50% for the lowest-$p_{\mathrm{T}}$ hadrons at large pseudorapidity, to greater than 90% for hadrons with $p_{\mathrm{T}} > 3$ GeV at mid-rapidity. Dependencies of the tracking efficiency on the event multiplicity and on particle species are ignored in the nominal analysis but considered in systematic uncertainties. The tracks used in this analysis are re-weighted, track-by-track, to corrected for their inefficiencies.

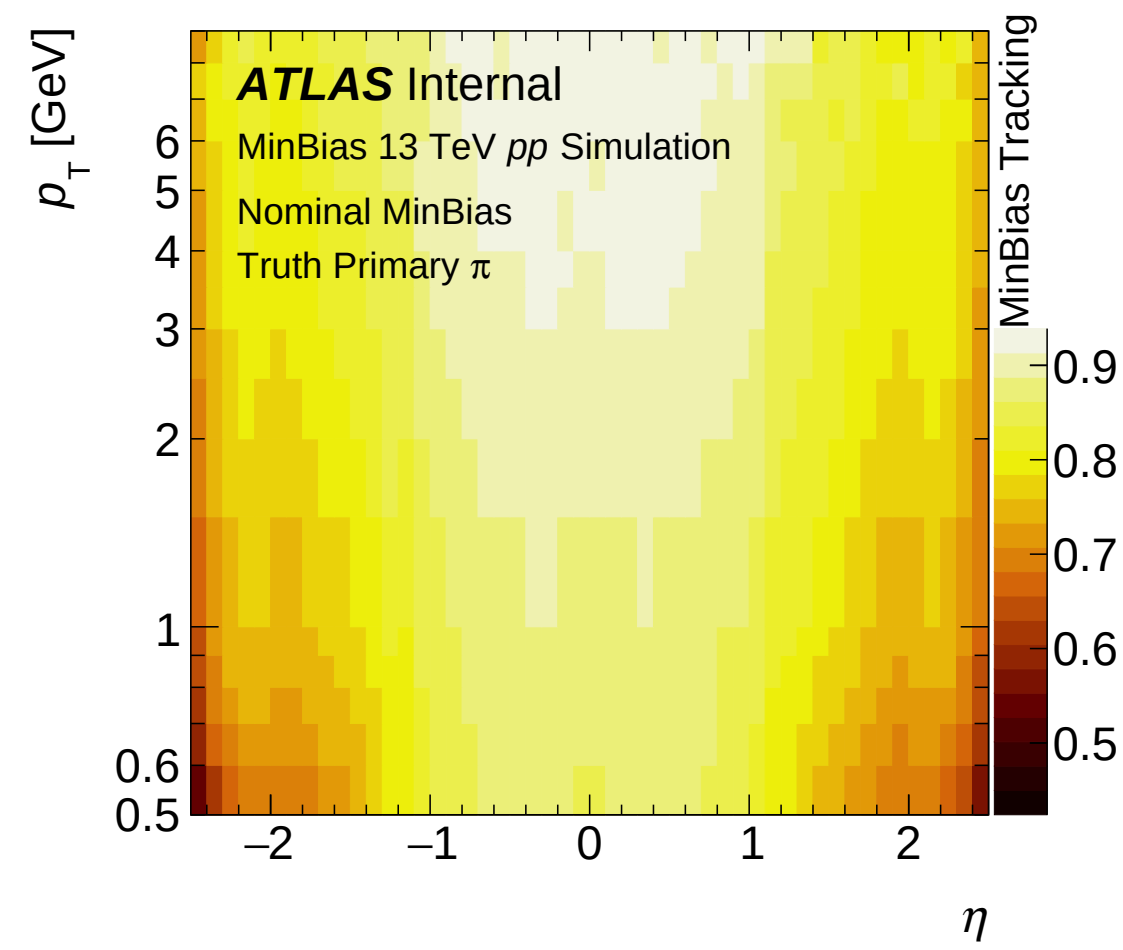

Figure 7.6: Two dimensional reconstruction efficiency of `MinBias` tracks as a function of track $p_{\mathrm{T}}$ and $\eta$ obtained from MC based prompt charged pions.

Jets are reconstructed using energy deposits in the calorimeter system, $|\eta| < 4.9$, closely following the procedure used in other measurements for Pb+Pb and *pp* collisions [149, 150]. Jets are measured by applying the anti-$k_t$ algorithm [151, 152] with radius parameter $R = 0.4$ to energy deposits in the calorimeter. No jets with $p_{\mathrm{T}} < 15$ GeV are considered. An iterative procedure is used to obtain an event-by-event estimate of the $\eta$-dependent underlying-event calorimetric energy density, while excluding jets from that estimate. The jet kinematics are corrected for this background and for the detector response using an $\eta$- and $p_{\mathrm{T}}$-dependent calibration derived from fully simulated and reconstructed Pythia [127] hard-scattering events configured with the NNPDF23LO parton distribution function set [128] and the A14 set of tuned parameters [129] to model non-perturbative effects. An additional, small correction, based on *in situ* studies of jets recoiling against photons, $Z$ bosons, and jets in other regions of the calorimeter, is applied [153, 154]. Simulation studies show that for jets with $p_{\mathrm{T}} > 75$ GeV, the average reconstructed jet $p_{\mathrm{T}}$ is within 1% of the generator level jet $p_{\mathrm{T}}$ and has a relative $p_{\mathrm{T}}$ resolution ($\sigma_{p_{\mathrm{T}}}/p_{\mathrm{T}}$) below 10% after the calibration procedure.

## 7.3 Methodology

The main technology employed in this analysis is that of two-particle correlations which has been used extensively within ATLAS [11, 12, 147, 155]. Due to the relatively larger contribution of non-flow correlations in $p$+Pb as compared to Pb+Pb, a non-flow subtraction procedure is used to extract the bulk flow correlations. A non-flow template fit method, first used in Ref. [12], is used to separate the flow from non-flow signal. In the jet events considered, the non-flow contribution is substantially larger than that of the MBT events, and the template method is found to be insufficient in describing the non-flow. It is found that by requiring the associated particles to be separated in $\eta$ from all jets with $p_{\mathrm{T}} > 15$ GeV, allows the template method to work as expected. Finally, an estimation of the jet particle yields is made in an effort to understand the differences between the results from MBT and jet events.

### 7.3.1 Two-Particle Correlations

Two-particle correlations are defined here as distributions of particle pair yields. The pairs are defined to be ordered A-B couples of particles-of-interest (particle A) and associate or reference particles (particle B). The particle pair yields are presented as distributions of relative azimuthal angle ($\Delta\phi = \phi^{\mathrm{A}} - \phi^{\mathrm{B}}$) and relative pseudorapidity ($\Delta\eta = \eta^{\mathrm{A}} - \eta^{\mathrm{B}}$), normalized by the number of A-particles. This is referred to as a per-particle yield (PPY), $Y(\Delta\phi, \Delta\eta)$, defined as:

$$Y(\Delta\phi, \Delta\eta) = \frac{1}{N_{\mathrm{A}}} \frac{\mathrm{d}^2 N_{\mathrm{pair}}}{\mathrm{d}\Delta\phi \cdot \mathrm{d}\Delta\eta}, \tag{7.1}$$

where $N_{\mathrm{pair}}$ is the event-wise yield of particle pairs integrated over all events, and $N_{\mathrm{A}}$ is the number of selected particles-of-interest integrated over all events. Thus, $Y$ is the average yield of pairs averaged over the number of A-particles and all events.

In practice, due to the limited coverage and imperfect performance of the real detector, there are acceptance effects that distort these distributions. These trivial distortions can be removed using the so-called mixed-event technique in which correlations are made between A particles from any given event with B-particles from different events. In this way, only trivial pair acceptance correlations will be present as there can be no physics correlation. The B-particles are drawn from events with similar characteristics so that the mixed event correlations faithfully represent the trivial geometric correlations in the signal distributions. The events are thus mixed in the following categories:

- collision beam configuration
- primary vertex position within 10 mm
- multiplicity ranges of $N_{\text{trk}}$ within 10 for $N_{\text{trk}} < 100$ and $N_{\text{trk}}$ within 20 for $N_{\text{trk}} > 100$
- centrality given by FCal $\Sigma E_{\text{T}}^{\text{Pb}}$ corresponding to 5% centrality ranges from 0-30%, 10% ranges from 30-70%, and the last range from 70-90%

To maximize the statistical weight of the mixed events, every signal jet event is mixed with five different qualified mix events, which is sufficiently many so that the introduced statistical uncertainty is significantly smaller to that of the signal distribution. However, for MBT events, the mixed event correction is applied only to estimate a systematic and is only mixed with a single event. Events are abandoned, in both same and mixed event distributions, if there are not exactly five events to mix with so that the events in the mixed distributions are correctly weighted to those of the signal distributions. After applying the mixed-event based acceptance correction, PPY becomes:

$$Y(\Delta\phi, \Delta\eta) = \frac{K}{N_{\text{A}}} \frac{\left\langle \mathrm{d}^2 N_{\text{same}} / (\mathrm{d}\Delta\phi \cdot \mathrm{d}\Delta\eta) \right\rangle_{\text{evt}}}{\left\langle \mathrm{d}^2 N_{\text{mixed}} / (\mathrm{d}\Delta\phi \cdot \mathrm{d}\Delta\eta) \right\rangle_{\text{evt}}}, \tag{7.2}$$

where $N_{\text{same}}$ is the yield of particle pairs with two-particles from the same event, $N_{\text{mixed}}$ is the yield of pair combinations of particles from mixed events, and $K$ is a scale factor to conserve the integral of PPY to what it should be before applying the mixed-event correction. Examples of two-particle correlations in same and mixed events are shown on the left and middle panels in Figure 7.7, respectively. Both the same and mixed event distributions have a trivial triangular shape in $\Delta\eta$ due to the detector acceptance. An example of PPY after correcting for this acceptance effect using the mixed event distribution is shown on the right panel in Figure 7.7.

Because the data were collected using triggers with different prescales and the object reconstruction is not fully efficient, luminosity weights and efficiency corrections are applied to approximate the genuine correlations

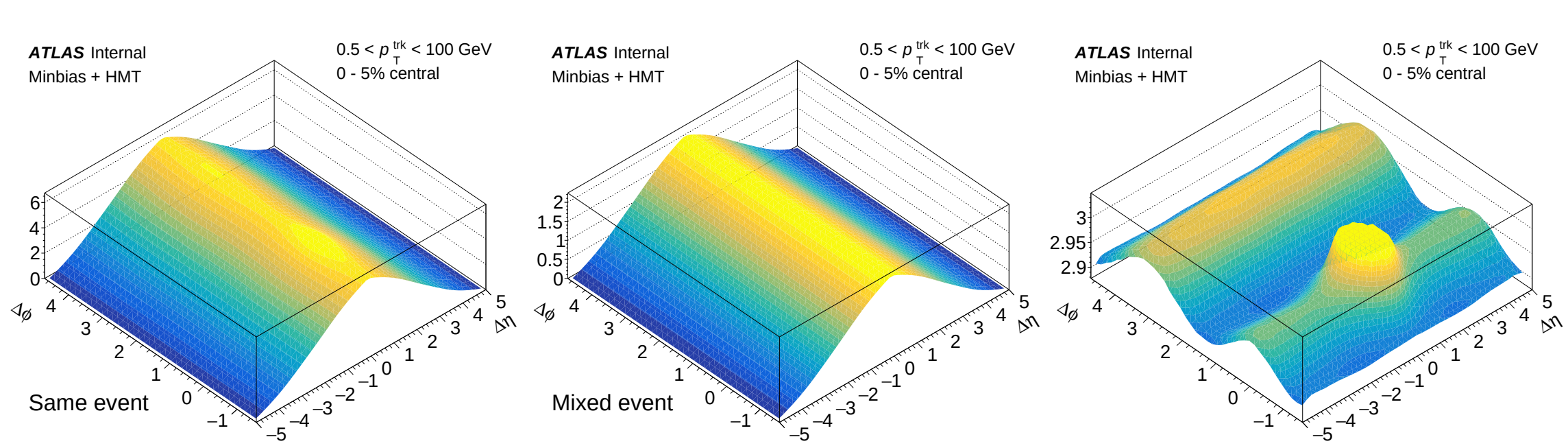

Figure 7.7: Example of same event two-particle correlation (left), mixed event two-particle correlation (middle) and per-particle-yield after correction for the acceptance effect.

with an ideal detector:

$$Y(\Delta\phi, \Delta\eta) \equiv \frac{K}{N_{\mathrm{A}}^{\mathrm{corr}}} \frac{S(\Delta\phi, \Delta\eta)}{B(\Delta\phi, \Delta\eta)}, \tag{7.3}$$

where

$$S(\Delta\phi, \Delta\eta) \equiv \left\langle \frac{\mathrm{d}^2 N_{\mathrm{signal}}^{\mathrm{corr}}}{\mathrm{d}\Delta\phi \cdot \mathrm{d}\Delta\eta} \right\rangle_{\text{evt, trigger}}, \tag{7.4}$$

$$B(\Delta\phi, \Delta\eta) \equiv \left\langle \frac{\mathrm{d}^2 N_{\mathrm{mixed}}^{\mathrm{corr}}}{\mathrm{d}\Delta\phi \cdot \mathrm{d}\Delta\eta} \right\rangle_{\text{evt, trigger}}, \tag{7.5}$$

$$N_{\mathrm{same(mixed)}}^{\mathrm{corr}} = \frac{1}{L_{\mathrm{int}}^{\mathrm{trigger}} \cdot \varepsilon_{\mathrm{trigger}}} \sum \frac{1}{\varepsilon^{\mathrm{A}} \cdot \varepsilon^{\mathrm{B}}}, \tag{7.6}$$

$$N_{\mathrm{A}}^{\mathrm{corr}} = \frac{1}{L_{\mathrm{int}}^{\mathrm{trigger}} \cdot \varepsilon_{\mathrm{trigger}}} \sum \frac{1}{\varepsilon^{\mathrm{A}}}, \tag{7.7}$$

where $L_{\mathrm{int}}^{\mathrm{trigger}}$ is the integrated luminosity of a single trigger as described in Sec. 7.2.1, and $\varepsilon_{\mathrm{trigger}}$ is the efficiency of the trigger. $L_{\mathrm{int}}^{\mathrm{trigger}}$ and $\varepsilon_{\mathrm{trigger}}$ are applied on a per-event basis. $\langle\rangle_{\mathrm{evt}}$ indicates the average over many events collected by different triggers. Each pair is also weighted by a factor to account for A and B particle reconstruction inefficiency on a per-particle-pair basis, and $\varepsilon^{\mathrm{A}}$ and $\varepsilon^{\mathrm{B}}$ are the $p_{\mathrm{T}}$ and $\eta$ dependent tracking reconstruction efficiencies for particle A and B respectively. Fig. 7.8 shows some selected $Y(\Delta\phi, \Delta\eta)$ correlations from MBT and jet events integrated over $p_{\mathrm{T}}$ for two different centrality classes.

These mixed event corrected two dimensional distributions in Fig. 7.8 are valuable because they provide a qualitative picture of the whole correlation space. One can see the strong correlation at small angles ($\Delta\eta \approx \Delta\phi \approx 0$) from decays and jets as well as the broad away-side ($\Delta\phi \approx \pi$) from the momentum balance of dijets.

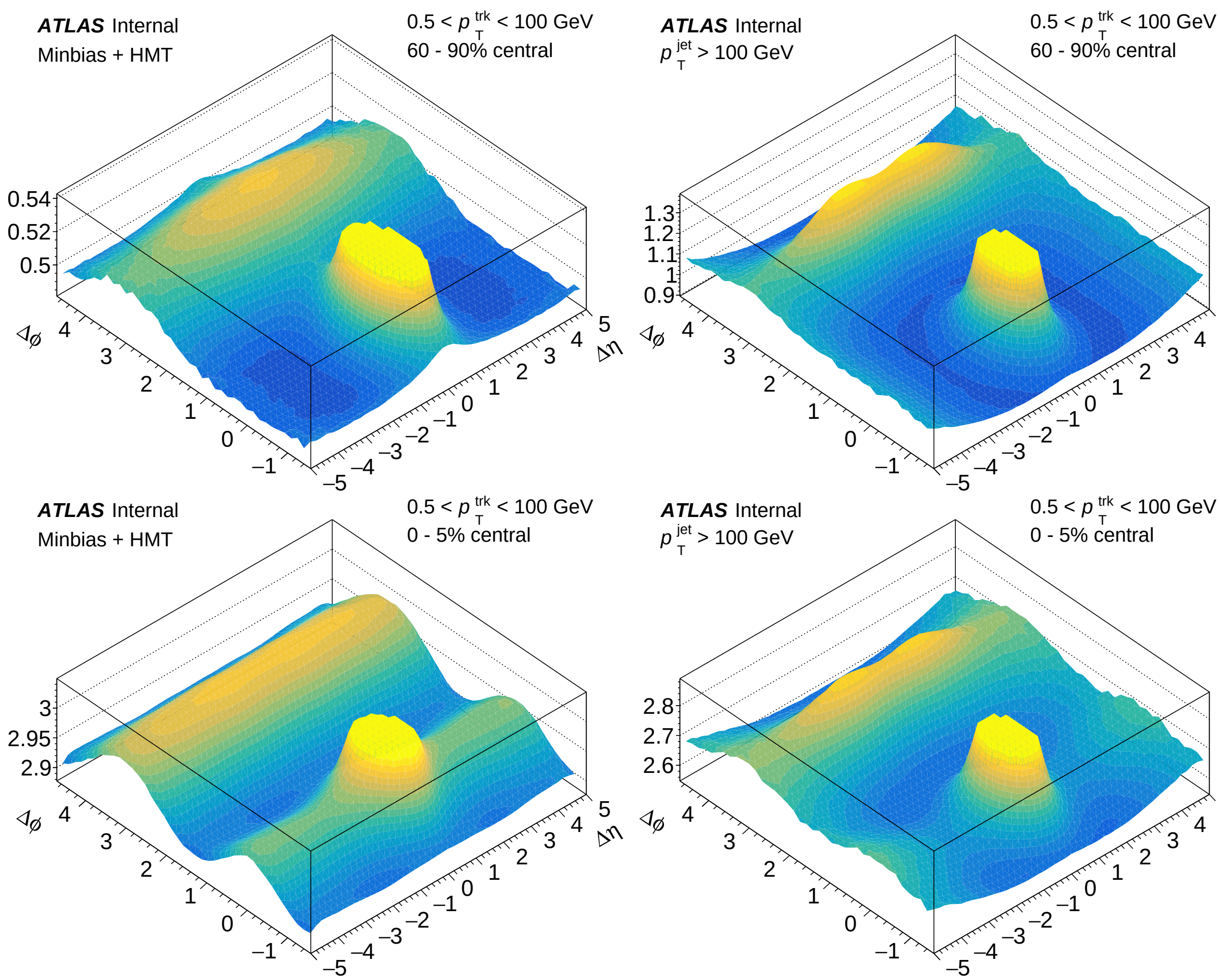

Figure 7.8: Example of fully corrected per-A-pair yields of two-particle correlations for MBT (left column) and jet (right column) events and peripheral (top row) and central (bottom row) selections.

It is also clear in the correlations from central events that there exist a $\cos(2\Delta\phi)$ modulation that extends the full range in $\Delta\eta$; these are the global flow correlations that are key to this measurement. To be quantitative, however, $\Delta\phi$ projections must be studied in detail. This is the subject of the next sections.

### 7.3.2 Signal Extraction

The global correlation signal is extracted from one-dimensional $\Delta\phi$ PPY distributions. To reduce the influence of short range non-flow correlations, the one dimensional distributions are made by integrating over pairs with large $\eta$ separation ($\Delta\eta > 2$). Long range non-flow is then accounted for via a template fitting procedure that leverages the difference in the relative non-flow contribution to the PPY between peripheral and central events. Within the fitting routine, the residual $\Delta\phi$ modulation is fit to a truncated Fourier series, the coefficients of which are the extracted signal. Under the assumption that the two-particle flow signal is the product of the signal from the A particle selection and that of the B particle (factorization), the $p_{\mathrm{T}}$ dependence can be extracted by making different A particle $p_{\mathrm{T}}$ selections.

#### 7.3.2.1 One Dimensional Correlation

The one-dimensional $\Delta\phi$ PPY, $Y(\Delta\phi)$, is constructed from the ratio of the same event distribution and mixed event distribution, both integrated over $|\Delta\eta|$ as:

$$Y(\Delta\phi) = \frac{\int_{\Delta_{\mathrm{l}}}^{\Delta_{\mathrm{h}}} S(\Delta\phi, |\Delta\eta|)\,\mathrm{d}|\Delta\eta|}{\int_{\Delta_{\mathrm{l}}}^{\Delta_{\mathrm{h}}} B(\Delta\phi, |\Delta\eta|)\,\mathrm{d}|\Delta\eta|}, \tag{7.8}$$

where $S(\Delta\phi, |\Delta\eta|)$ and $B(\Delta\phi, |\Delta\eta|)$ are the 2D yields in same and mixed events as defined in Eq. 7.3; $\Delta_{\mathrm{h}}$ is the phase space boundary of $\Delta\eta$ for the two-particle correlation, which is $\Delta_{\mathrm{h}} = 5.0$ in ATLAS; and $\Delta_{\mathrm{l}}$ is the lower boundary of the integration which is imposed for purpose of removing short range non-flow contributions. For this analysis, $\Delta_{\mathrm{l}} = 2.0$, which is set to eliminate particle pairs from a single jet or decays.

As indicated by Eq. 7.8, the 1D PPY is calculated by integrating $S(\Delta\phi, |\Delta\eta|)$ and $B(\Delta\phi, |\Delta\eta|)$ in $\Delta\eta$ first and then taking the ratio of the $S(\Delta\phi)$ and $B(\Delta\phi)$. The resulting PPY is dominated by contributions at $\Delta\eta \approx \Delta_{\mathrm{l}}$ due to larger pair yields and is less sensitive to fluctuations at the edge of the $\Delta\eta$ acceptance. This definition of PPY is used in previous ATLAS publications [12, 13]. However, one should note that there is a different way to define the PPY which treats all $\Delta\eta$ as having an equal weight. This is accomplished by first taking the ratio of $S(\Delta\phi, |\Delta\eta|)$ and $B(\Delta\phi, |\Delta\eta|)$, and then integrating over $\Delta\eta$. This alternative definition of PPY is used by the CMS Collaboration [**CMS-HIN-12-015**, 14, 156, 157]. The two definitions result in different average $\Delta\eta$ due to different relative weight at different $\Delta\eta$.

The one-dimensional $\Delta\phi$ PPY is characterized in terms of Fourier decomposition:

$$Y(\Delta\phi) = G\Big\{1 + 2\sum_{n=1}^{\infty} a_n \cos(n\Delta\phi)\Big\}, \tag{7.9}$$

where $a_n$ is the Fourier coefficient at order $n$ and $G$ is the normalization factor corresponding to the integral of $Y(\Delta\phi)$. Parameters $a_n$ and $G$ can be obtained from direct Fourier decomposition. As discussed in Section 7.1, the $a_n$ parameter extracted from Fourier decomposition contains both flow and non-flow contributions. Qualitatively, in events with low activity, PPY is driven by particles from dijets jets so $a_n$ is dominated by these non-flow contributions; in events with high activity, the flow contribution in $a_n$ becomes comparable with the non-flow contribution. Quantitatively, the $a_n$ parameters are not useful since the flow/non-flow contribution could vary with event activity and particle kinematic selections. So the main goal of two-particle correlation analysis is to systematically separate the flow from non-flow contributions.

The Fourier coefficients, $a_n$, are assumed to be separable into two linearly additive contributions $a_n = c_n + d_n$, where $c_n$ is the flow coefficient quantifying correlations related to the initial geometry and $d_n$ is the non-flow coefficient of pair correlations that are dominate by jet particle correlations. Then Eq. 7.9 can be rewritten as:

$$Y(\Delta\phi) = G\left\{1 + 2\sum_{n=1}^{\infty}(c_n + d_n)\cos(n\Delta\phi)\right\} \tag{7.10}$$

$$= G\left\{1 + 2\sum_{n=1}^{\infty} c_n \cos(n\Delta\phi)\right\} + G\left\{2\sum_{n=1}^{\infty} d_n \cos(n\Delta\phi)\right\} \tag{7.11}$$

$$= G\left\{1 + 2\sum_{n=1}^{\infty} c_n \cos(n\Delta\phi)\right\} + GJ(\Delta\phi). \tag{7.12}$$

where the first term in Eq. 7.12 is the flow contribution, and the second term, $J(\Delta\phi) = 2\sum_{n=1}^{\infty} d_n \cos(n\Delta\phi)$, is due to non-flow modulations in $\Delta\phi$. It will be shown in the following section that the flow term and non-flow term scale differently with event activity. By repeating two independent measurements at different event activity, one can separate the flow contribution from non-flow algebraically.

#### 7.3.2.2 Template Fit Method

The template fit procedure, as used here, has been applied in previous ATLAS measurements [13, 158] to extract $c_n$ as defined in Eq. 7.12. Flow and non-flow contributions are distinguished by comparing $Y(\Delta\phi)$ between two data selection samples: one at higher event activity, where flow is expected to have a larger influence, and one at lower event activity, where flow is expected to have a smaller influence. Previous measurements [12, 147] have studied this method using different proxies for event activity, namely, charged particle multiplicity, $N_{\rm trk}$, and the total transverse energy in the Pb-going forward calorimeter, $\Sigma E_{\rm T}^{\rm Pb}$. For reasons that will be discussed below 7.3.3, this analysis uses only $\Sigma E_{\rm T}^{\rm Pb}$. Parameters from central and peripheral correlations will be notated with superscript C and P respectively. Thus $\rho_n$ is defined as:

$$\rho_n = d_n^{\rm C}/d_n^{\rm P}, \tag{7.13}$$

$\rho_n$ is conjectured to be **almost** independent of its order $n$, i.e. $\rho_n = \varrho$ for all $n$. In other words, the shape of non-flow contribution in $Y(\Delta\phi)$ is the same in P and C samples as also discussed in Refs. [13] and [158], i.e. $J^{\rm C}(\Delta\phi) = \varrho J^{\rm P}(\Delta\phi)$. Two independent PPY in C events and in P events given by:

$$Y^{\rm P}(\Delta\phi) = G^{\rm P} J^{\rm P}(\Delta\phi) + G^{\rm P}\Big\{1 + 2\sum_{n=1}^{\infty} c_n^{\rm P}\cos(n\Delta\phi)\Big\}, \tag{7.14}$$

$$Y^{\rm C}(\Delta\phi) = G^{\rm C} J^{\rm C}(\Delta\phi) + G^{\rm C}\Big\{1 + 2\sum_{n=1}^{\infty} c_n^{\rm C}\cos(n\Delta\phi)\Big\}, \tag{7.15}$$

can be combined using $J^{\rm C}(\Delta\phi) = \varrho J^{\rm P}(\Delta\phi)$:

$$Y^{\rm C}(\Delta\phi) = \frac{G^{\rm C}\varrho}{G^{\rm P}} Y^{\rm P}(\Delta\phi) + G^{\rm C}(1-\varrho)\Big\{1 + 2\sum_{n=1}^{\infty}\Big(\frac{c_n^{\rm C} - \varrho c_n^{\rm P}}{1-\varrho}\Big)\cos(n\Delta\phi)\Big\} \tag{7.16}$$

One can re-parameterize Eq. 7.16 to match with the following form as used in Refs. [13] and [158]:

$$Y^{\rm C}(\Delta\phi) = F^{\rm temp} Y^{\rm P}(\Delta\phi) + G^{\rm temp}\Big\{1 + 2\sum_{n=1}^{\infty} c_n^{\rm temp}\cos(n\Delta\phi)\Big\}. \tag{7.17}$$

One then obtains the physical meanings of parameters $F^{\rm temp}$, $G^{\rm temp}$, and $c_n^{\rm temp}$ used in Eq. 7.16 in terms of $G^{\rm C}$, $G^{\rm P}$ and $\varrho$ as:

$$F^{\rm temp} = \frac{G^{\rm C}\varrho}{G^{\rm P}}, \tag{7.18}$$

$$G^{\rm temp} = G^{\rm C}(1-\varrho), \tag{7.19}$$

$$c_n^{\rm temp} = \frac{c_n^{\rm C} - \varrho c_n^{\rm P}}{1-\varrho}. \tag{7.20}$$

Motivated by previous measurements of the first-order flow correlation being small while the non-flow contribution at first-order is at its largest, the first order flow correlation, $c_1$, contribution is ignored. One can think of it being absorbed in $d_1' = d_1 + c_1 \sim d_1$. The C sample PPY is eventually fit as follows:

$$\begin{aligned} Y^{\mathrm{C}}(\Delta\phi) &= F^{\mathrm{temp}} Y^{\mathrm{P}}(\Delta\phi) + G^{\mathrm{temp}} \Big\{ 1 + 2\sum_{n=2}^{4} c_n^{\mathrm{temp}} \cos(n\Delta\phi) \Big\} \\ &= F^{\mathrm{temp}} Y^{\mathrm{P}}(\Delta\phi) + Y^{\mathrm{ridge}}(\Delta\phi), \end{aligned} \tag{7.21}$$

where $Y^{\mathrm{ridge}}(\Delta\phi) = G^{\mathrm{temp}} \Big\{ 1 + 2\sum_{n=2}^{4} c_n^{\mathrm{temp}} \cos(n\Delta\phi) \Big\}$ isolates the pure flow correlation. The parameters in Eq. 7.21, $F^{\mathrm{temp}}$, $G^{\mathrm{temp}}$, and $c_n^{\mathrm{temp}}$, are obtained from fitting $Y^{\mathrm{C}}(\Delta\phi)$ using $Y^{\mathrm{P}}(\Delta\phi)$ as a template via a global $\chi^2$ minimization. Thus, this approach is often referred to as the *template fit* method [12, 13]. The flow correlation coefficient of particles in C events is reported as:

$$v_{n,n} = c_n^{\mathrm{C}} = c_n^{\mathrm{temp}} - \varrho(c_n^{\mathrm{temp}} - c_n^{\mathrm{P}}) \tag{7.22}$$

where $c_n^{\mathrm{P}}$ is the correlation coefficient in the P sample. Here $c_n^{\mathrm{C}}$ is replaced by the symbol $v_{n,n}$ to be consistent with previous publications [12, 13].

A brief summary of the assumptions and caveats of using the template fit method is listed below:

- Assumption 1: The non-flow shape doesn't change with event activity, namely $\rho_n = \varrho$;
- Assumption 2: The first order flow contribution is negligible compared to non-flow contribution: $c_1 = 0$ or $a_1 = d_1$;
- If the flow correlation coefficient doesn't dependent on event activity, $v_{n,n} = c_n^{\mathrm{temp}}$, a *one-step* correction should be applied: $v_{n,n} = c_n^{\mathrm{temp}} - \varrho(c_n^{\mathrm{temp}} - c_n^{\mathrm{P}})$.

#### 7.3.2.3 $v_n$ Extraction Assuming Factorization

If the two-particle momentum correlations originate from a global field, as is the case for collective expansion, the $v_{n,n}$ will factorise such that

$$v_{n,n}(p_{\mathrm{T}}^{\mathrm{A}}, p_{\mathrm{T}}^{\mathrm{B}}) = v_n(p_{\mathrm{T}}^{\mathrm{A}}) \cdot v_n(p_{\mathrm{T}}^{\mathrm{B}}).$$

By assuming this relation and making specific $p_{\mathrm{T}}$ selections on A- and B-particles, the single-particle $v_n(p_{\mathrm{T}}^{\mathrm{A}})$ can be obtained from

$$v_n(p_{\mathrm{T}}^{\mathrm{A}}) = v_{n,n}(p_{\mathrm{T}}^{\mathrm{A}}, p_{\mathrm{T}}^{\mathrm{B}}) / \sqrt{v_{n,n}(p_{\mathrm{T}}^{\mathrm{B}}, p_{\mathrm{T}}^{\mathrm{B}})},$$

where $v_{n,n}(p_{\mathrm{T}}^{\mathrm{A}}, p_{\mathrm{T}}^{\mathrm{B}})$ is determined with A- and B-particles having $p_{\mathrm{T}}$ in range $p_{\mathrm{T}}^{\mathrm{A}}$ and $p_{\mathrm{T}}^{\mathrm{B}}$, respectively, and $v_{n,n}(p_{\mathrm{T}}^{\mathrm{B}}, p_{\mathrm{T}}^{\mathrm{B}})$ is determined with A- and B-particles both having $p_{\mathrm{T}}$ in range $p_{\mathrm{T}}^{\mathrm{B}}$. In this analysis, this range is nominally $p_{\mathrm{T}}^{\mathrm{B}} > 0.4$ GeV, although the dependence of the extracted anisotropy on this choice is explored in Chapter 8.

### 7.3.3 Event Activity Selection

Previous ATLAS measurements [12, 147] have studied template non-flow extraction using both charged particle multiplicity and $\Sigma E_{\mathrm{T}}^{\mathrm{Pb}}$ as proxies for event activity, and find similar results for $p$+Pb. This is also true of this analysis; similar results are found for each case when considering the MBT selection of events. However, in jet selected events, a bias on the jet shape is found when selecting low multiplicity events. Fig. 7.9 shows an example of the output of the template fitting procedure for MBT events. The left plot uses multiplicity for low and high event activity, and the right uses $\Sigma E_{\mathrm{T}}^{\mathrm{Pb}}$ centrality ranges. In each case, the P templates seem to be working well, and adding the harmonic modulation describes the C selection. The resulting $v_{n,n}$ are roughly consistent with each other.

In contrast, Fig. 7.10 shows the same comparison for jet events. On the left, for multiplicity selected P and C, the awayside structure of the P reference is much sharper than the C selection, and the fit is quite poor. On the right, the P and C selections are made with $\Sigma E_{\mathrm{T}}^{\mathrm{Pb}}$ centrality ranges. In this case, the template seems to do a better job describing the awayside shape of the C. The residual discrepancy in the awayside is discussed below in Sec. 7.3.4.

The problem when using multiplicity selections seems to come from a jet shape bias. In essence, the P selection is requiring a high $p_{\mathrm{T}}$ jet to be present in addition to there being a low number of particles. This selects jets that tend to fragment harder, i.e. into fewer, higher $p_{\mathrm{T}}$ particles that form more columnated shapes. Jets tend to be emitted at mid-rapidity, overlapping with the tracking acceptance, facilitating this correlation. When the selections are made using $\Sigma E_{\mathrm{T}}^{\mathrm{Pb}}$, the jet to $\Sigma E_{\mathrm{T}}^{\mathrm{Pb}}$ correlation is much weaker because the FCal is so far forward in rapidity. For this reason, the rest of this analysis will only make template event selections based on $\Sigma E_{\mathrm{T}}^{\mathrm{Pb}}$.

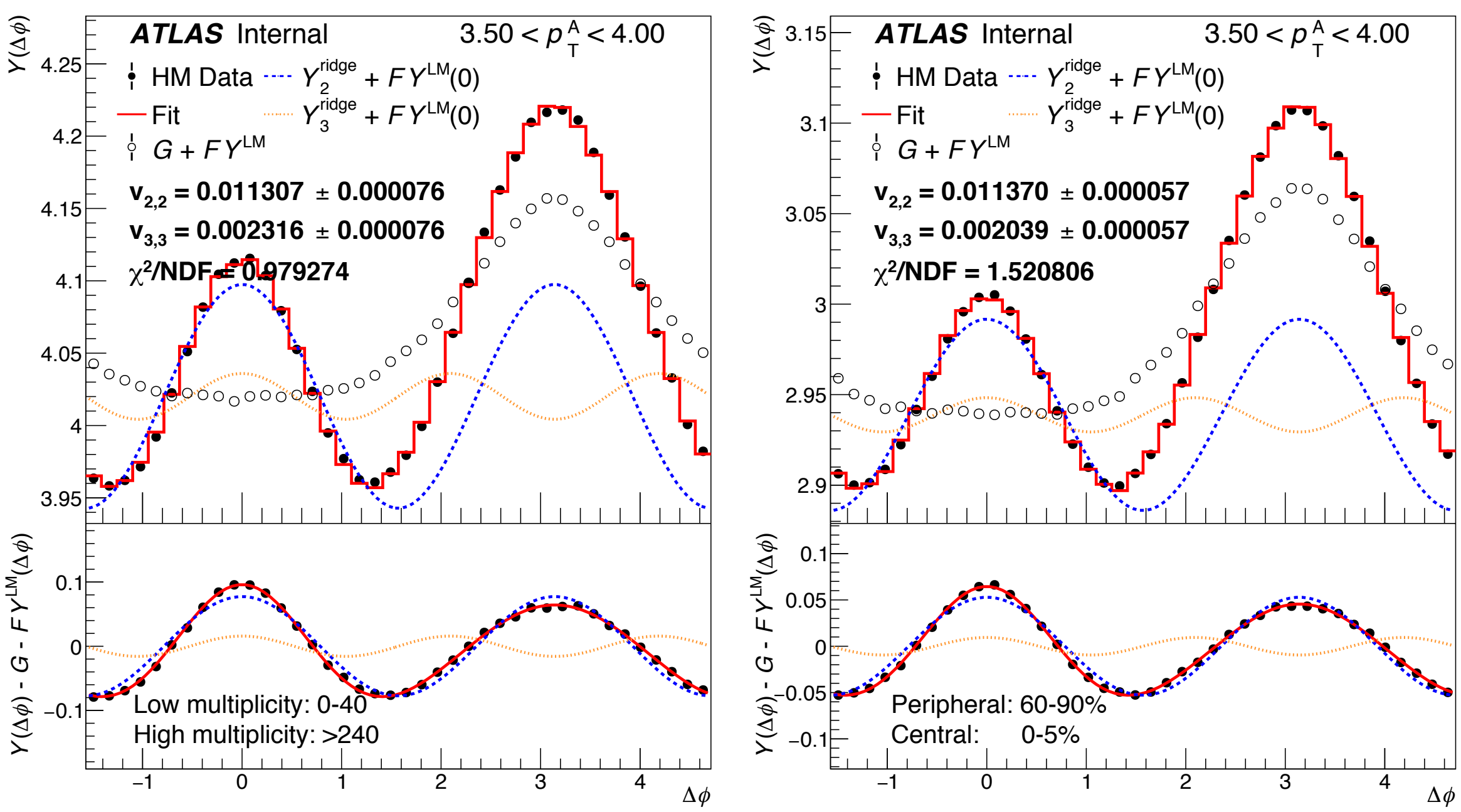

Figure 7.9: Template fitting output for MBT events. Both require $\Delta\eta > 2$ and the A particles to have $3.50 < p_{\mathrm{T}}^{\mathrm{trig}} < 4.00$ but no restriction on the B particles. The left plot is made selecting P in multiplicity range of $0 < N_{\mathrm{trk}} < 40$ and C $240 < N_{\mathrm{trk}}$. The right plot is made selecting P in $2.97 < \Sigma E_{\mathrm{T}}^{\mathrm{Pb}} < 17.04$ (60-90% central) and C in $79.03 < \Sigma E_{\mathrm{T}}^{\mathrm{Pb}}$ (0-5% central). In the upper panels, the open circles show the scaled and shifted P template with uncertainties omitted from the plot, the closed circles show the C data, and the red line shows the fit (template and harmonic functions). The blue dashed line shows the 2$^{\mathrm{nd}}$ order harmonic component and the yellow dashed line shows the total harmonic function. The lower panels show the difference between the C data and the scaled and shifted P template, showing that the residual looks harmonic.

### 7.3.4 Associated Particle Restriction

Fig. 7.10 shows that using $\Sigma E_{\mathrm{T}}^{\mathrm{Pb}}$ to select P and C allows the template fit to work more as expected from MBT events. However, there still exists an obvious discrepancy in the awayside correlation. This discrepancy is believed to be from a violation of the assumptions of the template method. The template method assumes that the shape of the non-flow does not change between the P and C selections, and free parameters $F^{\mathrm{temp}}$ and $G^{\mathrm{temp}}$ account for a scaling and pedestal shift respectively. This may be too simple in the case of jet events.

Consider a simple toy situation in which the particles in the event come from two processes: 1) soft underlying event interactions producing the bulk of the particles (UE), and 2) a hard scattering process producing, predominantly, dijet particle production (HS). MBT events rarely contain high $p_{\mathrm{T}}$ jets, and thus, the particles would be dominantly produced from 1). As can be seen from Fig. 7.9, a good fit is obtained in this case with a single non-flow scaling parameter. Jet events, on the other hand, would contain a mixture of particles produced from 1) and 2). It seems likely that the non-flow correlations from these two sets of particles would scale differently with event activity. In such a case, the template procedure would need to be modified to accommodate two different scaling behaviors.

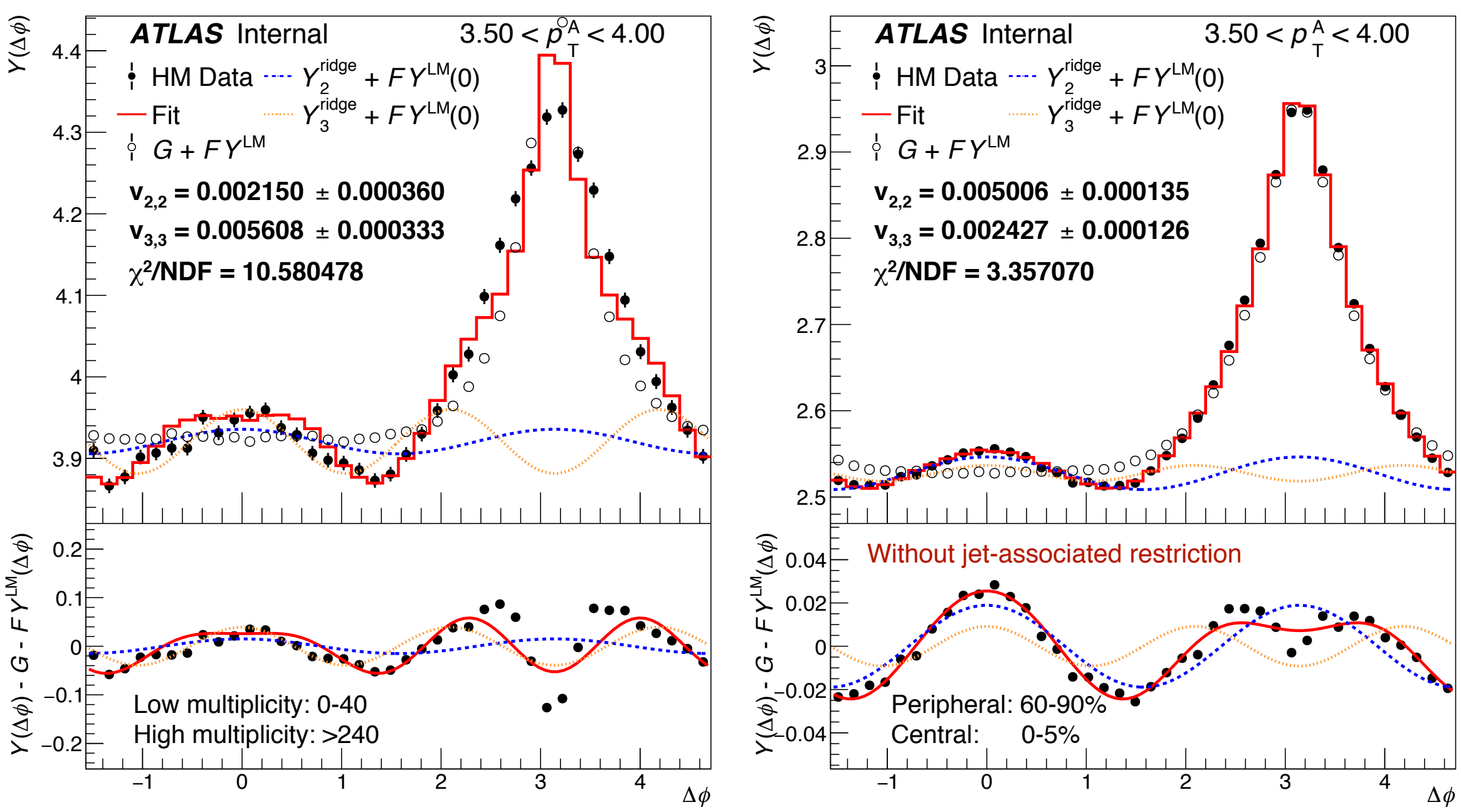

Figure 7.10: Template fitting output for jet events. Both require $\Delta\eta > 2$ and the A particles to have $3.50 < p_{\mathrm{T}}^{\mathrm{trig}} < 4.00$ but no restriction on the B particles. The left plot is made selecting P in multiplicity range of $0 < N_{\mathrm{trk}} < 40$ and C $240 < N_{\mathrm{trk}}$. The right plot is made selecting P in $2.97 < \Sigma E_{\mathrm{T}}^{\mathrm{Pb}} < 17.04$ (60-90% central) and C in $79.03 < \Sigma E_{\mathrm{T}}^{\mathrm{Pb}}$ (0-5% central). In the upper panels, the open circles show the scaled and shifted P template with uncertainties omitted from the plot, the closed circles show the C data, and the red line shows the fit (template and harmonic functions). The blue dashed line shows the 2$^{\mathrm{nd}}$ order harmonic component and the yellow dashed line shows the total harmonic function. The lower panels show the difference between the C data and the scaled and shifted P template, showing the problems in the template representing the awayside correlations.

Introducing additional parameters to the template method would undoubtedly produce better fits in these jet events; however, this comes at a cost of increased complexity of the flow extraction and moves away from a well vetted procedure. Therefore, the strategy is to make cuts in order to reduce the obvious jet component to the correlations while still benefiting from the increase in high $p_{\mathrm{T}}$ particles in these jet events.

Fig. 7.11 shows a distribution of track $\Delta\eta$ with respect to both the leading and sub-leading jets in the event. The quadratic fits (red dashed lines) are meant to guide the eye extrapolating into the short range. This motivates the nominal particle selection criteria as follows:

- No restriction on A particles
- B particles are required to have $|\Delta\eta| > 1$ with respect all jets in the event with $p_{\mathrm{T}}$ > 15 GeV.

In the simple two component model discussed above, this selection would keep correlations from UE-UE and HS-UE pair combinations but reduce the contribution from HS-HS correlations. Fig. 7.12 gives a comparison between the cases with (right) and without (left) this restriction. It is clear that the restriction gives a reduction in statistics. However, the obvious discrepancy on the awayside is gone, and the flow signal is a much greater fraction of the total correlation. Furthermore, the extracted parameters are significantly different; the $v_{3,3}$ is smaller because the fit is not trying to account for the depletion on the awayside, which allows $v_{2,2}$ to rise.

### 7.3.5 Estimation of Jet Particle Yields

In the previous section, a simple two component picture of a $p$+Pb jet event was discussed. The UE particles arise from soft interactions correlated with the overlapping nuclear geometry; a subset of the particles form jets emanating from a hard scattering (HS) of partons uncorrelated with the global geometry. In this case, the correlation functions are constructed from pairs pulled from this mixture. Under the two component assumption, pairs can be formed in the following four combinations:

- A: UE, B: UE (UE-UE)
- A: UE, B: HS (UE-HS)
- A: HS, B: UE (HS-UE)

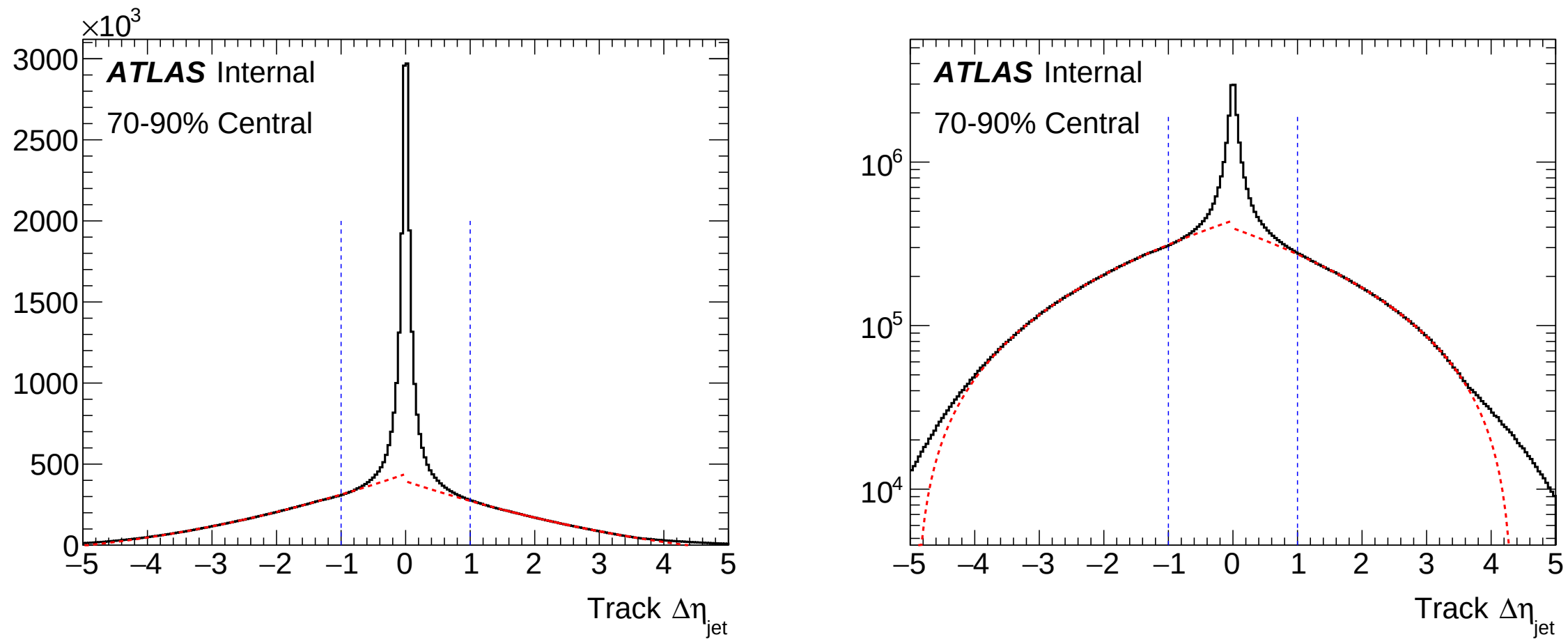

Figure 7.11: The distribution of track $\Delta\eta$ with respect to both the leading and sub-leading jets in 70-90% central jet events. To minimize the contamination from other jets in the event, the tracks are required to have $|\Delta\phi| < \pi/2$ with respect to the jet being compared to. The red dashed lines are quadratic fits to the histogram in the regions (-3.5,-1.5) and (1.5,3.5), respectively, and are included to guide the eye. The vertical blue dashed lines represent the nominal values of the B particle $\Delta\eta$ cut. The **left** plot is on a linear y scale and the **right** plot has a log y scale.

- A: HS, B: HS (HS-HS)

The following details the methodology used to estimated the relative particle pair yields of each contribution under minimal assumptions and taking inspiration from previous ATLAS measurements of the underlying in $pp$ collisions event [159–162].

Consider dividing the $\phi$ tracking acceptance into regions oriented by either the leading jet or leading track, in the case of MBT events that contain no jets with $p_{\mathrm{T}} > 15$ GeV. Define the following regions relative to this leading vector:

- towards: $(|\Delta\phi_{\mathrm{jet}}| < \frac{\pi}{4}) \cup (|\Delta\phi_{\mathrm{jet}}| > \frac{3\pi}{4})$
- transverse: $\frac{\pi}{4} < |\Delta\phi_{\mathrm{jet}}| < \frac{3\pi}{4}$

A graphical representations of these divisions is shown in Fig. 7.13, where the leading $p_{\mathrm{T}}$ object is represented as a yellow triangle. Then, under the assumption that 1) HS particles are completely contained in the towards region, and 2) the UE particles are uniformly distributed in the $\phi$ acceptance, the following relations follow:

$$N_{\mathrm{UE}} = 2N_{\mathrm{trans}}$$

$$N_{\mathrm{HS}} = N_{\mathrm{toward}} - N_{\mathrm{trans}},$$

where, $N_{\mathrm{UE}}$, and $N_{\mathrm{HS}}$ are the single particle yields from UE and HS processes respectively.

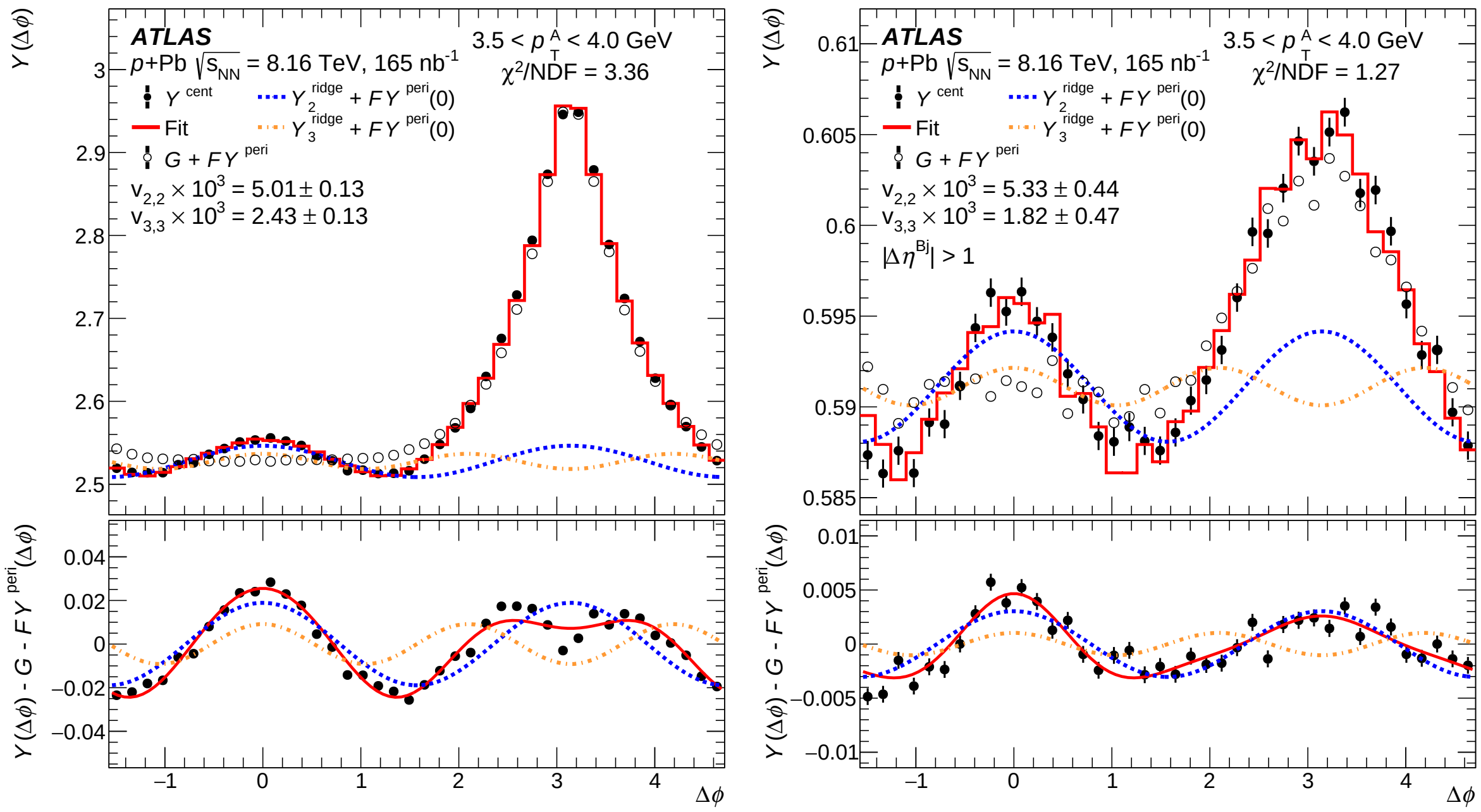

Figure 7.12: Template fitting output for jet events. Both require $\Delta\eta > 2$ and the A particles to have $3.50 < p_{\mathrm{T}}^{\mathrm{trig}} < 4.00$ and are made selecting P 60-90% central and C in 0-5% central events. The left plot is made with no selection on the B particles and the right plot is made requiring the B particles to have $\Delta\eta > 1$ with respect to all jets with $p_{\mathrm{T}}$ > 15 GeV in the event. In the upper panels, the open circles show the scaled and shifted P template with uncertainties omitted from the plot, the closed circles show the C data, and the red line shows the fit (template and harmonic functions). The blue dashed line shows the 2$^{\mathrm{nd}}$ order harmonic component and the yellow dashed line shows the total harmonic function. The lower panels show the difference between the C data and the scaled and shifted P template.

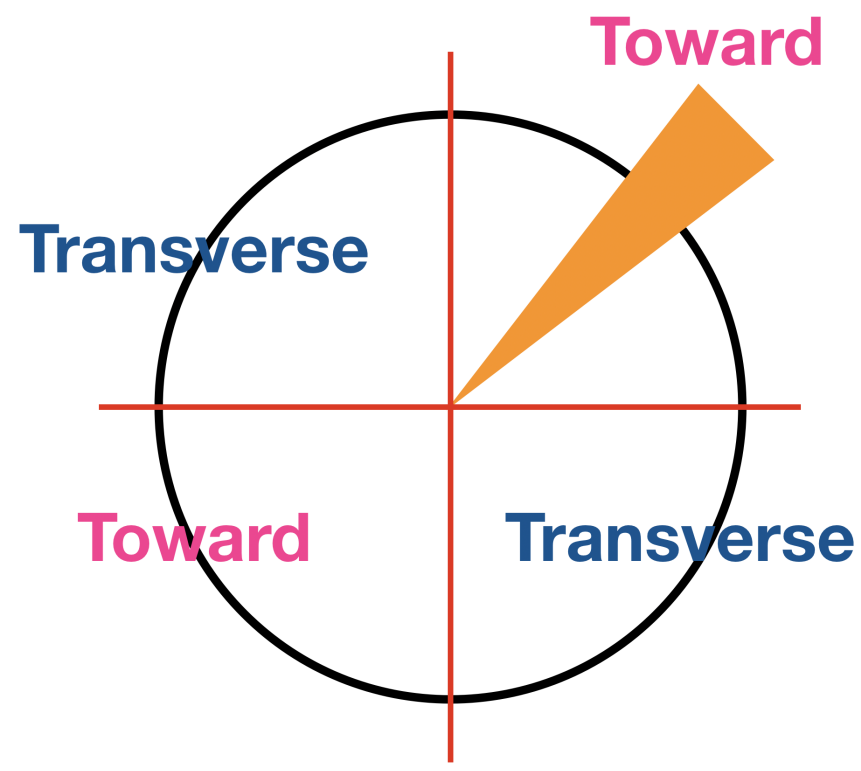

Figure 7.13: A graphical representation of the transverse and towards azimuthal regions relative to a high $p_{\mathrm{T}}$ object, shown as a yellow triangle.

The total yield of particle pairs can be decomposed in terms of products of the numbers of A ($N^{\mathrm{A}}$) and B ($N^{\mathrm{B}}$) particles in the following way:

$$P_{\text{total}} = N^{\mathrm{A}} \cdot N^{\mathrm{B}} \tag{7.23}$$

$$= (N^{\mathrm{A}}_{\mathrm{HS}} + N^{\mathrm{A}}_{\mathrm{UE}}) \cdot (N^{\mathrm{B}}_{\mathrm{HS}} + N^{\mathrm{B}}_{\mathrm{UE}}) \tag{7.24}$$

$$= N^{\mathrm{A}}_{\mathrm{HS}} \cdot N^{\mathrm{B}}_{\mathrm{HS}} + N^{\mathrm{A}}_{\mathrm{HS}} \cdot N^{\mathrm{B}}_{\mathrm{UE}} + N^{\mathrm{A}}_{\mathrm{UE}} \cdot N^{\mathrm{B}}_{\mathrm{HS}} + N^{\mathrm{A}}_{\mathrm{UE}} \cdot N^{\mathrm{B}}_{\mathrm{UE}}. \tag{7.25}$$

In the current case, because we enforce a $\Delta\eta$ gap and the detector acceptance is finite, $N^{\mathrm{A}} = N^{\mathrm{A}}(\eta^{\mathrm{A}})$ and $N^{\mathrm{B}} = N^{\mathrm{B}}(\eta^{\mathrm{A}}, \Delta\eta)$. That is, the yield of B particles, $N^{\mathrm{B}}$, depends on $\eta^{\mathrm{A}}$, and the particle separation, $\Delta\eta$. Thus, the products in 7.23 must take into account these dependencies:

$$N^{\mathrm{A}}_{\mathrm{X}} \cdot N^{\mathrm{B}}_{\mathrm{Y}} = \int_{-2.5}^{2.5} \frac{\mathrm{d}N^{\mathrm{A}}_{\mathrm{X}}(\eta^{\mathrm{A}})}{\mathrm{d}\eta^{\mathrm{A}}} \left[ \int_{2}^{5} \frac{\mathrm{d}^2 N^{\mathrm{B}}_{\mathrm{Y}}(\eta^{\mathrm{A}}, |\Delta\eta|)}{\mathrm{d}\eta^{\mathrm{A}}\mathrm{d}|\Delta\eta|} \mathrm{d}|\Delta\eta| \right] \mathrm{d}\eta^{\mathrm{A}} \tag{7.26}$$

Using the above assumptions and counting scheme, the pair yields of each combination may be estimated on a statistical basis, averaged over events. It is not possible to identify individual particles as originating from the UE or a HS using this method. It should be noted that the assumptions used in this derivation are likely not perfect; for example, the UE is not uniformly distributed in $\phi$, event by event, due to the presence of azimuthal anisotropy. The leading object may be more likely to be oriented with the anisotropy, in which case the UE yield would be underestimated and the HS yield overestimated. However, the analysis proceeds with the assumptions as given and includes no additional uncertainty for this potential effect.

## 7.4 Systematic Uncertainties

The sensitivity of each choice impacting the analysis is studied by making variations. The uncertainties are determined by assessing the difference between the nominal value of $v_2$ or $v_3$ and the value after a given variation. Unless otherwise stated, the uncertainties are defined as asymmetric one-standard-deviation errors. The final uncertainty is the quadrature sum of the uncertainty from each individual source. The variations are divided into three categories:

- Performance of the event triggers and tracking objects,
- Signal extraction, including choices made in the 2PC and template procedure,
- Jet selection.

In order to reduce the impact of statistical fluctuations, in most cases, the relative differences are smoothed using a bin-wise Gaussian convolution with width equal to one bin index. I.e. the value at any point is replaced by the Gaussian-weighted average of neighboring points. Below is a synopsis of the variations, as well as a comparison between the $v_2$ results versus $p_\mathrm{T}$. Comparisons with the $v_3$ results may be found in App. B.2, and a summary of the uncertainties in $v_2$ as a function of centrality may be found in App. B.3.

### 7.4.1 Performance

The sensitivity to the MBTS trigger efficiency correction and track efficiency corrections, is studied by checking the results with and without the corrections. Fig. 7.14 shows the results with and without the tracking and MBTS trigger (in the MBT case) corrections for both MBT and jet events. The effects of the correction are an almost negligible decrease in $v_2$ values. As a systematic variation on the corrections, the difference of the results with the corrections (nominal) to that without is taken as an uncertainty.

The 2016 $p$+Pb running period contained a bias to the tracking due to weak modes in the detector alignment. In these cases, the detector can be misshapen in a way that cannot be fully accounted for by the alignment procedure because they leave the global $\chi^2$ invariant. This leads to systematic problems with the track $q/p$ and sagitta that can be accounted for by applying bias corrections to tracks as a function of their angular position to simulate the impact of such weak modes. A comparison of the final results with and without this correction is shown in Fig 7.15. The MBT and 75 GeV jet events show a negligible difference, however, the 100 GeV jet events show a significant effect. Because it is not clear why the correction should affect the 100 GeV jet events and not the others, the difference is added as a systematic for the 100 GeVjet $v_2$ results only.

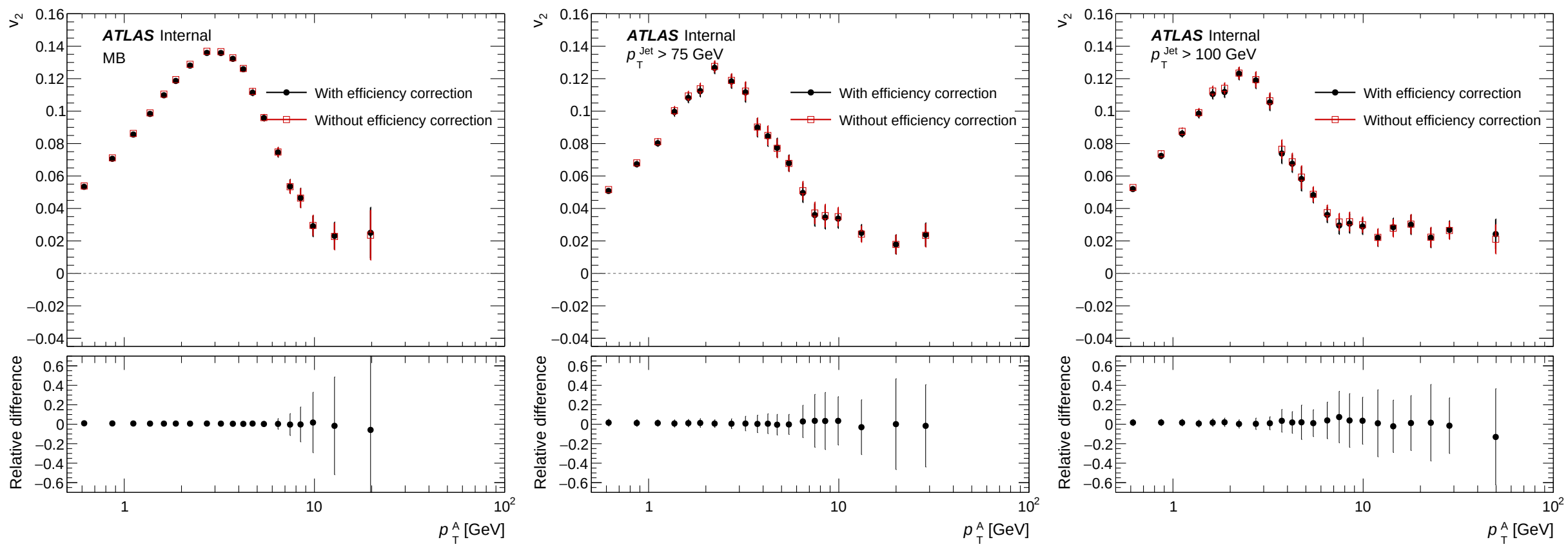

Figure 7.14: $v_2$ versus $p_T$ for MBT (left), 75 GeV jet (center), and 100 GeV jet (right) events with and without trigger and tracking efficiency corrections.

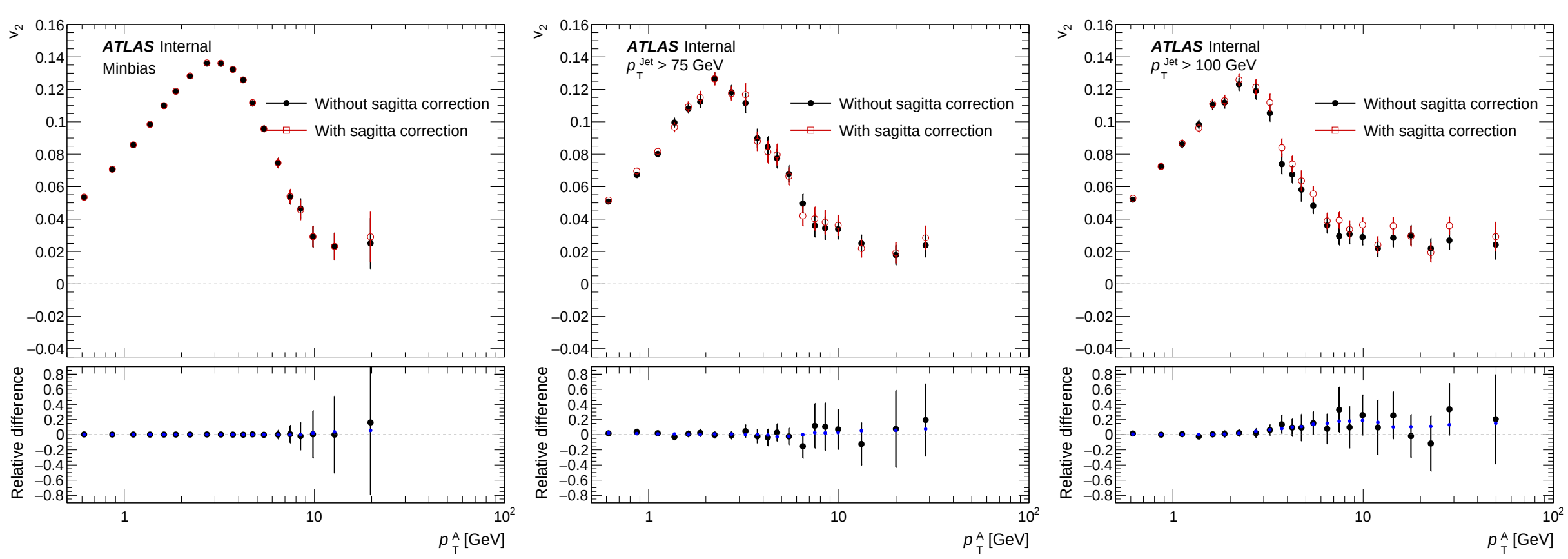

Figure 7.15: $v_2$ versus $p_T$ for MBT (left), 75 GeV jet (center), and 100 GeV jet (right) events with and without sagitta tracking correction.

The combined uncertainties for this category are plotted in Fig. 7.16.

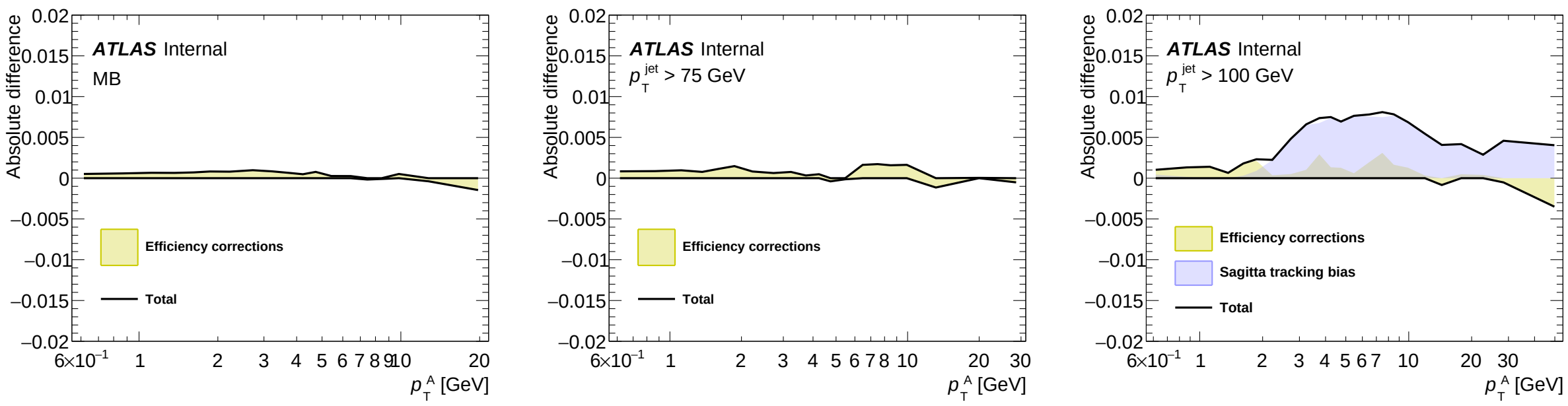

Figure 7.16: Combined performance uncertainty, plotted as the absolute difference in $v_2$ between the varied and nominal selections, for MBT (left), 75 GeV jet (center), and 100 GeV jet (right) events.

### 7.4.2 Signal Extraction

The following applies to the mixed event corrections described in Sec. 7.3.1. Because the 1D correlation functions are generated by first integrating over $\Delta\eta$ for the same and mixed events separately, and then dividing, the mixed event corrections do not correct for any $\Delta\eta$ acceptance effect in the final results. However, there are detector imperfections that are potentially being corrected for in $\Delta\phi$. To understand the magnitude of the effect of this correction on the results, the analysis is run with and without the correction. Fig. 7.17 gives the results of this test and shows a modest but systematic effect at high $p_{\mathrm{T}}$ in the case of the jet events, but an insignificant effect elsewhere. Because it is unclear what may cause the $p_{\mathrm{T}}$ dependence of the correction, the difference is taken as a systematic uncertainty.

The event mixing introduces additional statistical uncertainty in the results. Therefore, for the case of the MBT events, the results without the mixed event corrections are taken as nominal - adding any difference to the corrected as a systematic uncertainty. Conversely, due to the systematic difference observed for the jet events, the corrected are used as the nominal values and add the difference to the uncorrected as a systematic uncertainty.

The template method relies on there being a different fraction of flow and non-flow particles in the P and C selections. Here we test the sensitivity to this fraction by changing the P reference selection. The nominal $60-90\%$ reference is varied to use:

- $50-70\%$
- $70-90\%$

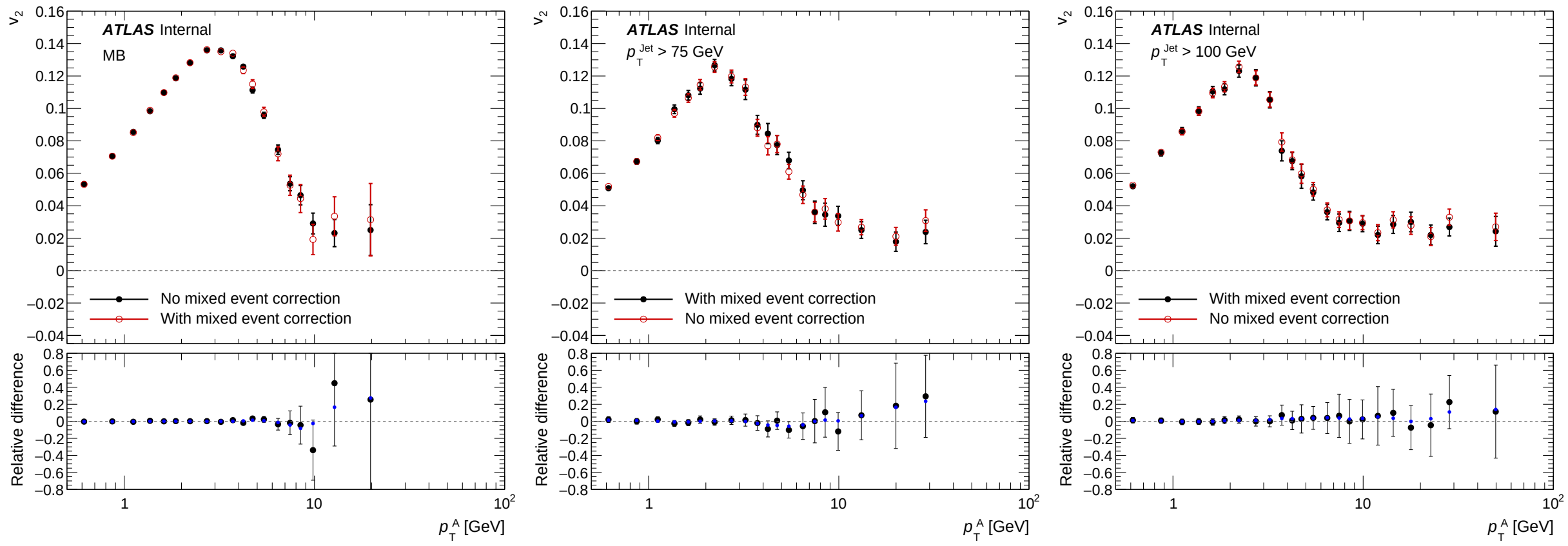

Figure 7.17: $v_2$ versus $p_T$ for both MBT (left), 75 GeV jet (center), and 100 GeV jet (right) events with the nominal values using the mixed event correction, and variation without the correction. The solid blue points in the sub-panels are the differences after smoothing.

Fig. 7.18 gives the result of this comparison. The maximum difference above and below the nominal case are taken as one standard deviation systematic uncertainties.

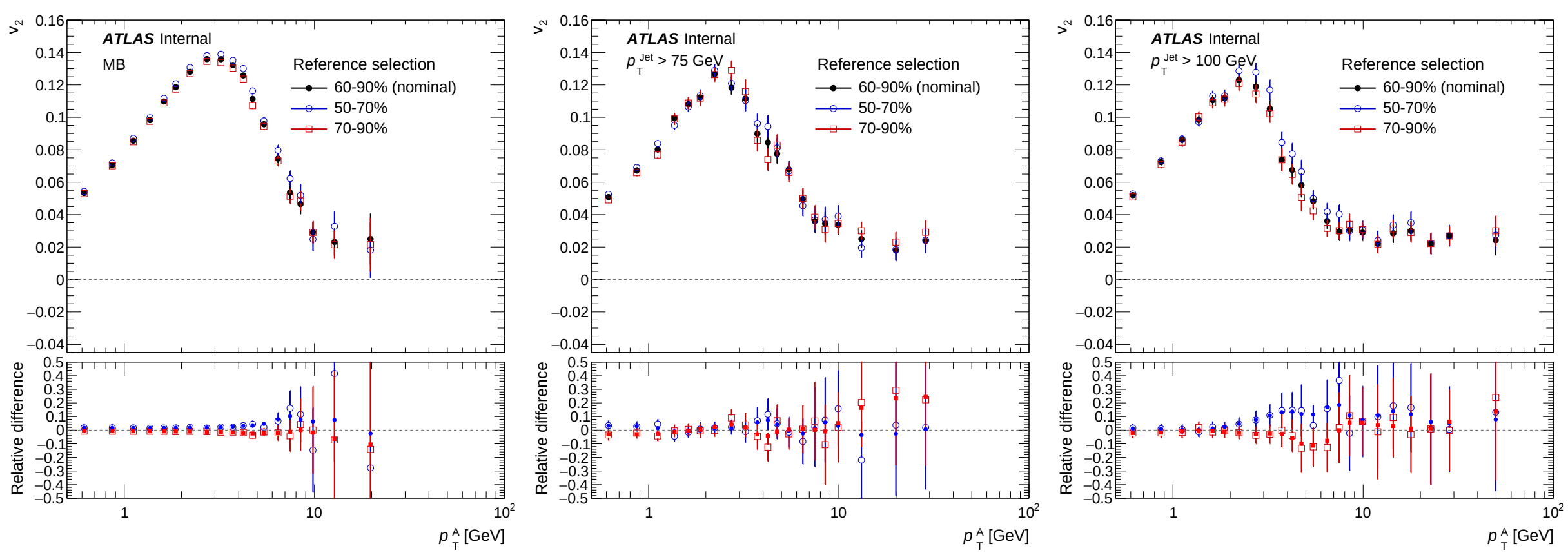

Figure 7.18: $v_2$ versus $p_T$ for both MBT (left), 75 GeV jet (center), and 100 GeV jet (right) events with the nominal and two varied P reference selections. The solid blue and red points in the sub-panels are the differences after smoothing.

The combined uncertainties for this category are plotted in Fig. 7.19.

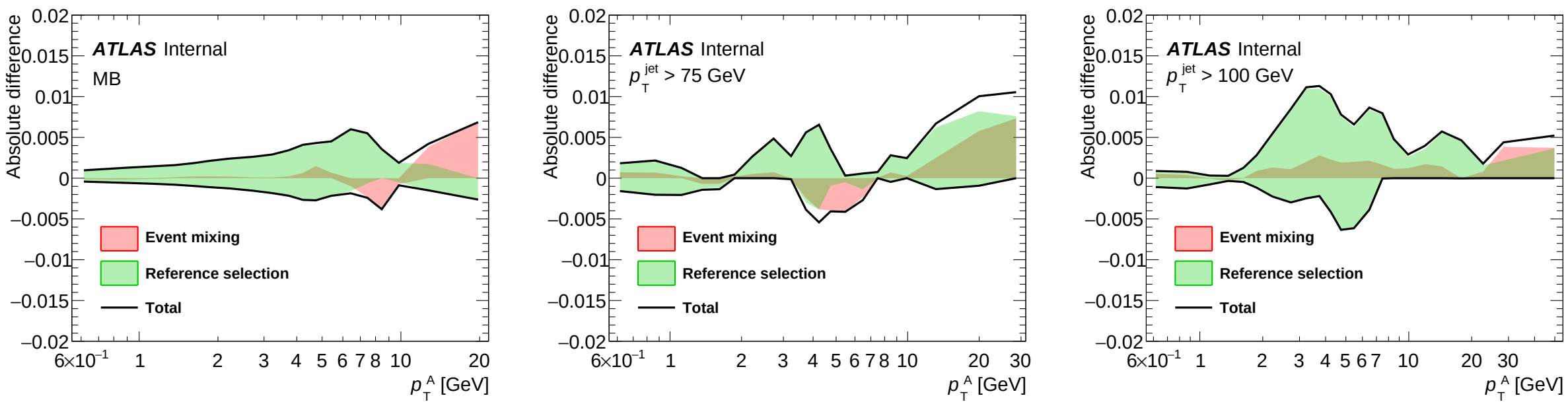

Figure 7.19: Combined signal extraction uncertainty, plotted as the absolute difference in $v_2$ between the varied and nominal selections, for MBT (left), 75 GeV jet (center), and 100 GeV jet (right) events.

### 7.4.3 Jet Selection

The reconstructed jets used in this analysis have been corrected for the average UE energy falling within the jet cone. However, this subtraction does not account for any azimuthal modulation of the UE. Therefore, the jets that happen to fall within the flow plane, will have, on average, more energy than those out of plane. This analysis selects jets based on their $p_{\mathrm{T}}$, thus, there will be more events with jets correlated to the UE flow plane passing the threshold. Because jets are partially composed of charged particles, this will enter the analysis as a positive contribution that will be strongest at high track $p_{\mathrm{T}}$. Additionally, the UE energy is not strongly dependent on jet energy, and thus, the $p_{\mathrm{T}}$ bias on the jets will be a larger percent for lower $p_{\mathrm{T}}$ jets. Therefore, this bias effect should be strongest for lower $p_{\mathrm{T}}$ jets.

To quantify this effect on the end results, we utilize the di-jet data overlay sample in which the jet processes are generated in Pythia, and are mixed with real data underlying events. If events are selected based on a $p_{\mathrm{T}}$ threshold made on the reconstructed jets (as in data) the sample jets will have been biased by the genuine flow of the data UE in a quantitatively similar way to jets in data. The magnitude of the effect is quantified by correlating tracks matched to truth Pythia particles with tracks coming from the data UE. If there is no bias, the correlation should be zero since the Pythia events are embedded randomly. Events are categorized by the presence of a reconstructed jet above either 75 or 100 GeV, as in data, and correlation functions are generated in exactly the same way as in data events except that the A-particles are required to be truth Pythia particles, and B-particles are required to be data particles. Fig. 7.20 gives the $v_2$ from this study, overlaid with the results from the jet datasets as a function of $p_{\mathrm{T}}$.

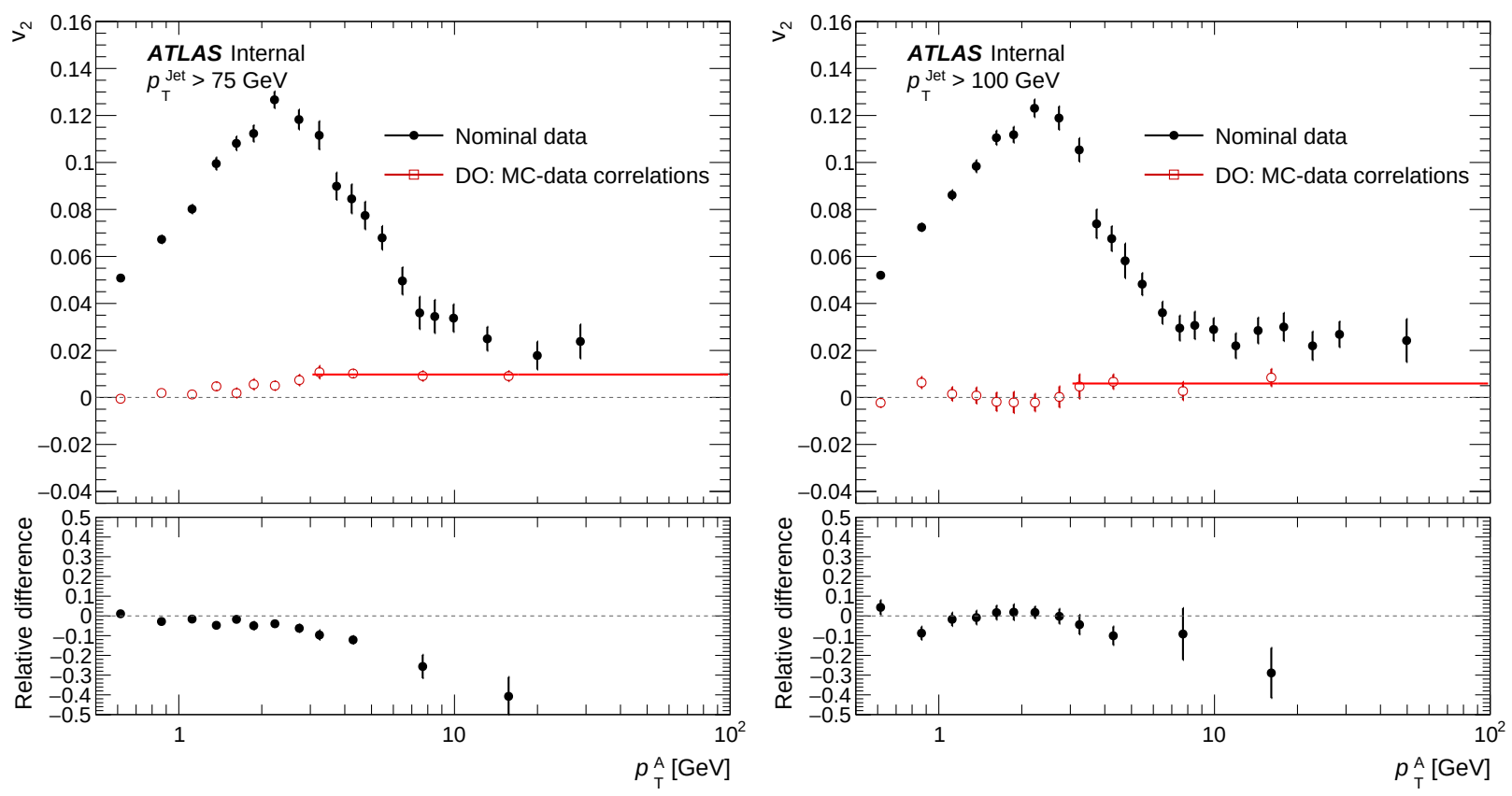

Figure 7.20: $v_2$ versus $p_{\mathrm{T}}$ for 75 GeV (left) and 100 GeV (right) jet events from data compared to those found using the MC data overlay sample. The red line is a fit to all points above 3 GeV, and acts as an estimate for the amount of the $v_2$ signal can be attributed to the jet-UE bias.

As expected, the effect is largest for the lower threshold jet events and at high track $p_{\mathrm{T}}$. The effect is significant in magnitude, however it accounts for only about 20% of the high $p_{\mathrm{T}}$ signal in 100 GeV events, and about 30% in the 70 GeV events. This is accounted for as a systematic uncertainty rather than a correction to the central values. The points are used as the uncertainty below 3 GeV and the constant fit is used above. For the centrality dependent $v_2$ results, the value determined in the 0-5% central case is conservatively used for all centralities.

The nominal selection criteria for jet events is an offline jet with $p_{\mathrm{T}} > 75$ GeV or $p_{\mathrm{T}} > 100$ GeV to be present in the region of $\eta < 1.16$. This $p_{\mathrm{T}}$ threshold is varied to $p_{\mathrm{T}} > 80$ GeV and $p_{\mathrm{T}} > 105$ GeV, respectively, to study the sensitivity to any slight jet trigger inefficiency or resolution effect. Fig. 7.21 shows the comparison of results with the varied jet threshold. The differences are taken as systematic uncertainties.

In jet events, the nominal B particle jet rejection cut made for all jets above 15 GeV. To understand the sensitivity to this choice, the threshold is varied to 20 GeV, and the difference to the nominal results is taken as a systematic uncertainty. The results using each cut value are compared in Fig. 7.22.

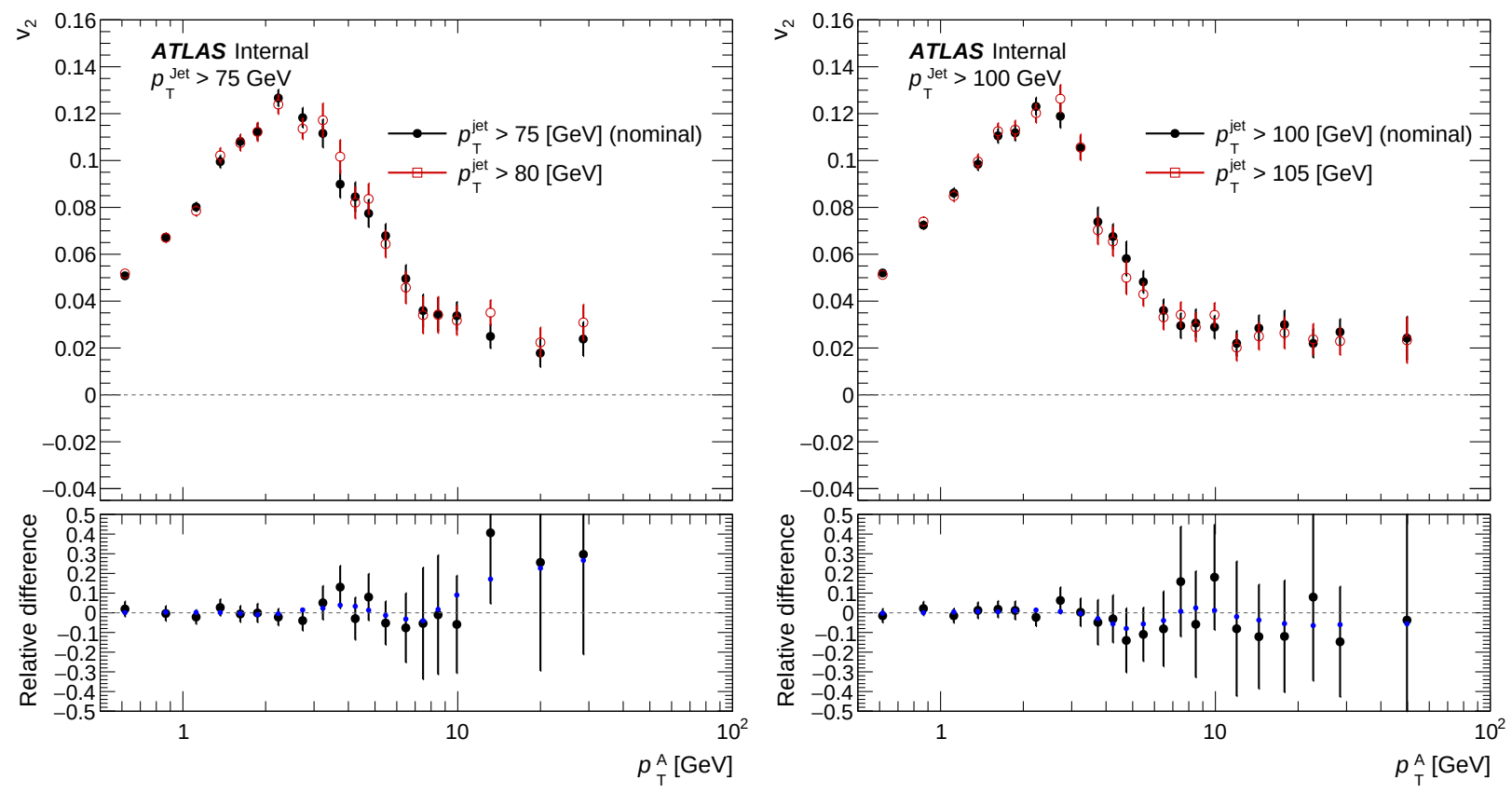

Figure 7.21: $v_2$ versus $p_T$ for 75 GeV (left) and 100 GeV (right) jet events with offline jet $p_T$ thresholds variations of +5 GeV. The solid blue points in the sub-panels are the differences after smoothing.

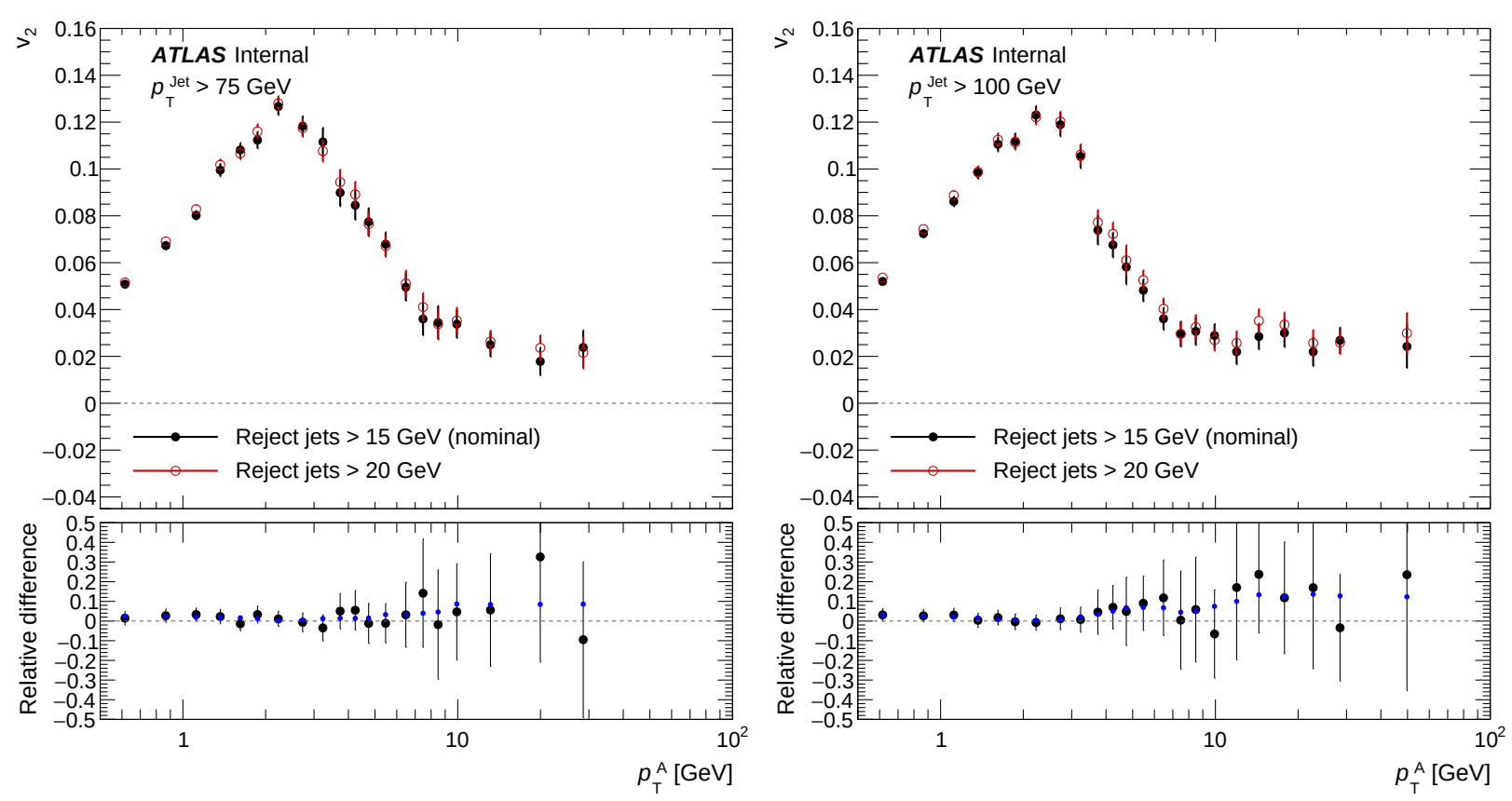

Figure 7.22: $v_2$ versus $p_T$ for 75 GeV (left) and 100 GeV (right) jet events with the nominal and varied $\Delta\eta_{\text{jet}}$ $p_T$ selection. The solid blue points in the sub-panels are the differences after smoothing.

The threshold $p_T$ for jets used in the rejection scheme is 15 GeV. Given this low threshold, it is conceivable that a significant portion of these jets are single hadrons on the tail of the UE particle distribution. To study this further, the number of charged tracks with $\Delta R < 0.4$ from the jet axis is determined. Figure 7.23 shows the multiplicity distributions of jets in different $p_T$ windows from 100 GeV jet events with two different centrality selections. The legends also report the fraction of these jets containing only a single track.

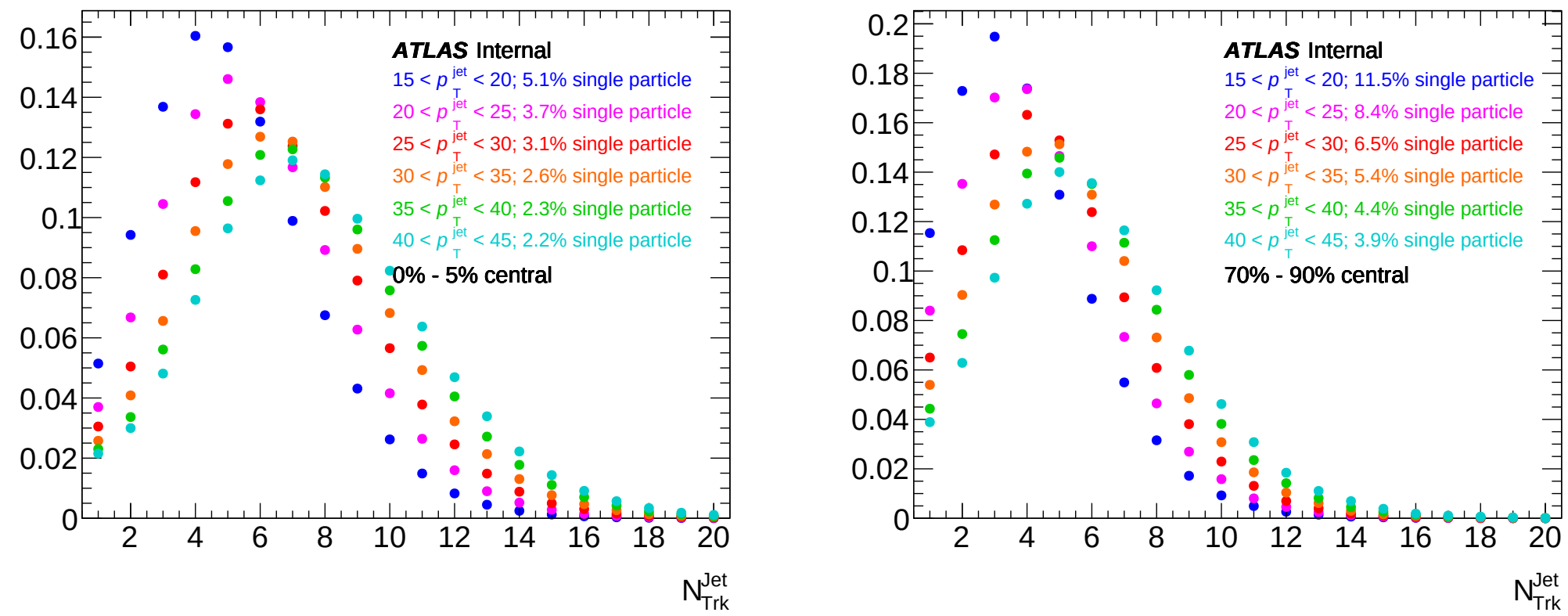

Figure 7.23: The number of charged tracks with $\Delta R < 0.4$ from the jet axis jets in different $p_{\mathrm{T}}$ windows from 100 GeVjet events. The left plot show the results from events from 0-5% central events and the right is for 70-90% central events. The solid blue points in the sub-panels are the differences after smoothing.

To estimate the sensitivity of the results to differences in jet multiplicity, all jets used in the jet rejection scheme are required to contain more than two tracks. The results of this test are plotted in Fig. 7.24. The difference from this test to the nominal scheme is taken as a systematic uncertainty on the final results.

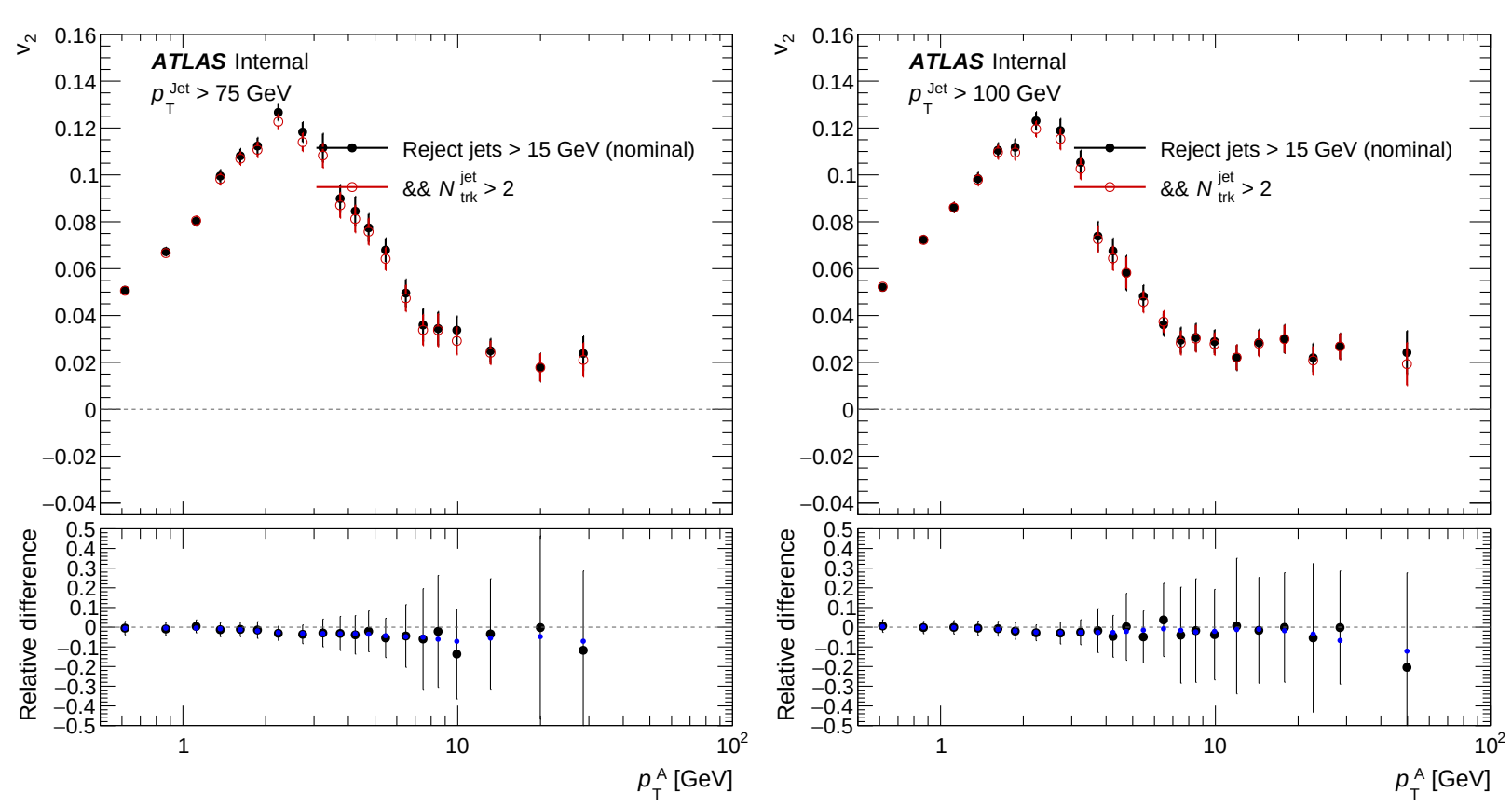

Figure 7.24: $v_2$ versus $p_{\mathrm{T}}$ for 75 GeV (left) and 100 GeV (right) jet events with the nominal and varied associated particle rejection jet multiplicity selection. The solid blue points in the sub-panels are the differences after smoothing.

As mentioned in Sec. 3.3, a sector of the HEC was disabled for this running period. Since the analysis uses calorimeter jet objects to reject B tracks from correlations, it is possible that the reduced response in this sector is allowing jets in that should be removed. To check for a bias, the analysis is run only using B tracks with $\eta < 1.56$ in the lab frame and compared with the nominal results. Fig. 7.25 gives the result of this comparison. There is a significant difference in each case, however, the trends at high $p_{\mathrm{T}}$ are opposite for the two jet samples. It is unclear what mechanism could have opposite effects for the two jet triggered events. As this difference could be due to jet bias, the difference is propagated as a systematic uncertainty.

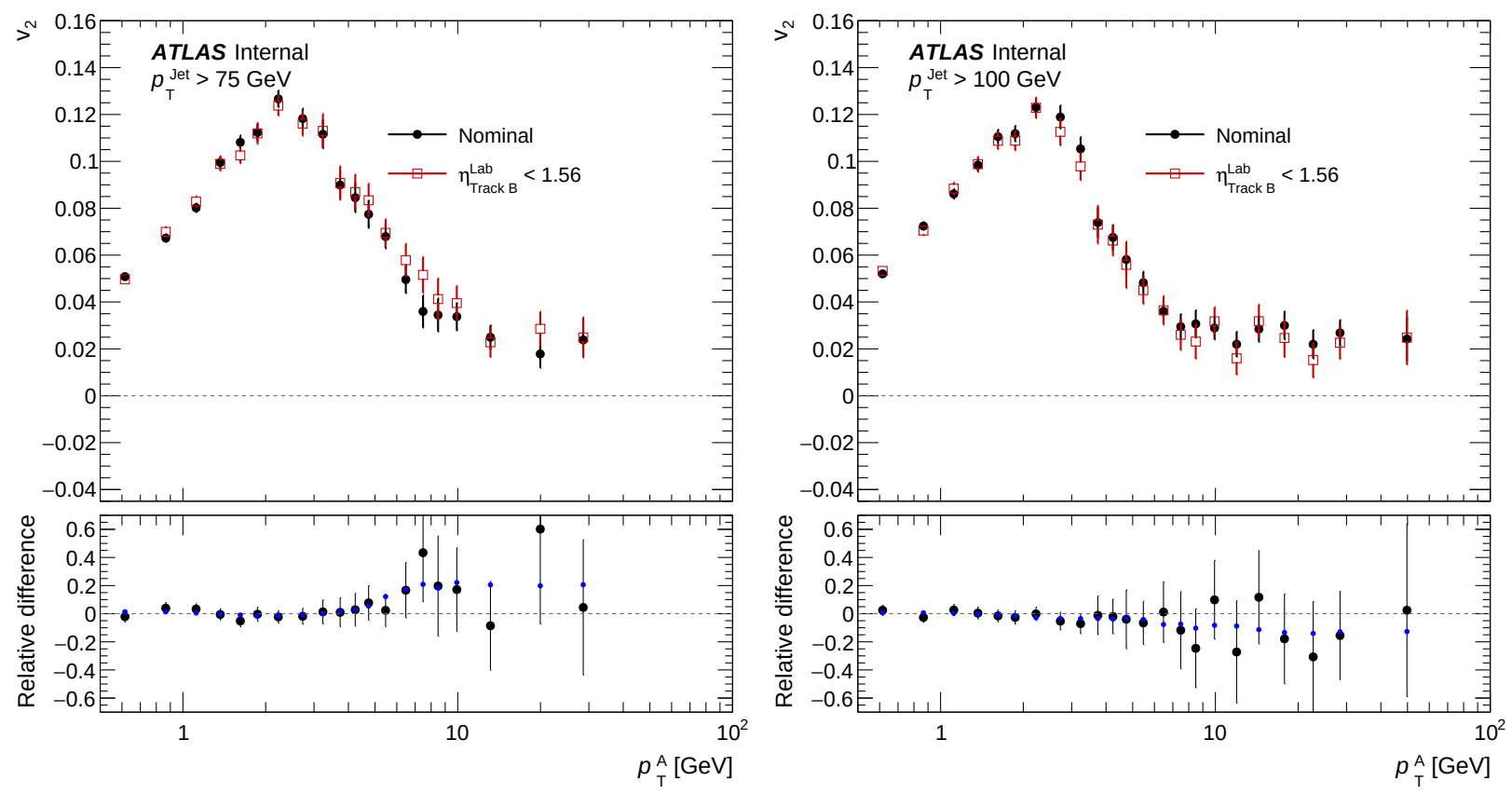

Figure 7.25: $v_2$ versus $p_{\mathrm{T}}$ for 75 GeV (left) and 100 GeV (right) events comparing the nominal results to those generated from tracks with an $\eta < 1.56$ in the lab frame. The solid blue points in the sub-panels are the differences after smoothing.

The combined uncertainties for this category are plotted in Fig. 7.26.

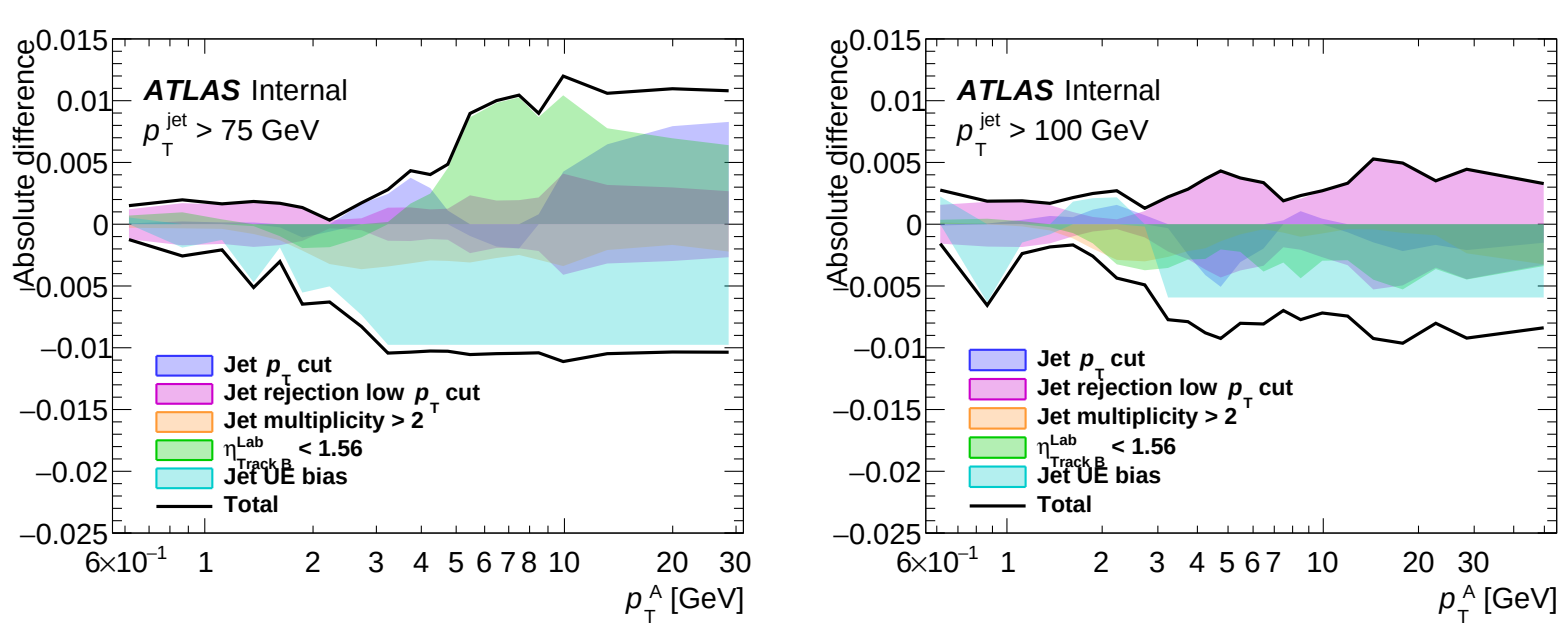

Figure 7.26: Combined jet selection uncertainty for MBT (left), 75 GeV jet (center), and 100 GeV jet (right) events.

### 7.4.4 Additional Checks

As mentioned in Sec. 7.3.2.1, a separation in $\eta$ between A and B particles is imposed to reduce the contamination from short range non-flow correlations. The nominal gap is $|\Delta\eta| > 2$, as has been used in numerous previous analyses. To understand the effect of this choice on the results, the flow extraction is made from correlations using different selections of $|\Delta\eta|$ for three different selections in A particle $p_{\mathrm{T}}$:

- Low $p_{\mathrm{T}}$: 0.5 < $p_{\mathrm{T}}$ < 2 GeV
- Mid $p_{\mathrm{T}}$: 2 < $p_{\mathrm{T}}$ < 9 GeV
- High $p_{\mathrm{T}}$: 9 < $p_{\mathrm{T}}$ < 100 GeV

Fig. 7.27 gives the result of this comparison, plotting $v_2$ vs. $|\Delta\eta|$ for the three $p_{\mathrm{T}}$ selections for all three datasets.

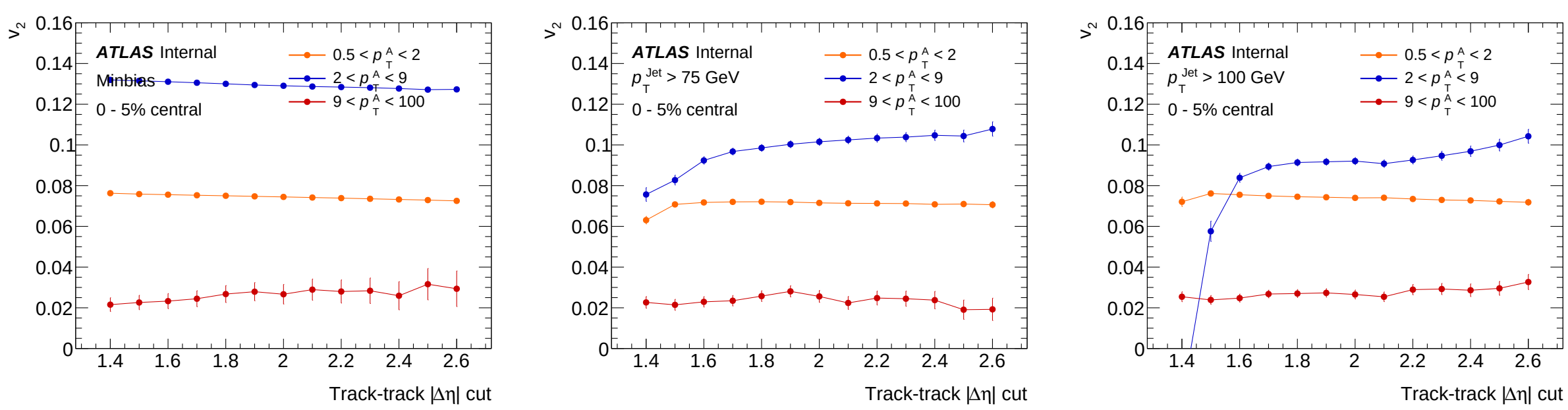

Figure 7.27: $v_2$ versus $|\Delta\eta|$ in low (orange), mid (blue), and high (red) $p_{\mathrm{T}}$ ranges for MBT (left), 75 GeV jet (center), and 100 GeV jet (right) events.

One sees in Fig. 7.27 a slow change in $v_2$ as the $|\Delta\eta|$ cut is varied around the nominal cut of 2. Because the separation cut is changing the $\eta$ distribution of tracks entering the correlations, as can be seen in Fig. 7.28, it is reasonable to expect this small change. However, when the cut becomes too small, one sees a change in the trend; this is particularly obvious in the jet events. This dramatic change is interpreted as a breakdown in the template assumptions and what we wish to avoid. The fact that the nominal cut value is well on the plateau gives confidence that the method is unbiased by the short-range correlations. The cut can be interpreted as a definition of the fiducial region of the measurement, and, therefore, do not include an uncertainty from this variation.

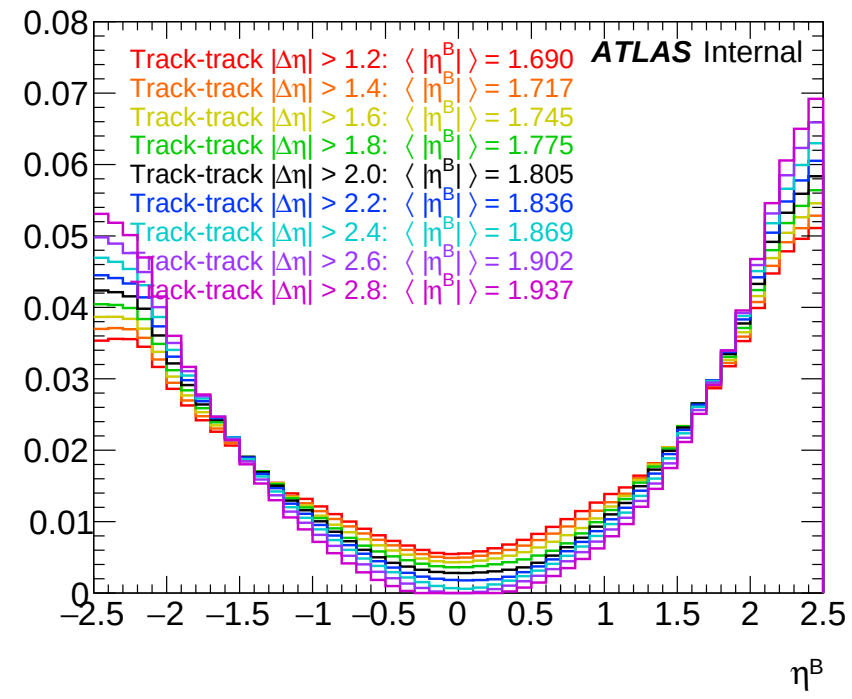

Figure 7.28: The $\eta$ distribution of associated (B) tracks used to generate correlation functions from 100 GeV jet events for several different values of track-track gap requirements. For each variation, the mean of $|\eta|$ is reported quantifying the change in the distributions.

The associated particle jet rejection scheme requires there to be a separation in $\eta$ of at least 1 unit between the B particle and all jets with $p_{\mathrm{T}} > 15$ GeV. Much like the previous check, we study the effect of this choice of separation on the results. The flow extraction is made from correlations using different selections of jet-track $|\Delta\eta|$ for three different selections in A particle $p_{\mathrm{T}}$:

- Low $p_{\mathrm{T}}$: $0.5 < p_{\mathrm{T}} < 2$ GeV
- Mid $p_{\mathrm{T}}$: $2 < p_{\mathrm{T}} < 9$ GeV
- High $p_{\mathrm{T}}$: $9 < p_{\mathrm{T}} < 100$ GeV

Fig. 7.29 gives the result of this comparison, plotting $v_2$ vs. $|\Delta\eta|$ for the three $p_{\mathrm{T}}$ selections for the two jet datasets.

Fig. 7.29 shows a slow change in $v_2$ as the $|\Delta\eta|$ cut is varied around the nominal cut of 1 unit. Because the separation cut is changing the $\eta$ distribution of tracks entering the correlations, as can be seen in Fig. 7.30, it is reasonable to expect this small change. Again the fact that the nominal cut is on a plateau gives confidence that the method is unbiased by the jet correlations. There is larger variation in the mid $p_{\mathrm{T}}$ transition region where the results are sensitive to changes in the UE-HS particle fractions. As before, this cut can be interpreted as a definition of the fiducial region of the measurement, and do not to incorporate a variation of it into the uncertainties.

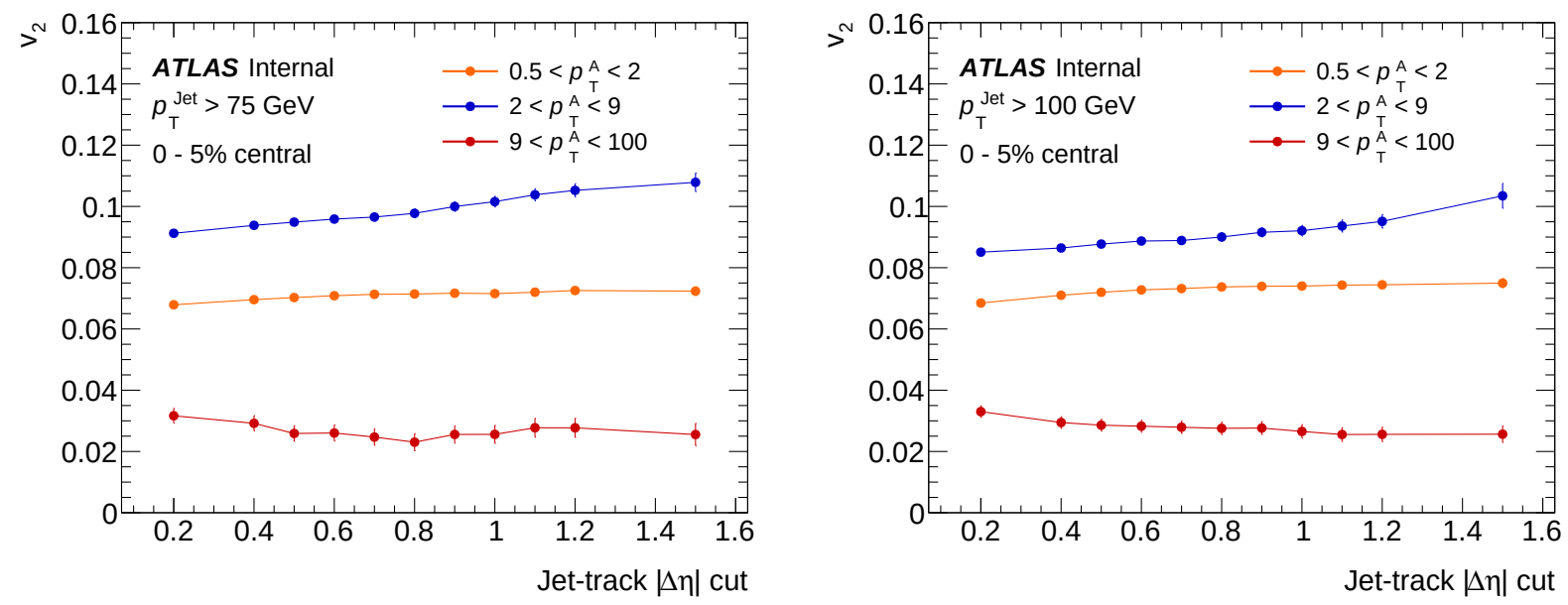

Figure 7.29: $v_2$ versus $|\Delta\eta|$ in low (orange), mid (blue), and high (red) $p_{\mathrm{T}}$ ranges for 75 GeV jet (left), and 100 GeV jet (right) events.

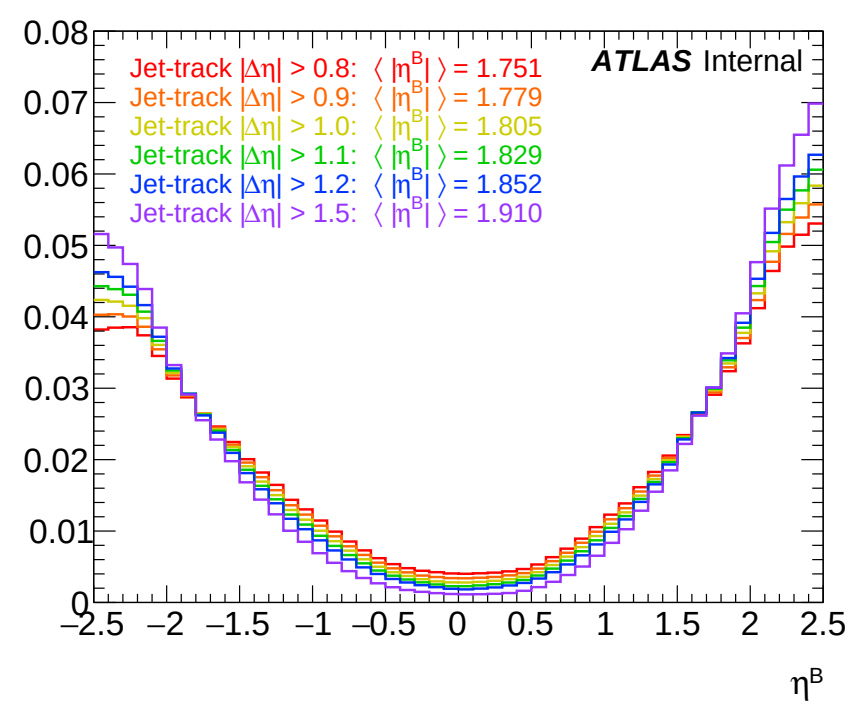

Figure 7.30: The $\eta$ distribution of associated (B) tracks used to generate correlation functions from 100 GeV jet events for several different values of jet-track gap requirements. For each variation, the mean of $|\eta|$ is reported quantifying the change in the distributions.

The results should be independent of the detector configuration, and therefore should be consistent between the two running periods (in which the $p$+Pb beam configurations were reversed). To check this, the results are determined for 100 GeV jet events in period 1 (run number < 313500) and period 2 (run number > 313500) separately. The left side of Fig. 7.31 shows a comparison between these two selections. There appears to be a slight bias towards higher values in period 1 vs. period 2 at high $p_{\mathrm{T}}$. Because the observed difference from the B particle disabled HEC rejection described in Sec. 7.4.3 is a selection made in the lab frame, the bias might be what is causing the slight period dependence. The right side of Fig. 7.31 shows the comparison again, but with the condition that B particles are restricted to have $\eta_{\mathrm{Lab}} < 1.56$. Since this check removes the apparent bias, the effect is attributed to the disabled HEC which is accounted for in Sec. 7.4.3. Therefore no additional systematic is applied.

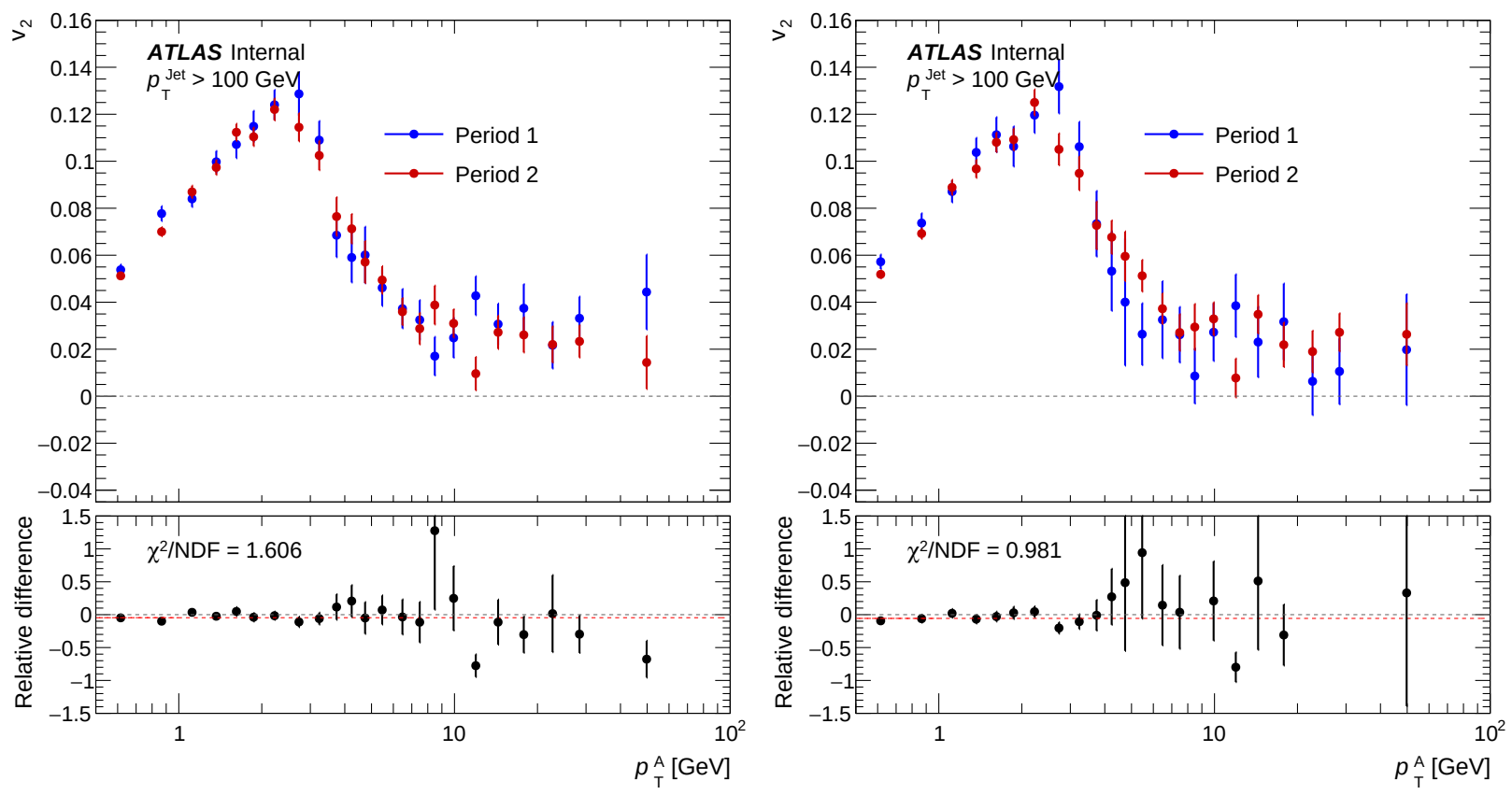

Figure 7.31: Comparison of $v_2$ results from data from period 1 and 2 separately. The right plot is after the associated particles are restricted to have $\eta_{Lab} < 1.56$, and the left is without restriction. The red dashed lines are the result of a constant fit for which the $\chi^2/NDF$ is quoted on the figure.

Fig. 7.3 shows small $\eta - \phi$ dependent tracking performance differences. To understand the sensitivity to any residual effects not removed by the event mixing scheme, $\phi$ flattening maps are generated using data and applied, particle-by-particle, when constructing the correlation functions. The corrections are derived by normalizing track $\phi$ distributions in $\Delta\eta = 0.5$ slices for tracks within a given $p_{\mathrm{T}}$ range. These correction maps are generated for MBT and jet events independently and are plotted in Figures B.26- B.28 in App. B.5. Fig. 7.32 gives the results with and without the corrections for the 75 GeV and 100 GeV jet events. As one would hope, the additional flattening yielded negligible differences to the nominal results, and no additional uncertainty is applied.

In addition to the nominal associated particle jet rejection scheme, others have been checked. Fig. 7.33 shows a comparison between the nominal "all" jet rejection, and one where only the leading and sub-leading (in $p_{\mathrm{T}}$) jets are rejected. There is a clear difference between 1.5 and 3 GeV where the all jet rejection case peaks at a higher value than the 2-jet case. This could be due to a higher concentration UE-UE correlations. At high $p_{\mathrm{T}}$, the 2-jet rejection yields a systematically higher value. It is not known if this is due to residual non-flow or a difference in the signal distribution being sampled. However, the all-jet rejection yields a larger suppression in the awayside non-flow peak in the correlation functions, making the extraction less sensitive to inconsistencies in the non-flow subtraction. Thus, we choose to use the all-jet scheme with no additional uncertainty.

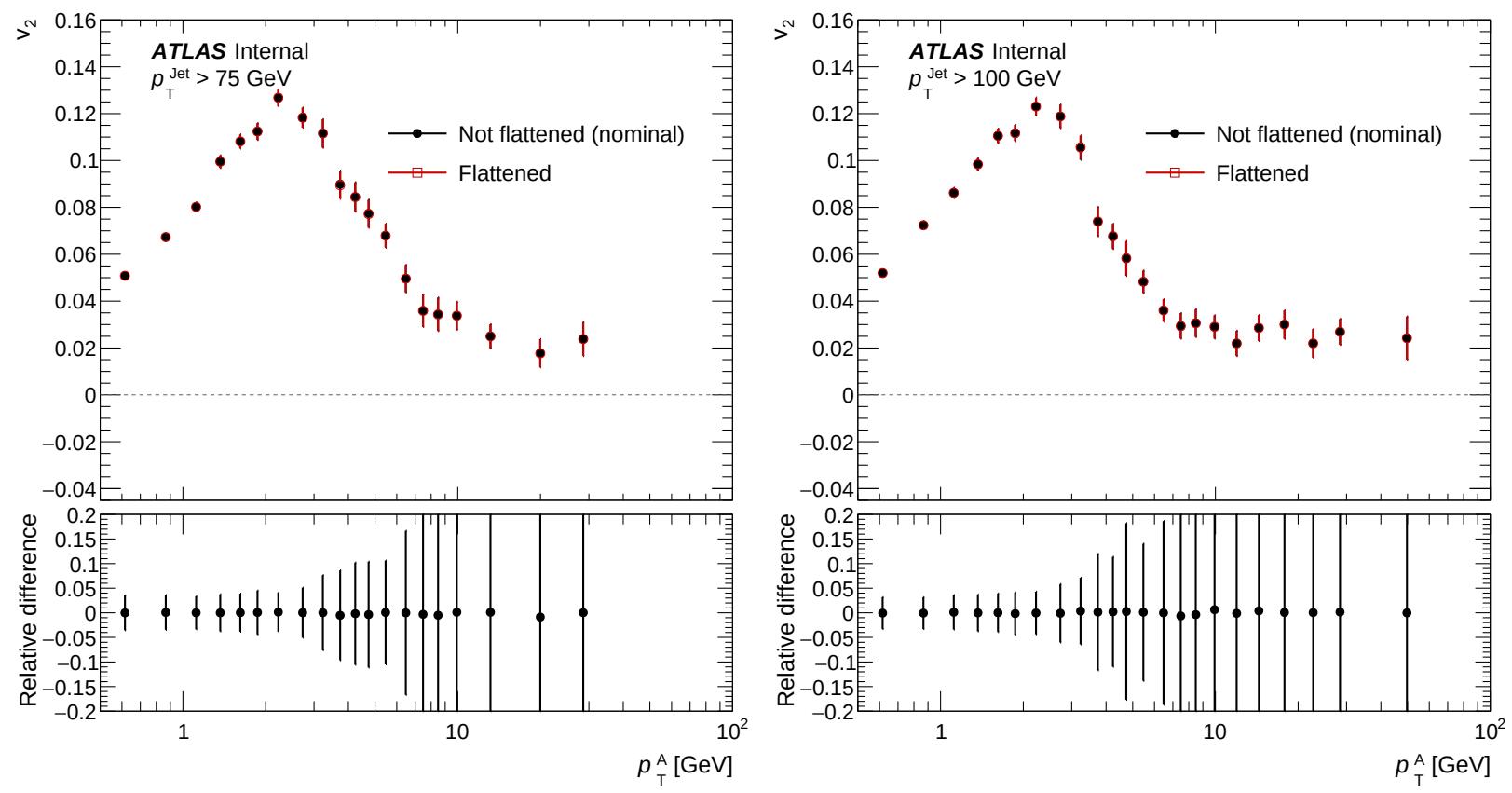

Figure 7.32: $v_2$ versus $p_T$ for both 75 GeV (left) and 100 GeV (right) jet events with the nominal values and those with flattening corrections applied.

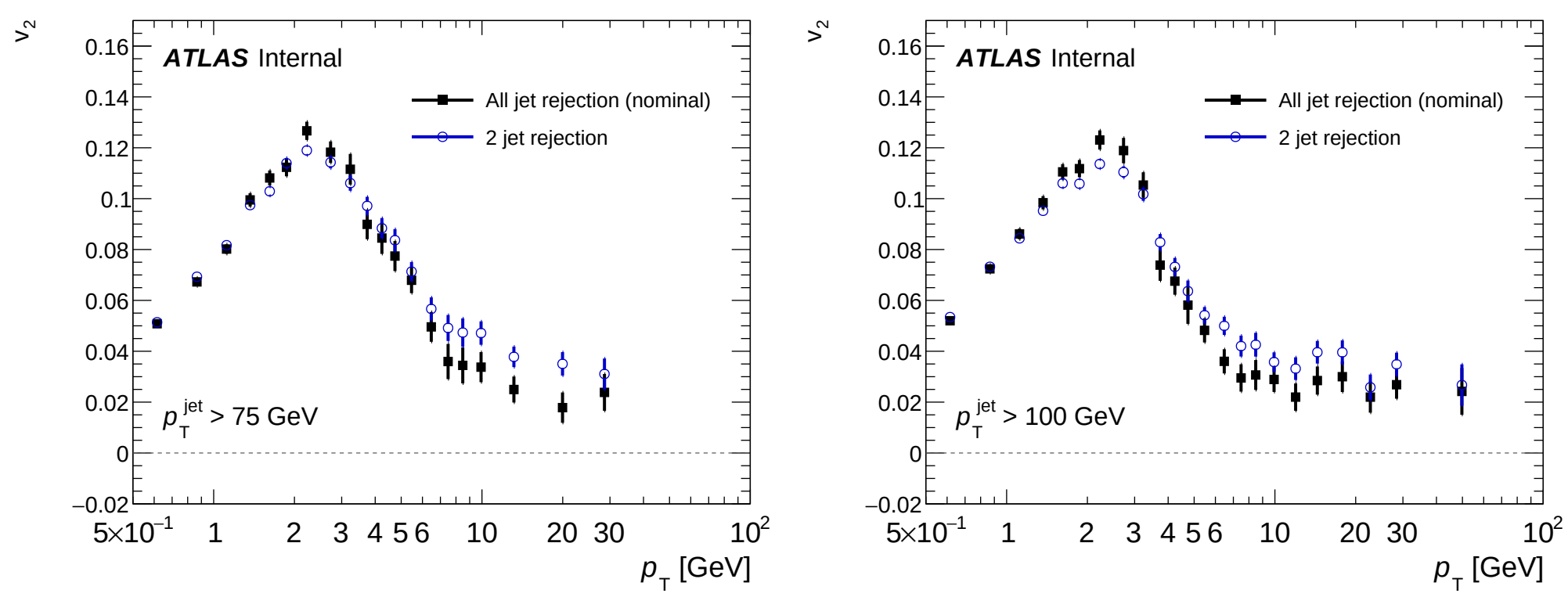

Figure 7.33: $v_2$ versus $p_T$ for 75 GeV (left) and 100 GeV (right) jet events with the nominal "all" jet rejection and those in which only the leading and sub-leading (in $p_T$) jets are used in the B particle rejection.

One might think that the simple two jet rejection scheme that rejects from the leading and sub-leading (in $p_T$) jets might not be as efficient in removing the jet non-flow as if you require the jets to be in opposite hemispheres. For this check, the basic two jet rejection is compared with a variant where the sub-leading jet must be in the opposite side in $\phi$ ($|\Delta\phi| > \pi/2$). Fig. 7.34 gives the result of this comparison showing almost no difference for 100 GeV jet events.

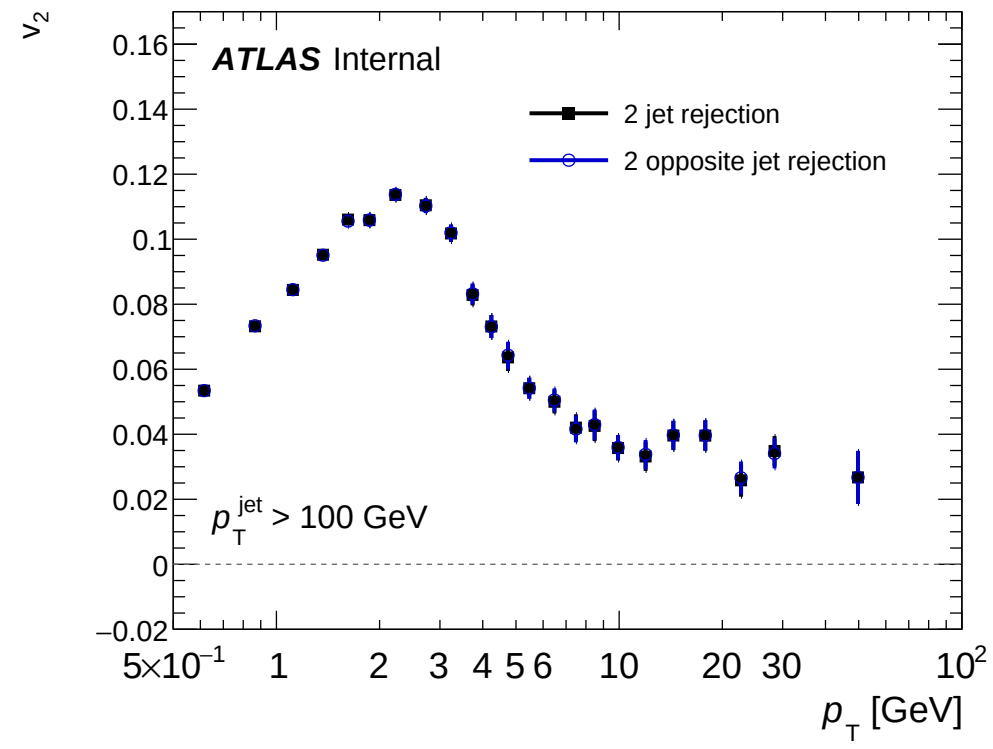

Figure 7.34: $v_2$ versus $p_\mathrm{T}$ 100 GeV jet events with the simple two jet rejection compared to those generated with the two opposite jet rejection.

As mentioned in Sec. 7.2.1, high multiplicity triggers were used to record events with multiplicities above a certain threshold. Because we are using the $\Sigma E_\mathrm{T}^\mathrm{Pb}$ to categorize events, if we use these triggers, our central and peripheral selections would be formed from a combination of triggers and the multiplicity distributions would be unnatural. Figure 7.35 gives a comparison of the results between the nominal selections from the MBT dataset and those from the MBT+HMT. If the MB+HMT is corrected for it's prescale differences, the results agree with just the MB. However comparing the MBT with the uncorrected MBT+HMT gives noticeable differences - particularly in the transition region between 2 GeV and 9 GeV. Since there is no statistical benefit in using the prescale corrected MBT+HMT, the MBT only is used for simplicity.

### 7.4.5 Total Systematic Uncertainty

The resultant variation from each source of uncertainty is added in quadrature asymmetrically. A summary of the sources as well as the total uncertainty in $v_2$ is plotted in Fig. 7.36. Likewise, the total uncertainty in the $v_3$ determination is plotted in Fig.7.37.

To determine the uncertainty in the particle pair yield fractions, the calculation is done after each variation, and each difference is considered an independent source. The variations relevant to the pair yield determination in MBT events are:

- Efficiency corrections

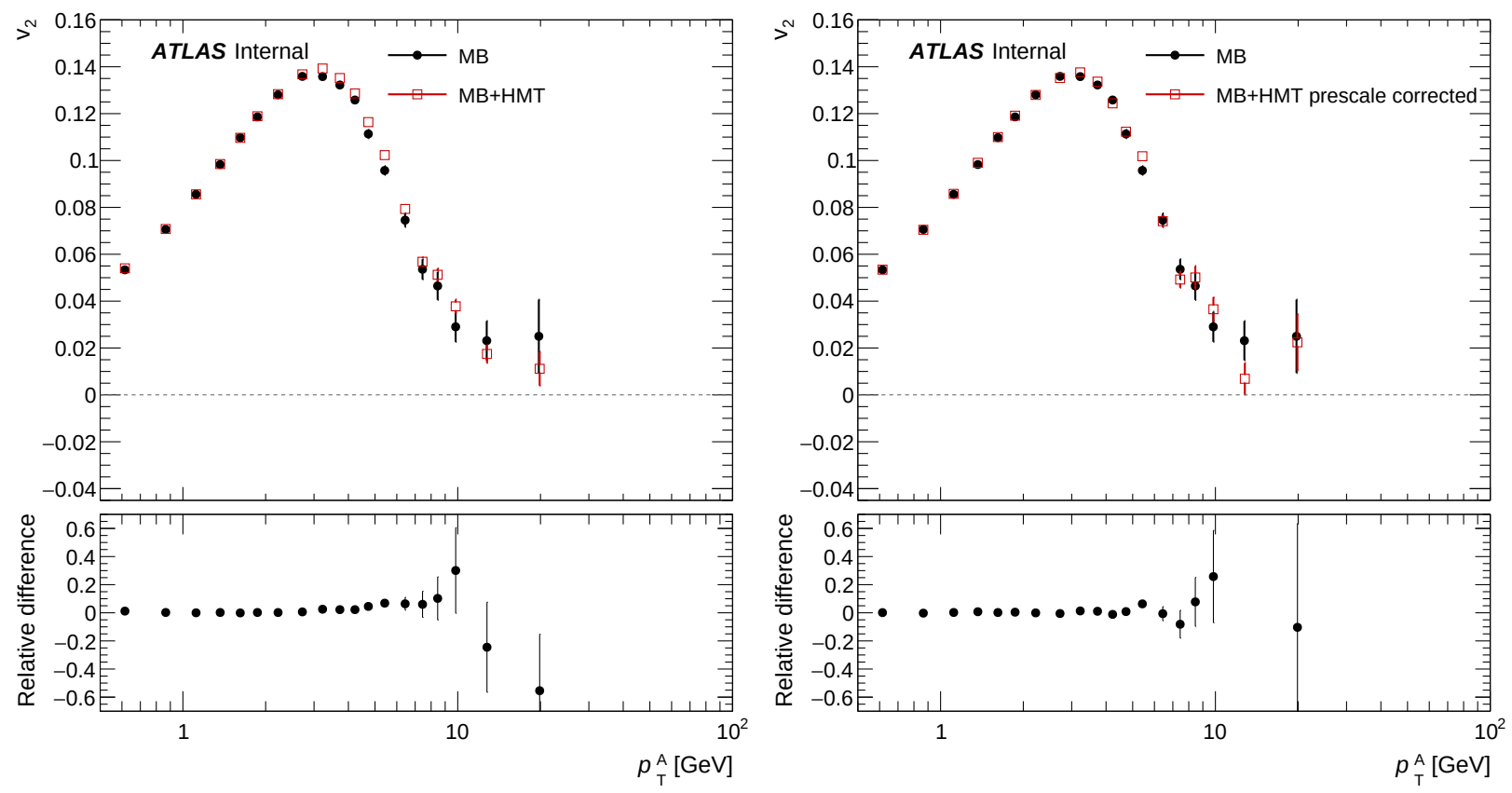

Figure 7.35: $v_2$ versus $p_T$ for MBT and MBT+HMT events. The right figure gives the MBT+HMT after trigger prescale correction, and the left is uncorrected.

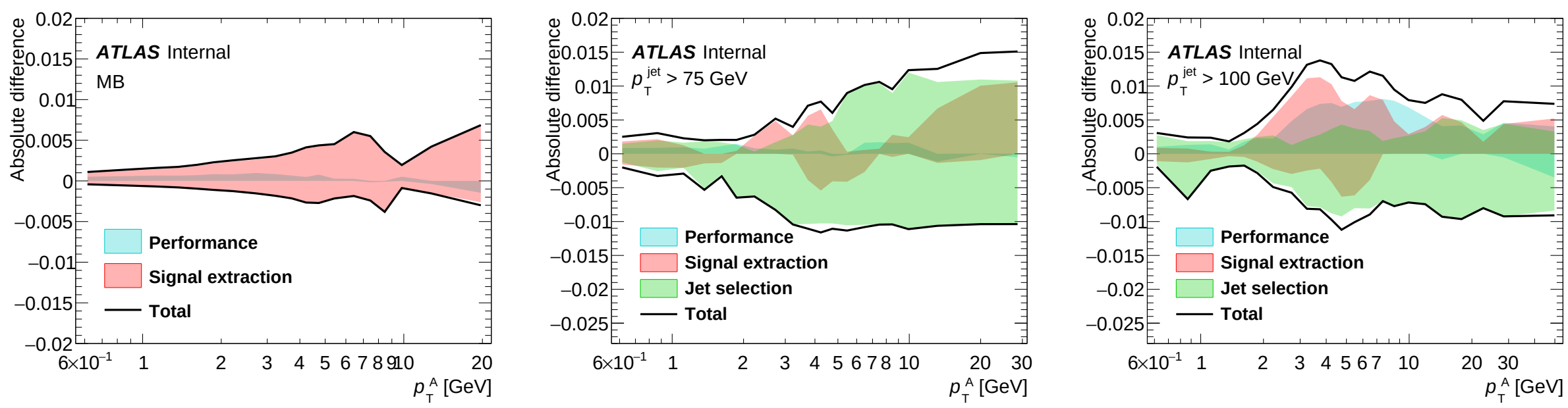

Figure 7.36: The relative uncertainty in $v_2$ from all sources versus $p_T$ for MBT (left), 75 GeV jet (center), and 100 GeV jet (right) events. The combined uncertainty is shown as the black curve.

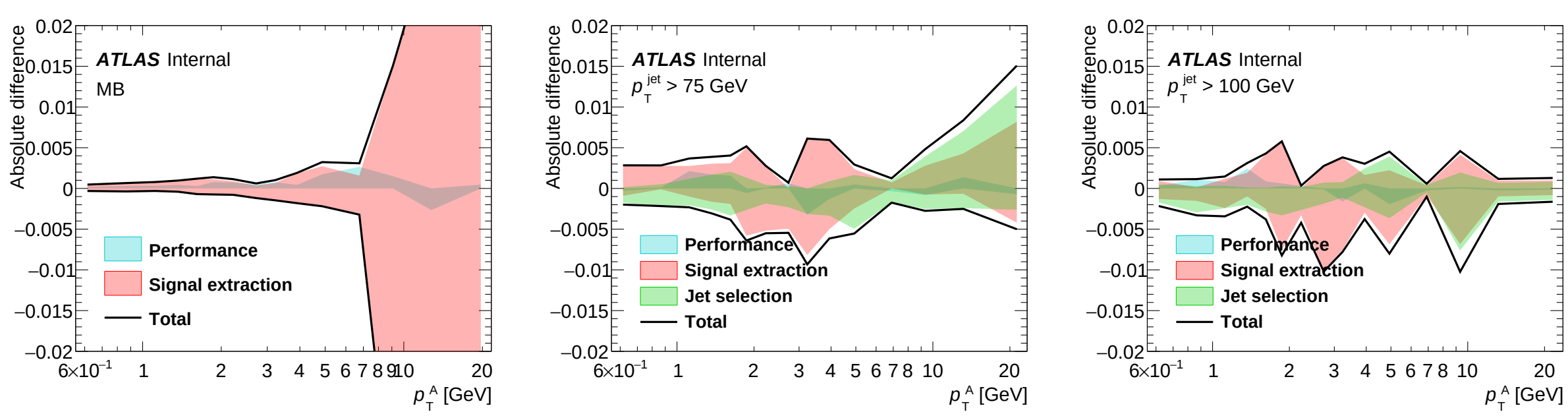

Figure 7.37: The relative uncertainty in $v_3$ from all sources versus $p_T$ for MBT (left), 75 GeV jet (center), and 100 GeV jet (right) events. The combined uncertainty is shown as the black curve.

and the variations relevant to the jet events are:

- Efficiency corrections
- Jet $p_{\mathrm{T}}$ cut
- B particle rejection jet low $p_{\mathrm{T}}$ cut
- B particle rejection jet multiplicity cut

Summary figures for UE-UE and HS-UE pair combinations are shown in Fig. 7.38.

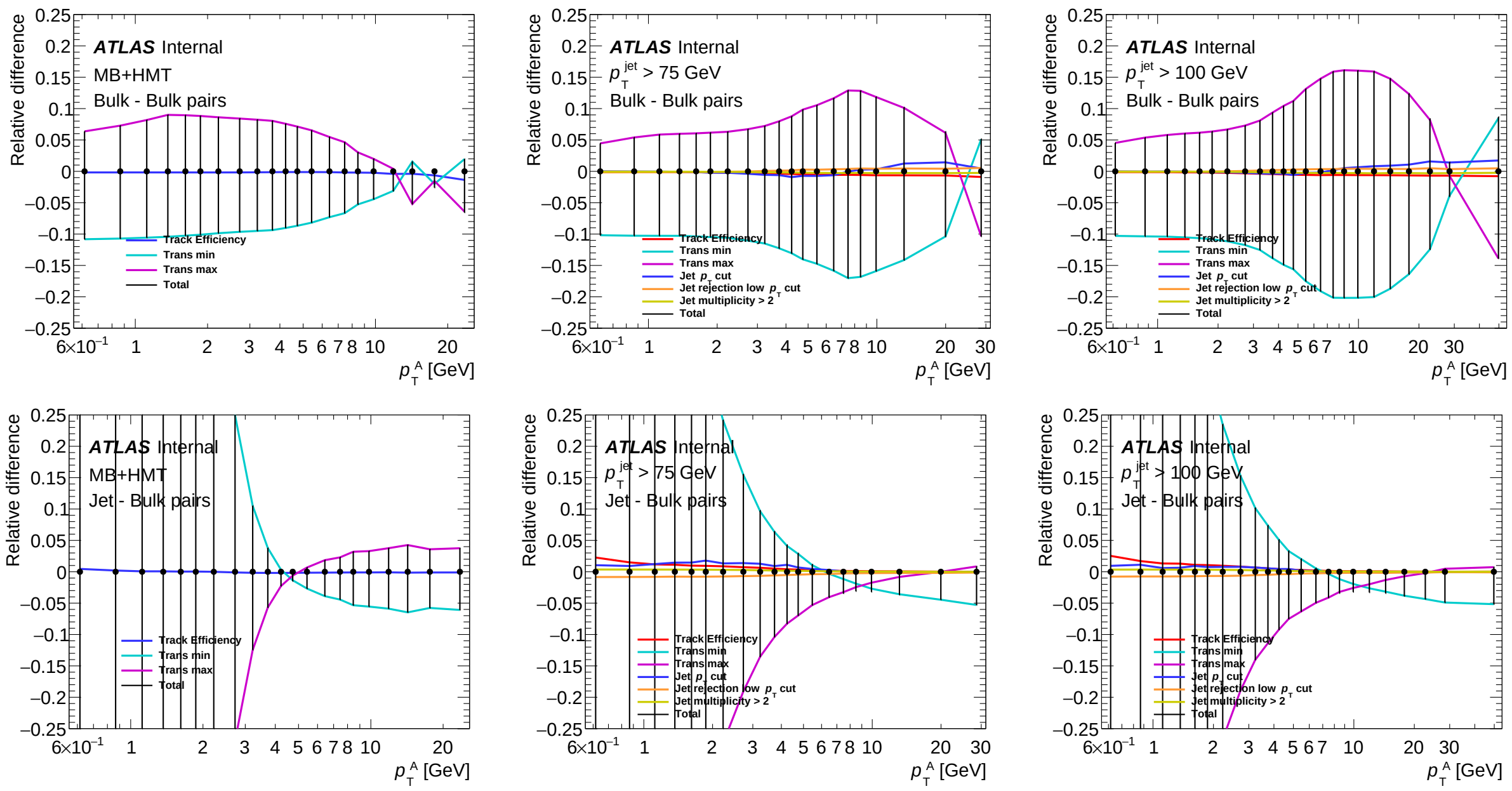

Figure 7.38: The relative uncertainty in UE-UE (top) and HS-UE (bottom) pair fractions in 0-5% central events from all sources versus $p_{\mathrm{T}}$ for MBT (left), 75 GeV jet (center), and 100 GeV jet (right) events. The combined uncertainty is shown as the black curve.

# Chapter 8

## Results of the Measurement of Azimuthal Anisotropy

Figure 8.1 shows the extracted second- ($v_2$) and third-order ($v_3$) anisotropy coefficients for the MBT events compared to those from both 75 and 100 GeV selections of jet events plotted as a function of A-particle $p_{\mathrm{T}}$ in the range $0.5 < p_{\mathrm{T}} < 100$ GeV. Each set of values is from events with the same 0–5% centrality selection. Points are located on the horizontal axis at the mean $p_{\mathrm{T}}$ of tracks within any given bin. The $v_2$ and $v_3$ coefficients increase as a function of $p_{\mathrm{T}}$ in the low $p_{\mathrm{T}}$ region ($p_{\mathrm{T}}$ < 2–3 GeV), then decrease (2–3 < $p_{\mathrm{T}}$ < 9 GeV), and finally plateau for high $p_{\mathrm{T}}$ ($p_{\mathrm{T}}$ > 9 GeV). The $v_2$ coefficients are consistent with being independent of $p_{\mathrm{T}}$ for $p_{\mathrm{T}}$ > 9 GeV, while the larger uncertainties in the values of $v_3$ preclude any strong conclusion.

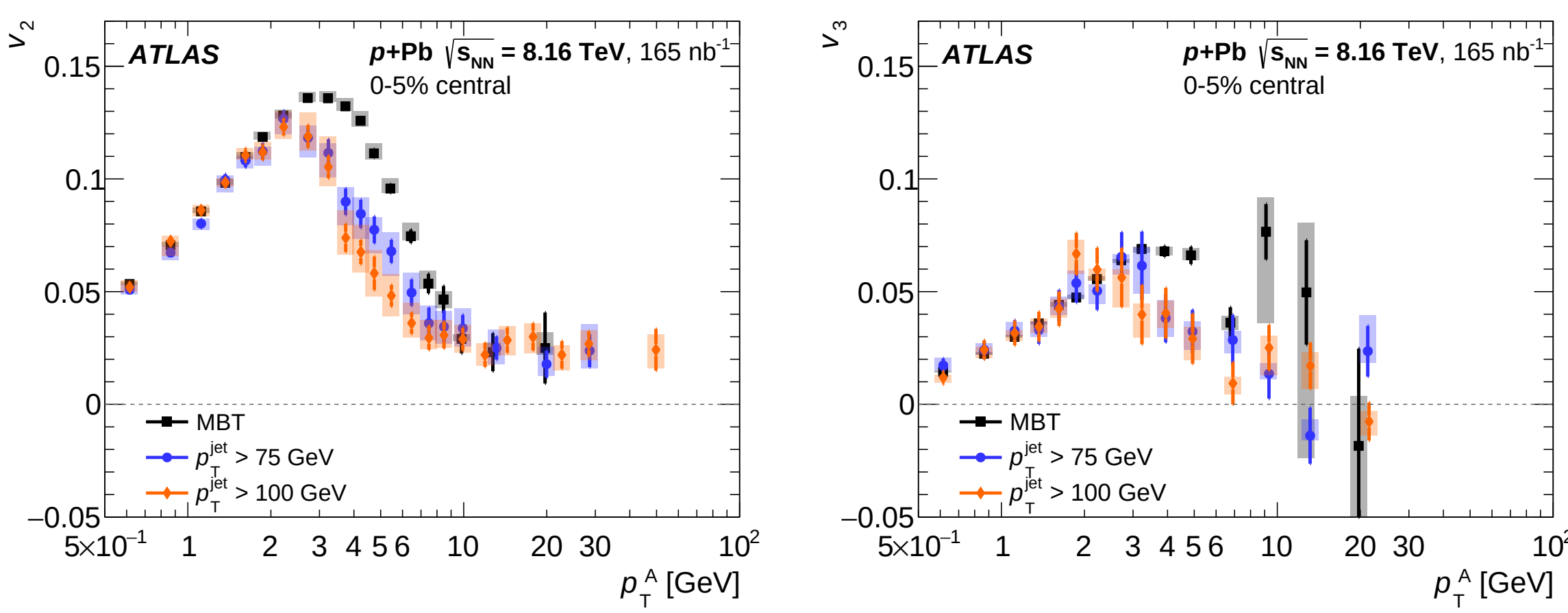

Figure 8.1: Distribution of $v_2$ (left) and $v_3$ (right) plotted as a function of the A-particle $p_{\mathrm{T}}$. Values from MBT events are plotted as black squares, and those from events with jet $p_{\mathrm{T}} > 75$ GeV and events with jet $p_{\mathrm{T}} > 100$ GeV are plotted as blue circles and orange diamonds respectively. Statistical uncertainties are shown as narrow vertical lines on each point, and systematic uncertainties are presented as colored boxes behind the points.

The $v_2$ results show agreement within uncertainties between the MBT and jet events for the low $p_{\mathrm{T}}$ ($p_{\mathrm{T}} \lesssim 2$ GeV) and high $p_{\mathrm{T}}$ ($p_{\mathrm{T}} \gtrsim 9$ GeV) regions. For the intermediate $p_{\mathrm{T}}$ region, the MBT events yield a higher $v_2$ value than jet events, although the trends are qualitatively similar. Similarly to $v_2$, the $v_3$ results show agreement between the MBT and jet events for $p_{\mathrm{T}} < 2$ GeV, and higher values from MBT events for $p_{\mathrm{T}} > 2$ GeV.

As mentioned in Section 7.3.2.3, if the measured anisotropy originates from a global momentum field, the $v_2$ and $v_3$ values, extracted for a given $p_{\mathrm{T}}^{\mathrm{A}}$ range, will be independent of B-particle selection. This assumption of factorization is explicitly tested by carrying out the analysis for different selections of $p_{\mathrm{T}}^{\mathrm{B}}$. Figure 8.2 shows the $v_2$ values, from each event trigger, for the nominal results using $p_{\mathrm{T}}^{\mathrm{B}} > 0.4$ GeV overlaid with results using $0.4 < p_{\mathrm{T}}^{\mathrm{B}} < 1$ GeV, $1 < p_{\mathrm{T}}^{\mathrm{B}} < 2$ GeV, and $2 < p_{\mathrm{T}}^{\mathrm{B}} < 3$ GeV. The test shows factorization breaking at the level of 5% for $p_{\mathrm{T}}^{\mathrm{A}} < 5$ GeV in MBT events. However, at higher $p_{\mathrm{T}}^{\mathrm{A}}$, the differences grow with $p_{\mathrm{T}}^{\mathrm{A}}$ to be 10–100% from the nominal values. For jet events, factorization holds within about 10–20% for all values of $p_{\mathrm{T}}^{\mathrm{A}}$, except for $4 < p_{\mathrm{T}}^{\mathrm{A}} < 9$ GeV in $p_{\mathrm{T}}^{\mathrm{jet}} > 100$ GeV events, where it is within about 30–40%. Although the large uncertainties prevent strong conclusions from being drawn, there is a hint of a difference in behavior at high $p_{\mathrm{T}}^{\mathrm{A}}$ where the factorization breaking is greater for MBT events than for jet events. This result could be due to the B-particle jet rejection scheme used for the jet events. Correlations resulting from hard-process, e.g. from back-to-back jets, are expected to specifically violate the factorization assumption, though Pythia studies have shown that it might be possible for hard processes to approximately pass factorization tests [163]. The B-particle jet rejection dramatically limits the contribution from these processes from entering the correlation functions in jet events. However, the correlations from MBT events have no such rejection, and could, therefore, be more susceptible to hard-process correlations at high $p_{\mathrm{T}}^{\mathrm{A}}$.

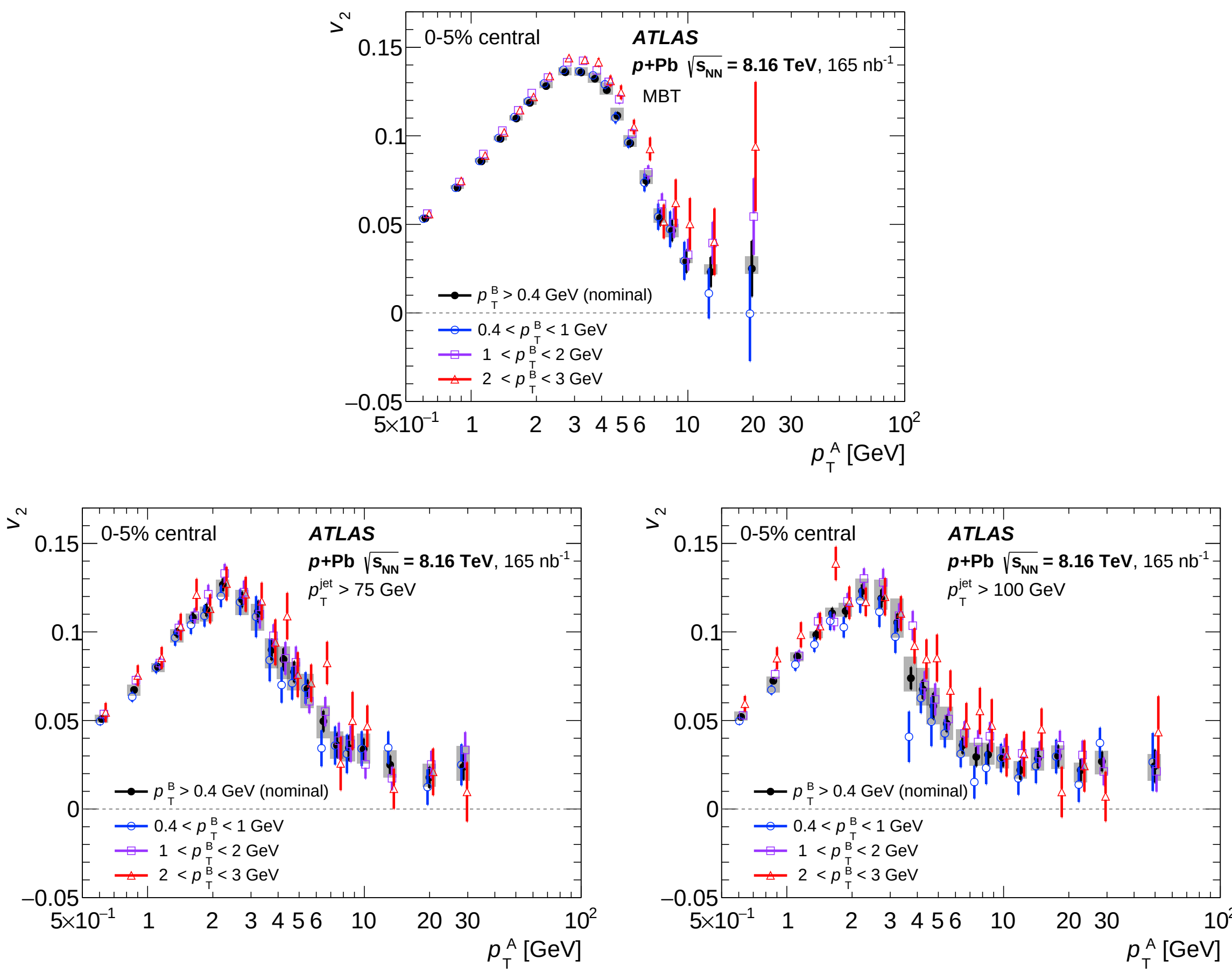

Figure 8.2: Measured $v_2$ values plotted as a function of the A-particle $p_{\mathrm{T}}$ for MBT events (top), events with jet $p_{\mathrm{T}} > 75$ GeV (bottom left), and events with jet $p_{\mathrm{T}} > 100$ GeV (bottom right). The nominal values (closed black circles) are overlaid with points generated by making different B-particle $p_{\mathrm{T}}$ selections: $0.4 < p_{\mathrm{T}}^{\mathrm{B}} < 1$ GeV (blue open circles), $1 < p_{\mathrm{T}}^{\mathrm{B}} < 2$ GeV (open violet squares), and $2 < p_{\mathrm{T}}^{\mathrm{B}} < 3$ GeV (open red triangles). The points with different B-particle $p_{\mathrm{T}}$ selections are offset slightly from the nominal horizontal-axis positions to make the uncertainties visible.

Figure 8.3 shows $v_2$ plotted as a function of centrality for MBT events and both classes of jet events. The results are divided into three regions in A-particle $p_{\mathrm{T}}$: $0.5 < p_{\mathrm{T}} < 2$ GeV, $2 < p_{\mathrm{T}} < 9$ GeV, and $9 < p_{\mathrm{T}} < 100$ GeV. The $v_2$ results show agreement, within uncertainties, between the MBT and jet events for $p_{\mathrm{T}}$ selections $0.5 < p_{\mathrm{T}} < 2$ GeV and $p_{\mathrm{T}} > 9$ GeV for all centralities and are found to be nearly independent of centrality. For $2 < p_{\mathrm{T}} < 9$ GeV, the MBT events give a higher $v_2$ value than the jet events, and all three sets show a trend to lower values of $v_2$ as the collisions become more peripheral.

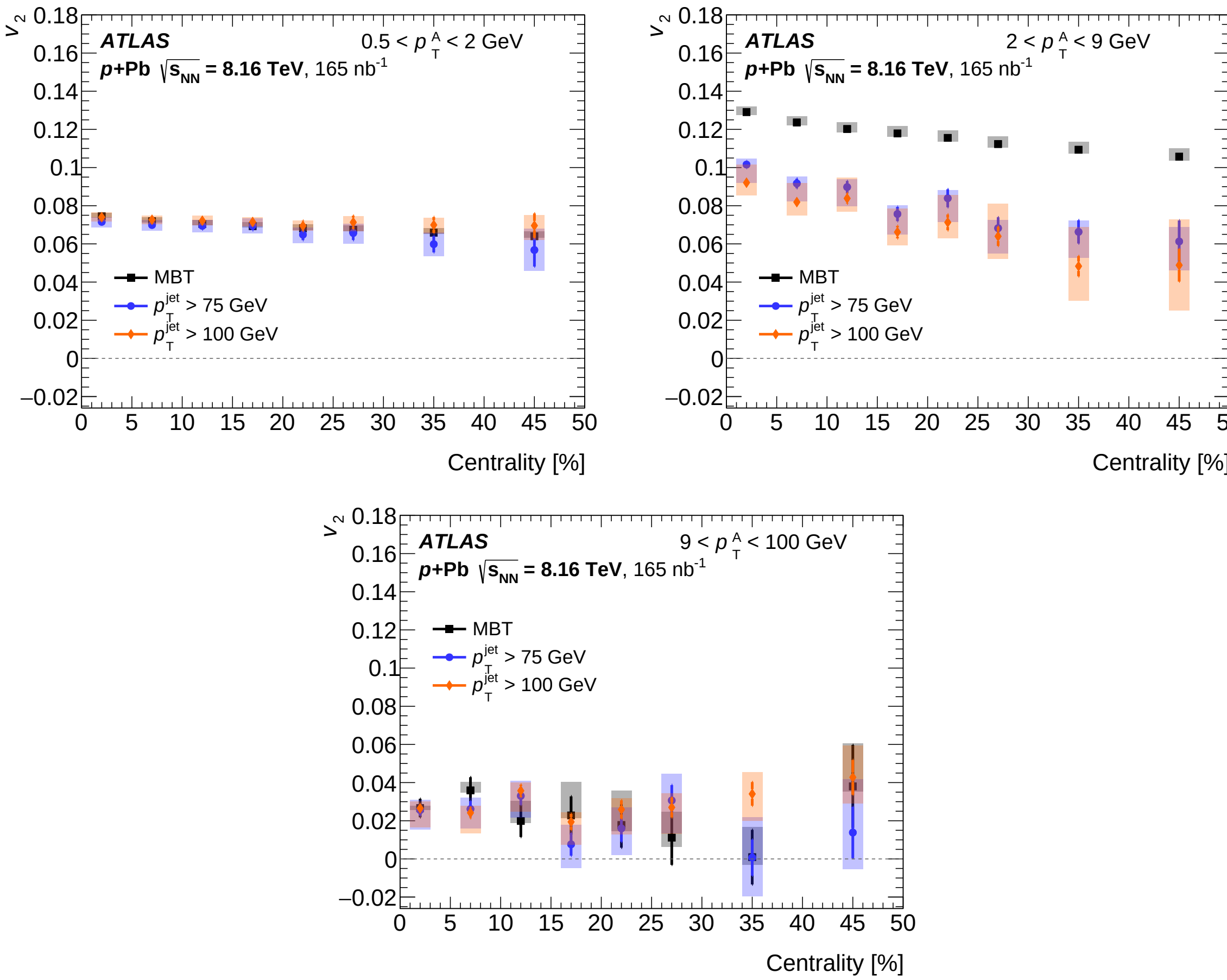

Figure 8.3: Distribution of $v_2$ plotted as a function of centrality for MBT events (black squares), events with jet $p_T > 75$ GeV (blue circles), and events with jet $p_T > 100$ GeV (orange diamonds). The results are obtained in three different selections of the A-particle $p_T$: $0.5 < p_T < 2$ GeV (top left), $2 < p_T < 9$ GeV (top right), and $9 < p_T < 100$ GeV (bottom). Statistical uncertainties are shown as narrow vertical lines on each point, and systematic uncertainties are presented as colored boxes behind the points.

## 8.1 Theory Comparisons

Focusing on the overall $p_T$ dependence of the anisotropies, Figure 8.4 (left panel) shows $v_2$ and $v_3$ coefficients from events with jet $p_T > 100$ GeV compared with theoretical calculations from Ref. [23]. This theoretical calculation, within the jet quenching paradigm, invokes a stronger parton coupling to the QGP near the transition temperature, which helps to reduce the tension in simultaneously matching the nucleus–nucleus high-$p_T$ hadron spectrum suppression and the azimuthal anisotropy $v_2$. The calculation tests two different initial $p$+Pb geometries referred to as 'size a' and 'size b', where the latter has a smaller initial QGP volume. The predictions are slightly lower than the data for both $v_2$ and $v_3$, and the 'size a' curve is within two standard deviations of all points. However, in the right panel of Figure 8.4, the same calculation predicts a substantial suppression of high-$p_\text{T}$ hadrons, as expressed by the quantity $R_{p\text{Pb}} = \text{d}^2N_{p\text{Pb}}/\text{d}p_\text{T}\text{d}y/(T_{p\text{Pb}} \times \text{d}^2\sigma_{pp}/\text{d}p_\text{T}\text{d}y)$ where $T_{p\text{Pb}}$ represents the nuclear thickness of the Pb nucleus, as determined via a Monte Carlo Glauber calculation [7]. Shown in comparison are published experimental results from ATLAS and ALICE for $R_{p\text{Pb}}$ in central events that are consistent with no nuclear suppression, i.e. $R_{p\text{Pb}} = 1$ [164, 165]. The ALICE experiment uses the notation $Q_{p\text{Pb}}$ for the same quantity to describe a bias that may exist due to the centrality categorization. There are uncertainties in the experimental measurements related to the centrality or multiplicity selection in $p$+Pb collisions, particularly in determining the nuclear thickness value $T_{p\text{Pb}}$. However, there is no indication of the large $R_{p\text{Pb}}$ suppression predicted by the jet quenching calculation. Thus, the jet quenching calculation is disfavored as it cannot simultaneously describe the non-zero high-$p_\text{T}$ azimuthal anisotropy and the lack of yield suppression.

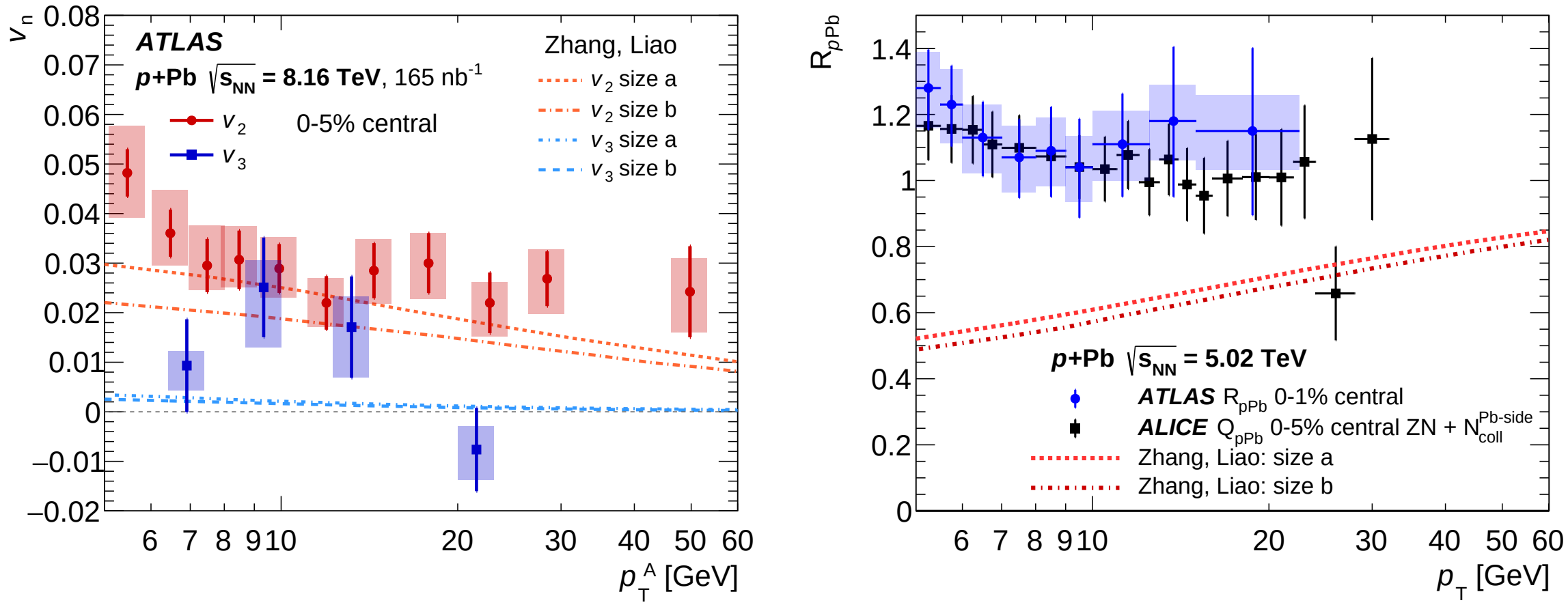

Figure 8.4: Coefficients $v_2$ and $v_3$ (left panel) and $R_{p\text{Pb}}$ (right panel) plotted as a function of particle $p_\text{T}$ for $p$+Pb collisions. The left panel is for central 0–5% events from the jet $p_\text{T} > 100$ GeV event sample. Statistical uncertainties are shown as narrow vertical lines on each point, and systematic uncertainties are presented as colored boxes behind the points. The left panel has two sets of curves showing theoretical predictions from a jet quenching framework with two different initial geometries in 0–4% central collisions [23]; the upper two (red/orange) are $v_2$ for 'size a' (dotted) and 'size b' (dash-dotted) configurations, and the lower two (blue) are $v_3$ where the 'size a' (dash-dotted) and 'size b' (dashed) curves are nearly indistinguishable from each other. The right panel shows $R_{p\text{Pb}}$ data from ATLAS [164] and $Q_{p\text{Pb}}$ data from ALICE [165]. Theoretical calculations (red/orange lines) from Ref. [23] are also shown in this panel; the dotted line gives the results of the 'size a' configuration and the dash-dotted line gives the results of the 'size b' configuration.

Figure 8.5 shows the MBT $v_2$ and $v_3$ coefficients compared with theoretical calculations from Ref. [5]. The calculations are derived from two opposite limits of kinetic theory. The low momentum bands represent zeroth-order hydrodynamic calculations for high-multiplicity $p$+Pb events that give quantitative agreement with $v_2$ up to $p_\mathrm{T} = 2$ GeV while predicting values of $v_3$ that are too high. Above some high $p_\mathrm{T}$ threshold, hadrons are expected to result, not from hydrodynamics, but instead from jets where the resulting partons have the opposite limit than in hydrodynamics, i.e. a large mean free path. To model this region, a non-hydrodynamic 'eremitic' expansion calculation (see Ref. [5] for the detailed calculation), shown as the bands at high $p_\mathrm{T}$, indicates slowly declining $v_2$ and $v_3$ coefficients. The dashed lines are a simple Padé-type fit connecting the two regimes [5]. The trends are qualitatively similar to those in the data, although there is not quantitative agreement. In particular, the calculation predicts values of $v_2$ and $v_3$ substantially below the experimental results for $p_\mathrm{T} = 4$–15 GeV. It should be noted that calculations presented in Ref. [5] are performed, consistently between the hydrodynamic and eremitic components, only for massless partons and with an ideal equation of state. Thus, one does not expect quantitative agreement and is looking for rather qualitative trends. More sophisticated treatments in the hydrodynamic regime result in better quantitative agreement with the anisotropy coefficients at low $p_{\mathrm{T}}$ [71, 166]. It is worth highlighting that traditional parton energy-loss calculations connect the high-$p_{\mathrm{T}}$ $v_2$ with a suppression in the overall yield of high-$p_{\mathrm{T}}$ particles. The same is true with this eremitic calculation, and thus, it should also be in contradistinction to $p$+Pb high $p_{\mathrm{T}}$ experimental data indicating almost no suppression, i.e. jet quenching.

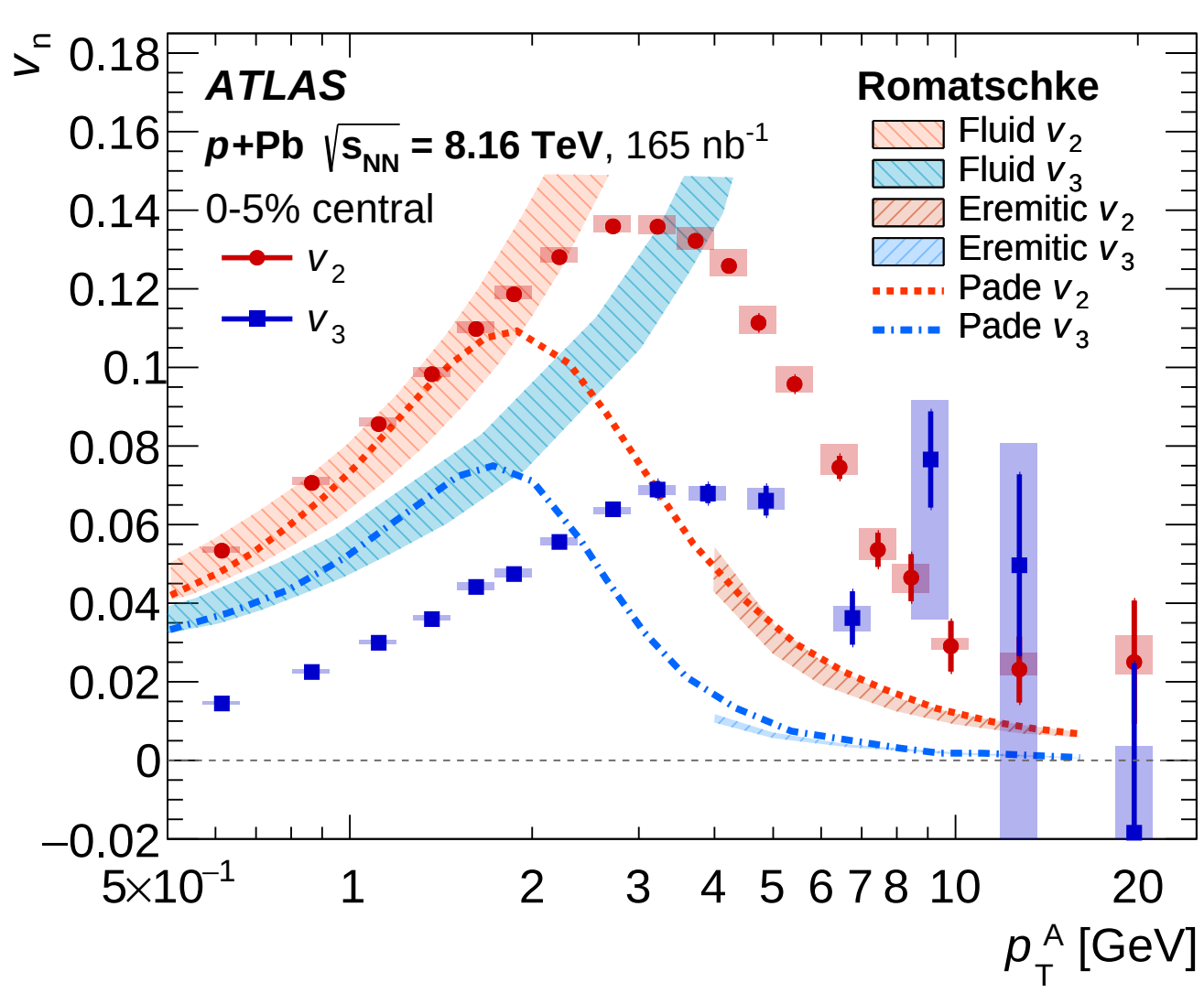

Figure 8.5: Coefficients $v_2$ and $v_3$ plotted as a function of $p_\mathrm{T}$ for central 0–5% $p$+Pb collisions from the MBT event sample. Theoretical calculations relevant to the low-$p_\mathrm{T}$ regime from hydrodynamics and to the high-$p_\mathrm{T}$ regime from an 'eremitic' framework from Romatschke [5] are also shown. The lines are Padé-type fits connecting the two regimes, where the red dotted line is for $v_2$ and the blue dash-dotted line is for $v_3$. Statistical uncertainties are shown as narrow vertical lines on each point, and systematic uncertainties are presented as colored boxes behind the points.

Another possible source of the high-$p_{\mathrm{T}}$ anisotropies could lie in an initial-state effect, potentially encoded in a model such as Pythia. Shown in Figure 8.6 is a Pythia calculation with hard[1] $pp$ events overlaid on minimum-bias $p$+Pb events generated in the default Angantyr framework [167]. It is emphasized that this version of Pythia does not include the recently developed string–string interaction, or so-called string shoving [168]. The generator-level charged particles are then processed with the entire analysis procedure, including the non-flow template fit. The result is a negative $v_{2,2}$ for all momenta, in contradistinction to the experimental data. Further investigation reveals that Pythia run in 'hard' scattering mode has correlations with large pseudorapidity separation between particle pairs as a result of the specific implementation of initial-state radiation. This correlation is reduced in high-multiplicity events because of the large number of uncorrelated UE particles, and thus results in a negative $v_{2,2}$ after subtracting the non-flow contribution.

[1] The term 'hard' refers to Pythia run with the following settings: `HardQCD:all=on`, `PartonLevel:MPI=off`, and containing a jet with $p_{\mathrm{T}} > 100$ GeV.

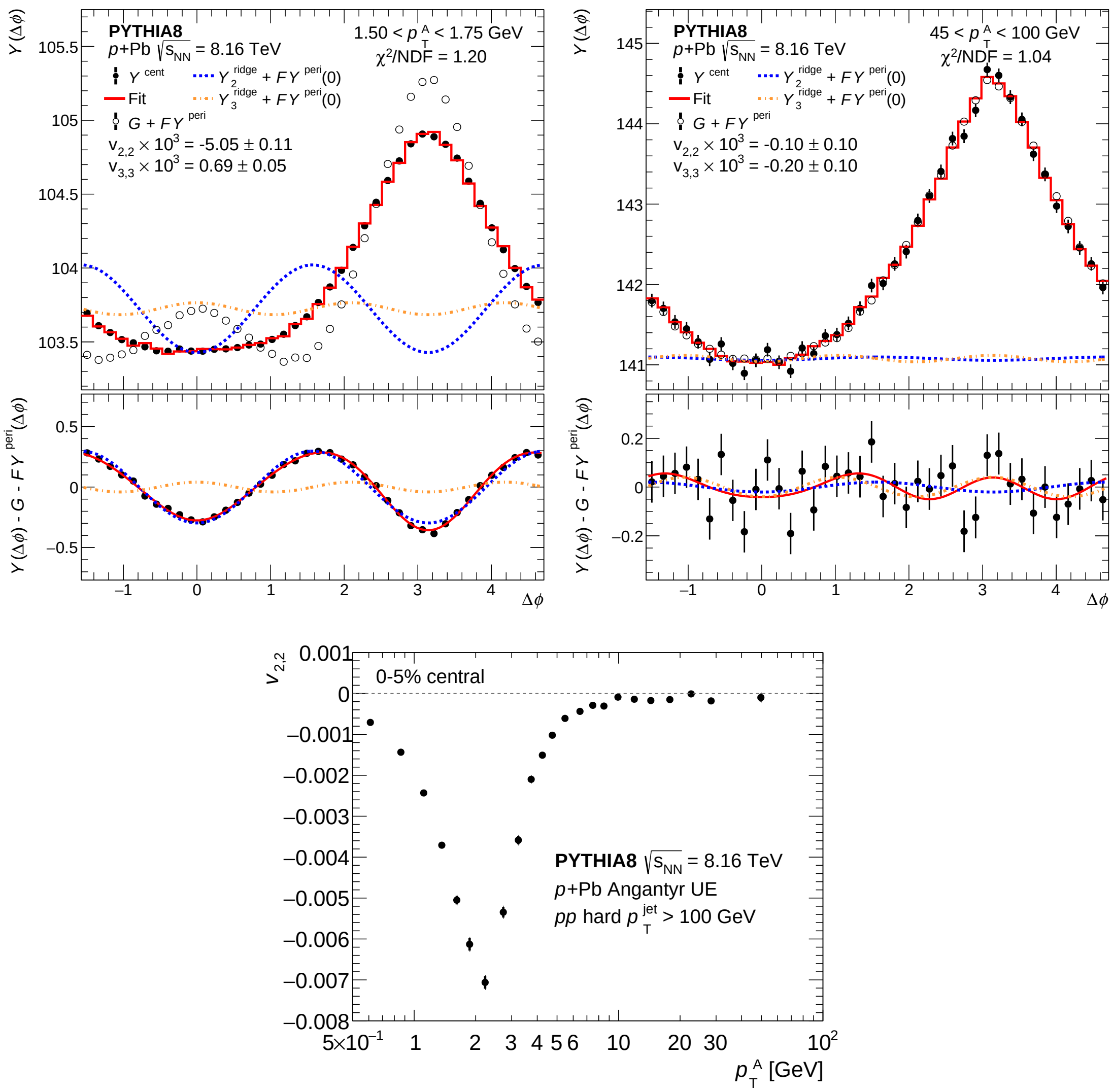

Figure 8.6: Predictions of azimuthal anisotropy from Pythia using the same two-particle formalism used for the data results. The events combine minimum-bias $p$+Pb underlying events generated in the Angantyr framework with hard $pp$ events that require the presence of a jet with $p_{\mathrm{T}} > 100$ GeV. The two top plots show example correlation functions, with template fits, from a low particle-$p_{\mathrm{T}}$ selection (top left) and a high particle-$p_{\mathrm{T}}$ selection (top right). In the upper panels of the two top plots, the open circles show the scaled and shifted peripheral template with uncertainties omitted, the closed circles show the central data, and the red histogram shows the fit (template and harmonic functions). The blue dashed line shows the second-order harmonic component, $Y_2^{\mathrm{ridge}}$, and the orange dashed line shows the third-order harmonic component, $Y_3^{\mathrm{ridge}}$ respectively). The lower panels show the difference between the central data and the peripheral template along with the second and third harmonic functions. The resulting $v_{2,2}$, $v_{3,3}$, and global fit $\chi^2$/NDF values are reported in the legends, where NDF $= 35$. The bottom plot shows the extracted $v_{2,2}$ values as a function of A-particle $p_{\mathrm{T}}$.

## 8.2 Comparison Between $p$+Pb and Pb+Pb Data

Figure 8.7 shows the published Pb+Pb results for $v_2$ as a function of $p_{\mathrm{T}}$ in the 20–30% centrality selection [80] compared to the $v_2$ from both the MBT $p$+Pb data and $p$+Pb containing a jet with $p_{\mathrm{T}} > 100$ GeV. This Pb+Pb centrality range is selected because the spatial elliptic eccentricity is approximately the same as in 0–5% centrality $p$+Pb collisions [169], despite having a much larger total particle multiplicity. The overall trends for Pb+Pb $v_2$ as a function of $p_{\mathrm{T}}$ are qualitatively similar to those presented here for $p$+Pb from MBT events and the jet events with jet $p_{\mathrm{T}} > 100$ GeV. Both sets of the $p$+Pb values are scaled by a single multiplicative factor (1.5) to match the Pb+Pb rise at low $p_{\mathrm{T}}$. The MBT $p$+Pb results quantitatively agree with those from the Pb+Pb system for $0.5 < p_{\mathrm{T}} < 8$ GeV, except for a slight difference in the peak value near $p_{\mathrm{T}} \approx 3$ GeV. For $p_{\mathrm{T}}$ above about 8 GeV, the Pb+Pb results indicate a slow decline of $v_2$ values with increasing $p_{\mathrm{T}}$, while the $p$+Pb results exhibit more of a plateau. Strikingly, the overall behavior of the $v_2$ values are quite similar.

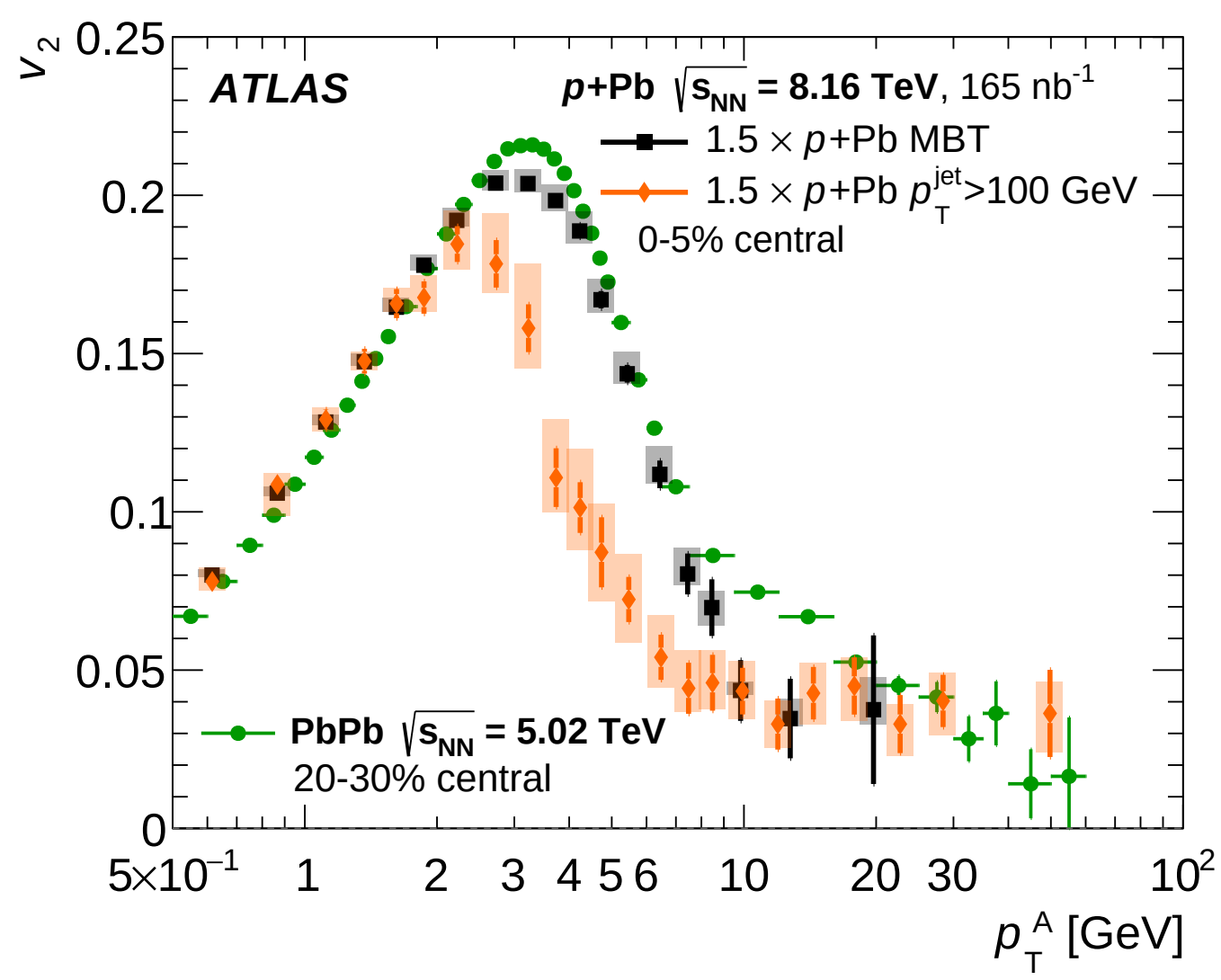

Figure 8.7: Scaled $p$+Pb $v_2$ values plotted as a function of the A-particle $p_{\mathrm{T}}$ overlaid with $v_2$ from 20–30% central Pb+Pb data at $\sqrt{s_{\mathrm{NN}}}$ = 5.02 TeV [80]. Results from MBT and jet $p_{\mathrm{T}} > 100$ GeV $p$+Pb events are plotted as black squares and orange diamonds, respectively, and those from Pb+Pb are plotted as green circles. Statistical uncertainties are shown as narrow vertical lines on each point, and systematic uncertainties are presented as colored boxes behind the points.

As detailed in Sections 2.4.3 and 2.4.4, the physics interpretations of the Pb+Pb elliptic anisotropies are hydrodynamic flow at low $p_{\mathrm{T}}$, differential jet quenching at high $p_{\mathrm{T}}$, and a transition between the two in the intermediate region of approximately $2 < p_{\mathrm{T}} < 10$ GeV. Since these effects all relate to the initial QGP geometric inhomogeneities, a common shape with a single scaling factor for $p$+Pb could indicate a common physics interpretation albeit with a different initial average geometry. This scaling factor of 1.5, as empirically determined, may be the result of slightly different initial spatial deformations, or from the much larger Pb+Pb overall multiplicity, which enables a stronger translation of spatial deformations into momentum space. For the high $p_{\mathrm{T}}$ region, this presents a conundrum in that it is difficult for differential jet quenching to cause the $v_2$ anisotropy in $p$+Pb collisions when there is no evidence for jet quenching overall. These measurements showing non-zero high $p_{\mathrm{T}}$ $v_2$ in $p$+Pb collisions in the absence the jet quenching observed in Pb+Pb collisions suggest there might be additional contributions to $v_2$ at high $p_{\mathrm{T}}$ in Pb+Pb collisions.

## 8.3 Particle Pair Yields

Returning to the issue of the difference in the intermediate $p_{\mathrm{T}}$ region between the $p$+Pb MBT and jet event results, the source of hadrons in this region should be considered. As detailed in Sec. 7.3.5, in a highly simplified picture one can classify hadrons as originating from hard scatterings (HS) or from the underlying event (UE). Thus, pairs of particles of A and B types can come from the combinations HS–HS, HS–UE, UE–HS, and UE–UE. Figure 8.8 presents the measured pair fractions for both MBT and jet, 0–5% central events plotted as a function of the A-particle $p_{\mathrm{T}}$. In each case UE–UE pairs dominate the correlation functions at low $p_{\mathrm{T}}$, and HS–UE combinations dominate at high $p_{\mathrm{T}}$. Combinations with HS B-particles are sub-dominant, because there are fewer jet particles than UE particles in central events; for the jet selected events, these combinations are further suppressed by the B-particle jet rejection condition. Figure 8.9 shows the dominant contributions from the MBT and jet events overlaid. Although the same qualitative behavior is found in each case, the point at which the HS–UE pairs become dominant over the other combinations is at a lower $p_{\mathrm{T}}$ for jet events than for MBT events.

This behavior can also be seen in Figure 8.10, in which the pair fractions are plotted as a function of centrality, and again, the values for MBT and jet events are overlaid. The centrality-dependent results are plotted for low, medium, and high A-particle $p_{\mathrm{T}}$ ranges in the same way as in Figure 8.3. At low $p_{\mathrm{T}}$, pair fractions from MBT and jet events agree, and in the mid-$p_{\mathrm{T}}$ transition region, MBT events have a larger UE–UE contribution and smaller HS–UE contribution compared to jet events. At high $p_{\mathrm{T}}$, central events show a difference between UE–UE and HS–UE that is reduced in more-peripheral events and absent for more peripheral than 25% centrality. The overall trend of the pair fractions with centrality is quite similar to that of $v_2$ shown in Figure 8.3, i.e. little centrality dependence for low and high $p_{\mathrm{T}}$ and significant centrality dependence in addition to MBT–jet event ordering in the mid-$p_{\mathrm{T}}$ transition region.

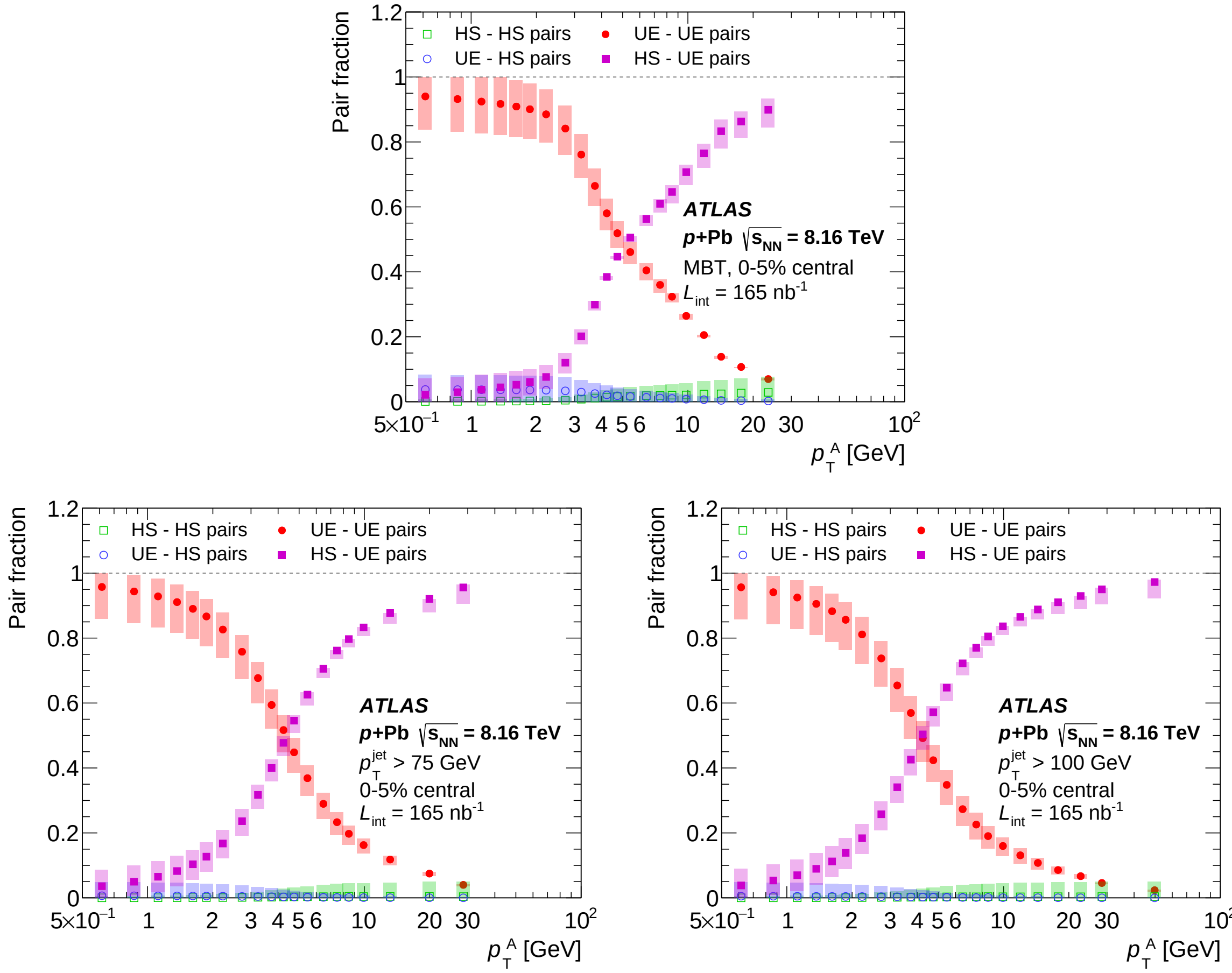

Figure 8.8: Particle pair yield composition fractions for MBT events (top), events with jet $p_{\mathrm{T}} > 75$ GeV (bottom left), and events with jet $p_{\mathrm{T}} > 100$ GeV (bottom right) plotted as a function of the A-particle $p_{\mathrm{T}}$. Green and blue open circles represent HS–HS and UE–HS pairs, respectively, and red and violet closed circles represent UE–UE and HS–UE pairs, respectively. Statistical uncertainties are shown as narrow vertical lines on each point, and systematic uncertainties are presented as colored boxes behind the points.

Thus, a potential explanation for the lower $v_2$ and $v_3$ in the intermediate $p_{\mathrm{T}}$ region is simply that, in that region, the HS particles have lower anisotropy coefficients than UE particles, and MBT events have a larger fraction of UE–UE pairs than jet-triggered events. In the low and high $p_T$ regions, the same types of pairs dominate in both the MBT and jet-triggered events, namely UE–UE and HS–UE respectively, and hence the anisotropy coefficients agree between the event samples. If this explanation is correct, it also aids in understanding Figure 8.7 in which there is a significant difference between the $p$+Pb jet event $v_2$ and the Pb+Pb $v_2$ in the intermediate $p_T$ region, because the relative pair fractions are potentially different.

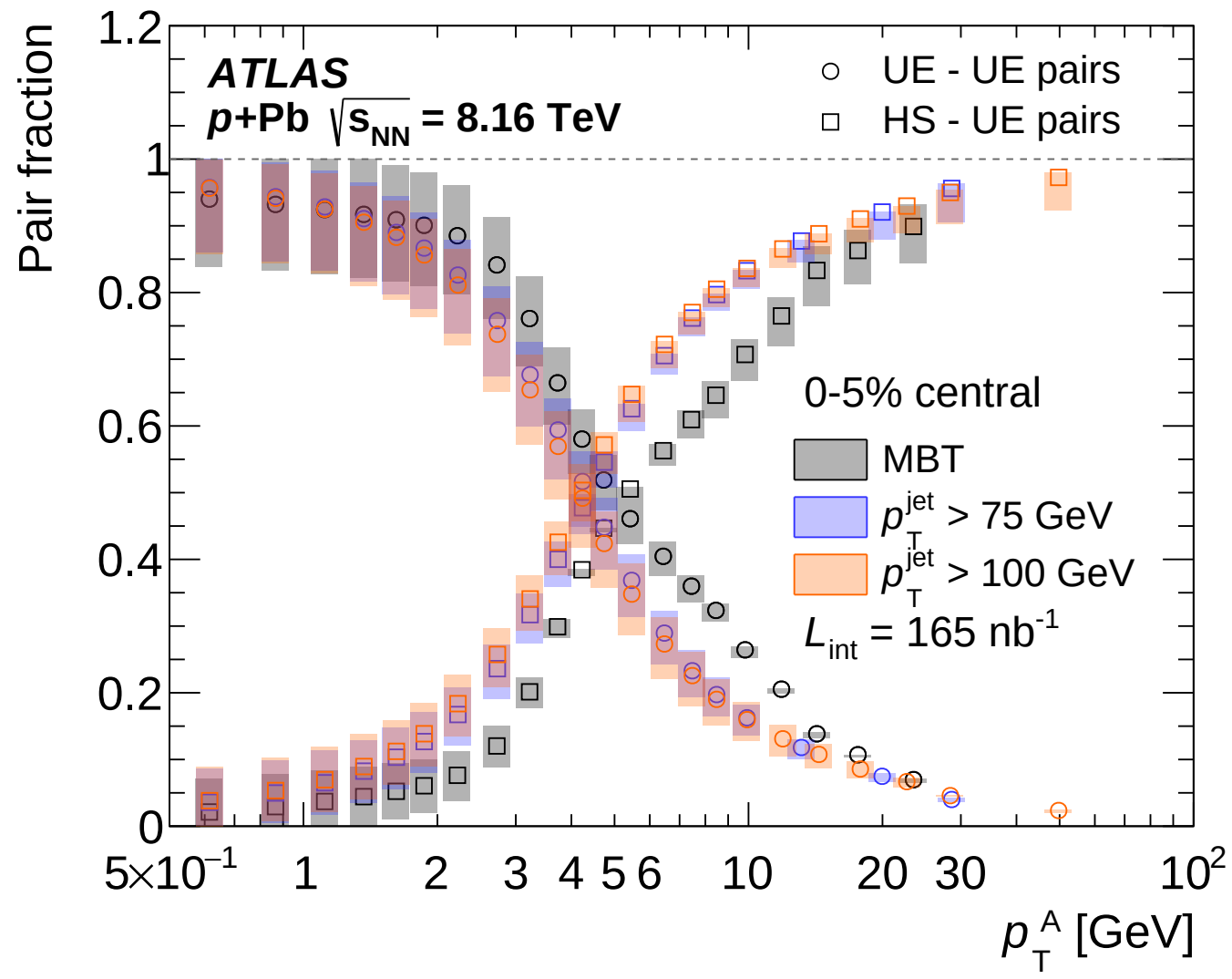

Figure 8.9: Underlying event–underlying event (UE–UE) (open circles) and hard scatter–underlying event (HS–UE) (open squares) particle-pair yield composition fractions for MBT events (black), events with jet $p_T > 75$ GeV (blue), and events with jet $p_T > 100$ GeV (orange) plotted as a function of the A-particle $p_T$. Statistical uncertainties are shown as narrow vertical lines on each point, and systematic uncertainties are presented as colored boxes behind the points.

This particle mixing picture is attractive in that it naturally explains the general shape of the $v_2(p_T)$ and $v_3(p_T)$ distributions as well as the ordering of the different event samples. However, it is noted that the correspondence between the differences in the flow coefficients and pair fractions is not quantitative; the differences in the flow coefficients are fractionally much larger than the differences in the pair fractions. Thus, there are either additional sources of correlation or our assumptions are violated in some way (e.g. the two assumed HS and UE sources are too simplistic or the measured pair fractions do not accurately represent the sources, as is discussed in Sec. 7.3.5). That said, for particle $p_T > 20$ GeV, where particle production in any model is thought to arise mainly from jet fragmentation, the non-zero $v_2$ demonstrates that a positive correlation exists between hard (high $p_{\mathrm{T}}$) and soft (low $p_{\mathrm{T}}$) particles, irrespective of the pair fractions.

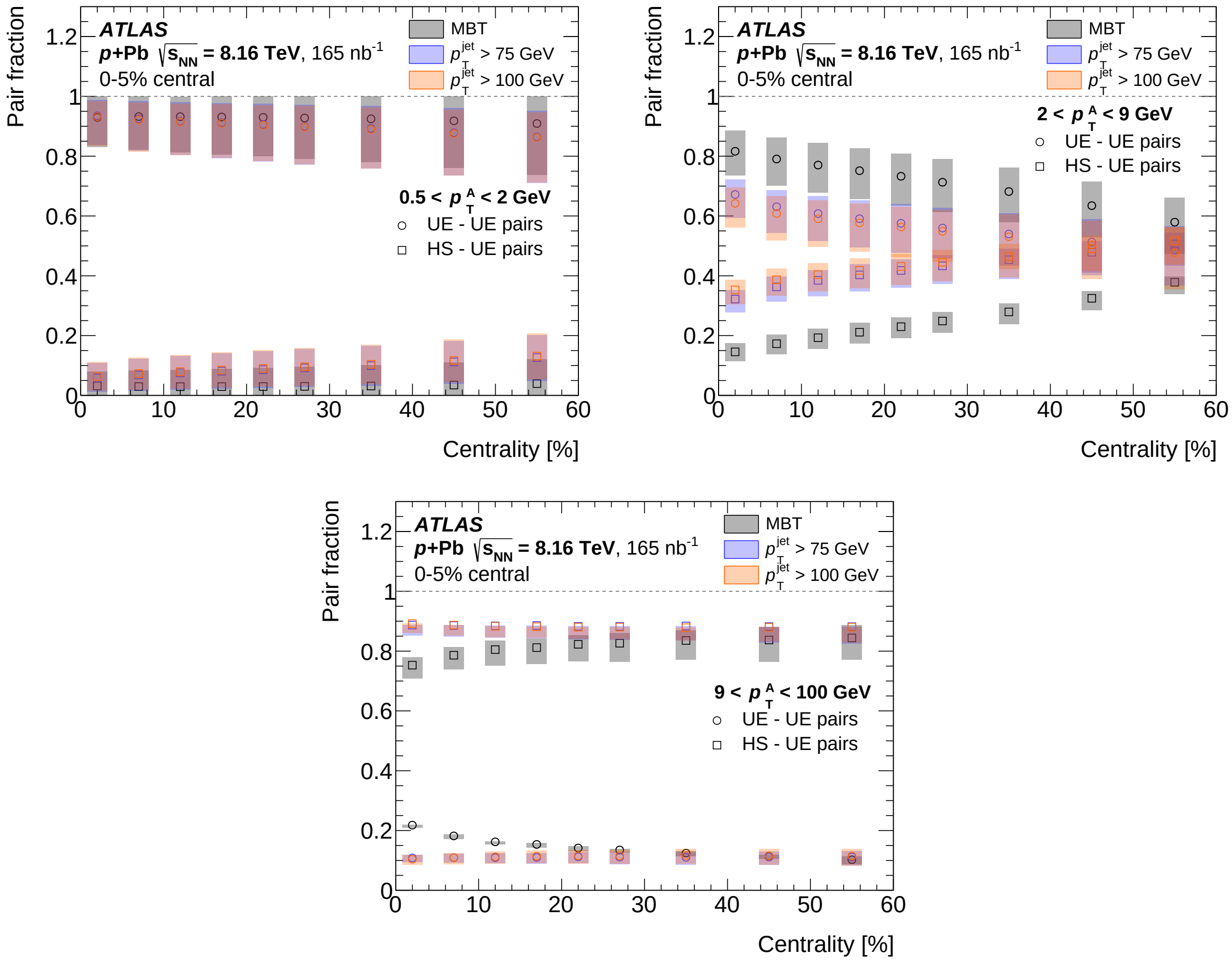

Figure 8.10: Underlying event–underlying event (UE–UE) (open circles) and hard scatter–underlying event (HS–UE) (open squares) particle-pair yield composition fractions for MBT events (black), events with jet $p_{\mathrm{T}} >$ 75 GeV (blue), and events with jet $p_{\mathrm{T}} > 100$ GeV (orange) plotted as a function of event centrality. The results are obtained in three different selections of the A-particle $p_{\mathrm{T}}$: $0.5 < p_{\mathrm{T}} < 2$ GeV (top left), $2 < p_{\mathrm{T}} < 9$ GeV (top right), and $9 < p_{\mathrm{T}} < 100$ GeV (bottom). Statistical uncertainties are shown as narrow vertical lines on each point, and systematic uncertainties are presented as colored boxes behind the points.

# Chapter 9

# Conclusions

This dissertation presented a study of 8.16 TeV $p$+Pb collisions generated at CERN's Large Hadron Collider and measured with the ATLAS detector. This study consisted of two novel measurements of particles over a broad range of energy scales, probing initial- and final-state nuclear phenomena; these results are published as Refs. [29, 30].

A measurement of the cross section and nuclear modification factor, $R_{p\mathrm{Pb}}$, of prompt and isolated photons was detailed in Chapters 5 and 6. These results span $20 < E_{\mathrm{T}}^{\gamma} < 500$ GeV and were presented from three nucleon–nucleon center-of-mass frame rapidity ranges: *forward* ($1.09 < \eta^{*} < 1.90$), *middle* ($-1.84 < \eta^{*} < 0.91$), and *backward* ($-2.83 < \eta^{*} < -2.02$). The cross sections, shown in Fig. 6.1, are compared to next-to-leading order pQCD calculations from JETPHOX using CT14 parton distribution functions [143] with EPPS16 nuclear modifications [58]. The theoretical predictions show a discrepancy of 20% at low $E_{\mathrm{T}}^{\gamma}$ that becomes negligible at high $E_{\mathrm{T}}^{\gamma}$. This discrepancy is of a similar level as has been observed from previous measurements in $pp$ collisions [116, 117].

To isolate the nuclear effects, $R_{p\mathrm{Pb}}$ was constructed in the same kinematic range using a $pp$ reference extrapolated from 8 TeV data [116]. The results, shown in Fig. 6.3 with forward-over-backward ratios in Fig. 6.4, are largely consistent with theoretical predictions using both free nucleon PDFs (including effects from the proton–neutron asymmetry of the Pb nucleus) and nuclear modified PDFs. At mid-rapidity and $E_{\mathrm{T}}^{\gamma} \approx 100$ GeV, corresponding to the shadowing range in nPDFs, the data hint at a preference for the slight modification predicted, but this is insignificant given the uncertainties in the data. The data are also compared to a prediction from a model of initial-state energy loss in which the colliding parton in the proton loses energy via gluon bremsstrahlung as it moves through the nucleus *before* interacting to create the photon [112, 113, 126]. The data show a preference for no energy loss, limiting the allowed amount of initial-state multiple scattering in theoretical models.

These photon results are now part of the world data that can be used in the global fit analyses to create future refinements of nPDFs. A precise understanding of the partonic structure of nucleons, and the initial state in general, is crucial to the ability to distinguish QGP-related final-state effects when studying strongly interacting probes. With this in mind, it would be interesting for future efforts to study prompt photon production differentially in centrality. Not only would this be an especially clean test of Glauber modeling and $T_{\mathrm{AB}}$ scaling, the results would be able to confirm that the centrality dependent modification measured in the jet spectra in Ref. [18] is indeed the result of an initial-state effect.

The second result, presented in Chapters 7 and 8, is a measurement of the azimuthal anisotropy coefficients, $v_2$ and $v_3$, from two-particle correlations of charged hadrons. By selecting events based on the presence of a high-$p_{\mathrm{T}}$ jet, this measurement is able to extend over an unprecedented range in particle $p_{\mathrm{T}}$, for small systems. Fig. 8.1 shows the observation of non-zero elliptic flow reaching $p_{\mathrm{T}} \approx 50$ GeV. In the context of the standard paradigm, in which anisotropy at low $p_{\mathrm{T}}$ is due to a hydrodynamic response to the nuclear geometry and the signal at high $p_{\mathrm{T}}$ is the result of path-length differential jet quenching, this is quite surprising because there is no observed energy loss in $p$+Pb. Furthermore, this does not appear to be a question of $R_{p\mathrm{Pb}}$ measurement sensitivity, since Fig. 8.4 shows that in order to have a level of anisotropy similar to the data, the $R_{p\mathrm{Pb}}$ must be suppressed to a level that is inconsistent with measurements (at least within this model).

This result casts doubt on the common explanation of the $v_n$ signal at high $p_{\mathrm{T}}$ in large systems [22–24]; at least in the sense that differential energy loss might not be the only component. Moreover, new questions are raised about the potential origin of the signal in $p$+Pb. Are final-state effects ruled out? It seems clear that droplets of QGP are being formed in small systems; is it possible that the signal is a result of an interaction with the QGP that does not manifest signs of energy loss? Measurements of jet fragmentation, such as that of Ref. [170], leave little room for the modification of the particle $p_{\mathrm{T}}$ and angular structure within jets. New precision measurements of jet or particle yields relative to photons, which are known to be unmodified, would provide the strongest limits to date on possible final-state effects. These measurements could also help to answer the, perhaps, more general but related question about whether it is possible to create droplets of QGP matter without any final state modification to the strongly interacting high-$p_{\mathrm{T}}$ objects.

On the other hand, the high-$p_{\mathrm{T}}$ anisotropy could be due to a speculative initial-state correlation between high- and low-$x$ partons in the colliding bodies. It is clear from Fig. 8.6 that there are mechanisms included in PYTHIA's implementation of initial-state radiation that produce long range azimuthal correlations; however, these correlations are diluted at higher centrality in disagreement with the data. Alternative approaches based on the CGC gluon saturation model have shown that, in $p$+Pb collisions, it's possible for initial-state momentum correlations to result in significant $v_2$ for heavy-flavor quarks that can only be produced through hard processes [171]. It would be interesting to know if a model similar to this is able to generate correlations of the same type and centrality scaling behavior as the present data.

Lastly, a simple two-component model of particle production was introduced to help explain the observed discrepancy in $v_n$ between MB and jet events in the mid-$p_{\mathrm{T}}$ transition region. In this model, particles originate from either the bulk underlying event or from jet-like hard scatterings. Then, the hypothesis is that the particles in any given range of $p_{\mathrm{T}}$ will have a certain fraction of UE and HS particles. Under the presumption of dijet-like HS particle production, the relative yields of particle pairs from each combination of the classes were estimated, as shown in Figures 8.8, 8.9, and 8.10. These pair fractions show a difference between MB and jet events in the mid-$p_{\mathrm{T}}$ range, qualitatively consistent with the behavior of the $v_n$ values. This raises a question: if the correlations are a linear combination of two components and the relative fractions of the components are known, can one extract the pure UE-UE and HS-UE correlations? In this case, the data are incompatible with this picture. As stated in Sec. 8.3, the differences in the $v_n$ are larger than the differences in the pair fractions, and thus, the pair fractions alone are not able to account for the differences in the flow coefficients. However, this model makes it clear that the high-$p_{\mathrm{T}}$ flow correlations that are reported in Fig. 8.1 are correlations between particles from hard jet fragmentation and the soft underlying event, and they are not correlations between hard particles themselves. As to what mechanism might cause such a relationship, it is hoped that future theoretical and experimental efforts will shed light and that this work will have contributed to a deeper understanding of the universe.

# Appendix A

# Measurement of Direct Photon Spectra

## A.1 MC Data

Table A.1: Overview of Pythia MC simulation samples used in this analysis.

| $E_{\mathrm{T}}$ slice [GeV] | $\sigma_{\mathrm{DP}}$ [nb] | $\epsilon_{\mathrm{DP}}$ | $N_{\mathrm{evt}}^{\mathrm{DP}}$ |
|---|---|---|---|
| $p$+Pb Data Overlay samples, 8.16 TeV($\Delta y = -0.465$ boost) | | | |
| mc15_pPb8TeV.423100.Pythia8EvtGen_A14NNPDF23LO_gammajet_DP17_35* | | | |
| 17-35 | $9.52\times10^{6}$ | $2.87\times10^{-5}$ | $321\times10^{3}$ |
| mc15_pPb8TeV.423101.Pythia8EvtGen_A14NNPDF23LO_gammajet_DP35_50* | | | |
| 35-50 | $7.26\times10^{5}$ | $2.62\times10^{-5}$ | $325\times10^{3}$ |
| mc15_pPb8TeV.423102.Pythia8EvtGen_A14NNPDF23LO_gammajet_DP50_70* | | | |
| 50-70 | $1.88\times10^{5}$ | $2.84\times10^{-5}$ | $328\times10^{3}$ |
| mc15_pPb8TeV.423103.Pythia8EvtGen_A14NNPDF23LO_gammajet_DP70_140* | | | |
| 70-140 | $5.01\times10^{4}$ | $4.21\times10^{-5}$ | $325\times10^{3}$ |
| mc15_pPb8TeV.423104.Pythia8EvtGen_A14NNPDF23LO_gammajet_DP140_280* | | | |
| 140-280 | $2.81\times10^{3}$ | $5.11\times10^{-5}$ | $325\times10^{3}$ |
| mc15_pPb8TeV.423105.Pythia8EvtGen_A14NNPDF23LO_gammajet_DP280_500* | | | |
| 280-500 | $1.24\times10^{2}$ | $5.21\times10^{-5}$ | $325\times10^{3}$ |
| Pb+$p$ Data Overlay samples, 8.16 TeV($\Delta y = +0.465$ boost) | | | |
| mc15_pPb8TeV.423100.Pythia8EvtGen_A14NNPDF23LO_gammajet_DP17_35* | | | |
| 17-35 | $9.52\times10^{6}$ | $2.87\times10^{-5}$ | $650\times10^{3}$ |
| mc15_pPb8TeV.423101.Pythia8EvtGen_A14NNPDF23LO_gammajet_DP35_50* | | | |
| 35-50 | $7.26\times10^{5}$ | $2.62\times10^{-5}$ | $671\times10^{3}$ |
| mc15_pPb8TeV.423102.Pythia8EvtGen_A14NNPDF23LO_gammajet_DP50_70* | | | |
| 50-70 | $1.88\times10^{5}$ | $2.84\times10^{-5}$ | $670\times10^{3}$ |
| mc15_pPb8TeV.423103.Pythia8EvtGen_A14NNPDF23LO_gammajet_DP70_140* | | | |
| 70-140 | $5.01\times10^{4}$ | $4.21\times10^{-5}$ | $549\times10^{3}$ |
| mc15_pPb8TeV.423104.Pythia8EvtGen_A14NNPDF23LO_gammajet_DP140_280* | | | |
| 140-280 | $2.81\times10^{3}$ | $5.11\times10^{-5}$ | $600\times10^{3}$ |
| mc15_pPb8TeV.423105.Pythia8EvtGen_A14NNPDF23LO_gammajet_DP280_500* | | | |
| 280-500 | $1.24\times10^{2}$ | $5.21\times10^{-5}$ | $600\times10^{3}$ |
| Generator-only $pp$ signal samples, 8 TeV($\Delta y = 0$, no boost) | | | |
| mc15_8TeV.423104.Pythia8EvtGen_A14NNPDF23LO_gammajet_DP17_35.evgen* | | | |
| 17-35 | $9.29\times10^{6}$ | $2.87\times10^{-5}$ | $0.2\times10^{6}$ |
| mc15_8TeV.423104.Pythia8EvtGen_A14NNPDF23LO_gammajet_DP35_50.evgen* | | | |
| 35-50 | $7.06\times10^{6}$ | $2.63\times10^{-5}$ | $0.2\times10^{6}$ |
| mc15_8TeV.423104.Pythia8EvtGen_A14NNPDF23LO_gammajet_DP50_70.evgen* | | | |
| 50-70 | $1.82\times10^{5}$ | $2.89\times10^{-5}$ | $0.2\times10^{6}$ |
| mc15_8TeV.423104.Pythia8EvtGen_A14NNPDF23LO_gammajet_DP70_140.evgen* | | | |
| 70-140 | $484\times10^{4}$ | $4.24\times10^{-5}$ | $0.2\times10^{6}$ |
| mc15_8TeV.423104.Pythia8EvtGen_A14NNPDF23LO_gammajet_DP140_280.evgen* | | | |
| 140-280 | $2.70\times10^{3}$ | $5.14\times10^{-5}$ | $0.2\times10^{6}$ |
| mc15_8TeV.423104.Pythia8EvtGen_A14NNPDF23LO_gammajet_DP280_500.evgen* | | | |
| 280-500 | $1.18\times10^{2}$ | $5.20\times10^{-5}$ | $0.2\times10^{6}$ |

Table A.2: Overview of Sherpa MC simulation samples used in this analysis.

| $E_T$ slice [GeV] | $\sigma$ [nb] | $\epsilon$ | $N_{evt}$ |
|---|---|---|---|
| $p$+Pb Data Overlay samples, 8.16 TeV($\Delta y = -0.465$ boost) | | | |
| mc15_pPb8TeV.420154.Sherpa_224_NNPDF30NNLO_SinglePhotonPt15_35_EtaFilter* | | | |
| 15-35 | $9.52\times10^6$ | $2.87\times10^{-5}$ | $312\times10^3$ |
| mc15_pPb8TeV.420155.Sherpa_224_NNPDF30NNLO_SinglePhotonPt35_50_EtaFilter* | | | |
| 35-50 | $7.26\times10^5$ | $2.62\times10^{-5}$ | $319\times10^3$ |
| mc15_pPb8TeV.420156.Sherpa_224_NNPDF30NNLO_SinglePhotonPt50_70_EtaFilter* | | | |
| 50-70 | $1.88\times10^5$ | $2.84\times10^{-5}$ | $320\times10^3$ |
| mc15_pPb8TeV.420157.Sherpa_224_NNPDF30NNLO_SinglePhotonPt70_140_EtaFilter* | | | |
| 70-140 | $5.01\times10^4$ | $4.21\times10^{-5}$ | $320\times10^3$ |
| mc15_pPb8TeV.420158.Sherpa_224_NNPDF30NNLO_SinglePhotonPt140_280_EtaFilter* | | | |
| 140-280 | $2.81\times10^3$ | $5.11\times10^{-5}$ | $316\times10^3$ |
| mc15_pPb8TeV.420159.Sherpa_224_NNPDF30NNLO_SinglePhotonPt280_500_EtaFilter* | | | |
| 280-500 | $1.24\times10^2$ | $5.21\times10^{-5}$ | $161\times10^3$ |
| Pb+$p$ Data Overlay samples, 8.16 TeV($\Delta y = +0.465$ boost) | | | |
| mc15_pPb8TeV.420154.Sherpa_224_NNPDF30NNLO_SinglePhotonPt15_35_EtaFilter* | | | |
| 15-35 | $9.52\times10^6$ | $2.87\times10^{-5}$ | $645\times10^3$ |
| mc15_pPb8TeV.420155.Sherpa_224_NNPDF30NNLO_SinglePhotonPt35_50_EtaFilter* | | | |
| 35-50 | $7.26\times10^5$ | $2.62\times10^{-5}$ | $657\times10^3$ |
| mc15_pPb8TeV.420156.Sherpa_224_NNPDF30NNLO_SinglePhotonPt50_70_EtaFilter* | | | |
| 50-70 | $1.88\times10^5$ | $2.84\times10^{-5}$ | $649\times10^3$ |
| mc15_pPb8TeV.420157.Sherpa_224_NNPDF30NNLO_SinglePhotonPt70_140_EtaFilter* | | | |
| 70-140 | $5.01\times10^4$ | $4.21\times10^{-5}$ | $657\times10^3$ |
| mc15_pPb8TeV.420158.Sherpa_224_NNPDF30NNLO_SinglePhotonPt140_280_EtaFilter* | | | |
| 140-280 | $2.81\times10^3$ | $5.11\times10^{-5}$ | $635\times10^3$ |
| mc15_pPb8TeV.420159.Sherpa_224_NNPDF30NNLO_SinglePhotonPt280_500_EtaFilter* | | | |
| 280-500 | $1.24\times10^2$ | $5.21\times10^{-5}$ | $329\times10^3$ |

## A.2 Cross section calculation data

Table A.3: Table of components in the $N_A^{sig}$ calculation in center of mass rapidity range $(1.90 < \eta* < 1.09)$. Yields are each quoted after prescale correction

| $E_T$ low | $E_T$ high | $N_A$ | $N_B$ | $N_C$ | $N_D$ | $f_B$ | $f_C$ | $f_D$ | $N_A^{sig}$ | Error (%) |
|---|---|---|---|---|---|---|---|---|---|---|
| 20 GeV | 25 GeV | 672821 | 124311 | 466354 | 153628 | 0.0018 | 0.0808 | 0.0003 | 317642 | 6.849 |
| 25 GeV | 35 GeV | 395205 | 109405 | 206615 | 123121 | 0.0029 | 0.0555 | 0.0002 | 223596 | 3.169 |
| 35 GeV | 45 GeV | 97645 | 37004 | 35378 | 37037 | 0.0027 | 0.0434 | 0.0002 | 65274 | 1.060 |
| 45 GeV | 55 GeV | 33865 | 15524 | 9627 | 16394 | 0.0030 | 0.0356 | 0.0002 | 25653 | 0.997 |
| 55 GeV | 65 GeV | 14147 | 7448 | 3284 | 8627 | 0.0029 | 0.0317 | 0.0002 | 11641 | 1.299 |
| 65 GeV | 75 GeV | 6629 | 3913 | 1419 | 4936 | 0.0040 | 0.0305 | 0.0002 | 5646 | 1.775 |
| 75 GeV | 85 GeV | 3471 | 2214 | 664 | 3112 | 0.0046 | 0.0289 | 0.0002 | 3064 | 2.294 |
| 85 GeV | 105 GeV | 3138 | 2178 | 510 | 3146 | 0.0050 | 0.0277 | 0.0003 | 2841 | 2.308 |
| 105 GeV | 125 GeV | 1222 | 936 | 193 | 1537 | 0.0050 | 0.0252 | 0.0003 | 1122 | 3.595 |
| 125 GeV | 150 GeV | 623 | 484 | 78 | 947 | 0.0073 | 0.0271 | 0.0008 | 591 | 4.758 |
| 150 GeV | 175 GeV | 278 | 217 | 34 | 458 | 0.0067 | 0.0262 | 0.0004 | 265 | 7.063 |
| 175 GeV | 200 GeV | 148 | 115 | 7 | 228 | 0.0077 | 0.0275 | 0.0006 | 146 | 9.213 |
| 200 GeV | 250 GeV | 94 | 95 | 5 | 211 | 0.0083 | 0.0269 | 0.0005 | 93 | 11.542 |
| 250 GeV | 350 GeV | 49 | 41 | 1 | 95 | 0.0092 | 0.0267 | 0.0009 | 49 | 15.643 |
| 350 GeV | 550 GeV | 7 | 2 | 3 | 19 | 0.0141 | 0.0309 | 0.0012 | 6 | 44.508 |

Table A.4: Table of components in the $N_A^{sig}$ calculation in center of mass rapidity range $(-1.84 < \eta* < 0.91)$. Yields are each quoted after prescale correction

| $E_T$ low | $E_T$ high | $N_A$ | $N_B$ | $N_C$ | $N_D$ | $f_B$ | $f_C$ | $f_D$ | $N_A^{sig}$ | Error (%) |
|---|---|---|---|---|---|---|---|---|---|---|
| 20 GeV | 25 GeV | 2417784 | 456241 | 1687869 | 549793 | 0.0026 | 0.0740 | 0.0002 | 1091930 | 3.363 |
| 25 GeV | 35 GeV | 1382959 | 331933 | 796584 | 403765 | 0.0028 | 0.0464 | 0.0002 | 760835 | 1.581 |
| 35 GeV | 45 GeV | 346858 | 110494 | 135962 | 125226 | 0.0029 | 0.0291 | 0.0001 | 233566 | 0.530 |
| 45 GeV | 55 GeV | 120035 | 46681 | 34341 | 49597 | 0.0031 | 0.0232 | 0.0001 | 89853 | 0.496 |
| 55 GeV | 65 GeV | 49222 | 22303 | 11437 | 23128 | 0.0034 | 0.0193 | 0.0001 | 38981 | 0.734 |
| 65 GeV | 75 GeV | 23190 | 11638 | 4694 | 12382 | 0.0033 | 0.0180 | 0.0001 | 19123 | 0.923 |
| 75 GeV | 85 GeV | 12160 | 6336 | 2044 | 7450 | 0.0038 | 0.0168 | 0.0001 | 10583 | 1.149 |
| 85 GeV | 105 GeV | 10890 | 6222 | 1695 | 7675 | 0.0037 | 0.0160 | 0.0001 | 9648 | 1.173 |
| 105 GeV | 125 GeV | 4192 | 2590 | 572 | 3424 | 0.0041 | 0.0159 | 0.0001 | 3807 | 1.808 |
| 125 GeV | 150 GeV | 2159 | 1458 | 247 | 2184 | 0.0044 | 0.0161 | 0.0001 | 2017 | 2.399 |
| 150 GeV | 175 GeV | 936 | 631 | 83 | 1072 | 0.0045 | 0.0156 | 0.0001 | 896 | 3.508 |
| 175 GeV | 200 GeV | 438 | 312 | 29 | 559 | 0.0055 | 0.0156 | 0.0001 | 426 | 5.013 |
| 200 GeV | 250 GeV | 362 | 270 | 18 | 519 | 0.0055 | 0.0148 | 0.0002 | 355 | 5.435 |
| 250 GeV | 300 GeV | 116 | 109 | 5 | 202 | 0.0063 | 0.0154 | 0.0004 | 114 | 9.443 |
| 300 GeV | 350 GeV | 46 | 45 | 2 | 78 | 0.0062 | 0.0153 | 0.0003 | 45 | 15.146 |
| 350 GeV | 400 GeV | 21 | 24 | 1 | 42 | 0.0069 | 0.0147 | 0.0003 | 21 | 22.602 |
| 400 GeV | 550 GeV | 13 | 21 | 1 | 23 | 0.0078 | 0.0167 | 0.0002 | 13 | 28.655 |

Table A.5: Table of components in the $N_A^{sig}$ calculation in center of mass rapidity range $(-2.83 < \eta* < -2.02)$. Yields are each quoted after prescale correction

| $E_{\mathrm{T}}$ low | $E_{\mathrm{T}}$ high | $N_A$ | $N_B$ | $N_C$ | $N_D$ | $f_B$ | $f_C$ | $f_D$ | $N_A^{sig}$ | Error (%) |
|---|---|---|---|---|---|---|---|---|---|---|
| 20 GeV | 25 GeV | 590254 | 129265 | 393165 | 140692 | 0.0043 | 0.0796 | 0.0003 | 249931 | 7.232 |
| 25 GeV | 35 GeV | 343385 | 97396 | 175942 | 108025 | 0.0044 | 0.0594 | 0.0002 | 196530 | 2.997 |
| 35 GeV | 45 GeV | 85099 | 32817 | 30019 | 34787 | 0.0050 | 0.0451 | 0.0004 | 59531 | 1.068 |
| 45 GeV | 55 GeV | 28773 | 14329 | 7915 | 14802 | 0.0063 | 0.0362 | 0.0002 | 21945 | 1.151 |
| 55 GeV | 65 GeV | 11520 | 6673 | 2776 | 7482 | 0.0056 | 0.0318 | 0.0004 | 9325 | 1.418 |
| 65 GeV | 75 GeV | 5164 | 3378 | 1141 | 4107 | 0.0065 | 0.0320 | 0.0004 | 4346 | 1.962 |
| 75 GeV | 85 GeV | 2521 | 1917 | 511 | 2520 | 0.0068 | 0.0305 | 0.0003 | 2186 | 2.637 |
| 85 GeV | 105 GeV | 2143 | 1740 | 378 | 2625 | 0.0066 | 0.0277 | 0.0004 | 1930 | 2.673 |
| 105 GeV | 125 GeV | 799 | 654 | 114 | 1116 | 0.0078 | 0.0295 | 0.0003 | 746 | 4.160 |
| 125 GeV | 150 GeV | 331 | 331 | 46 | 581 | 0.0086 | 0.0229 | 0.0009 | 309 | 6.405 |
| 150 GeV | 175 GeV | 118 | 128 | 29 | 269 | 0.0081 | 0.0245 | 0.0005 | 105 | 11.186 |
| 175 GeV | 200 GeV | 57 | 42 | 2 | 103 | 0.0088 | 0.0250 | 0.0004 | 57 | 14.639 |
| 200 GeV | 250 GeV | 28 | 43 | 1 | 71 | 0.0095 | 0.0221 | 0.0010 | 28 | 20.170 |
| 250 GeV | 350 GeV | 10 | 6 | 1 | 21 | 0.0087 | 0.0226 | 0.0010 | 10 | 33.534 |
| 350 GeV | 550 GeV | 0 | 1 | 1 | 1 | 0.0168 | 0.0247 | 0.0009 | 0 | 0 |

## A.3 Pythia- Efficiency data

Table A.6: Table of efficiency values in period A in center of mass rapidity range $(1.90 < \eta* < 1.09)$.

| $E_T$ low | $E_T$ high | $\epsilon_{Reco}$ | $\epsilon_{Reco \times ID}$ | $\epsilon_{Reco \times ID \times Iso}$ | Bin migration |
|---|---|---|---|---|---|
| 20 GeV | 25 GeV | 0.9647 | 0.7931 | 0.7845 | 1.0124 |
| 25 GeV | 35 GeV | 0.9721 | 0.8644 | 0.8542 | 1.0106 |
| 35 GeV | 45 GeV | 0.9730 | 0.8969 | 0.8848 | 1.0049 |
| 45 GeV | 55 GeV | 0.9743 | 0.9216 | 0.9073 | 0.9959 |
| 55 GeV | 65 GeV | 0.9749 | 0.9235 | 0.9093 | 0.9946 |
| 65 GeV | 75 GeV | 0.9771 | 0.9257 | 0.9090 | 0.9911 |
| 75 GeV | 85 GeV | 0.9760 | 0.9278 | 0.9093 | 0.9995 |
| 85 GeV | 105 GeV | 0.9743 | 0.9301 | 0.9121 | 0.9960 |
| 105 GeV | 125 GeV | 0.9706 | 0.9276 | 0.9097 | 1.0023 |
| 125 GeV | 150 GeV | 0.9800 | 0.9306 | 0.9137 | 0.9846 |
| 150 GeV | 175 GeV | 0.9765 | 0.9278 | 0.9035 | 0.9989 |
| 175 GeV | 200 GeV | 0.9768 | 0.9214 | 0.8954 | 1.0071 |
| 200 GeV | 250 GeV | 0.9792 | 0.9263 | 0.8984 | 0.9984 |
| 250 GeV | 350 GeV | 0.9771 | 0.9189 | 0.8874 | 0.9951 |
| 350 GeV | 550 GeV | 0.9812 | 0.9138 | 0.8758 | 1.0002 |

Table A.7: Table of efficiency values in period B in center of mass rapidity range $(1.90 < \eta* < 1.09)$.

| $E_T$ low | $E_T$ high | $\epsilon_{Reco}$ | $\epsilon_{Reco \times ID}$ | $\epsilon_{Reco \times ID \times Iso}$ | Bin migration |
|---|---|---|---|---|---|
| 20 GeV | 25 GeV | 0.9628 | 0.7847 | 0.7756 | 1.0124 |
| 25 GeV | 35 GeV | 0.9694 | 0.8552 | 0.8430 | 1.0106 |
| 35 GeV | 45 GeV | 0.9704 | 0.8906 | 0.8773 | 1.0049 |
| 45 GeV | 55 GeV | 0.9692 | 0.9129 | 0.8980 | 0.9959 |
| 55 GeV | 65 GeV | 0.9709 | 0.9117 | 0.8971 | 0.9946 |
| 65 GeV | 75 GeV | 0.9707 | 0.9113 | 0.8956 | 0.9911 |
| 75 GeV | 85 GeV | 0.9720 | 0.9198 | 0.9024 | 0.9995 |
| 85 GeV | 105 GeV | 0.9716 | 0.9230 | 0.9031 | 0.9960 |
| 105 GeV | 125 GeV | 0.9702 | 0.9214 | 0.9009 | 1.0023 |
| 125 GeV | 150 GeV | 0.9725 | 0.9241 | 0.8999 | 0.9846 |
| 150 GeV | 175 GeV | 0.9726 | 0.9201 | 0.8959 | 0.9989 |
| 175 GeV | 200 GeV | 0.9701 | 0.9159 | 0.8902 | 1.0071 |
| 200 GeV | 250 GeV | 0.9727 | 0.9157 | 0.8868 | 0.9984 |
| 250 GeV | 350 GeV | 0.9763 | 0.9124 | 0.8816 | 0.9951 |
| 350 GeV | 550 GeV | 0.9752 | 0.9068 | 0.8684 | 1.0002 |

Table A.8: Table of efficiency values in from both running periods as the luminosity weighted average in center of mass rapidity range $(1.90 < \eta* < 1.09)$.

| $E_T$ low | $E_T$ high | $\epsilon_{Reco}$ | $\epsilon_{Reco \times ID}$ | $\epsilon_{Reco \times ID \times Iso}$ | Bin migration |
|---|---|---|---|---|---|
| 20 GeV | 25 GeV | 0.9635 | 0.7876 | 0.7787 | 1.0150 |
| 25 GeV | 35 GeV | 0.9703 | 0.8584 | 0.8469 | 1.0034 |
| 35 GeV | 45 GeV | 0.9713 | 0.8928 | 0.8799 | 1.0077 |
| 45 GeV | 55 GeV | 0.9710 | 0.9159 | 0.9012 | 0.9899 |
| 55 GeV | 65 GeV | 0.9723 | 0.9158 | 0.9013 | 0.9910 |
| 65 GeV | 75 GeV | 0.9729 | 0.9162 | 0.9002 | 0.9976 |
| 75 GeV | 85 GeV | 0.9733 | 0.9226 | 0.9048 | 0.9988 |
| 85 GeV | 105 GeV | 0.9726 | 0.9255 | 0.9062 | 0.9917 |
| 105 GeV | 125 GeV | 0.9703 | 0.9235 | 0.9039 | 0.9932 |
| 125 GeV | 150 GeV | 0.9751 | 0.9263 | 0.9047 | 0.9899 |
| 150 GeV | 175 GeV | 0.9739 | 0.9228 | 0.8985 | 0.9958 |
| 175 GeV | 200 GeV | 0.9724 | 0.9178 | 0.8920 | 0.9938 |
| 200 GeV | 250 GeV | 0.9750 | 0.9193 | 0.8908 | 0.9944 |
| 250 GeV | 350 GeV | 0.9766 | 0.9146 | 0.8836 | 0.9846 |
| 350 GeV | 550 GeV | 0.9773 | 0.9092 | 0.8710 | 0.9888 |

Table A.9: Table of efficiency values in period A in center of mass rapidity range $(0.91 < \eta* < -1.84)$.

| $E_T$ low | $E_T$ high | $\epsilon_{Reco}$ | $\epsilon_{Reco \times ID}$ | $\epsilon_{Reco \times ID \times Iso}$ | Bin migration |
|---|---|---|---|---|---|
| 20 GeV | 25 GeV | 0.9812 | 0.8142 | 0.8039 | 1.0168 |
| 25 GeV | 35 GeV | 0.9858 | 0.8766 | 0.8660 | 1.0063 |
| 35 GeV | 45 GeV | 0.9884 | 0.9193 | 0.9078 | 1.0132 |
| 45 GeV | 55 GeV | 0.9894 | 0.9311 | 0.9202 | 0.9980 |
| 55 GeV | 65 GeV | 0.9903 | 0.9406 | 0.9280 | 1.0035 |
| 65 GeV | 75 GeV | 0.9902 | 0.9424 | 0.9303 | 0.9994 |
| 75 GeV | 85 GeV | 0.9897 | 0.9436 | 0.9295 | 1.0000 |
| 85 GeV | 105 GeV | 0.9901 | 0.9430 | 0.9300 | 0.9957 |
| 105 GeV | 125 GeV | 0.9896 | 0.9426 | 0.9281 | 1.0041 |
| 125 GeV | 150 GeV | 0.9895 | 0.9415 | 0.9274 | 0.9995 |
| 150 GeV | 175 GeV | 0.9909 | 0.9425 | 0.9288 | 0.9974 |
| 175 GeV | 200 GeV | 0.9910 | 0.9421 | 0.9261 | 0.9922 |
| 200 GeV | 250 GeV | 0.9911 | 0.9397 | 0.9241 | 0.9995 |
| 250 GeV | 300 GeV | 0.9915 | 0.9360 | 0.9187 | 0.9900 |
| 300 GeV | 350 GeV | 0.9928 | 0.9401 | 0.9244 | 0.9937 |
| 350 GeV | 400 GeV | 0.9926 | 0.9403 | 0.9222 | 0.9991 |
| 400 GeV | 550 GeV | 0.9926 | 0.9366 | 0.9179 | 0.9935 |

Table A.10: Table of efficiency values in period B in center of mass rapidity range $(0.91 < \eta* < -1.84)$.

| $E_{\mathrm{T}}$ low | $E_{\mathrm{T}}$ high | $\epsilon_{Reco}$ | $\epsilon_{Reco \times ID}$ | $\epsilon_{Reco \times ID \times Iso}$ | Bin migration |
|---|---|---|---|---|---|
| 20 GeV | 25 GeV | 0.9803 | 0.8045 | 0.7928 | 1.0168 |
| 25 GeV | 35 GeV | 0.9843 | 0.8690 | 0.8546 | 1.0063 |
| 35 GeV | 45 GeV | 0.9875 | 0.9164 | 0.9018 | 1.0132 |
| 45 GeV | 55 GeV | 0.9885 | 0.9311 | 0.9161 | 0.9980 |
| 55 GeV | 65 GeV | 0.9896 | 0.9382 | 0.9219 | 1.0035 |
| 65 GeV | 75 GeV | 0.9896 | 0.9405 | 0.9246 | 0.9994 |
| 75 GeV | 85 GeV | 0.9899 | 0.9420 | 0.9257 | 1.0000 |
| 85 GeV | 105 GeV | 0.9898 | 0.9434 | 0.9271 | 0.9957 |
| 105 GeV | 125 GeV | 0.9898 | 0.9433 | 0.9269 | 1.0041 |
| 125 GeV | 150 GeV | 0.9905 | 0.9418 | 0.9251 | 0.9995 |
| 150 GeV | 175 GeV | 0.9907 | 0.9427 | 0.9256 | 0.9974 |
| 175 GeV | 200 GeV | 0.9905 | 0.9387 | 0.9191 | 0.9922 |
| 200 GeV | 250 GeV | 0.9908 | 0.9407 | 0.9233 | 0.9995 |
| 250 GeV | 300 GeV | 0.9921 | 0.9401 | 0.9222 | 0.9900 |
| 300 GeV | 350 GeV | 0.9924 | 0.9380 | 0.9187 | 0.9937 |
| 350 GeV | 400 GeV | 0.9921 | 0.9375 | 0.9183 | 0.9991 |
| 400 GeV | 550 GeV | 0.9923 | 0.9339 | 0.9127 | 0.9935 |

Table A.11: Table of efficiency values in from both running periods as the luminosity weighted average in center of mass rapidity range $(0.91 < \eta* < -1.84)$.

| $E_{\mathrm{T}}$ low | $E_{\mathrm{T}}$ high | $\epsilon_{Reco}$ | $\epsilon_{Reco \times ID}$ | $\epsilon_{Reco \times ID \times Iso}$ | Bin migration |
|---|---|---|---|---|---|
| 20 GeV | 25 GeV | 0.9806 | 0.8079 | 0.7966 | 1.0203 |
| 25 GeV | 35 GeV | 0.9848 | 0.8717 | 0.8586 | 1.0092 |
| 35 GeV | 45 GeV | 0.9878 | 0.9174 | 0.9038 | 1.0112 |
| 45 GeV | 55 GeV | 0.9888 | 0.9311 | 0.9175 | 1.0032 |
| 55 GeV | 65 GeV | 0.9899 | 0.9390 | 0.9240 | 1.0047 |
| 65 GeV | 75 GeV | 0.9898 | 0.9411 | 0.9266 | 0.9984 |
| 75 GeV | 85 GeV | 0.9898 | 0.9426 | 0.9270 | 0.9989 |
| 85 GeV | 105 GeV | 0.9899 | 0.9433 | 0.9281 | 0.9982 |
| 105 GeV | 125 GeV | 0.9897 | 0.9431 | 0.9273 | 1.0015 |
| 125 GeV | 150 GeV | 0.9902 | 0.9417 | 0.9259 | 1.0007 |
| 150 GeV | 175 GeV | 0.9908 | 0.9427 | 0.9267 | 0.9969 |
| 175 GeV | 200 GeV | 0.9907 | 0.9399 | 0.9215 | 0.9984 |
| 200 GeV | 250 GeV | 0.9909 | 0.9404 | 0.9236 | 0.9984 |
| 250 GeV | 300 GeV | 0.9919 | 0.9387 | 0.9210 | 0.9909 |
| 300 GeV | 350 GeV | 0.9926 | 0.9387 | 0.9207 | 0.9924 |
| 350 GeV | 400 GeV | 0.9923 | 0.9384 | 0.9196 | 0.9979 |
| 400 GeV | 550 GeV | 0.9924 | 0.9348 | 0.9145 | 0.9931 |

Table A.12: Table of efficiency values in period A in center of mass rapidity range ($-2.02 < \eta * < -2.83$).

| $E_{\mathrm{T}}$ low | $E_{\mathrm{T}}$ high | $\epsilon_{Reco}$ | $\epsilon_{Reco \times ID}$ | $\epsilon_{Reco \times ID \times Iso}$ | Bin migration |
|---|---|---|---|---|---|
| 20 GeV | 25 GeV | 0.9641 | 0.7957 | 0.7834 | 1.0047 |
| 25 GeV | 35 GeV | 0.9651 | 0.8542 | 0.8406 | 1.0069 |
| 35 GeV | 45 GeV | 0.9692 | 0.8890 | 0.8723 | 0.9972 |
| 45 GeV | 55 GeV | 0.9710 | 0.9168 | 0.8966 | 0.9998 |
| 55 GeV | 65 GeV | 0.9725 | 0.9169 | 0.8990 | 0.9845 |
| 65 GeV | 75 GeV | 0.9737 | 0.9164 | 0.8965 | 1.0034 |
| 75 GeV | 85 GeV | 0.9715 | 0.9183 | 0.8939 | 0.9942 |
| 85 GeV | 105 GeV | 0.9706 | 0.9219 | 0.9005 | 0.9918 |
| 105 GeV | 125 GeV | 0.9778 | 0.9304 | 0.9082 | 0.9741 |
| 125 GeV | 150 GeV | 0.9723 | 0.9347 | 0.9034 | 1.0134 |
| 150 GeV | 175 GeV | 0.9695 | 0.9213 | 0.8954 | 0.9998 |
| 175 GeV | 200 GeV | 0.9762 | 0.9304 | 0.9067 | 0.9879 |
| 200 GeV | 250 GeV | 0.9764 | 0.9204 | 0.8868 | 0.9700 |
| 250 GeV | 350 GeV | 0.9807 | 0.9218 | 0.8992 | 0.9758 |
| 350 GeV | 550 GeV | 0.9692 | 0.9184 | 0.8850 | 0.9819 |

Table A.13: Table of efficiency values in period B in center of mass rapidity range ($-2.02 < \eta * < -2.83$).

| $E_{\mathrm{T}}$ low | $E_{\mathrm{T}}$ high | $\epsilon_{Reco}$ | $\epsilon_{Reco \times ID}$ | $\epsilon_{Reco \times ID \times Iso}$ | Bin migration |
|---|---|---|---|---|---|
| 20 GeV | 25 GeV | 0.9633 | 0.7774 | 0.7600 | 1.0047 |
| 25 GeV | 35 GeV | 0.9685 | 0.8483 | 0.8261 | 1.0069 |
| 35 GeV | 45 GeV | 0.9721 | 0.8916 | 0.8660 | 0.9972 |
| 45 GeV | 55 GeV | 0.9732 | 0.9159 | 0.8871 | 0.9998 |
| 55 GeV | 65 GeV | 0.9718 | 0.9168 | 0.8877 | 0.9845 |
| 65 GeV | 75 GeV | 0.9731 | 0.9182 | 0.8864 | 1.0034 |
| 75 GeV | 85 GeV | 0.9745 | 0.9253 | 0.8923 | 0.9942 |
| 85 GeV | 105 GeV | 0.9753 | 0.9307 | 0.8970 | 0.9918 |
| 105 GeV | 125 GeV | 0.9764 | 0.9324 | 0.8962 | 0.9741 |
| 125 GeV | 150 GeV | 0.9762 | 0.9415 | 0.9046 | 1.0134 |
| 150 GeV | 175 GeV | 0.9756 | 0.9324 | 0.8942 | 0.9998 |
| 175 GeV | 200 GeV | 0.9764 | 0.9287 | 0.8887 | 0.9879 |
| 200 GeV | 250 GeV | 0.9806 | 0.9374 | 0.8998 | 0.9700 |
| 250 GeV | 350 GeV | 0.9758 | 0.9297 | 0.8940 | 0.9758 |
| 350 GeV | 550 GeV | 0.9842 | 0.9346 | 0.8882 | 0.9819 |

Table A.14: Table of efficiency values in from both running periods as the luminosity weighted average in center of mass rapidity range $(-2.02 < \eta* < -2.83)$.

| $E_{\mathrm{T}}$ low | $E_{\mathrm{T}}$ high | $\epsilon_{Reco}$ | $\epsilon_{Reco \times ID}$ | $\epsilon_{Reco \times ID \times Iso}$ | Bin migration |
|---|---|---|---|---|---|
| 20 GeV | 25 GeV | 0.9636 | 0.7837 | 0.7680 | 1.0120 |
| 25 GeV | 35 GeV | 0.9673 | 0.8503 | 0.8311 | 1.0143 |
| 35 GeV | 45 GeV | 0.9711 | 0.8907 | 0.8682 | 0.9987 |
| 45 GeV | 55 GeV | 0.9725 | 0.9162 | 0.8903 | 1.0017 |
| 55 GeV | 65 GeV | 0.9720 | 0.9169 | 0.8916 | 0.9942 |
| 65 GeV | 75 GeV | 0.9733 | 0.9176 | 0.8899 | 1.0071 |
| 75 GeV | 85 GeV | 0.9735 | 0.9228 | 0.8929 | 0.9976 |
| 85 GeV | 105 GeV | 0.9737 | 0.9277 | 0.8982 | 0.9981 |
| 105 GeV | 125 GeV | 0.9769 | 0.9317 | 0.9003 | 0.9896 |
| 125 GeV | 150 GeV | 0.9749 | 0.9392 | 0.9042 | 1.0063 |
| 150 GeV | 175 GeV | 0.9735 | 0.9285 | 0.8946 | 1.0035 |
| 175 GeV | 200 GeV | 0.9763 | 0.9293 | 0.8949 | 0.9891 |
| 200 GeV | 250 GeV | 0.9792 | 0.9315 | 0.8953 | 0.9878 |
| 250 GeV | 350 GeV | 0.9775 | 0.9270 | 0.8958 | 0.9915 |
| 350 GeV | 550 GeV | 0.9791 | 0.9290 | 0.8871 | 0.9869 |

## A.4 Pythia- Purity data

Table A.15: Table of inputs for the purity calculation for the forward-rapidity bin, showing raw sideband yields $N_X$, sideband leakage fractions from MC $f_X$, and the final purity with asymmetrical errors.

| $E_T$ low | $E_T$ high | $N_A$ | $N_B$ | $N_C$ | $N_D$ | $f_B$ | $f_C$ | $f_D$ | P | Error low | Error high |
|---|---|---|---|---|---|---|---|---|---|---|---|
| 20 GeV | 25 GeV | 5889 | 1125 | 3971 | 1330 | 0.0012 | 0.0794 | 0.0000 | 0.46208 | 0.03633 | 0.03667 |
| 25 GeV | 35 GeV | 11080 | 3165 | 5674 | 3481 | 0.0022 | 0.0562 | 0.0002 | 0.56502 | 0.01927 | 0.01973 |
| 35 GeV | 45 GeV | 40297 | 15559 | 13560 | 16229 | 0.0020 | 0.0429 | 0.0003 | 0.70741 | 0.00666 | 0.00634 |
| 45 GeV | 55 GeV | 33865 | 15524 | 9627 | 16394 | 0.0030 | 0.0349 | 0.0002 | 0.75702 | 0.00577 | 0.00623 |
| 55 GeV | 65 GeV | 14147 | 7448 | 3284 | 8627 | 0.0026 | 0.0288 | 0.0000 | 0.82074 | 0.00699 | 0.00701 |
| 65 GeV | 75 GeV | 6629 | 3913 | 1419 | 4936 | 0.0040 | 0.0293 | 0.0001 | 0.85083 | 0.00908 | 0.00892 |
| 75 GeV | 85 GeV | 3471 | 2214 | 664 | 3112 | 0.0054 | 0.0309 | 0.0001 | 0.88414 | 0.00939 | 0.00961 |
| 85 GeV | 105 GeV | 3138 | 2178 | 510 | 3146 | 0.0049 | 0.0257 | 0.0001 | 0.90414 | 0.00839 | 0.00861 |
| 105 GeV | 125 GeV | 1222 | 936 | 193 | 1537 | 0.0042 | 0.0224 | 0.0003 | 0.91650 | 0.01225 | 0.01175 |
| 125 GeV | 150 GeV | 623 | 484 | 78 | 947 | 0.0053 | 0.0282 | 0.0012 | 0.94999 | 0.01174 | 0.01126 |
| 150 GeV | 175 GeV | 278 | 217 | 34 | 458 | 0.0065 | 0.0254 | 0.0006 | 0.95451 | 0.01626 | 0.01674 |
| 175 GeV | 200 GeV | 148 | 115 | 7 | 228 | 0.0080 | 0.0289 | 0.0002 | 0.99172 | 0.02047 | 0.01803 |
| 200 GeV | 250 GeV | 94 | 95 | 5 | 211 | 0.0076 | 0.0234 | 0.0006 | 0.98667 | 0.01492 | 0.01358 |
| 250 GeV | 350 GeV | 49 | 41 | 0 | 95 | 0.0101 | 0.0238 | 0.0013 | 1.00000 | 0.01325 | 0.00000 |
| 350 GeV | 550 GeV | 7 | 2 | 3 | 19 | 0.0142 | 0.0299 | 0.0012 | 0.95173 | 0.12248 | 0.04827 |

Table A.16: Table of inputs for the purity calculation for the mid-rapidity bin, showing raw sideband yields $N_X$, sideband leakage fractions from MC $f_X$, and the final purity with asymmetrical errors.

| $E_T$ low | $E_T$ high | $N_A$ | $N_B$ | $N_C$ | $N_D$ | $f_B$ | $f_C$ | $f_D$ | P | Error low | Error high |
|---|---|---|---|---|---|---|---|---|---|---|---|
| 20 GeV | 25 GeV | 20811 | 3883 | 14457 | 4723 | 0.0021 | 0.0726 | 0.0001 | 0.45889 | 0.01964 | 0.01936 |
| 25 GeV | 35 GeV | 38920 | 9694 | 21698 | 11671 | 0.0020 | 0.0462 | 0.0001 | 0.56032 | 0.01057 | 0.01043 |
| 35 GeV | 45 GeV | 143333 | 48218 | 52076 | 53646 | 0.0024 | 0.0294 | 0.0001 | 0.69322 | 0.00347 | 0.00353 |
| 45 GeV | 55 GeV | 120035 | 46681 | 34341 | 49597 | 0.0026 | 0.0241 | 0.0001 | 0.74893 | 0.00318 | 0.00282 |
| 55 GeV | 65 GeV | 49222 | 22303 | 11437 | 23128 | 0.0029 | 0.0191 | 0.0001 | 0.79154 | 0.00879 | 0.00921 |
| 65 GeV | 75 GeV | 23190 | 11638 | 4694 | 12382 | 0.0024 | 0.0184 | 0.0001 | 0.82471 | 0.00546 | 0.00554 |
| 75 GeV | 85 GeV | 12160 | 6336 | 2044 | 7450 | 0.0035 | 0.0163 | 0.0001 | 0.86987 | 0.00562 | 0.00538 |
| 85 GeV | 105 GeV | 10890 | 6222 | 1695 | 7675 | 0.0032 | 0.0169 | 0.0001 | 0.88651 | 0.00526 | 0.00574 |
| 105 GeV | 125 GeV | 4192 | 2590 | 572 | 3424 | 0.0031 | 0.0158 | 0.0000 | 0.90802 | 0.00727 | 0.00773 |
| 125 GeV | 150 GeV | 2159 | 1458 | 247 | 2184 | 0.0042 | 0.0168 | 0.0001 | 0.93449 | 0.00824 | 0.00776 |
| 150 GeV | 175 GeV | 936 | 631 | 83 | 1072 | 0.0040 | 0.0152 | 0.0001 | 0.95658 | 0.00883 | 0.00917 |
| 175 GeV | 200 GeV | 438 | 312 | 29 | 559 | 0.0052 | 0.0150 | 0.0001 | 0.97136 | 0.01111 | 0.01089 |
| 200 GeV | 250 GeV | 362 | 270 | 18 | 519 | 0.0056 | 0.0163 | 0.0002 | 0.98262 | 0.00887 | 0.00913 |
| 250 GeV | 300 GeV | 116 | 109 | 5 | 202 | 0.0062 | 0.0165 | 0.0004 | 0.98559 | 0.01384 | 0.01416 |
| 300 GeV | 350 GeV | 46 | 45 | 2 | 78 | 0.0054 | 0.0155 | 0.0002 | 0.98382 | 0.02757 | 0.01618 |
| 350 GeV | 400 GeV | 21 | 24 | 1 | 42 | 0.0067 | 0.0158 | 0.0003 | 0.98173 | 0.05298 | 0.01827 |
| 400 GeV | 550 GeV | 13 | 21 | 0 | 23 | 0.0071 | 0.0168 | 0.0002 | 1.00000 | 0.13625 | 0.00000 |

Table A.17: Table of inputs for the purity calculation for the backward-rapidity bin, showing raw sideband yields $N_X$, sideband leakage fractions from MC $f_X$, and the final purity with asymmetrical errors.

| $E_{\mathrm{T}}$ low | $E_{\mathrm{T}}$ high | $N_A$ | $N_B$ | $N_C$ | $N_D$ | $f_B$ | $f_C$ | $f_D$ | P | Error low | Error high |
|---|---|---|---|---|---|---|---|---|---|---|---|
| 20 GeV | 25 GeV | 5129 | 1077 | 3337 | 1210 | 0.0033 | 0.0754 | 0.0001 | 0.45545 | 0.03870 | 0.03830 |
| 25 GeV | 35 GeV | 9725 | 2766 | 4719 | 3061 | 0.0023 | 0.0565 | 0.0001 | 0.59365 | 0.01890 | 0.01910 |
| 35 GeV | 45 GeV | 34925 | 14443 | 11258 | 14714 | 0.0042 | 0.0449 | 0.0003 | 0.71708 | 0.00683 | 0.00717 |
| 45 GeV | 55 GeV | 28773 | 14329 | 7915 | 14802 | 0.0050 | 0.0327 | 0.0002 | 0.75954 | 0.00629 | 0.00671 |
| 55 GeV | 65 GeV | 11520 | 6673 | 2776 | 7482 | 0.0039 | 0.0300 | 0.0002 | 0.80766 | 0.00791 | 0.00809 |
| 65 GeV | 75 GeV | 5164 | 3378 | 1141 | 4107 | 0.0049 | 0.0326 | 0.0002 | 0.84178 | 0.01003 | 0.00997 |
| 75 GeV | 85 GeV | 2521 | 1917 | 511 | 2520 | 0.0056 | 0.0317 | 0.0003 | 0.86760 | 0.01235 | 0.01265 |
| 85 GeV | 105 GeV | 2143 | 1740 | 378 | 2625 | 0.0049 | 0.0287 | 0.0002 | 0.90088 | 0.01063 | 0.01037 |
| 105 GeV | 125 GeV | 799 | 654 | 114 | 1116 | 0.0070 | 0.0276 | 0.0000 | 0.93230 | 0.01305 | 0.01295 |
| 125 GeV | 150 GeV | 331 | 331 | 46 | 581 | 0.0104 | 0.0261 | 0.0014 | 0.93536 | 0.01961 | 0.01939 |
| 150 GeV | 175 GeV | 118 | 128 | 29 | 269 | 0.0065 | 0.0253 | 0.0006 | 0.89538 | 0.04013 | 0.03987 |
| 175 GeV | 200 GeV | 57 | 42 | 2 | 103 | 0.0065 | 0.0250 | 0.0000 | 0.99597 | 0.01522 | 0.00403 |
| 200 GeV | 250 GeV | 28 | 43 | 1 | 71 | 0.0103 | 0.0285 | 0.0007 | 0.99559 | 0.03534 | 0.00441 |
| 250 GeV | 350 GeV | 10 | 6 | 0 | 21 | 0.0062 | 0.0295 | 0.0003 | 1.00000 | 0.08575 | 0.00000 |
| 350 GeV | 550 GeV | 0 | 0 | 0 | 0 | 0.0117 | 0.0272 | 0.0009 | 0.00000 | 0.00000 | 0.00000 |

## A.5 Total systematic uncertainty data

Table A.18: Table of components of the total systematic uncertainty on the cross section shown as percents for each $p_T$ bin in center of mass rapidity range $(0.91 < \eta* < -1.84)$.

| $E_{\rm T}$ low | $E_{\rm T}$ high | E-scale low | E-scale high | Purity low | Purity high | Other low | Other high | Total low | Total high |
|---|---|---|---|---|---|---|---|---|---|
| 20 GeV | 25 GeV | 12.67 | 12.63 | 6.20 | 6.29 | 3.56 | 3.56 | 14.54 | 14.55 |
| 25 GeV | 35 GeV | 8.24 | 8.15 | 3.12 | 3.39 | 3.53 | 3.53 | 9.49 | 9.50 |
| 35 GeV | 45 GeV | 4.45 | 4.40 | 3.05 | 2.77 | 3.51 | 3.51 | 6.43 | 6.27 |
| 45 GeV | 55 GeV | 3.48 | 3.45 | 3.53 | 3.46 | 3.48 | 3.48 | 6.06 | 6.00 |
| 55 GeV | 65 GeV | 2.43 | 2.42 | 3.56 | 3.57 | 3.45 | 3.45 | 5.52 | 5.53 |
| 65 GeV | 75 GeV | 2.05 | 2.04 | 3.91 | 3.74 | 3.42 | 3.42 | 5.59 | 5.46 |
| 75 GeV | 85 GeV | 1.62 | 1.63 | 3.79 | 3.86 | 3.40 | 3.40 | 5.34 | 5.39 |
| 85 GeV | 105 GeV | 1.40 | 1.41 | 4.09 | 4.43 | 3.15 | 3.15 | 5.35 | 5.62 |
| 105 GeV | 125 GeV | 1.30 | 1.30 | 4.74 | 4.93 | 3.12 | 3.12 | 5.82 | 5.98 |
| 125 GeV | 150 GeV | 1.05 | 1.05 | 4.85 | 4.74 | 3.09 | 3.09 | 5.85 | 5.75 |
| 150 GeV | 175 GeV | 1.02 | 1.02 | 6.34 | 6.64 | 3.06 | 3.06 | 7.12 | 7.39 |
| 175 GeV | 200 GeV | 0.90 | 0.90 | 6.92 | 6.33 | 3.03 | 3.03 | 7.61 | 7.08 |
| 200 GeV | 250 GeV | 0.91 | 0.91 | 7.34 | 8.65 | 3.01 | 3.01 | 7.99 | 9.20 |
| 250 GeV | 350 GeV | 0.89 | 0.89 | 9.26 | 8.46 | 2.98 | 2.98 | 9.77 | 9.01 |
| 350 GeV | 550 GeV | 1.13 | 1.04 | 12.06 | 14.55 | 2.95 | 2.95 | 12.46 | 14.88 |

Table A.19: Table of components of the total systematic uncertainty on the cross section shown as percents for each $p_T$ bin in center of mass rapidity range $(0.91 < \eta* < -1.84)$.

| $E_{\rm T}$ low | $E_{\rm T}$ high | E-scale low | E-scale high | Purity low | Purity high | Other low | Other high | Total low | Total high |
|---|---|---|---|---|---|---|---|---|---|
| 20 GeV | 25 GeV | 15.67 | 15.52 | 3.08 | 3.07 | 2.98 | 2.98 | 16.25 | 16.10 |
| 25 GeV | 35 GeV | 11.10 | 11.02 | 1.64 | 1.67 | 2.95 | 2.95 | 11.60 | 11.53 |
| 35 GeV | 45 GeV | 7.33 | 7.31 | 1.52 | 1.54 | 2.92 | 2.92 | 8.03 | 8.02 |
| 45 GeV | 55 GeV | 5.58 | 5.58 | 1.54 | 1.47 | 2.90 | 2.90 | 6.47 | 6.46 |
| 55 GeV | 65 GeV | 4.28 | 4.26 | 1.63 | 1.54 | 2.87 | 2.87 | 5.40 | 5.36 |
| 65 GeV | 75 GeV | 3.23 | 3.23 | 1.61 | 1.52 | 2.84 | 2.84 | 4.59 | 4.56 |
| 75 GeV | 85 GeV | 2.22 | 2.21 | 1.75 | 1.62 | 2.81 | 2.81 | 3.99 | 3.93 |
| 85 GeV | 105 GeV | 1.59 | 1.58 | 1.48 | 1.46 | 2.79 | 2.79 | 3.54 | 3.52 |
| 105 GeV | 125 GeV | 1.36 | 1.37 | 2.06 | 1.63 | 2.76 | 2.76 | 3.70 | 3.49 |
| 125 GeV | 150 GeV | 1.14 | 1.14 | 2.08 | 2.19 | 2.74 | 2.74 | 3.62 | 3.68 |
| 150 GeV | 175 GeV | 1.01 | 1.01 | 2.74 | 2.45 | 2.71 | 2.71 | 3.98 | 3.79 |
| 175 GeV | 200 GeV | 0.95 | 0.95 | 2.61 | 2.53 | 2.68 | 2.68 | 3.86 | 3.80 |
| 200 GeV | 250 GeV | 0.92 | 0.91 | 2.84 | 2.60 | 2.66 | 2.66 | 3.99 | 3.83 |
| 250 GeV | 300 GeV | 0.91 | 0.91 | 3.29 | 3.26 | 2.63 | 2.63 | 4.31 | 4.28 |
| 300 GeV | 350 GeV | 0.91 | 0.91 | 3.36 | 2.95 | 2.60 | 2.60 | 4.35 | 4.04 |
| 350 GeV | 400 GeV | 0.92 | 0.92 | 3.71 | 3.21 | 2.58 | 2.58 | 4.61 | 4.22 |
| 400 GeV | 550 GeV | 0.90 | 0.90 | 5.91 | 5.48 | 2.55 | 2.55 | 6.50 | 6.12 |

Table A.20: Table of components of the total systematic uncertainty on the cross section shown as percents for each $p_T$ bin in center of mass rapidity range $(-2.02 < \eta* < -2.83)$.

| $E_T$ low | $E_T$ high | E-scale low | E-scale high | Purity low | Purity high | Other low | Other high | Total low | Total high |
|---|---|---|---|---|---|---|---|---|---|
| 20 GeV | 25 GeV | 13.31 | 13.09 | 5.18 | 5.20 | 3.90 | 3.90 | 14.80 | 14.62 |
| 25 GeV | 35 GeV | 7.35 | 7.29 | 2.88 | 2.89 | 3.87 | 3.87 | 8.79 | 8.74 |
| 35 GeV | 45 GeV | 4.34 | 4.21 | 3.16 | 3.29 | 3.83 | 3.83 | 6.60 | 6.58 |
| 45 GeV | 55 GeV | 3.43 | 3.42 | 3.17 | 3.37 | 3.80 | 3.80 | 6.02 | 6.12 |
| 55 GeV | 65 GeV | 2.66 | 2.62 | 3.46 | 3.71 | 3.77 | 3.77 | 5.77 | 5.90 |
| 65 GeV | 75 GeV | 2.17 | 2.18 | 3.82 | 3.38 | 3.73 | 3.73 | 5.77 | 5.49 |
| 75 GeV | 85 GeV | 1.81 | 1.81 | 4.10 | 4.00 | 3.70 | 3.70 | 5.81 | 5.74 |
| 85 GeV | 105 GeV | 1.44 | 1.46 | 3.91 | 4.14 | 3.46 | 3.46 | 5.41 | 5.59 |
| 105 GeV | 125 GeV | 1.15 | 1.16 | 3.47 | 4.32 | 3.42 | 3.42 | 5.01 | 5.63 |
| 125 GeV | 150 GeV | 1.15 | 1.13 | 4.89 | 4.35 | 3.39 | 3.39 | 6.06 | 5.63 |
| 150 GeV | 175 GeV | 1.48 | 1.51 | 5.46 | 5.86 | 3.35 | 3.35 | 6.58 | 6.92 |
| 175 GeV | 200 GeV | 0.88 | 0.88 | 6.24 | 5.85 | 3.32 | 3.32 | 7.12 | 6.78 |
| 200 GeV | 250 GeV | 0.88 | 0.88 | 6.46 | 7.56 | 3.28 | 3.28 | 7.30 | 8.29 |
| 250 GeV | 350 GeV | 0.88 | 0.88 | 9.04 | 8.28 | 3.24 | 3.24 | 9.64 | 8.94 |
| 350 GeV | 550 GeV | 0.88 | 0.88 | 10.35 | 12.53 | 3.21 | 3.21 | 10.87 | 12.96 |

Table A.21: Table of components of the total systematic uncertainty on $R_{p\mathrm{Pb}}$ shown as percents for each $p_T$ bin in center of mass rapidity range $(0.91 < \eta* < -1.84)$.

| $E_T$ low | $E_T$ high | p+Pb CS low | p+Pb CS high | p+p CS low | p+p CS high | Extrapolation low | Extrapolation high | Total low | Total high |
|---|---|---|---|---|---|---|---|---|---|
| 25 GeV | 35 GeV | 9.17 | 9.17 | 11.52 | 11.74 | 1.20 | 1.20 | 14.77 | 14.94 |
| 35 GeV | 45 GeV | 5.99 | 5.99 | 7.97 | 7.97 | 0.88 | 0.88 | 10.02 | 10.01 |
| 45 GeV | 55 GeV | 5.50 | 5.50 | 5.94 | 5.95 | 0.64 | 0.64 | 8.12 | 8.13 |
| 55 GeV | 65 GeV | 4.93 | 4.93 | 4.69 | 4.74 | 0.44 | 0.44 | 6.82 | 6.86 |
| 65 GeV | 75 GeV | 5.08 | 5.08 | 4.16 | 4.21 | 0.27 | 0.27 | 6.57 | 6.61 |
| 75 GeV | 85 GeV | 4.62 | 4.62 | 3.87 | 3.89 | 0.12 | 0.12 | 6.03 | 6.04 |
| 85 GeV | 105 GeV | 5.13 | 5.13 | 3.80 | 3.82 | 0.07 | 0.07 | 6.39 | 6.40 |
| 105 GeV | 125 GeV | 4.74 | 4.74 | 3.83 | 3.83 | 0.28 | 0.28 | 6.10 | 6.10 |
| 125 GeV | 150 GeV | 5.03 | 5.03 | 4.04 | 4.03 | 0.47 | 0.47 | 6.47 | 6.46 |
| 150 GeV | 175 GeV | 5.74 | 5.74 | 4.83 | 4.83 | 0.66 | 0.66 | 7.53 | 7.53 |
| 175 GeV | 200 GeV | 6.36 | 6.36 | 4.96 | 4.95 | 0.81 | 0.81 | 8.10 | 8.10 |
| 200 GeV | 250 GeV | 6.88 | 6.88 | 5.51 | 5.49 | 1.01 | 1.01 | 8.87 | 8.86 |
| 250 GeV | 350 GeV | 7.90 | 7.90 | 7.32 | 7.09 | 1.32 | 1.32 | 10.86 | 10.70 |
| 350 GeV | 550 GeV | 10.63 | 10.63 | 10.95 | 10.36 | 1.77 | 1.77 | 15.36 | 14.95 |

Table A.22: Table of components of the total systematic uncertainty on $R_{p\mathrm{Pb}}$ shown as percents for each $p_T$ bin in center of mass rapidity range $(0.91 < \eta* < -1.84)$.

| $E_T$ low | $E_T$ high | p+Pb CS low | p+Pb CS high | p+p CS low | p+p CS high | Extrapolation low | Extrapolation high | Total low | Total high |
|---|---|---|---|---|---|---|---|---|---|
| 25 GeV | 35 GeV | 11.34 | 11.34 | 12.13 | 12.22 | 0.53 | 0.53 | 16.61 | 16.68 |
| 35 GeV | 45 GeV | 7.71 | 7.71 | 8.21 | 8.24 | 0.64 | 0.64 | 11.28 | 11.30 |
| 45 GeV | 55 GeV | 6.11 | 6.11 | 5.82 | 5.83 | 0.73 | 0.73 | 8.47 | 8.48 |
| 55 GeV | 65 GeV | 5.00 | 5.00 | 4.37 | 4.38 | 0.81 | 0.81 | 6.69 | 6.70 |
| 65 GeV | 75 GeV | 4.15 | 4.15 | 3.48 | 3.48 | 0.87 | 0.87 | 5.48 | 5.49 |
| 75 GeV | 85 GeV | 3.48 | 3.48 | 3.02 | 3.03 | 0.92 | 0.92 | 4.70 | 4.71 |
| 85 GeV | 105 GeV | 3.06 | 3.06 | 2.66 | 2.66 | 0.99 | 0.99 | 4.18 | 4.18 |
| 105 GeV | 125 GeV | 3.18 | 3.18 | 2.52 | 2.51 | 1.07 | 1.07 | 4.19 | 4.19 |
| 125 GeV | 150 GeV | 3.09 | 3.09 | 2.53 | 2.54 | 1.14 | 1.14 | 4.16 | 4.16 |
| 150 GeV | 175 GeV | 3.53 | 3.53 | 2.61 | 2.62 | 1.21 | 1.21 | 4.55 | 4.56 |
| 175 GeV | 200 GeV | 3.41 | 3.41 | 2.70 | 2.70 | 1.27 | 1.27 | 4.53 | 4.53 |
| 200 GeV | 250 GeV | 3.57 | 3.57 | 2.79 | 2.76 | 1.34 | 1.34 | 4.72 | 4.71 |
| 250 GeV | 300 GeV | 3.93 | 3.93 | 2.92 | 2.89 | 1.42 | 1.42 | 5.10 | 5.08 |
| 300 GeV | 350 GeV | 3.99 | 3.99 | 3.15 | 3.11 | 1.49 | 1.49 | 5.29 | 5.27 |
| 350 GeV | 400 GeV | 4.28 | 4.28 | 3.40 | 3.39 | 1.55 | 1.55 | 5.69 | 5.67 |
| 400 GeV | 550 GeV | 6.28 | 6.28 | 3.73 | 3.75 | 1.64 | 1.64 | 7.49 | 7.49 |

Table A.23: Table of components of the total systematic uncertainty on $R_{p\mathrm{Pb}}$ shown as percents for each $p_T$ bin in center of mass rapidity range $(-2.02 < \eta* < -2.83)$.

| $E_{\mathrm{T}}$ low | $E_{\mathrm{T}}$ high | p+Pb CS low | p+Pb CS high | p+p CS low | p+p CS high | Extrapolation low | Extrapolation high | Total low | Total high |
|---|---|---|---|---|---|---|---|---|---|
| 25 GeV | 35 GeV | 8.62 | 8.62 | 11.52 | 11.74 | 2.14 | 2.14 | 14.55 | 14.72 |
| 35 GeV | 45 GeV | 6.33 | 6.33 | 7.97 | 7.97 | 2.22 | 2.22 | 10.42 | 10.42 |
| 45 GeV | 55 GeV | 5.89 | 5.89 | 5.94 | 5.95 | 2.28 | 2.28 | 8.67 | 8.68 |
| 55 GeV | 65 GeV | 5.53 | 5.53 | 4.69 | 4.74 | 2.33 | 2.33 | 7.61 | 7.65 |
| 65 GeV | 75 GeV | 5.51 | 5.51 | 4.16 | 4.21 | 2.37 | 2.37 | 7.30 | 7.33 |
| 75 GeV | 85 GeV | 5.67 | 5.67 | 3.87 | 3.89 | 2.40 | 2.40 | 7.27 | 7.28 |
| 85 GeV | 105 GeV | 5.48 | 5.48 | 3.80 | 3.82 | 2.45 | 2.45 | 7.11 | 7.12 |
| 105 GeV | 125 GeV | 5.55 | 5.55 | 3.83 | 3.83 | 2.50 | 2.50 | 7.19 | 7.20 |
| 125 GeV | 150 GeV | 6.13 | 6.13 | 4.04 | 4.03 | 2.55 | 2.55 | 7.77 | 7.76 |
| 150 GeV | 175 GeV | 7.52 | 7.52 | 4.83 | 4.83 | 2.59 | 2.59 | 9.30 | 9.30 |
| 175 GeV | 200 GeV | 7.80 | 7.80 | 4.96 | 4.95 | 2.63 | 2.63 | 9.61 | 9.60 |
| 200 GeV | 250 GeV | 8.39 | 8.39 | 5.51 | 5.49 | 2.68 | 2.68 | 10.39 | 10.38 |
| 250 GeV | 350 GeV | 11.37 | 11.37 | 7.32 | 7.09 | 2.75 | 2.75 | 13.80 | 13.68 |
| 350 GeV | 550 GeV | 12.89 | 12.89 | 10.95 | 10.36 | 2.86 | 2.86 | 17.15 | 16.78 |

Table A.24: Table of components of the total systematic uncertainty for the forward-to-backward $R_{p\mathrm{Pb}}$ ratio shown as percents for each $p_T$ bin.

| $E_{\mathrm{T}}$ low | $E_{\mathrm{T}}$ high | E-scale low | E-scale high | Purity low | Purity high | Other low | Other high | Extrapolation low | Extrapolation high | Total low | Total high |
|---|---|---|---|---|---|---|---|---|---|---|---|
| 20 GeV | 25 GeV | 3.43 | 5.01 | 1.72 | 1.67 | 1.07 | 1.07 | 2.56 | 2.56 | 4.74 | 5.97 |
| 25 GeV | 35 GeV | 2.15 | 1.95 | 1.18 | 1.07 | 1.07 | 1.07 | 2.46 | 2.46 | 3.63 | 3.48 |
| 35 GeV | 45 GeV | 1.62 | 1.64 | 1.37 | 1.21 | 1.07 | 1.07 | 2.39 | 2.39 | 3.37 | 3.32 |
| 45 GeV | 55 GeV | 1.58 | 1.57 | 1.11 | 1.17 | 1.07 | 1.07 | 2.37 | 2.37 | 3.23 | 3.26 |
| 55 GeV | 65 GeV | 1.59 | 1.63 | 1.07 | 1.12 | 1.06 | 1.06 | 2.37 | 2.37 | 3.23 | 3.26 |
| 65 GeV | 75 GeV | 1.59 | 1.58 | 1.05 | 1.21 | 1.06 | 1.06 | 2.38 | 2.38 | 3.23 | 3.28 |
| 75 GeV | 85 GeV | 1.59 | 1.59 | 1.11 | 1.47 | 1.06 | 1.06 | 2.41 | 2.41 | 3.27 | 3.41 |
| 85 GeV | 105 GeV | 1.58 | 1.58 | 0.81 | 0.27 | 1.06 | 1.06 | 2.45 | 2.45 | 3.21 | 3.11 |
| 105 GeV | 125 GeV | 1.60 | 1.59 | 1.02 | 1.31 | 1.05 | 1.05 | 2.51 | 2.51 | 3.32 | 3.42 |
| 125 GeV | 150 GeV | 1.59 | 1.59 | 0.79 | 1.20 | 1.05 | 1.05 | 2.59 | 2.59 | 3.31 | 3.43 |
| 150 GeV | 175 GeV | 1.73 | 1.73 | 1.57 | 2.40 | 1.05 | 1.05 | 2.67 | 2.67 | 3.70 | 4.12 |
| 175 GeV | 200 GeV | 1.57 | 1.57 | 0.93 | 1.63 | 1.05 | 1.05 | 2.75 | 2.75 | 3.46 | 3.71 |
| 200 GeV | 250 GeV | 1.58 | 1.58 | 3.90 | 1.67 | 1.05 | 1.05 | 2.86 | 2.86 | 5.20 | 3.82 |
| 250 GeV | 350 GeV | 1.57 | 1.57 | 2.08 | 4.26 | 1.04 | 1.04 | 3.05 | 3.05 | 4.15 | 5.57 |
| 350 GeV | 550 GeV | 1.72 | 1.66 | 4.72 | 2.39 | 1.04 | 1.04 | 3.36 | 3.36 | 6.13 | 4.57 |

## A.6 Pythia- Sherpa comparisons

This section contains comparisons of efficiencies and leakages from Pythia and Sherpa data overlay.

### A.6.1 Efficiencies

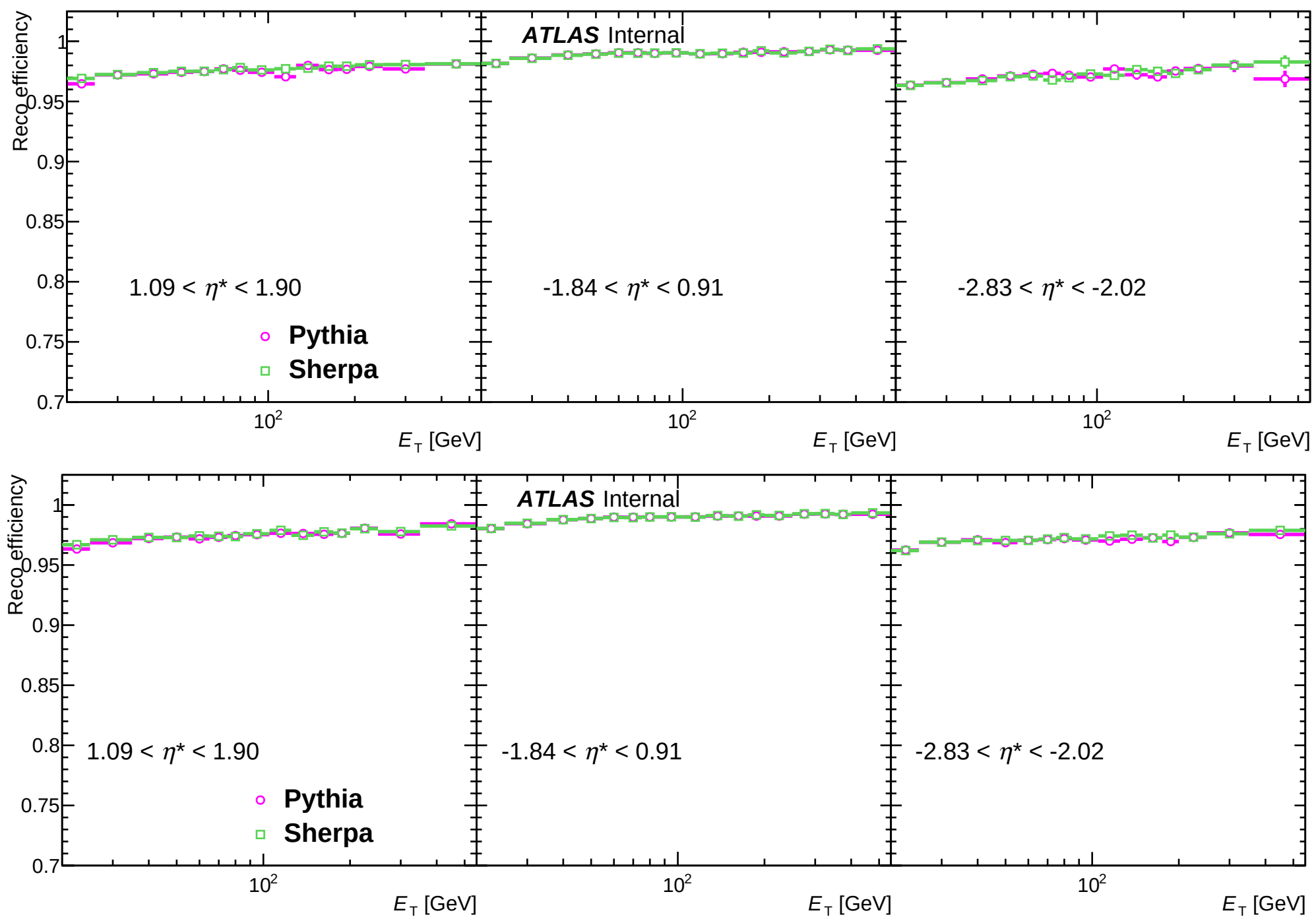

Figure A.1: Reconstruction efficiency comparison for period A (top) and period B (bottom).

### A.6.2 Leakages

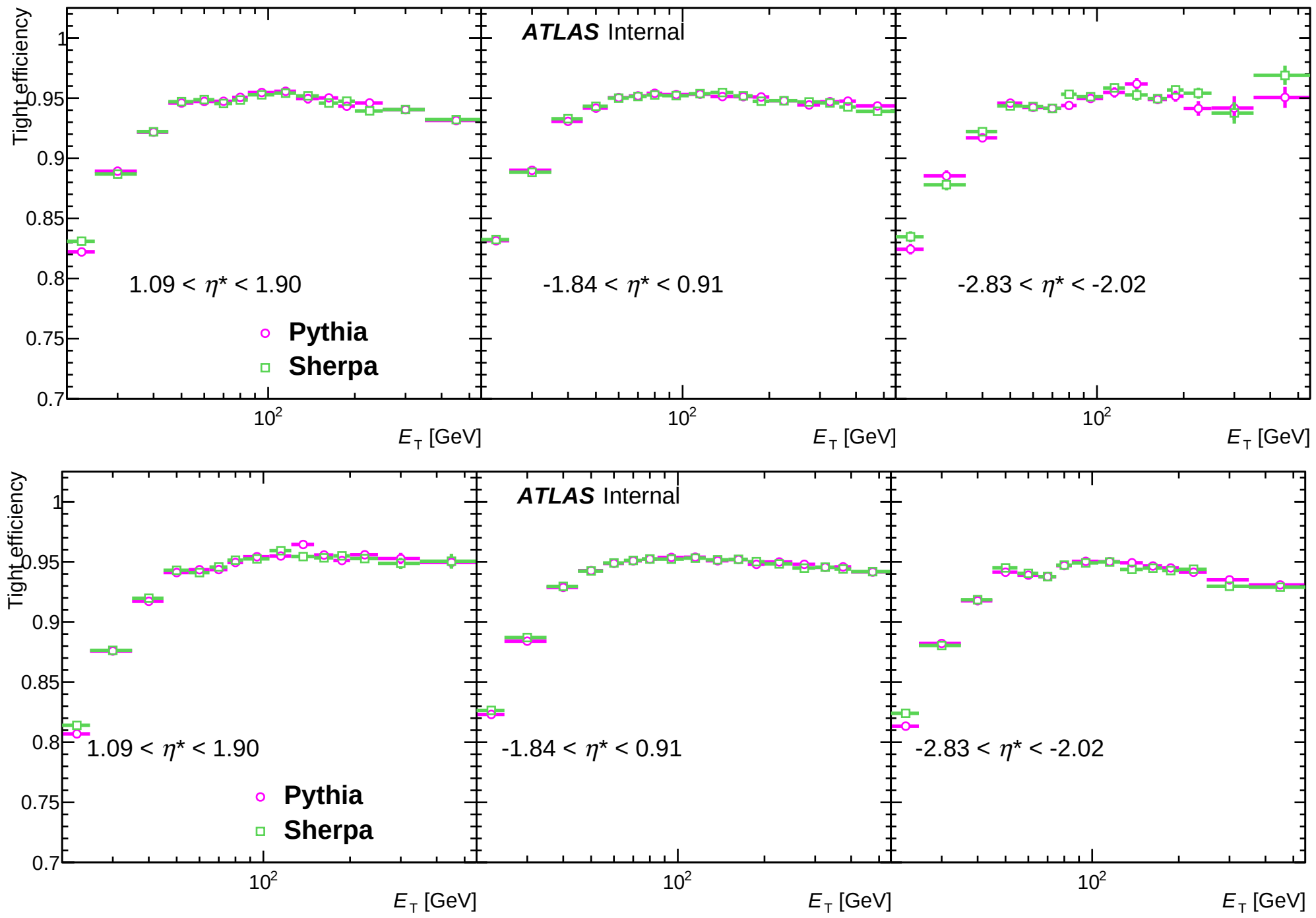

Figure A.2: Tight ID efficiency comparison for period A (top) and period B (bottom).

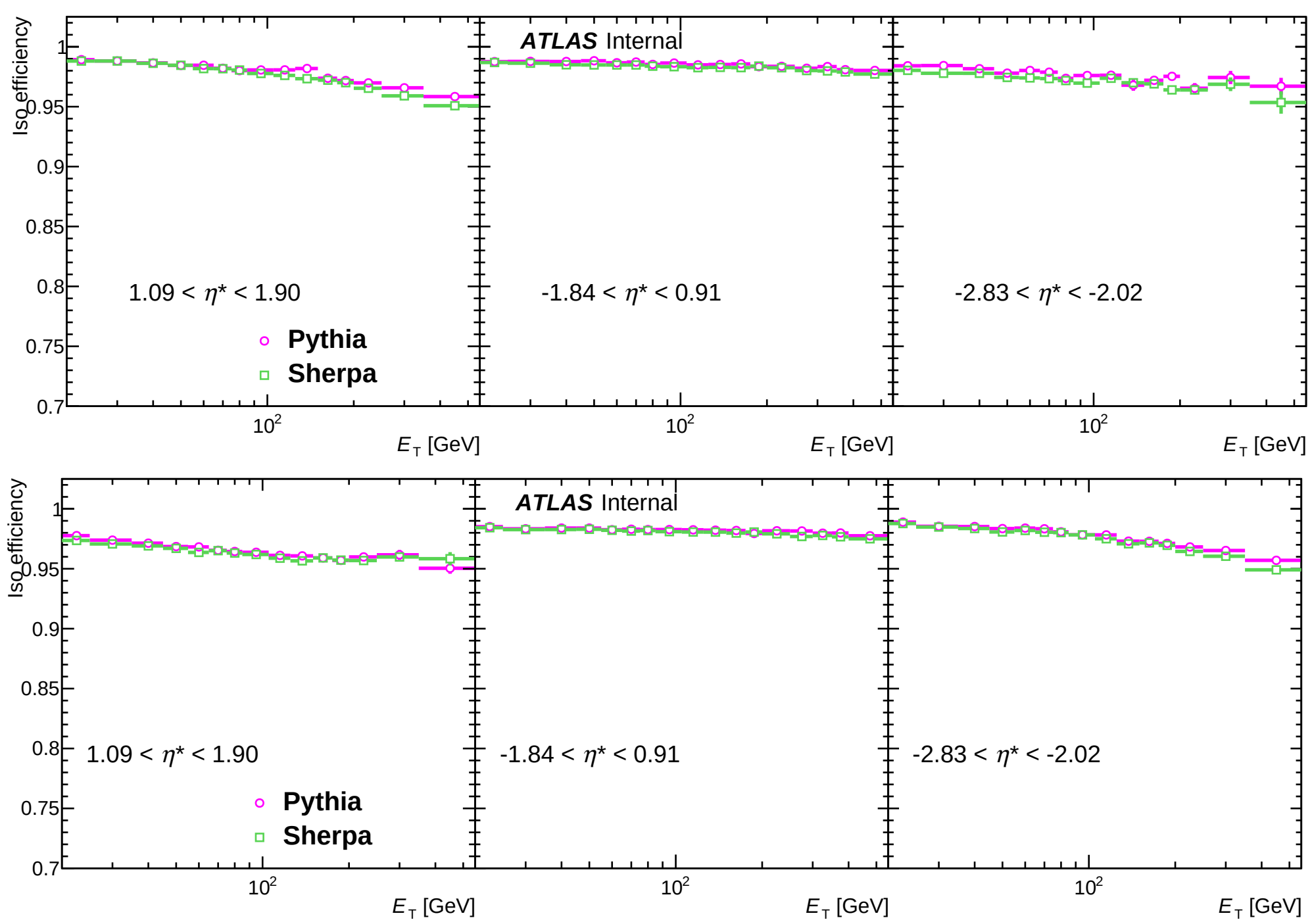

Figure A.3: Isolation efficiency comparison for period A (top) and period B (bottom).

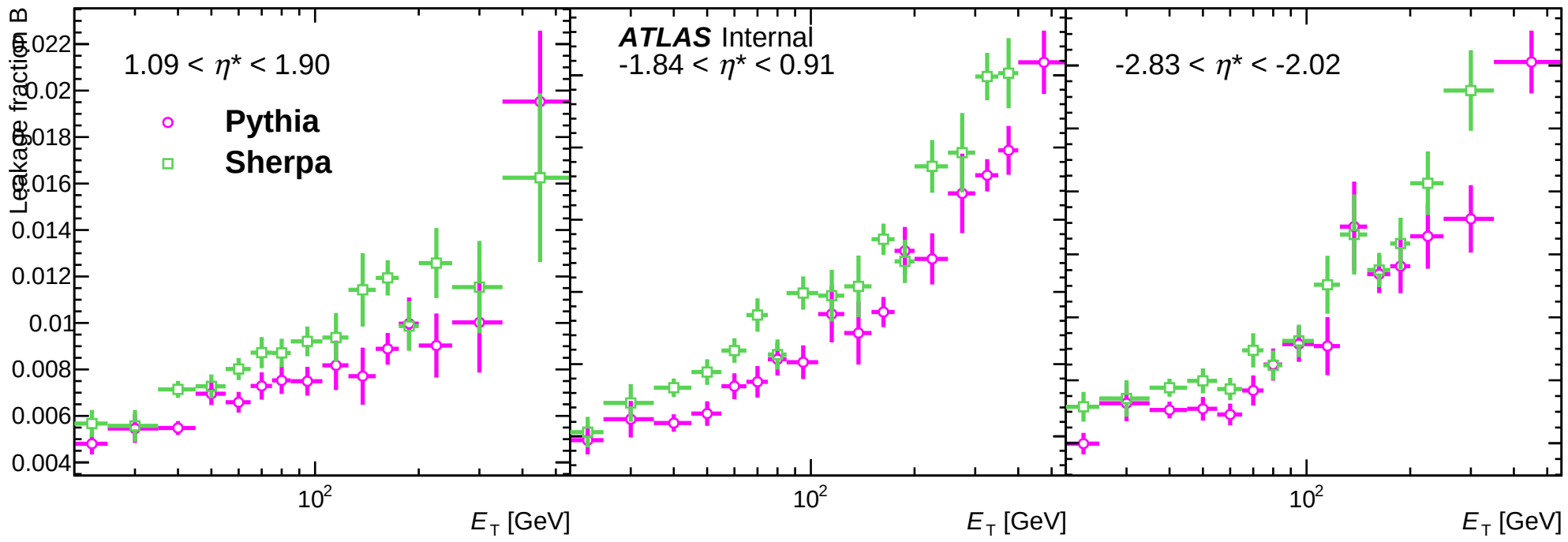

Figure A.4: Sideband B leakage fraction

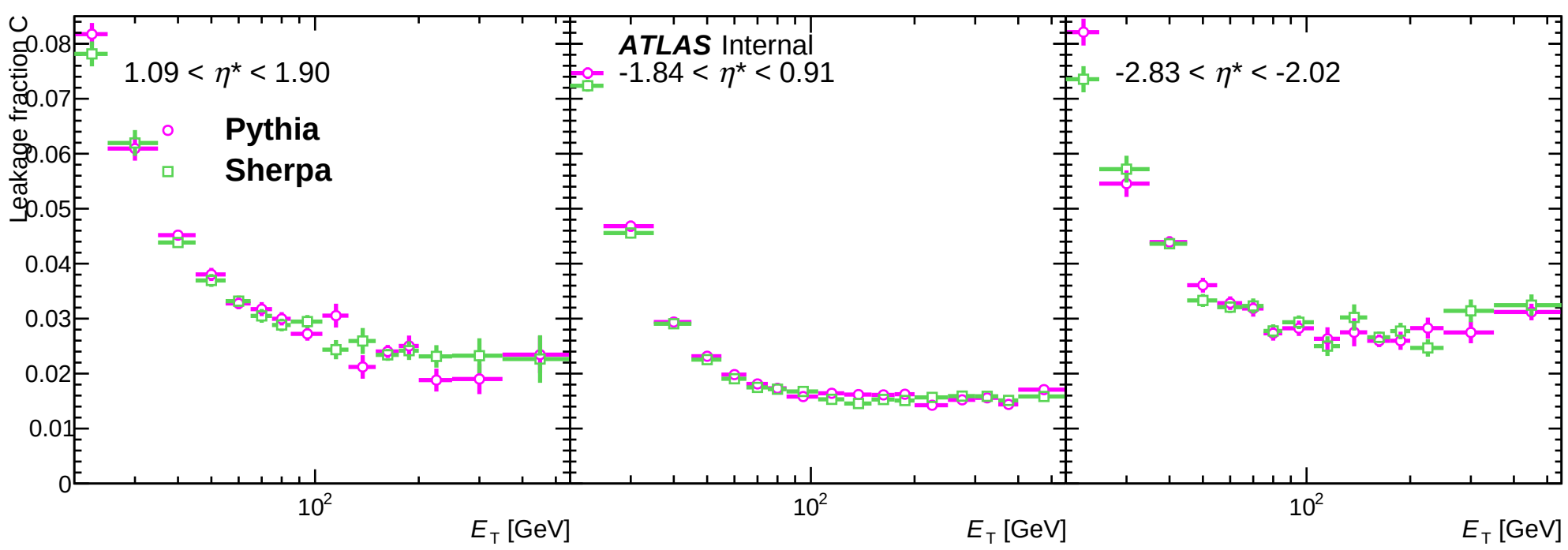

Figure A.5: Sideband C leakage fraction

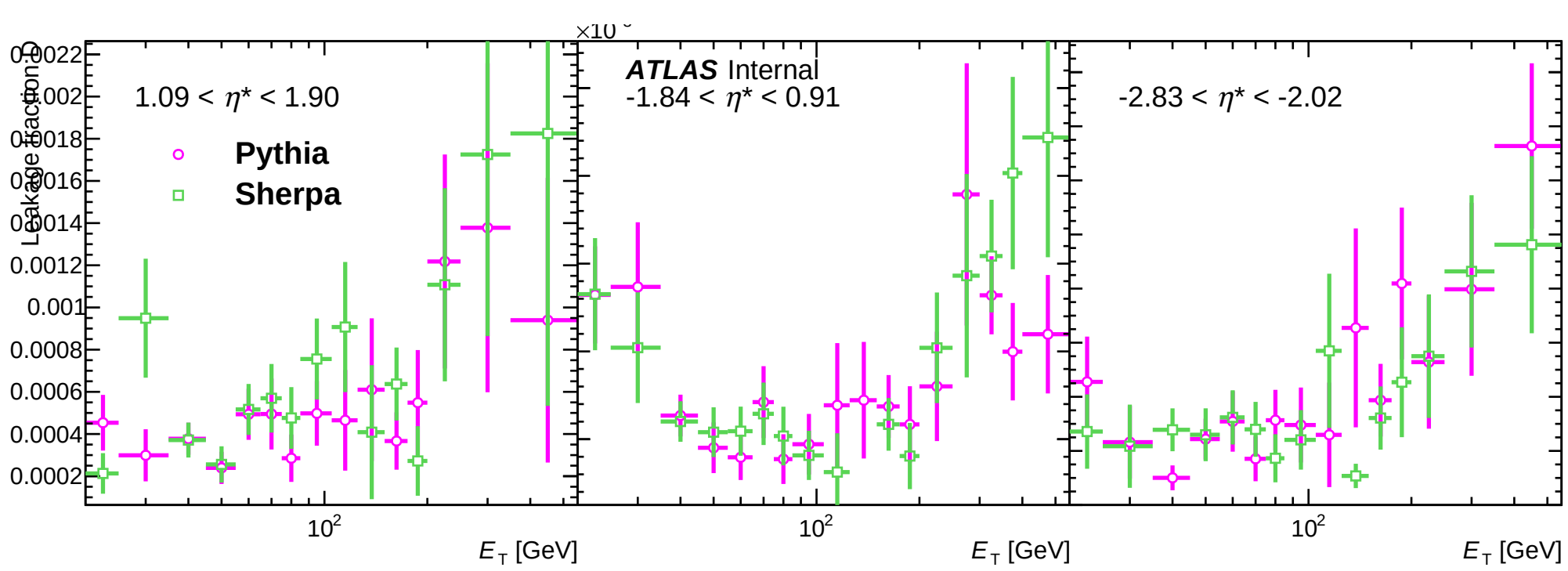

Figure A.6: Sideband D leakage fraction

## A.7 Shower shape comparisons

The following section contains comparisons for all ten shower shape variables. We show the distributions, compared to MC before and after applying fudge factors, in each $\eta$ slice and, for brevity, two representative $E_{\mathrm{T}}$ bins: $25\,\mathrm{GeV} < E_{\mathrm{T}} < 35\,\mathrm{GeV}$ and $105\,\mathrm{GeV} < E_{\mathrm{T}} < 125\,\mathrm{GeV}$.

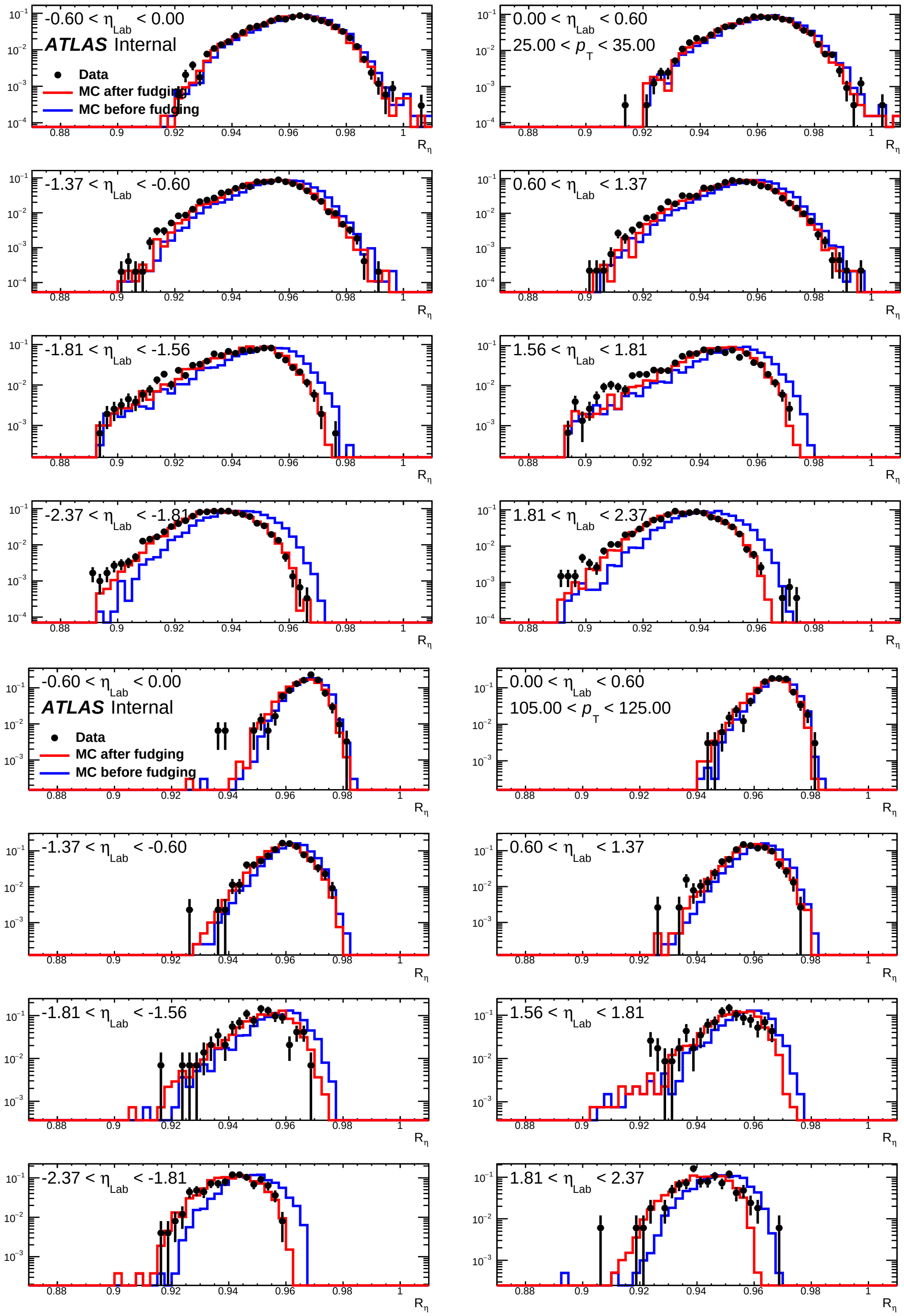

Figure A.7: Shower shape parameter, $R_\eta$, in each pseudorapidity slice from $E_{\mathrm{T}}$ bins 25 GeV $< E_{\mathrm{T}} <$ 35 GeV (above) and 105 GeV $< E_{\mathrm{T}} <$ 125 GeV (below). Reconstructed data plotted as black points overlaid with MC before (blue histogram) and after (red histogram) fudging.

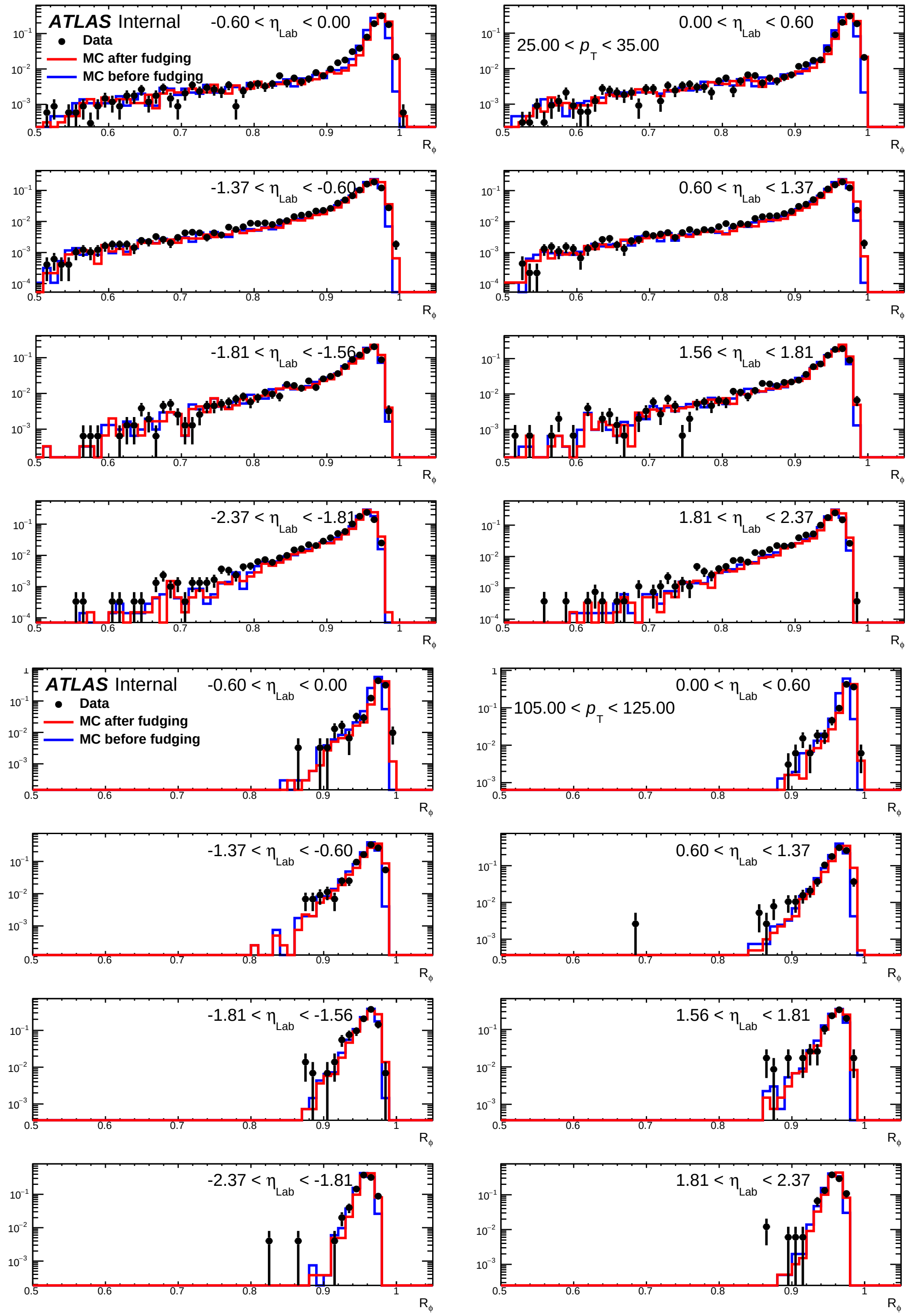

Figure A.8: Shower shape parameter, $R_{\phi}$, in each pseudorapidity slice from $E_{\mathrm{T}}$ bins 25 GeV $< E_{\mathrm{T}} <$ 35 GeV (above) and 105 GeV $< E_{\mathrm{T}} <$ 125 GeV (below). Reconstructed data plotted as black points overlaid with MC before (blue histogram) and after (red histogram) fudging.

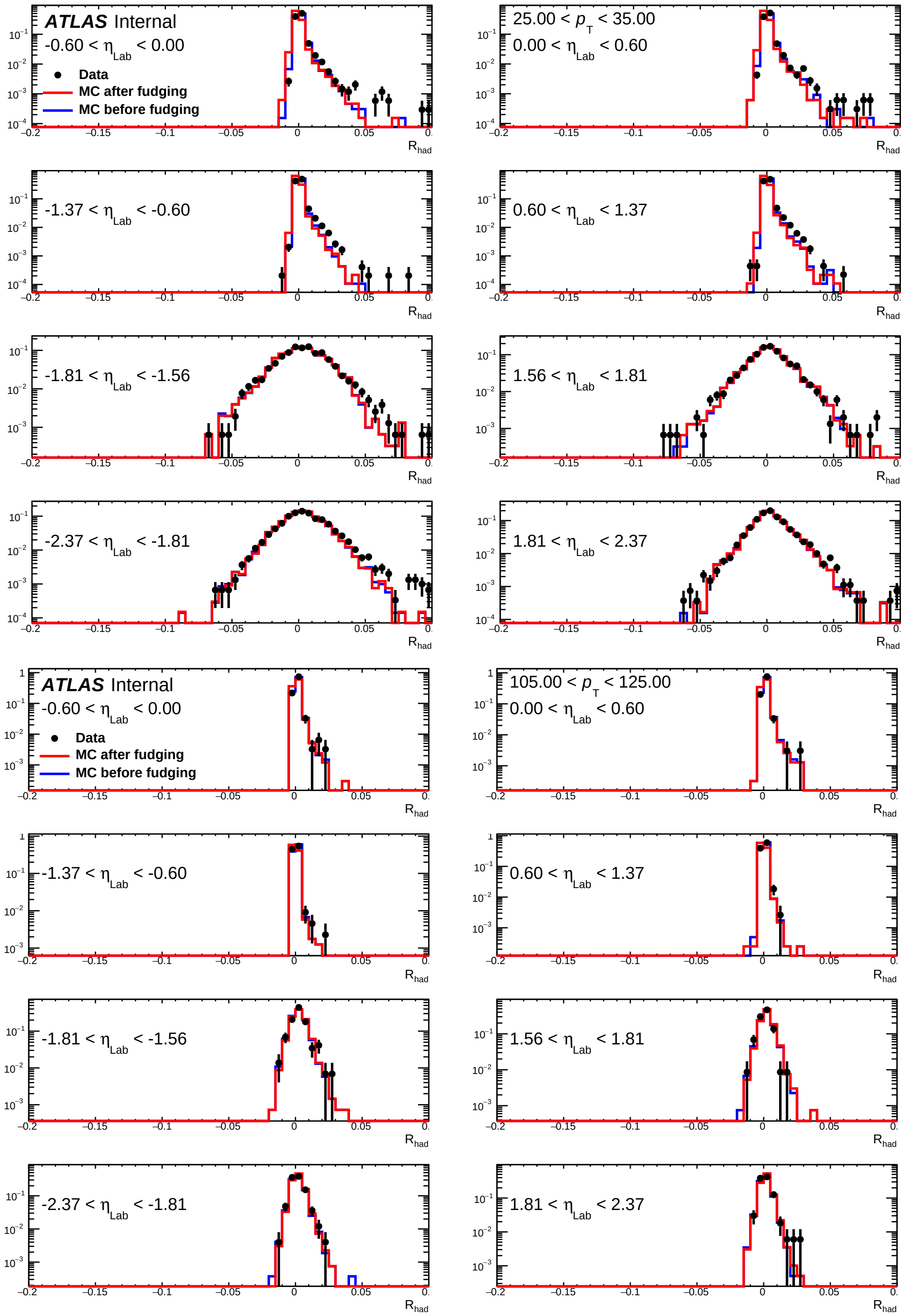

Figure A.9: Shower shape parameter, $R_{had}$, in each pseudorapidity slice from $E_{\mathrm{T}}$ bins 25 GeV $< E_{\mathrm{T}} <$ 35 GeV (above) and 105 GeV $< E_{\mathrm{T}} <$ 125 GeV (below). Reconstructed data plotted as black points overlaid with MC before (blue histogram) and after (red histogram) fudging.

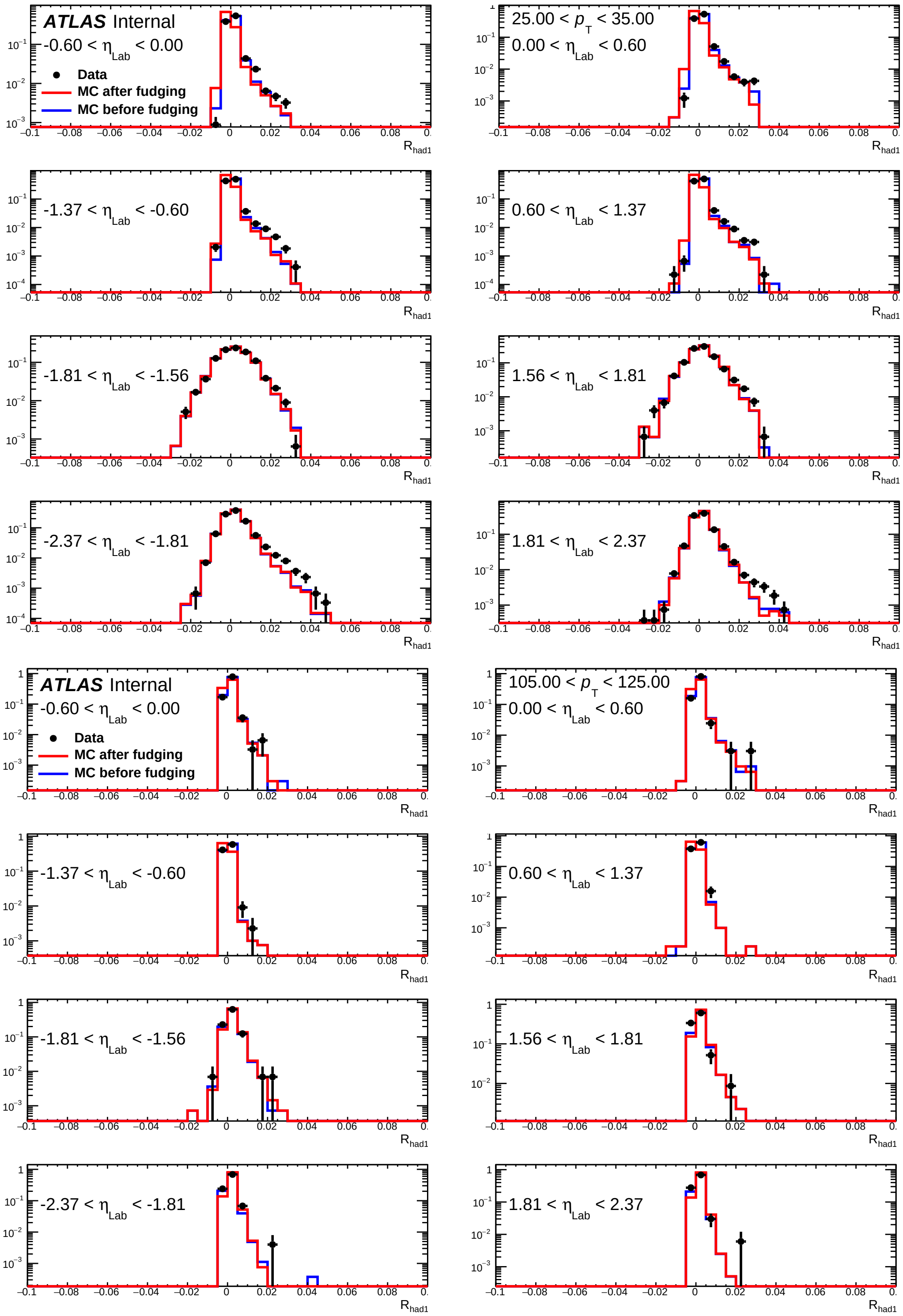

Figure A.10: Shower shape parameter, $R_{had1}$, in each pseudorapidity slice from $E_{\mathrm{T}}$ bins 25 GeV $< E_{\mathrm{T}} <$ 35 GeV (above) and 105 GeV $< E_{\mathrm{T}} <$ 125 GeV (below). Reconstructed data plotted as black points overlaid with MC before (blue histogram) and after (red histogram) fudging.

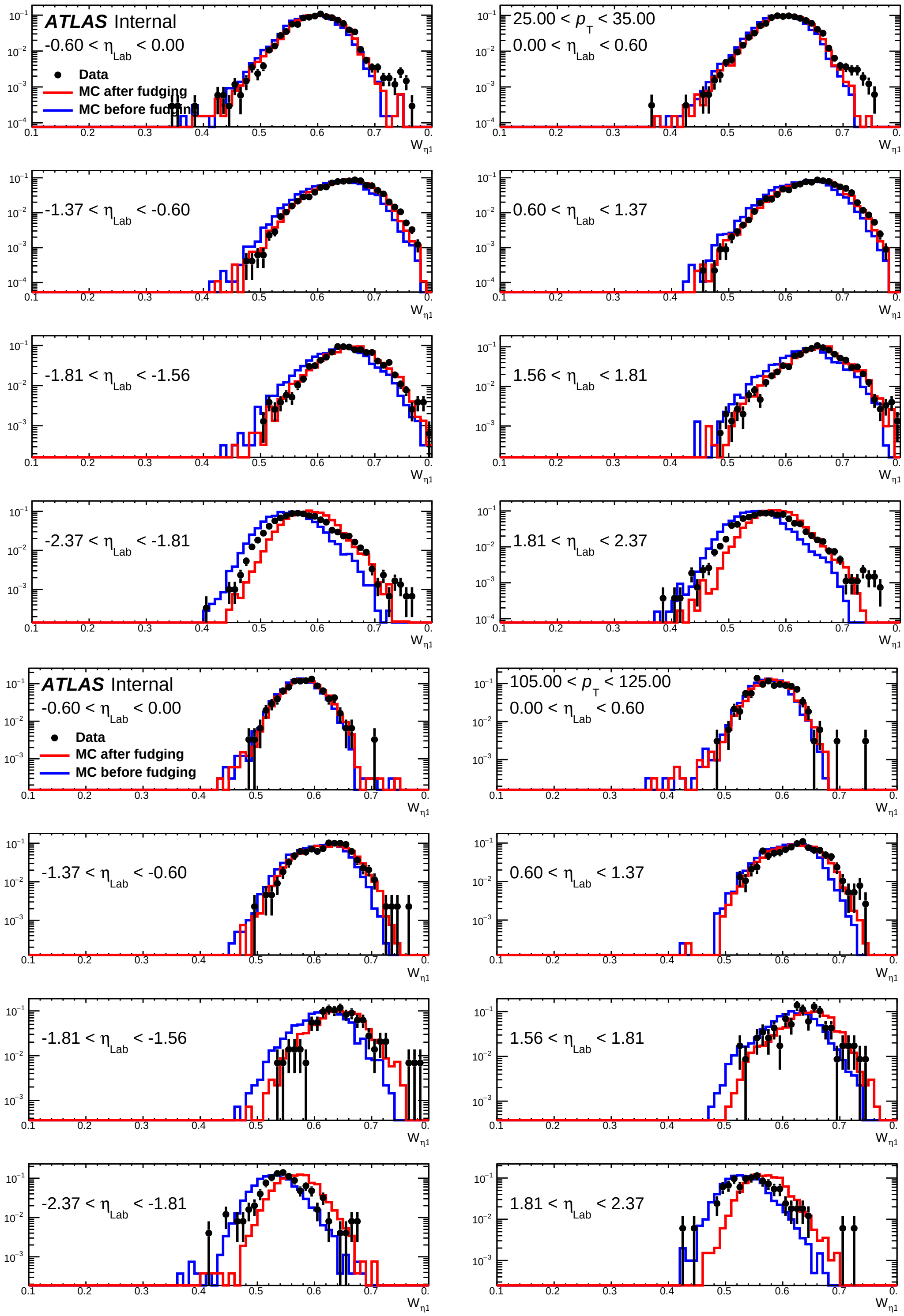

Figure A.11: Shower shape parameter, $W_{\eta 1}$, in each pseudorapidity slice from $E_{\mathrm{T}}$ bins 25 GeV $< E_{\mathrm{T}} <$ 35 GeV (above) and 105 GeV $< E_{\mathrm{T}} <$ 125 GeV (below). Reconstructed data plotted as black points overlaid with MC before (blue histogram) and after (red histogram) fudging.

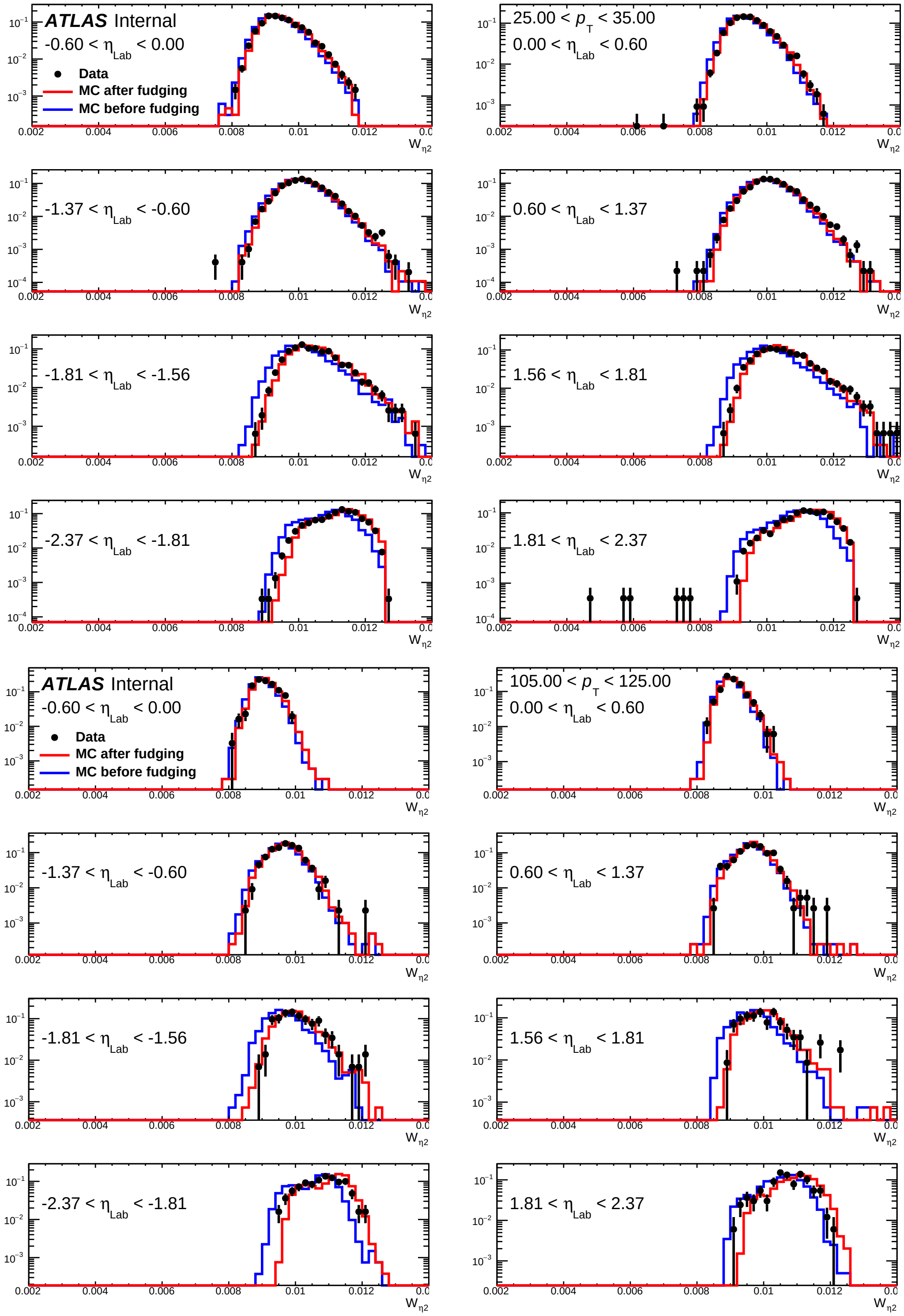

Figure A.12: Shower shape parameter, $W_{\eta 2}$, in each pseudorapidity slice from $E_{\mathrm{T}}$ bins 25 GeV $< E_{\mathrm{T}} <$ 35 GeV (above) and 105 GeV $< E_{\mathrm{T}} <$ 125 GeV (below). Reconstructed data plotted as black points overlaid with MC before (blue histogram) and after (red histogram) fudging.

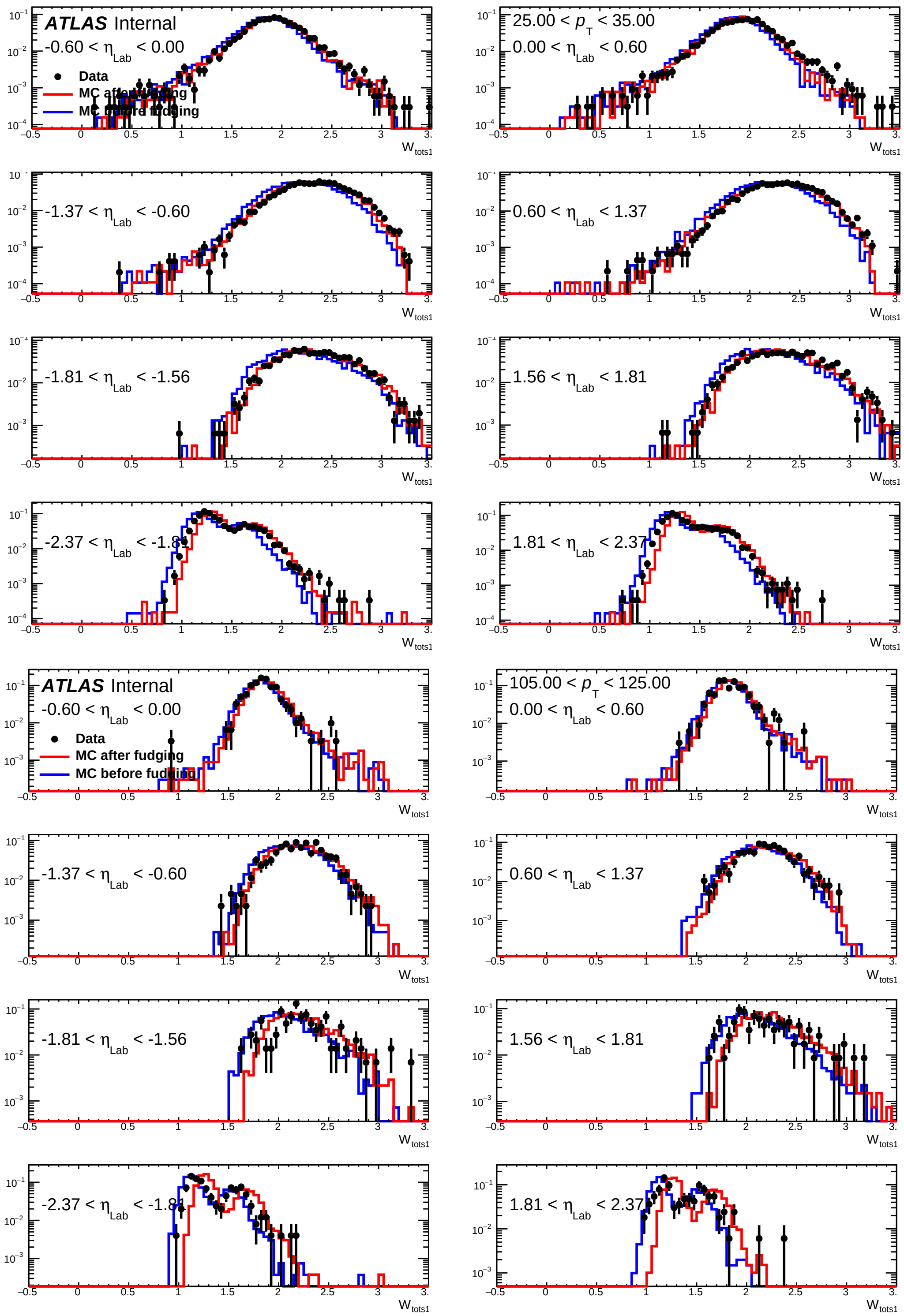

Figure A.13: Shower shape parameter, $W_{totes1}$, in each pseudorapidity slice from $E_{\mathrm{T}}$ bins 25 GeV $< E_{\mathrm{T}} <$ 35 GeV (above) and 105 GeV $< E_{\mathrm{T}} <$ 125 GeV (below). Reconstructed data plotted as black points overlaid with MC before (blue histogram) and after (red histogram) fudging.

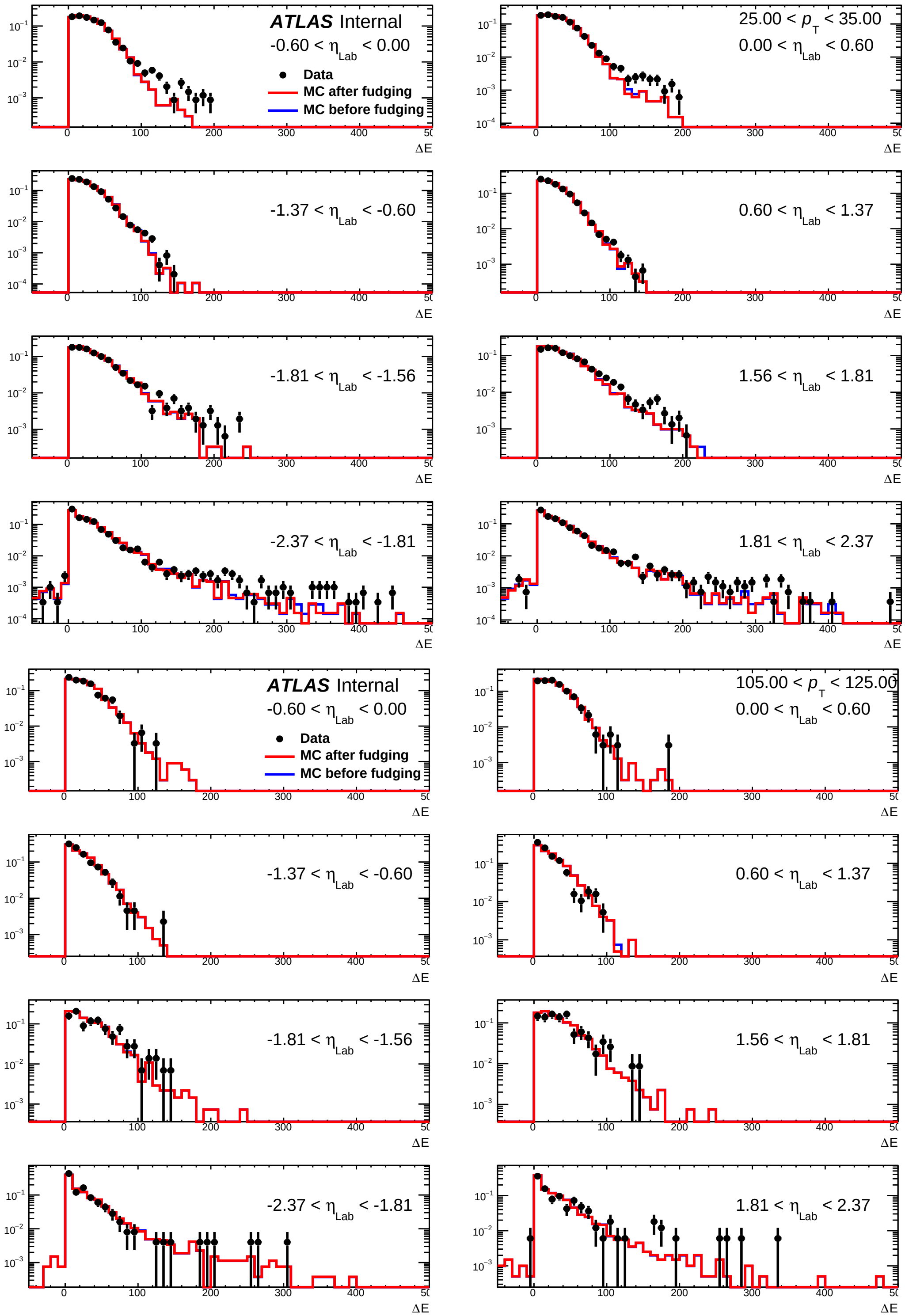

Figure A.14: Shower shape parameter, $\Delta E$, in each pseudorapidity slice from $E_{\mathrm{T}}$ bins 25 GeV $< E_{\mathrm{T}} <$ 35 GeV (above) and 105 GeV $< E_{\mathrm{T}} <$ 125 GeV (below). Reconstructed data plotted as black points overlaid with MC before (blue histogram) and after (red histogram) fudging.

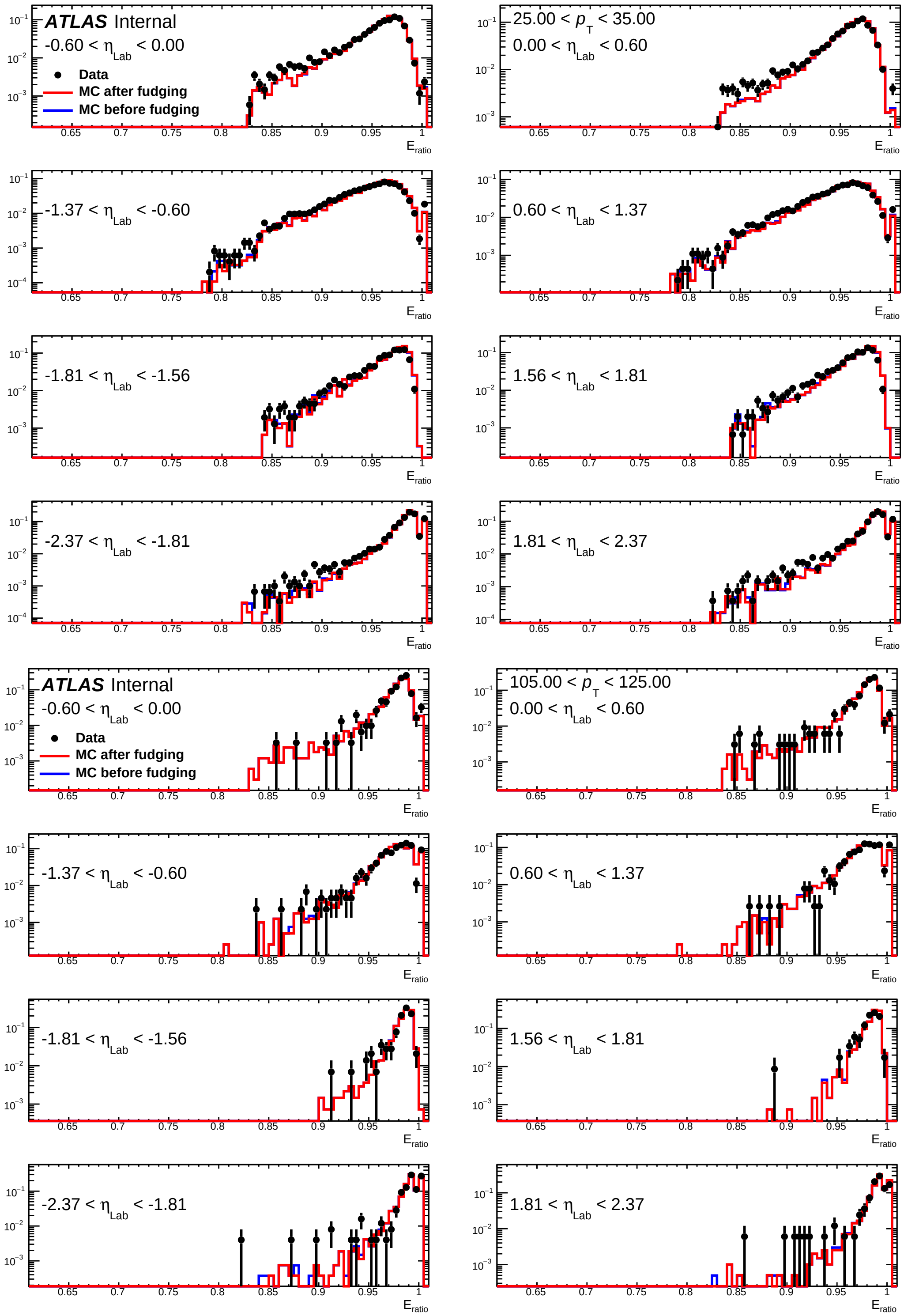

Figure A.15: Shower shape parameter, $E_{ratio}$, in each pseudorapidity slice from $E_{\mathrm{T}}$ bins $25\,\mathrm{GeV} < E_{\mathrm{T}} < 35\,\mathrm{GeV}$ (above) and $105\,\mathrm{GeV} < E_{\mathrm{T}} < 125\,\mathrm{GeV}$ (below). Reconstructed data plotted as black points overlaid with MC before (blue histogram) and after (red histogram) fudging.

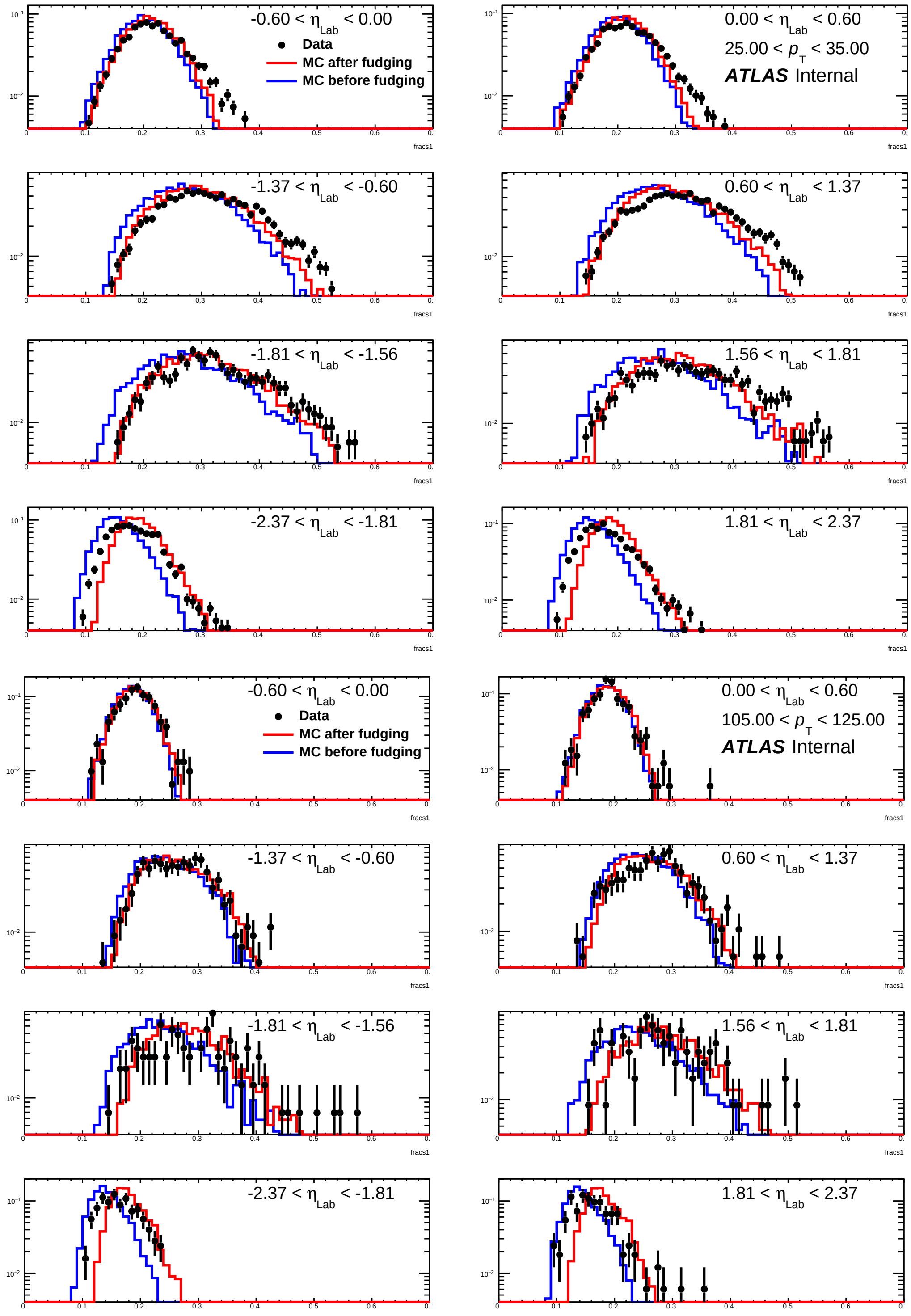

Figure A.16: Shower shape parameter, fracs1, in each pseudorapidity slice from $E_{\mathrm{T}}$ bins $25\,\mathrm{GeV} < E_{\mathrm{T}} < 35\,\mathrm{GeV}$ (above) and $105\,\mathrm{GeV} < E_{\mathrm{T}} < 125\,\mathrm{GeV}$ (below). Reconstructed data plotted as black points overlaid with MC before (blue histogram) and after (red histogram) fudging.

## A.8 Photon Isolation

This section contains the full set of isolation energy distributions for ”Tight” identified photons in data (black histogram) overlaid with ”Non-tight” designated photons (blue histogram) which are selected to increase backgrounds. The Non-tight histograms are scaled to match the integral of the tight histogram above 7 GeV. Additionally, the distribution of tight and truth-isolated reconstructed photons from MC are overlaid (red broken histogram). The second panel on each page compares the MC to the data isolation energy with the ”Non-Tight” subtracted from the ”Tight”. The distributions are plotted in their usual $\eta$ slices for each $E_{\mathrm{T}}$ bin.

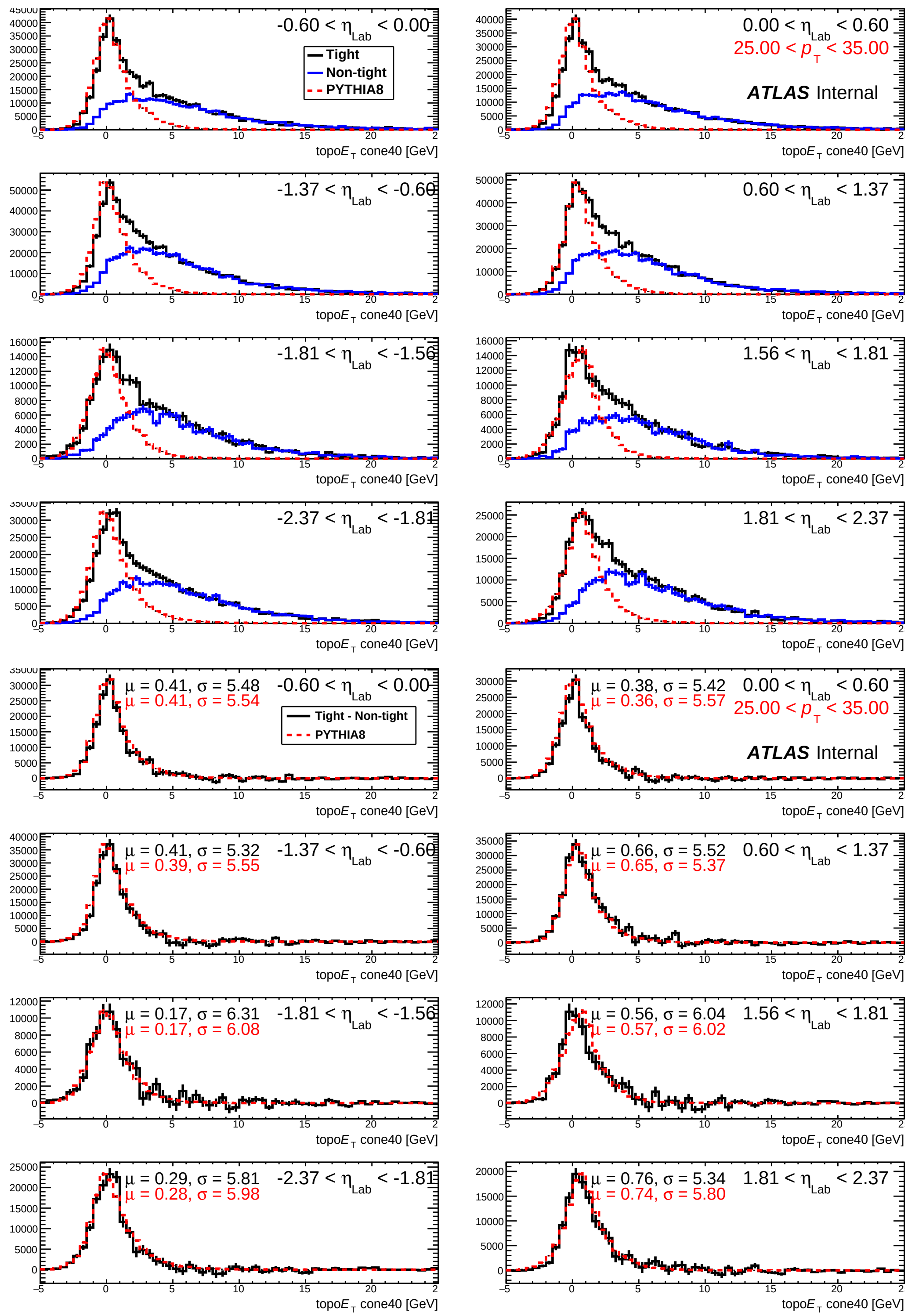
-0.60 < η_Lab < 0.00
Tight
Non-tight
PYTHIA8
0.00 < η_Lab < 0.60
25.00 < p_T < 35.00
ATLAS Internal
-1.37 < η_Lab < -0.60
0.60 < η_Lab < 1.37
-1.81 < η_Lab < -1.56
1.56 < η_Lab < 1.81
-2.37 < η_Lab < -1.81
1.81 < η_Lab < 2.37
topoE_T cone40 [GeV]
μ = 0.41, σ = 5.48
μ = 0.41, σ = 5.54
Tight - Non-tight
μ = 0.38, σ = 5.42
μ = 0.36, σ = 5.57
μ = 0.41, σ = 5.32
μ = 0.39, σ = 5.55
μ = 0.66, σ = 5.52
μ = 0.65, σ = 5.37
μ = 0.17, σ = 6.31
μ = 0.17, σ = 6.08
μ = 0.56, σ = 6.04
μ = 0.57, σ = 6.02
μ = 0.29, σ = 5.81
μ = 0.28, σ = 5.98
μ = 0.76, σ = 5.34
μ = 0.74, σ = 5.80

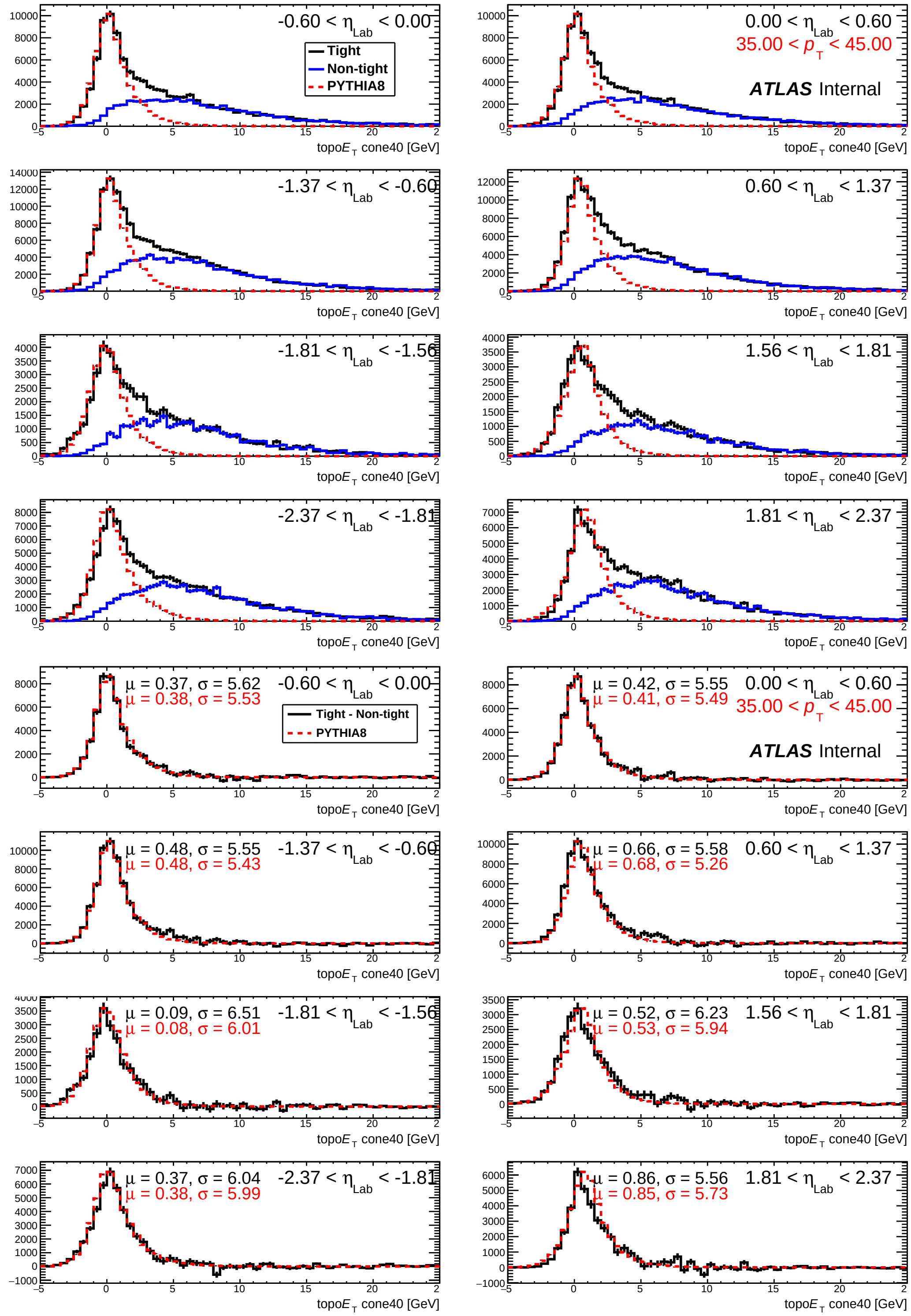
-0.60 < η_Lab < 0.00
Tight
Non-tight
PYTHIA8
0.00 < η_Lab < 0.60
35.00 < p_T < 45.00
ATLAS Internal
-1.37 < η_Lab < -0.60
0.60 < η_Lab < 1.37
-1.81 < η_Lab < -1.56
1.56 < η_Lab < 1.81
-2.37 < η_Lab < -1.81
1.81 < η_Lab < 2.37
μ = 0.37, σ = 5.62
μ = 0.38, σ = 5.53
Tight - Non-tight
μ = 0.42, σ = 5.55
μ = 0.41, σ = 5.49
μ = 0.48, σ = 5.55
μ = 0.48, σ = 5.43
μ = 0.66, σ = 5.58
μ = 0.68, σ = 5.26
μ = 0.09, σ = 6.51
μ = 0.08, σ = 6.01
μ = 0.52, σ = 6.23
μ = 0.53, σ = 5.94
μ = 0.37, σ = 6.04
μ = 0.38, σ = 5.99
μ = 0.86, σ = 5.56
μ = 0.85, σ = 5.73
topoE_T cone40 [GeV]

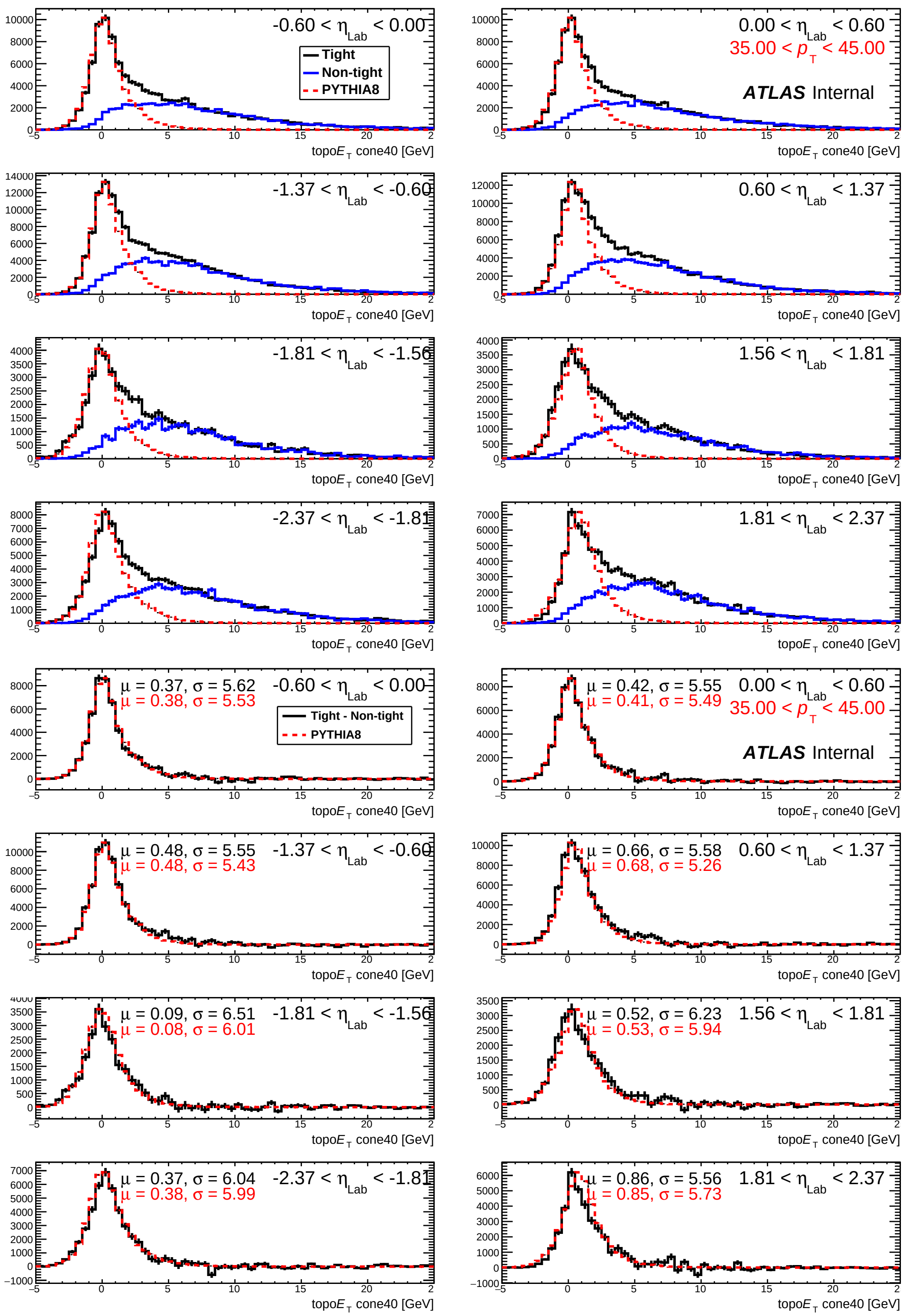
-0.60 < η_Lab < 0.00
Tight
Non-tight
PYTHIA8
0.00 < η_Lab < 0.60
35.00 < p_T < 45.00
ATLAS Internal
-1.37 < η_Lab < -0.60
0.60 < η_Lab < 1.37
-1.81 < η_Lab < -1.56
1.56 < η_Lab < 1.81
-2.37 < η_Lab < -1.81
1.81 < η_Lab < 2.37
μ = 0.37, σ = 5.62
μ = 0.38, σ = 5.53
Tight - Non-tight
μ = 0.42, σ = 5.55
μ = 0.41, σ = 5.49
μ = 0.48, σ = 5.55
μ = 0.48, σ = 5.43
μ = 0.66, σ = 5.58
μ = 0.68, σ = 5.26
μ = 0.09, σ = 6.51
μ = 0.08, σ = 6.01
μ = 0.52, σ = 6.23
μ = 0.53, σ = 5.94
μ = 0.37, σ = 6.04
μ = 0.38, σ = 5.99
μ = 0.86, σ = 5.56
μ = 0.85, σ = 5.73
topoE_T cone40 [GeV]

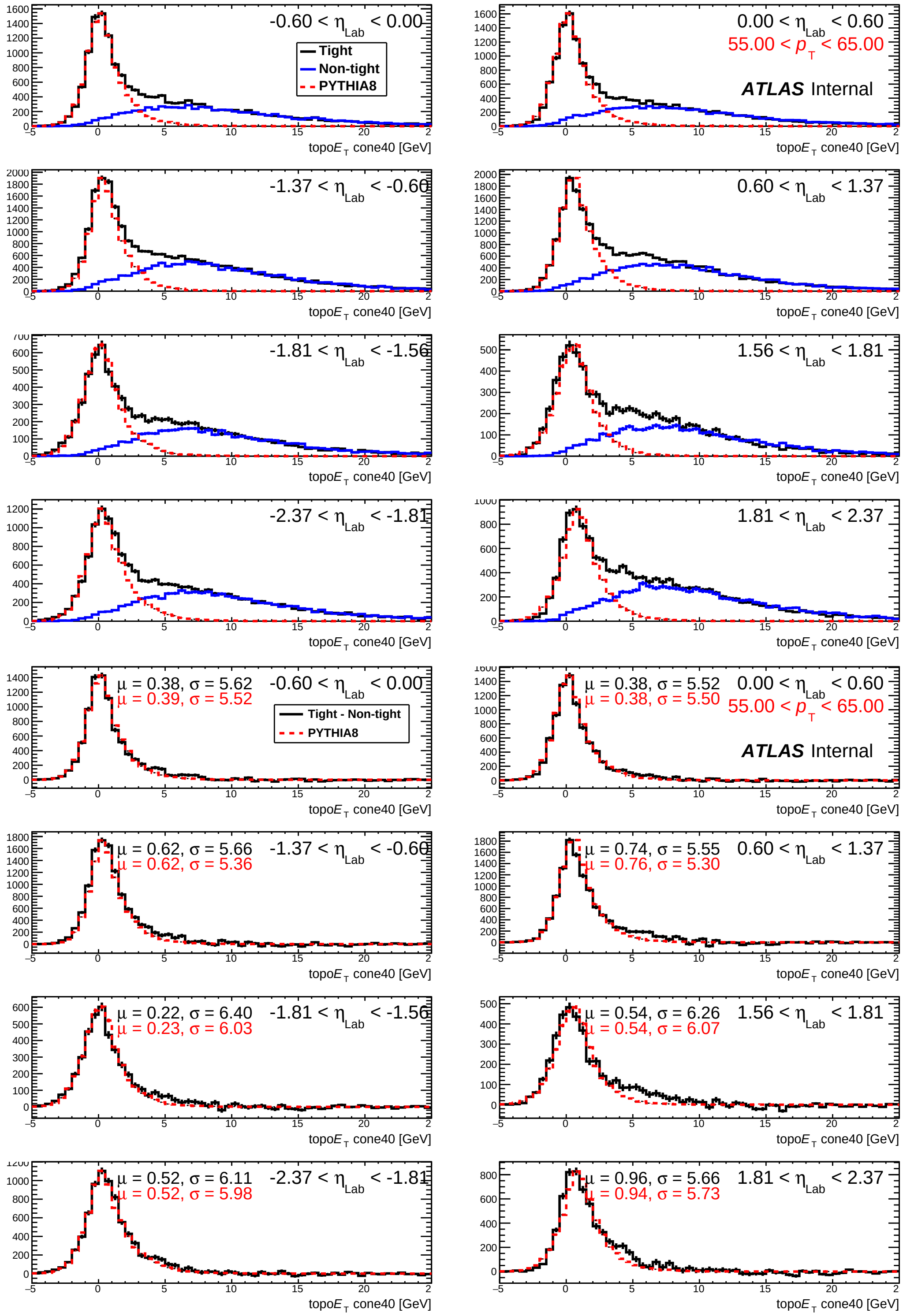

-0.60 < η_Lab < 0.00
Tight
Non-tight
PYTHIA8
0.00 < η_Lab < 0.60
55.00 < p_T < 65.00
ATLAS Internal
-1.37 < η_Lab < -0.60
0.60 < η_Lab < 1.37
-1.81 < η_Lab < -1.56
1.56 < η_Lab < 1.81
-2.37 < η_Lab < -1.81
1.81 < η_Lab < 2.37
μ = 0.38, σ = 5.62
μ = 0.39, σ = 5.52
Tight - Non-tight
μ = 0.38, σ = 5.52
μ = 0.38, σ = 5.50
μ = 0.62, σ = 5.66
μ = 0.62, σ = 5.36
μ = 0.74, σ = 5.55
μ = 0.76, σ = 5.30
μ = 0.22, σ = 6.40
μ = 0.23, σ = 6.03
μ = 0.54, σ = 6.26
μ = 0.54, σ = 6.07
μ = 0.52, σ = 6.11
μ = 0.52, σ = 5.98
μ = 0.96, σ = 5.66
μ = 0.94, σ = 5.73
topoE_T cone40 [GeV]

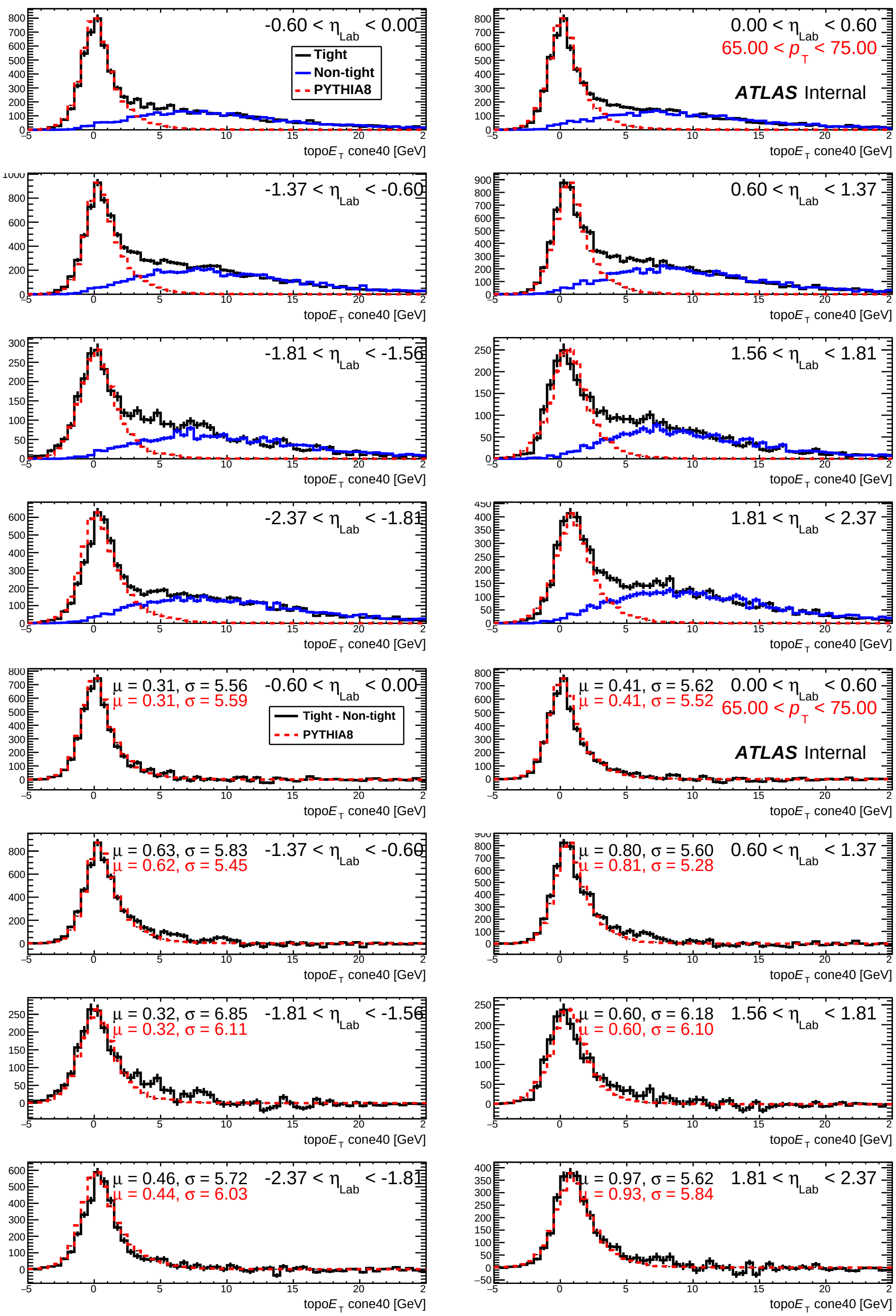
-0.60 < η_Lab < 0.00
Tight
Non-tight
PYTHIA8
0.00 < η_Lab < 0.60
65.00 < p_T < 75.00
ATLAS Internal
-1.37 < η_Lab < -0.60
0.60 < η_Lab < 1.37
-1.81 < η_Lab < -1.56
1.56 < η_Lab < 1.81
-2.37 < η_Lab < -1.81
1.81 < η_Lab < 2.37
μ = 0.31, σ = 5.56
μ = 0.31, σ = 5.59
Tight - Non-tight
PYTHIA8
μ = 0.41, σ = 5.62
μ = 0.41, σ = 5.52
μ = 0.63, σ = 5.83
μ = 0.62, σ = 5.45
μ = 0.80, σ = 5.60
μ = 0.81, σ = 5.28
μ = 0.32, σ = 6.85
μ = 0.32, σ = 6.11
μ = 0.60, σ = 6.18
μ = 0.60, σ = 6.10
μ = 0.46, σ = 5.72
μ = 0.44, σ = 6.03
μ = 0.97, σ = 5.62
μ = 0.93, σ = 5.84
topoE_T cone40 [GeV]

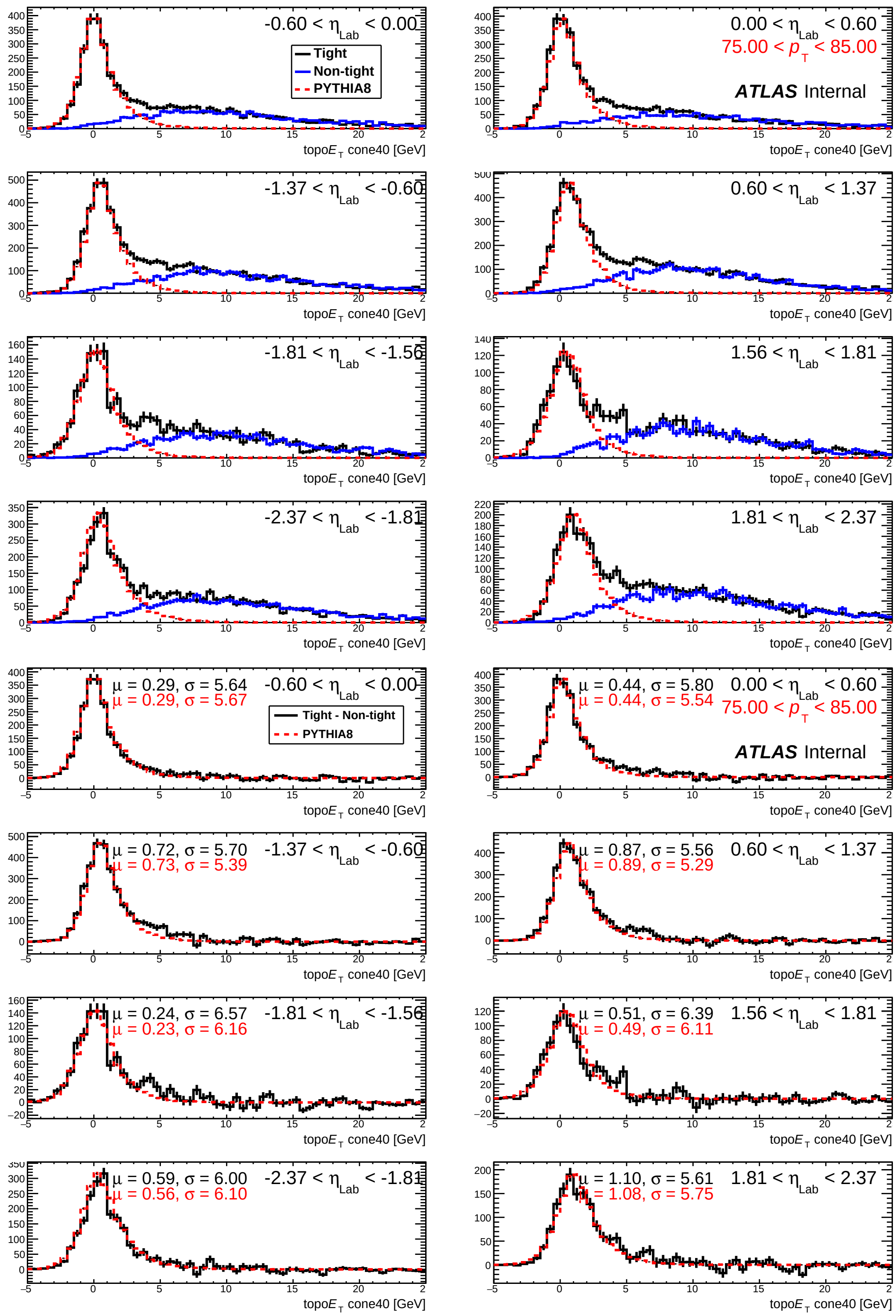
-0.60 < η_Lab < 0.00
Tight
Non-tight
PYTHIA8
0.00 < η_Lab < 0.60
75.00 < p_T < 85.00
ATLAS Internal
-1.37 < η_Lab < -0.60
0.60 < η_Lab < 1.37
-1.81 < η_Lab < -1.56
1.56 < η_Lab < 1.81
-2.37 < η_Lab < -1.81
1.81 < η_Lab < 2.37
μ = 0.29, σ = 5.64
μ = 0.29, σ = 5.67
Tight - Non-tight
μ = 0.44, σ = 5.80
μ = 0.44, σ = 5.54
μ = 0.72, σ = 5.70
μ = 0.73, σ = 5.39
μ = 0.87, σ = 5.56
μ = 0.89, σ = 5.29
μ = 0.24, σ = 6.57
μ = 0.23, σ = 6.16
μ = 0.51, σ = 6.39
μ = 0.49, σ = 6.11
μ = 0.59, σ = 6.00
μ = 0.56, σ = 6.10
μ = 1.10, σ = 5.61
μ = 1.08, σ = 5.75
topoE_T cone40 [GeV]

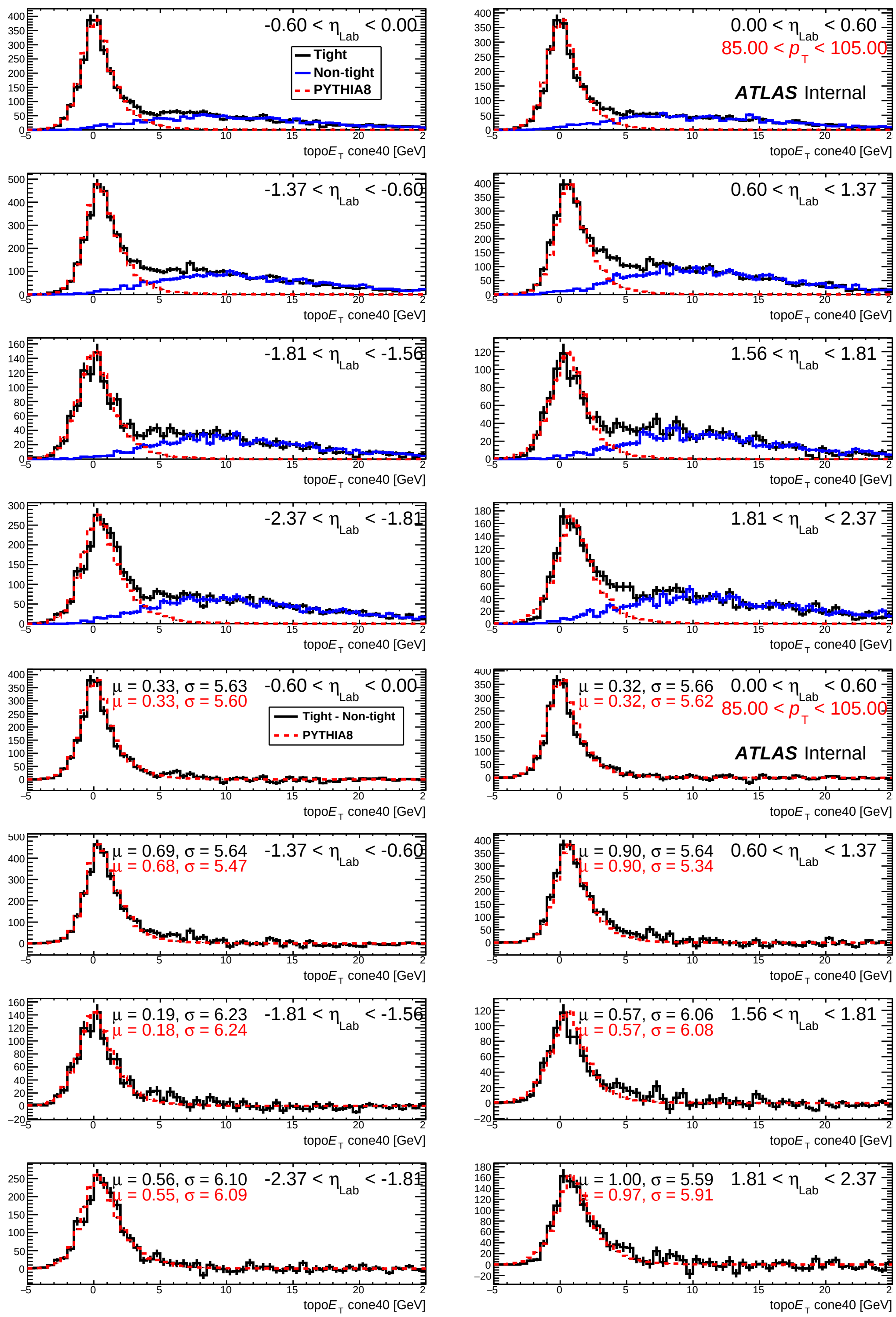
-0.60 < η_Lab < 0.00
Tight
Non-tight
PYTHIA8
0.00 < η_Lab < 0.60
85.00 < p_T < 105.00
ATLAS Internal
-1.37 < η_Lab < -0.60
0.60 < η_Lab < 1.37
-1.81 < η_Lab < -1.56
1.56 < η_Lab < 1.81
-2.37 < η_Lab < -1.81
1.81 < η_Lab < 2.37
μ = 0.33, σ = 5.63
μ = 0.33, σ = 5.60
Tight - Non-tight
PYTHIA8
μ = 0.32, σ = 5.66
μ = 0.32, σ = 5.62
μ = 0.69, σ = 5.64
μ = 0.68, σ = 5.47
μ = 0.90, σ = 5.64
μ = 0.90, σ = 5.34
μ = 0.19, σ = 6.23
μ = 0.18, σ = 6.24
μ = 0.57, σ = 6.06
μ = 0.57, σ = 6.08
μ = 0.56, σ = 6.10
μ = 0.55, σ = 6.09
μ = 1.00, σ = 5.59
μ = 0.97, σ = 5.91
topoE_T cone40 [GeV]

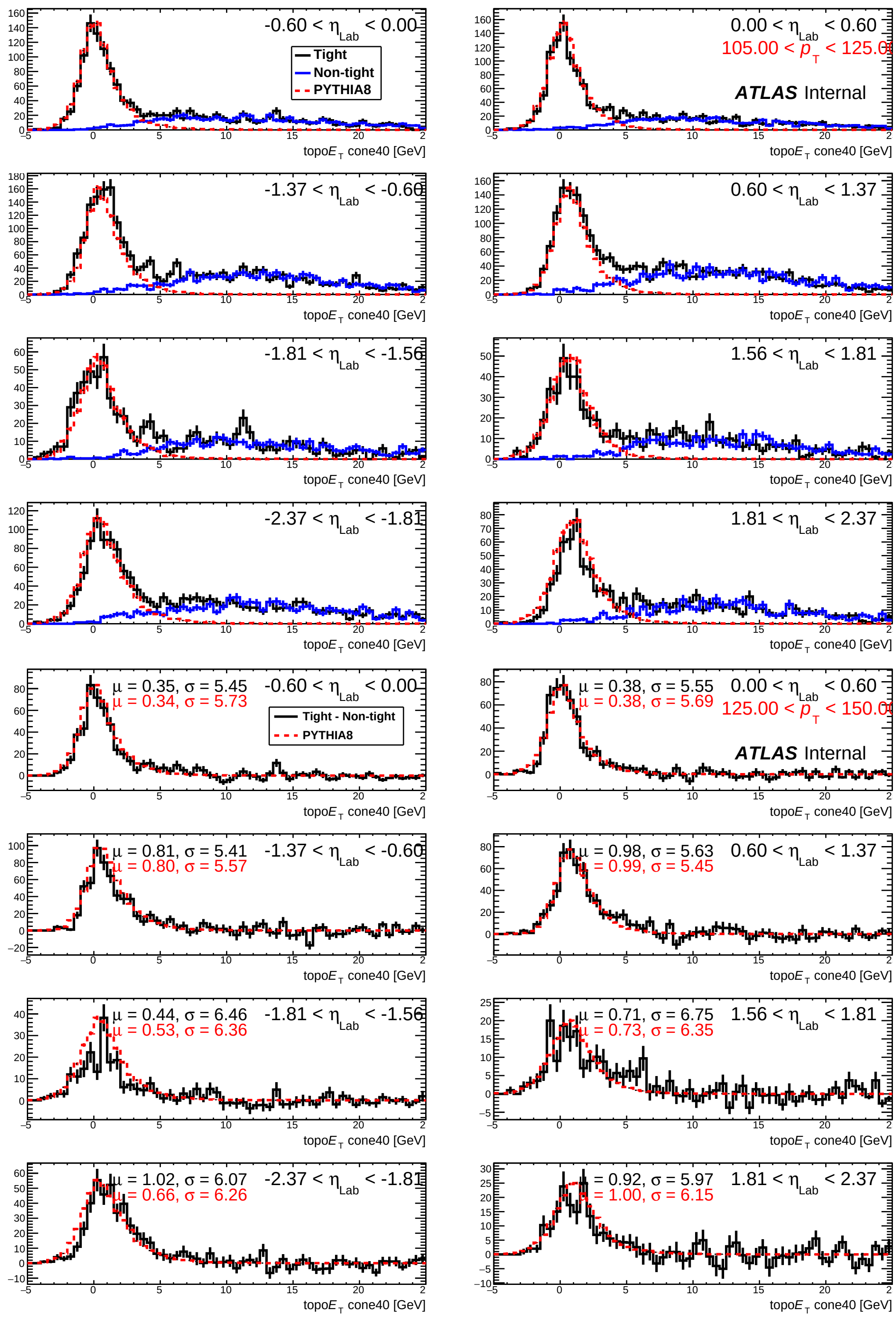
-0.60 < η_Lab < 0.00
Tight
Non-tight
PYTHIA8
0.00 < η_Lab < 0.60
105.00 < p_T < 125.0
ATLAS Internal
-1.37 < η_Lab < -0.60
0.60 < η_Lab < 1.37
-1.81 < η_Lab < -1.56
1.56 < η_Lab < 1.81
-2.37 < η_Lab < -1.81
1.81 < η_Lab < 2.37
μ = 0.35, σ = 5.45
μ = 0.34, σ = 5.73
Tight - Non-tight
PYTHIA8
μ = 0.38, σ = 5.55
μ = 0.38, σ = 5.69
125.00 < p_T < 150.0
μ = 0.81, σ = 5.41
μ = 0.80, σ = 5.57
μ = 0.98, σ = 5.63
μ = 0.99, σ = 5.45
μ = 0.44, σ = 6.46
μ = 0.53, σ = 6.36
μ = 0.71, σ = 6.75
μ = 0.73, σ = 6.35
μ = 1.02, σ = 6.07
μ = 0.66, σ = 6.26
μ = 0.92, σ = 5.97
μ = 1.00, σ = 6.15
topoE_T cone40 [GeV]

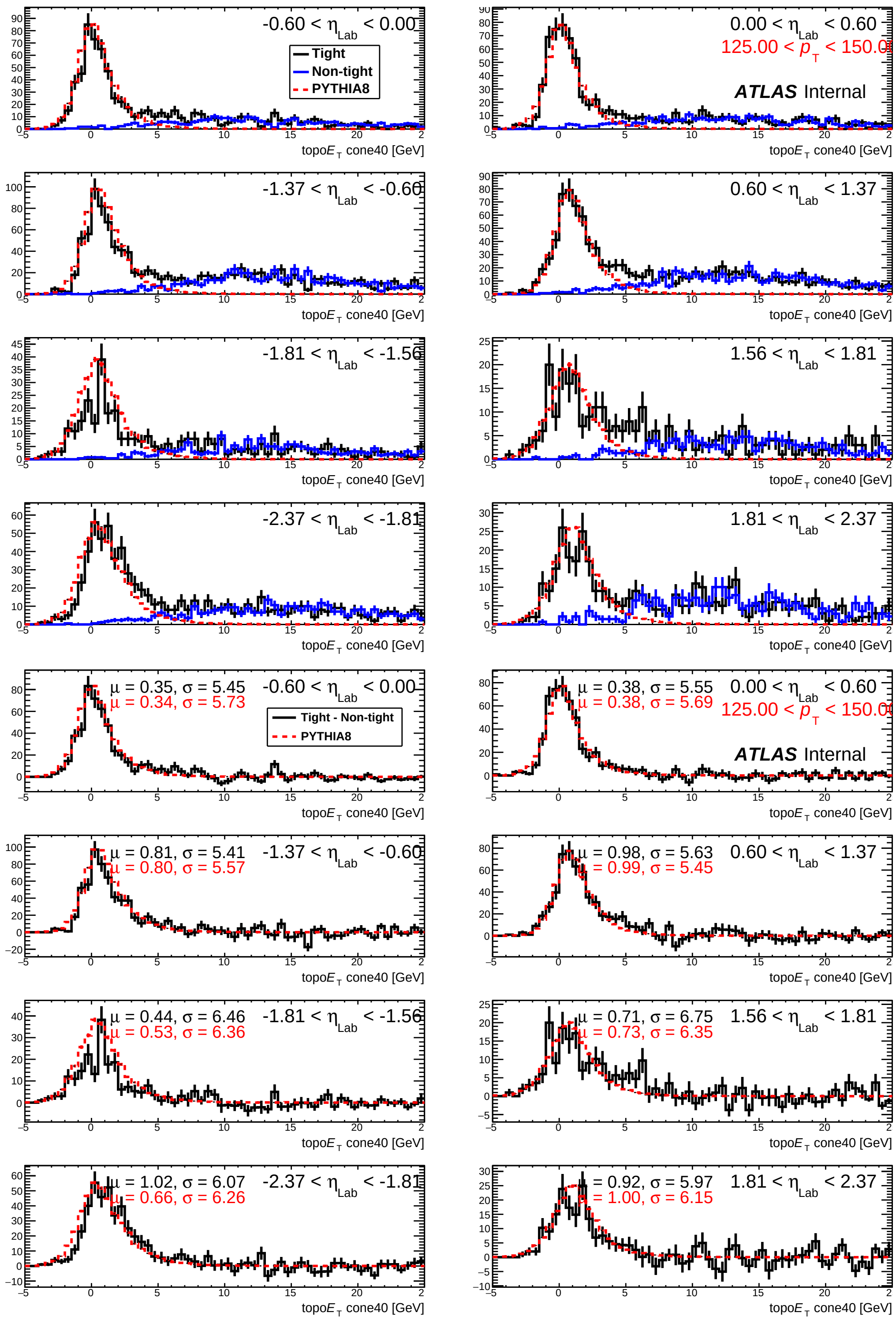
-0.60 < η_Lab < 0.00
Tight
Non-tight
PYTHIA8
0.00 < η_Lab < 0.60
125.00 < p_T < 150.00
ATLAS Internal
-1.37 < η_Lab < -0.60
0.60 < η_Lab < 1.37
-1.81 < η_Lab < -1.56
1.56 < η_Lab < 1.81
-2.37 < η_Lab < -1.81
1.81 < η_Lab < 2.37
μ = 0.35, σ = 5.45
μ = 0.34, σ = 5.73
Tight - Non-tight
μ = 0.38, σ = 5.55
μ = 0.38, σ = 5.69
μ = 0.81, σ = 5.41
μ = 0.80, σ = 5.57
μ = 0.98, σ = 5.63
μ = 0.99, σ = 5.45
μ = 0.44, σ = 6.46
μ = 0.53, σ = 6.36
μ = 0.71, σ = 6.75
μ = 0.73, σ = 6.35
μ = 1.02, σ = 6.07
μ = 0.66, σ = 6.26
μ = 0.92, σ = 5.97
μ = 1.00, σ = 6.15
topoE_T cone40 [GeV]

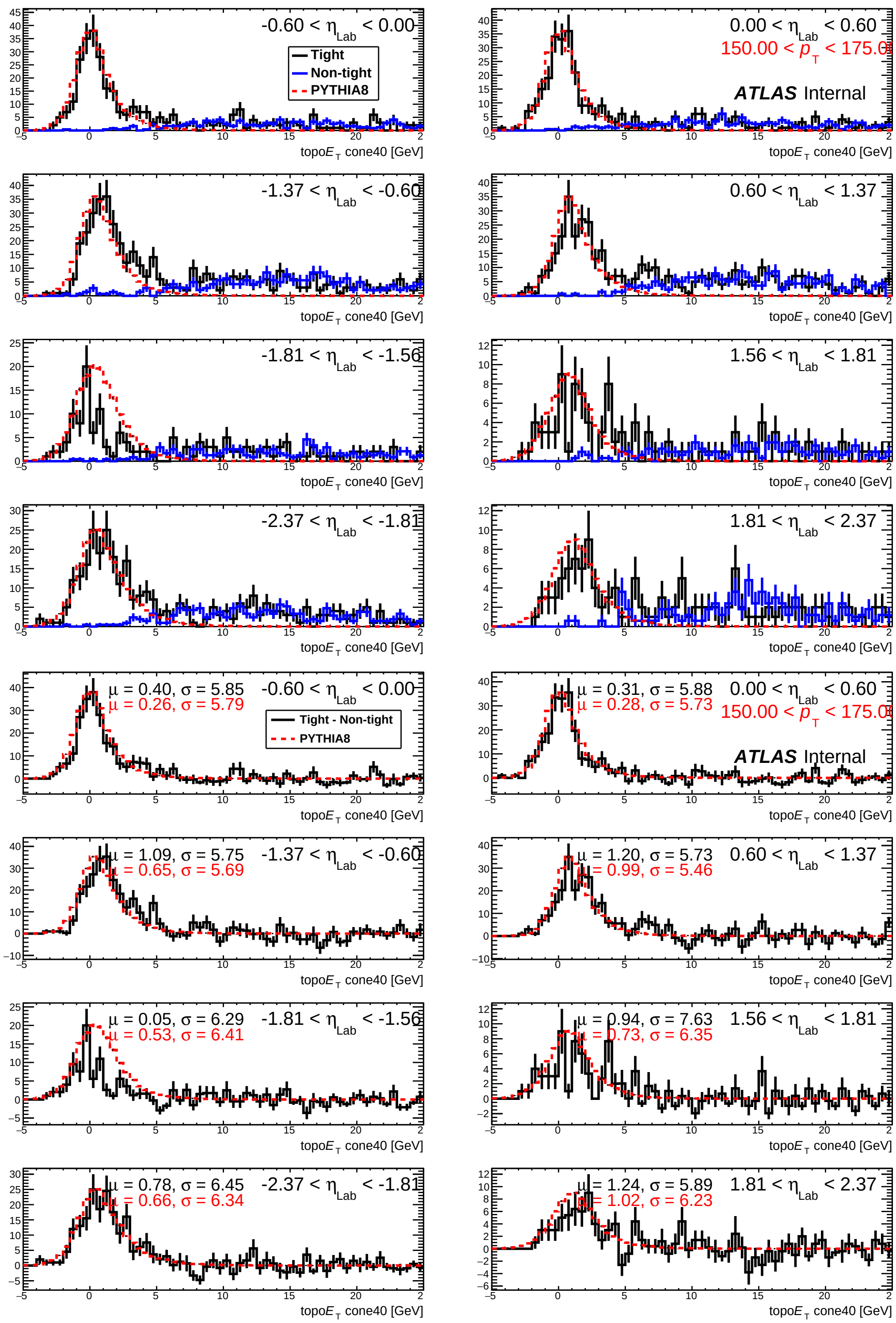
-0.60 < η_Lab < 0.00
Tight
Non-tight
PYTHIA8
topoE_T cone40 [GeV]
0.00 < η_Lab < 0.60
150.00 < p_T < 175.0
ATLAS Internal
-1.37 < η_Lab < -0.60
0.60 < η_Lab < 1.37
-1.81 < η_Lab < -1.56
1.56 < η_Lab < 1.81
-2.37 < η_Lab < -1.81
1.81 < η_Lab < 2.37
μ = 0.40, σ = 5.85
μ = 0.26, σ = 5.79
Tight - Non-tight
PYTHIA8
μ = 0.31, σ = 5.88
μ = 0.28, σ = 5.73
μ = 1.09, σ = 5.75
μ = 0.65, σ = 5.69
μ = 1.20, σ = 5.73
μ = 0.99, σ = 5.46
μ = 0.05, σ = 6.29
μ = 0.53, σ = 6.41
μ = 0.94, σ = 7.63
μ = 0.73, σ = 6.35
μ = 0.78, σ = 6.45
μ = 0.66, σ = 6.34
μ = 1.24, σ = 5.89
μ = 1.02, σ = 6.23

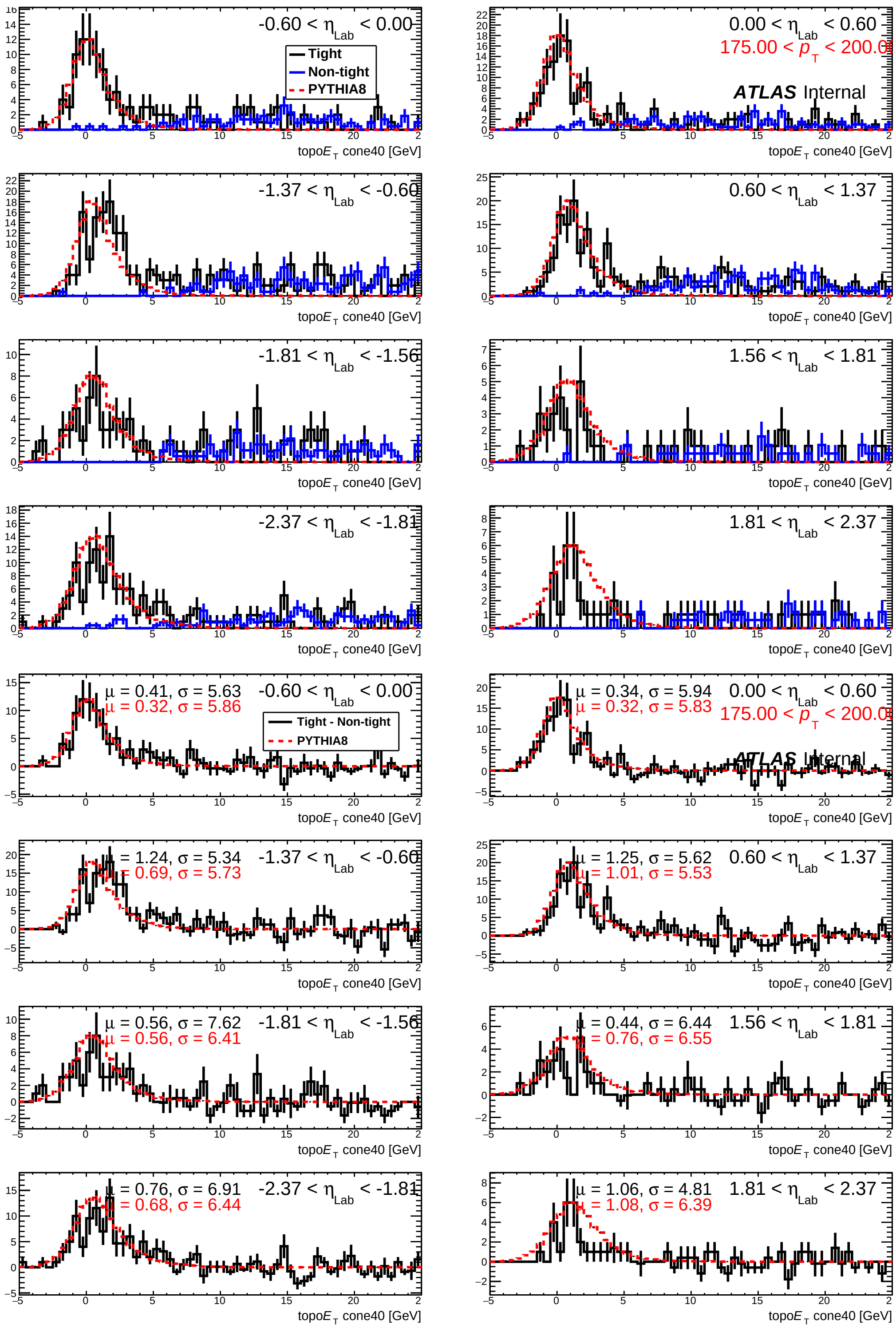
-0.60 < η_Lab < 0.00
Tight
Non-tight
PYTHIA8
0.00 < η_Lab < 0.60
175.00 < p_T < 200.00
ATLAS Internal
-1.37 < η_Lab < -0.60
0.60 < η_Lab < 1.37
-1.81 < η_Lab < -1.56
1.56 < η_Lab < 1.81
-2.37 < η_Lab < -1.81
1.81 < η_Lab < 2.37
μ = 0.41, σ = 5.63
μ = 0.32, σ = 5.86
Tight - Non-tight
μ = 0.34, σ = 5.94
μ = 0.32, σ = 5.83
μ = 1.24, σ = 5.34
μ = 0.69, σ = 5.73
μ = 1.25, σ = 5.62
μ = 1.01, σ = 5.53
μ = 0.56, σ = 7.62
μ = 0.56, σ = 6.41
μ = 0.44, σ = 6.44
μ = 0.76, σ = 6.55
μ = 0.76, σ = 6.91
μ = 0.68, σ = 6.44
μ = 1.06, σ = 4.81
μ = 1.08, σ = 6.39
topoE_T cone40 [GeV]

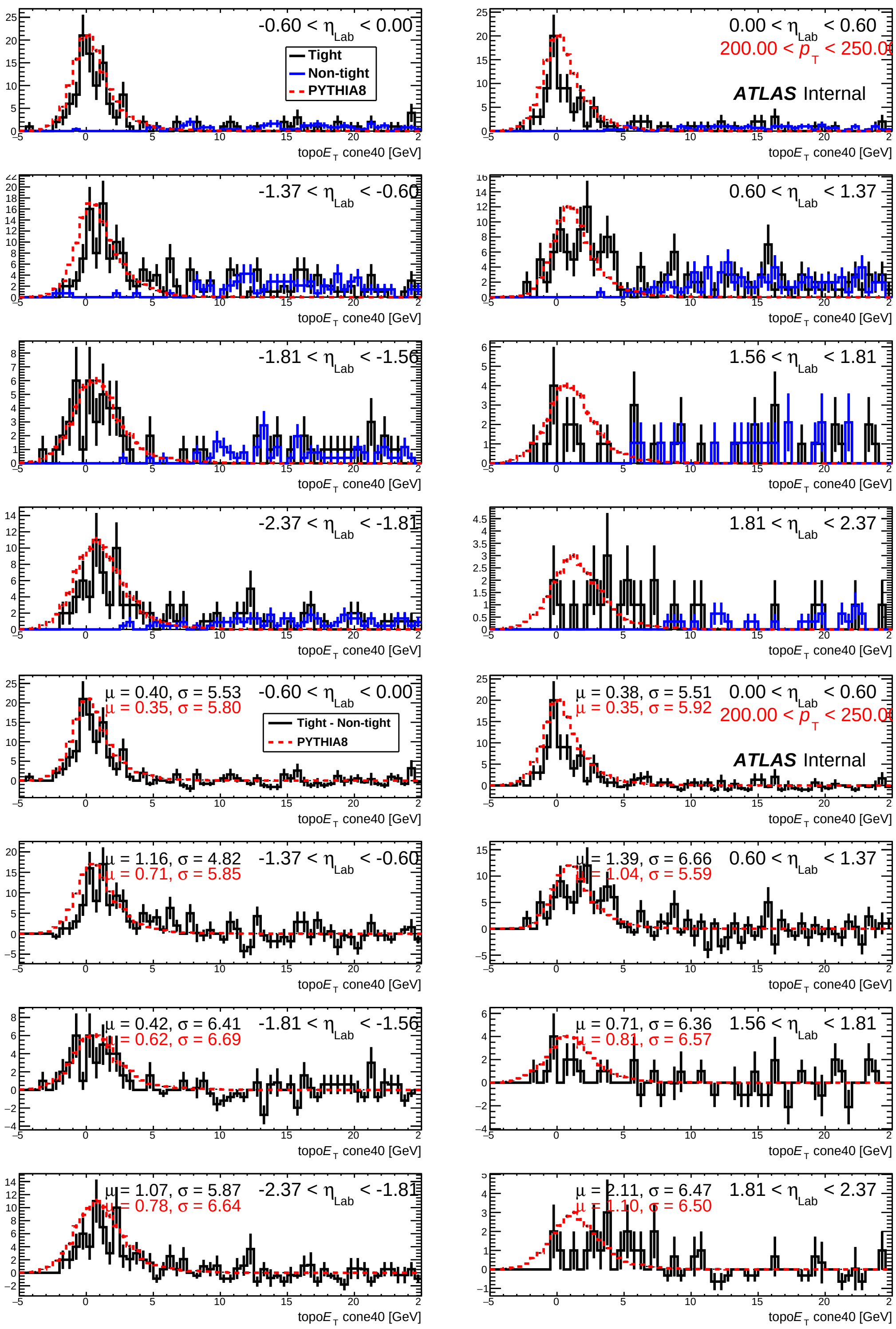
-0.60 < η_Lab < 0.00
Tight
Non-tight
PYTHIA8
0.00 < η_Lab < 0.60
200.00 < p_T < 250.00
ATLAS Internal
-1.37 < η_Lab < -0.60
0.60 < η_Lab < 1.37
-1.81 < η_Lab < -1.56
1.56 < η_Lab < 1.81
-2.37 < η_Lab < -1.81
1.81 < η_Lab < 2.37
μ = 0.40, σ = 5.53
μ = 0.35, σ = 5.80
Tight - Non-tight
μ = 0.38, σ = 5.51
μ = 0.35, σ = 5.92
μ = 1.16, σ = 4.82
μ = 0.71, σ = 5.85
μ = 1.39, σ = 6.66
μ = 1.04, σ = 5.59
μ = 0.42, σ = 6.41
μ = 0.62, σ = 6.69
μ = 0.71, σ = 6.36
μ = 0.81, σ = 6.57
μ = 1.07, σ = 5.87
μ = 0.78, σ = 6.64
μ = 2.11, σ = 6.47
μ = 1.10, σ = 6.50
topoE_T cone40 [GeV]

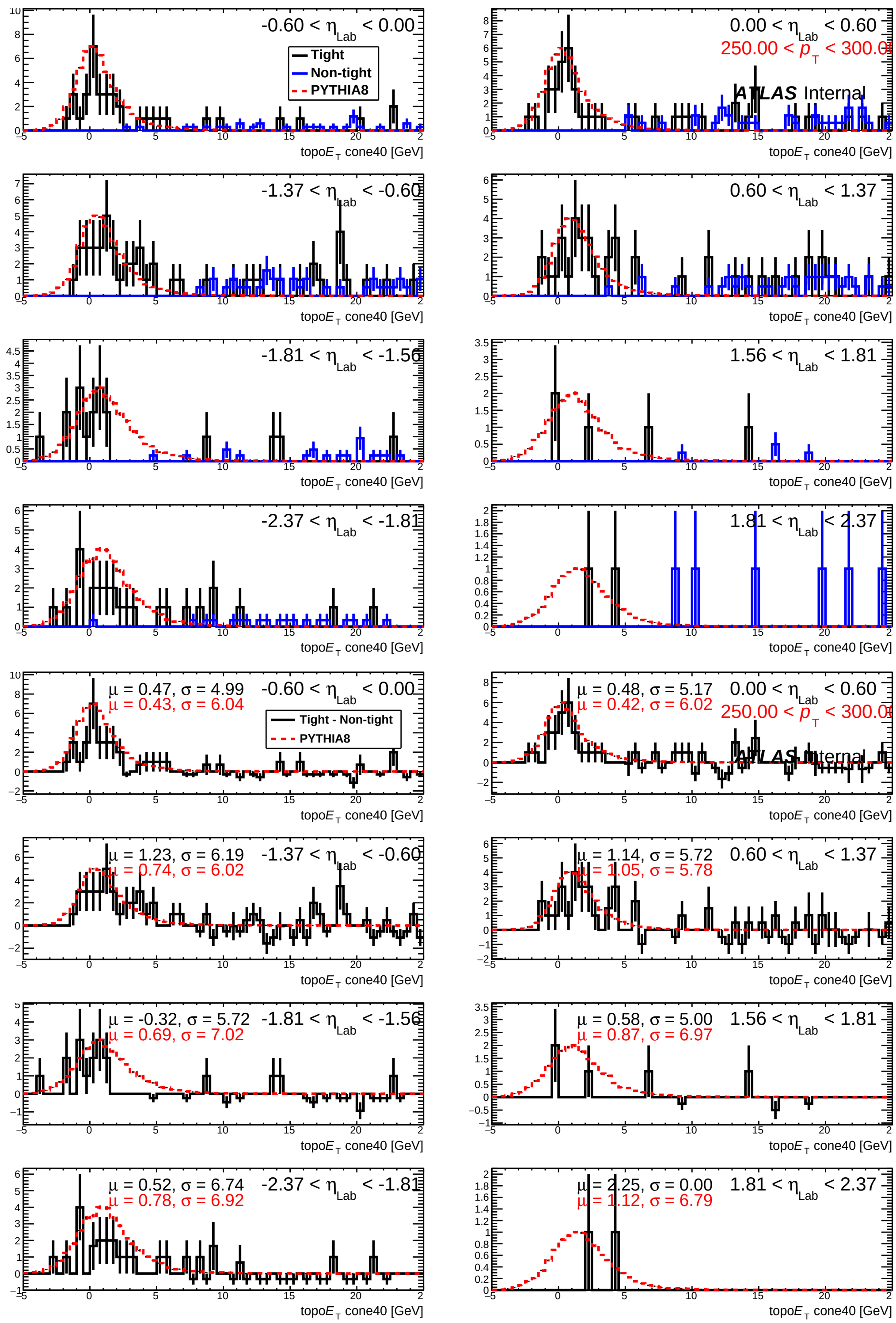
-0.60 < η_Lab < 0.00
Tight
Non-tight
PYTHIA8
0.00 < η_Lab < 0.60
250.00 < p_T < 300.00
ATLAS Internal
-1.37 < η_Lab < -0.60
0.60 < η_Lab < 1.37
-1.81 < η_Lab < -1.56
1.56 < η_Lab < 1.81
-2.37 < η_Lab < -1.81
1.81 < η_Lab < 2.37
μ = 0.47, σ = 4.99
μ = 0.43, σ = 6.04
Tight - Non-tight
μ = 0.48, σ = 5.17
μ = 0.42, σ = 6.02
μ = 1.23, σ = 6.19
μ = 0.74, σ = 6.02
μ = 1.14, σ = 5.72
μ = 1.05, σ = 5.78
μ = -0.32, σ = 5.72
μ = 0.69, σ = 7.02
μ = 0.58, σ = 5.00
μ = 0.87, σ = 6.97
μ = 0.52, σ = 6.74
μ = 0.78, σ = 6.92
μ = 2.25, σ = 0.00
μ = 1.12, σ = 6.79
topoE_T cone40 [GeV]

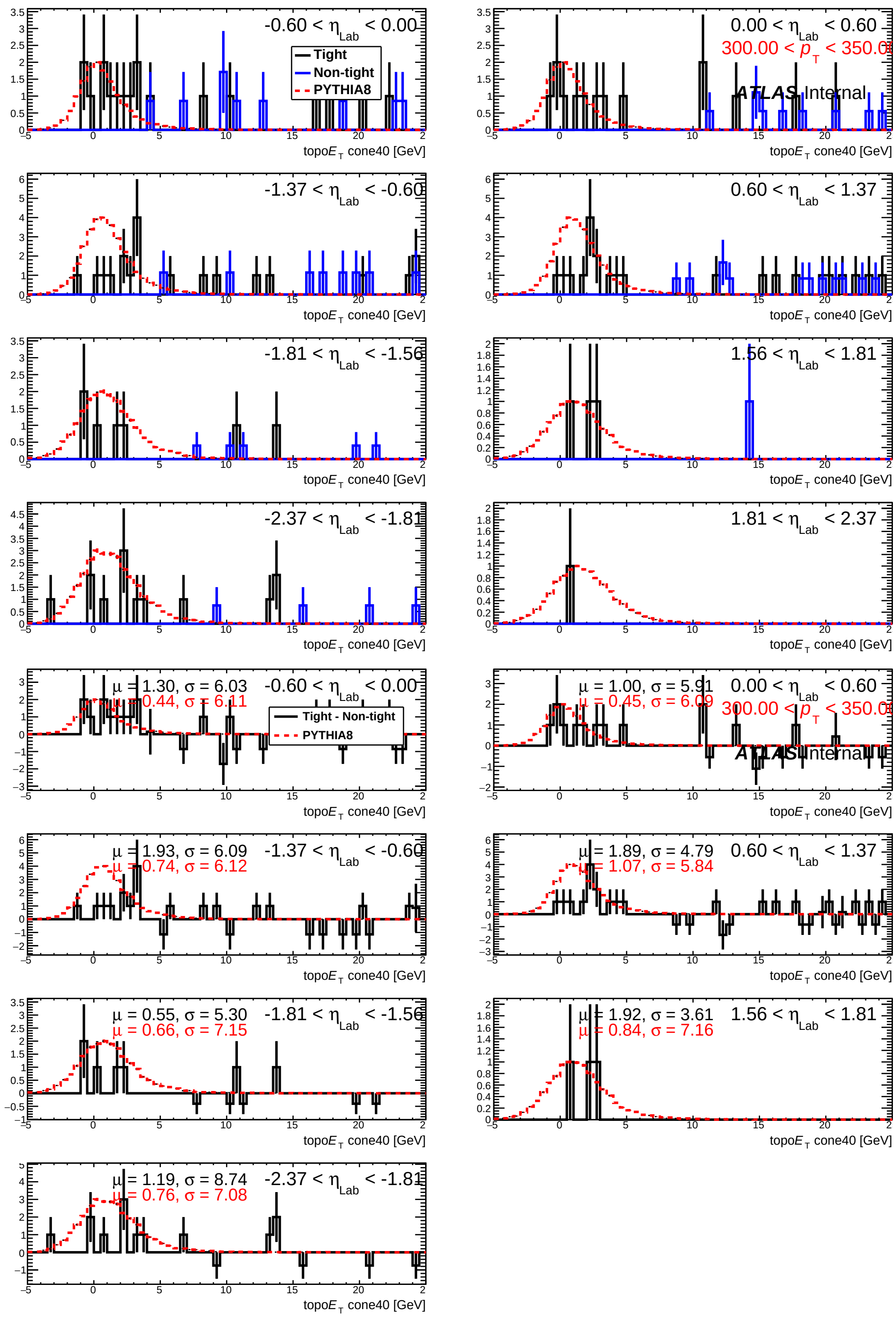
-0.60 < η_Lab < 0.00
Tight
Non-tight
PYTHIA8
0.00 < η_Lab < 0.60
300.00 < p_T < 350.00
ATLAS Internal
-1.37 < η_Lab < -0.60
0.60 < η_Lab < 1.37
-1.81 < η_Lab < -1.56
1.56 < η_Lab < 1.81
-2.37 < η_Lab < -1.81
1.81 < η_Lab < 2.37
μ = 1.30, σ = 6.03
μ = 0.44, σ = 6.11
Tight - Non-tight
μ = 1.00, σ = 5.91
μ = 0.45, σ = 6.09
μ = 1.93, σ = 6.09
μ = 0.74, σ = 6.12
μ = 1.89, σ = 4.79
μ = 1.07, σ = 5.84
μ = 0.55, σ = 5.30
μ = 0.66, σ = 7.15
μ = 1.92, σ = 3.61
μ = 0.84, σ = 7.16
μ = 1.19, σ = 8.74
μ = 0.76, σ = 7.08
topoE_T cone40 [GeV]

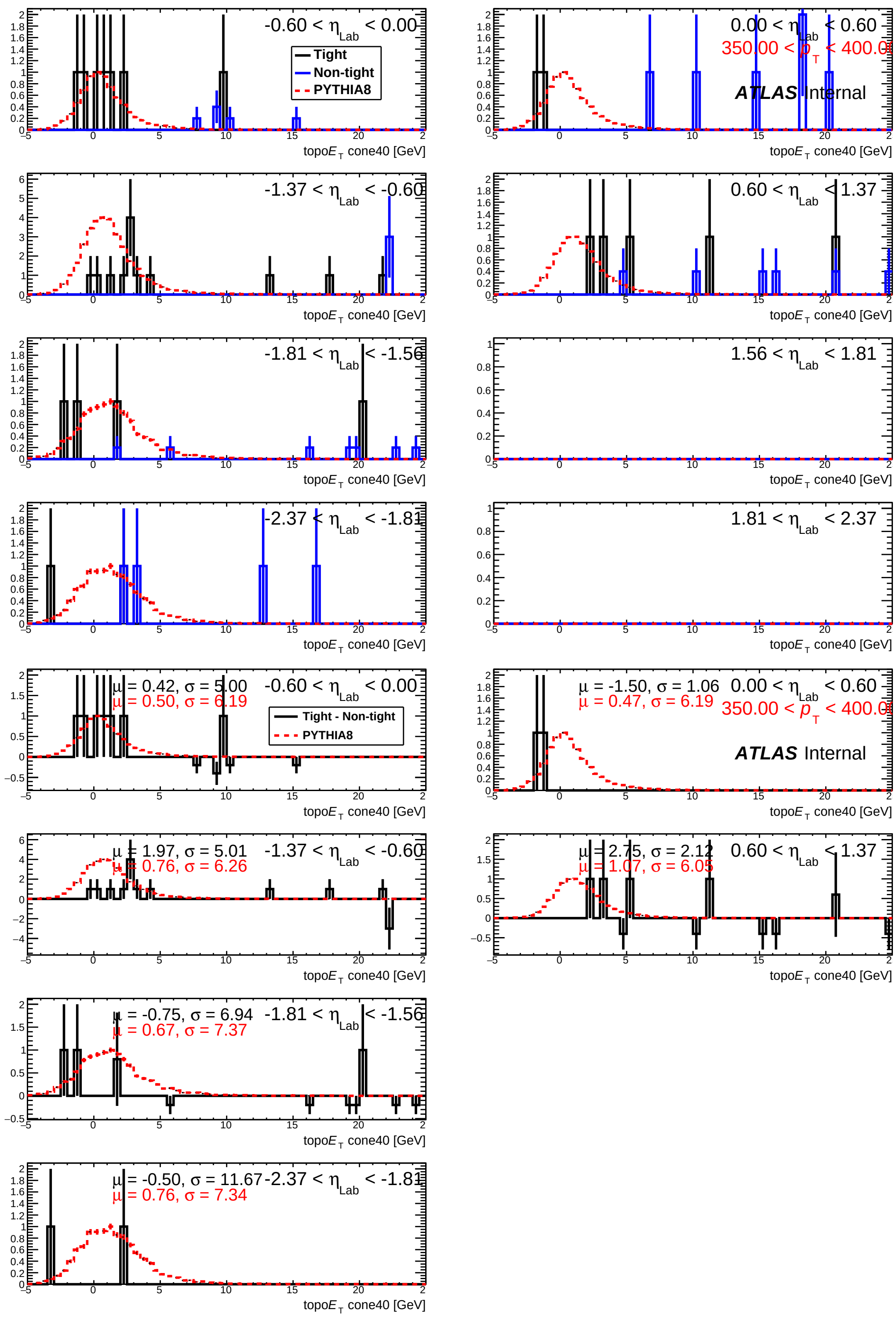
-0.60 < η_Lab < 0.00
Tight
Non-tight
PYTHIA8
0.00 < η_Lab < 0.60
350.00 < p_T < 400.00
ATLAS Internal
-1.37 < η_Lab < -0.60
0.60 < η_Lab < 1.37
-1.81 < η_Lab < -1.56
1.56 < η_Lab < 1.81
-2.37 < η_Lab < -1.81
1.81 < η_Lab < 2.37
μ = 0.42, σ = 5.00
μ = 0.50, σ = 6.19
Tight - Non-tight
μ = -1.50, σ = 1.06
μ = 0.47, σ = 6.19
μ = 1.97, σ = 5.01
μ = 0.76, σ = 6.26
μ = 2.75, σ = 2.12
μ = 1.07, σ = 6.05
μ = -0.75, σ = 6.94
μ = 0.67, σ = 7.37
μ = -0.50, σ = 11.67
μ = 0.76, σ = 7.34
topoE_T cone40 [GeV]

# Appendix B

## Measurement of Azimuthal Anisotropy

### B.1 Multiplicity Distributions

Figures B.1 and B.2 show the run-by-run multiplicity distributions for each trigger used to construct the minimum bias selection. The histograms contain raw counts and are not prescale corrected.

Figures B.3 and B.4 show the run-by-run multiplicity distributions summed over each trigger used to construct the minimum bias selection. The histograms contain raw counts and are not prescale corrected.

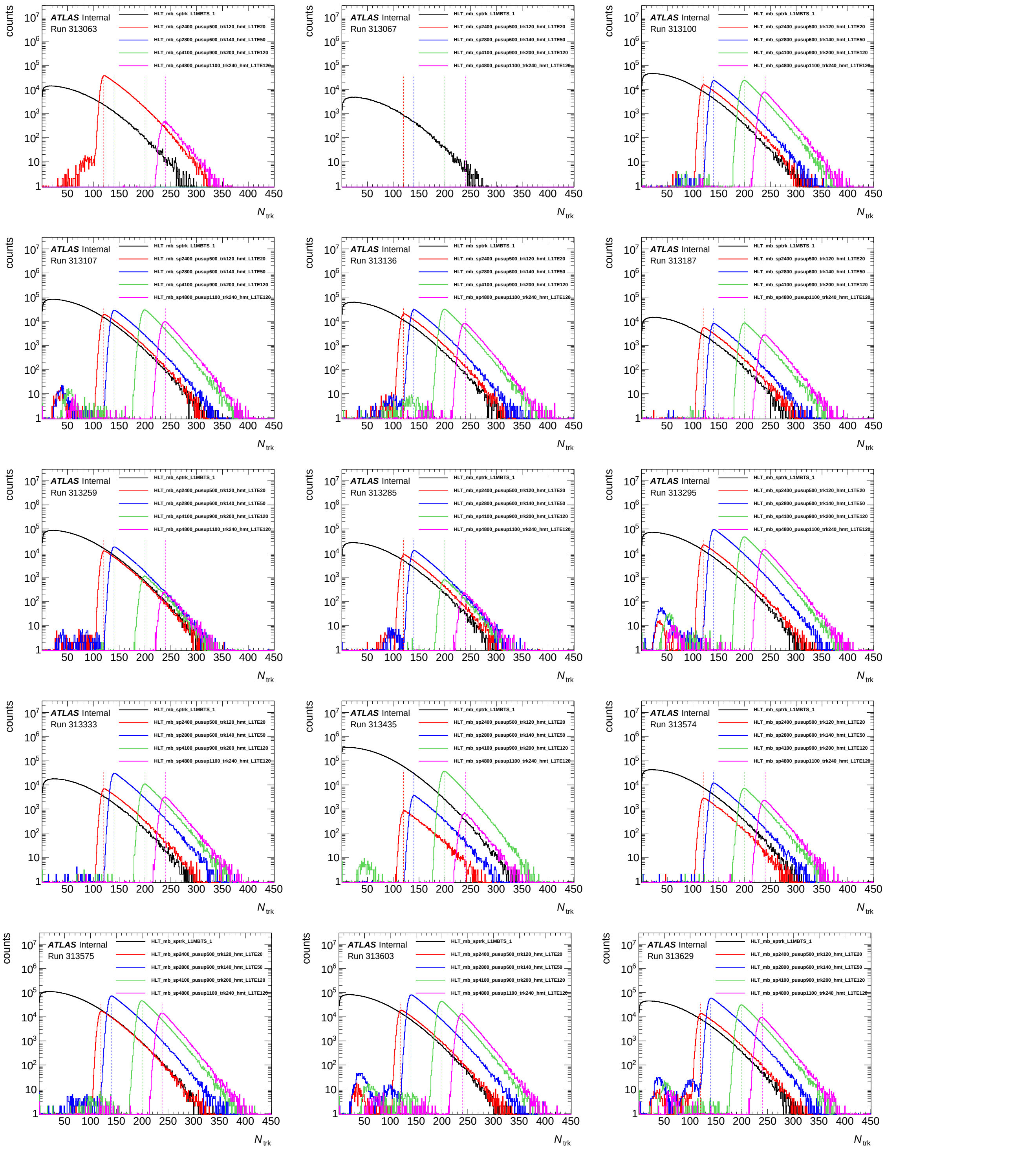

Figure B.1: Run-by-run multiplicity distributions for each trigger used to construct the minimum bias selection. Vertical lines are drawn to indicate the thresholds partitioning the $N_{\mathrm{trk}}$ range. In each region, only the trigger offering the largest number of events is used.

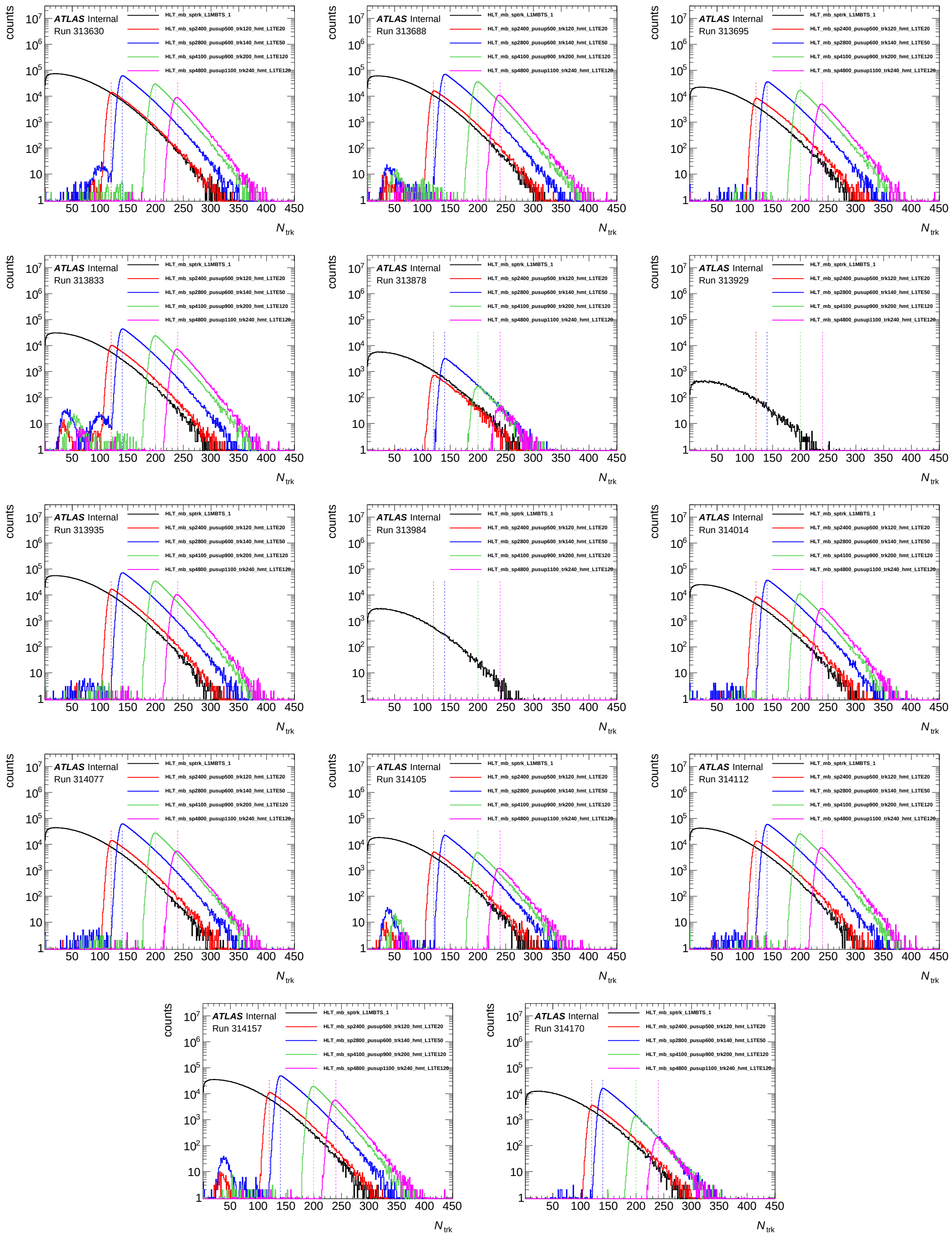

Figure B.2: Run-by-run multiplicity distributions for each trigger used to construct the minimum bias selection. Vertical lines are drawn to indicate the thresholds partitioning the $N_{\mathrm{trk}}$ range. In each region, only the trigger offering the largest number of events is used.

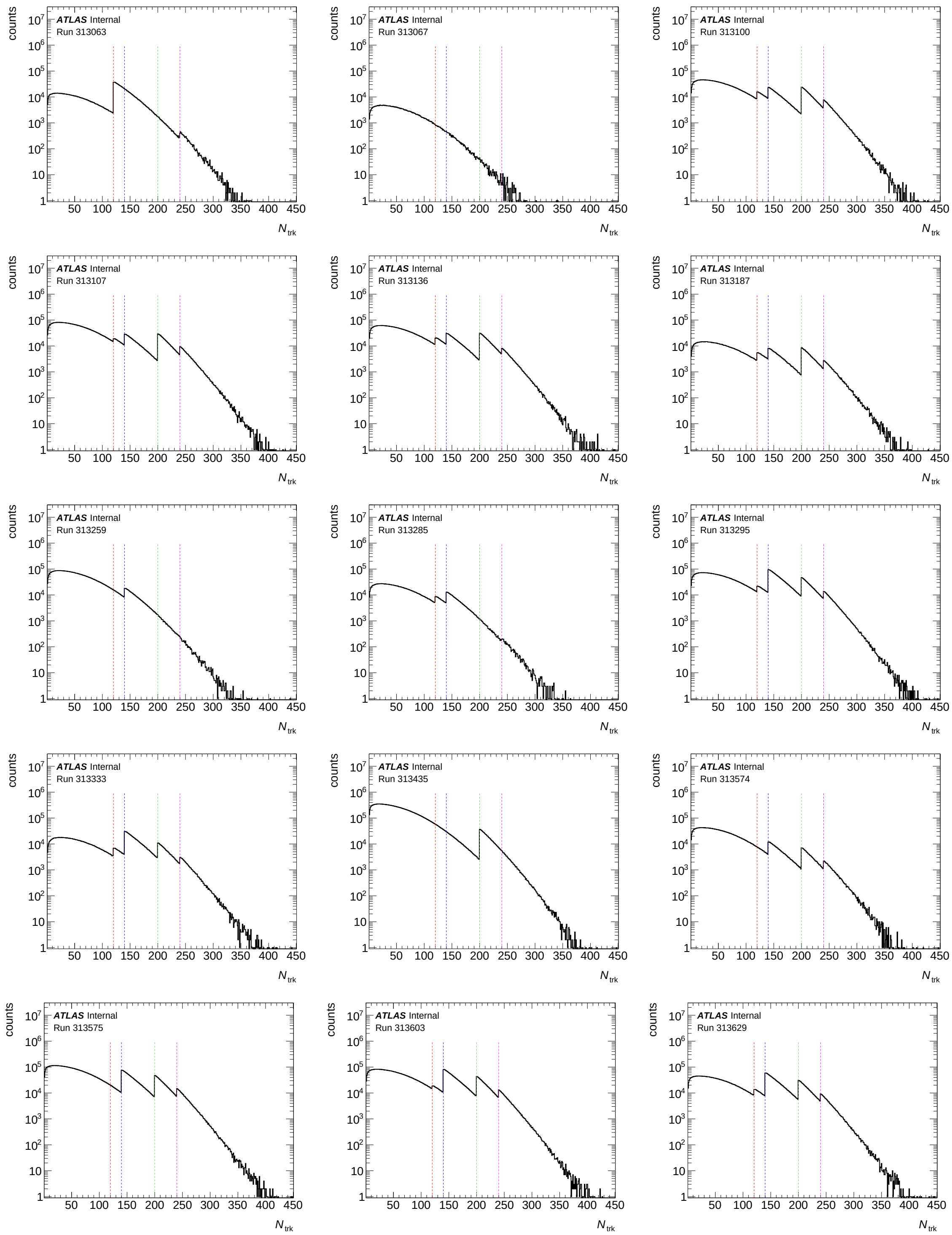

Figure B.3: Total run-by-run multiplicity distributions composing the minimum bias selection. Vertical lines are drawn to indicate the thresholds partitioning the $N_{\text{trk}}$ range. In each region, only the trigger offering the largest number of events is used.

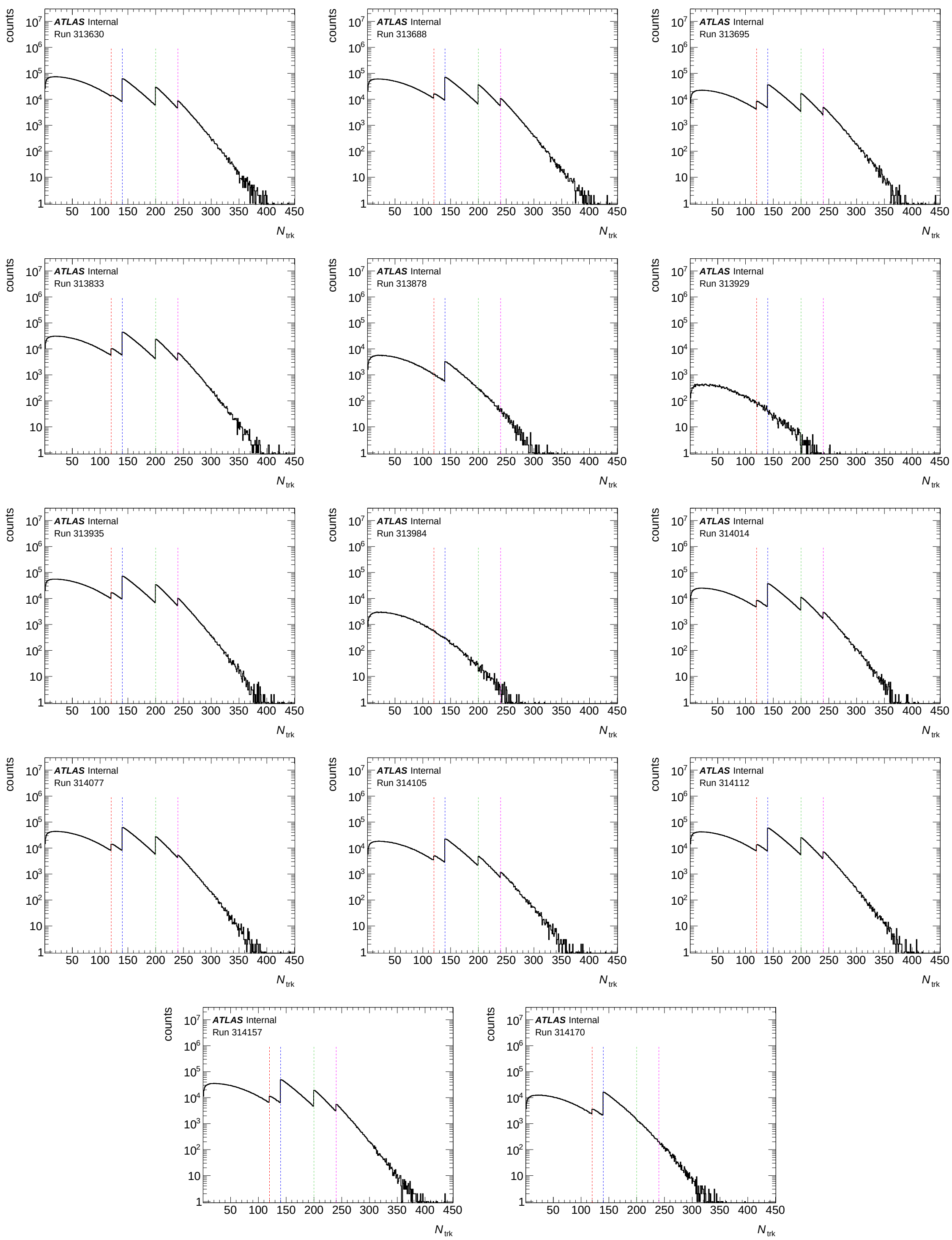

Figure B.4: Total run-by-run multiplicity distributions composing the minimum bias selection. Vertical lines are drawn to indicate the thresholds partitioning the $N_{\mathrm{trk}}$ range. In each region, only the trigger offering the largest number of events is used.

## B.2 Systematic uncertainties in $v_3$

### B.2.1 Performance

#### B.2.1.1 MinBias tracking efficiency

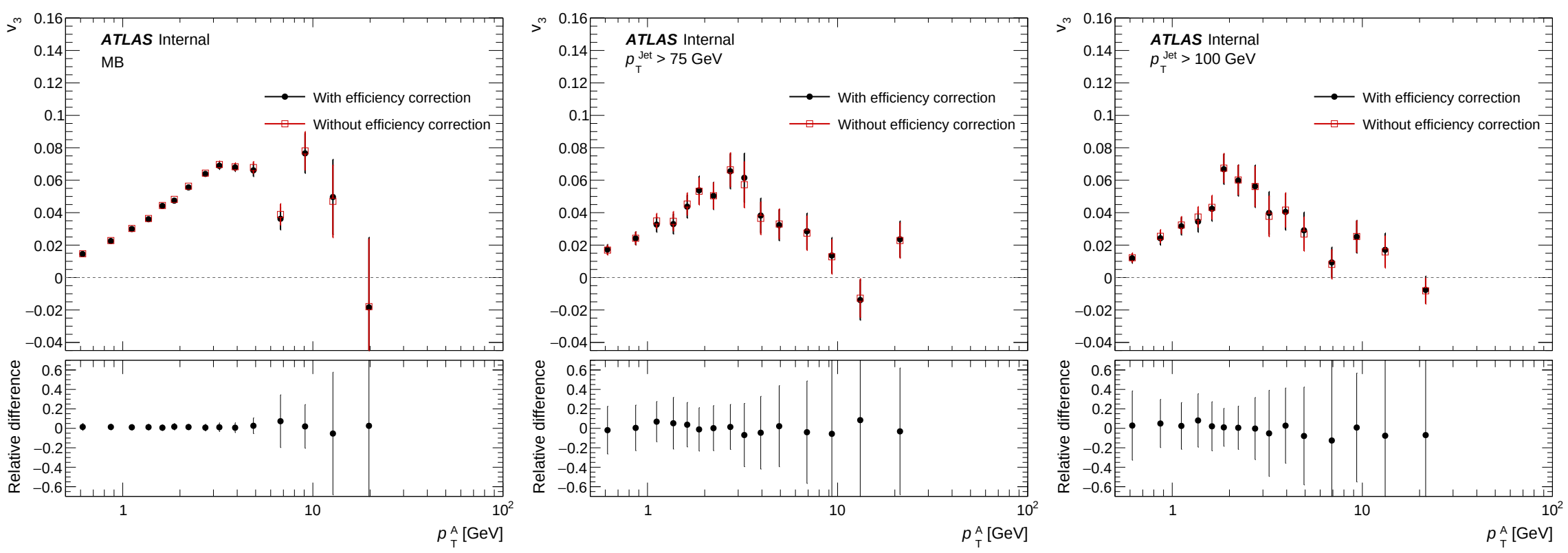

Figure B.5: $v_3$ versus $p_T$ for MB (left), 75 GeV jet (center), and 100 GeV jet (right) events with and without trigger and tracking efficiency corrections.

#### B.2.1.2 Total

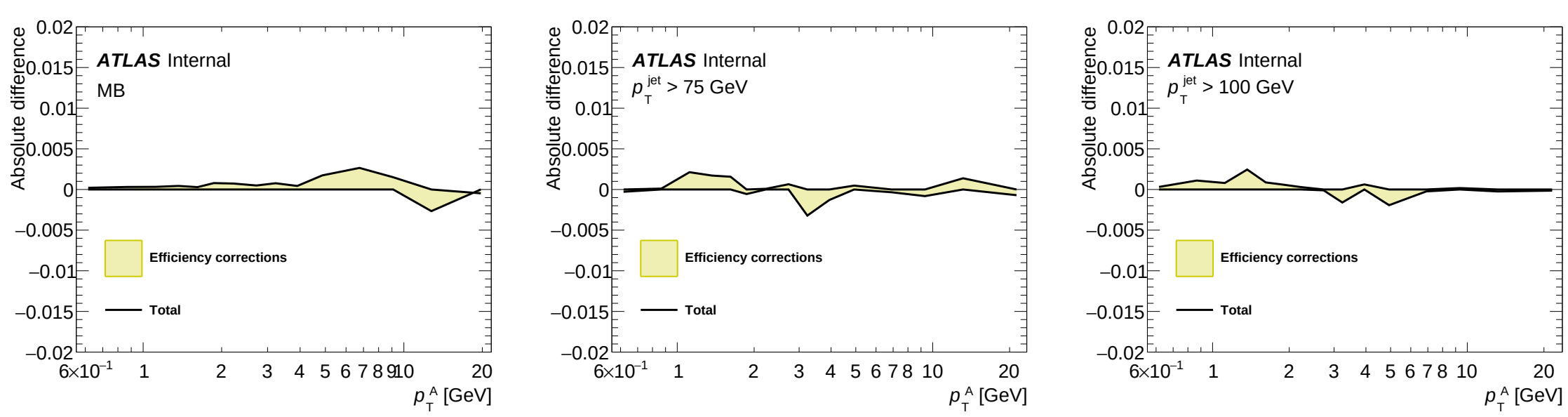

Figure B.6: Combined performance uncertainty, plotted as the absolute difference in $v_3$ between the varied and nominal selections, for MB (left), 75 GeV jet (center), and 100 GeV jet (right) events.

### B.2.2 Signal extraction

#### B.2.2.1 Event Mixing

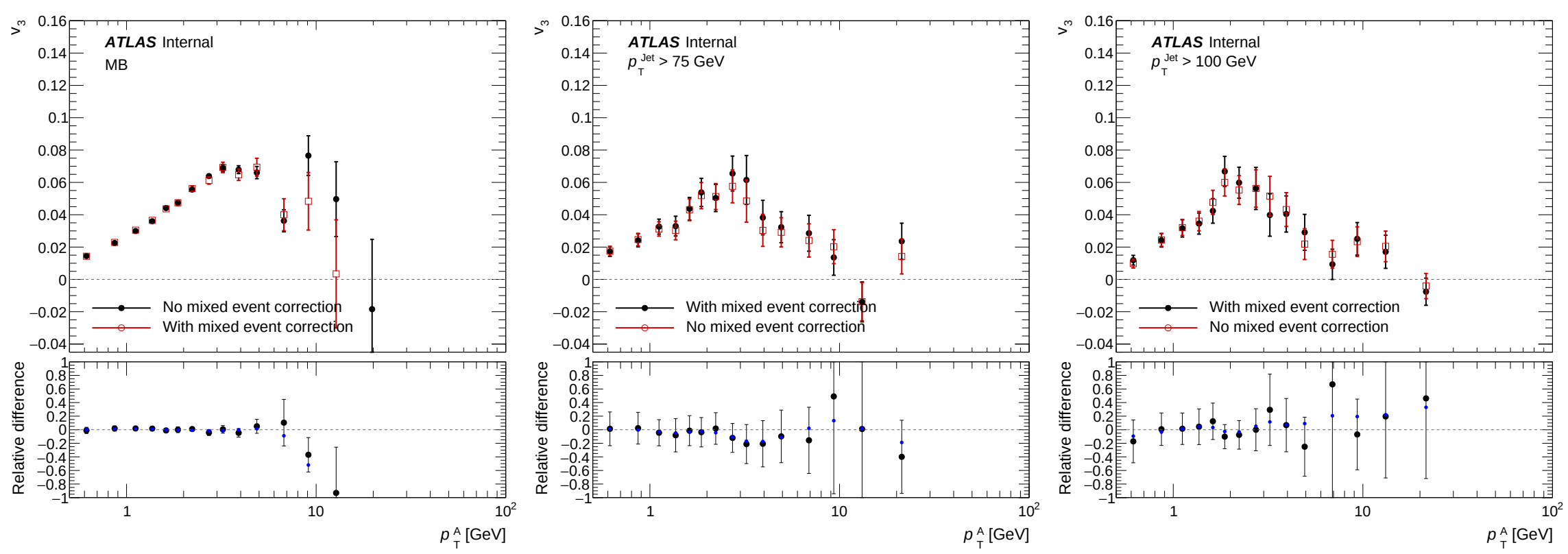

Figure B.7: $v_3$ versus $p_T$ for both MB (left), 75 GeV jet (center), and 100 GeV jet (right) events with the nominal values using the mixed event correction, and variation without the correction.

#### B.2.2.2 Template fitting reference selection

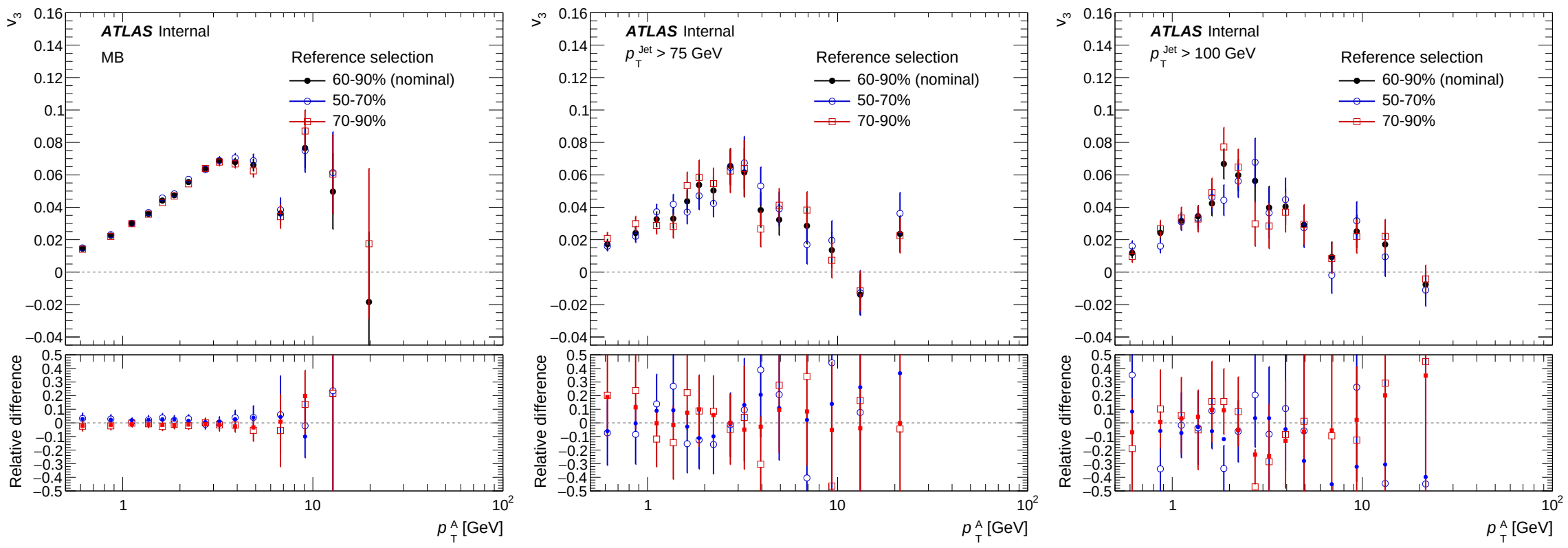

Figure B.8: $v_3$ versus $p_T$ for both MB (left), 75 GeV jet (center), and 100 GeV jet (right) events with the nominal and two varied P reference selections.

#### B.2.2.3 Total

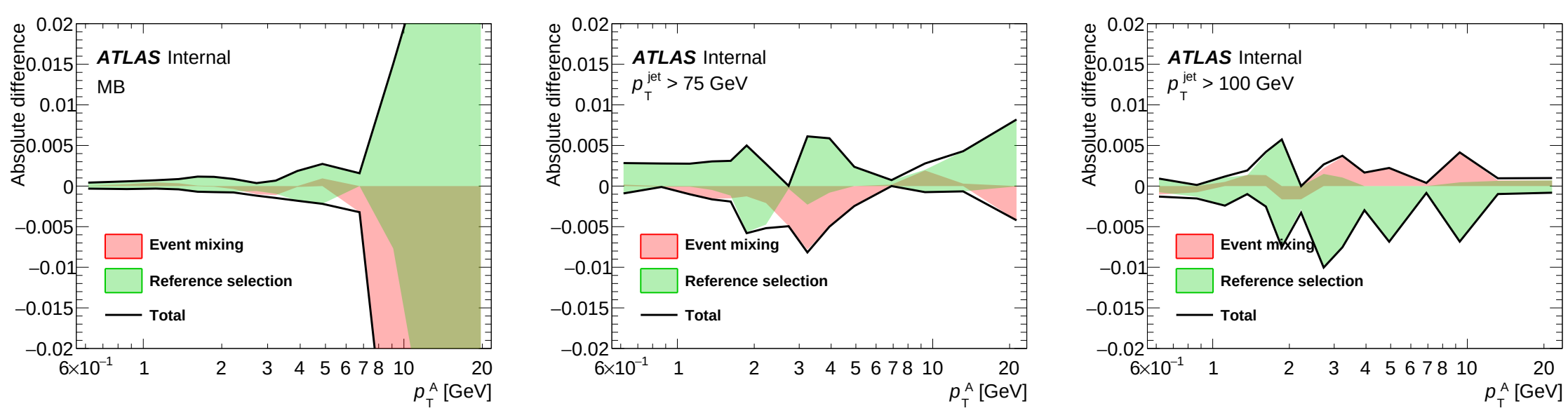

Figure B.9: Combined signal extraction uncertainty, plotted as the absolute difference in $v_3$ between the varied and nominal selections, for MB (left), 75 GeV jet (center), and 100 GeV jet (right) events.

### B.2.3 Jet selection

#### B.2.3.1 Jet $p_T$ threshold

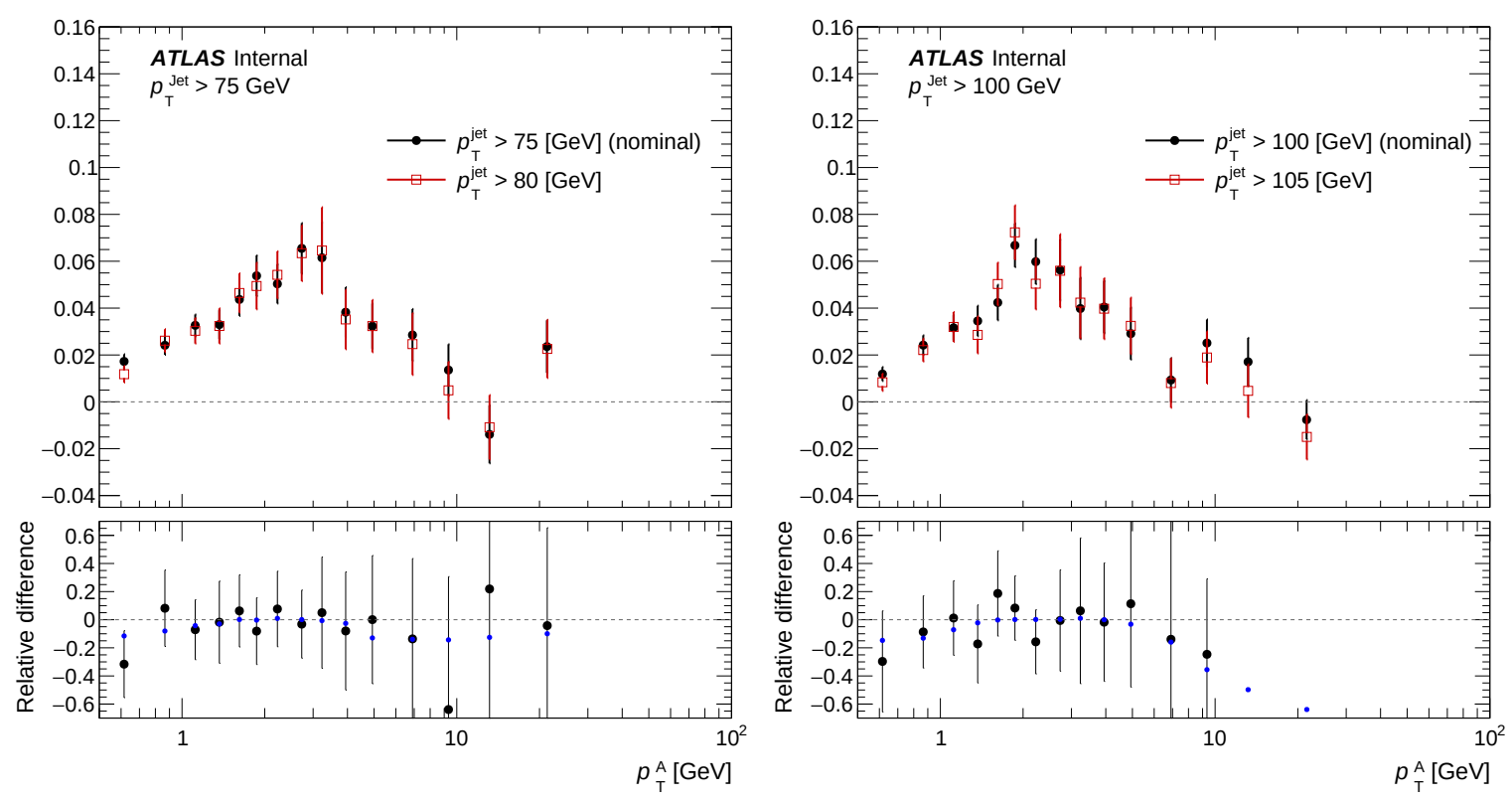

Figure B.10: $v_3$ versus $p_T$ for 75 GeV (left) and 100 GeV (right) jet events with offline jet $p_T$ thresholds variations of +5 GeV.

#### B.2.3.2 Associated particle jet rejection min jet $p_{\mathrm{T}}$

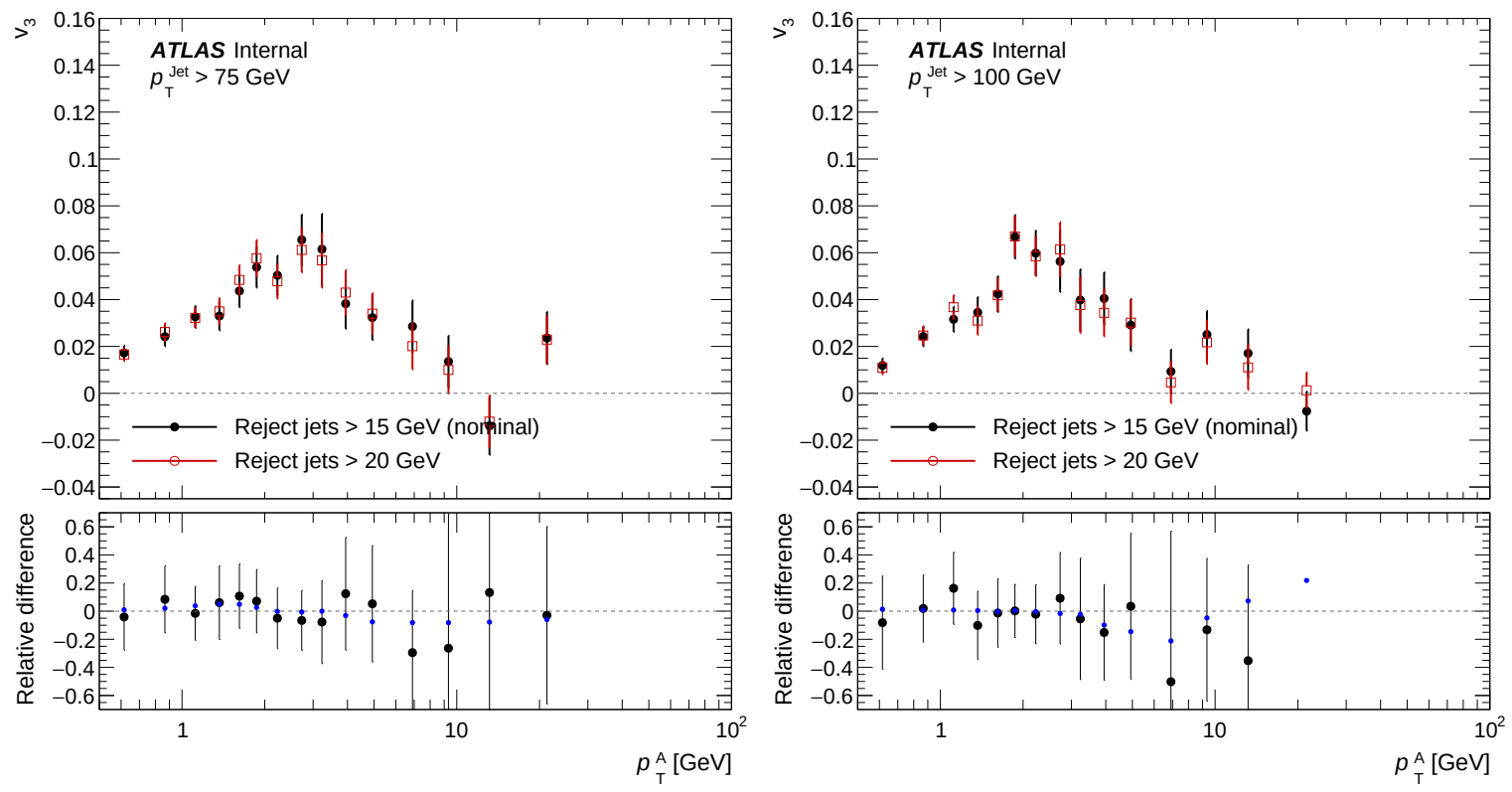

Figure B.11: $v_3$ versus $p_{\mathrm{T}}$ for 75 GeV (left) and 100 GeV (right) jet events with the nominal and varied $\Delta\eta_{\mathrm{jet}}$ $p_{\mathrm{T}}$ selection.

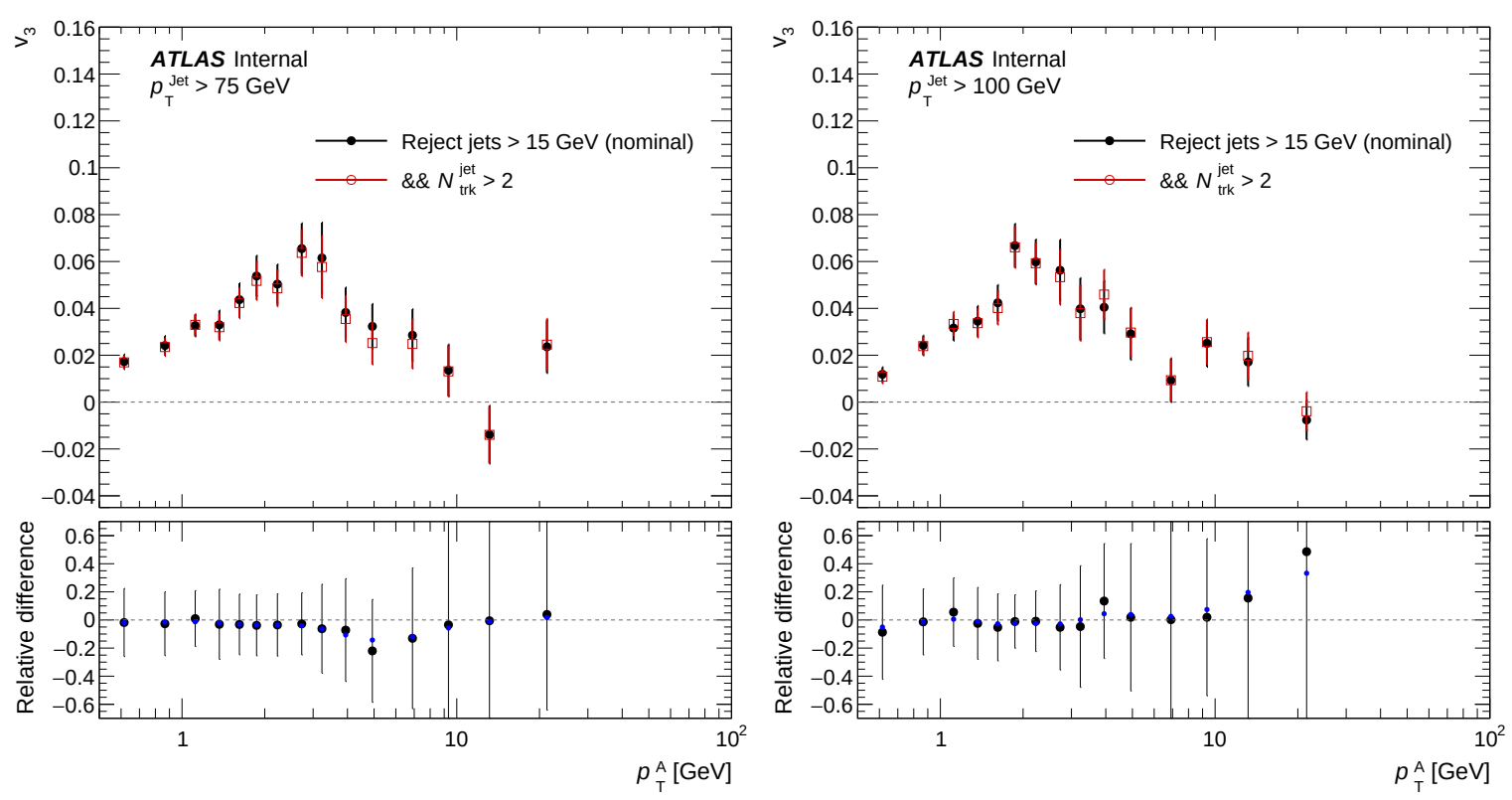

Figure B.12: $v_2$ versus $p_{\mathrm{T}}$ for 75 GeV (left) and 100 GeV (right) jet events with the nominal and varied associated particle rejection jet multiplicity selection.

#### B.2.3.3 Total

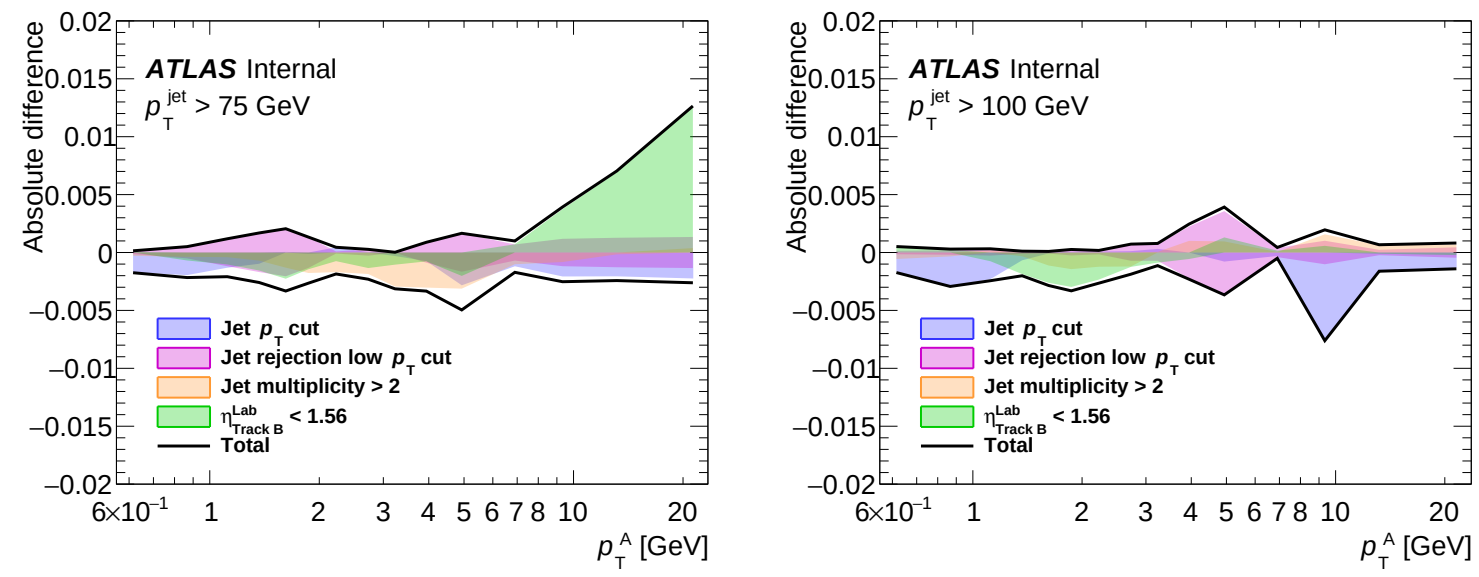

Figure B.13: Combined jet selection uncertainty for MB (left), 75 GeV jet (center), and 100 GeV jet (right) events.

## B.3 Systematic uncertainties in $v_2$ vs. centrality

### B.3.1 Performance

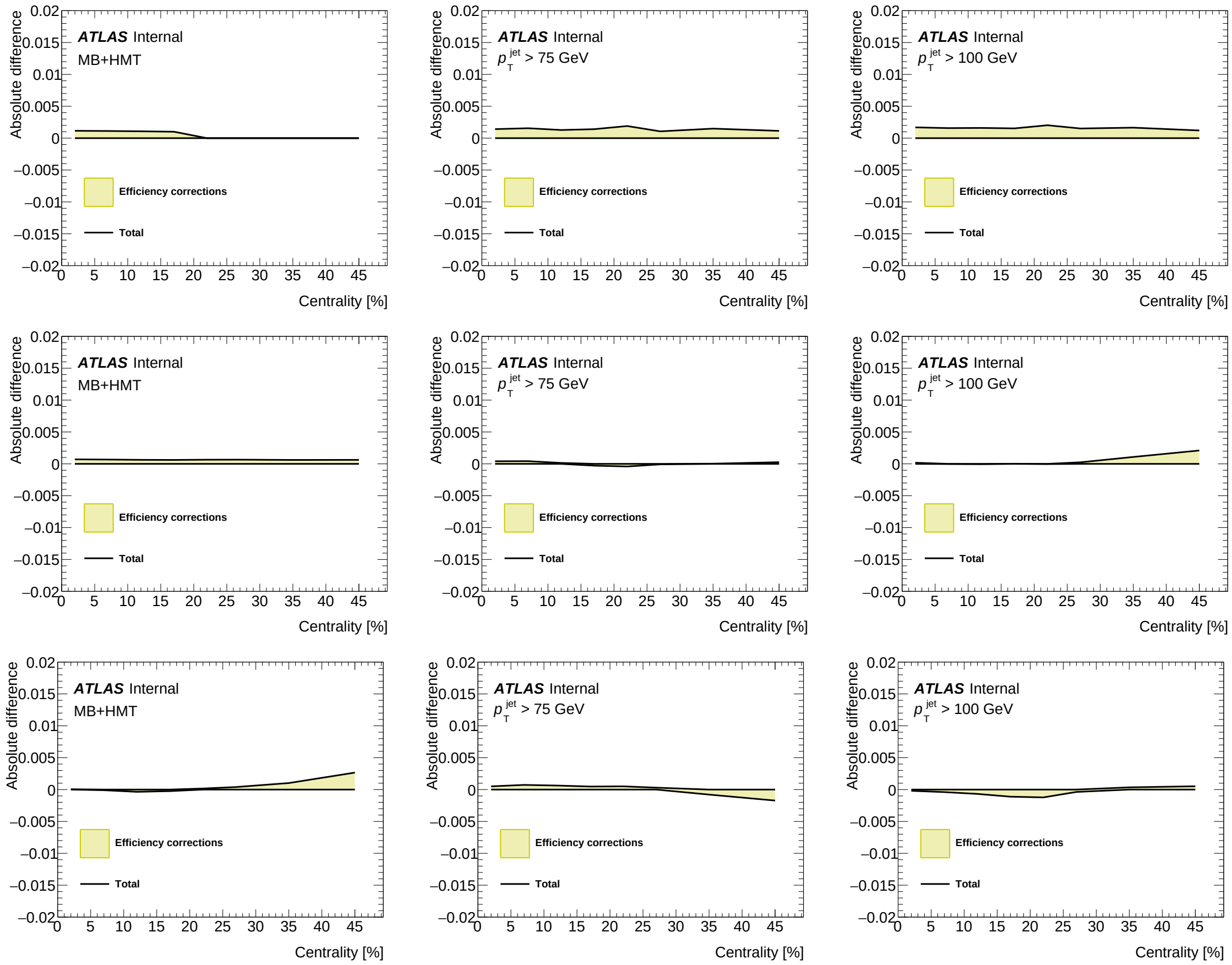

Figure B.14: Combined performance uncertainty in $v_2$ as a function of centrality, plotted as the absolute difference between the varied and nominal selections, for MB (left), 75 GeV jet (center), and 100 GeV jet (right) events. The uncertainties are determined separately for $0.5 < p_{\mathrm{T}}^{\mathrm{A}} < 2$ GeV (top row), $2 < p_{\mathrm{T}}^{\mathrm{A}} < 9$ GeV (middle row), and $9 < p_{\mathrm{T}}^{\mathrm{A}} < 100$ GeV (bottom row).

### B.3.2 Signal extraction

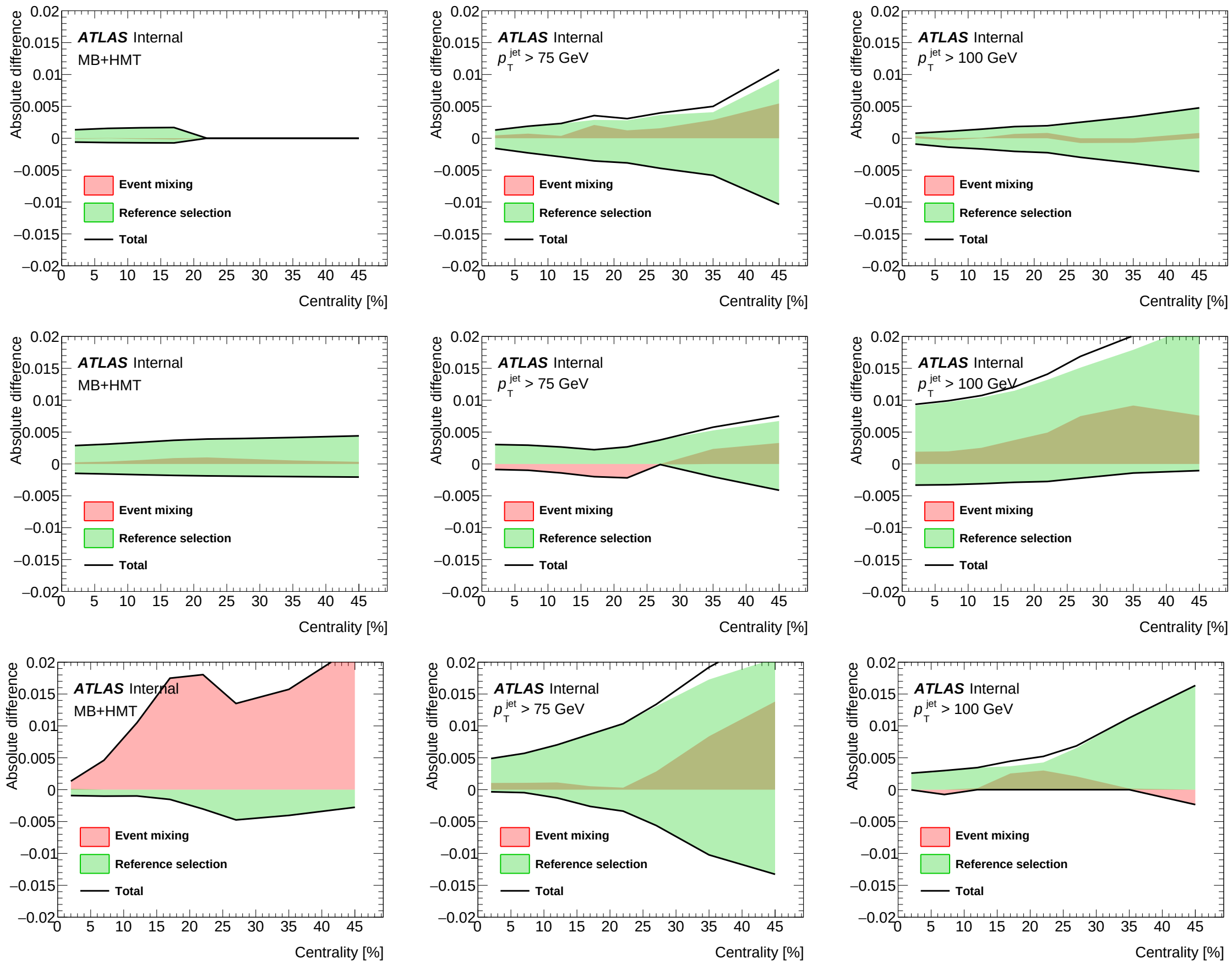

Figure B.15: Combined signal extraction uncertainty in $v_2$ as a function of centrality, plotted as the absolute difference between the varied and nominal selections, for MB (left), 75 GeV jet (center), and 100 GeV jet (right) events. The uncertainties are determined separately for $0.5 < p_T^A < 2$ GeV (top row), $2 < p_T^A < 9$ GeV (middle row), and $9 < p_T^A < 100$ GeV (bottom row).

### B.3.3 Jet selection

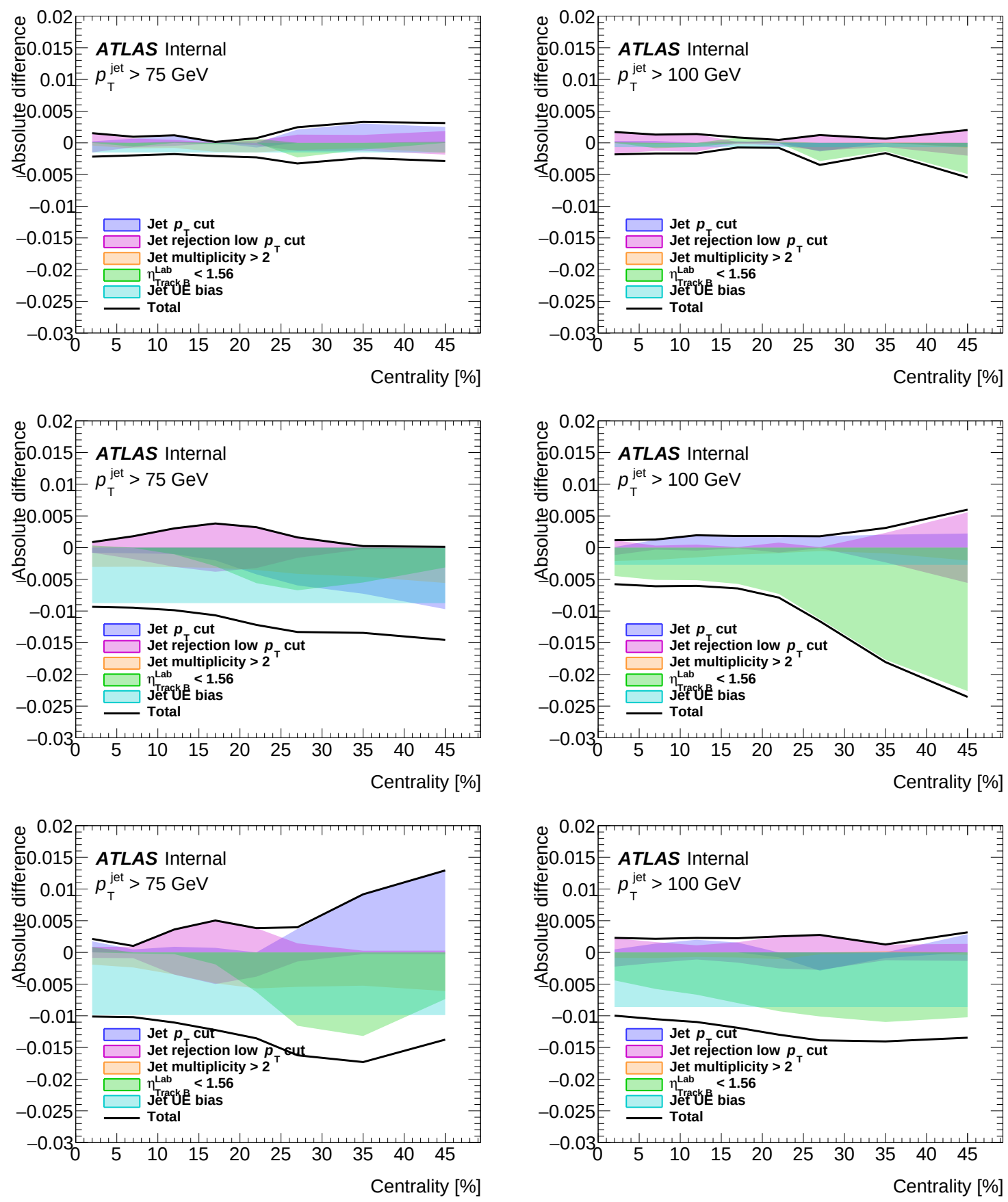

Figure B.16: Combined jet selection uncertainty in $v_2$ as a function of centrality, plotted as the absolute difference between the varied and nominal selections, for 75 GeV (left) and 100 GeV jet (right) events. The uncertainties are determined separately for $0.5 < p_{\mathrm{T}}^{\mathrm{A}} < 2$ GeV (top row), $2 < p_{\mathrm{T}}^{\mathrm{A}} < 9$ GeV (middle row), and $9 < p_{\mathrm{T}}^{\mathrm{A}} < 100$ GeV (bottom row).

### B.3.4 Signal extraction

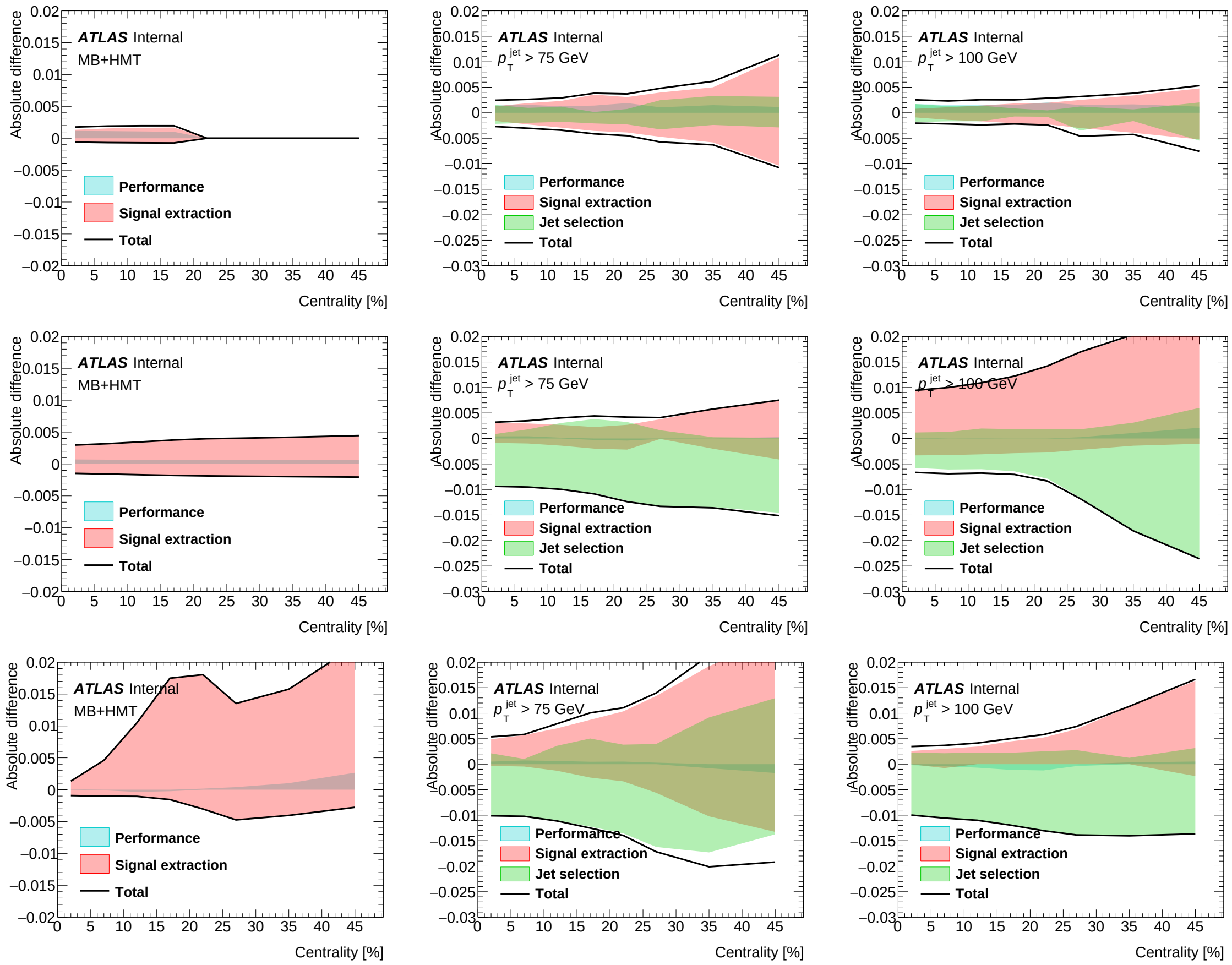

Figure B.17: Total combined uncertainty in $v_2$ as a function of centrality, plotted as the absolute difference between the varied and nominal selections, for MB (left), 75 GeV jet (center), and 100 GeV jet (right) events. The uncertainties are determined separately for $0.5 < p_T^A < 2$ GeV (top row), $2 < p_T^A < 9$ GeV (middle row), and $9 < p_T^A < 100$ GeV (bottom row).

## B.4 Template fits

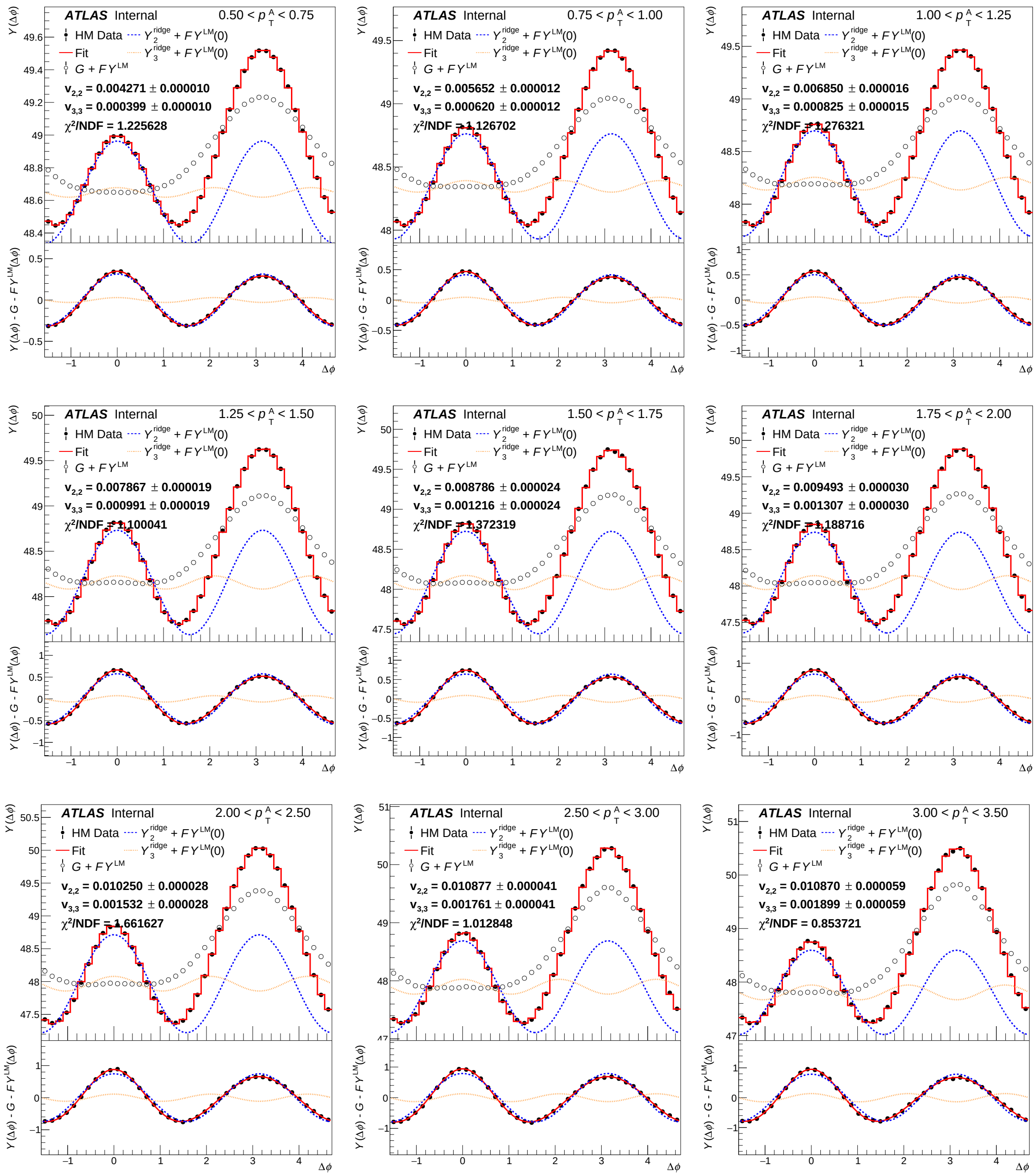

Figure B.18: Template fits to correlation functions from 0-5% central events using 60-90% central events as peripheral reference from the MB dataset. Each figure shows a different $p_T$ range.

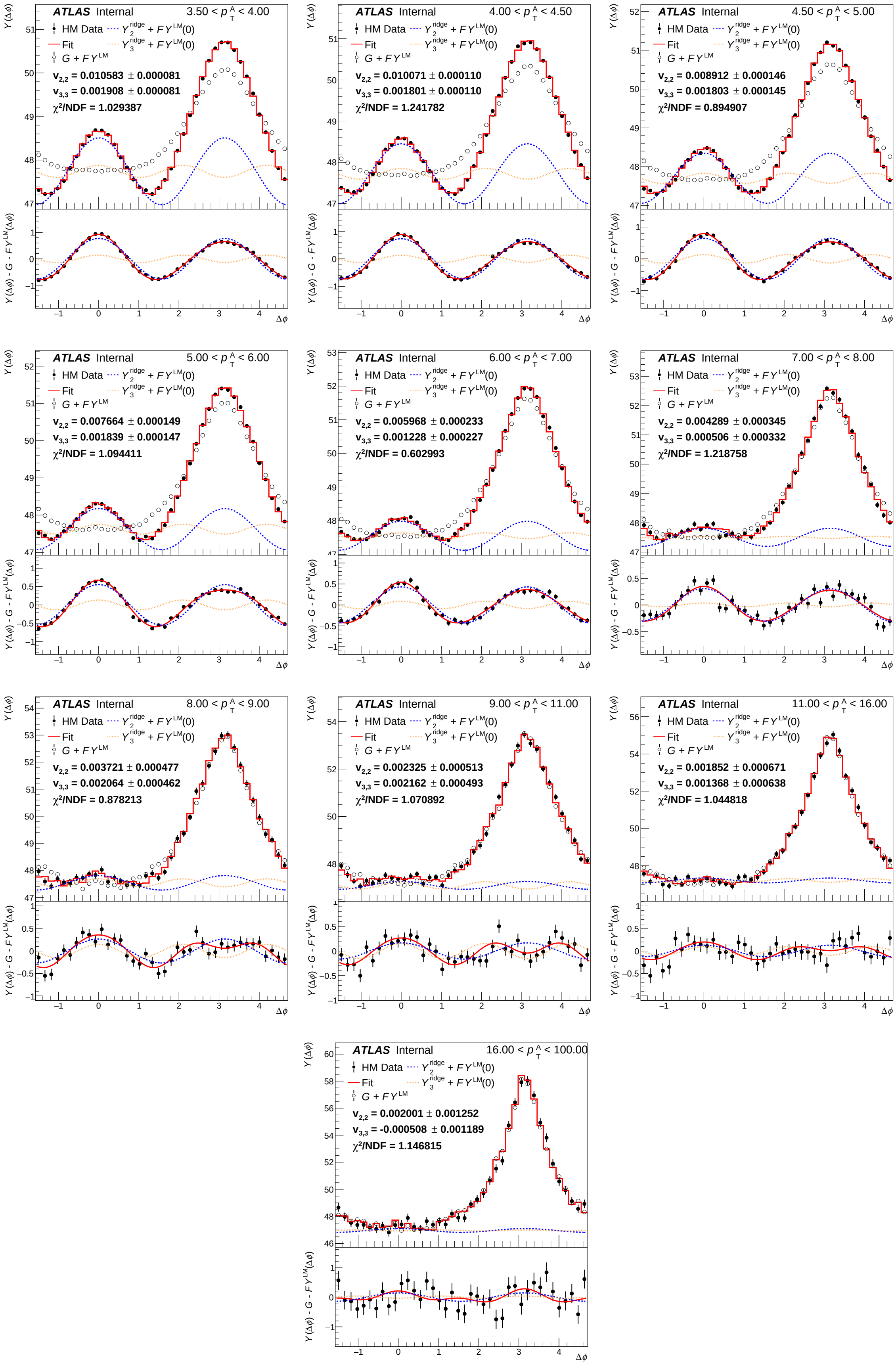

Figure B.19: Template fits to correlation functions from 0-5% central events using 60-90% central events as peripheral reference from the MB dataset. Each figure shows a different $p_{\mathrm{T}}$ range.

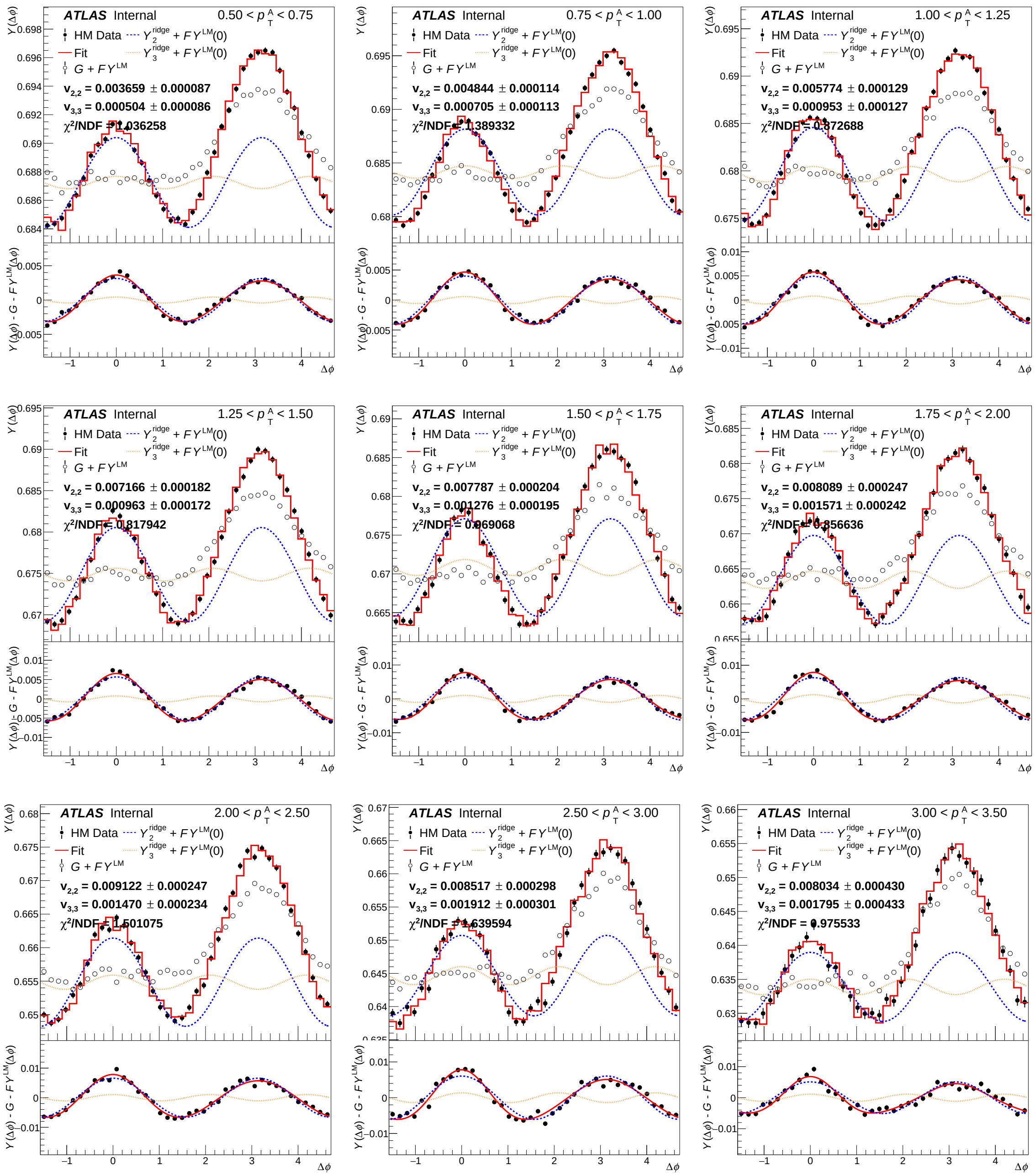

Figure B.20: Template fits to correlation functions from 0-5% central events using 60-90% central events as peripheral reference from the 75 GeV jet dataset. Each figure shows a different $p_{\mathrm{T}}$ range.

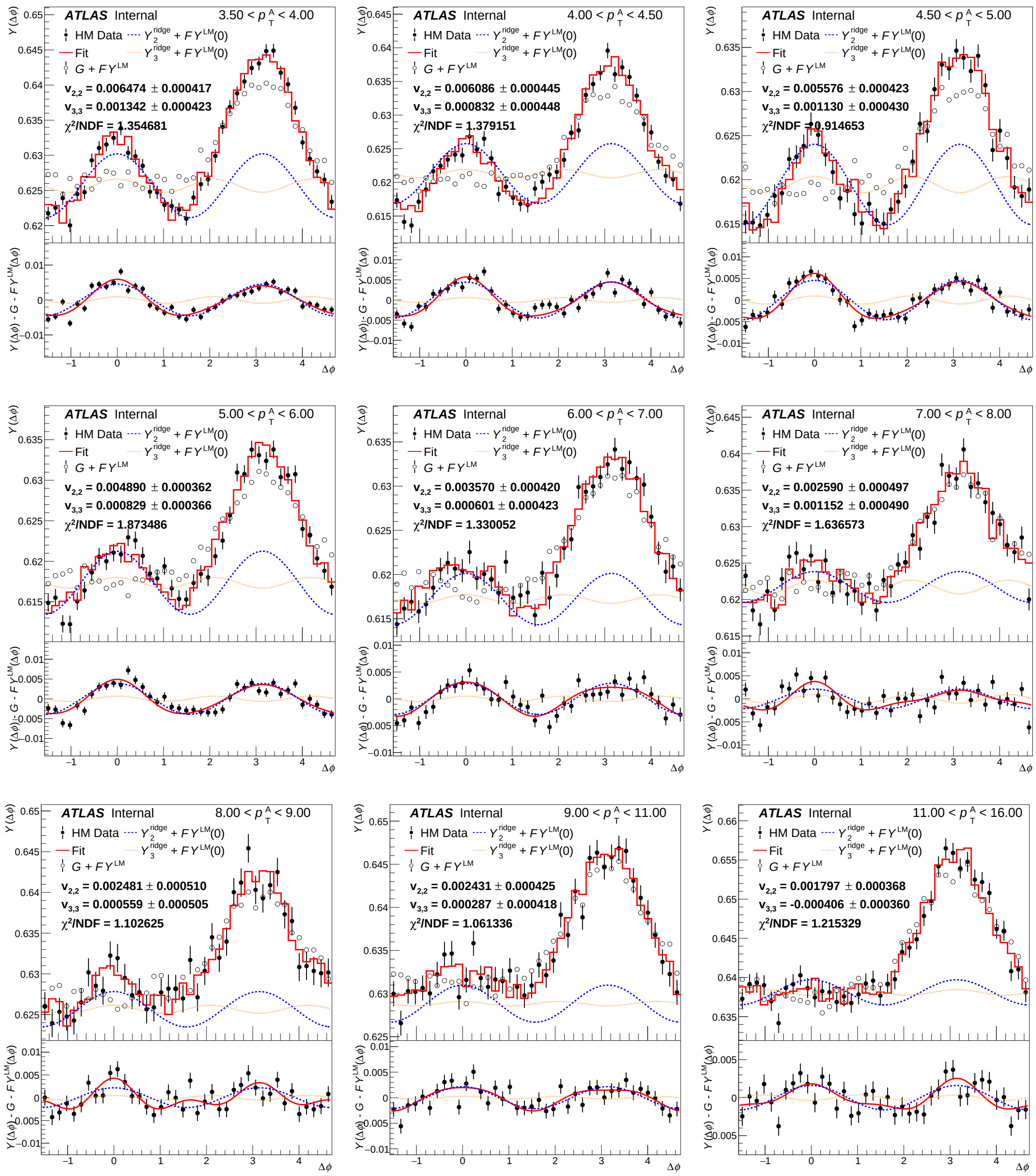

Figure B.21: Template fits to correlation functions from 0-5% central events using 60-90% central events as peripheral reference from the 75 GeV jet dataset. Each figure shows a different $p_{\mathrm{T}}$ range.

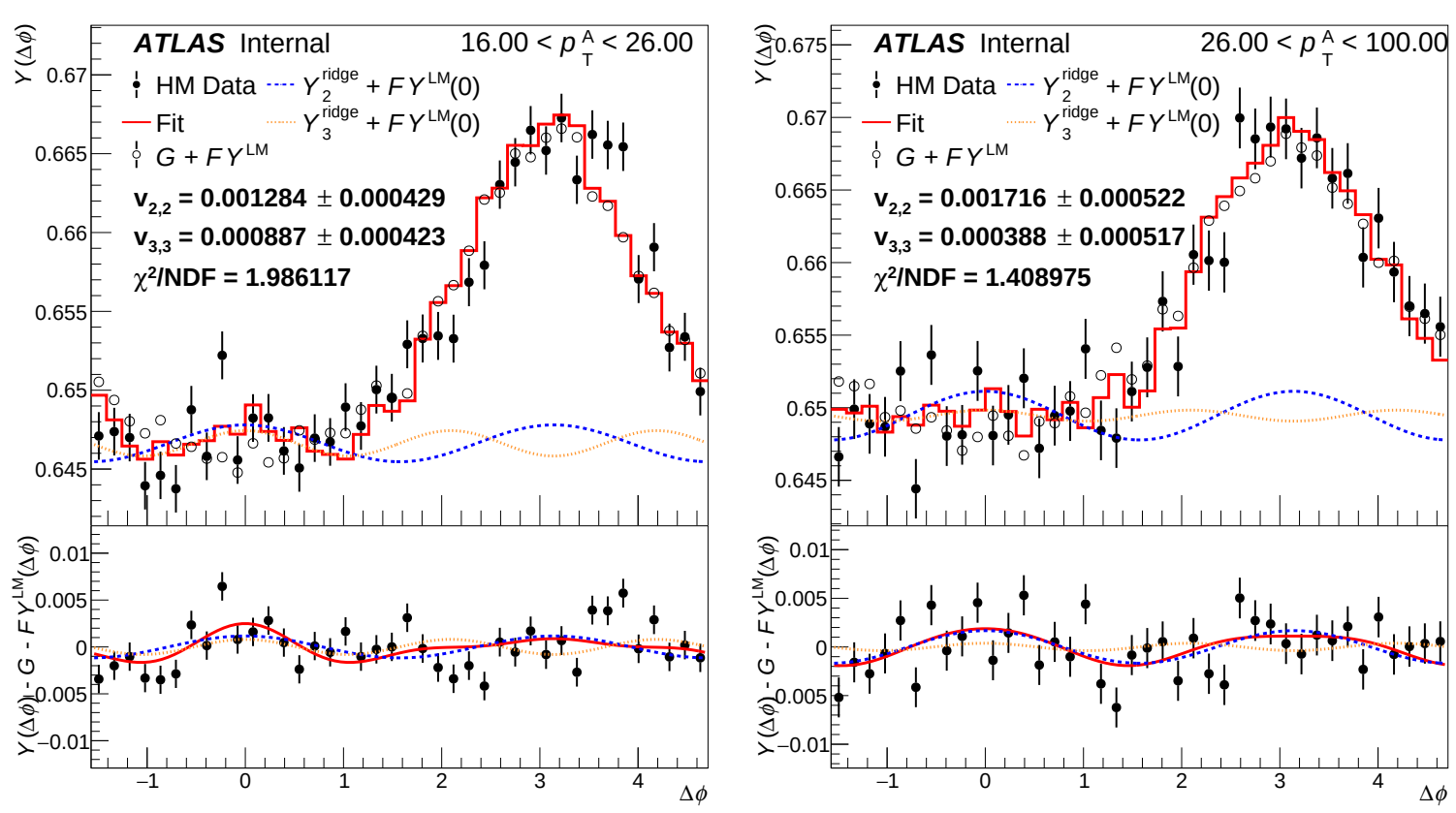

Figure B.22: Template fits to correlation functions from 0-5% central events using 60-90% central events as peripheral reference from the 75 GeV jet dataset. Each figure shows a different $p_{\mathrm{T}}$ range.

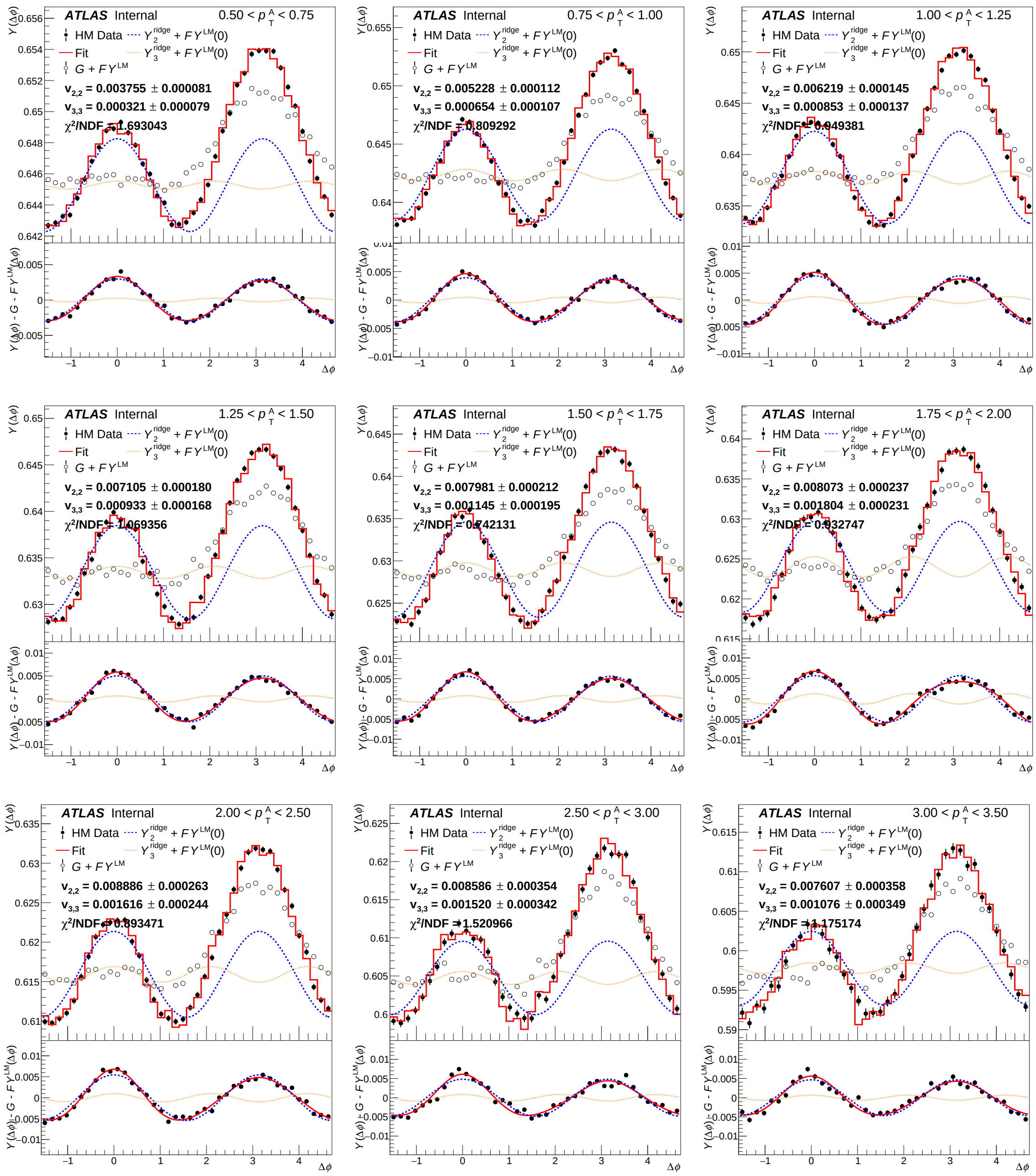

Figure B.23: Template fits to correlation functions from 0-5% central events using 60-90% central events as peripheral reference from the 100 GeV jet dataset. Each figure shows a different $p_{\mathrm{T}}$ range.

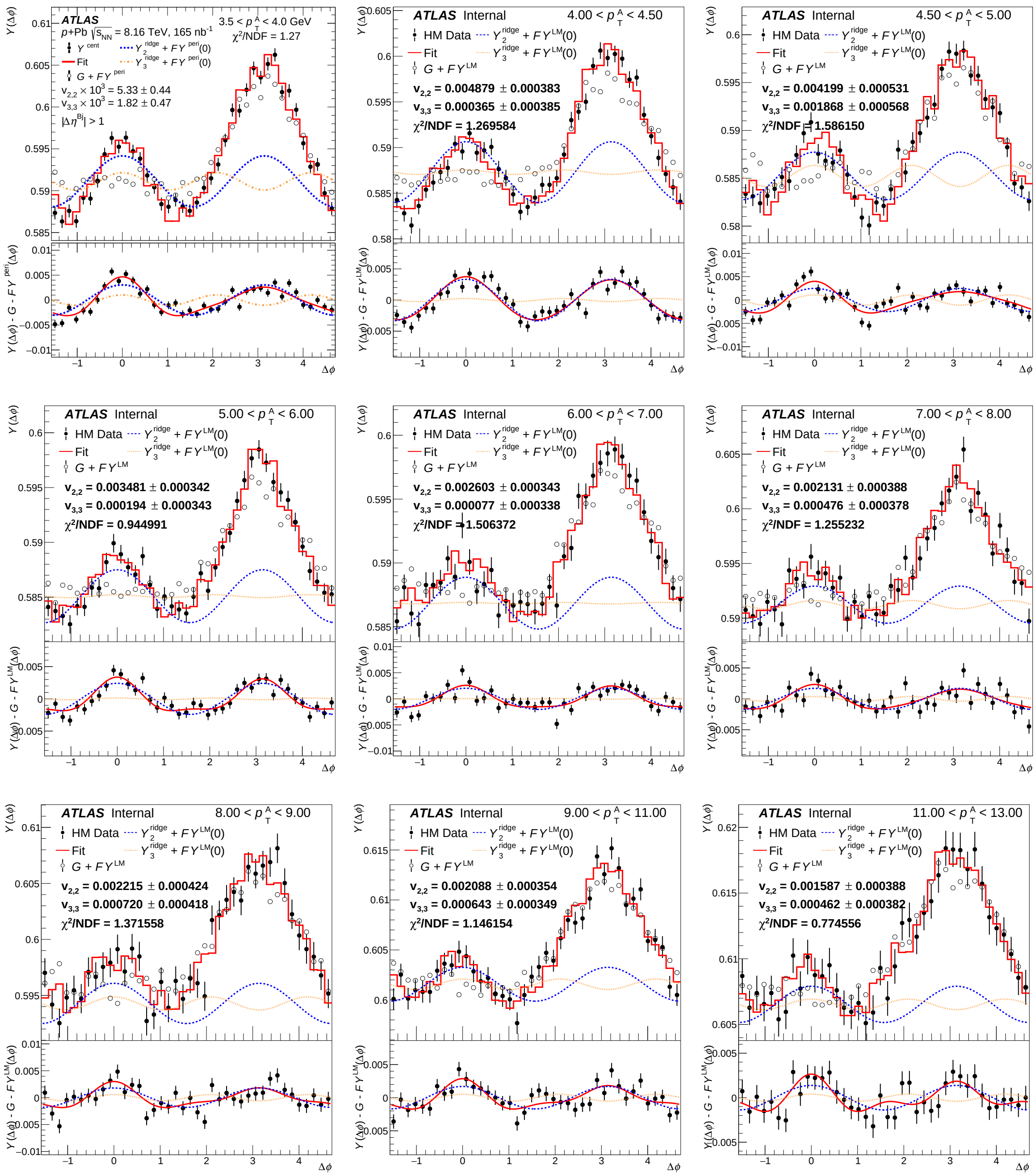

Figure B.24: Template fits to correlation functions from 0-5% central events using 60-90% central events as peripheral reference from the 100 GeV jet dataset. Each figure shows a different $p_T$ range.

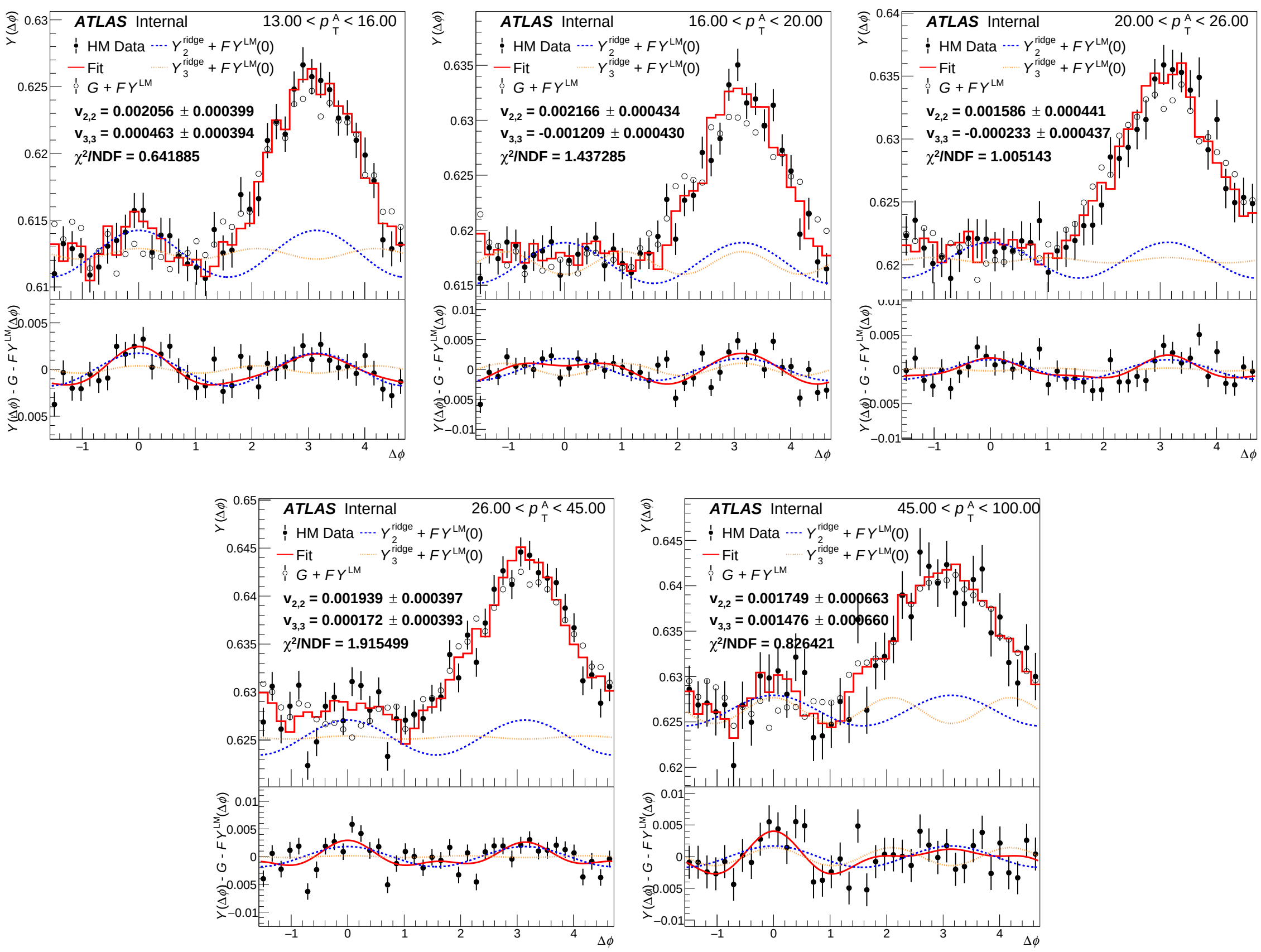

Figure B.25: Template fits to correlation functions from 0-5% central events using 60-90% central events as peripheral reference from the 100 GeV jet dataset. Each figure shows a different $p_\mathrm{T}$ range.

## B.5 Track $\phi$ flattening maps

Track $\phi$ flattening corrections are derived by normalizing track $\phi$ distributions in $\Delta\eta = 0.5$ slices for tracks within a given $p_{\mathrm{T}}$ range. These correction maps are generated for MB and jet events independently.

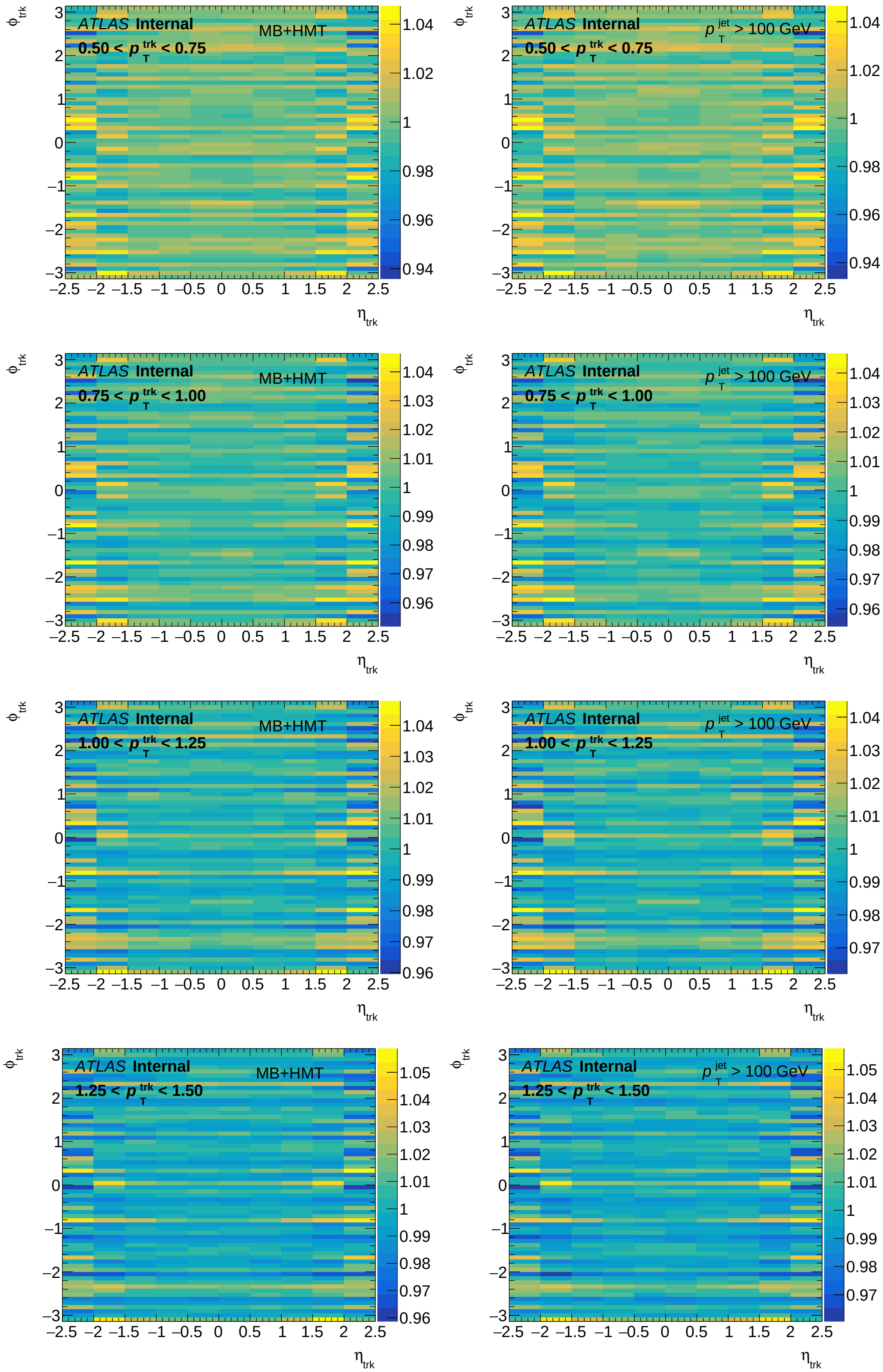

Figure B.26: Track $\phi$ flattening corrections for MB (left) and jet (right) events for tracks in various $p_{\mathrm{T}}$ ranges.

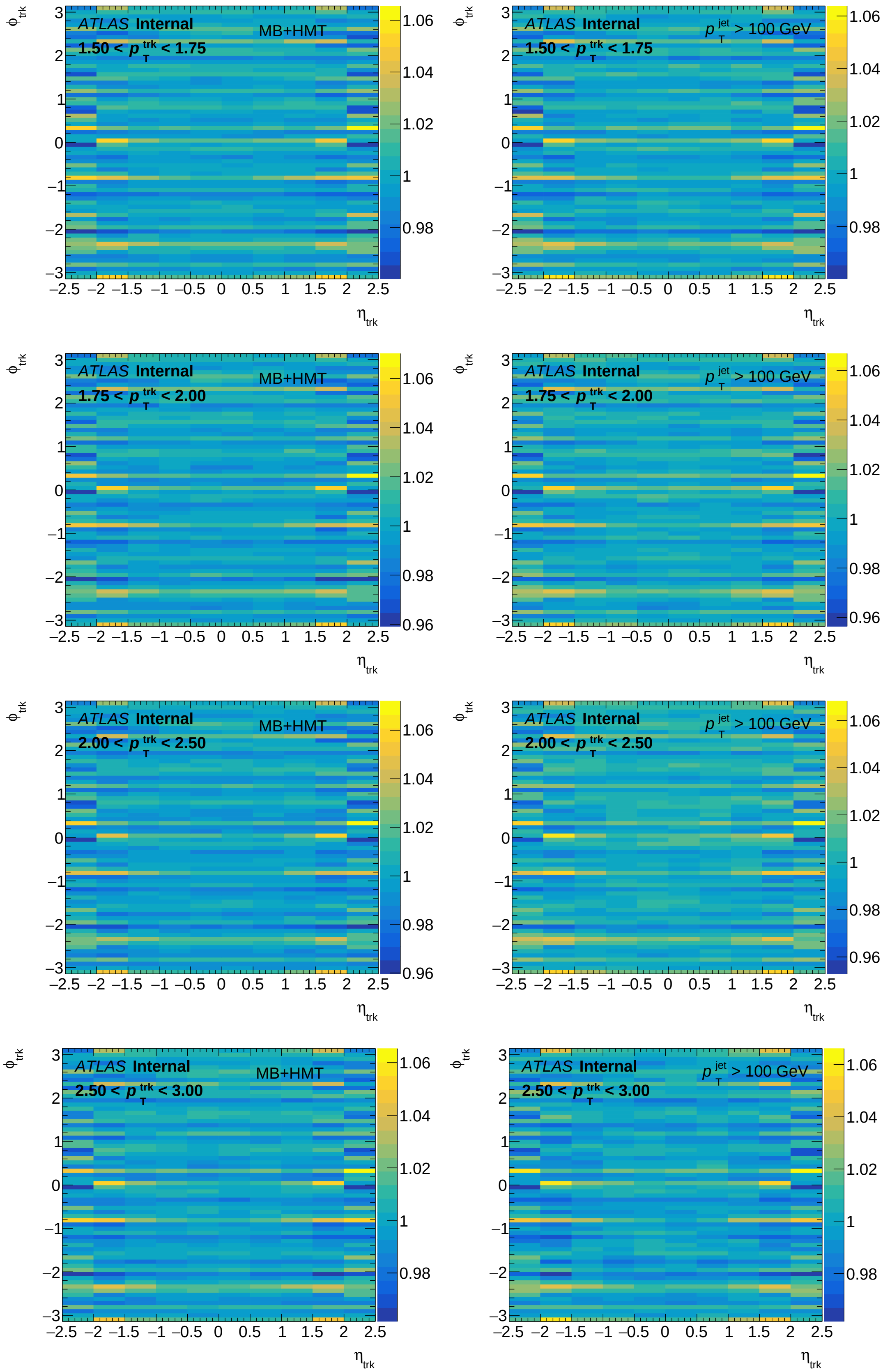

Figure B.27: Track $\phi$ flattening corrections for MB (left) and jet (right) events for tracks in various $p_{\mathrm{T}}$ ranges.

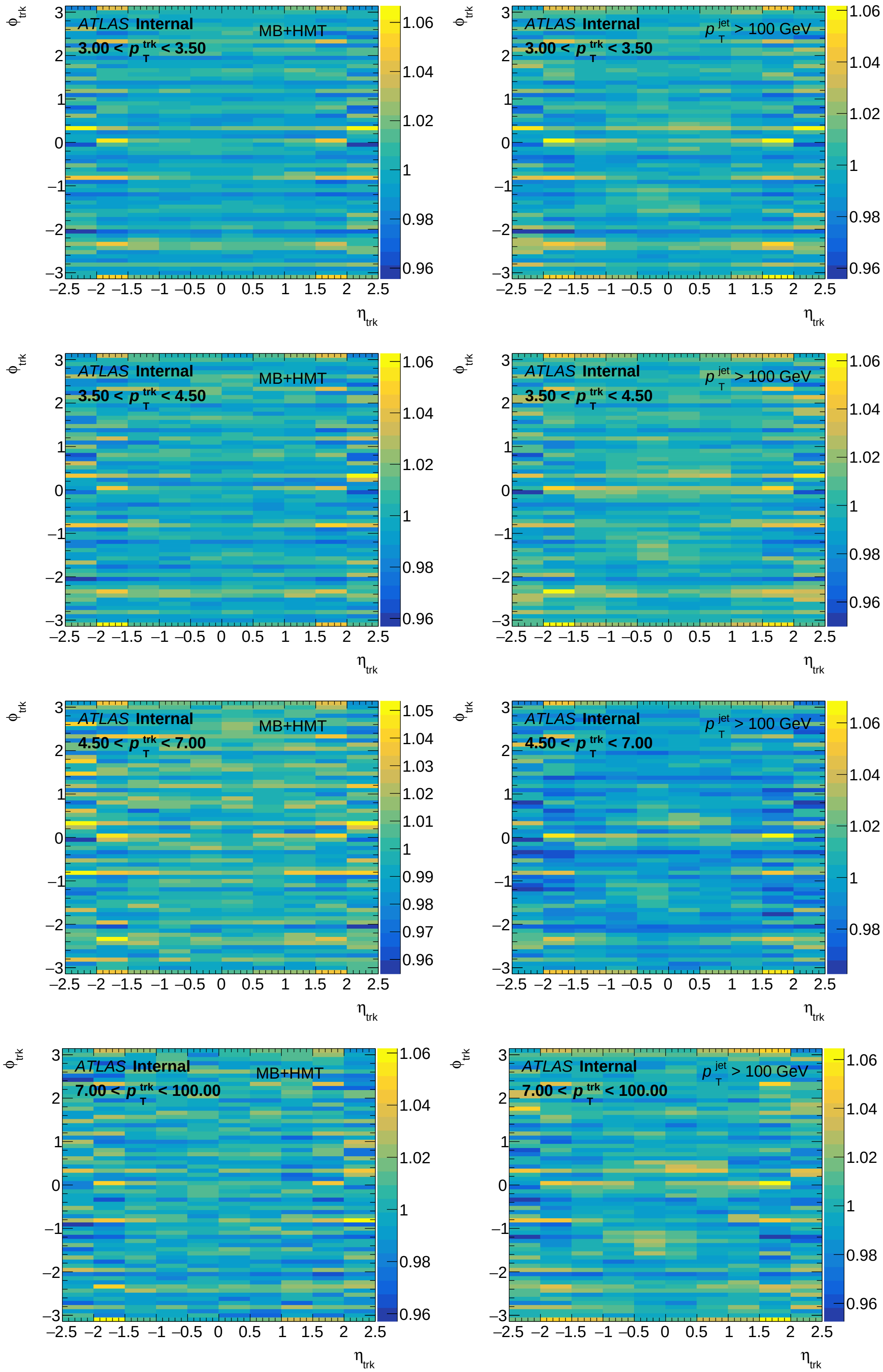

Figure B.28: Track $\phi$ flattening corrections for MB (left) and jet (right) events for tracks in various $p_T$ ranges.

# Bibliography

[1] Particle Data Group. "Review of Particle Physics". In: **Phys. Rev. D** 98 (3 Aug. 2018), p. 030001. DOI: 10.1103/PhysRevD.98.030001.

[2] E. V. Shuryak. "Theory of Hadronic Plasma". In: **Sov. Phys. JETP** 47 (1978). [Zh. Eksp. Teor. Fiz.74,408(1978)], pp. 212–219.

[3] P. Amaro-Seoane et al. "Laser Interferometer Space Antenna". In: (2017). arXiv: 1702.00786 [astro-ph.IM].

[4] C. Gale, S. Jeon, and B. Schenke. "Hydrodynamic Modeling of Heavy-Ion Collisions". In: **Int. J. Mod. Phys.** A28 (2013), p. 1340011. arXiv: 1301.5893 [nucl-th].

[5] P. Romatschke. "Azimuthal Anisotropies at High Momentum from Purely Non-Hydrodynamic Transport". In: **Eur. Phys. J. C** 78.8 (2018), p. 636. arXiv: 1802.06804 [nucl-th].

[6] P. Huovinen and P. V. Ruuskanen. "Hydrodynamic Models for Heavy Ion Collisions". In: **Ann. Rev. Nucl. Part. Sci.** 56 (2006), pp. 163–206. arXiv: nucl-th/0605008 [nucl-th].

[7] M. L. Miller et al. "Glauber modeling in high energy nuclear collisions". In: **Ann. Rev. Nucl. Part. Sci.** 57 (2007), pp. 205–243. arXiv: nucl-ex/0701025 [nucl-ex].

[8] G.-Y. Qin and X.-N. Wang. "Jet quenching in high-energy heavy-ion collisions". In: **Int. J. Mod. Phys. E** 24.11 (2015), p. 1530014. arXiv: 1511.00790 [hep-ph].

[9] Y. Mehtar-Tani, J. G. Milhano, and K. Tywoniuk. "Jet physics in heavy-ion collisions". In: **Int. J. Mod. Phys. A** 28 (2013), p. 1340013. arXiv: 1302.2579 [hep-ph].

[10] J.-P. Blaizot and Y. Mehtar-Tani. "Jet structure in heavy ion collisions". In: **Int. J. Mod. Phys. E** 24.11 (2015), p. 1530012. arXiv: 1503.05958 [hep-ph].

[11] ATLAS Collaboration. "Observation of Associated Near-Side and Away-Side Long-Range Correlations in $\sqrt{s_{\mathrm{NN}}} = 5.02\,\mathrm{TeV}$ Proton–Lead Collisions with the ATLAS Detector". In: **Phys. Rev. Lett.** 110 (2013), p. 182302. arXiv: `1212.5198 [hep-ex]`.

[12] ATLAS Collaboration. "Measurements of long-range azimuthal anisotropies and associated Fourier coefficients for $pp$ collisions at $\sqrt{s} = 5.02$ and 13 TeV and $p$+Pb collisions at $\sqrt{s_{\mathrm{NN}}} = 5.02$ TeV with the ATLAS detector". In: **Phys. Rev. C** 96 (2017), p. 024908. arXiv: `1609.06213 [hep-ex]`.

[13] ATLAS Collaboration. "Observation of Long-Range Elliptic Azimuthal Anisotropies in $\sqrt{s} = 13$ and 2.76 TeV $pp$ Collisions with the ATLAS Detector". In: **Phys. Rev. Lett.** 116 (2016), p. 172301. arXiv: `1509.04776 [hep-ex]`.

[14] CMS Collaboration. "Observation of long-range, near-side angular correlations in proton–proton collisions at the LHC". In: **JHEP** 09 (2010), p. 091. arXiv: `1009.4122 [hep-ex]`.

[15] CMS Collaboration. "Evidence for Collective Multiparticle Correlations in $p$Pb Collisions". In: **Phys. Rev. Lett.** 115 (2015), p. 012301. arXiv: `1502.05382 [hep-ex]`.

[16] ALICE Collaboration. "Multiplicity dependence of the average transverse momentum in pp, p-Pb, and Pb-Pb collisions at the LHC". In: **Phys. Lett.** B727 (2013), pp. 371–380. arXiv: `1307.1094 [nucl-ex]`.

[17] PHENIX Collaboration. "Creation of quark–gluon plasma droplets with three distinct geometries". In: **Nature Phys.** 15.3 (2019), pp. 214–220. arXiv: `1805.02973 [nucl-ex]`.

[18] ATLAS Collaboration. "Centrality and rapidity dependence of inclusive jet production in $\sqrt{s_{\mathrm{NN}}} = 5.02$ TeV proton-lead collisions with the ATLAS detector". In: **Phys. Lett. B** 748 (2015), p. 392. arXiv: `1412.4092 [hep-ex]`.

[19] CMS Collaboration. "Measurement of inclusive jet production and nuclear modifications in $p$Pb collisions at $\sqrt{s_{\mathrm{NN}}} = 5.02$ TeV". In: **Eur. Phys. J. C** 76 (2016), p. 372. arXiv: `1601.02001 [hep-ex]`.

[20] ALICE Collaboration. "Transverse Momentum Distribution and Nuclear Modification Factor of Charged Particles in $p$+Pb Collisions at $\sqrt{s_{NN}}$=5.02 TeV". In: **Phys. Rev. Lett.** 110 (8 Feb. 2013), p. 082302. arXiv: `1210.4520 [nucl-ex]`.

[21] PHENIX Collaboration. "Centrality-Dependent Modification of Jet-Production Rates in Deuteron-Gold Collisions at $\sqrt{s_{NN}} = 200$ GeV". In: **Phys. Rev. Lett.** 116 (12 Mar. 2016), p. 122301. arXiv: `1509.04657 [nucl-ex]`.

[22] J. Noronha-Hostler et al. "Event-by-Event Hydrodynamics + Jet Energy Loss: A Solution to the $R_{AA} \bigotimes v_2$ Puzzle". In: **Phys. Rev. Lett.** 116 (25 June 2016), p. 252301. DOI: `10.1103/PhysRevLett.116.252301`.

[23] X. Zhang and J. Liao. "Jet Quenching and Its Azimuthal Anisotropy in AA and possibly High Multiplicity pA and dA Collisions". In: (2013). arXiv: `1311.5463 [nucl-th]`.

[24] J. Liao and E. Shuryak. "Angular Dependence of Jet Quenching Indicates Its Strong Enhancement near the QCD Phase Transition". In: **Phys. Rev. Lett.** 102 (20 May 2009), p. 202302. DOI: `10.1103/PhysRevLett.102.202302`.

[25] ATLAS Collaboration. "Centrality, rapidity and transverse momentum dependence of isolated prompt photon production in lead–lead collisions at $\sqrt{s_{\mathrm{NN}}} = 2.76$ TeV measured with the ATLAS detector". In: **Phys. Rev. C** 93 (2016), p. 034914. arXiv: `1506.08552 [hep-ex]`.

[26] ALICE Collaboration. "Direct photon production in Pb-Pb collisions at $\sqrt{s_{\mathrm{NN}}} = 2.76$ TeV". In: **Phys. Lett. B** 754 (2016), pp. 235–248. arXiv: `1509.07324 [nucl-ex]`.

[27] PHENIX Collaboration. "Measurement of Direct Photons in Au+Au Collisions at $\sqrt{s_{NN}} = 200$ GeV". In: **Phys. Rev. Lett.** 109 (2012), p. 152302. arXiv: `1205.5759 [nucl-ex]`.

[28] T. A. Collaboration. "$Z$ boson production in Pb+Pb collisions at $\sqrt{s_{\mathrm{NN}}}$= 5.02 TeV measured by the ATLAS experiment". In: **Phys. Lett.** B802 (2020), p. 135262. arXiv: `1910.13396 [nucl-ex]`.

[29] ATLAS Collaboration. "Measurement of prompt photon production in $\sqrt{s_{\mathrm{NN}}} = 8.16$ TeV $p$+Pb collisions with ATLAS". In: **Phys. Lett.** (2019). arXiv: `1903.02209 [hep-ex]`.

[30] ATLAS Collaboration. "Transverse momentum and process dependent azimuthal anisotropies in $\sqrt{s_{\rm NN}} =$ 8.16 TeV $p$+Pb collisions with the ATLAS detector". In: **Eur. Phys. J. C** 80.1 (2020), p. 73. arXiv: 1910.13978 [nucl-ex].

[31] Wikipedia contributors. **Standard Model — Wikipedia, The Free Encyclopedia**. [Online; accessed January-2020]. 2020. URL: https://https://en.wikipedia.org/wiki/Standard_Model.

[32] M. Thomson. **Modern particle physics**. New York: Cambridge University Press, 2013. ISBN: 9781107034266.

[33] M. Sachs. "On the Origin of Spin in Relativity*". In: **The British Journal for the Philosophy of Science** 40.3 (Sept. 1989), pp. 409–412. ISSN: 0007-0882. DOI: 10.1093/bjps/40.3.409.

[34] J. P. Costella, B. H. J. McKellar, and A. A. Rawlinson. "Classical anti-particles". In: **Am. J. Phys.** 65 (1997), pp. 835–841. arXiv: hep-ph/9704210 [hep-ph].

[35] NobelPrize.org. Nobel Media AB 2020. **The Nobel Prize in Physics 2004**. 2004. URL: https://www.nobelprize.org/prizes/physics/2004/summary/.

[36] G. S. Bali. "QCD forces and heavy quark bound states". In: **Phys. Rept.** 343 (2001), pp. 1–136. arXiv: hep-ph/0001312 [hep-ph].

[37] A. Deur, S. J. Brodsky, and G. F. de Teramond. "The QCD Running Coupling". In: **Prog. Part. Nucl. Phys.** 90 (2016), pp. 1–74. arXiv: 1604.08082 [hep-ph].

[38] L. Susskind. "Harmonic-Oscillator Analogy for the Veneziano Model". In: **Phys. Rev. Lett.** 23 (10 Sept. 1969), pp. 545–547. DOI: 10.1103/PhysRevLett.23.545.

[39] C. D. Roberts. "Hadron Physics and QCD: Just the Basic Facts". In: **J. Phys. Conf. Ser.** 630.1 (2015), p. 012051. arXiv: 1501.06581 [nucl-th].

[40] R. N. Mohapatra. "From old symmetries to new symmetries: Quarks, leptons and B-L". In: **International Journal of Modern Physics A** 29.29 (2014), p. 1430066. DOI: 10.1142/S0217751X1430066X.

[41] F. Halzen and A. D. Martin. **Quarks And Leptons: An Introductory Course In Modern Particle Physics**. 1984. ISBN: 0471887412, 9780471887416.

[42] B. Andersson, G. Gustafson, and B. Soderberg. "A General Model for Jet Fragmentation". In: **Z. Phys.** C20 (1983), p. 317. DOI: `10.1007/BF01407824`.

[43] T. Sjöstrand et al. "An introduction to PYTHIA 8.2". In: **Computer Physics Communications** 191 (2015), pp. 159–177. ISSN: 0010-4655. DOI: `https://doi.org/10.1016/j.cpc.2015.01.024`.

[44] B. Andersson, S. Mohanty, and F. Soderberg. "Recent developments in the Lund model". In: 2002. arXiv: `hep-ph/0212122 [hep-ph]`.

[45] M. Cacciari, G. P. Salam, and G. Soyez. "The anti-kt jet clustering algorithm". In: **Journal of High Energy Physics** 2008.04 (Apr. 2008), pp. 063–063. DOI: `10.1088/1126-6708/2008/04/063`.

[46] N. F. Mott and N. H. D. Bohr. "The scattering of fast electrons by atomic nuclei". In: **Proceedings of the Royal Society of London. Series A, Containing Papers of a Mathematical and Physical Character** 124.794 (1929), pp. 425–442. DOI: `10.1098/rspa.1929.0127`.

[47] J. C. Collins, D. E. Soper, and G. F. Sterman. "Factorization of Hard Processes in QCD". In: **Adv. Ser. Direct. High Energy Phys.** 5 (1989), pp. 1–91. arXiv: `hep-ph/0409313 [hep-ph]`.

[48] C. A. G. Canal and R. Sassot. "Deep inelastic scattering, diffraction and all that". In: **AIP Conference Proceedings** 531.1 (2000), pp. 199–246. DOI: `10.1063/1.1315039`.

[49] V. N. Gribov and L. N. Lipatov. "Deep inelastic e p scattering in perturbation theory". In: **Sov. J. Nucl. Phys.** 15 (1972). [Yad. Fiz.15,781(1972)], pp. 438–450.

[50] L. N. Lipatov. "The parton model and perturbation theory". In: **Sov. J. Nucl. Phys.** 20 (1975). [Yad. Fiz.20,181(1974)], pp. 94–102.

[51] G. Altarelli and G. Parisi. "Asymptotic Freedom in Parton Language". In: **Nucl. Phys.** B126 (1977), pp. 298–318. DOI: `10.1016/0550-3213(77)90384-4`.

[52] Y. L. Dokshitzer. "Calculation of the Structure Functions for Deep Inelastic Scattering and e+ e- Annihilation by Perturbation Theory in Quantum Chromodynamics." In: **Sov. Phys. JETP** 46 (1977). [Zh. Eksp. Teor. Fiz.73,1216(1977)], pp. 641–653.

[53] R. D. Ball et al. “Parton distributions for the LHC Run II”. In: **JHEP** 04 (2015), p. 040. arXiv: 1410.8849 [hep-ph].

[54] E. Iancu, A. Leonidov, and L. McLerran. “The Color glass condensate: An Introduction”. In: **QCD perspectives on hot and dense matter. Proceedings, NATO Advanced Study Institute, Summer School, Cargese, France, August 6-18, 2001**. 2002, pp. 73–145. arXiv: hep-ph/0202270 [hep-ph].

[55] H. Kowalski and D. Teaney. “Impact parameter dipole saturation model”. In: **Phys. Rev. D** 68 (11 Dec. 2003), p. 114005. DOI: 10.1103/PhysRevD.68.114005.

[56] J. Aubert et al. “The ratio of the nucleon structure functions F2N for iron and deuterium”. In: **Physics Letters B** 123.3 (1983), pp. 275–278. ISSN: 0370-2693. DOI: https://doi.org/10.1016/0370-2693(83)90437-9.

[57] J. Ashman et al. “Measurement of the ratios of deep inelastic muon-nucleus cross sections on various nuclei compared to deuterium”. In: **Physics Letters B** 202.4 (1988), pp. 603–610. ISSN: 0370-2693. DOI: https://doi.org/10.1016/0370-2693(88)91872-2.

[58] K. J. Eskola et al. “EPPS16: Nuclear parton distributions with LHC data”. In: **Eur. Phys. J.** C77.3 (2017), p. 163. arXiv: 1612.05741 [hep-ph].

[59] R. Hagedorn. “Thermodynamics of strong interactions”. In: (1971). DOI: 10.5170/CERN-1971-012.

[60] J. C. Collins and M. J. Perry. “Superdense Matter: Neutrons or Asymptotically Free Quarks?” In: **Phys. Rev. Lett.** 34 (21 May 1975), pp. 1353–1356. DOI: 10.1103/PhysRevLett.34.1353.

[61] N. Cabibbo and G. Parisi. “Exponential hadronic spectrum and quark liberation”. In: **Physics Letters B** 59.1 (1975), pp. 67–69. ISSN: 0370-2693. DOI: https://doi.org/10.1016/0370-2693(75)90158-6.

[62] P. de Forcrand. “Simulating QCD at finite density”. In: **PoS** LAT2009 (2009), p. 010. arXiv: 1005.0539 [hep-lat].

[63] W. Busza, K. Rajagopal, and W. van der Schee. “Heavy Ion Collisions: The Big Picture, and the Big Questions”. In: **Ann. Rev. Nucl. Part. Sci.** 68 (2018), pp. 339–376. arXiv: 1802.04801 [hep-ph].

[64] M. G. Alford et al. "Color superconductivity in dense quark matter". In: **Rev. Mod. Phys.** 80 (2008), pp. 1455–1515. arXiv: 0709.4635 [hep-ph].

[65] J. D. Bjorken. "Highly relativistic nucleus-nucleus collisions: The central rapidity region". In: **Phys. Rev. D** 27 (1 Jan. 1983), pp. 140–151. DOI: 10.1103/PhysRevD.27.140.

[66] C. Loizides. "Glauber modeling of high-energy nuclear collisions at the subnucleon level". In: **Phys. Rev.** C94.2 (2016), p. 024914. arXiv: 1603.07375 [nucl-ex].

[67] C. Loizides, J. Nagle, and P. Steinberg. "Improved version of the PHOBOS Glauber Monte Carlo". In: (2014). [SoftwareX1-2,13(2015)]. arXiv: 1408.2549 [nucl-ex].

[68] B. Alver and G. Roland. "Collision-geometry fluctuations and triangular flow in heavy-ion collisions". In: **Phys. Rev. C** 81 (5 May 2010), p. 054905. DOI: 10.1103/PhysRevC.81.054905.

[69] S. Jeon and U. Heinz. "Introduction to Hydrodynamics". In: **Int. J. Mod. Phys.** E24.10 (2015), p. 1530010. arXiv: 1503.03931 [hep-ph].

[70] B. Schenke, S. Jeon, and C. Gale. "Elliptic and triangular flow in event-by-event (3+1)D viscous hydrodynamics". In: **Phys. Rev. Lett.** 106 (2011), p. 042301. arXiv: 1009.3244 [hep-ph].

[71] R. D. Weller and P. Romatschke. "One fluid to rule them all: viscous hydrodynamic description of event-by-event central p+p, p+Pb and Pb+Pb collisions at $\sqrt{s} = 5.02$ TeV". In: **Phys. Lett.** B774 (2017), pp. 351–356. arXiv: 1701.07145 [nucl-th].

[72] B. Schenke, P. Tribedy, and R. Venugopalan. "Fluctuating Glasma initial conditions and flow in heavy ion collisions". In: **Phys. Rev. Lett.** 108 (2012), p. 252301. arXiv: 1202.6646 [nucl-th].

[73] B. Schenke, P. Tribedy, and R. Venugopalan. "Event-by-event gluon multiplicity, energy density, and eccentricities in ultrarelativistic heavy-ion collisions". In: **Phys. Rev.** C86 (2012), p. 034908. arXiv: 1206.6805 [hep-ph].

[74] J. M. Maldacena. "The Large N limit of superconformal field theories and supergravity". In: **Int. J. Theor. Phys.** 38 (1999). [Adv. Theor. Math. Phys.2,231(1998)], pp. 1113–1133. arXiv: hep-th/9711200 [hep-th].

[75] P. Romatschke. "Light-Heavy Ion Collisions: A window into pre-equilibrium QCD dynamics?" In: **Eur. Phys. J.** C75.7 (2015), p. 305. arXiv: 1502.04745 [nucl-th].

[76] F. Cooper and G. Frye. "Single-particle distribution in the hydrodynamic and statistical thermodynamic models of multiparticle production". In: **Phys. Rev. D** 10 (1 July 1974), pp. 186–189. DOI: 10.1103/PhysRevD.10.186.

[77] B. Muller, J. Schukraft, and B. Wyslouch. "First Results from Pb+Pb collisions at the LHC". In: **Ann. Rev. Nucl. Part. Sci.** 62 (2012), pp. 361–386. arXiv: 1202.3233 [hep-ex].

[78] N. Borghini, P. M. Dinh, and J.-Y. Ollitrault. "Flow analysis from multiparticle azimuthal correlations". In: **Phys. Rev.** C64 (2001), p. 054901. arXiv: nucl-th/0105040 [nucl-th].

[79] J.-Y. Ollitrault, A. M. Poskanzer, and S. A. Voloshin. "Effect of flow fluctuations and nonflow on elliptic flow methods". In: **Phys. Rev.** C80 (2009), p. 014904. arXiv: 0904.2315 [nucl-ex].

[80] ATLAS Collaboration. "Measurement of the azimuthal anisotropy of charged particles produced in $\sqrt{s_{\mathrm{NN}}} = 5.02$ TeV Pb+Pb collisions with the ATLAS detector". In: **Eur. Phys. J. C** 78 (2018), p. 997. arXiv: 1808.03951 [hep-ex].

[81] CMS Collaboration. "Measurement of the elliptic anisotropy of charged particles produced in PbPb collisions at $\sqrt{s_{\mathrm{NN}}} = 2.76$ TeV". In: **Phys. Rev. C** 87 (2013), p. 014902. arXiv: 1204.1409 [hep-ex].

[82] C. Gale et al. "Event-by-event anisotropic flow in heavy-ion collisions from combined Yang-Mills and viscous fluid dynamics". In: **Phys. Rev. Lett.** 110.1 (2013), p. 012302. arXiv: 1209.6330 [nucl-th].

[83] ALICE Collaboration. "Higher harmonic flow coefficients of identified hadrons in Pb-Pb collisions at $\sqrt{s_{\mathrm{NN}}}$ = 2.76 TeV". In: **JHEP** 09 (2016), p. 164. arXiv: 1606.06057 [nucl-ex].

[84] W.-T. Deng, X.-N. Wang, and R. Xu. "Hadron production in p+p, p+Pb, and Pb+Pb collisions with the HIJING 2.0 model at energies available at the CERN Large Hadron Collider". In: **Phys. Rev. C** 83 (2011), p. 014915. arXiv: 1008.1841 [hep-ph].

[85] D. Kharzeev, E. Levin, and M. Nardi. "Color glass condensate at the LHC: Hadron multiplicities in pp, pA and AA collisions". In: **Nucl. Phys.** A747 (2005), pp. 609–629. arXiv: hep-ph/0408050 [hep-ph].

[86] ALICE Collaboration. “Centrality determination of Pb-Pb collisions at $\sqrt{s_{NN}}$ = 2.76 TeV with ALICE”. In: **Phys. Rev.** C88.4 (2013), p. 044909. arXiv: `1301.4361 [nucl-ex]`.

[87] P. Aurenche et al. “A New critical study of photon production in hadronic collisions”. In: **Phys. Rev.** D73 (2006), p. 094007. arXiv: `hep-ph/0602133 [hep-ph]`.

[88] C. Collaboration. “Nuclear modification factor of isolated photons in PbPb collisions at $\sqrt{s_{\rm NN}} = 5.02$ TeV”. In: (2019).

[89] ATLAS Collaboration. “Measurement of charged-particle spectra in Pb+Pb collisions at $\sqrt{s_{\rm NN}} = 2.76$ TeV with the ATLAS detector at the LHC”. In: **JHEP** 09 (2015), p. 050. arXiv: `1504.04337 [hep-ex]`.

[90] ATLAS Collaboration. “Measurement of the nuclear modification factor for inclusive jets in Pb+Pb collisions at $\sqrt{s_{\rm NN}} = 5.02$ TeV with the ATLAS detector”. In: **Phys. Lett. B** 790 (2019), p. 108. arXiv: `1805.05635 [hep-ex]`.

[91] CMS Collaboration. “Observation and studies of jet quenching in PbPb collisions at $\sqrt{s_{\rm NN}} = 2.76$ TeV”. In: **Phys. Rev. C** 84 (2011), p. 024906. arXiv: `1102.1957 [hep-ex]`.

[92] ATLAS Collaboration. “Measurement of jet $p_{\rm T}$ correlations in Pb+Pb and *pp* collisions at $\sqrt{s_{\rm NN}} = 2.76$ TeV with the ATLAS detector”. In: **Phys. Lett. B** 774 (2017), p. 379. arXiv: `1706.09363 [hep-ex]`.

[93] M. Gyulassy, I. Vitev, and X. N. Wang. “High p(T) azimuthal asymmetry in noncentral A+A at RHIC”. In: **Phys. Rev. Lett.** 86 (2001), pp. 2537–2540. arXiv: `nucl-th/0012092 [nucl-th]`.

[94] M. Gyulassy et al. “Transverse expansion and high p(T) azimuthal asymmetry at RHIC”. In: **Phys. Lett.** B526 (2002), pp. 301–308. arXiv: `nucl-th/0109063 [nucl-th]`.

[95] J. L. Nagle and W. A. Zajc. “Small System Collectivity in Relativistic Hadronic and Nuclear Collisions”. In: **Ann. Rev. Nucl. Part. Sci.** 68 (2018), pp. 211–235. arXiv: `1801.03477 [nucl-ex]`.

[96] L. Evans and P. Bryant. “LHC Machine”. In: **Journal of Instrumentation** 3.08 (Aug. 2008), S08001–S08001. DOI: `10.1088/1748-0221/3/08/s08001`.

[97] E. Mobs. “The CERN accelerator complex - 2019. Complexe des accélérateurs du CERN - 2019”. In: (July 2019). General Photo. URL: `https://cds.cern.ch/record/2684277`.

[98] CERN. **LHC report: make way for the heavy ions**. [Online; accessed 2020-03-15]. 2018. URL: https://home.cern/news/news/accelerators/lhc-report-make-way-heavy-ions.

[99] ATLAS Collaboration. "The ATLAS Experiment at the CERN Large Hadron Collider". In: **JINST** 3 (2008), S08003. DOI: 10.1088/1748-0221/3/08/S08003.

[100] ATLAS Collaboration. **ATLAS Insertable B-Layer Technical Design Report**. ATLAS-TDR-19. 2010. URL: https://cds.cern.ch/record/1291633. Addendum: ATLAS-TDR-19-ADD-1. 2012. URL: https://cds.cern.ch/record/1451888.

[101] B. Abbott et al. "Production and integration of the ATLAS Insertable B-Layer". In: **JINST** 13 (2018), T05008. arXiv: 1803.00844 [physics.ins-det].

[102] ATLAS Collaboration. "Performance of the ATLAS track reconstruction algorithms in dense environments in LHC Run 2". In: **Eur. Phys. J. C** 77 (2017), p. 673. arXiv: 1704.07983 [hep-ex].

[103] ATLAS Collaboration. **Track Reconstruction Performance of the ATLAS Inner Detector at $\sqrt{s}$ = 13 TeV**. ATL-PHYS-PUB-2015-018. 2015. URL: https://cds.cern.ch/record/2037683.

[104] A. Sidoti. "Minimum Bias Trigger Scintillators in ATLAS Run II". In: **JINST** 9.10 (2014), p. C10020. DOI: 10.1088/1748-0221/9/10/C10020.

[105] X.-N. Wang and M. Gyulassy. "HIJING: A Monte Carlo model for multiple jet production in p p, p A and A A collisions". In: **Phys. Rev.** D44 (1991), pp. 3501–3516. DOI: 10.1103/PhysRevD.44.3501.

[106] S. Agostinelli et al. "GEANT4—a simulation toolkit". In: **Nucl. Instrum. Meth. A** 506 (2003), pp. 250–303. DOI: 10.1016/S0168-9002(03)01368-8.

[107] ATLAS Collaboration. "The ATLAS Simulation Infrastructure". In: **Eur. Phys. J. C** 70 (2010), p. 823. arXiv: 1005.4568 [physics.ins-det].

[108] TOTEM Collaboration. "Luminosity-Independent Measurement of the Proton-Proton Total Cross Section at $\sqrt{s} = 8$ TeV". In: **Phys. Rev. Lett.** 111 (1 July 2013), p. 012001. DOI: 10.1103/PhysRevLett.111.012001.

[109] C. A. Salgado et al. "Proton-nucleus collisions at the LHC: scientific opportunities and requirements". In: **J. Phys. G** 39 (2012), p. 015010. arXiv: 1105.3919 [hep-ph].

[110] F. Arleo et al. "Inclusive prompt photon production in nuclear collisions at RHIC and LHC". In: **JHEP** 04 (2011), p. 055. arXiv: 1103.1471 [hep-ph].

[111] I. Helenius, K. J. Eskola, and H. Paukkunen. "Probing the small-$x$ nuclear gluon distributions with isolated photons at forward rapidities in p+Pb collisions at the LHC". In: **JHEP** 09 (2014), p. 138. arXiv: 1406.1689 [hep-ph].

[112] I. Vitev and B.-W. Zhang. "A Systematic study of direct photon production in heavy ion collisions". In: **Phys. Lett.** B669 (2008), pp. 337–344. arXiv: 0804.3805 [hep-ph].

[113] Z.-B. Kang, I. Vitev, and H. Xing. "Effects of cold nuclear matter energy loss on inclusive jet production in p+A collisions at energies available at the BNL Relativistic Heavy Ion Collider and the CERN Large Hadron Collider". In: **Phys. Rev.** C92 (2015), p. 054911. arXiv: 1507.05987 [hep-ph].

[114] ATLAS Collaboration. "Measurements of the Nuclear Modification Factor for Jets in Pb+Pb Collisions at $\sqrt{s_{\mathrm{NN}}} = 2.76\,\mathrm{TeV}$ with the ATLAS Detector". In: **Phys. Rev. Lett.** 114 (2015), p. 072302. arXiv: 1411.2357 [hep-ex].

[115] ATLAS Collaboration. "Measurement of the inclusive isolated prompt photon cross section in *pp* collisions at $\sqrt{s} = 7\,\mathrm{TeV}$ with the ATLAS detector". In: **Phys. Rev. D** 83 (2011), p. 052005. arXiv: 1012.4389 [hep-ex].

[116] ATLAS Collaboration. "Measurement of the inclusive isolated prompt photon cross section in *pp* collisions at $\sqrt{s} = 8\,\mathrm{TeV}$ with the ATLAS detector". In: **JHEP** 08 (2016), p. 005. arXiv: 1605.03495 [hep-ex].

[117] ATLAS Collaboration. "Measurement of the cross section for inclusive isolated-photon production in *pp* collisions at $\sqrt{s} = 13\,\mathrm{TeV}$ using the ATLAS detector". In: **Phys. Lett. B** 770 (2017), p. 473. arXiv: 1701.06882 [hep-ex].

[118] CMS Collaboration. "Measurement of the Differential Cross Section for Isolated Prompt Photon Production in $pp$ Collisions at 7 TeV". In: **Phys. Rev. D** 84 (2011), p. 052011. arXiv: 1108.2044 [hep-ex].

[119] CMS Collaboration. "Measurement of the Production Cross Section for Pairs of Isolated Photons in $pp$ collisions at $\sqrt{s} = 7$ TeV". In: **JHEP** 01 (2012), p. 133. arXiv: 1110.6461 [hep-ex].

[120] STAR Collaboration. "Inclusive $\pi^0$, $\eta$, and direct photon production at high transverse momentum in $p + p$ and $d$+Au collisions at $\sqrt{s_{NN}} = 200$ GeV". In: **Phys. Rev. C** 81 (2010), p. 064904. arXiv: 0912.3838 [hep-ex].

[121] PHENIX Collaboration. "Direct photon production in $d$+Au collisions at $\sqrt{s_{NN}} = 200$ GeV". In: **Phys. Rev.** C87 (2013), p. 054907. arXiv: 1208.1234 [nucl-ex].

[122] ATLAS Collaboration. "$Z$ boson production in $p + Pb$ collisions at $\sqrt{s_{\mathrm{NN}}} = 5.02$ TeV measured with the ATLAS detector". In: **Phys. Rev. C** 92 (2015), p. 044915. arXiv: 1507.06232 [hep-ex].

[123] CMS Collaboration. "Study of $Z$ boson production in $p$Pb collisions at $\sqrt{s_{\mathrm{NN}}} = 5.02$ TeV". In: **Phys. Lett. B** 759 (2016), p. 36. arXiv: 1512.06461 [hep-ex].

[124] CMS Collaboration. "Study of $W$ boson production in $p$Pb collisions at $\sqrt{s_{\mathrm{NN}}}$ 5.02 TeV". In: **Phys. Lett. B** 750 (2015), p. 565. arXiv: 1503.05825 [hep-ex].

[125] K. Kovarik et al. "nCTEQ15 - Global analysis of nuclear parton distributions with uncertainties in the CTEQ framework". In: **Phys. Rev.** D93.8 (2016), p. 085037. arXiv: 1509.00792 [hep-ph].

[126] Y.-T. Chien et al. "Jet Quenching from QCD Evolution". In: **Phys. Rev.** D93 (2016), p. 074030. arXiv: 1509.02936 [hep-ph].

[127] T. Sjostrand, S. Mrenna, and P. Z. Skands. "A Brief Introduction to PYTHIA 8.1". In: **Comput. Phys. Commun.** 178 (2008), pp. 852–867. arXiv: 0710.3820 [hep-ph].

[128] R. D. Ball et al. "Parton distributions with LHC data". In: **Nucl. Phys. B** 867 (2013), pp. 244–289. arXiv: 1207.1303 [hep-ph].

[129] ATLAS Collaboration. **ATLAS Pythia 8 tunes to 7 TeV data**. ATL-PHYS-PUB-2014-021. 2014. URL: https://cds.cern.ch/record/1966419.

[130] T. Gleisberg et al. "Event generation with SHERPA 1.1". In: **JHEP** 02 (2009), p. 007. arXiv: `0811.4622 [hep-ph]`.

[131] M. Cacciari, G. P. Salam, and S. Sapeta. "On the characterisation of the underlying event". In: **JHEP** 04 (2010), p. 065. arXiv: `0912.4926 [hep-ph]`.

[132] ATLAS Collaboration. "Measurement of the photon identification efficiencies with the ATLAS detector using LHC Run-1 data". In: **Eur. Phys. J. C** 76 (2016), p. 666. arXiv: `1606.01813 [hep-ex]`.

[133] ATLAS Collaboration. "Measurement of the photon identification efficiencies with the ATLAS detector using LHC Run 2 data collected in 2015 and 2016". In: **Eur. Phys. J. C** 79 (2019), p. 205. arXiv: `1810.05087 [hep-ex]`.

[134] ATLAS Collaboration. "Electron and photon energy calibration with the ATLAS detector using 2015-2016 LHC proton-proton collision data". In: (2018). arXiv: `1812.03848 [hep-ex]`.

[135] ATLAS Collaboration. "Topological cell clustering in the ATLAS calorimeters and its performance in LHC Run 1". In: **Eur. Phys. J. C** 77 (2017), p. 490. arXiv: `1603.02934 [hep-ex]`.

[136] ATLAS Collaboration. "Measurement of the inclusive isolated prompt photon cross-section in $pp$ collisions at $\sqrt{s} = 7\,\text{TeV}$ using $35\,\text{pb}^{-1}$ of ATLAS data". In: **Phys. Lett. B** 706 (2011), p. 150. arXiv: `1108.0253 [hep-ex]`.

[137] S. Manzoni, L. Carminatia, and G. Marchiorib. **Electron-to-photon fake rate measurements: Supporting documentation for the Photon identification in 2015+2016 ATLAS data**. ATL-COM-PHYS-2017-1277. 2018. URL: `https://cds.cern.ch/record/2280801`.

[138] A. Cueto et al. **First measurement of the cross section for inclusive isolated-photon production in pp collisions at $\sqrt{s} = 13$TeV using the ATLAS detector**. ATL-COM-PHYS-2016-449. 2016. URL: `https://cds.cern.ch/record/2151096`.

[139] ATLAS Collaboration. "Measurement of the inclusive isolated prompt photons cross section in $pp$ collisions at $\sqrt{s} = 7\,\text{TeV}$ with the ATLAS detector using $4.6\,\text{fb}^{-1}$". In: **Phys. Rev. D** 89 (2014), p. 052004. arXiv: `1311.1440 [hep-ex]`.

[140] ATLAS Collaboration. "High-$E_\mathrm{T}$ isolated-photon plus jets production in $pp$ collisions at $\sqrt{s} = 8\,\mathrm{TeV}$ with the ATLAS detector". In: **Nucl. Phys. B** 918 (2017), p. 257. arXiv: `1611.06586 [hep-ex]`.

[141] ATLAS Collaboration. "Luminosity determination in $pp$ collisions at $\sqrt{s} = 8\,\mathrm{TeV}$ using the ATLAS detector at the LHC". In: **Eur. Phys. J. C** 76 (2016), p. 653. arXiv: `1608.03953 [hep-ex]`.

[142] G. Avoni et al. "The new LUCID-2 detector for luminosity measurement and monitoring in ATLAS". In: **JINST** 13.07 (2018), P07017. DOI: `10.1088/1748-0221/13/07/P07017`.

[143] S. Dulat et al. "New parton distribution functions from a global analysis of quantum chromodynamics". In: **Phys. Rev.** D93.3 (2016), p. 033006. arXiv: `1506.07443 [hep-ph]`.

[144] L. Bourhis, M. Fontannaz, and J. P. Guillet. "Quarks and gluon fragmentation functions into photons". In: **Eur. Phys. J. C** 2 (1998), pp. 529–537. arXiv: `hep-ph/9704447 [hep-ph]`.

[145] X. Chen et al. "Isolated photon and photon+jet production at NNLO QCD accuracy". In: **Submitted to: J. High Energy Phys.** (2019). arXiv: `1904.01044 [hep-ph]`.

[146] H. Paukkunen and C. A. Salgado. "Constraints for the nuclear parton distributions from Z and W production at the LHC". In: **JHEP** 03 (2011), p. 071. arXiv: `1010.5392 [hep-ph]`.

[147] ATLAS Collaboration. "Measurement of long-range pseudorapidity correlations and azimuthal harmonics in $\sqrt{s_\mathrm{NN}} = 5.02\,\mathrm{TeV}$ proton–lead collisions with the ATLAS detector". In: **Phys. Rev. C** 90 (2014), p. 044906. arXiv: `1409.1792 [hep-ex]`.

[148] ATLAS Collaboration. "Study of the material of the ATLAS inner detector for Run 2 of the LHC". In: **JINST** 12 (2017), P12009. arXiv: `1707.02826 [hep-ex]`.

[149] ATLAS Collaboration. "Measurement of the jet radius and transverse momentum dependence of inclusive jet suppression in lead–lead collisions at $\sqrt{s_\mathrm{NN}} = 2.76\,\mathrm{TeV}$ with the ATLAS detector". In: **Phys. Lett. B** 719 (2013), p. 220. arXiv: `1208.1967 [hep-ex]`.

[150] ATLAS Collaboration. "Measurement of jet fragmentation in Pb+Pb and $pp$ collisions at $\sqrt{s_\mathrm{NN}} = 5.02\,\mathrm{TeV}$ with the ATLAS detector". In: **Phys. Rev. C** 98 (2018), p. 024908. arXiv: `1805.05424 [hep-ex]`.

[151] M. Cacciari, G. P. Salam, and G. Soyez. "The anti-kt jet clustering algorithm". In: **JHEP** 04 (2008), p. 063. arXiv: `0802.1189 [hep-ph]`.

[152] M. Cacciari, G. P. Salam, and G. Soyez. "FastJet user manual". In: **Eur. Phys. J. C** 72.3 (Mar. 2012), p. 1896. DOI: `10.1140/epjc/s10052-012-1896-2`.

[153] ATLAS Collaboration. "Jet energy scale measurements and their systematic uncertainties in proton–proton collisions at $\sqrt{s} = 13$ TeV with the ATLAS detector". In: **Phys. Rev. D** 96 (2017), p. 072002. arXiv: `1703.09665 [hep-ex]`.

[154] ATLAS Collaboration. "Measurement of photon-jet transverse momentum correlations in 5.02 TeV Pb+Pb and $pp$ collisions with ATLAS". In: **Phys. Lett. B** 789 (2019), p. 167. arXiv: `1809.07280 [hep-ex]`.

[155] ATLAS Collaboration. "Measurement of the azimuthal anisotropy for charged particle production in $\sqrt{s_{\mathrm{NN}}} = 2.76$ TeV lead–lead collisions with the ATLAS detector". In: **Phys. Rev. C** 86 (2012), p. 014907. arXiv: `1203.3087 [hep-ex]`.

[156] CMS Collaboration. "Measurement of long-range near-side two-particle angular correlations in $pp$ collisions at $\sqrt{s} = 13$ TeV". In: **Phys. Rev. Lett.** 116 (2016), p. 172302. arXiv: `1510.03068 [hep-ex]`.

[157] CMS Collaboration. "Multiplicity and transverse momentum dependence of two- and four-particle correlations in $p$Pb and PbPb collisions". In: **Phys. Lett. B** 724 (2013), p. 213. arXiv: `1305.0609 [hep-ex]`.

[158] ATLAS Collaboration. **Measurement of the long-range pseudorapidity correlations between muons and charged particles in $\sqrt{s_{\mathrm{NN}}} = 8.16$ TeV proton–lead collisions with the ATLAS detector**. ATLAS-CONF-2017-006. 2017. URL: `https://cds.cern.ch/record/2244808`.

[159] ATLAS Collaboration. "Underlying event characteristics and their dependence on jet size of charged-particle jet events in $pp$ collisions at $\sqrt{s} = 7$ TeV with the ATLAS detector". In: **Phys. Rev. D** 86 (2012), p. 072004. arXiv: `1208.0563 [hep-ex]`.

[160] ATLAS Collaboration. "Measurement of the underlying event in jet events from 7 TeV proton–proton collisions with the ATLAS detector". In: **Eur. Phys. J. C** 74 (2014), p. 2965. arXiv: 1406.0392 [hep-ex].

[161] ATLAS Collaboration. "Measurement of distributions sensitive to the underlying event in inclusive $Z$-boson production in $pp$ collisions at $\sqrt{s} = 7$ TeV with the ATLAS detector". In: **Eur. Phys. J. C** 74 (2014), p. 3195. arXiv: 1409.3433 [hep-ex].

[162] ATLAS Collaboration. "Measurement of charged-particle distributions sensitive to the underlying event in $\sqrt{s} = 13$ TeV proton–proton collisions with the ATLAS detector at the LHC". In: **JHEP** 03 (2017), p. 157. arXiv: 1701.05390 [hep-ex].

[163] D. Kikola et al. "Nonflow 'factorization' and a novel method to disentangle anisotropic flow and non-flow". In: **Phys. Rev.** C86 (2012), p. 014901. arXiv: 1110.4809 [nucl-ex].

[164] ATLAS Collaboration. "Transverse momentum, rapidity, and centrality dependence of inclusive charged-particle production in $\sqrt{s_{\text{NN}}} = 5.02$ TeV $p$+Pb collisions measured by the ATLAS experiment". In: **Phys. Lett. B** 763 (2016), p. 313. arXiv: 1605.06436 [hep-ex].

[165] ALICE Collaboration. "Centrality dependence of particle production in p-Pb collisions at $\sqrt{s_{\text{NN}}}$= 5.02 TeV". In: **Phys. Rev. C** 91.6 (2015), p. 064905. arXiv: 1412.6828 [nucl-ex].

[166] H. Mäntysaari et al. "Imprints of fluctuating proton shapes on flow in proton-lead collisions at the LHC". In: **Phys. Lett. B** 772 (2017), pp. 681–686. arXiv: 1705.03177 [nucl-th].

[167] C. Bierlich et al. "The Angantyr model for Heavy-Ion Collisions in PYTHIA8". In: **JHEP** 10 (2018), p. 134. arXiv: 1806.10820 [hep-ph].

[168] C. Bierlich. "Microscopic collectivity: The ridge and strangeness enhancement from string–string interactions". In: **Nucl. Phys. A** 982 (2019). The 27th International Conference on Ultrarelativistic Nucleus-Nucleus Collisions: Quark Matter 2018, pp. 499–502. DOI: https://doi.org/10.1016/j.nuclphysa.2018.07.015.

[169] C. Loizides, J. Kamin, and D. d'Enterria. "Improved Monte Carlo Glauber predictions at present and future nuclear colliders". In: **Phys. Rev. C** 97.5 (2018), p. 054910. arXiv: 1710.07098 [nucl-ex].

[170] ATLAS Collaboration. "Measurement of jet fragmentation in 5.02 TeV proton–lead and proton–proton collisions with the ATLAS detector". In: **Nucl. Phys.** A978 (2018), p. 65. arXiv: 1706.02859 [hep-ex].

[171] C. Zhang et al. "Elliptic Flow of Heavy Quarkonia in $pA$ Collisions". In: **Phys. Rev. Lett.** 122.17 (2019), p. 172302. arXiv: 1901.10320 [hep-ph].